\documentclass[a4paper,12pt]{book}

\usepackage[utf8x]{inputenc}

\usepackage[left=2cm,right=2cm,top=2cm,bottom=3cm]{geometry}

\usepackage[linktoc=all,hidelinks]{hyperref}
\usepackage{cite}
\usepackage{amsmath}
\usepackage{amsfonts}
\usepackage{amssymb}
\usepackage{graphicx}
\usepackage{afterpage}
\usepackage[nottoc,numbib]{tocbibind}
\usepackage{enumitem}

\usepackage[toc,page]{appendix}
\usepackage{tikz}
\usetikzlibrary{cd}
\usepackage[compat=1.1.0]{tikz-feynman} \tikzfeynmanset{warn luatex=false}
\usepackage{mathtools}
\usepackage{amsthm,amsmath,latexsym,amssymb,amsfonts,amscd}
\usepackage{graphics,lscape,fancyhdr,array,stmaryrd,euscript,wrapfig}

\usepackage{empheq,slashed}
\usepackage{verbatim,slashed}
\numberwithin{equation}{section}
\usepackage{hyperref,setspace}
\usepackage{dsfont}
\usepackage{mathrsfs}

\usepackage[numbers,sort&compress]{natbib}
\usepackage[nottoc]{tocbibind}

\makeatletter
\renewcommand\paragraph{\@startsection{paragraph}{4}{0pt}%
  {0.2ex plus 0.1ex minus 0.1ex} 
  {-1em} 
  {\normalfont\normalsize\bfseries}}
\makeatother

\usepackage{float} 

\newcommand{\pl}{\partial}

\newcommand{\be}{\begin{equation}}
\newcommand{\ee}{\end{equation}}

\newcommand{\fdu}[2]{{}_{#1}{}^{#2}\,}

\newcommand{\besubeqs}{\begin{subequations}}
\newcommand{\esubeqs}{\end{subequations}}

\newcommand{\zb}{{\bar{z}}}

\newcommand{\pb}{{\bar{p}}}
\newcommand{\qb}{{\bar{q}}}

\newcommand{\jb}{{\bar{j}}}

\newcommand{\pfrac}[1]{{\frac{\pl}{\pl #1}}}

\newcommand{\deltas}[1]{{\delta\left(#1\right)}}
\newcommand{\PP}{{\mathbb{P}}}

\newcommand{\PPb}{{\bar{\mathbb{P}}}}
\newcommand{\NPP}{\mathbb{N}_{\PP}}
\newcommand{\NPPb}{\mathbb{\bar N}_{\PPb}}

\newcommand{\fA}{\mathfrak{f}}   
\newcommand{\fB}{\mathbf{f}}     

\newcommand{\mA}{\mathcal{A}}
\newcommand{\mtA}{\tilde{\mathcal{A}}}

\newcommand{\definition}{\mathrel{\mathop:}=}

\begin{document}
\raggedbottom

\thispagestyle{empty}

\includegraphics[width=30mm]{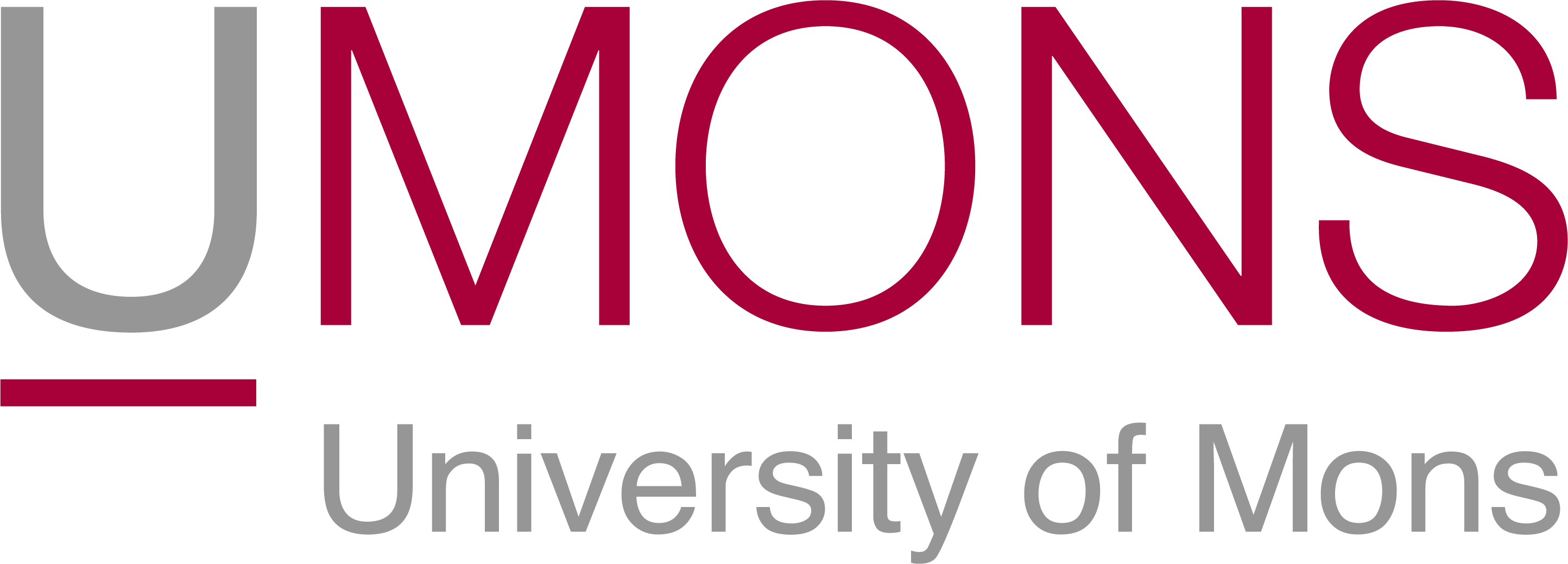}
\hfill
\includegraphics[width=30mm]{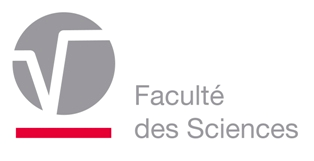}

\begin{center}

\vspace*{5mm}
\hrule
\vspace{10mm}
{\Large \sc Light-Front approach\\
[5mm]
to 4d massless Higher-Spin interactions}
\vspace{10mm}
\hrule

\vspace{15mm}
{\large {Mattia Serrani}}

\vspace{5mm}
{Service de Physique de l’Univers, Champs et Gravitation\\
Facult\'e des Sciences\\
Université de Mons}

\vspace{5mm}
{27 August 2025}

\vspace{20mm}
{\sc Th\`ese pr\'esent\'ee en vue de l’obtention\\
du grade de Doctorat en sciences}

\vspace{20mm}
{\sc Promoteur de th\`ese}
\begin{align}
    \text{Evgeny D. Skvortsov} &\quad \text{Universit\'e de Mons, Belgique} \nonumber
\end{align}

\vspace{5mm}
\centering{\sc Membres du jury}
\begin{align}
\text{Nicolas Boulanger} &\quad \text{Universit\'e de Mons, Belgique} \nonumber\\
\text{Andrea Campoleoni} &\quad \text{Universit\'e de Mons, Belgique} \nonumber\\
\text{Dario Francia} &\quad \text{Universit\`a Roma Tre, Italie} \nonumber\\
\text{Dmitry S. Ponomarev} &\quad \text{Lomonosov Moscow State University, Russie} \nonumber
\end{align}

\end{center}

\clearpage
\thispagestyle{empty}
\vspace*{6cm}
\begin{center}
    \emph{A Maria e Franco,\\
    e alla mia famiglia.\\
    Vi voglio bene.}
\end{center}
\clearpage

\clearpage
\begin{flushright}
    \textit{``The greatest quality of humans lies in their ability to recognize patterns and solve problems. These problems are sometimes artificial, as in much of mathematics, while others arise from our effort to understand nature and, by extension, the universe itself. I believe this is the natural direction we should follow, rather than using our intelligence to wage war against one another, treating peace merely as a pause while preparing for the next conflict. We share the same world and we should take care of it, together.''}\\
    --- A path to harmony and well-being
\end{flushright}
\vspace{5mm}
\begin{flushright}
    \textit{``Il fine di una rivolta, che sia il calcio in bocca ad un maschio violento o i tre giorni di un rave party, é la sospensione del tempo e l'apertura di squarci nel quotidiano dentro la quale sperimentare gioiosamente avventure di libertá reale e collettiva... la bellezza di riappropiarsi di una parte di ciò che ti viene sottratto ogni giorno... un atto d'amore spontaneo verso sé stessi e verso gli altri.''}\\
    --- Un articolo
\end{flushright}
\vspace{80mm}
\begin{flushright}
    \textit{``The mass of men lead lives of quiet desperation.''}\\
    --- Henry David Thoreau, Walden.
\end{flushright}
\vspace{5mm}
\begin{flushright}
    \textit{``Aveva quella bellezza di cui solo i vinti sono capaci. E la limpidezza delle cose deboli. E la solitudine, perfetta, di ciò che si è perduto.''}\\
    --- Alessandro Baricco, Oceano Mare.
\end{flushright}
\vspace{5mm}
\begin{flushright}
    \textit{``Je ne puis vivre personnellement sans mon art. Mais je n'ai jamais placé cet art au-dessus de tout.''}\\
    --- Albert Camus, Discours de Suède.
\end{flushright}

\clearpage

\chapter*{Acknowledgements}
\addcontentsline{toc}{chapter}{Acknowledgements}
\pagenumbering{roman}

This thesis presents part of the research I carried out during my doctoral studies at the University of Mons under the supervision of Zhenya.

As tradition requires, I begin by thanking my supervisor, Zhenya. I am grateful to you --- and to your ERC grant --- for giving me the opportunity to pursue a PhD in theoretical physics, one of my greatest dreams. As is perhaps inevitable in any PhD journey, there were ups and downs. We probably both made mistakes along the way, but the last two years --- and especially this final one --- gave me the confidence to defend this thesis. Thank you again; I admire your approach to physics and your interest in mathematics. I only regret not learning enough from you during my early years. I am especially grateful for your constant support throughout this final year of my PhD.

Thanks to the friendly atmosphere and the coffee room (yet for some, a tea room) of the Service de Physique de l’Univers, Champs et Gravitation. I am grateful to Nicolas, the head of the unit, and I also warmly thank the other two professors, Andrea and Maxime.

Thanks to Nicolas and Andrea, members of my jury, for their helpful comments.

I would also like to thank the other two members of the jury, Mitya and Dario, for their help in improving this thesis and for their valuable comments and discussions.

Now the hardest part; thanking all the people who shared moments with me along the way while hoping not to forget anyone. As a proper theoretical physicist, I will attempt to classify them into inequivalent classes.

When I arrived in Mons in August 2021, there were five of us starting our PhDs. Along the way, we lost two (Kamil and Akshay), gained one (Shailesh), and ended up with four. Thanks to Josh, with whom I shared a flat for three whole years --- it wasn’t easy, as we both know, but now I certainly have stories to tell. 
Thanks to Shailesh, we may never know whether Zhenya is really in the unit or not. I also thank his wife, Swati, for her wonderful hospitality and all the delicious food she prepared for us.
Thanks to Richard, whose rare appearances in the unit left some students wondering who he really was.

Thanks also to Ismael, whose love for mathematics inspired many of us, and to Arsenii, Misha and Vanya --- the Russians. Together with Josh, thank you for all the beers, and the many great moments we shared together.

I thank all other members of the unit, past and present, listed here in reverse alphabetical order: Yannick, William, Victor, Vasileios, Tung, Thomas, Sylvain, Simon, Robin, Noémie, Nicolas Mdx, Matthieu, Lea, Joachim, Ivano, Guillaume, Felipe, Clara, Chrysoula, Arnauld, and Arghya. 

I thank everyone from the Solvay School 2022 --- in Brussels, Paris, and Geneva. It was an amazing experience for me, and you were wonderful company. 

I also thank my friends from Vesuvius: Javi, Lorenzo, Andrea, Tiago, Ismail and Hugo.

Along the way, I was fortunate to form some special connections with wonderful people. I will never forget my Greek, Italian, and Bulgarian friends: $K\!\alpha\tau\epsilon\rho\iota\nu\alpha$, $A\sigma\pi\alpha$, Michela, Isabella, and $A\scalebox{0.8}{HH}a$. You made the end of my second year in Mons different and memorable.

Thanks to Jacobo and Chloé. Gracias and Merci, I will never forget our trio. Thanks also to Ludo and his amazing coffee (Joliecœur).

It is now the turn of my Italian friends from Italy, some of whom even came to visit me here in Mons. In alphabetical order, grazie Ari, Diego, Doda, Fede, Gabbo, Gabry, Giovi, Leo, Marcy, Matte, Meli, Mile, Rachi, Samu, Sara, Staffo, Ste and Vasia. Siete i migliori amici di sempre. Quando torno a casa, a Marina di Montemarciano, mi sento sempre così bene.

Thanks also to my friends from my loved Pisa: grazie Vasco, Dani, Deppo, Ventu, Ricca, Ale, Leo, and Enrico.

I will conclude by thanking my family. Grazie, mamma, paccio e fiola. Vi voglio un mondo di bene. Mi sembra ovvio dirlo, ma senza di voi nulla di tutto questo avrebbe lo stesso senso. Grazie nonna Maria, garzie nonno Franco e grazie nonno Fernando. Vi penso. Grazie anche a nonna Flavia.

Et Margaux, je suis tellement heureux qu'on se soit trouvés. Miam :)

\vfill

\paragraph{Funding.} This project has received funding from the European Research Council (ERC) under the European Union’s Horizon 2020 research and innovation programme (grant agreement No 101002551).

\chapter*{Abstract}
\addcontentsline{toc}{chapter}{Abstract}

By higher-spin gravities, we refer to any extension of gravity that includes at least one massless field of spin greater than two, interacting non-trivially with the spin-2 graviton. These theories are expected to provide a pathway toward a deeper understanding --- and possibly a solution --- to quantum gravity. A central feature of such models is the presence of an enhanced gauge symmetry, known as higher-spin symmetry, which is often (though not always) infinite-dimensional. This symmetry plays a crucial role in constraining the dynamics and may provide a mechanism to address the ultraviolet (UV) non-renormalisability of gravity by restricting the set of allowed counterterms --- potentially reducing them to a finite set or eliminating them all --- thereby rendering the quantum theory effectively classical.

Among the various ways to study interactions in higher-spin gravity --- such as the Noether procedure, the BV-BRST formalism, Unfolding, Twistors, and a few others --- this thesis focuses on the light-front approach. In particular, most of the results are derived from the analysis of the holomorphic and non-holomorphic quartic constraints, which arise from requiring the closure of the non-linearly realised Poincaré algebra --- thereby ensuring the Poincaré invariance of the theory.

Recently, it was discovered --- quite surprisingly --- that a particular type of theory, known as chiral higher-spin theory, can be fully consistent while involving only cubic interaction vertices, provided that all such vertices are included (except simple truncations, such as restricting to only even spins). This remarkable feature is closely tied to deep concepts such as integrability, twistor theory, deformation quantisation, and to the chirality of the theory.

To date, this represents the only example of a perturbatively local higher-spin extension of gravity that propagates the correct degrees of freedom expected for a gravitational theory --- apart from including only one chirality of the graviton couplings. All other known examples are either not perturbatively local (e.g. the collective-dipole approach or not making specific well-defined predictions for interactions as in Vasiliev’s equations) or exhibit non-unitarity already at the free level (as in conformal higher-spin gravity) or do not have propagating degrees of freedom as in $3d$-examples of matter-free higher-spin theories.

This theory, already compelling in its own right, may serve as a promising starting point for the search for a complete, unitary theory of higher-spin gravity --- potentially achievable by extending it while preserving a controlled or mild form of non-locality. Importantly, any higher-spin gravity in $4d$ has to contain the (anti)-chiral theory as a subsector.

This thesis explores three main directions. First, it examines the possibility of truncating the chiral higher-spin theory, considering whether it can be isolated as a special case within a broader class of consistent chiral higher-spin theories, and initiates a systematic program for classifying all possible chiral higher-spin gravities. Second, it aims to shed light on the connection between the quartic holomorphic constraint and a constraint that arises in the context of $2d$ Celestial CFT --- specifically, OPE associativity --- which is conjectured to be dual to asymptotically flat $4d$ bulk gravity. Third, it explores potential no-go and yes-go results by attempting to solve the non-holomorphic quartic constraint and by searching for possible four-point interaction vertices that ensure the consistency of the theory up to fourth order, for both lower- and higher-spin fields.

\chapter*{Outline and main results of the Thesis}
The present work is a natural continuation of the study initiated by Bengtsson, Bengtsson, and Brink in 1983 \cite{Bengtsson:1983pd,Bengtsson:1983pg,Bengtsson:1986kh}, and later developed by Metsaev in 1991 \cite{Metsaev:1991mt,Metsaev:1991nb}. More recently, in 2017, Ponomarev and Skvortsov revisited and extended the formalism \cite{Ponomarev:2016lrm}. This thesis builds upon their approach and aims to explore new results within that framework.

The thesis is based on the following three research papers:
\begin{enumerate}[itemsep=0.6em]
    \item \cite{Serrani:2025owx} \textbf{M.~Serrani, ``{On classification of (self-dual) higher-spin gravities in flat space},'' {\em JHEP} {\bf 08} (2025) 032, \href{http://arXiv.org/abs/2505.12839}{{\tt arXiv:2505.12839 [hep-th]}}.}

    \item \cite{Serrani:2025oaw} \textbf{M.~Serrani, ``{Associativity of celestial OPE, higher spins and self-duality},'' {\em JHEP} {\bf 04} (2026) 047, \href{http://arXiv.org/abs/2508.16804}{{\tt arXiv:2508.16804 [hep-th]}}.}

    \item \cite{Serrani:2026dbs} \textbf{M.~Serrani, ``{Massless spinning fields on the Light-Front: quartic vertices and amplitudes},'' \href{http://arXiv.org/abs/2602.12826}{{\tt arXiv:2602.12826 [hep-th]}}.}
    
\end{enumerate}
The first two papers have been published in JHEP, while the third is currently available as an arXiv preprint and has been submitted to JHEP. The thesis begins with an introductory chapter, followed by one chapter corresponding to each of the three papers.

\vspace{0.5 cm}

\paragraph{Chapter 1: Introduction.} The first part of the introduction is intended to guide the reader smoothly into the concepts and ideas of higher-spin gravity. The second part provides a brief historical overview of the light-front approach to higher-spin interactions, along with essential background material to facilitate the understanding of the core of the thesis.

We emphasise that the particular focus on certain topics within higher-spin gravity reflects the author's personal choice and subjective experience, and is by no means intended to provide a comprehensive overview of the full literature on the subject. Rather, it should be viewed as a personal account. The field is rapidly evolving, and for a more extensive and up-to-date review of higher-spin gravity, covering both massless and massive fields, we refer the reader to \cite{Bekaert:2022poo}.

\vspace{0.5 cm}

\paragraph{Chapter 2: Self-dual higher-spin theories in flat
space.} This chapter is devoted to the study of the light-cone quartic holomorphic constraint in $4d$ higher-spin theory. The main results are:
\begin{itemize}
\item A complete solution to the light-cone holomorphic constraint for both even- and odd-derivative vertices, in the presence or absence of a gauge group $G$. 

\item We provide a systematic framework for classifying all consistent lower- and higher-derivative chiral higher-spin theories.

\item A full classification of chiral higher-spin theories with one- and two-derivative vertices, with and without a gauge group $G$.
\end{itemize}

\vspace{0.5 cm}

\paragraph{Chapter 3: Associativity of celestial OPE, higher-spins and self-duality.} This chapter is dedicated to exploring the connection between the light-cone holomorphic constraint in $4d$ higher-spin theory and the OPE associativity constraint in $2d$ Celestial CFT. The main results are:

\begin{itemize}
\item Finding an explicit solution to the OPE associativity constraint in the most general case, including higher-spin fields and higher-derivative vertices.

\item  We highlight and clarify the connection between several ideas: (a)  the operator product expansion (OPE) associativity in celestial conformal field theory (CCFT); (b) the vanishing of tree-level amplitudes; (c) the Jacobi identity for the kinematic algebra; (d) the light-cone holomorphic constraints. Naturally, (b), (c) or (d) are closely related to self-duality.
\end{itemize}

\vspace{0.5 cm}

\paragraph{Chapter 4: Massless spinning fields on the Light-Front (quartic vertices and amplitudes).} This chapter is devoted to the study of the full light-cone quartic constraint in $4d$ higher-spin theory and on the construction of all local four-point higher-spin amplitudes in the spinor-helicity formalism. The main results are:

\begin{itemize}

\item We develop a systematic method to determine the Hamiltonian four-point density, either by explicitly constructing or by systematically checking the existence (via Frobenius integrability) of local quartic vertices solving the quartic light-cone constraint.

\item By computing local quartic vertices with increasing numbers of derivatives that solve the quartic light-cone constraint, we conjecture the complete set of quartic vertices satisfying the light-cone consistency conditions. 

\item We reproduce known no-go and yes-go results for lower-spin interactions.

\item We find all allowed unitary local higher-spin theories and identify new families of local quasi-chiral higher-spin theories.

\item We determine all local higher-spin four-point amplitudes satisfying factorisation in the spinor-helicity formalism.

\end{itemize}

\setcounter{tocdepth}{2}
\tableofcontents
\addcontentsline{toc}{chapter}{Contents}
\pagenumbering{arabic}

\chapter{Introduction}

The greatest quality of humans lies in their ability to recognize patterns and solve problems. These problems are sometimes artificial\footnote{Here, by ``artificial'' I mean ``made by humans''. I do not intend to delve into the philosophical debate over whether mathematics is invented or discovered --- a question which, in any case, I believe is not well defined.}, as in much of mathematics, while others arise from our effort to understand nature and, by extension, the universe itself.

Theoretical physics perfectly embodies this concept: it begins by studying structures provided by nature\footnote{The first people we can call ``physicists'' are those who looked to the stars and uncovered patterns through astronomy and plane geometry. From Thales of Miletus (c. 624–546 BCE), who is credited with predicting a solar eclipse and recognizing that the Moon shines by reflecting the Sun’s light, to Galileo Galilei (1564–1642), who ushered in modern physics through the scientific method, combining careful experimentation with mathematical reasoning.}, passes through abstract mathematical constructions, and leads to a deeper understanding of observed phenomena. The ultimate aim of physics is to develop a consistent mathematical framework that can describe all possible observations as accurately as possible. 

This is a challenging task for several reasons. First, nature’s complexity often makes phenomena appear chaotic\footnote{Here, by ``chaotic'', we mean systems in which, without a solid theoretical prediction, it becomes extremely difficult to identify any patterns. There exist processes --- even fully deterministic ones --- that are chaotic by definition. A well-known example is the so-called “butterfly effect”, where tiny variations in initial conditions can lead to dramatically different outcomes.}, and recognizing meaningful patterns within them is far from trivial. Second, there is the fundamental difficulty of defining what we mean by ``observation'' or, in the language of modern physics, ``observables''\footnote{These are physical quantities that can, in principle, be measured experimentally. In classical physics, they can correspond to the position of a particle, its energy, angular momentum, or also the electric or magnetic field strength. In quantum physics the most famous ones are scattering amplitudes that give rise to cross sections; a quantitative measures of the probability for a given physical process to occur.}. And even more subtle is the question of what it means to describe phenomena ``as accurately as possible''. As we now understand, certain aspects of reality may be intrinsically unpredictable, as in quantum mechanics, and we cannot expect mathematics to reveal what nature itself conceals.

Let us also keep in mind the key distinction between physics and mathematics; the role played by experiments. In the end, physics must be guided by experimental evidence, and we must be prepared to abandon even the most mathematically elegant theory if it fails to match observations.

At the time of writing this thesis, there is a broad consensus within the scientific community that our current understanding of the physical world rests on two well-established and experimentally confirmed theoretical frameworks. Historically, the first is General Relativity (GR), formulated by Albert Einstein in 1915 \cite{Einstein:1916vd}, which provides a classical description of gravity as the curvature of spacetime. The second is Quantum Field Theory (QFT), whose development began in the 1920s through the contributions of many physicists and unifies quantum mechanics with special relativity. By the late 1970s, it was established as the theoretical framework that underpins the Standard Model (SM) of particle physics \cite{Weinberg:2004kv}.

General Relativity describes the classical evolution of space-time and its classical interaction with matter fields. Aside from being a beautiful mathematical construction rooted in pseudo-Riemannian geometry, it has, to date, successfully reproduced all known physical observations. From its early predictions, such as the perihelion precession of Mercury and the deflection of light by the Sun \cite{NYT1919,Dyson:1920cwa}, to the groundbreaking discovery of gravitational waves in 2015 \cite{LIGOScientific:2016aoc,LIGOScientific:2017vwq}, the theory has proven remarkably accurate.

The Standard Model of particle physics describes the quantum interactions of matter (spin-$\tfrac12$ fermions) and gauge bosons (spin-$1$) with unprecedented precision, as confirmed by high-energy collisions at the LHC and numerous other experiments around the world. Theoretically, it describes three of the four known forces of nature, Quantum Electrodynamics (QED)\footnote{The theory describing the interactions of matter via the exchange of photons (the quanta of lights), massless spin-$1$ particles, and which classical counterpart is standard electrodynamics.}, the Weak force\footnote{The theory responsable for the radioactive decay of atoms and which mediators are the $W^+$, $W^-$ and $Z$ massive spin-$1$ particles.} and the Strong force\footnote{The theory governed by Quantum Chromodynamics (QCD), responsable for the existence of atoms and allowing the confinements of quarks into protons, neutrons, and other hadron particles via the exchange of gluons, massless spin-$1$ particles.}. To date, the most accurate prediction is in the context of QED via the measurements of the electron magnetic dipole moment $a_e^{(exp)}=\tfrac{g-2}{2}=0.00115965218059(13)$ \cite{Fan:2022eto} and $a_e^{(theo)}=0.001 159 652 182032(745)$ \cite{Aoyama:2017uqe}, where $g=2$ corresponds to the value predicted by the Dirac equation, and the rest is called anomalous magnetic moment of the electron and comes from loop corrections $a_e=\frac{\alpha}{2\pi}+ O(\alpha^2)$, where $\alpha=\frac{e^2}{4\pi}\sim \frac{1}{137}$ is the fine structure constant of QED. 

In the Standard Model, the theoretical idea that stands out is the concept of symmetries and a special thanks goes to Emmy Noether \cite{Noether:1918zz} for introducing this concept. In particular, it is built upon the concept of gauge symmetries\footnote{Gauge symmetries are local symmetries, in contrast to global symmetries. Global symmetries are expected to be absent in any consistent theory of quantum gravity, as suggested by one of the Swampland conjectures \cite{Palti:2019pca}. On the other hand, gauge symmetries are just redundancies of description (as is manifested in the light-cone gauge, for example). Nevertheless, they can be extremely useful, e.g. it would be painful to explain GR without using differential geometry where diffeomorphisms play the role of such gauge symmetries.}. It is a QFT with gauge symmetry $SU(3)_C\times SU(2)_L\times U(1)_Y$ and it is precisely this gauge structure that ensures the renormalisability\footnote{Renormalisability is synonymous with predictivity at all energy scales. Strictly speaking, a theory is called (non-)renormalisable if it allows for predictions at both low- and high-energies using a (in-)finite number of free parameters. In the Standard Model, there are $19$ such parameters, which must be determined experimentally. An ambiguity in the choice of the gauge group is present as well as in the choice of the matter content, which is one of the open problems. } of the theory. Another key concept is that of spontaneous symmetry breaking (SSB), the phenomenon by which a global or a gauge symmetry is broken at low energies. In the SM, this breaking is triggered by the Higgs field\footnote{The Higgs field is a spin-$0$ complex scalar field.} \cite{Englert:1964et, Higgs:1964pj, Guralnik:1964eu}, which acquires a non-zero vacuum expectation value. This mechanism has far-reaching consequences, such as giving mass to the  $W^{\pm}$ and $Z$ bosons, as well as to fermions via Yukawa interactions.

Let us also note that the SM is built upon a set of fundamental particles\footnote{In principle, there exist many other massive excited states that can have arbitrarily large spin, both integer and half-integer. However, these are not considered fundamental for two main reasons. First, fundamental particles are those that participate in interactions in an essential way, meaning they explicitly appear in the action of the fundamental theory. Second, fundamental particles are stable and do not decay, unlike excited states. The only possible exception might be the proton, whose absolute stability remains unconfirmed, although its lifetime is extremely long.} with spins $0,\tfrac12,1$. As we will explore in this thesis, our main focus is on investigating the possibility of constructing a fundamental theory that includes higher-spin particles $(s>2)$, alongside the familiar lower-spin fields.

While some might say that the golden age of 20th-century physics --- characterized by both theoretical breakthroughs and experimental excitement --- is coming to an end, this view is probably too pessimistic. Rather than a decline, it seems we are in a transitional phase, and we need to recalibrate our approaches. A new era of discovery could well be on the horizon.

On the experimental side, growing attention is being directed toward the possibility of observing the universe through entirely new eyes. In particular, there is significant excitement surrounding the upcoming LISA mission \cite{Barausse:2020rsu} and the Einstein Telescope \cite{Abac:2025saz}, which are expected to give groundbreaking results and place stringent constraints on potential deviations from General Relativity.

On the theoretical side, it is increasingly more important to develop models and theories well in advance of experimental verification. Since the phenomena we are probing are becoming ever more subtle and elusive --- though not necessarily less important, and often more fundamental --- we are now in a position where we must know the answer before we see it in the data\footnote{Possibilities of discovering something due to serendipity, like happened for the discovery of radioactivity or the Cosmic Microwave Background (CMB) radiation is less and less probable nowadays. Though, at least in the theoretical side, this can still happen.}. This is notably the case with the detection of gravitational waves: identifying the signal required using filters\footnote{Extracting gravitational waves (GW) signals from the (typically much bigger) detector noise is both a technological and a theoretical challenge. The idea of matched filtering is to enhance the signal-to-noise ratio through smart integration techniques. This is possible only if the waveform is predicted in advance with high precision, for more details, see Chapter $7$ of \cite{Maggiore:2007ulw}.} specifically tuned to theoretically predicted waveforms. Indeed, there is currently a major effort devoted to predicting all possible gravitational waveforms \cite{Buonanno:1998gg}, arising from a wide variety of processes --- from binary systems with diverse mass ratios, to black hole mergers, or intriguing phenomena such as superradiance \cite{Tomaselli:2024dbw}. This exemplifies the nature of modern physics. Therefore, more than ever, we must focus our efforts on the theoretical side --- building as many consistent models and frameworks as possible
to guide current and future experiments in their search for new physics.

Interestingly, contemporary physics still faces a range of unresolved challenges, both phenomenological and theoretical.

On the phenomenological side, key open problems include the nature and origin of dark matter, dark energy, the Hubble tension, the matter–antimatter asymmetry, and the non-zero neutrino masses, among others.

On the theoretical side, major puzzles persist, such as the strong CP problem, the hierarchy problem, the black hole information paradox, the trans-Planckian problem in inflationary cosmology, and the challenge of formulating a consistent theory of quantum gravity.

This thesis is devoted to the problem of quantum gravity. A resolution of this issue could also shed light on several of the other problems mentioned above, most notably, the black hole information paradox.

Quantum gravity is the effort to reconcile General Relativity with the principles of quantum mechanics. The natural starting point is to study gravity within the framework of quantum field theory. While this is indeed possible, it turns out that General Relativity is perturbatively non-renormalizable as a QFT \cite{tHooft:1974toh,Goroff:1985th}.

The motivation behind the search for a theory of quantum gravity can be several. First, since the other fundamental interactions are all described by quantum field theories, unifying all forces of nature into a single consistent framework strongly suggests that gravity, too, must be quantised.  Second, General Relativity predicts its breakdown through the singularity theorems \cite{Penrose:1964wq,Hawking:1967ju}, which indicate the inevitability of singularities such as those inside black holes and at the origin of the universe. These are regions where the classical description of spacetime ceases to be valid, and a quantum theory of gravity becomes essential. Quantum gravitational effects are expected to become significant at energy scales close to the Planck mass, $M_pc^2=\sqrt{\frac{\hbar c^5}{G}}\sim 10^{19}GeV$, which is far beyond the reach of any current or foreseeable experiments on Earth (at LHC we ``only'' reach $\sim 10^4 GeV$). At present, the most promising avenues for probing quantum gravity effects come from astrophysical observations, such as the study of gravitational waves generated by black hole or neutron star mergers. Alternatively, the only option is to turn to Gedanken (thought) experiments.

Another possibility is to study quantum gravity at low energies. While this approach may not directly help with the problem of quantum gravity, it could nonetheless provide crucial evidence that gravity --- like all other known forces --- obeys the laws of quantum mechanics. Several proposed tabletop experiments aim to test this, based on the idea of preparing a massive object in a non-classical state (e.g. an entangled state). The object can then be probed to determine whether it exhibits genuine quantum behavior. Although quantum decoherence makes such experiments extremely challenging, recent technological advances suggest that they may become feasible in the near future, see \cite{Carney:2018ofe}. Importantly, it remains an open question whether gravity is fundamentally quantum or purely classical. These types of experiments could offer a definite answer to this issue.

Let us recall that, even though General Relativity is non-renormalisable, it is still possible to extract meaningful and nontrivial information from its quantisation. Two important examples are the following.

First, General Relativity can be treated as an effective field theory (EFT) \cite{Weinberg:1978kz,Georgi:1993mps,Donoghue:1994dn}. Within this framework, it remains a well-defined QFT at energies well below the Planck scale, allowing for the computation of quantum corrections to classical results. Notably, this includes quantum corrections to the Newtonian potential \cite{Burgess_2004}.

Second, one can study quantum fields propagating on a fixed classical curved background \cite{Birrell:1982ix}. This semiclassical approach has led to one of the most profound theoretical predictions of the past century: the discovery that black holes radiate as black bodies, emitting radiation with a specific temperature, now known as the Hawking temperature $T_H$ \cite{Hawking:1974rv,Hawking:1975vcx}, and entropy $S_{BH}\sim A_H$, where $A_H$ is the area of their event horizon. Together, these quantities obey the four laws of black hole thermodynamics \cite{Bekenstein:1973ur,Bardeen:1973gs}. This provided the first insight that the degrees of freedom of a quantum gravity theory may be stored on a codimension-one hypersurface. This holographic perspective foreshadows the AdS/CFT correspondence, which was first hinted at in the work of 't Hooft \cite{tHooft:1973alw, tHooft:1993dmi}, then Flato and Fronsdal \cite{Flato:1978qz}, then Brown and Henneaux \cite{Brown:1986nw}\footnote{The contributions of Flato and Fronsdal, along with those of Brown and Henneaux, were acknowledged as early precursors to the AdS/CFT correspondence only after the correspondence had been firmly established.}, then Susskind \cite{Susskind:1994vu}, then Polyakov \cite{Polyakov:1997tj}, and later established by Maldacena \cite{Maldacena:1997re}.

Up to date, several theories have been proposed for a theory of quantum gravity: String theory, Higher-spin gravity, Loop quantum gravity, Asymptotic Safety, Higher-derivative gravity, Lorentz violating gravity, Group field theory, Causal dynamical triangulation, Causal Set Theory, Emergent Gravity, Non-Commutative Geometry, Matrix models, and even more unconventional ideas such as p-adic String Theory. 

Each approach distinguishes itself by the way it attempts to solve the problem of quantum gravity. Some introduce new frameworks, such as the worldsheet formulation in String Theory, which also requires higher dimensions and an infinite tower of massive states. Others, like Lorentz-violating gravity theories, relax some of the fundamental principles of quantum field theory by breaking Lorentz invariance at high energy and in specific ways. More conservative approaches, such as Asymptotic Safety, aim to show that gravity itself remains well-behaved at high energies due to a yet-to-be-established UV fixed point. Moreover, these theories have different goals: while some focus on quantising gravity, others, like String Theory, are more ambitious, attempting not only to quantise gravity but to unify it with the Standard Model of particle physics and more.

This thesis is devoted to the study of higher-spin gravity, with a particular focus on the search for consistent interactions between higher-spin fields using a specific framework known as the light-front approach. Higher-spin gravity is a proposed extension of Einstein's general relativity, which includes --- alongside the spin-$2$ graviton --- at least one interacting massless field of spin greater than two. While less ambitious than string theory in scope\footnote{String theory, in addition to addressing the ultraviolet problem of gravity, aims to unify all four fundamental interactions and to provide a framework capable of explaining all possible phenomenological processes. In contrast, higher-spin gravity takes a more modest approach and is best regarded as a toy model for exploring aspects of quantum gravity within a mathematically consistent framework. While there are strong reasons --- discussed below --- to believe that higher-spin gravity is unlikely to be the ultimate theory of nature, it nonetheless offers a valuable and mathematically elegant first step toward a deeper understanding of quantum gravity.}, its goal is to address the ultraviolet (UV) divergences of gravity by introducing a finite, or more likely infinite, tower of higher-spin fields. 

Higher-spin theories\footnote{Note that by higher-spin gravity we refer to a theory in which higher-spin fields interact among themselves as well as with a spin-$2$ graviton, whereas higher-spin theories more generally refer to models where the last requirement can be relaxed.} are gauge theories, and they are fundamentally tied to the concept of gauge symmetry manifesting in the form of a Lie algebra. The algebra associated with a higher-spin theory is called higher-spin algebra and is usually, but not always, infinite dimensional. This is a consequence of the fact that, usually, we need to deal with arbitrary higher-spin fields, and thus infinitely many of them. 

To introduce and motivate the study of such a theory, it is useful to draw a parallel with supersymmetry. Supersymmetry (SUSY), first introduced in \cite{Wess:1974tw}, is based on the idea of a symmetry that relates fermions and bosons. This symmetry offers several theoretical advantages: it partially addresses the hierarchy problem, improves the ultraviolet behavior of gravity, and plays a central role in the formulation of superstring theory \cite{Green:1987sp,Green:1987mn}, a better-behaved, supersymmetric extension of bosonic string theory. These improvements are largely due to the enhanced cancellation of divergences in supersymmetric models.

Examples of such cancellations include the finiteness to all orders in perturbation theory of $\mathcal{N}=4$ Super Yang-Mills (SYM) theory \cite{Mandelstam:1982cb,Brink:1982wv}, and the improved UV behavior observed in supergravity (SUGRA) models \cite{Grisaru:1976vm,Deser:1977yyz,VanNieuwenhuizen:1981ae}.\footnote{Supergravity refers to any gravitational theory with a local supersymmetric algebra and can be systematically derived by gauging a global supersymmetric theory.} A direct group-theoretical consequence of supersymmetry is that the field content (the spectrum) must organize into so-called supermultiplets, which allows for representations of the supersymmetry algebra \cite{Wess:1992cp}.

The idea of higher-spin gravity is similar in spirit, though it ask for a symmetry that relates massless fields of all spins. As a consequence, one would naively expect that consistent representations of the higher-spin symmetry require multiplets that include an infinite tower of massless fields, with some notable exceptions, such as consistent truncations to only even-spin fields, or to only odd-spin fields in the presence of a $U(N)$ gauge algebra \cite{Fradkin:1986ka}. However, as we will explore in this thesis, there also exist constructions with a finite number of higher-spin fields, at least when the theory is restricted to a chiral sector.

The presence of higher-spin symmetry should make the UV divergences of gravity disappear, by restricting the allowed counterterms — potentially reducing them to a finite set or eliminating them all --- thereby rendering the genuine quantum problems to the classical ones (indeed, if the higher-spin symmetry kills all but finitely many counterterms, the theory is renormalizable, at least). This could also be the key mechanism behind the improved UV behavior observed in string theory. Then supersymmetry might play a secondary role: rather than being the primary reason for UV finiteness, it could simply facilitate the construction of consistent models\footnote{Note that there exist consistent non-supersymmetric tachyon-free models \cite{Angelantonj:2002ct,Sugimoto:1999tx,Baykara:2024tjr}.}. The true origin of finiteness might instead lie in the presence of an infinite tower of higher-spin states, which become massless in the high-energy (tensionless) limit of string theory, $\alpha' \to \infty$ \cite{Gross:1987ar, Gross:1987kza, Gross:1988ue}.

From this perspective, our physical world could correspond to a spontaneously broken phase of an underlying higher-spin theory\footnote{ it is important to emphasize that, unlike the internal gauge symmetries of the Standard Model, higher-spin symmetry is a spacetime gauge symmetry. This distinction significantly complicates both the formulation of consistent interactions and the description of symmetry breaking, making the study of higher-spin theories more subtle.} \cite{Vasiliev:1986td,Vasiliev:1988sa}, where the symmetry is no longer manifest, and the higher-spin fields acquire masses. This mechanism could also provide a pathway for a phenomenological theory of higher-spin, maybe pointing towards string theory, or maybe not.

\newpage

\section{Higher-spin gravity}

The first account on higher-spin fields dates back to Majorana’s proposal \cite{Majorana:1932chs} where he generalised the Dirac relativistic invariant wave equations for the spin-$\tfrac12$ particle (the electron) to arbitrary higher-spin fields, both integer and half-integer. This was then further revisited by Dirac himself \cite{Dirac:1936tg}. A few years later, in a seminal work of Wigner \cite{Wigner:1939cj}, all unitary irreducible representations UIRs of the inhomogeneous Lorentz group $ISO(3,1)$ in $4d$ were classified, via the novel method of induced representations.\footnote{For a modern and comprehensive review of UIRs of the Poincaré group in arbitrary spacetime dimensions and their relation to relativistic field equations, see \cite{Bekaert:2006py}.} 

This method reduces the problem of classifying the representation of a group (in this case, the Poincaré group $ISO(d-1,1)\equiv\mathbb{R}^{d-1,1}\rtimes SO(d-1,1)$) to that of classifying representations of its stability subgroups (``little group''). In particular, the little group of massless particles ($p^2=0$) is $ISO(d-2)$, which in turn admits two stability subgroups. The subgroup $SO(d-2)$, which stabilises the origin in $\mathbb{R}^{d-2}$, gives rise to finite-dimensional representations describing particles with fixed, arbitrary helicity $\lambda$. The subgroup $SO(d-3)$ , which stabilises a $(d-3)$-sphere of radius $\mu>0$ in $\mathbb{R}^{d-2}$, leads to infinite-dimensional representations known as continuous (or infinite) spin representations, see \cite{Bekaert:2005in,Bekaert:2017khg}. 

Higher-spin theory could be understood as the program aimed at constructing consistent interacting field theories\footnote{Free particles are of limited physical interest. Even if they exist, they would neither be directly detected nor have any observable effect, making them essentially a theoretical abstraction.} for all massless UIRs of the Poincaré group, with the possible exception of the continuous spin ones, and with particular emphasis on the spin-$2$ graviton. It remains an open question whether all UIRs of Poincaré correspond to physical particles. To date, only massive UIRs and massless ones with low-spin ($\lambda\leq 2$) are known to exist in nature.

The first step in this direction was already taken by Bargmann and Wigner \cite{Bargmann:1948ck}, trying to associate, to any UIRs of Poincaré, a manifestly covariant differential equation whose (positive-energy) solutions transform accordingly to the corresponding UIR. The restricted program of searching them only for massive and massless representations in $d=4$ was completed in \cite{Bargmann:1948ck}. For a generalisation to any dimensions, see \cite{Bekaert:2006py}. In $3d$ the complete programme was completed in \cite{Binegar:1981gv}.

An important step toward a modern field-theoretical description of free higher-spin fields was made by Singh and Hagen \cite{Singh:1974qz,Singh:1974rc}, who constructed a Lagrangian formulation for massive fields of arbitrary spin. Their work built upon earlier results, notably the Lagrangian descriptions of massive spin-$2$ and spin-$\frac{3}{2}$ by Fierz and Pauli \cite{Fierz:1939ix}, and by Rarita and Schwinger \cite{Rarita:1941mf}, respectively. Furthermore, in \cite{Fierz:1939ix}, the field equations for arbitrary massive spin were written down, and these are the same as those reproduced by the Singh–Hagen Lagrangian description.

A few years later, Fronsdal, by taking the massless limit of the Singh–Hagen Lagrangian, managed to realise an action principle for free massless integer helicity fields \cite{Fronsdal:1978rb} and, in collaboration with Fang \cite{Fang:1978wz}, extended it to half-integer helicities. Fronsdal subsequently generalized the formulation to (A)dS backgrounds \cite{Fronsdal:1978vb}. The resulting field equations are two-derivative and are built out from totally symmetric tensor fields, $\varphi_{\mu_1\cdots\mu_s}$ for a spin-$s$ field. A notable feature of the resulting Lagrangian is the emergence of a novel gauge symmetry $\delta\varphi_{\mu_1\cdots\mu_s}=\partial_{(\mu_1}\xi_{\mu_2\cdots\mu_s)}$, accompanied by a peculiar double-tracelessness condition on the fields $\varphi_{\mu_1\cdots\mu_{s-4}\nu\rho}^{\phantom{\mu_1\cdots\mu_{s-4}\nu\rho}\nu\rho}=0$ and traceless gauge parameters $\xi_{\mu_1\cdots\mu_{s-3}\nu}^{\phantom{\mu_1\cdots\mu_{s-3}\nu}\nu}=0$, which becomes essential starting from spin-$3$ and above. 

We notice that in $d\geq 5$, the classification of UIRs of the Poincaré group, while qualitatively similar to the $4d$ case, becomes more intricate. In higher dimensions, there are additional ways to embed these UIRs into mixed-symmetry tensor representations, with the first covariant formulation of mixed-symmetry fields developed by Labastida \cite{Labastida:1987kw}. Such descriptions may offer practical advantages and, notably, they naturally arise in string theory spectra, though for massive fields. 

We also point out that the Fronsdal formulation and, in general, the metric-like one, are not the only available descriptions of massless higher-spin fields. As we will describe in detail, the light-cone has a very simple free Lagrangian for arbitrary massless higher-spin fields, albeit at the cost of manifest Lorentz covariance. Another alternative approach is offered by the frame-like formulation of higher-spin fields, which generalizes the vielbein formalism of gravity; see \cite{Didenko:2014dwa} for a comprehensive review.

The idea of the frame-like approach to higher-spin fields is to emulate the Cartan formulation of gravity \cite{Hehl:1976kj}\footnote{Notice that in the literature this is sometimes referred to as the first-order formulation of gravity, or the tetrad formulation. To be precise also Palatini formulation of gravity \cite{Palatini:1919ffw} is first-order but as independent variables use the metric $g_{\mu\nu}$ and the connection $\Gamma_{\mu\nu}^{\rho}$.} via the vielbein $e^a_\mu$ and the spin-connection $\omega^{a,b}_{\mu}$.\footnote{Here $a,b,...=0,1,...,d-1$ are $so(d-1,1)$ frame indices, referring to the local Minkowski frame, while $\mu,\nu,...=0,1,...,d-1$ are spacetime coordinate indices.} By studying the Unfolding equations of higher-spin fields, it is possible to find the proper generalisation of the Cartan formulation for a spin-$s$ particle. This corresponds to a set of one-forms $\omega^{a(s-1),b(k)}$ (Weyl modules) for $k\leq s-1$ and zero-forms $C^{a(s+k),b(s)}$ (gauge modules) for $k\geq 0$. Therefore, all spins together cover all two-row Young diagrams of the Lorentz group, each appearing in one copy. Roughly speaking, the zero-forms encode the physical degrees of freedom of the theory and the one-forms (the gauge parameters associated with them) the gauge symmetries.

One of the deep earlier results is the Flato-Fronsdal theorem \cite{Flato:1978qz} for $so(3,2)$. The theorem says, roughly speaking, that the tensor product of two singletons decomposes into a direct sum of UIRs of the $4d$ anti-de Sitter group that are massless fields of all spins, each appearing in one copy. More precisely, $D\left(\tfrac{1}{2},0\right)\otimes D\left(\tfrac{1}{2},0\right)=\sum_{s=0}^{\infty}D(s+1,s)$, where $D\left(\tfrac{1}{2},0\right)$ is the $Rac$ singleton representation\footnote{Same as for the flat case, where free fields are associated to UIRs of the Poincaré, in $AdS_d$ free fields can be identified with UIRs of its isometry algebra, the non-compact group $SO(d-1,2)$. Analysing all the UIRs, there is a peculiar representation. Representations that saturate the so-called unitarity bounds, on the field theory side they coincide with fields that acquire a gauge symmetry. Surprisingly, this also happens for the scalar representation. This representation is so short that it does not have enough degrees of freedom to live in the bulk, but it can correspond to a scalar field in $d-1$ dimension Minkowski space, with $SO(d-1,2)$ playing the role of the $(d-1)$-dimensional conformal group. There is also an analogue for the fermionic representation. These ``remarkable'' representations where discovered in $AdS_4$ by Dirac \cite{Dirac:1963ta} and called $Di$, for the fermionic singleton $D\left(1,\tfrac{1}{2}\right)$, and $Rac$, for the scalar singleton $D\left(\tfrac{1}{2},0\right)$.} of $so(3,2)$ and $\sum_{s=0}^{\infty}D(s+1,s)$ are the UIRs corresponding to massless fields of spin-$s$. They are closely related to the same ``all two-row Young diagrams'' representations of the Lorentz algebra one finds in the spectrum of zero-forms. This fact is related to the concept of higher-spin algebra.

Even before Fronsdal, in $d=4$ another (twistor-inspired) approach to massless fields with spin was proposed by Penrose starting from \cite{Penrose:1965am}, see also \cite{Hughston:1979tq,Eastwood:1981jy,Woodhouse:1985id}, which allows one to bypass some of the restrictions of the Fronsdal description, e.g. to construct gravitational interactions in flat space (more generally, any self-dual spacetime), and is the way to covariantize chiral higher-spin theory. 
When going back from twistor space to the usual spacetime, it employs a ``very chiral'' spinorial representation of $so(3,1)\sim sl(2,\mathbb{C})$ Lorentz fields.

Once the free description, in the metric-like, or the frame-like, or the chiral, or the light-front approach is found, we need to search for possible consistent interactions. By consistent, we mean Lorentz invariant and perturbatively local.

In this regard, a fundamental concept for a theory of higher-spin fields is the existence of a higher-spin algebra\footnote{For a formal definition and an introduction see \cite{Joung:2014qya}.}, just as for supergravity, it is the existence of a supersymmetry algebra. This is defined as a (typically infinite-dimensional) Lie algebra that arises from the global symmetries of a higher-spin theory. The higher-spin algebra is usually the starting point for the construction of a fully non-linear higher-spin theory. Indeed, to search directly for a consistent theory of higher spins is complicated, but it can be easier to construct first higher-spin algebras and then try to consider interactions order by order. The first higher-spin algebra in $4d$, $shs(1,\mathbb{C})$, containing $AdS_4$ as a subalgebra, was found by Fradkin and Vasiliev in \cite{Fradkin:1986ka}.

The higher-spin algebra can be seen as the first step towards a complete higher-spin theory. We could say that it corresponds to the first cubic deformation, like for the Lie algebra in Yang-Mills theory. Though, as for Yang-Mills, we need to add higher-order terms in order to make the theory consistent. 

In the following, we explain in detail the light-front approach, as one of the methods to introduce consistent interactions in higher-spin theories. There are some pros and cons. The most important ``pro'' is that one deals with physical degrees of freedom only. All redundancies due to gauge symmetry are eliminated. The outcome does not depend on which particular Lorentz tensor one has (luckily or unluckily) chosen to embed the physical degrees of freedom into (modulo gauge redundancies). Therefore, one gets unambiguous results concerning (non)-existence of a certain type of interactions (this is to be contrasted with the non-existence of gravitational interactions in flat space within the Fronsdal approach, which is simply a drawback of this particular covariant approach and has no fundamental meaning). There is no higher-spin algebra in the light-cone gauge. Instead, it gets transferred (if the covariant description is available at all) to the form of the generators of the Poincaré algebra. We should also point out that the most minimal QFT setup is to have the Hamiltonian, the time-ordered exponent of which gives the S-matrix, and the existence of other generators of the Poincaré algebra allows one to prove the Poincaré invariance of the S-matrix. There is no theorem that would guarantee the existence of a manifestly covariant description that gives the same S-matrix. The most important ``con'' is the lack of the manifest Lorentz symmetry, which is the price to pay.

\newpage

\section{Higher-spin interactions in the light-front}

The standard approach in contemporary physics is to formulate theories using Lorentz covariant fields, so that the equations are manifestly Poincaré invariant. However, particularly in the case of higher-spin fields, there exist multiple ways to embed the physical degrees of freedom --- two in four dimensions --- into tensorial fields. As a result, when searching for interactions, their presence or absence may strongly depend on the specific field realisation chosen. In this context, the light-front approach provides a more reliable framework: by directly working with the physical degrees of freedom, it offers a systematic and complete method to explore consistent interactions without the ambiguities introduced by covariant fields. Therefore, it stands out as the safest and most exhaustive method for analysing interactions of (higher-spin) theories.

The light-front approach to quantum field theory is a way to work only with the physical degrees of freedom of the theory. It can either be reached from the gauge fixing of a covariant formulation or --- and it is important to stress --- we can start directly with a theory in the light-front, with no reference to any covariant description --- such a covariant formulation may not even exist. This method is generally considered as the most effective for establishing rigorous no-go results or identifying promising yes-go results. However, to construct a complete theory, to simplify computations, to improve its geometrical interpretation, to better exploit all the symmetries and to study its quantum properties, one typically needs a covariant formulation. We will elaborate more on these points in the following sections.

The light-front quantisation of relativistic quantum field theory was introduced in Dirac’s seminal 1949 paper \cite{Dirac:1949cp}. In that work, Dirac proposed three distinct forms of Hamiltonian dynamics based on different choices of the time coordinate: the instant form (the conventional one), the point form, and the front form --- now commonly referred to as light-front or light-cone quantisation\footnote{The term light-cone can be somewhat misleading. Quantisation does not occur on the light-cone itself, but rather on a light-like codimension-one hypersurface defined by constant $x^+$. Geometrically, this corresponds to the wavefront of a plane light wave, which is why Dirac referred to it as the front form. Nevertheless, the terminology light-cone is widely used in the literature, including in the higher-spin context.}. These formulations differ in the number of kinematical versus dynamical Poincaré generators, as well as in the structure of the Cauchy initial-value problem (to be fair, the light front is not a Cauchy surface and there are some subtle global issues related to that).

Given a particular foliation of spacetime, let $\Sigma$ be the codimension-one hypersurface on which quantisation is performed. The subset of Poincaré generators that leave $\Sigma$ invariant forms a subgroup of the Poincaré group, known as the stability group $G_{\Sigma}$ of $\Sigma$. In \cite{Leutwyler:1977vy}, Leutwyler and Stern, under the assumption that the stability group $G_{\Sigma}$ acts transitively on $\Sigma$, proved that there are exactly five inequivalent classes of hypersurfaces. Dirac had originally identified only three of them, missing two cases which, however, are less significant since their stability groups are rather small.

The light-front approach has been applied in a wide range of problems since Dirac's work. It was rediscovered in the context of current algebra, as the ``infinite momentum frame'' in \cite{Fubini:1964boa}. It was used to study quantum electrodynamics in a laser beam \cite{Neville1968,Neville:1971uc}. It was applied to gauge theories, notably in QED \cite{Kogut:1969xa,Bjorken:1970ah}, and later in QCD \cite{Lepage:1980fj}. The motivation was to provide a more robust theoretical framework that could reconcile the constituent quark model (CQM) and the quark parton model, originally proposed by Richard Feynman \cite{Feynman:1969ej} to describe hadrons. The importance of the light-front approach in the context of QCD is also closely tied to the strongly coupled nature of the theory. Indeed, in the infrared (IR) regime, QCD exhibits confinement and becomes strongly coupled, in contrast to its behavior in the UV, where asymptotic freedom holds \cite{Gross:1973id,Politzer:1973fx}. Being strongly coupled in the IR means that standard QFT techniques --- such as perturbation theory and Feynman diagrams --- are no longer directly applicable. The light-front approach offers a powerful framework for studying QFT in such non-perturbative regimes and for studying relativistic bound states.  For an extensive review on the subject, along with an interesting introduction to light-front field theory for lower-spin fields, see \cite{Brodsky:1997de}.

The light-front approach has played a crucial role in both string theory and supersymmetry. Notably, it was instrumental in the first proof of the finiteness of $4d$ $\mathcal{N}=4$ super Yang-Mills (SYM) theory \cite{Mandelstam:1982cb,Brink:1982wv}. It was also used for the initial quantisation of bosonic string theory \cite{Goddard:1973qh} and later extended to superstring theory \cite{Schwarz:1982jn}. Moreover, for certain supersymmetric theories with extended supersymmetry, the light-cone superspace remains the only known consistent off-shell superspace formulation. Examples include $4d$ $\mathcal{N}=4$ super Yang-Mills (SYM) theory \cite{Mandelstam:1982cb,Brink:1982wv}, $11d$ $\mathcal{N}=1$ SUGRA \cite{Ananth:2005vg,Metsaev_2005}, $10d$ $\mathcal{N}=1$ SYM \cite{Ananth:2004es}.
Additionally, the light-cone gauge provides one of the most effective methods to extract physical states in string theory \cite{Goddard:1973qh}.

It is important to distinguish two kinds of ``light-cone'' approach. One is just to take an advantage of the light-cone gauge as a manifestly unitary gauge (the one that keeps only the physical degrees of freedom). Another one is the actual light-cone/front quantization where $x^+$ is chosen as ``time'' and one looks for the deformation of the generators of the Poincare algebra. 

More recently, the light-front approach has also been applied to higher-spin interactions, as we will now see and describe in detail. The use of the light-front approach to study possible interactions in higher-spin theories began in 1983 with the work of Bengtsson, Bengtsson, and Brink \cite{Bengtsson:1983pd,Bengtsson:1983pg,Bengtsson:1986kh}, where cubic interactions in flat space were found by solving the cubic light-cone constraint, and the complete classification of cubic vertices was obtained in \cite{Bengtsson:1986kh}. Just a few months later, the first covariant cubic vertex was constructed in \cite{Berends:1984wp} via the Noether procedure, starting from a free massless spin-$3$ Fronsdal field $\phi^{\mu\nu\lambda}$ and trying to make it self-interacting\footnote{The same analysis was performed in \cite{Deser:1990bk} for a cubic coupling of the form $4-2-2$ but the same conclusion was found. Consistency at the first non-trivial order, the cubic, but impossibility to extend it further.}. The idea is to search for a Yang–Mills-type theory only involving a higher-spin field with an odd spin, in this case spin-$3$\footnote{For Yang-Mills, starting from a Maxwell-type action $S_2=-\int d^4x F_a^{\mu\nu}F^a_{\mu\nu}$ and gauge symmetry $\delta_0A^a_{\mu}=\partial_{\mu}\xi^a$, the first non-trivial order give the vertex $S_3=\int d^4x f_{abc}A^a_{\mu}\partial^{\mu}A^b_{\nu}A^{c\nu}$ and the gauge transformation $\delta_1A^a_{\mu}=-f^a_{\phantom{a}bc}\xi^bA^c_{\mu}$, where $f_{abc}$ is a structure constant. A similar result can also be found for odd higher-spin fields. Though, for Yang-Mills, at the next order we find $S_4=-\frac{1}{4}\int d^4x f^e_{\phantom{e}ac}f_{ebd}A^a_{\mu}A^{b\mu}A^c_{\nu}A^{d\nu}$ and $\delta_2=0$, with the additional requirement that the structure constant $f_{abc}$ have to satisfy the Jacobi identity, and the procedure terminates. For higher-spin fields, the Noether procedure usually fails at the quartic order, see e.g. \cite{Bekaert:2010hp}.}. Although a non-abelian algebra emerged at the first non-trivial order, it was obstructed at the next order \cite{Berends:1984rq,Bengtsson:1983bp}\footnote{A more recent analysis \cite{Bekaert:2010hp}, using the antifield formalism to search for consistent deformations, assuming arbitrary symmetric Fronsdal gauge fields in dimensions strictly greater than three, showed that this deformation is obstructed even when an infinite tower of higher-spin fields is introduced.}. This led to the realization that a consistent interacting theory would likely require an infinite tower of massless higher-spin fields. A related paper where conserved currents for massless fields of arbitrary spin were explicitly constructed is \cite{Berends:1985xx}.
Moreover, some years before, it was realized in \cite{Aragone:1979hx,Berends:1979kg} that the minimal coupling of gravity to higher-spin particles is problematic. In particular, such couplings introduce terms proportional to the Riemann tensor, which cannot be cancelled by the variations of the metric in flat space. This observation motivated the study of higher-spin interactions in $(A)dS$ backgrounds, especially after the positive results obtained in \cite{Fradkin:1986qy,Fradkin:1987ks}, where the earlier no-go results could be avoided thanks to the presence of the cosmological constant $\Lambda$, which is produced together with the Riemann tensor from the commutator of two covariant derivatives $[\nabla,\nabla]\xi\sim R\xi +\Lambda \xi$.

It is useful to note that the same year the existence of the gravitational two-derivative interactions of massless higher-spin fields in flat space was discovered in the light-cone gauge \cite{Bengtsson:1986kh}, which went almost unnoticed, even by the authors. One can also show that there is a one-to-one correspondence between interactions of massless higher-spin fields in flat and anti-de Sitter spacetimes \cite{Boulanger:2008tg, Metsaev:2018xip}. As discussed in \cite{Krasnov:2021nsq}, what happens is that the two-derivative gravitational interaction cannot be written (locally) within the Fronsdal approach in flat space. Instead, one can construct a $(2s-2)$-derivative vertex. In anti-de Sitter space, a specific linear combination of the $2$- and $(2s-2)$-derivative vertices of type $s-s-2$ can be written down in terms of Fronsdal fields (this is the Fradkin-Vasiliev vertex \cite{Fradkin:1986qy, Fradkin:1987ks}). Therefore, what breaks down when going to flat space is the Fronsdal approach, while the gravitational and all other interactions are smooth!

One of the cornerstones of the light-front approach to higher-spin interactions are the two papers by Metsaev \cite{Metsaev:1991mt,Metsaev:1991nb}, where, for the first time, the quartic light-cone constraint was systematically analyzed, leading to, what we will call, the Metsaev solution. This result completely determines the values of the coupling constants for all cubic vertices.

It is known that light-cone perturbation theory proceeds as follows: the $n$-order constraint fixes the form of the $n$-point vertices up to some free coefficients, while the $(n+1)$-order constraint determines those coefficients, which we will apply at the lowest cubic and quartic orders. Therefore, the importance of the quartic constraint lies not only in its role in constructing the quartic interaction vertices themselves, but, more fundamentally, in its ability to fix the free parameters of the cubic theory --- thereby determining the consistent higher-spin spectrum.

Moreover, as we will see, it is sufficient to consider only a specific sector of the quartic constraint --- namely, the (anti-)holomorphic part --- to fix these free coefficients and, with them, the spectrum of higher-spin fields required for consistency. The (anti)-holomorphic part is a subset of the quartic consistency relations to which the quartic interaction vertices make no contribution, which is a ``miracle'' that happens for massless fields in $4d$. 

25 years later, the results of Metsaev were reviewed in \cite{Ponomarev:2016lrm}, where new features were uncovered, most notably, the possibility of consistently truncating the theory at the cubic level, including only the holomorphic vertices. The resulting fully consistent theory was named the chiral higher-spin theory.

Let us emphasize that by the chiral theory\footnote{Chirality in the light-cone formulation is expected to correspond to self-duality in the covariant approach. This correspondence is well established in lower-spin theories, such as self-dual Yang–Mills (SDYM) and self-dual gravity (SDGR). See \cite{Ponomarev:2017nrr} for the formulation of Chiral theory as a self-duality constraint. A part of this thesis is also devoted to make the connection between what was initially called chiral higher-spin theories and self-dual theories more transparent. }, we refer to a theory in which positive and negative helicities are treated in a specific --- but different --- way. It is important to clarify a common misconception: chirality does not mean that one helicity completely decouples from the interactions. Rather, both helicities participate, but the interactions are structured in a chiral way. In the light-cone gauge, this is very easy to describe, as we will review below.

A very important remark to make about chiral (or self-dual) theories is that even though they are non-unitary and not parity invariant theories, they are consistent subsectors of the complete theories \cite{Krasnov:2016emc}. By consistent subsectors, we mean that at the level of scattering amplitudes, all amplitudes we can compute in a self-dual theory exactly coincide with the ones of the complete theory. It is also true that all solutions of self-dual theories are solutions of the full ones. However, due to the chirality of the interactions, not all of the amplitudes can be reproduced by the chiral theory. For instance, both SDYM and SDGR are consistent subsectors of YM and GR. Moreover, they are one-loop exact and finite theories, with no quantum divergences. Therefore, the chiral higher-spin theory also represents a consistent subsector of the putative full complete higher-spin theory (which is not available at present and is likely non-local).

We now turn to an introduction to the basics of the light-front approach to quantum field theory. We stress that in this thesis, the light-front formalism is primarily employed as a tool to introduce interactions in a consistent and the most general way. Although we make use of canonical quantisation, our focus is not on studying quantum properties such as loop corrections or renormalisation. Consequently, potential issues related to the quantisation of light-cone theories are not of concern in our treatment. For further discussion on these aspects, see \cite{Bassetto:1991ue,Mannheim:2020rod}. It remains an open question whether quantum field theories formulated in different forms of dynamics (e.g. instant form vs. light-front form) are physically equivalent.

We now introduce the light-front approach to higher-spin theories to settle our notations and make the presentation of the following three chapters more accessible. A more detailed discussion can be found in the first sections of \cite{Ponomarev:2016lrm,Ponomarev:2016cwi,Ponomarev:2017nrr}, in \cite{Metsaev:2005ar,Bengtsson:2012jm}, and also in the lecture notes \cite{Ponomarev:2022vjb}.

The light-front deformation procedure relies on two key ingredients: the Hamiltonian formulation of quantum field theory in the light-cone gauge, and the non-linear realization of the Poincaré algebra via higher-order corrections to the quadratic free Hamiltonian. As Dirac said in his work \cite{Dirac:1949cp}, the ``real difficulty in the construction of a theory of a relativistic dynamical system'' lies in determining the interaction terms (the potentials) that are consistent with the commutation relations of the Poincaré algebra.

While the problem is conceptually straightforward, the challenge lies in performing a significant amount of computation to determine the correct form of the interactions. Once the free field realization for massless higher-spin fields is in place, the analysis of the cubic constraints gives a number of cubic interactions vertices. The coupling constants in front of them remain free at this order, as well as the spectrum of fields cannot be fixed yet. As we will see, it is only at the level of the quartic constraints that the spectrum of a theory begins to be constrained.

For simplicity, we will work in four dimensions. Although the assumption of $4d$ will not affect the early stages of the analysis, it will become important later on. Spinning massless fields in $d$-dimensional Minkowski space are described by irreps of the $SO(d-2)$ little group. Totally symmetric fields carry $\frac{(d+s-5)!}{s! (d-4)!}(d+2s-4)$ propagating degrees of freedom. For $d=4$ and $s\geq 1$, we have only $2$ degrees of freedom, and this simplifies their description. For the analysis of both massive and massless higher-spin fields in $d$-dimensions using oscillators, see \cite{Metsaev:2005ar}.

We refrain from discussing no-go results in the light-cone approach in this section as they are delegated to Chapter 4.

\paragraph{Notations.} We use both light-cone coordinates (in fact, double-null) and the light-cone gauge.
In flat spacetime, we adopt the $4d$ Minkowski metric with ``mostly plus'' signature
\begin{align}
    &x^{\mu}=(x^0,x^1,x^2,x^3)\,,&
    &\eta^{\mu\nu}=\text{diag}(-,+,+,+)\,,\\
    &ds^2=-(dx^0)^2+(dx^1)^2+(dx^2)^2+(dx^3)^2\,.
\end{align}
We define light-cone coordinates as
\begin{align}
    x^+&=\frac{x^3+x^0}{\sqrt{2}}\,,&
    x^-&=\frac{x^3-x^0}{\sqrt{2}}\,,&
    x&=\frac{x^1-ix^2}{\sqrt{2}}\,,&
    \bar{x}&=\frac{x^1+ix^2}{\sqrt{2}}\,,\\
    \partial_-=\partial^+ &= \frac{\partial^3+\partial^0}{\sqrt{2}}\,, &
    \partial_+=\partial^- &= \frac{\partial^3-\partial^0}{\sqrt{2}}\,, &
    \partial &= \frac{\partial^1+i\partial^2}{\sqrt{2}}\,, &
    \bar{\partial} &= \frac{\partial^1-i\partial^2}{\sqrt{2}}\,,
\end{align}
where the metric and the wave-operator become
\begin{align}
    &x^{\mu}=(x^+,x^-,x,\bar{x})\,,\qquad
    ds^2=2\,dx^+ dx^- + dx^a dx_a=2\,dx^+ dx^- + 2\,dx d\bar{x}\,,\\
    &\Box=\partial_{\mu}\partial^{\mu}=(2\partial^+\partial^-+\partial^a\partial_a)=2(\partial^+\partial^-+\partial\bar{\partial})\,.
\end{align}
The light-front scalar product becomes
\begin{align}
    A_{\mu}B^{\mu}=A_+B^++A_-B^-+A\bar{B}+\bar{A}B=A^-B^++A^+B^-+A\bar{B}+\bar{A}B\,,
\end{align}
and our definitions for the derivatives imply
\begin{align}
    \partial^+ x^-=\partial^-x^+=\partial x=\bar{\partial}\bar{x}=1\,.
\end{align}
In the light-front, $x^+$ is taken to be the time variable and $\partial^-$ is the time derivative. Moreover, one assumes that $\partial^+$ is always non-zero and therefore can be inverted as an operator. The coordinate $x^-$ defines the longitudinal direction, whereas $x^1$ and $x^2$, or equivalently $x$ and $\bar{x}$, are the transverse directions.
We also use the following shorthand notation:
\begin{align}
    &d^2x=dxd\bar{x}\,,&
    &d^2p=dpd\bar{p}\,,\\
    &d^3x=dx^-d^2x\,,&
    &d^3p=dp^+d^2p\,,\\
    &\delta^2(x-y)=\delta(x-y)\delta(\bar{x}-\bar{y})\,,&
    &\delta^2(p+q)=\delta(p+q)\delta(\bar{p}+\bar{q})\,,\\
    &\delta^3(x-y)=\delta(x^--y^-)\delta^2(x-y)\,,&
    &\delta^3(p+q)=\delta(p^++q^+)\delta^2(p+q)\,.
\end{align}
In the following part of the introduction, Greek indices $\mu,\nu,...=0,1,2,3$ label $4d$ spacetime coordinates; Latin indices $a,b,...=1,2$ are reserved for the transverse directions; when needed, $i,j,k,...=1,2,3$ denote the standard spatial coordinates, with $x^0$ as the time variable.

\paragraph{Massless spin-1 field.} Before studying higher-spin fields in the light-front formulation, we consider a simple example of a spin‑$1$ field, which highlights three key features characteristic of this approach. Starting from the Maxwell Lagrangian for a free spin-$1$ field
\begin{align}\label{spin1_eom}
    &S=-\frac{1}{4}\int d^4 x F_{\mu\nu}F^{\mu\nu}\,,&
    &F^{\mu\nu}=\partial^{\mu}A^{\nu}-\partial^{\nu}A^{\mu}\,,&
    &\delta A^{\mu}=\partial^{\mu}\xi\,,
\end{align}
whose equations of motion are
\begin{equation}\label{EOMspin1}
    \partial_{\nu}F^{\nu\mu}=\Box A^{\mu}-\partial^{\mu}\partial_{\nu}A^{\nu}=0\,.
\end{equation}
There are various ways to gauge fix the theory; two standard gauges are the Coulomb gauge $\partial^iA_i=0$ and the Lorentz gauge $\partial^{\mu}A_{\mu}=0$. Here we employ the light-cone gauge that consists of setting $A^+=0$. This directly eliminates one component of the gauge field and can be reached by a suitable gauge transformation
\begin{align}
    &A^+\rightarrow A^++\partial^+\xi&
    &\implies&
    &\partial^+\xi=-A^+
    &\implies&
    &\xi=-\frac{1}{\partial^+}A^+\,.
\end{align}
The first remark is that this gauge can be achieved once we assume that the operator $\partial^+$ is invertible. Indeed, in the presence of zero modes ($p^+=0$), when $\phi$ satisfies $\partial^+\phi=0$, this gauge cannot be reached. Therefore, when working in the light-cone gauge, we need to assume $\partial^+\phi\neq0$ ($p^+\neq 0$). Notice that the problem of physical zero modes or topological obstructions is typical of the gauge fixing procedure in general. Zero modes are subtle and can affect the vacuum structure, the topological sectors, and quantisation. For a further look at the subtleties with zero modes in the light-cone gauge, some discussion can be found in \cite{Perry:1994kp,Burkardt:1995ct,6db1d5c6aea346a68475d79dd4ab667e,Harindranath:1996hq,Heinzl:2000ht,Mannheim:2020rod}. From now on, we will assume the operator $\partial^+$ to be always invertible.

Let us look at the equations of motion \eqref{EOMspin1} in the ``+'' direction 
\begin{align}
    &\partial^+(\partial^+A^-+\partial_aA^a)=0&
    &\implies&
    &A^-=-\frac{\partial_a}{\partial^+}A^a\,.
\end{align}
As we see, no time derivatives are involved; hence the equation for $A^-$ acts as a constraint rather than a dynamical equation, implying that $A^-$ is not a dynamical field. It is also important that the action is quadratic in $A^-$ (i.e. the path integral is Gaussian in $A^-$). Therefore, solving the classical equations of motion for $A^-$, as above, is equivalent to performing the path integral over $A^-$ exactly. 

The second remark is that the operator $\frac{1}{\partial^+}$ has to be interpreted as the Green’s function of the derivative operator $\partial^+$. Explicitly, one can define \cite{Mandelstam:1982cb,Bjorken:1970ah}
\begin{align}
    &\partial^+\left(\frac{1}{\partial^+}\phi(x,x^-)\right)=\phi(x,x^-)&
    &\implies&
    &\frac{1}{\partial^+}\phi(x,x^-)=\int dy^- \epsilon(y^--x^-)\phi(x,y^-)\,,
\end{align}
where $\epsilon(x)$ is the sign function, and, in this context, $x=\{x^+,x,\bar{x}\}$. The non-local form of the constraint is not problematic: the operator $\frac{1}{\partial^+}$ is well-defined as an integral operator.

Third and final remark is that the appearance of terms such as $\frac{1}{\partial^+}$ originates from the elimination of non-dynamical fields and does not indicate any genuine physical non-locality. They arise naturally when solving the constraint equations. While $\partial^+$ is formally non-local in $x^-$, it remains causal in light-front time $x^+$, and therefore does not violate physical locality. Equivalently, this can be seen from momentum space: for any physical field, locally, one can always choose a frame where $p^+>0$, so the operator $\frac{1}{p^+}$ is well-defined.\footnote{This is entirely analogous to Coulomb gauge ($\partial^iA_i=0$), where solving Gauss’s law expresses the electric field (and the Coulomb potential) in terms of an instantaneous non-local kernel $\frac{1}{|\vec{x}-\vec{y}|}$.} In contrast, terms involving $\frac{1}{\partial_a}$, or equivalently $\frac{1}{\partial}$ or $\frac{1}{\bar{\partial}}$ do indicate genuine non-localities, even in light-cone gauge, and would correspond to true non-local behavior when interpreted in a covariant framework.

Let us look at the equations of motion \eqref{EOMspin1} in the transverse ``$a$'' direction 
\begin{align}\label{KGspin1}
    &\Box A^a-\partial^a(\partial_bA^b+\partial^+A^-)=0&
    &\implies&
    &\Box A^a \approx 0\,,
\end{align}
whereby $\approx$ we indicate that the equation is valid on the solution of the equations of motion for auxiliary fields ($A^-$ in our case is an auxiliary field). Then the two components of $A^a$ describe a massless vector field. Finally, looking at the equations of motion \eqref{EOMspin1} in the ``$-$'' direction 
\begin{align}
    &\Box A^--\partial^-(\partial_bA^b+\partial^+A^-)=0&
    &\implies&
    &\Box A^-\approx \frac{\partial_a}{\partial^+}\Box A^a\approx 0\,,
\end{align}
we see that this condition is automatically satisfied. Thus, in the light-front formulation, a spin‑$1$ field is described by a massless vector $A^i$ satisfying the Klein–Gordon equation \eqref{KGspin1}. Equivalently, it can be represented by two complex conjugate scalar fields $A$ and $\bar{A}$, corresponding to helicities $+1$ and $-1$. The action can then be rewritten in the form
\begin{equation}
    S=-\frac{1}{2}\int d^4x \partial_{\mu}A^a\partial^{\mu}A_a=-\int d^4x \bar{A}\Box A\,.
\end{equation}

\paragraph{Light-front gauge from Fronsdal fields.} We start from the off-shell Fronsdal massless spin-$s$ field
\begin{align}
    &\phi^{\mu(s)}\,,&
    &\phi_{\nu\rho}^{\phantom{\nu\rho}\nu\rho\mu(s-4)}=0\,,&
    &\delta\phi^{\mu(s)}=\partial^\mu\xi^{\mu(s-1)}\,,&
    &\xi_{\nu}^{\phantom{\nu}\nu\mu(s-3)}=0\,,
\end{align}
and its equations of motion
\begin{equation}\label{FronsdalEOM}
    F^{\mu(s)}=\Box\phi^{\mu(s)}-s\partial^{\mu}\left(\partial_{\nu}\phi^{\nu\mu(s-1)}-\frac{s-1}{2}\partial^{\mu}\phi_{\nu}^{\phantom{\nu}\nu\mu(s-2)}\right)=0\,,
\end{equation}
where we used the standard notation 
\begin{align}
    &\phi^{\mu(s)}\equiv \phi^{(\mu_1\mu_2...\mu_s)}\,,&
    &(\partial_x)_{\mu(s)}\equiv \frac{\partial}{\partial x^{(\mu_1}}\frac{\partial}{\partial x^{\mu_2}}\cdots\frac{\partial}{\partial x^{\mu_n)}}\,,
\end{align}
to denote symmetric tensors. Fixing the light-cone gauge amounts to setting to zero all off-shell fields $\phi^{\mu(s)}$ that have at least one upper ``$+$'' index
\begin{equation}\label{LCgauge}
    \phi^{+\mu(s-1)}=0\,.
\end{equation}
First, let us observe that the light-cone gauge above can always be reached. Indeed, it amounts to finding a $\xi$ such that
\begin{equation}
    \partial^+\xi^{\mu(s-1)}+(s-1)\partial^{\mu}\xi^{+\mu(s-2)}=-\phi^{+\mu(s-1)}\,.
\end{equation}
This can be reached step by step, starting by fixing $\xi^{+...+}$ up to $\xi^{\mu(s-1)}$. For example, for the spin-$1$ case, we find as before 
\begin{equation}
\xi=-\frac{1}{\partial^+}\phi^+\,,
\end{equation}
while for spin-$2$, we first set to zero $\phi^{++}$ by using $\xi^+$ as
\begin{equation}
    \xi^+=-\frac{1}{2\partial^+}\phi^{++}\,,
\end{equation}
and then set to zero $\phi^{+\mu}$ by using $\xi^\mu$ as
\begin{equation}
    \xi^{\mu}=-\frac{1}{\partial^+}\left(\phi^{+\mu}-\frac{\partial^{\mu}}{2\partial^+}\phi^{++}\right)\,,
\end{equation}
and so on for the generic spin-$s$ massless field. As before, this gauge can be reached only if we assume the operator $\partial^+$ to be invertible everywhere; then we assume there are no zero modes.

We know that in four dimensions, massless fields propagate only two physical degrees of freedom --- the two helicity states, corresponding to positive and negative helicity. In the light-cone gauge, not all components of the field are set to zero explicitly. Many field components are eliminated via their equations of motion. Only two components remain dynamical. As we now discuss, all other components can be expressed in terms of the two physical helicity states by solving the non-dynamical equations of motion, i.e. the constraints.

The components of the equation of motions \eqref{FronsdalEOM} with two ``+'' after imposing the gauge \eqref{LCgauge} give
\begin{align}
    &\partial^+\partial^+\phi_{\nu}^{\phantom{\nu}\nu\mu(s-2)}\approx 0&
    &\implies&
    &\phi_{\nu}^{\phantom{\nu}\nu\mu(s-2)}\approx 0&
    &\implies&
    &\phi_a^{\phantom{a}ab(s-2)}\approx 0\,.
\end{align}
Traceless symmetric two-dimensional tensors, such as $\phi^{a(s)}$, have only $2$ non-vanishing components, which we denote 
\begin{align}\label{twoscalars}
    &\phi^s=\phi^{x(s)}\,,&
     &\phi^{-s}=\phi^{\bar{x}(s)}\,,
\end{align}
where we use the indices $x$ and $\bar{x}$ to denote the transverse directions in terms of the two complex conjugate variables. For real Fronsdal fields, these are complex conjugate to each other, and they correspond to the two physical helicity $+s$ and $-s$ states. The components of \eqref{FronsdalEOM} with only one ``$+$'' give
\begin{align}
    &\partial^+\partial_{\nu}\phi^{\nu\mu(s-1)}\approx 0&
    &\implies&
    &\partial_{\nu}\phi^{\nu\mu(s-1)}\approx 0\,,
\end{align}
so the field is also transverse. Focusing on $\mu$ with values in the transverse directions, we obtain
\begin{align}\label{first-}
    &\partial^+\phi^{-a(s-1)}+\partial_b\phi^{ba(s-1)}\approx 0
    &\implies&
    &\phi^{-a(s-1)}\approx -\frac{\partial_b}{\partial^+}\phi^{ba(s-1)}\,.
\end{align}
This implies that on-shell $\phi$ with a single ``$-$'' can be expressed in terms of the two independent fields we introduce in \eqref{twoscalars}. We can continue considering one more ``$-$'' index as
\begin{align}
    &\partial^+\phi^{--a(s-2)}+\partial_b\phi^{-ba(s-2)}\approx 0
    &\implies&
    &\phi^{--a(s-2)}\approx -\frac{\partial_b}{\partial^+}\phi^{-ba(s-2)}\,.
\end{align}
Then using \eqref{first-} we get
\begin{equation}
    \phi^{--a(s-2)}\approx \left(\frac{\partial_b}{\partial^+}\right)^2\phi^{bba(s-2)}\,,
\end{equation}
then can be expressed in terms of $\phi^{\pm s}$. We can proceed like this iteratively, and we find in general that
\begin{equation}
    \phi^{-(n)a(s-n)}\approx \left(-\frac{\partial_b}{\partial^+}\right)^n\phi^{b(n)a(s-n)}\,.
\end{equation}
This finally shows that each component of the Fronsdal field that was not gauged away by \eqref{LCgauge} can be expressed in terms of the two dynamical fields $\phi^{\pm s}$. The dynamical fields satisfy the Klein-Gordon equations
\begin{align}
    \square \phi^{\pm s}=0\,,
\end{align}
which is the easiest way to see what the Fronsdal action reduces to. Finally, if we substitute the terms $\phi^{\pm s}$ into the Fronsdal action
\begin{align}
    &S=\frac{1}{2}\int d^4x\,\phi_{\mu(s)}G^{\mu(s)}[\phi]\,,&
    &G^{\mu(s)}[\phi]=F^{\mu(s)}-\frac{s(s-1)}{4}\eta^{\mu\mu}F_{\nu}^{\phantom{\nu}\nu\mu(s-2)}\,,
\end{align}
we get
\begin{equation}
    S=-\tfrac{1}{2}\int d^4x \partial_{\mu}\phi^{-s}\partial^{\mu}\phi^{+s}\,.
\end{equation}
Let us emphasize that, although the fields $\phi^{-s}$ and $\phi^{+s}$ resemble scalar fields\footnote{For our purposes, we will treat them as if they are scalar fields, complex conjugate of each other.}, only $\phi^0$ transforms as a scalar under Lorentz transformations. The other fields transform differently, as we will briefly discuss in the following section.

Notice that starting from any covariant formulation, we should find the same light-front formulation. For instance, by starting from the chiral formulation used in \cite{Krasnov:2021nsq}, we find the same result.

\paragraph{Basics of the light-front approach.} Quantum field theory in flat spacetime has a rigorous definition and requires a Hilbert space endowed with a unitary action of the Poincaré algebra. The generators of the $4d$ Poincaré algebra $iso(3,1)$ are the four momenta $P^{\mu}$ and the six operators $J^{\mu\nu}$ which combine angular momentum $L^i$ and Lorentz boost $K^i$ accordingly to
\begin{align}
    &L^i=\frac{1}{2}\epsilon^{ijk}J^{jk}\,,&
    &K^i=J^{0i}\,.
\end{align}
The commutation relations (to be realized via Poisson brackets in classical theory and via commutators in quantum theory) defining the algebra are
\begin{subequations}
\begin{align}\label{Poincare}
    [P^{\mu},P^{\nu}]=&\,0\,,\\
    [J^{\mu\nu},P^{\rho}]=&\,P^{\mu}\eta^{\nu\rho}-P^{\nu}\eta^{\mu\rho}\,,\\
    [J^{\mu\nu},J^{\rho\sigma}]=&\,J^{\mu\sigma}\eta^{\nu\rho}-J^{\nu\sigma}\eta^{\mu\rho}-J^{\mu\rho}\eta^{\nu\sigma}+J^{\nu\rho}\eta^{\mu\sigma}\,.
\end{align}
\end{subequations}

There are two main frameworks to quantise a field theory:
\begin{itemize}
    \item The Lagrangian formalism is a manifestly covariant approach. Starting from a Lagrangian density, and consequently an action functional, quantisation can be performed covariantly via the path integral formulation. In the case of gauge theories, gauge fixing is required to define the functional integral properly. To preserve covariance and systematically handle gauge symmetries, one typically employs BRST-BV quantisation methods.
    \item The Hamiltonian formalism \cite{Henneaux:1992ig,Gitman:1990qh} breaks manifest covariance due to the choice of a hypersurface on which quantisation is performed. Quantisation proceeds via the canonical quantisation. This framework is particularly well-suited to the light-front quantisation. Moreover, the Hamiltonian formalism offers greater control over constrained systems and provides an explicit description of time evolution. In what follows, we adopt this Hamiltonian perspective.
\end{itemize}
 Canonical quantisation consists of postulating equal-time commutation relations between the fields and their conjugate momenta. This procedure requires selecting a hypersurface $\Sigma$ (at fixed time) on which the initial data are defined and from which the evolution of the system is determined. This choice induces a foliation of spacetime into space-like or light-like (as in the light-front approach) hypersurfaces, each characterized by a normal vector $\vec{n}$ that is time-like or light-like, respectively.

To ensure causality, the hypersurface must intersect every physical world-line once and only once, so as to guarantee both the existence and uniqueness of the evolution. The particular choice of this surface leads to different quantisation schemes \cite{Dirac:1949cp}, each associated with a distinct stability subgroup of the Poincaré group, i.e. the subgroup of symmetries that leaves the hypersurface invariant. The generators associated with the stability group are called kinematical (K) and do not receive any corrections due to interactions, while the others are called dynamical (D) generators.

The canonical equal-time quantisation uses $t=t_0$ as a choice for the space-like hypersurface $\Sigma_t$ upon which to quantise. The stability group of a space-like hypersurface consists of spatial rotations $L^i$ and translations $P^i$. Instead, all boosts $K^i$ and time translations $P^0=H$ affect time; as a consequence, they do not preserve the surface, and they receive contributions from interactions. Therefore, out of the $10$ Poincaré generators, we have $6$ kinematical and $4$ dynamical ones.  

The light-front quantisation instead uses a light-like hypersurface, given by the choice of $x^+=\text{const}$ that becomes our ``time'' direction, and the new Hamiltonian is $H=P^-$. This, of all five possible quantisations \cite{Leutwyler:1977vy}, is the one with the least number of dynamical generators. There are just three of them. Below is the list of the kinematical and dynamical generators of $iso(3,1)$ we will have to work with:
\begin{align}\label{KD}
    &\text{Kinematical:}&
    & P^+,P^a,J^{a+},J^{+-},J^{ab}&
    &:7\,.\\
    &\text{Dynamical:}&
    & P^-,J^{a-}&
    &:3\,.
\end{align}

The time evolution of any operator $G$ is determined by the Hamiltonian as $\dot{G}=\partial^-G=i[H,G]$. Therefore, if the Poincaré algebra relations are satisfied at the initial light-cone time $x^+=0$, then they are satisfied at all times. As a consequence, generators that have an explicit $x^+$ dependence like
\begin{align}
    &J^{-+}=x^-\partial^+-x^+P^-\,,&
    &J^{a+}=x^a\partial^+-x^+\partial^a\,,
\end{align}
should be declared to be kinematical, as we indeed have done above, since the dynamical part vanishes at $x^+=0$. The $x^+$ dependence can then be reconstructed by using the equations of motion.

Each generator can be decomposed as $G=G_2+G_{int}$, where $G_2$ denotes the free part (quadratic in the fields) and $G_{int}$ encodes the interactions (higher-order in the fields). By definition, the kinematical generators are fixed to their free expressions and do not receive any corrections from interactions (they do not evolve with time). In contrast, the dynamical generators do get $G_{int}$ contributions, but not arbitrary ones. Their form is highly constrained by the requirement that the full algebra of generators --- kinematical and dynamical --- must be preserved at all orders in the interaction expansion.

The explicit form of all commutation relations in the light-front can be seen in \cite{Ponomarev:2016lrm}. The simplest type of commutator is
\begin{equation}
    [K,K]=K\,,
\end{equation}
and they are automatically satisfied at the non-linear level. Two more types of commutators are
\begin{align}
    &[K,D]=K\quad\implies\quad[K,D_{int}]=0\,,&
    &[K,D]=D\quad\implies\quad[K,D_{int}]=D_{int}\,.
\end{align}
Both of them result in linear differential equations on the deformation $D_{int}$, and will be called kinematical constraints. These can be easily solved exactly for $D_{int}$ at any order, and they will restrict the form of the dynamical generators. The last type of commutator is
\begin{equation}\label{dynamical_constraints}
    [D,D]=0\quad\implies \quad\text{explicitly: }\quad
    [J^{a-},J^{c-}]=0\,,\quad
    [J^{a-},P^-]=0\,,
\end{equation}
and are the most difficult to solve, and the main challenge of the light-front deformation procedure. These are called dynamical constraints.

\paragraph{Free action and canonical quantisation.} 
In four dimensions, all massless spinning fields have two degrees of freedom, i.e. effectively they are made by two scalar fields --- except for the scalar field itself. We denote $\phi^{\lambda}$ and $\phi^{-\lambda}$, the helicity $+\lambda$ and $-\lambda$ scalar fields\footnote{In the light-front literature, the helicity is more commonly denoted by $\lambda$. We will adopt this notation from now on.}, respectively. As we have seen, this can be achieved starting from a covariant formulation, such as the Fronsdal action, and then going to the light-cone gauge. However, we must remember that the light-front formulation exists on its own. We then start from the free action.

The free action for a generic helicity $\lambda$ massless field is (here, it is convenient to abbreviate $\phi\equiv \phi^{+s}$ and $\bar{\phi}\equiv\phi^{-s}$)
\begin{equation}\label{freeAction}
    S_2\equiv \int d^4x L_2=-\tfrac{1}{2}\int\partial_{\mu}\bar\phi\partial^{\mu}\phi\, d^4x=-\int\left(\partial^+\bar\phi\dot{\phi}+ \pl^+\phi\dot{\bar{\phi}}+V(\phi)\right)d^4x\,,
\end{equation}
where $V(\phi)=(\partial_a\bar{\phi}\partial^a\phi)$. 
Since the Lagrangian is linear in the velocities, we cannot solve for the momenta $\Pi$ and $\bar\Pi$. At this point, an even simpler model is to take a single real scalar field
\begin{equation}\label{freeActionA}
    S_2\equiv \int d^4x L_2=-\tfrac12\int\partial_{\mu}\phi\partial^{\mu}\phi\, d^4x=-\int\left(\partial^+\phi\dot{\phi}+V(\phi)\right)d^4x\,,
\end{equation}
where $V(\phi)=\tfrac12(\partial_a\phi\partial^a\phi)$. Since we cannot solve for the velocity $\dot\phi$, we find a primary constraint:
\begin{align}
    \Phi(x)&= \Pi(x)- \frac{\delta L_2}{\delta (\dot\phi(x))}=\Pi(x)+\pl^+{\phi}(x)\,,
\end{align}
and the ``naive'' canonical Poisson brackets are then 
\begin{equation}
    [\phi(x,x^+),\Pi(y,y^+)]_{x^+=y^+}=\delta^3(x-y)\,,
\end{equation}
where we have defined $x=\{x^-,x,\bar{x}\}$. The (extended) Hamiltonian is given by 
\begin{equation}
    H_2=\int d^3x \, [V(\phi) + u(x)\Phi(x)]\,,
\end{equation}
where one needs to integrate over the equal-time $x^+$ hypersurfaces. In fact, we do not have a single constraint, but infinitely many of them, parameterized by the space point $x$. The (now in plural) constraints are second class 
\begin{align}
    [\Phi(x),\Phi(y)]=[\Pi(x), \pl^+ \phi(y)]+[\pl^+\phi(x),\Pi(y)]&= 2\pl^+_x \delta^3(x-y)\,.
\end{align}
The factor of two above is crucial! Basically, the constraints ``want'' us to identify $\Pi$ with $-\pl^+ {\phi}$. A way to do it is to reduce everything to the constraint surface and to compute the Dirac bracket. For a general system of second class constraints $\Phi_\alpha$ one has
\begin{align}
    [F,G]_D&=[F,G]-[F,\Phi_\alpha ] C^{\alpha,\beta}[\Phi_\beta,G]\,,
\end{align}
where $C^{\alpha,\beta}$ is the inverse of $[\Phi_\alpha,\Phi_\beta]$. In our case, the inverse has the kernel $\tfrac12 (\pl^+)^{-1} \delta(x-y)$, with $(\pl^+)^{-1}$ understood as discussed above. The net effect is that the Dirac bracket differs from the naive one by a factor of two
\begin{equation}
    [\phi(x,x^+),\Pi(y,y^+)]_{x^+=y^+}=\tfrac12\delta^3(x-y)\,,
\end{equation}
and we can interpret $\Phi$ as $-\pl^+ \phi$. Curiously enough, one finds also
\begin{equation}
[\phi(x,x^+),\phi(y,y^+)]_{x^+=y^+}=\frac1{2\pl^+}\delta^3(x-y)\,,
\end{equation}
while in the initial system $\phi(x)$ were commuting to zero, of course. Coming back to the higher-spin fields, we find
\begin{align}
    [\partial^+_x\phi^{\lambda}(x,x^+),\phi^{s}(y,y^+)]_{x^+=y^+}=\frac{1}{2}\delta^{\lambda,-s}\delta^3(x-y)\,,
\end{align}
or, equivalently
\begin{align}
    [\phi^{\lambda}(x,x^+),\phi^{s}(y,y^+)]_{x^+=y^+}=\frac{1}{\partial^+_x-\partial^+_y}\delta^{\lambda,-s}\delta^3(x-y)\,.
\end{align}
As standard in the Hamiltonian formulation, the Dirac bracket above gives the equation of motion, and then the time evolution of the system via
\begin{equation}
    \dot{F}(\phi)\equiv\partial^-F(\phi)=[F(\phi),H_2]\,.
\end{equation}
In particular, for $F(\phi)=\phi^{\lambda}(x)$ we have 
\begin{equation}
    \partial^-\phi^{\lambda}(x)=[\phi^{\lambda}(x),H_2]=-\frac{\partial\bar{\partial}}{\partial^+}\phi^{\lambda}(x)\,.
\end{equation}
The Poisson bracket with the factor of $\tfrac12$ just derived will be used many times in the text below.

\paragraph{Free field realisations.} At this point, we can find a field realisation of the Poincaré algebra on the ``scalar'' field $\phi^{\lambda}$, see e.g. \cite{Siegel:1999ew} for an excellent review of many aspects of the light-cone formulation. The standard linear free field realisation of the Poincaré algebra \eqref{Poincare} is
\begin{align}\label{FreeRealisation1}
    &[x^{\mu},\partial^{\nu}]=-\eta^{{\mu}{\nu}}\,,&
    &P^{\mu}\cdot\phi^{\lambda}=\partial^{\mu}\phi^{\lambda}\,,&
    &J^{{\mu}{\nu}}\cdot\phi^{\lambda}=(x^{\mu}\partial^{\nu}-x^{\nu}\partial^{\mu}+S^{\mu\nu})\phi^{\lambda}\,,
\end{align}
where $S^{\mu\nu}$ is the spin part of the angular momentum. It is straightforward to check that this realisation does satisfy the Poincaré algebra. In the light-cone gauge, we require
\begin{align}\label{FreeRealisation2}
    &S^{+\mu}\cdot\phi^{\lambda}=0\,,&
    &S^{\mu\nu}\cdot\partial_{\mu}\phi^{\lambda}=0\,,
\end{align}
where the first condition tells us that the only non-zero components of $S^{\mu\nu}$ are $S^{x\bar{x}}$, $S^{x-}$, and $S^{\bar{x}-}$. The second condition tells us that we can express the components of $S^{x-}$ and $S^{\bar{x}-}$ in terms of $S^{x\bar{x}}$ as
\begin{align}\label{FreeRealisation3}
    &S^{x-}\cdot\phi^{\lambda}=-S^{x\bar{x}}\cdot\frac{\partial}{\partial^+}\phi^{\lambda}\,,&
    &S^{\bar{x}-}\cdot\phi^{\lambda}=S^{x\bar{x}}\cdot\frac{\bar{\partial}}{\partial^+}\phi^{\lambda}\,.
\end{align}
Then, the helicity representation is specified by the action of $S^{x\bar{x}}$ generating the Wigner little group
\begin{equation}\label{FreeRealisation4}
    S^{x\bar{x}}\cdot\phi^{\lambda}=-\lambda\phi^{\lambda}\,,
\end{equation}
where $\lambda$ is the helicity of the field. There is a formal way to prove that $S^{x\bar{x}}$ does really correspond to the helicity operator and then $\lambda$ to the helicity, defined as the proportionality coefficients between the Pauli-Lubanski pseudovector $W^{\mu}$ and the momenta $P^{\mu}$; see Appendix $B$ of \cite{Ponomarev:2022vjb} for a clear explanation. Notice that the fact that the field transforms non-trivially under $S^{x\bar{x}}$ is what distinguishes a scalar field from a spinning field. We can go through the same argument starting from the complex conjugate field $\phi^{-\lambda}$.

The free action \eqref{freeAction} is Poincaré invariant, by checking it is invariant under the transformations \eqref{FreeRealisation1}. We can read the Noether current associated with them:
\begin{align}
    &P^{\mu}:&
    &T^{{\mu},\nu}=\sum_{\lambda}\frac{\delta L_2}{\delta(\partial_{\nu}\phi^{\lambda})}\partial^{\mu}\phi^{\lambda}-\eta^{\mu\nu}L_2\,,\\
    &J^{\mu\nu}:&
    &L^{\mu\nu,\rho}=x^{\mu}T^{\nu,\rho}-x^{\nu}T^{\mu,\rho}+R^{\mu\nu,\rho}\,,
\end{align}
where $T^{\mu,\nu}$ is the energy-momentum tensor and $L^{\mu\nu,\rho}$ the angular-momentum tensor, with $R^{\mu\nu,\rho}$ the spin part
\begin{equation}
    R^{\mu\nu,\rho}=\sum_{\lambda}\frac{\delta L_2}{\delta(\partial_{\rho}\phi^{\lambda})}S^{\mu\nu}\phi^{\lambda}\,.
\end{equation}
The Noether charges can be found by integrating over a fixed-time hypersurface as
\begin{align}
    &P^{\mu}_2=\int d^3x T^{\mu,+}\,,&
    &J_2^{\mu\nu}=\int d^3x L^{\mu\nu,+}\,,
\end{align}
and explicitly, they read as
\begin{align}\label{Noether_currents}
    &P^{\mu}_2=-\sum_{\lambda}\int d^3x \partial^+\phi^{-\lambda}p^{\mu}_2\phi^{\lambda}\,,&
    &J_2^{\mu\nu}=-\sum_{\lambda}\int d^3x \partial^+\phi^{-\lambda}j^{\mu\nu}_2\phi^{\lambda}\,.
\end{align}
The terms $p^{\mu}_2$ and $j^{\mu\nu}_2$ above correspond to the free field realisation \eqref{FreeRealisation1} of the Poincaré algebra rewritten in light-cone coordinates. The kinematical operators are
\begin{subequations}
\begin{align}
    &p_2^+=\partial^+\,,&
    &p_2=\partial\,,&
    &\bar{p}_2=\bar{\partial}\,,\\
    &j_2^{x+}=x\partial^+\,,&
    &j_2^{\bar{x}+}=\bar{x}\partial^+\,,&
    &j_2^{-+}=-x^-\partial^+\,,\\
    &j_2^{x\bar{x}}=x\bar{\partial}+\bar{x}\partial-\lambda\,,
\end{align}
\end{subequations}
and the dynamical generators are
\begin{align}
    &p_2^-=h_2=-\frac{\partial\bar{\partial}}{\partial^+}\,,&
    &j_2^{x-}=-x\frac{\partial\bar{\partial}}{\partial^+}-x^-\partial+\lambda\frac{\partial}{\partial^+}\,,&
    &j_2^{\bar{x}-}=-\bar{x}\frac{\partial\bar{\partial}}{\partial^+}-x^-\bar{\partial}-\lambda\frac{\bar{\partial}}{\partial^+}\,.
\end{align}
As it should be, the charges generate the algebra via the commutator (still the Poisson bracket, in fact):
\begin{align}
    &[\phi^{\lambda},P^{\mu}_2]=p^{\mu}_2\phi^{\lambda}\,,&
    &[\phi^{\lambda},J^{\mu\nu}_2]=j^{\mu\nu}_2\phi^{\lambda}\,,
\end{align}
where for $p^-_2=h_2$ and $P^-_2=H_2$ we correctly reproduce the Hamiltonian evolution.

\paragraph{Momentum space.} In the core of the thesis, we will work with fields in the momentum space. We introduce here the notation and conventions for the Fourier transformations with respect to $x^-$ and the transverse directions $x$ and $\bar{x}$ as
{\allowdisplaybreaks
\begin{align}
    \phi(x,x^+)&=\frac{1}{(2\pi)^{\frac{3}{2}}}\int \phi(p,x^+)e^{i(x^-p^++x\cdot p)}d^3 p\,,\\
    \phi(p,x^+)&=\frac{1}{(2\pi)^{\frac{3}{2}}}\int \phi(x,x^+)e^{-i(x^-p^++p\cdot x)}d^3 x\,,
\end{align}}%
\noindent followed by a change of variables $p=iq$, in order to avoid complex $i$ factors around. Effectively, the Fourier transform amounts to the replacement
\begin{align}
    &\frac{\partial}{\partial x^{\mu}}\rightarrow q_{\mu}\,,&
    &x^{\mu}\rightarrow -\frac{\partial}{\partial q_{\mu}}\,,&
\end{align}
and by analogy with the coordinate space, we denote $q=\{q^+,q,\bar{q}\}$. We also adopt the notation
\begin{equation}
    \phi_q^{\lambda}\equiv\phi^{\lambda}(q^+,q,\bar{q},x^+=0)\,. 
\end{equation}
The free field realisation of the kinematical Poincaré generators becomes\footnote{Following standard notations in the light-cone, we rename $\beta=p^+$.}
\begin{subequations}
\begin{align}
    &p_2^+=\beta\,,&
    &p_2=q\,,&
    &\bar{p}_2=\bar{q}\,,\\
    &j_2^{x+}=-\beta\frac{\partial}{\partial \bar{q}}\,,&
    &j_2^{\bar{x}+}=-\beta\frac{\partial}{\partial q}\,,&
    &j_2^{-+}=-N_{\beta}-1=-\frac{\partial}{\partial\beta}\beta\,,\\
    &j_2^{x\bar{x}}=N_q-N_{\bar{q}}-\lambda\,,
\end{align}
\end{subequations}
where $N_q=q\partial_q$ and $N_{\bar{q}}=\bar{q}\partial_{\bar{q}}$ are the Euler operators.
The dynamical generators are
\begin{align}\label{dynamical_genera}
    &h_2=-\frac{q\bar{q}}{\beta}\,,&
    &j_2^{x-}=\frac{\partial}{\partial\bar{q}}\frac{q\bar{q}}{\beta}+q\frac{\partial}{\partial\beta}+\lambda\frac{q}{\beta}\,,&
    &j_2^{\bar{x}-}=\frac{\partial}{\partial q}\frac{q\bar{q}}{\beta}+\bar{q}\frac{\partial}{\partial\beta}-\lambda\frac{\bar{q}}{\beta}\,.
\end{align}
The equal-time commutator in terms of the Fourier transformed fields becomes 
\begin{align}\label{DiracBracket}
    [\phi_q^{\lambda},\phi_p^{s}]=\delta^{\lambda,-s}\frac{\delta^3(q+p)}{2q^+}\,,
\end{align}
and the Noether charges, by the Fourier transform of \eqref{Noether_currents}, take the form 
\begin{align}
    &P^{\mu}_2=-\sum_{\lambda}\int d^3q\,d^3p\,\delta^3(q+p)p^+\phi_p^{-\lambda}p^{\mu}_2(q,\partial_q)\phi_q^{\lambda}\,,\\
    &J_2^{\mu\nu}=-\sum_{\lambda}\int d^3q\,d^3p\,\delta^3(q+p)p^+\phi_p^{-\lambda}j^{\mu\nu}_2(q,\partial_q)\phi_q^{\lambda}\,,
\end{align}
where once we solve for the delta function $\delta^3(q+p)$ we get
\begin{equation}\label{charges}
    Q_{\xi}=\int d^3q\, q^+\phi^{-\lambda}_{-q}\mathcal{O}_{\xi}(q,\partial_q)\phi^{\lambda}_q\,.
\end{equation}
This is the general form of the Poincaré charges, where $\xi$ is the generator of the Poincaré algebra associated with a Killing vector $\xi$. For $\xi^\mu=a^{\mu}$ and $\xi^\mu=b^{\mu\nu}x_\nu$, we recover the standard free charges.

The Poincaré algebra is realised via the commutator in the usual way
\begin{equation}
    \delta_{\xi}\phi_q^{\lambda}=[\phi_q^{\lambda},Q_{\xi}]\,.
\end{equation}
Notice that the action of the free charges differs from \eqref{FreeRealisation1}-\eqref{FreeRealisation4}. This is the case because we realised the symmetries on the hypersurface at $x^+=0$, then the $x^+$ dependence disappears. The action of a generator realised on an equal-time surface with $x^+\neq 0$  is related to its realization at $x^+=0$ through the Hamiltonian equations of motion.

Notice that due to the nontrivial integration measure $q^+$ in \eqref{charges}, the conjugate operators are defined as
\begin{equation}
    \mathcal{O}^{\dagger}=-\frac{1}{q^+}\mathcal{O}^T(-q)q^+\,,
\end{equation}
where the transposed operator is defined as usual via integration by parts, e.g. $q^T=q$, $\partial_q^T=-\partial_q$. In particular, the generators of the Poincaré algebra are Hermitian $\mathcal{O}^{\dagger}=\mathcal{O}$, like $p^{\dagger}=p$. Using the Dirac bracket \eqref{DiracBracket} and 
\begin{equation}
    \delta_{\xi}\phi_q^{\lambda}=\frac{1}{2}\mathcal{O}_{\xi}\phi_q^{\lambda}+\frac{1}{2}\mathcal{O}^{\dagger}_{\xi}\phi_q^{\lambda}=\mathcal{O}_{\xi}\phi_q^{\lambda}\,,
\end{equation}
for Hermitian operators, one can verify that the following is satisfied:
\begin{align}
    &[Q_{\xi},Q_{\eta}]=Q_{[\xi,\eta]}\,,&
    &[\delta_{\xi},\delta_{\eta}]=\delta_{[\xi,\eta]}\,.
\end{align}
This relation implies that the algebra satisfied by the charges is the same as the one realised by the symmetry generators (i.e. the Killing vectors $\xi$).
A useful formula that is used to compute the constraints is
\begin{equation}
    [F(\phi),Q_{\xi}]=\int dv\,\mathcal{O}_{\xi}(v)\phi^{\lambda}_v\frac{\partial}{\partial\phi^{\lambda}_v}F(\phi)\,.
\end{equation}
This section provided the essential toolkit needed to understand, compute, and analyse both the kinematical and dynamical constraints. However, we will not solve these constraints explicitly here. Instead, our focus will be on examining the results and offering a series of observations. For a detailed treatment of the explicit solutions, we refer the reader to \cite{Ponomarev:2016lrm,Ponomarev:2017nrr,Ponomarev:2022vjb}, as well as to the introduction of Chapter 4.

\paragraph{Deformation procedure.} To start the light-front deformation procedure, we need a general ansatz for the dynamical generators at the interacting level. The three dynamical generators take the form
\begin{align}
    &H=P^-=H_2+\sum_n H_n\,,&
    &J^{x-}=J^{x-}_2+\sum_n J^{x-}_n\,,&
    &J^{\bar{x}-}=J^{\bar{x}-}_2+\sum_n J^{\bar{x}-}_n\,,&
\end{align}
where 
\begin{align}\label{hamiltonian_intro}
    H_n=&\,\sum_{\lambda_i}\int d^{3n}q\;\delta^3\Big(\sum_i q_i\Big)h^{q_1,...,q_n}_{\lambda_1,...,\lambda_n}\phi^{\lambda_1}_{q_1}\cdots\phi^{\lambda_n}_{q_n}\,,\\\label{boostz_intro}
    J^{x-}_n=&\,\sum_{\lambda_i}\int d^{3n}q\;\delta^3\Big(\sum_i q_i\Big)\Big[j^{q_1,...,q_n}_{\lambda_1,...,\lambda_n}-\frac{1}{n}\,h^{q_1,...,q_n}_{\lambda_1,...,\lambda_n}\Big(\sum_j\frac{\partial}{\partial \bar{q}_j}\Big)\Big]\phi^{\lambda_1}_{q_1}\cdots\phi^{\lambda_n}_{q_n}\,,\\ \label{boostzbar_intro}
    J^{\bar{x}-}_n=&\,\sum_{\lambda_i}\int d^{3n}q\;\delta^3\Big(\sum_i q_i\Big)\Big[\bar{j}^{q_1,...,q_n}_{\lambda_1,...,\lambda_n}-\frac{1}{n}\,h^{q_1,...,q_n}_{\lambda_1,...,\lambda_n}\Big(\sum_j\frac{\partial}{\partial q_j}\Big)\Big]\phi^{\lambda_1}_{q_1}\cdots\phi^{\lambda_n}_{q_n}\,.
\end{align}
To simplify notation, we denote the densities\footnote{We call them densities because we integrate them.} as
\begin{align}
    &h_n=h^{q_1,...,q_n}_{\lambda_1,...,\lambda_n}\,,&
    &j_n=j^{q_1,...,q_n}_{\lambda_1,...,\lambda_n}\,,&
    &\bar{j}_n=\bar{j}^{q_1,...,q_n}_{\lambda_1,...,\lambda_n}\,.
\end{align}
All the effort is now devoted to constraining --- and, when possible, determining --- the precise structure of these densities.

Let us now make a few remarks concerning this specific choice of generators:
\begin{itemize}
    \item The delta function imposes the conservation of the total momenta $q^+$ and transverse momenta $q$ and $\bar{q}$ as a consequence of the translation invariance imposed by $P^+$ and $P^A$. 
    
    \item The operators $h_n$, $j_n$ and $\bar{j}_n$ do all depend on transversal momenta $q_i,\bar{q_i}$ and $q_i^+$, but not on the time $x^+$ and, hence, its momenta $q_i^-$ since we need to construct generators at $x^+=0$ only. 
    
    \item For cubic vertices, the combination $q_i\bar{q}_i$ can be eliminated via a field redefinition
\begin{align}\label{redefintion_cubic}
    &h_2(q_i)\equiv -\frac{q_i\bar{q}_i}{\beta_i}\,.
\end{align}
At the cubic level, a field redefinition is closely related to going on-shell (i.e. setting $H_2=0$). However, this equivalence no longer holds at higher orders. For a detailed discussion of field redefinitions within the Hamiltonian light-front framework, see Appendix B of \cite{Metsaev:2005ar}.

\item The explicit $h_n$ correction to both $J^{x-}$ and $J^{\bar{x}-}$ is a trick which slightly simplifies the remaining kinematical constraints for $j_n$ and $\bar{j}_n$.

\item The form of the three densities $h_n$, $j_n$, and $\bar{j}_n$ is not only tightly constrained by both kinematical and dynamical constraints, but, most importantly, is constrained by locality. The locality requirement rules out any terms involving $\frac{1}{q}$ or/and $\frac{1}{\bar{q}}$, which would introduce genuine non-localities. However, as previously noted, terms like $\frac{1}{\beta}$ are allowed, since they naturally arise from solving the equations of motion and do not signal physical non-locality. Furthermore, permitting more severe forms of non-locality --- such as dividing by the free Hamiltonian $\frac{1}{H_2}$ --- would trivialise the deformation procedure, in the sense that any interaction would appear to be consistent. 
\end{itemize}

We can rewrite the dynamical constraints \eqref{dynamical_constraints} as
\begin{align}
    &[J^{a-},J^{c-}]=0&
    &\iff&
    &[J^{a-}_2,J^{c-}_n]-[J^{c-}_2,J^{a-}_n]=\sum_{\substack{i,j>2\\i+j=n+2}}[J^{c-}_i,J^{a-}_j]\,,\\\label{dynamical_constraints2}
    &[J^{a-},H_2]=0&
    &\iff&
    &[J^{a-}_2,H_n]-[H_2,J^{a-}_n]=\sum_{\substack{i,j>2\\i+j=n+2}}[H_i,J^{a-}_j]\,.
\end{align}
Interestingly, in Appendix B of \cite{Ponomarev:2016lrm}, it was proved that the $[J^{a-},J^{c-}]=0$ constraint is automatically satisfied provided that the $[J^{a-},H_2]=0$ constraint is solved and $J^{a-}$ is solved for in terms of $H$ at any given order.

\paragraph{Kinematical constraints.} These constraints can be solved exactly. Those involving $P^+$ and $P^A$ are solved by the delta functions enforcing momentum conservation. The analysis of the remaining kinematical constraints can be found in \cite{Ponomarev:2016lrm}. The outcome is that all densities must depend on specific combinations of momenta, denoted $\PP_{ij}$, defined as
\begin{align}
    &\PP_{ij}=q_i\beta_j-q_j\beta_i\,,&
    &\PPb_{ij}=\bar{q}_i\beta_j-\bar{q}_j\beta_i\,,
\end{align}
where $\PPb_{ij}=-\PPb_{ji}$ and $\PP_{ij}=-\PP_{ji}$. In four dimensions, there are $N-2$ such independent variables for an $N$-point function. Once these $N-2$ independent variables are chosen, the kinematical constraints require that the densities be expressed in the following form:
\begin{align}
    h_n&=h_n(\PP_{ij},\PPb_{ij},\beta_k)\,,\\
    j_n&=j_n(\PP_{ij},\PPb_{ij},\beta_k)\,,\qquad \text{same for }\bar{j}_n\,.
\end{align}
We also find the following homogeneity conditions
\begin{align}
    &h_n:&
    &\#\PPb-\#\PP=\sum_k\lambda_k\,,&
    &\#\PPb+\#\PP=-\#\beta\,,\\
    &j_n:&
    &\#\PPb-\#\PP=\sum_k\lambda_k-1\,,&
    &\#\PPb+\#\PP=-\#\beta\,,\\
    &\bar{j}_n:&
    &\#\PPb-\#\PP=\sum_k\lambda_k+1\,,&
    &\#\PPb+\#\PP=-\#\beta\,.
\end{align}

\paragraph{Cubic Dynamical constraints.} The cubic dynamical constraints determine the form of the cubic vertices, up to an overall coefficient --- the coupling constant. For three-point vertices, with $N=3$, there is only one independent variable $\PP$ and one $\PPb$. In particular, after imposing momentum conservation, we find
\begin{align}
    &\PP_{12}=\PP_{23}=\PP_{31}=\PP=\frac{1}{3}\,\Big[(\beta_1-\beta_2)q_3+(\beta_2-\beta_3)q_1+(\beta_3-\beta_1)q_2\Big]\,,
\end{align}
and the same for $\PPb$. Both $\PP$ and $\PPb$ are cyclic invariant, then $\sigma_{123}\PP=\PP$, and $\sigma_{123}\PPb=\PPb$.

The restriction to the cubic order of the dynamical constraint \eqref{dynamical_constraints2} is
\begin{equation}\label{cubic_constraint}
    [H,J^{a-}]\Big|_3=[H_3,J^{a-}_2]-[J^{a-}_3,H_2]=0\,.
\end{equation}
This can be solved, and leads to the classification of cubic vertices in the light-cone gauge:
\begin{align}\label{cubic_vertices}
h_{\lambda_1,\lambda_2,\lambda_3}=&\,C^{\lambda_1,\lambda_2,\lambda_3}\frac{\PPb^{\lambda_{123}}}{\beta_1^{\lambda_1}\beta_2^{\lambda_2}\beta_3^{\lambda_3}}+\bar{C}^{-\lambda_1,-\lambda_2,-\lambda_3}\frac{\PP^{-\lambda_{123}}}{\beta_1^{-\lambda_1}\beta_2^{-\lambda_2}\beta_3^{-\lambda_3}}\,,\\\label{cubic_vertices_j}
    j_{\lambda_1,\lambda_2,\lambda_3}=&\,\frac{2}{3}\,C^{\lambda_1,\lambda_2,\lambda_3}\frac{\PPb^{\lambda_{123}-1}}{\beta_1^{\lambda_1}\beta_2^{\lambda_2}\beta_3^{\lambda_3}}\Lambda^{\lambda_1,\lambda_2,\lambda_3}\,,\\\label{cubic_vertices_jbar}
    \bar{j}_{\lambda_1,\lambda_2,\lambda_3}=&\,-\frac{2}{3}\,\bar{C}^{-\lambda_1,-\lambda_2,-\lambda_3}\frac{\PP^{-\lambda_{123}-1}}{\beta_1^{-\lambda_1}\beta_2^{-\lambda_2}\beta_3^{-\lambda_3}}\Lambda^{\lambda_1,\lambda_2,\lambda_3}\,,
\end{align}
with $\lambda_{123}=\lambda_1+\lambda_2+\lambda_3$ and we defined
\begin{align}
    &\Lambda^{\lambda_1,\lambda_2,\lambda_3}=\,\beta_1(\lambda_2-\lambda_3)+\beta_2(\lambda_3-\lambda_1)+\beta_3(\lambda_1-\lambda_2)\,,
\end{align}
while $C$ and $\bar{C}$ are the holomorphic and anti-holomorphic coupling constants, respectively. Some observations are in order:
\begin{itemize}
    \item An interesting feature of the cubic interactions, which is hard to see in any covariant description, is that the vertices split into holomorphic, $\PP$-dependent, and anti-holomorphic, $\PPb$-dependent. Indeed, field redefinition at the cubic level \eqref{redefintion_cubic} allows one to eliminate any power of $\PP\PPb\sim H_2$ (via an appropriate canonical transformation) but not $\PP$ or $\PPb$ separately. Covariant descriptions usually lead to some $\PP\PPb$-terms by default. This split is fundamental for identifying the chiral higher-spin theory, as it isolates the two distinct chiral sectors of the cubic vertices.
    
    \item The dynamical information about the theory is encoded in the coupling constants $C^{\lambda_1,\lambda_2,\lambda_3}$ and $\bar{C}^{-\lambda_1,-\lambda_2,-\lambda_3}$. Note that $\lambda_1+\lambda_2+\lambda_3>0$ for $C^{\lambda_1,\lambda_2,\lambda_3}$ unless $\lambda_i=0$ and $\lambda_1+\lambda_2+\lambda_3<0$ for $\bar{C}^{-\lambda_1,-\lambda_2,-\lambda_3}$ unless $\lambda_i=0$\footnote{These constraints on the helicities arise from the light-front locality condition, which forbids inverse powers of $q$ and $\bar{q}$. In terms of $\PP$ and $\PPb$, this translates into the requirement that only non-negative powers of $\PP$ and $\PPb$ are allowed.}. There is a unique $\lambda_i=0$ vertex, which corresponds to $(\phi^0)^3$. It is a matter of choice where to put this scalar cubic self-interaction, into $C$ or $\bar{C}$. 
    
    \item To have the right dimension, we should set 
\begin{align}
    C^{\lambda_1,\lambda_2,\lambda_3}&= (\ell_P)^{\lambda_1+\lambda_2+\lambda_3-1} c^{\lambda_1,\lambda_2,\lambda_3}\,,
\end{align}
where $c$ are dimensionless and $\ell_P$ is some constant of length. Any choice of $c$ leads to a theory that is consistent up to the cubic order. At this order, all interactions do not feel each other's presence since the constraint \eqref{cubic_constraint} is a linear equation with respect to $H_3$, $J_3$, i.e. any linear combination of solutions is a solution again.
\item Let us notice that even if in \eqref{cubic_vertices}-\eqref{cubic_vertices_jbar} we displayed both the value of the Hamiltonian and the boost densities, the latter are not independent from the former, and we can always find $j_n$ and $\bar{j}_n$ starting from $h_n$. Conceptually, this is clear; the theory must be completely specified by the form of the Hamiltonian $H$, and the boosts $J^{x-}$ and $J^{\bar{x}-}$ must follow from it.\footnote{The procedure to extract $j_n$ and $\bar{j}_n$ once $h_n$ is known is described in section 3.1 of \cite{Ponomarev:2016cwi}.} 
\end{itemize}

\paragraph{Light-front vs covariant cubic vertices.} The classification of light-front cubic vertices is given in equation ~\eqref{cubic_vertices}. A parallel classification exists in covariant approaches; for our purposes, we refer the reader to \cite{Bekaert:2010hp,Boulanger:2008tg}. The literature on the subject is extensive --- for a slightly broader overview, see also the references in \cite{Ponomarev:2016lrm}. In the covariant (Fronsdal) formalism, one generically has two cubic vertices (this holds specifically in $4d$) for any triplet of spins in $4d$.  As we can see from \eqref{cubic_vertices}, there exists a specific vertex $h_{\lambda_1,\lambda_2,\lambda_3}$ for every possible combination of helicities $(\lambda_1,\lambda_2,\lambda_3)$. Therefore, the two classifications do not coincide: the light-front approach allows for a broader set of vertices. This mismatch can be attributed to two reasons. First, in covariant approaches, the existence of interactions heavily depends on the field content; for example, the local electromagnetic interaction mediated by $A_{\mu}$ would look non-local in terms of the field strength $F_{\mu\nu}$, i.e. we cannot realize the minimal electromagnetic interaction with $F_{\mu\nu}$ provided the locality is under control. Second, the notion of locality between the two formalisms can differ. Indeed, some light-front vertices can be written in covariant form only at the cost of introducing some non-localities \cite{Conde:2016izb}.

To clearly illustrate the difference between the two approaches, we can list, for each triplet of spins\footnote{By spin we refer to the modulus of the helicity, $s=|\lambda|$. Hence, $s$ is always positive, with $\lambda=\pm s$.} $(s_1,s_2,s_3)$, the possible number of interaction vertices in both frameworks. In the light-front approach, one can distinguish between different types of vertices using the total helicity sum $|\lambda_1+\lambda_2+\lambda_3|$ , which corresponds to the total power of $\PP$ or $\PPb$. In coordinate space, this translates to the total number of derivatives appearing in the vertex (whenever it has a realization via the Fronsdal field). Instead, in the covariant approach, we can write the total number of derivatives. In the light-front approach, we have the following independent vertices with $\PPb$
\begin{align}
    \begin{split}
    &(+s_1,+s_2,+s_3)\,,\qquad
    (+s_1,+s_2,-s_3)\,,\qquad
    (+s_1,-s_2,+s_3)\,,\\
    &(-s_1,+s_2,+s_3)\,,\qquad
    (+s_1,-s_2,-s_3)\,,
    \end{split}
\end{align}
and similarly for the $\PP$-vertices, where all signs have to be flipped. Let us order spins as $s_1\geq s_2\geq s_3$. It is important that the last $(+--)$-configuration is only available when $s_1>s_2+s_3$, while the second to last is when $s_1<s_2+s_3$. Therefore, we always have four $\PPb$-vertices. These are, however, not real and are not parity invariant on their own. Remember that parity has the following effect
\begin{align}
    &P(\phi^{\lambda})=\phi^{-\lambda}\,,&
    &P(C^{\lambda_i})=C^{-\lambda_i}\,,&
    &P(\PP)=\PPb\,.
\end{align}
Therefore, to have parity invariant vertices, the independent vertices with opposite helicities must both be present and with the same coupling constant. This yields $4$ parity invariant vertices.

In contrast, in the covariant formulation, there exist only two vertices. With $s_1\geq s_2\geq s_3$ we find the following helicity configurations:
\begin{align}
    &(+s_1,+s_2,+s_3)\oplus(-s_1,-s_2,-s_3) \,,&
    &(+s_1,+s_2,-s_3)\oplus (-s_1,-s_2,+s_3)\,.
\end{align}
For a more detailed account of the various types of vertices --- such as current interactions, abelian vertices, and, most notably, the non-abelian ones --- see \cite{Boulanger:2008tg,Ponomarev:2016lrm}. In particular, one can see the two-derivative gravitational interactions corresponding to $(+s,-s,+2)$ and $(-s,+s,-2)$ helicity configurations. 

Let us also point out that the classification of the vertices coincides with the one that can be quite easily obtained in the spinor-helicity formalism. However, there is an important distinction. In the light-front approach, the vertices \eqref{cubic_vertices} are actual densities, used to construct the interacting Hamiltonian of the system. As such, they carry off-shell information --- something not accessible in the purely on-shell spinor-helicity framework \cite{Benincasa:2007xk,Benincasa:2011pg}. This justifies the more involved procedure required to derive them.

\paragraph{Quartic-order, Metsaev solution and Chiral higher-spin theory.} The first challenging constraint comes at the quartic order. It is at the quartic order that the spectrum of a theory gets fixed, as well as all the cubic couplings. The constraint \eqref{dynamical_constraints2} at the quartic order is
\begin{equation}\label{quartic_constraint}
    [H,J^{a-}]\Big|_4=[H_4,J^{a-}_2]+[H_3,J^{a-}_3]-[J^{a-}_4,H_2]=0\,.
\end{equation}
A remarkable property of \eqref{quartic_constraint} observed in \cite{Metsaev:1991mt} is that the chiral sectors receive no contribution from $H_4$ and $J^{a-}_4$.\footnote{This becomes clear by analyzing the homogeneity degree in $q$ and $\bar{q}$ of the various subsectors of the commutators, and observing that the chiral sectors decouple and must be consistent independently of the quartic generators.} In other words, there are two closed subsystems of equations, for $CC$ and for $\bar C\bar C$, that are independent of higher orders:
\begin{align}
     &[H_3(\PPb),J^{x-}_3]=0\,,&
    &[H_3(\PP),J^{\bar{x}-}_3]=0\,.
\end{align}
These equations are very constraining, especially for higher-spin fields. Solving the holomorphic ($CC$) or anti-holomorphic ($\bar C\bar C$) constraint allows one to find the allowed spectrum and cubic vertices and to fix all the couplings $C^{\lambda_1,\lambda_2,\lambda_3}$. This is the Metsaev solution, originally discovered by Metsaev in \cite{Metsaev:1991mt,Metsaev:1991nb}. However, it was only in \cite{Ponomarev:2016lrm} that it was realized that, by truncating the theory to a single chiral sector --- either holomorphic or anti-holomorphic --- the resulting theory becomes consistent at all orders on its own, without the need for higher-order interaction vertices.

The Metsaev solution completely determines all cubic couplings to take the following form:
\begin{align}
&C^{\lambda_1,\lambda_2,\lambda_3}=\frac{(\ell_p)^{\lambda_{123}-1}}{\Gamma(\lambda_1+\lambda_2+\lambda_3)}\,,&
\bar C^{\lambda_1,\lambda_2,\lambda_3}=\frac{(\bar{\ell}_p)^{\lambda_{123}-1}}{\Gamma(\lambda_1+\lambda_2+\lambda_3)}\,.&
\end{align}
Moreover, if one requires the theory to be unitary and parity invariant, the coupling constants above must satisfy the relation:
\begin{equation}
    \bar C^{\lambda_1,\lambda_2,\lambda_3}=(C^{\lambda_1,\lambda_2,\lambda_3})^*\,.
\end{equation} 
This condition then requires the inclusion of higher-point vertices, in addition to the cubic ones, in order to ensure the consistency of the theory. In the final chapter, we will analyse the full quartic constraint and study the quartic vertices.

Chiral higher-spin theory is defined as the maximal consistent theory that includes all cubic vertices of a single chirality (practically speaking, we mean that it includes only $C$ or only $\bar{C}$ couplings, i.e. the total helicity in every vertex is either positive, for the chiral theory, or negative, for the anti-chiral one) --- either holomorphic or anti-holomorphic. When fields are assumed to transform in a specific representation (which we will discuss in more detail later in the thesis) of a gauge group $G$, all such chiral vertices are activated. In the absence of a gauge group, only the vertices with an even number of derivatives --- i.e. those for which $|\lambda_1+\lambda_2+\lambda_3|$ is even --- are allowed, as a consequence of the symmetry property of the cubic vertices.

\paragraph{Simple actions for chiral theories.} 
Using the cubic vertices \eqref{cubic_vertices} we can write down in a simple way some consistent to all orders cubic actions. The action for SDYM in the light-front, which was first given by Chalmers and Siegel in \cite{Chalmers:1996rq}, is 
\begin{align}\label{SDYM_action}
    \begin{split}
    S=&\int d^3q_1\, d^3q_2\,\delta^3(q_1+q_2)(\phi^{-1}_{q_1})^a\frac{q_2\bar{q}_2}{\beta_2}(\phi^1_{q_2})_a\\
    &+\int d^3q_1\, d^3q_2\, d^3q_3\,\delta^3(q_1+q_2+q_3)gf_{abc}\frac{\PPb_{12}}{\beta_1\beta_2}\beta_3(\phi^1_{q_1})^a(\phi^1_{q_2})^b(\phi^{-1}_{q_3})^c\,,
    \end{split}
\end{align}
where $g\sim C^{1,1,-1}$ is the SDYM coupling constant and $f_{abc}$ is the structure constant of some Lie algebra $\mathfrak{g}$. By varying \eqref{SDYM_action} with respect to $\phi^{-1}$, we obtain the SDYM equations of motion for the helicity $+1$ field in the light-front. Once rewritten in the coordinate representation, we get
\begin{align}
    &\Box (\phi^1(x))_a-2f_{abc}\partial^+\left(\frac{\bar{\partial}}{\partial^+}(\phi^1(x))^b(\phi^1(x))^c\right)=0\,,
\end{align}
where we used the antisymmetry of $f_{abc}$ to simplify the result. 

The action for SDGR in the light-front was also given by Chalmers and Siegel in \cite{Chalmers:1996rq}, is 
\begin{align}\label{SDGR_action}
    \begin{split}
    S=&\int d^3q_1\, d^3q_2\,\delta^3(q_1+q_2)\phi^{-2}_{q_1}\frac{q_2\bar{q}_2}{\beta_2}\phi^2_{q_2}\\
    &+\int d^3q_1\, d^3q_2\, d^3q_3\,\delta^3(q_1+q_2+q_3)g\frac{\PPb_{12}^2}{\beta_1^2\beta_2^2}\beta_3^2\phi^2_{q_1}\phi^2_{q_2}\phi^{-2}_{q_3}\,,
    \end{split}
\end{align}
where $g\sim C^{2,2,-2}$ is the SDGR coupling constant. By varying \eqref{SDGR_action} with respect to $\phi^{-2}$, we obtain the SDGR equations of motion for the helicity $+2$ field in the light-front. Once rewritten in the coordinate representation, we get
\begin{align}
    &\Box \phi^2(x)+2g(\partial^+)^2\left(\frac{\bar{\partial}^2}{(\partial^+)^2}\phi^2(x)\phi^2(x)-\frac{\bar{\partial}}{\partial^+}\phi^2(x)\frac{\bar{\partial}}{\partial^+}\phi^2(x)\right)=0\,.
\end{align}
The action for a general chiral higher-spin theory in the light-front is 
\begin{align}\label{Chiral_action}
    \begin{split}
    S=&\frac{1}{2}\sum_{\lambda}\int d^3q_1\, d^3q_2\,\delta^3(q_1+q_2)\phi^{-\lambda}_{q_1}\frac{q_2\bar{q}_2}{\beta_2}\phi^{\lambda}_{q_2}\\
    &+\sum_{\lambda_1,\lambda_2,\lambda_3}\int d^3q_1\, d^3q_2\, d^3q_3\,\delta^3(q_1+q_2+q_3)C^{\lambda_1,\lambda_2,\lambda_3}\frac{\PPb_{12}^{\lambda_{123}}}{\beta_1^{\lambda_1}\beta_2^{\lambda_2}\beta_3^{\lambda_3}}\phi^{\lambda_1}_{q_1}\phi^{\lambda_2}_{q_2}\phi^{\lambda_3}_{q_3}\,,
    \end{split}
\end{align}
where $\lambda_{123}=\lambda_1+\lambda_2+\lambda_3$ and $C^{\lambda_1,\lambda_2,\lambda_3}$ denote the coupling constants of the theory, which must be properly fixed in the case of a specific chiral higher-spin theory. A more detailed discussion of this point will be provided in the core of the thesis.

\clearpage
\chapter[On classification of (self-dual) higher-spin gravities in flat space]
        {On classification of (self-dual) higher-spin gravities in flat space}
\chaptermark{Self-dual higher-spin theories in flat space}

\vspace{5mm}

\paragraph{Abstract:} There\footnote{The content of this chapter is identical to the content of the paper.} is a great number of higher-spin gravities in $3d$ that can have both finite and infinite spectra of fields and can be formulated as Chern-Simons theories. It was believed that this is impossible in higher dimensions, where higher-spin fields do have propagating degrees of freedom. We show that there are infinitely many higher-spin theories in the $4d$ flat space featuring nontrivial local interactions that can have either a finite or infinite number of fields. We classify all one- and two-derivative (i.e. with gauge and gravitational interactions) higher-spin theories by solving the holomorphic constraint in the light-cone gauge obtained by Metsaev. Therefore, these theories are consistent subsectors of the higher-spin extensions of self-dual Yang-Mills/gravity, which in turn are truncations of the chiral higher-spin gravity.

\clearpage

\section{Introduction}\label{Paper1-section1}
The basic idea of looking for extensions of gravity by massless fields with higher spins dates back to at least as early as \cite{Fronsdal:1978rb}. The first candidate spectrum of a higher-spin theory was proposed in \cite{Flato:1978qz} and contains massless fields of all spins from $0$ to $\infty$. It was obtained via the representation theory of the $AdS_4$ symmetry group, which, in modern terms, corresponds to working out the spectrum of single-trace operators in the free vector model and using the AdS/CFT dictionary to read off the spectrum of the dual theory. It also came with the idea \cite{Flato:1980zk} that interactions of higher-spin fields have something to do with the scattering of dipoles --- bilinears in the free fields --- on the boundary of $AdS_4$, which was a precursor to the modern AdS/CFT duality with vector models on the boundary, see e.g. \cite{Sezgin:2002rt,Klebanov:2002ja,Sezgin:2003pt,Leigh:2003gk,Giombi:2011kc}. 

Despite the above mentioned initial input, where it is $AdS_4$ and not the flat space that played a role, the systematic study of possible interactions had first been started in the light-front approach\footnote{Introduced in a seminal paper by Dirac \cite{Dirac:1949cp}.} by Bengtsson, Bengtsson and Brink \cite{Bengtsson:1983pg,Bengtsson:1983pd,Bengtsson:1986kh} in the flat space with the complete classification of cubic vertices obtained in \cite{Bengtsson:1986kh}. Around the same time, the first covariant cubic vertices had been found in \cite{Berends:1984wp,Berends:1984rq}, where it also became evident that the smallest spectrum of fields to admit consistent interactions has to be infinite and unbounded in spin. One cannot help mentioning numerous no-go theorems, most notably the Weinberg low energy theorem \cite{Weinberg:1964ew} and the Coleman-Mandula theorem \cite{Coleman:1967ad} that seem to shut down the whole higher-spin idea. Analogues of both Weinberg's and Coleman-Mandula theorems can be extended to $AdS_d$ \cite{Maldacena:2011jn,Fitzpatrick:2012cg,Boulanger:2013zza,Alba:2013yda,Alba:2015upa,Sleight:2021iix}.\footnote{The initial impulse towards anti-de Sitter space \cite{Fradkin:1986qy} was triggered by the fact that these analogues had not yet been available and by the encouraging result \cite{Fradkin:1986qy} on the construction of ``gravitational interactions'' of higher-spin fields that avoid the no-go of \cite{Aragone:1979hx}. However, the cubic gravitational interactions in the flat space were also constructed with the help of the light-front approach, the same year, as a part of the complete classification \cite{Bengtsson:1986kh}. As discussed in \cite{Krasnov:2021nsq}, the ``non-existence'' of the gravitational interactions for higher-spin fields in flat space \cite{Aragone:1979hx} is related to a somewhat degenerate nature of the description based on symmetric (Fronsdal) fields and is not an invariant statement about higher-spin fields. In particular, the ``gravitational'' interaction of \cite{Fradkin:1986qy} is a linear combination of several independent (the minimal and nonminimal) vertices in the light-cone gauge. } However, these results mostly indicate that higher-spin theories should be integrable in some sense, in particular, the S-matrix must be trivial or simple enough, rather than telling anything about the (non)-existence of these theories as field theories.

The present understanding of the problem of higher-spin interactions is that it is difficult for the masslessness of higher-spin fields to coexist with the usual requirements of locality within the field theory approach.\footnote{One of the possible applications of higher-spin gravities, apart from the general search for fundamental interactions, is to provide simple models of quantum gravity where the extended symmetry associated with massless fields should eliminate counterterms. One would not expect a quantum gravity model to be a local field theory, which can partially explain why it is difficult to stick to the usual definitions of locality when higher-spin fields are involved. Therefore, the higher-spin problem can also be rephrased as how to deal with nonlocalities that higher-spin fields usually lead to. } This problem is insensitive to the sign of the cosmological constant. However, it takes a rather technical argument to see how the construction of interactions order by order breaks down in the flat space, see e.g. \cite{Bekaert:2010hp,Ponomarev:2017nrr,Roiban:2017iqg}, while it is much easier to arrive at the same conclusion for the negative cosmological constant by using AdS/CFT \cite{Bekaert:2015tva,Maldacena:2015iua,Sleight:2017pcz,Ponomarev:2017qab}, see, however, \cite{Neiman:2023orj}. 

There are a number of well-defined, i.e. local or perturbatively local, higher-spin gravities known to date. In lower dimensions,  $d<4$, the graviton and massless higher-spin fields do not have propagating degrees of freedom, which neutralizes the locality problem. Also, the famous no-go theorems do not apply. For example, there is a great number of (matter-free) higher-spin gravities in $3d$ (purely massless, conformal and partially-massless) that can be formulated with the help of the Chern-Simons theory with a Lie algebra into which the gravitational $sl_2$ subalgebra is embedded in a sufficiently nontrivial way, \cite{Blencowe:1988gj,Bergshoeff:1989ns,Campoleoni:2010zq,Henneaux:2010xg,Grigoriev:2020lzu,Pope:1989vj,Fradkin:1989xt,Grigoriev:2019xmp}. An example of a matter-coupled theory has recently been constructed in \cite{Sharapov:2024euk}. In $d=2$ one can use the BF-theory in a similar way \cite{Alkalaev:2020kut}.

In $d\geq4$ the situation has been more complicated and, for that reason, less rich in models. The first perturbatively local theory has been the conformal higher-spin gravity of Segal and Tseytlin \cite{Segal:2002gd,Tseytlin:2002gz,Bekaert:2010ky, Basile:2022nou}, which is a higher-spin extension of conformal gravity. The local Weyl symmetry keeps interactions perturbatively local. A recent example is chiral higher-spin gravity \cite{Metsaev:1991mt,Metsaev:1991nb,Ponomarev:2016lrm,Ponomarev:2017nrr}, which can be thought of as a higher-spin extension of self-dual Yang-Mills theory and of self-dual gravity at the same time. The theory admits two simple contractions that are higher-spin extensions either of self-dual Yang-Mills (HS-SDYM) or of self-dual gravity (HS-SDGR) \cite{Ponomarev:2017nrr,Krasnov:2021nsq}. Chiral theory and its contractions can also be double-copied \cite{Ponomarev:2024jyg} to give theories with even bigger spectra. Recently, a quasi-chiral theory has also been found \cite{Adamo:2022lah}. All of these theories have an infinite, unbounded in spin, spectrum of fields.\footnote{One can also mention a higher-spin extension of the IKKT matrix model \cite{Sperling:2017dts}, which is a noncommutative field theory with higher-spin fields in the spectrum. Also, one can try to reverse-engineer the AdS/CFT duals of the vector models, see e.g. \cite{deMelloKoch:2018ivk}. } 

Let us stress that the light-front approach has always played an important role in the study of higher-spin interactions, see e.g. \cite{Metsaev:1993ap,Metsaev:2005ar,Metsaev:2007rn,Metsaev:2018xip}.\footnote{To be fair, one should also mention covariant results, e.g. \cite{Boulanger:2008tg,Manvelyan:2010je,Boulanger:2012dx,Francia:2016weg}, see also a very general review \cite{Bekaert:2022poo}. } Its main appealing feature is that all unphysical degrees of freedom are eliminated and one can get unambiguous results concerning (non)existence of certain interactions.\footnote{Any covariant approach picks a certain embedding of the physical degrees of freedom into a Lorentz-covariant field and, depending on the embedding, may miss important types of interactions. For example, the conclusion of \cite{Aragone:1979hx} that there are no gravitational interactions of higher-spin fields relied on a particular choice of a Lorentz-covariant field, while gravitational two-derivative interactions are present on the list \cite{Bengtsson:1986kh} within the light-front approach. Similarly, while the graviton described by the symmetric tensor allows one to construct the usual gravitational interactions, it is impossible to do so if a different Lorentz tensor is chosen \cite{Bekaert:2002uh}. } In the present paper, we study the consistency condition that appears at the quartic order in the light-cone gauge. It was obtained and analysed by Metsaev in \cite{Metsaev:1991mt,Metsaev:1991nb}. An immediate consequence \cite{Ponomarev:2016lrm} of this result is the existence of chiral higher-spin gravity.\footnote{Note that the light-cone approach is instrumental in getting sharp no-go's \cite{Ponomarev:2017nrr} or existence statements \cite{Ponomarev:2016lrm}. However, once some positive result is established, it is also helpful to have a Lorentz-invariant formulation, e.g. to study solutions, compute amplitudes, extend the results into (anti)-de Sitter space, some of which was achieved for chiral higher-spin gravity in \cite{Sharapov:2022faa,Sharapov:2022wpz,Sharapov:2022awp,Sharapov:2022nps,Sharapov:2023erv,Skvortsov:2024rng,Tran:2025yzd}. Still, the first quantum corrections have been computed in the light-cone gauge \cite{Skvortsov:2018jea,Skvortsov:2020wtf,Skvortsov:2020gpn,Tsulaia:2022csz,Neiman:2024vit}. Another interesting direction is AdS/CFT duality for chiral higher-spin gravity \cite{Skvortsov:2018uru,Sharapov:2022awp,Jain:2024bza,Aharony:2024nqs}. } What we show is that there are more sufficiently interesting solutions hidden in there.  

In a few words, the analysis of the quartic consistency proceeds as follows. One begins with some cubic interaction vertex $V_{\lambda_1,\lambda_2,\lambda_3}$, which also implies the presence of fields with helicities $\lambda_{1,2,3}$ in the spectrum. Every light-cone vertex leads to the standard spinor-helicity amplitude, e.g.
\begin{align}\label{Paper1-genericV}
   V_{\lambda_1,\lambda_2,\lambda_3}\Big|_{\text{on-shell}} \sim 
        [12]^{\lambda_1+\lambda_2-\lambda_3}[23]^{\lambda_2+\lambda_3-\lambda_1}[13]^{\lambda_1+\lambda_3-\lambda_2}\,, \qquad \sum\lambda_i>0
\end{align}
and there is a similar expression with $\langle | \rangle$ when $\sum\lambda_i<0$. The propagator connects $+\lambda$ to $-\lambda$. The quartic light-cone constraint is equivalent to the Poincaré invariance of the S-matrix, see e.g. \cite{Ponomarev:2016cwi}. To have a nontrivial quartic constraint, one either has $\lambda_i=-\lambda_j$ among the already introduced vertices or has to add more fields/interactions to ``make them talk'' to each other.\footnote{Of course, one can always introduce abelian interactions, which correspond to all $\lambda_{1,2,3}>0$ or $\lambda_{1,2,3}<0$ in a vertex or have a non-abelian interaction that is, nevertheless, consistent on its own because one cannot even form an exchange diagram, e.g. one can take vertex $V_{666,-13,-42}$ and it does not lead to any nontrivial quartic constraint.} However, the quartic constraint may not be satisfied as is and may force one to introduce more exchange diagrams and, hence, more fields/interactions. 

Quite often, the process does not stop till the chiral higher-spin gravity is reached. One can confine oneself to gauge and gravitational interactions (i.e. one- and two-derivative ones) that are present in the contractions --- HS-SDYM and HS-SDGR --- of the chiral higher-spin gravity. As we show, there are still many nontrivial solutions in this subclass of interactions, and we classify them all. Therefore, all the theories we found belong to the class of higher-spin extensions of self-dual Yang-Mills and of self-dual gravity. In a bit more detail, we have found $9$ families of HS-SDGR-like theories with finitely many fields (plus $4$ particular cases) and $3$ families with infinitely many fields. As for HS-SDYM-like theories, we have found $8$ families with finitely many fields (plus $2$ particular cases) and $2$ families with infinitely many fields.

It is tempting to make a parallel with the matter-free higher-spin gravities in $3d$, which can always be formulated as Chern-Simons theories \cite{Grigoriev:2020lzu}. They provide a huge variety of options to have a finite number of higher-spin fields. Our results reveal that the case of $4d$ is somewhat similar if one restricts oneself to holomorphic/anti-holomorphic interactions/amplitudes.  

Among other highlights, one can mention that there are particular low-spin couplings, e.g. $V_{-1,0,2}$, that induce higher-spin amplitudes from the Poincaré invariance of the S-matrix, which is surprising. There are also consistent solutions with both a finite spectrum and higher-derivative nonabelian interactions. We also find that it is possible for the graviton to have colour (multi-graviton theories) once we restrict to the self-dual subsector. Another curiosity is that there are solutions with fractional helicities, e.g. $\lambda=2/3,4/3,...$, which, at least at present, are more of an oddity.  

The outline of the paper is as follows:
In Section \ref{Paper1-section2}, we briefly summarise the derivation of the holomorphic quartic constraint, along with some basics of the light-front approach to higher-spin interactions.
In Section \ref{Paper1-section3}, we review and extend the solution of the holomorphic constraint \cite{Metsaev:1991mt,Metsaev:1991nb,Ponomarev:2016lrm,Ponomarev:2017nrr,Monteiro:2022xwq} to all integer spins, allowing for arbitrary couplings involving both even- and odd-derivative interactions. We also analyse the constraint in the presence of a $U(N)$ gauge group.
In Section \ref{Paper1-section4}, we present our main results: the complete classification of two-derivative chiral higher-spin theories, all of which are contained within HS-SDGR, and the classification of all one-derivative chiral higher-spin theories in the presence of a gauge group, contained within HS-SDYM. We also discuss the higher-derivative case and highlight some interesting features. Additionally, we provide explicit solutions for the couplings in certain chiral higher-spin theories.
In Section \ref{Paper1-section5}, we compute the $4$-pt amplitudes of a generic chiral higher-spin theory.
Finally, in Section \ref{Paper1-section6}, we conclude with some comments and outline possible directions for future work.

We also include five appendices. In Appendix \ref{Paper1-AppendixA}, we settle our notation and review standard results of the light-front approach to massless higher-spin fields. In Appendix \ref{Paper1-AppendixB}, we provide some simple yet useful formulas used in the main text. In Appendix \ref{Paper1-AppendixC}, we solve the holomorphic constraint for the low-derivative cases. In Appendix \ref{Paper1-AppendixD}, we consider the holomorphic constraint for the gauge groups $SO(N)$ and $USp(N)$. In Appendix \ref{Paper1-AppendixE}, we consider the holomorphic constraint for the case where all fields take values in the adjoint representation of some Lie algebra of internal symmetry.

\section{Quartic higher-spin Lorentz constraints}\label{Paper1-section2}
We begin by reviewing the derivation of the quartic holomorphic constraint.
To construct higher-spin vertices in $4d$ flat space using the light-cone gauge, one begins by expressing the Poincaré algebra generators in terms of free fields. These are initially quadratic in the fields. The idea is to deform them by adding higher-order local terms and require that the modified generators continue to satisfy the Poincaré algebra, order by order in the fields.

An extensive study of this approach to construct consistent interactions for massless higher-spin fields, along with various results, can be found in \cite{Ponomarev:2016lrm}. We review some relevant notation and results in Appendix \ref{Paper1-AppendixA}.
Here, we begin by recalling the results for the cubic vertices and boost generators, which are fixed up to field redefinitions to be
\begin{align}\label{Paper1-CubicVertices}
    h_{\lambda_1,\lambda_2,\lambda_3}&=C^{\lambda_1,\lambda_2,\lambda_3}\frac{\PPb^{\lambda_{123}}}{\beta_1^{\lambda_1}\beta_2^{\lambda_2}\beta_3^{\lambda_3}}+\bar{C}^{-\lambda_1,-\lambda_2,-\lambda_3}\frac{\PP^{-\lambda_{123}}}{\beta_1^{-\lambda_1}\beta_2^{-\lambda_2}\beta_3^{-\lambda_3}}\,,\\
    j_{\lambda_1,\lambda_2,\lambda_3}&=\frac{2}{3}\,C^{\lambda_1,\lambda_2,\lambda_3}\frac{\PPb^{\lambda_{123}-1}}{\beta_1^{\lambda_1}\beta_2^{\lambda_2}\beta_3^{\lambda_3}}\Lambda^{\lambda_1,\lambda_2,\lambda_3}\,,\\
    \bar{j}_{\lambda_1,\lambda_2,\lambda_3}&=-\frac{2}{3}\,\bar{C}^{-\lambda_1,-\lambda_2,-\lambda_3}\frac{\PP^{-\lambda_{123}-1}}{\beta_1^{-\lambda_1}\beta_2^{-\lambda_2}\beta_3^{-\lambda_3}}\Lambda^{\lambda_1,\lambda_2,\lambda_3}\,,
\end{align}
where $\lambda_{123}=\lambda_1+\lambda_2+\lambda_3$. Here, $C^{\lambda_1,\lambda_2,\lambda_3}$ are coupling constants and the main task is to determine which ones lead to consistent theories, i.e. which couplings may not be zero and what the relations between various couplings are. Written in this form, the cubic vertices exhibit a clear separation between holomorphic and anti-holomorphic components, corresponding respectively to the terms involving $\PPb$ and $\PP$ in the equations above.
These expressions can be used to construct the Hamiltonian $H$ and the dynamical boost generators $J^{z-}$ and $J^{\bar{z}-}$ via Eqs.~\eqref{Paper1-hamiltonian} and \eqref{Paper1-boostz}.

By examining the expressions for $H$ and the dynamical boosts $J^{z-}, J^{\bar{z}-}$, we can assume certain symmetry properties of the couplings $C^{\lambda_1,\lambda_2,\lambda_3}$ to simplify our computations.\footnote{This assumption is harmless, as it can be made without loss of generality.}
Indeed, one can observe that the parity under the exchange $(\lambda_i, q_i) \leftrightarrow (\lambda_j, q_j)$ of both $h^{q_1,q_2,q_3}_{\lambda_1,\lambda_2,\lambda_3}$ and $j^{q_1,q_2,q_3}_{\lambda_1,\lambda_2,\lambda_3}$ is given by $(-)^{\lambda_{123}}$.
Indeed, we have\footnote{For simplicity, we focus only on the holomorphic part, but the same reasoning applies more generally.}
\begin{alignat}{2}
&H_3^{\lambda_1,\lambda_2,\lambda_3}&&=C^{\lambda_1,\lambda_2,\lambda_3}\int d^9q\;\delta\Big(\sum_i q_i\Big)\frac{\PPb^{\lambda_{123}}}{\beta_1^{\lambda_1}\beta_2^{\lambda_2}\beta_3^{\lambda_3}}\phi^{\lambda_1}_{q_1}\phi^{\lambda_2}_{q_2}\phi^{\lambda_3}_{q_3}\,,\\
\nonumber
&H_3^{\lambda_2,\lambda_1,\lambda_3}&&=C^{\lambda_2,\lambda_1,\lambda_3}\int d^9q\;\delta\Big(\sum_i q_i\Big)\frac{\PPb^{\lambda_{123}}}{\beta_1^{\lambda_2}\beta_2^{\lambda_1}\beta_3^{\lambda_3}}\phi^{\lambda_2}_{q_1}\phi^{\lambda_1}_{q_2}\phi^{\lambda_3}_{q_3}\\
    & &&\overset{q_1\leftrightarrow q_2}{=}(-)^{\lambda_{123}}C^{\lambda_2,\lambda_1,\lambda_3}\int d^9 q\;\delta\Big(\sum_i q_i\Big)\frac{\PPb^{\lambda_{123}}}{\beta_1^{\lambda_1}\beta_2^{\lambda_2}\beta_3^{\lambda_3}}\phi^{\lambda_1}_{q_1}\phi^{\lambda_2}_{q_2}\phi^{\lambda_3}_{q_3}\,,
\end{alignat}
and the same symmetry considerations apply to the boost generators. Therefore, we can assume that the coupling constants are symmetric for even-derivative interactions and antisymmetric for odd-derivative ones:
\begin{equation}\label{Paper1-nocolour_coupling_sym}
    C^{\lambda_1,\lambda_2,\lambda_3}=(-)^{\lambda_{123}}C^{\lambda_{\sigma_1},\lambda_{\sigma_2},\lambda_{\sigma_3}}\,,
\end{equation}
where $\sigma\in\Sigma_3$ represents an odd permutation. Then we can use the following form for the generators:
\begin{align}\label{Paper1-generatorssym}
    H_3&=\sum_{\lambda_1,\lambda_2,\lambda_3}\int d^9 q\;\delta\Big(\sum_i q_i\Big)h^{q_1,q_2,q_3}_{\lambda_1,\lambda_2,\lambda_3}\,\phi^{\lambda_1}_{q_1}\phi^{\lambda_2}_{q_2}\phi^{\lambda_3}_{q_3}\,,\\
    J^{z-}_3&=\sum_{\lambda_1,\lambda_2,\lambda_3}\int d^9q\;\delta\Big(\sum_i q_i\Big)\Big[j^{q_1,q_2,q_3}_{\lambda_1,\lambda_2,\lambda_3}-\frac{1}{3}\,h^{q_1,q_2,q_3}_{\lambda_1,\lambda_2,\lambda_3}\Big(\sum_j\frac{\partial}{\partial \bar{q}_j}\Big)\Big]\phi^{\lambda_1}_{q_1}\phi^{\lambda_2}_{q_2}\phi^{\lambda_3}_{q_3}\,,
\end{align}
i.e. to sum over all triplets of helicities instead of all distinct (up to permutation) triplets. 
This observation also implies that odd-derivative couplings involving at least two identical fields\footnote{In this context, ``identical'' refers to fields with the same helicity, since no additional quantum numbers are yet considered.} vanish by symmetry  $C^{\lambda,\lambda,\lambda'}\equiv 0$.

The quartic dynamical constraint takes the simple form
\begin{equation}\label{Paper1-quartic_constarint}
    H_2j_4^{z-}=J_2^{z-}h_4+[H_3,J_3^{z-}]\,.
\end{equation}
We can observe, by examining the degree of homogeneity of $q$, that these two conditions must be satisfied independently
\begin{align}
    &[H_3(\PPb),J_3]=0,&
    &[H_3(\PP),\bar{J}_3]=0\,.
\end{align}
We refer to them, respectively, as holomorphic and anti-holomorphic quartic constraints.

In particular, it is worth noting that if we assume the theory to be purely holomorphic (or anti-holomorphic) and to satisfy the holomorphic quartic constraints, it will be a well-defined theory at any order and will contain only cubic interactions. These are the theories we will study. We can explicitly write down the holomorphic constraint as
\begin{align}\label{Paper1-HoloexpressionGR}
\begin{split}
[H_3,J_3^{z-}]=&\sum_{\lambda_i,\alpha_j}\int d^9p\;d^9q\; \delta\left(\sum_i q_i\right) \left[j_3^{\lambda_i}(q_i)-\frac{h_3^{\lambda_i}(q_i)}{3}\left(\sum_{k}\frac{\partial}{\partial \bar{q}_k}\right)\right] \times \\
&\delta\left(\sum_j p_j\right)h_3^{\alpha_j}(p_j)\left[\prod_{i=1}^3\phi_{q_i}^{\lambda_i},\prod_{j=1}^3\phi_{p_j}^{\alpha_j}\right]\,.
\end{split}
\end{align}
In computing the Poisson bracket, we get
\begin{equation}
    \sum_{\lambda_1,\lambda_2,\lambda_3}\sum_{\alpha_1,\alpha_2,\alpha_3}[\phi^{\lambda_1}_{q_1}\phi^{\lambda_2}_{q_2}\phi^{\lambda_3}_{q_3},\phi^{\alpha_1}_{p_1}\phi^{\alpha_2}_{p_2}\phi^{\alpha_3}_{p_3}]=9\sum_{\lambda_1,\lambda_2}\sum_{\alpha_1,\alpha_2}\sum_{\omega}\phi^{\lambda_1}_{q_1}\phi^{\lambda_2}_{q_2}\phi^{\alpha_1}_{p_1}\phi^{\alpha_2}_{p_2}[\phi^{\omega}_{q_3},\phi^{-\omega}_{p_3}]\,,
\end{equation}
and the constraint can be brought to the form
\begin{align}
\begin{split}
[H_3,J_3^{z-}]= &\sum_{\lambda_i,\alpha_j}\int d^9p\;d^9q\; \delta\left(\sum_i q_i\right) \delta\left(\sum_j p_j\right)9\,\delta^{\lambda_3,-\alpha_3}\frac{\delta(q_3+p_3)}{2q_3^+}\,\phi^{\lambda_1}_{q_1}\phi^{\lambda_2}_{q_2}\phi^{\alpha_1}_{p_1}\phi^{\alpha_2}_{p_2}\,\times\\
&\left(j_3^{\lambda_i}(q_i)+\sum_{k\neq 3}\frac{\partial}{\partial \bar{q}_{k}}\frac{h_3^{\lambda_i}(q_i)}{3}\right)h_3^{\alpha_j}(p_j)\,.
\end{split}
\end{align}
Once we substitute the explicit form of the generators in \eqref{Paper1-generatorssym}, we get
\begin{align}\label{Paper1-holoGR}
    \begin{split}
    [H_3,J_3^{z-}]=&\sum_{\lambda_i,\omega}\int d^{12}q\;\delta \left(\sum_i q_i\right)\frac{9}{2}\Big[(-)^{\omega}\frac{(\lambda_1+\omega-\lambda_2)\beta_1-(\lambda_2+\omega-\lambda_1)\beta_2}{(\beta_1+\beta_2)\beta_1^{\lambda_1}\beta_2^{\lambda_2}\beta_3^{\lambda_3}\beta_4^{\lambda_4}}\,\times\\
    &C^{\lambda_1,\lambda_2,\omega}C^{-\omega,\lambda_3,\lambda_4}\PPb_{12}^{\lambda_{12}+\omega-1}\PPb_{34}^{\lambda_{34}-\omega}\,\phi^{\lambda_1}_{q_1}\phi^{\lambda_2}_{q_2}\phi^{\lambda_3}_{q_3}\phi^{\lambda_4}_{q_4}\Big]\,,
    \end{split}
\end{align}
where $\lambda_{ij}\equiv\lambda_i+\lambda_j$ and we have to remember that the couplings $C^{\lambda_1,\lambda_2,\lambda_3}$ have the symmetry \eqref{Paper1-nocolour_coupling_sym}.

In general, we can also assume that the fields live in some representation of a gauge group $G$. The simplest option is to take the fields to be $N\times N$ matrices, i.e. $\phi\in\text{Mat}_N$ and make the replacement:
\begin{align}
    &\phi^{\lambda}_{q}\quad\rightarrow\quad (\phi^{\lambda}_q)_a T^a\equiv (\phi^{\lambda}_{q})^A_{\;B}\,,&
    &H_3\sim \phi^{\lambda_1}_{q_1}\phi^{\lambda_2}_{q_2}\phi^{\lambda_3}_{q_3}\quad\rightarrow\quad H_3\sim\mathrm{Tr}(\phi^{\lambda_1}_{q_1}\phi^{\lambda_2}_{q_2}\phi^{\lambda_3}_{q_3})\,.
\end{align}
In the presence of a gauge group the holomorphic constraint takes the form
\begin{equation}
\begin{split}
[H_3,J_3^{z-}]= & \sum_{\lambda_i,\alpha_j}\int d^9p\;d^9q\;\delta\left(\sum_i q_i\right) \left[j_3^{\lambda_i}(q_i)-\frac{h_3^{\lambda_i}(q_i)}{3}\left(\sum_{k}\frac{\partial}{\partial \bar{q}_k}\right)\right] \times \\
 & \delta\left(\sum_j p_j\right)h_3^{\alpha_j}(p_j)\left[\mathrm{Tr}\prod_{i=1}^3\phi_{q_i}^{\lambda_i},\mathrm{Tr}\prod_{j=1}^3\phi_{p_j}^{\alpha_j}\right]\,,
 \end{split}
\end{equation}
and once we compute the Poisson bracket and substitute the explicit form of the generators, we get
\begin{align}\label{Paper1-holocolour}
    \begin{split}
    [H_3,J_3^{z-}]=&\sum_{\lambda_i,\omega}\int d^{12}q\,\delta\left(\sum_i q_i\right)\frac{9}{2}\Big[(-)^{\omega}\frac{(\lambda_1+\omega-\lambda_2)\beta_1-(\lambda_2+\omega-\lambda_1)\beta_2}{(\beta_1+\beta_2)\beta_1^{\lambda_1}\beta_2^{\lambda_2}\beta_3^{\lambda_3}\beta_4^{\lambda_4}}\times\\
    &C^{\lambda_1,\lambda_2,\omega}C^{-\omega,\lambda_3,\lambda_4}\PPb_{12}^{\lambda_{12}+\omega-1}\PPb_{34}^{\lambda_{34}-\omega}\,\mathrm{Tr}(\phi^{\lambda_1}_{q_1}\phi^{\lambda_2}_{q_2}\phi^{\lambda_3}_{q_3}\phi^{\lambda_4}_{q_4})\Big]\,.
    \end{split}
\end{align}
In this case, we have used the following form for the generators:
\begin{align}\label{Paper1-generatorscyc}
    H_3=&\;\sum_{\lambda_1,\lambda_2,\lambda_3}\int d^9 q\;\delta\Big(\sum_i q_i\Big)h^{q_1,q_2,q_3}_{\lambda_1,\lambda_2,\lambda_3}\mathrm{Tr}(\phi^{\lambda_1}_{q_1}\phi^{\lambda_2}_{q_2}\phi^{\lambda_3}_{q_3})\,,\\
    J^{z-}_3=&\;\sum_{\lambda_1,\lambda_2,\lambda_3}\int d^9q\;\delta\Big(\sum_i q_i\Big)\Big[j^{q_1,q_2,q_3}_{\lambda_1,\lambda_2,\lambda_3}-\frac{1}{3}h^{q_1,q_2,q_3}_{\lambda_1,\lambda_2,\lambda_3}\Big(\sum_j\frac{\partial}{\partial \bar{q}_j}\Big)\Big]\mathrm{Tr}(\phi^{\lambda_1}_{q_1}\phi^{\lambda_2}_{q_2}\phi^{\lambda_3}_{q_3})\,.
\end{align}
Note that both  $h^{q_1,q_2,q_3}_{\lambda_1,\lambda_2,\lambda_3}$ and $j^{q_1,q_2,q_3}_{\lambda_1,\lambda_2,\lambda_3}$ are symmetric under the cyclic permutation of $(\lambda_i,q_i)$. By applying the same argument as before, we can assume $C^{\lambda_1,\lambda_2,\lambda_3}$ to be cyclic symmetric.

We will now proceed to solve the holomorphic constraint explicitly, both with and without a gauge group.

\section{Quartic holomorphic constraint}\label{Paper1-section3}
In this section, we solve the light-cone holomorphic constraint for any integer helicity. We begin by studying the case without a gauge group, and then extend the analysis to several cases where the fields take values in certain representations of  $G=U(N), SO(N),$ and $USp(N)$. Our primary focus will be on the $U(N)$ case, as the others follow a similar pattern. Additional comments on the other cases are provided in Appendix \ref{Paper1-AppendixD}.
\subsection{Light-cone holomorphic constraint}
We solve the holomorphic constraint \eqref{Paper1-holoGR} for fixed integer helicities $\lambda_{1,2,3,4}$ on the external legs, considering both even- and odd-derivative interactions. As we will see, however, Eq.~\eqref{Paper1-holoGR} may require additional couplings to be nonzero on top of the initial input. These, in turn, can trigger further instances of Eq.~\eqref{Paper1-holoGR} involving different helicity configurations, leading to an iterative process where new constraints arise. Consequently, finding a consistent solution to the cubic deformation involves more than solving the constraint for fixed external helicities. Complete solutions to \eqref{Paper1-holoGR} will be presented in the next section.

The analysis begins by following Appendix A of \cite{Ponomarev:2016lrm}. We rewrite the light-cone holomorphic constraint as\footnote{Note that we use a different notation compared to \cite{Ponomarev:2016lrm}, writing the pair of couplings with the exchanged helicity in the middle, as in $C^{\lambda_1,\lambda_2,\omega}C^{-\omega,\lambda_3,\lambda_4}$. This is equivalent, since, as we have seen above, the couplings are always cyclic symmetric.}
\begin{equation}\label{Paper1-holo2}
    \sum_{\omega}\text{Sym}\left[(-)^{\omega}\frac{(\lambda_1+\omega-\lambda_2)\beta_1-(\lambda_2+\omega-\lambda_1)\beta_2}{\beta_1+\beta_2}\,C^{\lambda_1,\lambda_2,\omega}C^{-\omega,\lambda_3,\lambda_4}\PPb_{12}^{\lambda_{12}+\omega-1}\PPb_{34}^{\lambda_{34}-\omega}\right]=0\,.
\end{equation}
Here, Sym denotes the sum over $6$ distinct contributions
\begin{equation}
    (1234)=\{1234\}+\{1324\}+\{1423\}+\{3412\}+\{2413\}+\{2314\}\,.
\end{equation}
Starting from the variables $\PPb_{ij}$ and $\beta_i$, using momentum conservation, only $5$ of them are independent. For instance, we can choose $\PPb_{12}$, $\PPb_{34}$ and three of the $\beta_i$'s. The various relations can be found in Appendix \ref{Paper1-AppendixB}.

Now we can sum the two contributions $(1234)\leftrightarrow (3412)$ and $\omega\leftrightarrow-\omega$, and after defining new independent variables
\begin{equation}
    2A=\PPb_{12}+\PPb_{34}=\PPb_{23}-\PPb_{14}\,,\quad
    2B=\PPb_{13}-\PPb_{24}=\PPb_{34}-\PPb_{12}\,,\quad
    2C=\PPb_{14}+\PPb_{23}=-\PPb_{13}-\PPb_{24}\,,
\end{equation}
and using the relations in Appendix \ref{Paper1-AppendixB}, the $\beta$ dependence disappears
\begin{align}
    \begin{split}
    &\frac{(\lambda_1+\omega-\lambda_2)\beta_1-(\lambda_2+\omega-\lambda_1)\beta_2}{\beta_1+\beta_2}\,\PPb_{34}+\frac{(\lambda_3-\omega-\lambda_4)\beta_3-(\lambda_4-\omega-\lambda_3)\beta_4}{\beta_3+\beta_4}\,\PPb_{12}\\
    &=(\lambda_1-\lambda_2)\PPb_{34}+(\lambda_3-\lambda_4)\PPb_{12}+\frac{\omega}{2}(\PPb_{13}-\PPb_{23}+\PPb_{24}-\PPb_{14})\\
    &=(\lambda_1-\lambda_2+\lambda_3-\lambda_4)A+(\lambda_1-\lambda_2-\lambda_3+\lambda_4)B-2\omega C\,.
    \end{split}
\end{align}
The constraint \eqref{Paper1-holo2} can then be rewritten as a polynomial in just $3$ independent variables
\begin{align}\label{Paper1-LCholo}
\nonumber
    &\sum_{\omega}(-)^{\omega}\big[((\lambda_{13}-\lambda_{24})A+(\lambda_{14}-\lambda_{23})B-2\omega C)\,
    \mathcal{C}^{1234\omega}(A-B)^{\lambda_{12}+\omega-1}(A+B)^{\lambda_{34}-\omega-1}\\
    \nonumber
    &+(-)^{\lambda_{24}+\omega}((\lambda_{14}-\lambda_{23})B+(\lambda_{12}-\lambda_{34})C-2\omega A)\,
    \mathcal{C}^{1324\omega}(B-C)^{\lambda_{13}+\omega-1}(B+C)^{\lambda_{24}-\omega-1}\\
    &+((\lambda_{12}-\lambda_{34})C+(\lambda_{13}-\lambda_{24})A-2\omega B)\,
     \mathcal{C}^{1423\omega}(C-A)^{\lambda_{14}+\omega-1}(C+A)^{\lambda_{23}-\omega-1}\big]=0\,,
\end{align}
where $\lambda_{ij}\equiv\lambda_i+\lambda_j$ and $\mathcal{C}^{1234\omega}\equiv C^{\lambda_1,\lambda_2,\omega}C^{-\omega,\lambda_3,\lambda_4}$. We will use the fact that $\mathcal{C}^{1234\omega}$ is a product of two couplings only at the end; for now, we will treat it as a generic function. 

To avoid unnecessary minus signs between products of couplings and extra $i$ factors between couplings, we assign an additional factor of $i$ to odd-helicity fields. This is equivalent to taking even-helicity fields to be Hermitian and odd-helicity fields to be anti-Hermitian matrices (with singlets treated as $1\times 1$ matrices). With this convention, the factor $(-)^{\omega}$ in \eqref{Paper1-LCholo} is removed.

Before we begin solving the constraint, we set the following definitions:
\begin{align}\label{Paper1-definitions1}
\nonumber
&k^{1234}_+\equiv(-)^{\lambda_{12}}\sum_{\omega}(-)^{\omega}\mathcal{C}^{1234\omega}\,,\quad
    k^{1234}_-\equiv\sum_{\omega}\mathcal{C}^{1234\omega}\,,\;\;\;\;\;\;\;\quad
    k^{1234}_{\omega+}\equiv(-)^{\lambda_{12}}\sum_{\omega}(-)^{\omega}\omega\,\mathcal{C}^{1234\omega}\,,\\
    &k^{1234}_{\omega-}\equiv\sum_{\omega}\omega\, \mathcal{C}^{1234\omega}\,,\quad
    k^{1234}_{\omega^2+}\equiv(-)^{\lambda_{12}}\sum_{\omega}(-)^{\omega}\omega^2\, \mathcal{C}^{1234\omega}\,,\quad
    k^{1234}_{\omega^2-}\equiv\sum_{\omega}\omega^2\, \mathcal{C}^{1234\omega}\,,\\ \label{Paper1-definitions2}
    \begin{split}
    &f_+^{1234}(A,B)\equiv(-)^{\lambda_{34}}\sum_{\omega}(-)^{\omega}\mathcal{C}^{1234\omega}(A-B)^{\lambda_{12}+\omega-1}(A+B)^{\lambda_{34}-\omega-1}\,,\\
    &f^{1234}_{\omega+}(A,B)\equiv(-)^{\lambda_{34}}\sum_{\omega}(-)^{\omega}\omega\,\mathcal{C}^{1234\omega}(A-B)^{\lambda_{12}+\omega-1}(A+B)^{\lambda_{34}-\omega-1}\,,\\
    &f_-^{1234}(A,B)\equiv\sum_{\omega}\,\mathcal{C}^{1234\omega}(A-B)^{\lambda_{12}+\omega-1}(A+B)^{\lambda_{34}-\omega-1}\,,\\
    &f^{1234}_{\omega-}(A,B)\equiv\sum_{\omega}\omega\,\mathcal{C}^{1234\omega}(A-B)^{\lambda_{12}+\omega-1}(A+B)^{\lambda_{34}-\omega-1}\,,
     \end{split}
\end{align}
where the upper index $1234$ indicates the order of the external helicities. We can then rewrite the constraint \eqref{Paper1-LCholo} using the functions \eqref{Paper1-definitions2} just defined as
\begin{align}\label{Paper1-short_LCholo}
    \begin{split}
    &\;\;\;\;((\lambda_{13}-\lambda_{24})A+(\lambda_{14}-\lambda_{23})B)f^{1234}_-(A,B)-2\omega Cf^{1234}_{\omega -}(A,B)\\
    &+((\lambda_{14}-\lambda_{23})B+(\lambda_{12}-\lambda_{34})C)f^{1324}_+(B,C)-2\omega Af^{1324}_{\omega+}(B,C)\\
    &+((\lambda_{12}-\lambda_{34})C+(\lambda_{13}-\lambda_{24})A)f^{1423}_-(C,A)-2\omega Bf^{1423}_{\omega-}(C,A)=0\,.
    \end{split}
\end{align}
We divide the solution into four distinct cases, which differ based on the form of the constraint once specific relations among the external helicities are imposed. The general approach proceeds as follows:

\begin{itemize}
\item First, we determine the polynomial form of the functions appearing in \eqref{Paper1-short_LCholo} --- up to some free coefficients (later identified with those in \eqref{Paper1-definitions1}) --- by setting to zero all monomials except those that can be generated by at least two different functions $f^{\cdots\cdot}_{\pm}$.

\item Second, we substitute these polynomial expressions back into the constraint \eqref{Paper1-short_LCholo} to derive relations among the coefficients in \eqref{Paper1-definitions1}. 

\item Third, once all coefficients in \eqref{Paper1-definitions1} are fixed, the equation \eqref{Paper1-LCholo} is solved and we can extract the products of couplings $\mathcal{C}^{1234\omega}$ from the functions $f^{\cdots\cdot}_{\pm}$ via a simple change of variables. 
\end{itemize}
As we will see, we always end up with a system of the general form
\begin{equation}
    \mathcal{C}^{1234\omega}\sim\frac{(\Lambda-2)!}{2^{\Lambda-2}(\lambda_{12}+\omega-1)!(\lambda_{34}-\omega-1)!}\qquad 
    \forall\;\omega\,,\qquad
    \Lambda\equiv \lambda_1+\lambda_2+\lambda_3+\lambda_4\,.
\end{equation}
Assuming this form and using the formulas collected in Appendix \ref{Paper1-AppendixB}, we obtain the following simple relations, which will be useful later:
\begin{align}\label{Paper1-usefulrelations}
\nonumber
 &(\lambda_{34}-\lambda_{12})k^{1234}_{\omega+}-2k^{1234}_{\omega^2+}=\frac12\,(2-\Lambda)k^{1234}_+\,,&
   &(\lambda_{34}-\lambda_{12})k^{1234}_{\omega-}-2 k^{1234}_{\omega^2-}=\frac12\,(2-\Lambda)k^{1234}_-\,,\\
    &k^{1234}_{\omega+}=\frac12\,(\lambda_{34}-\lambda_{12})k^{1234}_+\,,&
   &k^{1234}_{\omega-}=\frac12\,(\lambda_{34}-\lambda_{12})k^{1234}_-\,,\\
\nonumber
    &k^{1212}_{\omega^2+}=\frac{\Lambda-2}{4}\,k^{1212}_{+}\,,&
    &k^{1212}_{\omega^2-}=\frac{\Lambda-2}{4}\,k^{1212}_{-}\,.
\end{align}
These relations allow us to solve $k^{\cdots\cdot}_{\omega\pm}$ and $k^{\cdots\cdot}_{\omega^2\pm}$ in terms of $k^{\cdots\cdot}_{\pm}$.
Given the symmetry properties of the couplings \eqref{Paper1-nocolour_coupling_sym}, we obtain the relations
\begin{align}\label{Paper1-useful_relations}
    &k^{1234}_{\mp}=k^{2134}_{\pm}=(-)^{\Lambda}k^{1243}_{\pm}\,,&
    &k^{3412}_-=k^{1234}_-\,,&
    &k^{3412}_+=(-)^{\Lambda}k^{1234}_+\,.
\end{align}
In all cases, we will present both the general solution and the solution assuming only even-derivative vertices (i.e. $k^{\cdots\cdot}_+=k^{\cdots\cdot}_-$).  In the following, we focus on $\Lambda\geq 4$; for lower-derivative cases, fewer constraints arise due to the somewhat degenerate structure of the monomials. The cases $\Lambda=2,3$ are analysed separately in Appendix \ref{Paper1-AppendixC}.

\paragraph{Case 1.} Here, we assume that all helicities are equal, i.e.  $\lambda_1=\lambda_2=\lambda_3=\lambda_4$. The constraint \eqref{Paper1-LCholo} becomes
\begin{align}\label{Paper1-case1}
    \begin{split}
    \sum_{\omega}\omega\, \mathcal{C}^{1111\omega}&\big[ C
    (A-B)^{2\lambda+\omega-1}(A+B)^{2\lambda-\omega-1}+(-)^{\omega}A
    (B-C)^{2\lambda+\omega-1}(B+C)^{2\lambda-\omega-1}\\
    &+B (C-A)^{2\lambda+\omega-1}(C+A)^{2\lambda-\omega-1}\big]=0\,.
    \end{split}
\end{align}
First, we determine the polynomial form of the functions \eqref{Paper1-definitions2} as
\begin{subequations}\label{Paper1-case11}
\begin{align}
    f^{1111}_{\omega-}(A,B)&=2(k^{1111}_{\omega^2+} AB^{\Lambda-3}-k^{1111}_{\omega^2-}A^{\Lambda-3}B)\,,\\
    f^{1111}_{\omega+}(B,C)&=2(k^{1111}_{\omega^2-} BC^{\Lambda-3}-k^{1111}_{\omega^2+}B^{\Lambda-3}C)\,,\\
    f^{1111}_{\omega-}(C,A)&=2(k^{1111}_{\omega^2+} CA^{\Lambda-3}-k^{1111}_{\omega^2-}C^{\Lambda-3}A)\,.
\end{align}
\end{subequations}
By changing variables and using $S=A+B$, $D=A-B$, we can derive a system of equations for the couplings by solving \eqref{Paper1-case11} for fixed $\omega$. The resulting system is
\begin{equation}
    \mathcal{C}^{1111\omega}=\frac{k^{1111}_{\omega^2-}+(-)^{\omega}k^{1111}_{\omega^2+}}{2^{\Lambda-4}}  \begin{pmatrix}
        \Lambda-3\\
        2\lambda-\omega-1
    \end{pmatrix}\qquad
    \forall\,\omega\neq 0\,.
\end{equation}
Now, substituting \eqref{Paper1-case11} into \eqref{Paper1-case1}, we obtain an additional constraint
\begin{align}
    \begin{split}
   &(k^{1111}_{\omega^2-}-k^{1111}_{\omega^2+})A^{\Lambda-3}BC=0\quad
    \Rightarrow\quad
    k^{1111}_{\omega^2-}=k^{1111}_{\omega^2+}\quad
    \Rightarrow\quad 
    k^{1111}_{\omega^2E}\neq 0,\; k^{1111}_{\omega^2O}=0\,,\\
    &k^{1111}_{\omega^2E}\equiv\sum_{\omega\in\text{even}}\omega^2\, \mathcal{C}^{1111\omega}\,,\qquad
    k^{1111}_{\omega^2O}\equiv\sum_{\omega\in\text{odd}}\omega^2\, \mathcal{C}^{1111\omega}\,.
    \end{split}
\end{align}
Only the solutions with even $\omega$ survive. This was expected, since by symmetry, odd-derivative vertices with at least two equal helicities vanish. The final solution is
\begin{equation}
    \mathcal{C}^{1111\omega}=\frac{k^{1111}_{\omega^2E}(\Lambda-3)!}{2^{\Lambda-5}(2\lambda+\omega-1)!(2\lambda-\omega-1)!}\qquad
    \forall\,\omega\neq 0\,\text{even}\,.
\end{equation}
Consistency is guaranteed by the relation
\begin{equation}
    \sum_{\omega\in\text{even}}\frac{\omega^2(\Lambda-3)!}{2^{\Lambda-5}(2\lambda+\omega-1)!(2\lambda-\omega-1)!}=1\,.
\end{equation}
Note that the product $C^{\lambda,\lambda,0}C^{0,\lambda,\lambda}$ decouples from the system; indeed, it does not contribute to the constant $k^{1111}_{\omega^2E}$ and does not appear in the constraint \eqref{Paper1-case1}; it is then unconstrained. As we will see, this has important consequences, such as allowing for non-vanishing amplitudes and ensuring the correct relations among the couplings.

\paragraph{Case 2.}
Here, we assume $\lambda_{12}=\lambda_{34}$ and $\lambda_{14}=\lambda_{23}$, which implies $\lambda_1=\lambda_3$ and $\lambda_2=\lambda_4$. The constraint \eqref{Paper1-LCholo} becomes
\begin{align}\label{Paper1-case2}
    \begin{split}
    \sum_{\omega}&\big[((\lambda_1-\lambda_2)A-\omega C)\,
    \mathcal{C}^{1212\omega}(A-B)^{\lambda_{12}+\omega-1}(A+B)^{\lambda_{12}-\omega-1}\\
    &-(-)^{\omega}\omega A\,
    \mathcal{C}^{1122\omega}(B-C)^{2\lambda_1+\omega-1}(B+C)^{2\lambda_2-\omega-1}\\
    &+((\lambda_1-\lambda_2)A-\omega B)
     \,\mathcal{C}^{1221\omega}(C-A)^{\lambda_{12}+\omega-1}(C+A)^{\lambda_{12}-\omega-1}\big]=0\,.
     \end{split}
\end{align}
First, we determine the polynomial form of the functions \eqref{Paper1-definitions2} as
\begin{subequations}\label{Paper1-case2_funct}
\begin{align}
    \nonumber
    f_-^{1212}(A,B)=&k^{1212}_-A^{\Lambda-2}-2k^{1212}_{\omega-}A^{\Lambda-3}B-k^{1212}_+ B^{\Lambda-2}\\ \label{Paper1-case2f1}
    =&k^{1212}_-A^{\Lambda-2}-k^{1212}_+ B^{\Lambda-2}\,,\\
    \nonumber
    f^{1212}_{\omega-}(A,B)=&k^{1212}_{\omega-}A^{\Lambda-2}-2k^{1212}_{\omega^2-}A^{\Lambda-3}B+2k^{1212}_{\omega^2+}AB^{\Lambda-3}\\\label{Paper1-case2f2}
    =&2(k^{1212}_{\omega^2+}AB^{\Lambda-3}-k^{1212}_{\omega^2-}A^{\Lambda-3}B)\,,\\ 
    \nonumber
    f^{1122}_{\omega+}(B,C)=&k^{1122}_{\omega+}B^{\Lambda-2}+2((\lambda_2-\lambda_1)k^{1122}_{\omega+}-k^{1122}_{\omega^2+})B^{\Lambda-3}C\\ \label{Paper1-case2f3}
    &-2((\lambda_2-\lambda_1)k^{1122}_{\omega-}-k^{1122}_{\omega^2-})BC^{\Lambda-3}-k^{1122}_{\omega-}C^{\Lambda-2}\,.
\end{align}
\end{subequations}
In this case $k^{1212}_{\omega+}$ and $k^{1212}_{\omega-}$ are zero because the sum runs over both $\omega$ and $-\omega$, then
\begin{align}
    &k^{1212}_{\omega+}=(-)^{\lambda_{12}}\sum_{\omega}\omega\, \mathcal{C}^{1212\omega}=0\,,&
    k^{1212}_{\omega-}=\sum_{\omega}(-)^{\omega}\omega\, \mathcal{C}^{1212\omega}=0\,.
\end{align}
The same happens to the coefficients $k^{1221}_{\omega+}$ and $k^{1221}_{\omega-}$.
The functions $f^{1221}_-(C,A)$ and $f^{1221}_{\omega-}(C,A)$ have the same form as $f^{1212}_-(A,B)$ and $f^{1212}_{\omega-}(A,B)$.

Now, by using \eqref{Paper1-useful_relations}, we have $k_{\pm}^{1212}=k_{\mp}^{1221}$ and by substituting the functions back into \eqref{Paper1-case2}, we obtain the constraints
\begin{equation}\label{Paper1-case2conditions}
    k^{1122}_{\omega-}=(\lambda_2-\lambda_1)k^{1212}_-=k^{1122}_{\omega+}\,,
\end{equation}
and
\begin{align}\label{Paper1-case2further}
    &k^{1122}_{\omega^2-}+(\lambda_1-\lambda_2)k^{1122}_{\omega-}=k^{1212}_{\omega^2-}\,,&
    &k^{1122}_{\omega^2+}+(\lambda_1-\lambda_2)k^{1122}_{\omega+}=k^{1212}_{\omega^2+}\,.
\end{align}
Notice that the only nontrivial constraints are those in \eqref{Paper1-case2conditions}, while \eqref{Paper1-case2further} are automatically satisfied, as can be seen using \eqref{Paper1-usefulrelations}.
From the relations above, we can rewrite \eqref{Paper1-case2_funct} as
\begin{subequations}
\begin{align}\label{Paper1-case2f1new2}
    f_-^{1212}(A,B)&=k^{1212}_-A^{\Lambda-2}-k^{1212}_+ B^{\Lambda-2}\,,\\\label{Paper1-case2f2new2}
    f^{1212}_{\omega-}(A,B)&=2(k^{1212}_{\omega^2+}AB^{\Lambda-3}-k^{1212}_{\omega^2-}A^{\Lambda-3}B)\,,\\\label{Paper1-case2f3new2}
    f^{1122}_{\omega+}(B,C)&=(\lambda_2-\lambda_1)k^{1212}_-(B^{\Lambda-2}-C^{\Lambda-2})+2(k^{1212}_{\omega^2-}BC^{\Lambda-3}-k^{1212}_{\omega^2+}B^{\Lambda-3}C)\,.
\end{align}
\end{subequations}
From the functions \eqref{Paper1-case2f1new2} and \eqref{Paper1-case2f2new2}, we obtain
\begin{align}\label{Paper1-case2_1212}
     \mathcal{C}^{1212\omega}&=\frac{k^{1212}_-+(-)^{\lambda_{12}+\omega}k^{1212}_+}{2^{\Lambda-2}}\begin{pmatrix}
         \Lambda-2\\
         \lambda_{12}-\omega-1
     \end{pmatrix}\qquad
     \forall\,\omega\,,\\
    \mathcal{C}^{1212\omega}&=\frac{k^{1212}_{\omega^2-}+(-)^{\lambda_{12}+\omega}k^{1212}_{\omega^2+}}{2^{\Lambda-4}(\Lambda-2)}
   \begin{pmatrix}
        \Lambda-2\\
        \lambda_{12}-\omega-1
    \end{pmatrix}
    \qquad
    \forall\,\omega\,.
\end{align}
These systems are, in fact, equivalent due to the relation given in \eqref{Paper1-usefulrelations}. From the function \eqref{Paper1-case2f3new2}, we obtain
\begin{equation}\label{Paper1-case2_1122}
    \mathcal{C}^{1122\omega}=\frac{k^{1212}_-+(-)^{\omega}k^{1212}_+}{2^{\Lambda-2}}
   \begin{pmatrix}
        \Lambda-2\\
        2\lambda_2-\omega-1
    \end{pmatrix}
    \qquad
    \forall\,\omega\neq 0\,.
\end{equation}
Also in this case, the product $C^{\lambda_1,\lambda_1,0}C^{0,\lambda_2,\lambda_2}$ remains unconstrained. Now, assuming we only have even-derivative vertices, we can rewrite \eqref{Paper1-case2_1212} and \eqref{Paper1-case2_1122} as
\begin{align}
     \mathcal{C}^{1212\omega}&=\frac{k^{1212}_-(\Lambda-2)!}{2^{\Lambda-3}(\lambda_{12}+\omega-1)!(\lambda_{12}-\omega-1)!}\qquad
     \forall\,\omega\,,\\
     \mathcal{C}^{1122\omega}&=\frac{k^{1212}_-(\Lambda-2)!}{2^{\Lambda-3}(2\lambda_1+\omega-1)!(2\lambda_2-\omega-1)!}\qquad
     \forall\,\omega\neq 0\,.
\end{align}
Here, the sum runs over odd or even values of $\omega$, depending on the parity of the external helicities. Everything discussed above holds for $\Lambda \geq 5$. For $\Lambda = 4$, although some monomial powers become degenerate, the conditions in \eqref{Paper1-case2conditions} are still satisfied, leading to the same solution.

\paragraph{Case 3.}
Here, we assume $\lambda_{12}=\lambda_{34}$, in which case the constraint becomes
\begin{align}\label{Paper1-case3}
\begin{split}
    \sum_{\omega}&\big[((\lambda_{13}-\lambda_{24})A+(\lambda_{14}-\lambda_{23})B-2\omega C)\,
    \mathcal{C}^{1234\omega}(A-B)^{\lambda_{12}+\omega-1}(A+B)^{\lambda_{12}-\omega-1}\\
    &+(-)^{\lambda_{24}+\omega}((\lambda_{14}-\lambda_{23})B-2\omega A)\,
    \mathcal{C}^{1324\omega}(B-C)^{\lambda_{13}+\omega-1}(B+C)^{\lambda_{24}-\omega-1}\\
    &+((\lambda_{13}-\lambda_{24})A-2\omega B)\,
     \mathcal{C}^{1423\omega}(C-A)^{\lambda_{14}+\omega-1}(C+A)^{\lambda_{23}-\omega-1}\big]=0\,.
     \end{split}
\end{align}
First, we determine the polynomial form of the functions \eqref{Paper1-definitions2} as
\begin{subequations}\label{Paper1-case_3func}
\begin{align}\label{Paper1-case3f1}
   f_-^{1234}(A,B)&=k^{1234}_-A^{\Lambda-2}-k^{1234}_+B^{\Lambda-2}\,,\\ \label{Paper1-case3f2}
   f_+^{1324}(B,C)&=(-)^{\Lambda}(k^{1324}_+B^{\Lambda-2}+((\lambda_{24}-\lambda_{13})k^{1324}_+-2k^{1324}_{\omega+})CB^{\Lambda-3}-k^{1324}_-C^{\Lambda-2})\,,\\ \label{Paper1-case3f3}
   f_-^{1423}(C,A)&=k^{1423}_-C^{\Lambda-2}-((\lambda_{23}-\lambda_{14})k^{1423}_+-2k^{1423}_{\omega+})CA^{\Lambda-3}-k^{1423}_+A^{\Lambda-2}\,,\\ \label{Paper1-case3f4}
   \nonumber
    f^{1234}_{\omega-}(A,B)&=k^{1234}_{\omega-}A^{\Lambda-2}+((\lambda_{34}-\lambda_{12})k^{1234}_{\omega-}-2k^{1234}_{\omega^2-})A^{\Lambda-3}B\\
    &-((\lambda_{34}-\lambda_{12})k^{1234}_{\omega+}-2k^{1234}_{\omega^2+})AB^{\Lambda-3}-k^{1234}_{\omega+}B^{\Lambda-2}\,.
\end{align}
\end{subequations}
The functions $f_{\omega+}^{1324}$ and $f_{\omega-}^{1423}$ have the same form as $f_{\omega-}^{1234}$. We can now substitute these functions into \eqref{Paper1-case3}, leading to the constraints
\begin{align}\label{Paper1-case3conditions}
    \begin{split}
    &k^{1234}_-=k^{1423}_+,\qquad
    k^{1324}_+=(-)^{\Lambda}k^{1234}_+,\qquad
    k^{1234}_{\omega+}=k^{1234}_{\omega-}=0\,,\\
    &k^{1324}_{\omega-}=\frac{1}{2}(-)^{\Lambda}(\lambda_{24}-\lambda_{13})k^{1423}_-\,,\qquad
    k^{1423}_{\omega-}=\frac{1}{2}(-)^{\Lambda}(\lambda_{23}-\lambda_{14})k^{1324}_-\,,\\
    &((\lambda_{23}-\lambda_{14})k^{1423}_{\omega-}-2k^{1423}_{\omega^2-})=(-)^{\Lambda}((\lambda_{24}-\lambda_{13})k^{1324}_{\omega-}-2k^{1324}_{\omega^2-})\,,\\
    \end{split}
\end{align}
and
\begin{equation}
\begin{aligned}\label{Paper1-case3further}
    &k^{1324}_{\omega+}=\frac{1}{2}(\lambda_{24}-\lambda_{13})k^{1324}_+\,,\qquad
    &&k^{1423}_{\omega+}=\frac{1}{2}(\lambda_{23}-\lambda_{14})k^{1423}_+\,,\\
    &k^{1234}_{\omega^2-}=\frac{1}{2}((\lambda_{14}-\lambda_{23})k^{1423}_{\omega+}+2k^{1423}_{\omega^2+})\,,
    &&k^{1234}_{\omega^2+}=\frac{1}{2}(-)^{\Lambda}((\lambda_{23}-\lambda_{14})k^{1324}_{\omega+}+2k^{1324}_{\omega^2+})\,.
\end{aligned}
\end{equation}
Using the relations above, we can rewrite \eqref{Paper1-case_3func} as
\begin{subequations}
\begin{align}\label{Paper1-case3f1new}
    f_-^{1234}(A,B)&=k^{1234}_-A^{\Lambda-2}-k^{1234}_+B^{\Lambda-2}\,,\\ \label{Paper1-case3f2new}
    f^{1324}_+(B,C)&=(-)^{\Lambda}(k^{1324}_+B^{\Lambda-2}-k^{1324}_-C^{\Lambda-2})\,,\\ \label{Paper1-case3f3new}
    f^{1423}_-(C,A)&=k^{1423}_-C^{\Lambda-2}-k^{1423}_+A^{\Lambda-2}\,,\\ \label{Paper1-case3f4new}
    f_{\omega-}^{1234}(A,B)&=2(k_{\omega^2+}^{1234}AB^{\Lambda-3}-k_{\omega^2-}^{1234}A^{\Lambda-3}B)\,. 
\end{align}
\end{subequations}
Notice that the only nontrivial constraints are those in \eqref{Paper1-case3conditions}, while \eqref{Paper1-case3further} are automatically
satisfied, as can be seen using \eqref{Paper1-usefulrelations}.
From the function \eqref{Paper1-case3f1new}, we find
\begin{equation}
    \mathcal{C}^{1234\omega}=\frac{(k^{1234}_- +(-)^{\omega+\lambda_{12}}k^{1234}_+)(\Lambda-2)!}{2^{\Lambda-2}(\lambda_{12}+\omega-1)!(\lambda_{34}-\omega-1)!}\qquad 
    \forall\;\omega\,,
\end{equation}
and an identical expression holds for the external helicity configurations $(1324)$ and $(1423)$.

Now, assuming only even-derivative vertices, from the constraints \eqref{Paper1-case3conditions} we obtain
\begin{equation}
    k^{1234}_+=k^{1324}_+=k^{1423}_+\,,
\end{equation}
that leads to the solution
\begin{equation}
    \mathcal{C}^{1234\omega}=\frac{k^{1234}_+(\Lambda-2)!}{2^{\Lambda-3}(\lambda_{12}+\omega-1)!(\lambda_{34}-\omega-1)!}\qquad 
    \forall\;\omega\,.
\end{equation}
The same solution applies to the other two orderings of external helicities $(1324)$ and $(1423)$. Everything discussed above holds for $\Lambda \geq 5$. For $\Lambda = 4$, although some monomial powers become degenerate, the conditions in \eqref{Paper1-case3conditions} are still satisfied, leading to the same solution.

\paragraph{Case 4.}
Here, we solve \eqref{Paper1-LCholo} assuming generic helicities. First, we determine the polynomial form of the functions \eqref{Paper1-definitions2} as
\begin{subequations}\label{Paper1-case4_func}
\begin{align}\label{Paper1-case4f1}
   f_-^{1234}(A,B)=&k^{1234}_-A^{\Lambda-2}-k^{1234}_+B^{\Lambda-2}\,,\\ \label{Paper1-case4f2}
   \nonumber
    f^{1234}_{\omega-}(A,B)=&k^{1234}_{\omega-}A^{\Lambda-2}+((\lambda_{34}-\lambda_{12})k^{1234}_{\omega-}-2k^{1234}_{\omega^2-})A^{\Lambda-3}B\\
    &+((\lambda_{12}-\lambda_{34})k^{1234}_{\omega+}+2k^{1234}_{\omega^2+})AB^{\Lambda-3}-k^{1234}_{\omega+}B^{\Lambda-2}\,.
\end{align}
\end{subequations}
The functions $f_+^{1324}(B,C)$ and $f_-^{1423}(C,A)$ have the same form as $f_-^{1234}(A,B)$, and the functions $f_{\omega+}^{1324}(B,C)$ and $f_{\omega-}^{1423}(C,A)$ have the same form as $f^{1234}_{\omega-}(A,B)$.

From equation \eqref{Paper1-case4f1}, we can extract the system for the couplings and by substituting the functions \eqref{Paper1-case4_func} into \eqref{Paper1-LCholo}, also the relations among them:
\begin{align}\label{Paper1-case4system1}
    \begin{split}
&\mathcal{C}^{1234\omega}=\frac{(k^{1234}_- +(-)^{\omega+\lambda_{12}}k^{1234}_+)(\Lambda-2)!}{2^{\Lambda-2}(\lambda_{12}+\omega-1)!(\lambda_{34}-\omega-1)!}\qquad 
    \forall\;\omega\,,\quad \text{same for $(1324)$ and $(1423)$}\,,\\
     &k^{1234}_-=k^{1423}_+\,,\qquad
    k^{1324}_+=(-)^{\Lambda}k^{1234}_+\,,\qquad
   k^{1423}_-=(-)^{\Lambda}k^{1324}_-\,.
   \end{split}
\end{align}
This gives the most general solution to the constraint \eqref{Paper1-LCholo} in the presence of both even- and odd-derivative couplings. We also find these further constraints
\begin{align}\label{Paper1-case4further}
    \begin{split}
   &k^{1234}_{\omega-}=\frac{1}{2}(\lambda_{34}-\lambda_{12})k^{1234}_-\,,\quad
   k^{1234}_{\omega+}=\frac{1}{2}(\lambda_{34}-\lambda_{12})k^{1234}_+\,,\quad
   k^{1324}_{\omega+}=\frac{1}{2}(\lambda_{24}-\lambda_{13})k^{1324}_+\,,\\
   &k^{1324}_{\omega-}=\frac{1}{2}(\lambda_{24}-\lambda_{13})k^{1324}_-\,,\quad
   k^{1423}_{\omega-}=\frac{1}{2}(\lambda_{23}-\lambda_{14})k^{1423}_-\,,\quad
   k^{1423}_{\omega+}=\frac{1}{2}(\lambda_{23}-\lambda_{14})k^{1423}_+\,,\\
   &(\lambda_{34}-\lambda_{12})k^{1234}_{\omega-}-2 k^{1234}_{\omega^2-}=((\lambda_{23}-\lambda_{14})k^{1423}_{\omega+}-2 k^{1423}_{\omega^2+})\,,\\
   &(\lambda_{24}-\lambda_{13})k^{1324}_{\omega+}-2 k^{1324}_{\omega^2+}=(-)^{\Lambda}((\lambda_{34}-\lambda_{12})k^{1234}_{\omega+}-2 k^{1234}_{\omega^2+})\,,\\
   &(\lambda_{23}-\lambda_{14})k^{1423}_{\omega-}-2 k^{1423}_{\omega^2-}=(-)^{\Lambda}((\lambda_{24}-\lambda_{13})k^{1324}_{\omega-}-2 k^{1324}_{\omega^2-})\,.
   \end{split}
\end{align}
However, these are automatically satisfied, as can be seen using \eqref{Paper1-usefulrelations}. Everything discussed above holds for $\Lambda \geq 5$. For $\Lambda = 4$, although some monomial powers become degenerate, the conditions in \eqref{Paper1-case4system1} are still satisfied, leading to the same solution.

The next problem is to find a meaningful solution\footnote{By this, we mean a solution that involves several couplings, perhaps, including the gravitational ones.} to \eqref{Paper1-case4system1}. This can be done by examining each case individually; however, finding a general procedure to classify all solutions is not straightforward.

For instance, we observe that a solution involving only odd-derivative vertices (i.e. $k^{\cdots\cdot}_+=-k^{\cdots\cdot}_-$) is inconsistent with \eqref{Paper1-case4system1}. Consequently, any consistent solution must involve a nontrivial combination of both even- and odd-derivative couplings (or only even).

Let us rewrite the coefficients in a more convenient form, which will facilitate the identification of additional constraints:
\begin{align}\label{Paper1-new_variablesE}
    k^{1234}_E=\frac{1}{2}(k^{1234}_-+k^{1234}_+)&=\sum_{(\lambda_{12}+\omega)\in\text{even}}\mathcal{C}^{1234\omega}\,,\\ \label{Paper1-new_variablesO}
    k^{1234}_O=\frac{1}{2}(k^{1234}_--k^{1234}_+)&=\sum_{(\lambda_{12}+\omega)\in\text{odd}}\mathcal{C}^{1234\omega}\,.
\end{align}
We can then rewrite the system governing the couplings in the form
\begin{align}\label{Paper1-even_first}
    \mathcal{C}^{1234\omega}&=\frac{k^{1234}_E (\Lambda-2)!}{2^{\Lambda-3}(\lambda_{12}+\omega-1)!(\lambda_{34}-\omega-1)!}\qquad 
    \forall\;(\lambda_{12}+\omega)\in\text{even}\,,\\ \label{Paper1-odd_first}
    \mathcal{C}^{1234\omega}&=\frac{k^{1234}_O (\Lambda-2)!}{2^{\Lambda-3}(\lambda_{12}+\omega-1)!(\lambda_{34}-\omega-1)!}\qquad 
    \forall\;(\lambda_{12}+\omega)\in\text{odd}\,.
\end{align}
Recall that $\mathcal{C}^{1234\omega}$ is a product of two terms. Therefore, following the procedure outlined in Appendix A of \cite{Ponomarev:2016lrm}, assuming that all even couplings are non-vanishing, and using \eqref{Paper1-even_first} we find that the unique solution (Metsaev solution) for the couplings is
\begin{equation}\label{Paper1-alleven}
    C^{\lambda_1,\lambda_2,\lambda_3}\sim \frac{1}{(\lambda_1+\lambda_2+\lambda_3-1)!}\,,\qquad
    \sum_{i=1}^3\lambda_i\in\text{even}\,.
\end{equation}
Similarly, assuming that all odd-derivative vertices are activated and using \eqref{Paper1-odd_first}, we obtain
\begin{equation}\label{Paper1-allodd}
    C^{\lambda_1,\lambda_2,\lambda_3}\sim \frac{\epsilon^{\lambda_1\lambda_2\lambda_3}}{(\lambda_1+\lambda_2+\lambda_3-1)!}\,,\qquad
    \sum_{i=1}^3\lambda_i\in\text{odd}\,,
\end{equation}
where $\epsilon^{\lambda_1\lambda_2\lambda_3}$ is a totally antisymmetric tensor. Using \eqref{Paper1-new_variablesE} and \eqref{Paper1-new_variablesO}, the constraint \eqref{Paper1-case4system1} becomes
\begin{align}
    \begin{split}
    k^{1234}_E+k^{1234}_O&=k^{1423}_E-k^{1423}_O\,,\qquad
   k^{1324}_E-k^{1324}_O=(-)^{\Lambda}(k^{1234}_E-k^{1234}_O)\,,\\
   k^{1423}_E+k^{1423}_O&=(-)^{\Lambda}(k^{1324}_E+k^{1324}_O)\,.
   \end{split}
\end{align}
Now, assuming \eqref{Paper1-alleven}, which implies $k^{1234}_E=k^{1324}_E=k^{1423}_E$, we arrive at
\begin{align}
    &k^{1234}_O=-k^{1423}_O\,,&
    &k^{1234}_O=(-)^{\Lambda}k^{1324}_O=k^{1423}_O\,,
\end{align}
that implies $k^{\cdots\cdot}_O=0$. Therefore, in the presence of all even-derivative couplings, no consistent solution can accommodate the inclusion of any odd-derivative interactions.

On the other hand, assuming \eqref{Paper1-allodd}, we have $k^{1234}_O=k^{1324}_O=k^{1423}_O=k_O$, which leads to
\begin{align}
\begin{split}
&k^{1234}_E-k^{1423}_E+2k_O=0,\qquad
k^{1324}_E+(-)^{\Lambda+1}k^{1234}_E+(-)^{\Lambda}k_O-k_O=0\,,\\
&k^{1423}_E+(-)^{\Lambda+1}k^{1324}_E+(-)^{\Lambda+1}k_O+k_O=0\,.
\end{split}
\end{align}
Attempting to find a solution for even or odd $\Lambda$ leads to $k_O=0$. Therefore, this case is also ruled out.

We conclude that any viable solution must involve a truncated spectrum --- i.e. one that cannot contain all even- and/or odd-derivative couplings --- and must feature nontrivial relations among the couplings. A detailed investigation of this possibility is left for future work. From this point onward, we restrict our analysis to the case in which only even-derivative vertices are present.

When only even-derivative vertices are present (i.e. $k^{\cdots\cdot}_+=k^{\cdots\cdot}_-)$, the solution to \eqref{Paper1-case4system1} yields
\begin{equation}\label{Paper1-unique_even_solution}
    k^{1234}_+=k^{1324}_+=k^{1423}_+=k^{1234}_-=k^{1324}_-=k^{1423}_-\,,
\end{equation}
and these are the solutions we will study in detail.

\paragraph{Summary.} The general solution to the holomorphic constraint \eqref{Paper1-LCholo} is given by
\begin{equation}
\boxed{
\begin{aligned}\label{Paper1-case4system}
&\mathcal{C}^{1234\omega}=\frac{(k^{1234}_- +(-)^{\omega+\lambda_{12}}k^{1234}_+)(\Lambda-2)!}{2^{\Lambda-2}(\lambda_{12}+\omega-1)!(\lambda_{34}-\omega-1)!}\quad 
    \forall\;\omega\,,\quad \text{same for $(1324)$ and $(1423)$}\,,\\ 
    &k^{1234}_-=k^{1423}_+\,,\;\;
    k^{1324}_+=(-)^{\Lambda}k^{1234}_+\,,\;\;
   k^{1423}_-=(-)^{\Lambda}k^{1324}_-\,,\;\;C^{\lambda_1,\lambda_1,0}C^{0,\lambda_2,\lambda_2}=\;\text{generic}\,.
\end{aligned}
}
\end{equation}
Assuming only even-derivative vertices, the solution to the holomorphic constraint is given by
\begin{equation}
\boxed{
\begin{aligned}\label{Paper1-symfinalsystem}
    &\mathcal{C}^{1234\omega}=\frac{k^{1234}_+(\Lambda-2)!}{2^{\Lambda-3}(\lambda_{12}+\omega-1)!(\lambda_{34}-\omega-1)!}\quad 
    \forall\;\omega\,,\quad \text{same for $(1324)$ and $(1423)$}\,,\\
    &k_+^{1234}= k_+^{1324}= k_+^{1423}\,,\qquad
    C^{\lambda_1,\lambda_1,0}C^{0,\lambda_2,\lambda_2}=\;\text{generic}\,.
\end{aligned}
}
\end{equation}
Let us stress that $\mathcal{C}^{1234\omega}$ is not assumed to factorize into $C^{\lambda_1,\lambda_2,\lambda_3}$.

\paragraph{Lower-spin analysis.}
We begin by examining solutions for lower helicities, specifically $0,1,2$. The corresponding even-derivative couplings are given by
\begin{equation}
    \{C^{-2,2,2},C^{-1,1,2},C^{0,0,2},C^{0,1,1},C^{0,2,2},C^{1,1,2},C^{2,2,2}\}\,.
\end{equation}
In total, we have $7$ couplings: $5$ abelian and $2$ non-abelian.\footnote{The non-abelian vertices are the most interesting ones, as they deform the gauge algebra in a covariant formulation. The abelian ones do not. Nevertheless, it is worth noting that, in general, abelian vertices can also be constrained and may play an important role in ensuring the consistency of the theory. Since we work in the light-cone gauge and with holomorphic theories, we define abelian couplings as those with $\lambda_i\geq0$ and $\sum_i\lambda_i>0$. Note that $(0,s,s)$ is still considered abelian this way, which is justified since the spin-zero exchange disappears from the constraint. Such couplings are, obviously, consistent on their own unless there are non-abelian couplings. } We can now explore all possible chiral theories that can be constructed from these couplings.

We begin with two-derivative theories, involving the couplings $\{C^{-2,2,2},C^{-1,1,2},C^{0,0,2},C^{0,1,1}\}$.
All possible two-derivative chiral theories with lower-spin fields that solve the constraints are
\begin{subequations}\label{Paper1-lower_der_lower_spin}
\begin{align}
        &\{C^{-2,2,2}\}\,,&&\text{graviton coupling}\,,\\
        &\{C^{1,1,0}\}\,,&&\text{photons coupled to scalars}\,,\\
        &\{C^{-2,2,2}=C^{0,0,2}\}\,,&&\text{graviton coupled to scalars}\,,\\\label{Paper1-case4lowerspin}
        &\{C^{-2,2,2}=C^{-1,1,2}\}\,,&&\text{graviton coupled to photons}\,,\\\label{Paper1-case5lowerspin}
        &\{C^{-2,2,2}=C^{-1,1,2}=C^{0,0,2},C^{0,1,1}\}\,,&& \text{graviton, photons and scalars}\,.
\end{align}
\end{subequations}
We use the following notation: within $\{\cdots\}$, we list the active couplings. When a coupling appears alone, it is considered unconstrained --- that is, it can appear in the theory’s action with a free coefficient. However, the constraints may sometimes impose nontrivial relations among these couplings. It is important to specify the active couplings before solving the constraints. Notably, the resulting theories do not always correspond to truncations of the larger one. For instance, the above theories do not always correspond to a truncation of the larger one \eqref{Paper1-case5lowerspin} by setting to zero certain couplings, but rather to distinct theories. For example, to obtain \eqref{Paper1-case4lowerspin} from \eqref{Paper1-case5lowerspin}, we remove $C^{0,0,2}$, which is not equivalent to setting it to zero in \eqref{Paper1-case5lowerspin}.
Note that the relations above follow from the well-known universality of gravitational interactions.

If we allow for the higher-derivative terms, the most general theory we obtain is
\begin{equation}\label{Paper1-HDlowerspin}
    \{C^{-2,2,2}=C^{-1,1,2}=C^{0,0,2},C^{0,1,1},C^{0,2,2},C^{1,1,2},C^{2,2,2}\}\,.
\end{equation}
The fact that the couplings satisfy the constraints implies that a consistent action can be written in the light-cone gauge. In principle,\footnote{In some cases, a covariant formulation of the theory may not exist or may require introducing non-localities. However, a covariant description can often be found, as is the case here.} it may be possible to find a covariant formulation of the theory with an appropriate gauge symmetry.

Let us consider, for example, the largest two-derivative theory \eqref{Paper1-case5lowerspin}. In this case, the covariant (parity completed) action corresponds to the truncation to the cubic order of the Einstein-Maxwell-scalar theory
\begin{equation}
    S=\int d^4x\sqrt{-g}\left(\frac{1}{2k}R-\frac{1}{2}\nabla^{\mu}\varphi\nabla_{\mu}\varphi-\frac{1}{4}F^{\mu\nu}F_{\mu\nu}+\frac{1}{4}a\,\varphi F^{\mu\nu}F_{\mu\nu}\right),
\end{equation}
where $\nabla$ is the gravitational covariant derivative, $g$ is the determinant of the metric, $k = 8\pi G$ is the gravitational coupling, $R$ is the Ricci scalar, $\varphi$ is the scalar  field, and $F_{\mu\nu} = \pl_{\mu} A_{\nu} - \pl_{\nu} A_{\mu}$ is the Maxwell–Faraday $2$-form. The parameter $a$ is the coupling constant between the scalar and the vector field. The values of the coefficients in the action are directly related to those appearing in light-cone gauge: we have $k \sim C^{-2,2,2}$, while the relations among the other couplings are determined by minimal couplings imposed by diffeomorphism invariance, and $a \sim C^{0,1,1}$.
Note that we can always include a scalar cubic coupling $C^{0,0,0}$, corresponding to a potential term of the form $V(\varphi) \sim \varphi^3$.
As expected from the covariant perspective, most of the abelian couplings remain unconstrained.\footnote{This is a typical feature of lower-spin theories. As we introduce higher-spin interactions, we will see that many abelian couplings become constrained as well.}

The covariant (parity completed) higher-derivative action of \eqref{Paper1-HDlowerspin} corresponds to the truncation to the cubic order of 
\begin{align}
\nonumber
    S&=\int d^4x\sqrt{-g}\left(\frac{1}{2k}R-\frac{1}{2}\nabla^{\mu}\varphi\nabla_{\mu}\varphi-\frac{1}{4}F^{\mu\nu}F_{\mu\nu}+\frac{1}{4}a\,\varphi F^{\mu\nu}F_{\mu\nu}+b R^3 +c RF^2+dR^2\varphi\right)\,,\\
    \nonumber
    &R^3=R_{\mu\nu\rho\sigma}R^{\mu\nu}_{\phantom{\mu\nu}\alpha\beta}R^{\rho\sigma\alpha\beta}\,,\qquad
    RF^2=R_{\mu\nu\rho\sigma}F^{\mu\nu}F^{\rho\sigma}\,,\qquad
    R^2=R_{\mu\nu\rho\sigma}R^{\mu\nu\rho\sigma}\,,\\
    & k\sim C^{-2,2,2}\,,\qquad 
    a\sim C^{0,1,1}\,,\qquad 
    b\sim C^{2,2,2}\,,\qquad 
    c\sim C^{1,1,2}\,,\qquad 
    d\sim C^{0,2,2}\,.
\end{align}
It is worth emphasising the importance of the freedom in the product $C^{\lambda_1,\lambda_1,0}C^{0,\lambda_2,\lambda_2}$. Without this freedom, we would obtain the constraint (as in the case of chiral higher-spin gravity)
\begin{align}\label{Paper1-wrong_condition}
    &C^{2, 2, 2}=\frac{3}{10}\frac{(C^{0, 2, 2})^2}{C^{-2, 2, 2}}\,,&
    &C^{1, 1, 2}=\frac{C^{0, 1, 1} C^{0, 2, 2}}{C^{-2, 2, 2}}\,.
\end{align}
This would be in contrast with the covariant framework, where it is perfectly possible to introduce both $R^3$ and $FR^2$ terms with arbitrary coefficients in a Lorentz-invariant way. This is a further consistency check for our solutions.

We also point out that a method for constructing covariant chiral Lagrangians was recently developed in \cite{Krasnov:2021nsq}, using the spinorial representation of the Lorentz group in four dimensions, $SO(3,1)$. This formalism appears to be the most suitable for formulating Lagrangian descriptions of all consistent chiral higher-spin theories identified in the light-cone approach.

\paragraph{Higher-spin analysis.}
We now begin to search for complete solutions involving higher-spin fields. In particular, we need to require that $\mathcal{C}^{1234\omega}$ factorizes into the product of two couplings. While it is commonly believed that the presence of higher-spin fields imposes strong constraints on the theory, requiring an infinite spectrum of fields, we will show that this is not always the case. In fact, we can find many interesting families of solutions, some with finite spectra and others with infinite ones. Well-known examples such as HS-SDGR, HS-SDYM, and chiral higher-spin gravity will appear as special cases.

To better understand the underlying logic, particularly for higher-derivative interactions, we will sometimes label the couplings by their number of derivatives, rather than by the specific helicities. We denote a generic $n$-derivative coupling as $C^n$, allowing us to schematically rewrite the system of couplings \eqref{Paper1-symfinalsystem} as
\begin{equation}\label{Paper1-LCcouplingder}
C^nC^m=C^mC^n=\frac{k_{\Lambda}(\Lambda-2)!}{2^{\Lambda-3}(n-1)!(m-1)!}\,,\qquad
k_{\Lambda}=\sum_{\omega}\mathcal{C}^{1234\omega}\,,\qquad
\Lambda=n+m\,.
\end{equation}
Remember, we consider $n,m$ even and $n,m\geq 2$.
The key observation is that \eqref{Paper1-LCholo} decomposes into independent constraints at fixed total derivative order $\Lambda$; the same can be seen from the system above. Let us assume a chiral theory that includes $2$- and $4$-derivative interactions. Using \eqref{Paper1-LCcouplingder}, we can write the system (very schematically)\footnote{We emphasize that this is just a sketch to understand the main logic behind the existence of a truncated spectrum, especially for higher derivative theories. To identify consistent theories, one must solve all the constraints explicitly. This will be carried out in the next section for lower-derivative theories.} as
\begin{equation}
    \begin{cases}
    C^2C^2=C^2C^2\,,\\
    C^2C^4=C^2C^4\,,\\
    C^4C^4=\frac{5}{8}(C^4C^4+2\,C^2C^6)\,.
    \end{cases}
\end{equation}
The first line corresponds to the $2$-derivative sector, which is consistent on its own and includes HS-SDGR and other theories all contained in the former. However, problems arise when we try to go beyond $2$-derivative interactions.

If we include $4$-derivative interactions, the second constraint above is consistent and admits solutions, but the third equation becomes inconsistent unless $6$-derivative couplings are introduced.
Proceeding to the next order, we can resolve the previous inconsistency, but another one will arise:
\begin{equation}
    \begin{cases}
    C^4C^4=\frac{5}{8}(C^4C^4+2\,C^2C^6)\,,\\
    C^2C^6=\frac{3}{16}(C^4C^4+2\,C^2C^6)\,,\\
    C^4C^6=\frac{7}{16}(2\,C^4C^6+2\,C^2C^8)\,,\\
    C^6C^6=\frac{63}{128}(C^6C^6+2\,C^4C^8+2\,C^2C^{10})\,.\\
    \end{cases}
\end{equation}
The first two equations are consistent and fix the previous problem. However, truncating the theory at $6$ derivatives makes the remaining equations inconsistent, thereby requiring the inclusion of terms up to  $10$-derivative couplings. This structure arises from the identity
\begin{equation}
    \sum_{\omega}\frac{(\Lambda-2)!}{2^{\Lambda-3}(\lambda_{12}+\omega-1)!(\lambda_{34}-\omega-1)!}=1\qquad 
    \forall\;\omega\,,
\end{equation}
which underlies the consistency of the system \eqref{Paper1-symfinalsystem}.

This pattern persists at all orders: truncating the theory at any finite number of derivatives generally introduces inconsistencies at higher orders, pointing toward a unique solution with all couplings turned on, which is the full chiral higher-spin theory.

However, there is an alternative. One can impose that lower-derivative couplings do not generate higher-order constraints. For instance, a theory involving only $C^2$ and $C^4$  couplings must be arranged so that no $C^4C^4$ term ever arises. Surprisingly, this condition can be satisfied, and it provides a pathway for classifying all consistent chiral higher-spin theories in $4d$ flat space.

\subsection{Light-cone holomorphic constraint with U(N) gauge group}
Here, we solve the holomorphic constraint in the presence of a $U(N)$ gauge group in the most general setting, allowing for both even- and odd-derivative vertices. The generators of the $u(N)$ algebra satisfy the Poisson bracket
\begin{equation}\label{Paper1-U(N)poisson_bracket}
  [(\phi^{\lambda}_p)^A_{\;B},(\phi^{\mu}_q)^C_{\;D}]=\frac{\delta^{\lambda,-\mu}\delta^3(p+q)}{2p^+}\,\theta_{\lambda}\delta^C_{\;B}\delta^A_{\;D}\,,
\end{equation}
where $\theta_{\lambda}=e^{ix\lambda}$ is a phase factor reflecting a possible ambiguity in the commutator.

The constraint is given in \eqref{Paper1-holocolour}. Due to the trace, we can decompose the contributions into several colour-ordered terms, as
\begin{align}\label{Paper1-colour-ordered_terms}
    \begin{split}
    (1234)=&\;[1234]+[1342]+[1423]+[1324]+[1243]+[1432]\,,\\
    [1234]=&\;\{1234\}+\{2341\}+\{3412\}+\{4123\}\,.
    \end{split}
\end{align}
We will focus on a single colour-ordered term, $[1234]$, as the same will follow for the other orders. The constraint for $[1234]$ takes the form
\begin{equation}\label{Paper1-Unconstraint}
\sum_{\omega}\text{Cycl}\Big[(-)^{\omega}\theta_{\omega}\frac{(\lambda_1+\omega-\lambda_2)\beta_1-(\lambda_2+\omega-\lambda_1)\beta_2}{\beta_1+\beta_2}\,
\mathcal{C}^{1234\omega}\PPb_{12}^{\lambda_{12}+\omega-1}\PPb_{34}^{\lambda_{34}-\omega}\Big]=0\,.
\end{equation}
As before, we can apply the trick of summing the terms $(1234)\leftrightarrow (3412)$ and $\omega\leftrightarrow -\omega$, followed by a change of variables to $A,B,C$. However, we must now account for the phase $\theta_{\omega}$, and the simplification works only if we have $\theta_{\omega}=\theta_{-\omega}$. This is the case due to the antisymmetry of the Poisson bracket.\footnote{Using the antisymmetric property of the Poisson bracket, we have
\begin{equation}
    [(\phi^{\lambda}_p)^A_{\;B},(\phi^{\mu}_q)^C_{\;D}]=\frac{\delta^{\lambda,-\mu}\delta^3(p+q)}{2p^+}\,\theta_{\lambda}\delta^C_{\;B}\delta^A_{\;D}=-[(\phi^{\mu}_q)^C_{\;D},(\phi^{\lambda}_p)^A_{\;B}]=\frac{\delta^{\mu,-\lambda}\delta^3(p+q)}{2p^+}\,\theta_{-\lambda}\delta^A_{\;D}\delta^C_{\;B}\,,
\end{equation}
that implies $\theta_{\lambda}=\theta_{-\lambda}$.}
The constraint then becomes
\begin{align}\label{Paper1-LCholocolour}
\nonumber
    &\sum_{\omega} \big[(-)^{\omega}\theta_{\omega}((\lambda_{13}-\lambda_{24})A+(\lambda_{14}-\lambda_{23})B-2\omega C)\,
    \mathcal{C}^{1234\omega}(A-B)^{\lambda_{12}+\omega-1}(A+B)^{\lambda_{34}-\omega-1}\\
    &+(-)^{\lambda_{14}}\theta_{\omega}((\lambda_{12}-\lambda_{34})C+(\lambda_{13}-\lambda_{24})A-2\omega B)\,
    \mathcal{C}^{4123\omega}(C-A)^{\lambda_{14}+\omega-1}(C+A)^{\lambda_{23}-\omega-1}\big]=0\,.
\end{align}
Following the same conventions used in \cite{Skvortsov:2020wtf} and in analogy with the previous case, we fix $\theta_{\omega}=(-)^{\omega}$. This choice is again equivalent to taking even-helicity fields to be Hermitian and odd-helicity fields to be anti-Hermitian matrices.

As in the previous case, we divide the solution into several subcases. However, we note that the constraints \eqref{Paper1-LCholo} and \eqref{Paper1-LCholocolour} are structurally very similar. In fact, the only difference lies in the form of the functions, which, in this case, include only half of the terms present in the previous one.
Therefore, we focus on two representative cases, as the others follow straightforwardly.

In the following, we assume $\Lambda \geq 3$. Indeed, as in the previous case, for lower derivatives, fewer constraints arise due to the partially degenerate structure of the monomials. Therefore, the case $\Lambda = 2$ is treated separately in Appendix \ref{Paper1-AppendixC}.

\paragraph{Case 1.}
Here, we assume all helicities are equal, i.e. $\lambda_1=\lambda_2=\lambda_3=\lambda_4$. In this case, the constraint \eqref{Paper1-LCholocolour} becomes
\begin{align}
\begin{split}
    \sum_{\omega}\omega\,\mathcal{C}^{1111\omega}\big(&C
    (A-B)^{2\lambda+\omega-1}(A+B)^{2\lambda-\omega-1}\\
    &+(-)^{\omega}B
(C-A)^{2\lambda+\omega-1}(C+A)^{2\lambda-\omega-1}\big)=0\,.
\end{split}
\end{align}
First, we determine the polynomial form of the functions \eqref{Paper1-definitions2} as
\begin{align}\label{Paper1-U(1)_case1}
    &f^{1111}_{\omega-}(A,B)=-2k^{1111}_{\omega^2_-}A^{\Lambda-3}B\,,&
    &f^{1111}_{\omega+}(C,A)=2k^{1111}_{\omega^2_-}CA^{\Lambda-3}\,.
\end{align}
Upon substituting them back into the constraint, we find it is trivially satisfied. From the functions \eqref{Paper1-U(1)_case1}, we find the system
\begin{align}
\begin{split}
    \mathcal{C}^{1111\omega}=&\;\frac{k^{1111}_{\omega^2-}(\Lambda-3)!}{2^{\Lambda-4}(2\lambda+\omega-1)!(2\lambda-\omega-1)!}\qquad
    \forall\;\omega\neq 0\,,\\
    k^{1111}_{\omega^2-}=&\;\sum_{\omega}\,\omega^2\,\mathcal{C}^{1111\omega}\,.
    \end{split}
\end{align}
As in the previous analysis, here too, just by looking at \eqref{Paper1-LCholocolour}, one can see that the product $C^{\lambda_1,\lambda_1,0}C^{0,\lambda_2,\lambda_2}$ decouples from the system; therefore, it is unconstrained. 

\paragraph{Case 2.} Here, we solve \eqref{Paper1-LCholocolour} assuming generic helicities. First, we determine the polynomial form of the functions \eqref{Paper1-definitions2} as
\begin{subequations}
\begin{align}\label{Paper1-case4colourf1}
   f_-^{1234}(A,B)=&\;k^{1234}_-A^{\Lambda-2}\,,\\ \label{Paper1-case4colour2}
   f^{1234}_{\omega-}(A,B)=&\;k^{1234}_{\omega-}A^{\Lambda-2}+((\lambda_{34}-\lambda_{12})k^{1234}_{\omega-}-2k^{1234}_{\omega^2-})A^{\Lambda-3}B\,.
\end{align}
\end{subequations}
The functions $f^{4123}_+$ and $f^{4123}_{\omega+}$ have the same form as $f_-^{1234}(A,B)$ and $f^{1234}_{\omega-}(A,B)$, respectively. Upon substituting these into \eqref{Paper1-LCholocolour} gives
\begin{equation}
    k^{1234}_-=k^{4123}_-\,.
\end{equation}
From the function \eqref{Paper1-case4colourf1}, we find the system
\begin{align}
    \begin{split}
    \mathcal{C}^{1234\omega}=&\;\frac{k^{1234}_-(\Lambda-2)!}{2^{\Lambda-2}(\lambda_{12}+\omega-1)!(\lambda_{34}-\omega-1)!}\qquad 
    \forall\;\omega\,,\\
    k^{1234}_-=&\;\sum_{\omega}\,\mathcal{C}^{1234\omega}\,.
    \end{split}
\end{align}
and an identical expression holds for the external helicity configurations $(4123)$. In all other cases, when particular relations among the couplings are imposed, the same pattern follows.

The same procedure used to find a solution to the constraint for $U(N)$ can be applied to other gauge groups, such as $G = SO(N),\,USp(N)$, where the case of $SO(N)$ was first studied in \cite{Metsaev:1991nb}. We comment on them in Appendix \ref{Paper1-AppendixD}. 

Another interesting case is when all fields are taken to live in the adjoint representation of a Lie algebra. This scenario is studied in Appendix \ref{Paper1-AppendixE}.

\paragraph{Summary.} A solution to the holomorphic constraint in the presence of $U(N)$ gauge group is given by even- and odd-derivative vertices that satisfy 
\begin{equation}
\boxed{
\begin{aligned}\label{Paper1-symfinalsystem_U(N)}
    &\mathcal{C}^{1234\omega}=\frac{k^{1234}_-(\Lambda-2)!}{2^{\Lambda-2}(\lambda_{12}+\omega-1)!(\lambda_{34}-\omega-1)!}\quad 
    \forall\;\omega\,,\quad\text{same for $(4123)$}\,,\\
    &k_-^{1234}= k_-^{4123}\,,\qquad
     C^{\lambda_1,\lambda_1,0}C^{0,\lambda_2,\lambda_2}=\;\text{generic}\,.
\end{aligned}
}
\end{equation}

\paragraph{Lower-spin analysis.}
We now examine all solutions in the presence of a gauge group, focusing on lower helicities $0,1,2$. The complete set of even- and odd-derivative couplings that can be constructed is given by
\begin{align}\label{Paper1-all_colour_couplings}
\begin{split}
    \{&C^{-2,1,2},\textcolor{blue}{C^{-2,2,2}},\textcolor{blue}{C^{-1,0,2}},C^{-1,1,1},\textcolor{blue}{C^{-1,1,2}},\textcolor{red}{C^{-1,2,2}},C^{0,0,1},\textcolor{blue}{C^{0,0,2}},\\
&C^{0,1,1},\textcolor{blue}{C^{0,1,2}},\textcolor{blue}{C^{0,2,2}},C^{1,1,1},\textcolor{red}{C^{1,1,2}},\textcolor{red}{C^{1,2,2}},\textcolor{red}{C^{2,2,2}}\}\,.
\end{split}
\end{align}
In total, we have $15$ couplings: $9$ abelian and $6$ non-abelian. We now search for possible consistent chiral theories.

We highlight the lower-spin couplings as follows: blue indicates those which, on their own, require higher-spin couplings for consistency; red denotes those which, when combined with any other coupling, also require higher-spin ones; uncoloured couplings are, as we will show in more detail below, the only consistent lower-spin couplings.

We begin by considering one-derivative interactions, involving the subset of couplings
\begin{equation}\label{Paper1-one_derivative_colour}
    \{C^{-2,1,2},C^{-1,0,2},C^{-1,1,1},C^{0,0,1}\}\,.
\end{equation}
Consistent theories are given by
\begin{subequations}
\begin{align}
    &\{C^{-1,1,1}\}\,, && \text{Yang-Mills coupling}\,,\\
    &\{C^{-1,1,1}=C^{0,0,1}\}\,,&& \text{Yang-Mills coupled to scalars}\,,\\
    &\{C^{-2,1,2}=C^{-1,1,1}\}\,,&& \text{coloured graviton}\,.
\end{align}
\end{subequations}
Note that including $C^{-1,0,2}$ and at least one other of the couplings in \eqref{Paper1-one_derivative_colour} leads, by consistency, to an infinite tower of higher-spin fields. 

If instead we look for the minimal consistent theory that includes $C^{-1,0,2}$, we still necessarily generate at least one higher-spin field. Indeed, $C^{-1,0,2}$ gives rise to an exchange via the scalar field, which is proportional to $C^{-1,0,2}C^{-1,0,2}$. Next, the minimal consistent set of couplings is given by\footnote{We are making use of the concept of the small crystal \eqref{Paper1-small_crystal_1d}, which will be introduced and justified in the next section.}  
\begin{equation}
    \begin{aligned}
     &C^{2,-1,0}C^{0,2,-1},&&\;\;&& C^{2,2,-3}C^{3,-1,-1},&&\;\;&&C^{2,-1,0}C^{0,-1,2}\,.
     \end{aligned}
\end{equation}
In particular, the solutions for the couplings give
\begin{equation}
    \{C^{-1,0,2},C^{-3, 2, 2},C^{-1, -1, 3}=\frac{(C^{-1, 0, 2})^2}{C^{-3, 2, 2}}\}\,.
\end{equation}
This is a surprising result on its own: a one-derivative lower-spin coupling induces higher-spin couplings!

Another interesting option is the coloured graviton. It is well known that, under the usual assumptions, gravitons cannot carry colour \cite{Boulanger:2000rq}. However, we find that it is possible to have multi-graviton theories provided we restrict to the self-dual case, which is thanks to $C^{-2,1,2}$. However, we expect that once the unitary completion is considered --- via the inclusion of its parity-related anti-self-dual counterpart $C^{-2,-1,2}$ --- the general quartic constraint will lead to non-localities.

Now we can start to include higher-derivative couplings, then the most general consistent theory is
\begin{equation}\label{Paper1-HDcolourtheory}
\{C^{-1,1,1}=C^{-2,1,2}=C^{0,0,1},C^{0,1,1},C^{1,1,1}\}\,.
\end{equation}
Indeed, any attempt to include one of the couplings highlighted in blue or red in \eqref{Paper1-all_colour_couplings} leads to inconsistencies, forcing us to introduce higher-spin fields.
This is consistent with expectations from the covariant formalism. For example, the coupling $C^{-2,2,2}$ would lead to ``colour gravity'', which is known not to be consistent \cite{Boulanger:2000rq}, at least including only lower spins.

Save for the $C^{-2,1,2}$ coupling, the covariant (parity completed) action, corresponding to \eqref{Paper1-HDcolourtheory}, is the truncation to the cubic order of
\begin{align}
    \begin{split}
    S=&\;\int d^4x\left(-\frac{1}{2\alpha^2}\mathrm{Tr}(F_{\mu\nu}F^{\mu\nu})-\frac{1}{2}D^{\mu}\varphi D_{\mu}\varphi+aF^3+b\,\varphi F^2\right)\,,\\
    F^3=&\;\mathrm{Tr}(F_{\mu\nu}F^{\mu}_{\phantom{\mu}\alpha}F^{\nu\alpha})\,,\qquad
    \varphi F^2=\mathrm{Tr}(\varphi F_{\mu\nu}F^{\mu\nu})\,,\\
    \alpha\sim&\;C^{-1,1,1}\,,\qquad 
    a\sim C^{1,1,1}\,,\qquad 
    b\sim C^{0,1,1}\,.
    \end{split}
\end{align}
Here, $D$ denotes the covariant derivative, $F^a_{\mu\nu}=\pl_\mu A^{a}_\nu-\pl_\nu A^{a}_\mu+f^{abc}A_{\mu}^bA_{\nu}^c$ is the YM field strength and $\alpha\sim C^{-1,1,1}$ indicates the YM coupling. The relations among the couplings are determined by the minimal coupling, required by the gauge invariance.

\paragraph{Higher-spin analysis.}The study of solutions involving higher-spin fields follows the same logic as in the case without a gauge group.

\section{Low-derivative chiral higher-spin theories}\label{Paper1-section4}
In this section, we provide a complete classification of chiral higher-spin theories with one- and two-derivatives. In particular, we identify new families of theories beyond the well-known HS-SDYM, HS-SDGR, and full chiral higher-spin gravity. These novel higher-spin theories feature a reduced spectrum, which in turn allows more freedom to the couplings.\footnote{Let us note that the problem of solving the light-cone constraints should be closely related to the problem of classifying higher-spin algebras, as in \cite{Fradkin:1986ka,Boulanger:2013zza}, see also \cite{Ponomarev:2017nrr} for the steps towards establishing such a relation.}

In the following, we always consider that at least two cubic couplings ``talk'' with each other (i.e. they share at least one pair of opposite helicities.). In fact, it is always possible to construct well-defined theories with isolated couplings that do not ``talk'', but such cases are rather trivial.
\subsection{Two-derivative chiral higher-spin theories}
Here, we classify all inequivalent solutions of chiral higher-spin theories involving two-derivative interactions.
As previously established, satisfying the holomorphic constraint \eqref{Paper1-holo2} requires respecting two basic rules:
\begin{itemize}
    \item Starting from a single pair of two-derivative couplings $C^2C^2$, four additional ones are required. In particular, the following set of couplings must be present together:
    \begin{equation}\label{Paper1-small_crystal_2d}
        C^{\lambda_1,\lambda_2,2-\lambda_{12}}C^{\lambda_{12}-2,\lambda_3,\lambda_4},\;\;C^{\lambda_1,\lambda_3,2-\lambda_{13}}C^{\lambda_{13}-2,\lambda_2,\lambda_4},\;\;C^{\lambda_1,\lambda_4,2-\lambda_{14}}C^{\lambda_{14}-2,\lambda_2,\lambda_3}\,.
    \end{equation}
    We also impose  $\sum_{i=1}^4\lambda_i=4$ to ensure that only two-derivative interactions are present. We refer to the collection of the $6$ couplings above as a small crystal. An exception is the pair $C^{\lambda_1,\lambda_1,0}C^{0,\lambda_2,\lambda_2}$, which is excluded from the small crystal, as it does not participate in the constraint \eqref{Paper1-LCholo}.
    \item These couplings have to satisfy the following system of equations:
    \begin{align}\label{Paper1-coupling2d}
    \begin{split}
    &C^{\lambda_1,\lambda_2,2-\lambda_{12}}C^{\lambda_{12}-2,\lambda_3,\lambda_4}= C^{\lambda_1,\lambda_3,2-\lambda_{13}}C^{\lambda_{13}-2,\lambda_2,\lambda_4}=C^{\lambda_1,\lambda_4,2-\lambda_{14}}C^{\lambda_{14}-2,\lambda_2,\lambda_3}\,,\\
    &C^{\lambda_1,\lambda_1,0}C^{0,\lambda_2,\lambda_2}=\text{generic}\,.
     \end{split}
    \end{align}
    While this system does not affect the classification below, it plays a role in determining relations among the couplings in any given theory. Notably, a solution is always guaranteed to exist --- namely the one corresponding to HS-SDGR --- in which all two-derivative couplings in the theory take the same value. However, in general, for smaller theories than HS-SDGR, some couplings remain unconstrained and correspond to free parameters. In what follows, we assume that all such free parameters are kept different from zero.
\end{itemize}
We begin by outlining the method used to identify solutions and introducing some key terminology. We define a crystal as a self-consistent set of couplings that, taken together, satisfy the condition \eqref{Paper1-small_crystal_2d}. Each crystal is uniquely determined by the $C^2C^2$ pair of couplings from which it originates.

In particular, starting from a set of external helicities $(\lambda_1, \lambda_2, \lambda_3, \lambda_4)$, they uniquely identify a crystal. We refer to this configuration as the seed.

For instance, the seed $(2, 2 - 2\lambda, \lambda, \lambda)$ generates the crystal
\begin{equation}
\begin{aligned}\label{Paper1-example}
     &C^{2,2-2 \lambda,-2+2 \lambda}C^{2-2 \lambda,\lambda,\lambda},&&\;\;&& C^{2,\lambda,-\lambda}C^{\lambda,2-2 \lambda,\lambda},&&\;\;&&C^{2,\lambda,-\lambda}C^{\lambda,2-2 \lambda,\lambda}\\
     &C^{2,2,-2}C^{2,-2,2},&&\;\;&&C^{2,-2,2}C^{-2,2,2},&&\;\;&&C^{2,2,-2}C^{2,2,-2}\\
     &C^{2,2,-2}C^{2,2-2 \lambda,-2+2 \lambda},&&\;\;&&C^{2,2-2 \lambda,-2+2 \lambda}C^{2-2 \lambda,2,-2+2 \lambda},&&\;\;&&C^{2,-2+2 \lambda,2-2 \lambda}C^{-2+2 \lambda,2,2-2 \lambda}\\
     &C^{2,\lambda,-\lambda}C^{\lambda,2,-\lambda},&&\;\;&&C^{2,2,-2}C^{2,\lambda,-\lambda},&&\;\;&&C^{2,-\lambda,\lambda}C^{-\lambda,\lambda,2}\,.
\end{aligned}
\end{equation}
To construct the crystal above, we do as follows. The first line represents the small crystal generated by the seed. We then search for all possible pairs that we can connect (via fields with opposite helicity) to generate other small crystals. We repeat this process till it terminates. Note that all couplings involved are two-derivative ones.

The crystal above defines a specific chiral higher-spin theory with two-derivative interactions. It contains a finite amount of couplings, a finite spectrum, and includes the SDGR interaction $C^{2,-2,2}$.
We can observe that starting from a different seed, corresponding to a small crystal other than the initial one, would generate a different crystal. For instance, starting from $(2,2,-2,2)$ leads to SDGR, producing only the second line of \eqref{Paper1-example}. Starting from $(2,\lambda,2,-\lambda)$ or $(2,2,2-2\lambda,-2+2\lambda)$ would still generate smaller solutions.

We say that two crystals are equivalent if their seeds can be mapped into each other via a combination of the following two operations:
\begin{itemize}
    \item A general permutation of the external legs:
    \begin{align}\label{Paper1-eq_rel_permutations}
   &(\lambda_1,\lambda_2,\lambda_3,\lambda_4)\rightarrow (\lambda_{\sigma_1},\lambda_{\sigma_2},\lambda_{\sigma_3},\lambda_{\sigma_4})\,,& 
    &\sigma\in S_4\,.
\end{align}
This is a simple consequence of the symmetry of the corresponding amplitude, which is also seen from the three products of couplings in \eqref{Paper1-small_crystal_2d} that contribute to the same holomorphic constraint.
    \item An affine transformation of the helicities:
    \begin{align}\label{Paper1-eq_rel_affine}
    &(\lambda_1,\lambda_2,\lambda_3,\lambda_4)\rightarrow
    (\lambda'_1,\lambda'_2,\lambda'_3,\lambda'_4)\,,&
    &\vec{{\lambda'}}_i=\hat{A}\vec{\lambda_i}+\vec{b}\,.
    \end{align}
    What we mean is that a solution can have some of the helicities $\lambda_i$ as free parameters and two solutions that differ by an affine integer-valued transformation of these free parameters should be considered equivalent.\footnote{We just mean that, for example, if the classification contains theories with helicities $\lambda,\lambda+1,\lambda+2$ and $2\lambda'-1,2\lambda',2\lambda'+1$ for any $\lambda$, $\lambda'$, then this is the same solution (parameterised by $\lambda=2\lambda'-1$).}
\end{itemize}

\paragraph{Observations and simple solutions.} Here, we present some general observations that enable us to readily recover known solutions and identify new ones.

First, HS-SDGR is a well-defined theory. It contains all higher-spin fields and all possible chiral two-derivative couplings, and is thus defined by the largest possible crystal.
Moreover, starting from a seed containing only even helicities, we will never generate odd ones. Conversely, it is not possible to construct a theory with only odd helicities. This tells us that HS-SDGR admits a consistent truncation to the even-helicity sector.

Some solutions have only (++$-$) cubic couplings. This possibility was first pointed out in \cite{Ponomarev:2017nrr} while exploring possible subalgebras of the ``gauge algebra'' of HS-SDGR (see also \cite{Monteiro:2022xwq}). A covariant action for the case including all (++$-$) couplings and no scalar fields was constructed in \cite{Krasnov:2021nsq}. 

To see the existence of such theories using our method, we have to start from a small crystal \eqref{Paper1-small_crystal_2d} and impose that the newly generated couplings never produce any ($--$+) terms. We can achieve this by requiring
\begin{equation}
    \lambda_1,\lambda_2,\lambda_3\geq n\geq 2\,,\qquad n\in\mathbb{Z}\,.
\end{equation}
As a consequence
\begin{align}
    \begin{split}
    \lambda_4=4-\lambda_{123}\leq -n\leq -2\,,&\\
    2-\lambda_{ij}\leq -n\leq -2,&\qquad i,j=\{1,2,3\}\,,\\
    \lambda_{ij}-2\geq n\geq 2,&\qquad i,j=\{1,2,3\}\,. 
    \end{split}
\end{align}
These define consistent truncations of HS-SDGR: we have a truncation for every $n\geq 2$ where all couplings will only contain fields of helicities $|\lambda|\geq n$.
We can also observe that there exist solutions containing all possible cubic couplings constructed from the truncated spectrum.\footnote{This is, in fact, a more general statement. As we will see, all consistent two-derivative chiral higher-spin theories are entirely determined by their spectrum.}
To see it, we need to prove that any coupling with $|\lambda|\geq n$ can be reached by a small crystal generated by two couplings with helicities $|\lambda|\geq n$. To do this, we simply use that pair of couplings as the seed of the small crystal. These are exactly the truncations found in \cite{Ponomarev:2017nrr}.

Note that this does not prevent the existence of consistent ``smaller''\footnote{Here by smaller we mean theories that contain fewer fields and/or fewer cubic couplings.} theories. Indeed, as we will see later, this is the case. We should interpret the solutions we just described as a collection (finite or infinite) of crystals, nested one inside the other. This idea is related to the notion of subalgebras, see \cite{Ponomarev:2017nrr}.

An additional interesting observation arises when we attempt to include a scalar field. A two-derivative cubic coupling involving a scalar takes the form $C^{\lambda,2-\lambda,0}$ and generates (even on its own, but here we assume the presence of two such couplings) the small crystal
\begin{equation}
    C^{\lambda_1,2-\lambda_1,0}C^{0,\lambda_2,2-\lambda_2},\;\;
    C^{\lambda_1,\lambda_2,2-\lambda_{12}}C^{\lambda_{12}-2,2-\lambda_1,2-\lambda_2},\;\;
    C^{\lambda_1,2-\lambda_2,\lambda_2-\lambda_1}C^{\lambda_1-\lambda_2,2-\lambda_1,\lambda_2}\,.
\end{equation}
This crystal contains both (++$-$) and ($--$+) couplings if at least one of the helicities satisfies $\lambda\geq 3$ or $\lambda\leq -1$. Therefore, it is not possible to consistently add a scalar field and retain only (++$-$) couplings while preserving Lorentz invariance; adding a scalar necessarily introduces ($--$+) couplings as well.

The only exceptions occur for $\lambda=1$, which gives a crystal with a single coupling $C^{1,1,0}$, or $\lambda=0,2$, yielding the two couplings $C^{2,0,0}$ and $C^{-2,2,2}$.

We can now ask which is the simplest theory that includes the SDGR coupling $C^{-2,2,2}$ and a generic coupling containing at least one scalar field $C^{\lambda,2-\lambda,0}$. The smallest crystal containing them is generated by the seed $(2,\lambda,0,2-\lambda)$, which gives the spectrum
\begin{equation}\label{Paper1-crystal_sdgr_scalar}
    \pm\{0,2,\lambda,\lambda-2,2\lambda-2\}\,.
\end{equation}
The cubic couplings are then all possible ones constructed from the spectrum above:
\begin{equation}
\{C^{-2,2,2},C^{2,-\lambda ,\lambda},C^{2,2-\lambda ,\lambda -2},C^{2,2-2 \lambda ,2 \lambda -2},C^{0,0,2},C^{2-2 \lambda ,\lambda ,\lambda},C^{0,2-\lambda ,\lambda},C^{2-\lambda ,2-\lambda ,2 \lambda -2}\}\,.
\end{equation}
One final observation is that we can easily recover the minimal coupling\footnote{For massless fields, the minimal coupling corresponds to the cubic interaction with the lowest number of derivatives, two for gravitational interactions and one for gauge interactions. Note also that the mass dimension of the coupling scales as $1-|\lambda_1+\lambda_2-\lambda_3|$.  Consequently, the gravitational minimal coupling has mass dimension $-1$ and is proportional to $1/M_P$, while the gauge (spin-$1$) coupling is dimensionless and corresponds to the electric charge $e$ for photons or the gauge coupling $g$ for gluons. For a more detailed explanation, including a generalization to massive higher-spin fields, we refer to \cite{Arkani-Hamed:2017jhn}. There, all cubic interactions --- not only the minimal ones --- are classified using the spinor-helicity formalism, which is also adapted to the massive case.} to gravity for higher-spin fields, thanks to the small crystal 
\begin{align}\label{Paper1-sdgr_universality}
    &C^{2,2,-2}C^{2,\lambda,-\lambda},&
    &C^{2,\lambda,-\lambda}C^{\lambda,2,-\lambda},&
    &C^{2,-\lambda,\lambda}C^{-\lambda,2,\lambda}\,.
\end{align}
This configuration yields the following solution to the system of equations:
\begin{align}
    &C^{2,2,-2}C^{2,\lambda,-\lambda}=C^{2,\lambda,-\lambda}C^{\lambda,2,-\lambda}&
    &\Rightarrow&
    C^{2,2,-2}=C^{2,\lambda,-\lambda}\,.
\end{align}
This result demonstrates the universality of gravitational interactions, even for higher-spin fields. An expression of what is commonly referred to as the equivalence principle.\\
Both the crystal \eqref{Paper1-crystal_sdgr_scalar} and \eqref{Paper1-sdgr_universality} will appear again in the classification we present below.

\paragraph{Spectrum $\Rightarrow$ couplings.} There is an interesting ``experimental fact'' that all two- and one-derivative theories we classify below have all possible cubic couplings that can be constructed from the spectrum of the helicities that participate in the interactions, which we can call ``interacting spectrum''. In other words, suppose that the free approximation contains fields with helicities $S'=\{\lambda_i'\}$, while the cubic couplings contain a subset of these helicities $S'\ni S=\{\lambda_i\}$. Then, the theory contains all cubic vertices $V_{\lambda_i,\lambda_j,\lambda_k}$, $\lambda_i+\lambda_j+\lambda_k=1 \text{ or } 2$. Some theories contain free parameters, and for the above to hold, we assume these parameters are generic --- that is, they do not vanish. The fact that the (interacting) spectrum determines the couplings is nontrivial, and as we will show later, this property no longer holds in the higher-derivative case.

\paragraph{All two-derivative solutions.}
To construct inequivalent crystals, we begin with a seed containing generic helicities $(\lambda_1,\lambda_2,\lambda_3,\lambda_4)$, with the condition $\sum_{i=1}^4\lambda_i=4$. We then proceed as follows:
\begin{itemize}
    \item First, we construct the small crystal generated by the seed and identify all possible relations among the helicities that would generate new small crystals. This is done by fixing two generic helicities to be opposite to each other.
    \item For each such relation, once imposed, we compute the newly generated crystals and then look for further constraints they induce.
    \item Since the initial small crystal contains three independent helicities --- taken to be $(\lambda_1,\lambda_2,\lambda_3)$ --- we iterate this process up to three times and then stop.
    \item Finally, we identify inequivalent crystals, modulo the equivalence relations \eqref{Paper1-eq_rel_permutations} and \eqref{Paper1-eq_rel_affine}.
\end{itemize}
In general, the solutions may also include non-integer helicities,\footnote{In particular, most of them do not contain half-integer spins, which should be interpreted as (bosonic) fermions (note, that the generators for fermions are slightly different and are not just obtained but letting $\lambda$ be half-integer). Instead, they involve helicities like $\lambda=\frac{2}{3},\frac{4}{3}$, or other generic rational values. These configurations, even though they may be mathematically interesting they are not physically relevant. For this reason, we decided to omit them in the classification. Nevertheless, it is tempting to consider $\lambda \in \mathbb{R}$ since the light-cone generators tolerate that. One can also speculate that the light-cone gauge allows one to define ``fractional spin'' in four dimensions. } that would correspond to non-physical configurations. In the classification below, we only report the physical solutions.

We present the solutions starting from the three-parameter family, then proceed to the two- and one-parameter families, followed by particular cases. However, it is important to interpret the classification in the opposite direction. The solution with the greatest helicity freedom remains valid until an additional relation is imposed, reducing the number of free parameters.

As all solutions are completely determined by the (interacting) spectrum, for each inequivalent crystal, we only write down the interacting spectrum in the following way: $\pm\{ \lambda_1, ...\}\cup\{s_1,...\}$. The spectrum of a theory consists of fields with helicity $\pm \lambda_1$, ..., $\pm s_1$, ...\footnote{The free Hamiltonian always contains pairs of fields with opposite helicities.}. The cubic vertices are given by all possible triplets of fields with helicities $\pm \lambda_1$, ... and $+s_1$, ... (no $-s_1$!) such that the total helicity is $(1)2$ for (one-)two-derivative theories.

Note that a crystal can exhibit two possible behaviours when a new relation among the helicities is imposed: 
\begin{itemize}
    \item The new relation may trigger the formation of new and different small crystals. This leads to an inequivalent crystal compared to the original one, but with one less degree of freedom in the helicities. 
    \item The new relation does not result in any new small crystals. In this case, the resulting crystal is simply a contraction of the original one, in the sense that, although it contains fewer fields and couplings, it is not independent of the original. 
\end{itemize}

We begin by describing inequivalent crystals with a finite spectrum of fields and a finite number of cubic couplings. The first case is the simplest and least interesting, involving only a single small crystal. This case gives a three-parameter family of solutions, with the following seed $(\lambda_1,\lambda_2,\lambda_3,4-\lambda_{123})$ and the crystal contains $10$ fields:
\begin{equation}\label{Paper1-2dgeneral}
    \pm\{2-\lambda_{12},2-\lambda_{13},2-\lambda_{23}\}\cup \{\lambda_1,\lambda_2,\lambda_3,4-\lambda_{123}\}\,.
\end{equation}
There are $4$ two-parameter families of finite crystals:
\begin{enumerate}
    \item The seed $(\lambda_1,\lambda_2,0,4-\lambda_{12})$ generates a crystal with $22$ fields:
    \begin{align}\label{Paper1-2pcase1}
        \begin{split}
        \pm\{&0,\lambda_1-2,\lambda_2-2,\lambda_1-\lambda_2,2\lambda_1-2,2\lambda_2-2,6-2 \lambda_{12},\\
        &2-\lambda_{12},4-\lambda_1-2 \lambda_2,4-2 \lambda_1-\lambda_2\}\cup\{\lambda_1,\lambda_2,4-\lambda_{12}\}\,.
         \end{split}
    \end{align}
    It is obtained by imposing $\lambda_3=-\lambda_3\Rightarrow\lambda_3=0$.
    \item The seed $(\lambda_1,\lambda_2,2,2-\lambda_{12})$ generates a crystal with $8$ fields:
    \begin{equation}\label{Paper1-2pcase2}
        \pm\{2,\lambda_1,\lambda_2,2-\lambda_{12}\}\,.
    \end{equation}
    It is obtained by imposing $\lambda_{13}-2=\lambda_1\Rightarrow\lambda_3=2$.
    \item The seed $(\lambda_1,\lambda_2,2-2 \lambda_1,2+\lambda_1-\lambda_2)$ generates a crystal with $13$ fields:
    \begin{equation}\label{Paper1-2pcase3}
        \pm\{\lambda_1,2\lambda_2-2,2\lambda_1-\lambda_2,2-\lambda_{12},2+2 \lambda_1-2 \lambda_2\}\cup \{\lambda_2,2-2 \lambda_1,2+\lambda_1-\lambda_2\}\,.
    \end{equation}
    It is obtained by imposing $2-\lambda_{13}=\lambda_1\Rightarrow\lambda_3=2-2\lambda_1$.
    \item The seed $(\lambda_1,\lambda_2,2-\lambda_1,2-\lambda_2)$ generates a crystal with $13$ fields:
    \begin{equation}\label{Paper1-2pcase4}
        \pm\{0,2\lambda_1-2,2\lambda_2-2,\lambda_1-\lambda_2,2-\lambda_{12}\}\cup\{\lambda_1,\lambda_2,2-\lambda_1,2-\lambda_2\}\,.
    \end{equation}
    It is obtained by imposing $2-\lambda_{13}=\lambda_{13}-2\Rightarrow\lambda_3=2-\lambda_1$.
\end{enumerate}
There are $5$ one-parameter families of finite crystals:
\begin{enumerate}
    \item The seed $(0,6-4 \lambda ,\lambda ,3 \lambda -2)$ generates a crystal with $24$ fields:
   \begin{align}
        \begin{split}
       \pm&\{0,\lambda-2,2\lambda-2,3 \lambda -4,4 \lambda -4,5 \lambda -6,6 \lambda -6,\\
       &7 \lambda -8, 8 \lambda -10,9 \lambda -10,12 \lambda -14\} \cup\{\lambda,3\lambda-2,6-4\lambda\}\,.
        \end{split}
   \end{align}
   It comes from \eqref{Paper1-2pcase1} by imposing $\lambda_2-\lambda_1=2\lambda_1-2\Rightarrow\lambda_2=3\lambda_1-2$.
   \item The seed $(0,4-3 \lambda ,\lambda ,2 \lambda)$ generates a crystal with $31$ fields:
   \begin{align}
       \begin{split}
       \pm&\{0,\lambda,\lambda-2,2\lambda-2,3 \lambda -2,4 \lambda -4,4 \lambda -2,5 \lambda -4,6 \lambda -6,6 \lambda -4,\\
       &7 \lambda -6,8 \lambda -6,9\lambda -8,10 \lambda -8,12 \lambda -10\}\cup\{2\lambda,4-3 \lambda\}\,.
       \end{split}
   \end{align}
    It comes from \eqref{Paper1-2pcase1} by imposing $\lambda_2-\lambda_1=\lambda_1\Rightarrow\lambda_2=2\lambda_1$.
     \item The seed $(4-3 \lambda,2-2\lambda,\lambda,4 \lambda-2)$ generates a crystal with $16$ fields:
    \begin{equation}
    \pm\{12 \lambda-10,9 \lambda-8,8 \lambda-6,6 \lambda-6,5\lambda-4,2 \lambda-2, \lambda\}\cup\{4-3 \lambda,4 \lambda-2\}\,.
    \end{equation}
    It comes from \eqref{Paper1-2pcase3} by imposing $\lambda_1=2-2\lambda_2$.
    \item The seed $(1,2-2 \lambda ,\lambda ,\lambda +1)$ generates a crystal with $16$ fields:
   \begin{equation}
       \pm\{0,\lambda,\lambda-1,2\lambda,2 \lambda -1,3 \lambda -1,4 \lambda -2\}\cup\{1,\lambda+1,2-2\lambda\}\,.
   \end{equation}
   It comes from \eqref{Paper1-2pcase3} by imposing $2\lambda_2-2=2-2\lambda_2\Rightarrow\lambda_2=1$.
   \item The seed $(6 \lambda -6,\lambda ,4-3 \lambda ,6-4 \lambda)$ generates a crystal with $13$ fields:
   \begin{equation}
       \pm\{2 \lambda -2,3 \lambda -4,7 \lambda -8,8 \lambda -10,12 \lambda -14\}\cup\{\lambda,6 \lambda -6,6-4 \lambda\}\,.
   \end{equation}
   It comes from \eqref{Paper1-2pcase3} by imposing $\lambda_1+\lambda_2-2=2-2\lambda_2\Rightarrow\lambda_1=4-3\lambda_2$.
\end{enumerate}
There are $4$ particular finite crystals:
\begin{enumerate}
    \item The seed $(0,1,1,2)$ generates a crystal with $5$ fields:
    \begin{equation}\label{Paper1-2d_lower_spin}
        \pm\{0,1,2\}\,.
    \end{equation}
    This corresponds to SDGR coupled to the photon and a scalar field. 
    \item The seed $(0,0,2,2)$ generates a crystal with $3$ fields:
    \begin{equation}
        \pm\{0,2\}\,.
    \end{equation}
    This corresponds to SDGR with a scalar field.
    \item The seed $(-2,2,2,2)$ generates a crystal with $2$ fields:
    \begin{equation}
        \pm\{2\}\,.
    \end{equation}
    This corresponds to SDGR.
    \item The seed $(1,1,1,1)$ generates a crystal with $2$ fields:
    \begin{equation}
        \{0,1\}\,.
    \end{equation}
    This corresponds to the abelian vertex $\varphi F^2$ alone.
\end{enumerate}
Below, we present the crystals with infinite spectra and infinitely many cubic couplings, which we refer to as infinite crystals. We explicitly indicate whether helicities $0$ and $\pm 2$ are present in the theory, as these fields are of particular relevance.  There is $1$ two-parameter family of infinite crystals:
\begin{enumerate}
    \item The seed $(-\lambda_1,\lambda_1,4-\lambda_2,\lambda_2)$ generates a crystal with the following fields:
    \begin{equation}\label{Paper1-2inf}
        \pm\{2,(2-\lambda_2)k\pm 2,(2-\lambda_2)k\pm\lambda_1\}\,,\qquad
        k\in \mathbb{Z}_{\geq 0}\,.
    \end{equation}
    It is obtained by imposing $\lambda_3=-\lambda_1$.
\end{enumerate}
There are $2$ one-parameter families of infinite crystals:
\begin{enumerate}
    \item The seed $(-2,6,-\lambda,\lambda)$ generates a crystal with the following fields:
    \begin{subequations}
    \begin{align}
        &\pm\{2,4k\}\,,&&k\in \mathbb{Z}_{\geq 0}\,,\qquad
        \lambda\equiv 2\;(\text{mod}\;4)\,,\\
        &\pm\{0,2,2k\}\,,&&k\in \mathbb{Z}_{\geq 0}\,,\qquad
        \lambda\equiv 0\;(\text{mod}\;4)\,,\\
        &\pm\{0,2,k\}\,,&&k\in \mathbb{Z}_{\geq 0}\,,\qquad
        \lambda\;\text{odd}\,.
    \end{align}
    \end{subequations}
    It comes from \eqref{Paper1-2pcase3} by imposing $\lambda_2=\lambda_2-\lambda_1-2\Rightarrow\lambda_1=-2$ or from \eqref{Paper1-2inf} by imposing $\lambda_2=-2$.
    \item The seed $(0,6-2 \lambda,\lambda-2,\lambda)$ generates a crystal with the following fields:
    \begin{equation}
        \pm\{0,2,(2-\lambda)k,(2-\lambda)k\pm 2\}\,,\qquad k\in \mathbb{Z}_{\geq 0}\,.
    \end{equation}
    It comes from \eqref{Paper1-2pcase1} by imposing $\lambda_1=\lambda_2-2$.
\end{enumerate}
There are some particular infinite solutions:\footnote{
We include these cases because they arise by fixing specific helicities in the finite crystals discussed above.}
\begin{itemize}
    \item The following seeds generate all integer helicities:
    \begin{align}\label{Paper1-exceptions1}
         \begin{split}
        \{&(0,0,1,3),(0,1,-1,4),(0,3,-3,4),(1,2,-2,3),\\
        &(1,-1,-2,6),(1,4,-4,3),(1,-1,1,3)\}\,.
        \end{split}
    \end{align}
    \item The following seeds generate all even helicities:
    \begin{align}\label{Paper1-exceptions2}
        \begin{split}
        \{&(0,2,-2,4),(0,0,0,4),(0,4,-4,4),(0,0,-2,6),\\
        &(0,6,-6,4),(0,10,-10,4),(4,-4,-2,6),(0,10,-4,-2)\}\,.
        \end{split}
    \end{align}
    \item The following seeds generate all helicities $\lambda\equiv 2$ (mod $4$):
    \begin{equation}
        \{(2,-2,-2,6),(10,-10,-2,6),(6,-6,-2,6)\}\,.
    \end{equation}
\end{itemize}
In the following, we present additional crystals which, as described above, are merely contractions of larger crystals, even though they exhibit a different spectrum. Nevertheless, they are relevant, and we include them here as well.
We also indicate whether the crystal coincides with its small crystal, and whether it contains only (++$-$) couplings. There is $1$ two-parameter family of finite crystals:
\begin{enumerate}
    \item The seed $(\lambda_1,\lambda_2,\lambda_1,4-2 \lambda_1-\lambda_2)$ generates a crystal with $7$ fields:
    \begin{equation}\label{Paper1-2pcase5}
        \pm\{2 \lambda_1-2,2-\lambda_{12}\}\cup\{\lambda_1,\lambda_2,4-2 \lambda_1-\lambda_2\}\,.
    \end{equation}
    It is a contraction of \eqref{Paper1-2dgeneral} by imposing $2-\lambda_{12}=2-\lambda_{23}\Rightarrow\lambda_3=\lambda_1$. It coincides with the small crystal.
\end{enumerate}
There are $13$ one-parameter families of finite crystals:
\begin{enumerate}
   \item The seed $(0,1,3-\lambda ,\lambda)$ generates a crystal with $21$ fields:
   \begin{equation}
       \pm\{0,1,\lambda-1,\lambda-2,2 \lambda -4,2 \lambda
   -3,2 \lambda -2,3 \lambda -5,3 \lambda -4,4 \lambda-6\}\cup\{\lambda,3-\lambda\}\,.
   \end{equation}
   It is a contraction of \eqref{Paper1-2pcase1} by imposing $\lambda_2=2-\lambda_2\Rightarrow\lambda_2=1$.
   \item The seed $(2,\lambda,0,2-\lambda)$ generates a crystal with $9$ fields:
   \begin{equation}\label{Paper1-example3}
       \pm\{0,2,\lambda,\lambda -2,2 \lambda -2\}\,.
   \end{equation}
   It is a contraction of \eqref{Paper1-2pcase1} by imposing $2\lambda_2-2=\lambda_2\Rightarrow\lambda_2=2$.
   \item The seeds $(0,\lambda ,\lambda ,4-2 \lambda)$ and $(2 \lambda -2,\lambda ,2-\lambda ,4-2 \lambda)$ generate a crystal with $11$ fields:
   \begin{equation}
       \pm\{0,\lambda-2,2 \lambda -2,3 \lambda -4,4 \lambda -6\}\cup\{\lambda,4-2 \lambda\}\,.
   \end{equation}
   It is a contraction of \eqref{Paper1-2pcase1} by imposing $\lambda_1-\lambda_2=\lambda_2-\lambda_1\Rightarrow\lambda_2=\lambda_1$ and of \eqref{Paper1-2pcase3} by imposing $2-2\lambda_1=2\lambda_2-2\Rightarrow\lambda_1=2-\lambda_2$, respectively.
   \item The seed $(0,6-3 \lambda ,\lambda ,2 \lambda -2)$ generates a crystal with $17$ fields:
   \begin{equation}
       \pm\{0,2-\lambda,2 \lambda -4,2 \lambda -2,3 \lambda -4,4 \lambda
   -6,5 \lambda -8,6 \lambda -10\}\cup\{\lambda,6-3\lambda\}\,.
   \end{equation}
   It is a contraction of \eqref{Paper1-2pcase1} by imposing $\lambda_2=2\lambda_1-2$.
   \item The seed $(2,2,-\lambda ,\lambda)$ generates a crystal with $4$ fields:
   \begin{equation}\label{Paper1-simplestcase}
       \pm\{2,\lambda \}\,.
   \end{equation}
   It is a contraction of \eqref{Paper1-2pcase2} by imposing $\lambda_2=2$. This crystal contains only (++$-$) couplings. It coincides with the small crystal.
   \item The seed $(2,2-2 \lambda ,\lambda ,\lambda)$ generates a crystal with $6$ fields:
   \begin{equation}
       \pm\{2,\lambda ,2 \lambda -2\}\,.
   \end{equation}
   It is a contraction of \eqref{Paper1-2pcase2} by imposing $\lambda_2=\lambda_1$. This crystal contains only (++$-$) couplings.
   \item The seed $(6-4 \lambda ,\lambda ,\lambda ,2 \lambda -2)$ generates a crystal with $6$ fields:
   \begin{equation}
       \pm\{3 \lambda -4,2 \lambda -2\}\cup\{\lambda,6-4 \lambda\}\,.
   \end{equation}
   It is a contraction of \eqref{Paper1-2pcase3} by imposing $\lambda_1=2\lambda_2-2$. This crystal contains only (++$-$) couplings. It coincides with the small crystal.
   \item The seed $(2-2 \lambda ,2-2 \lambda ,\lambda ,3 \lambda)$ generates a crystal with $8$ fields:
   \begin{equation}
       \pm\{\lambda,4 \lambda -2,6 \lambda -2\}\cup\{2-2\lambda,3\lambda\}\,.
   \end{equation}
   It is a contraction of \eqref{Paper1-2pcase3} by imposing  $\lambda_2=2-2\lambda_1$.
   \item The seed $(10-6 \lambda ,\lambda ,2 \lambda -2,3 \lambda -4)$ generates a crystal with $10$ fields:
   \begin{equation}
       \pm\{2 \lambda -2,3 \lambda -4,4 \lambda -6,5 \lambda -8\}\cup\{\lambda,10-6 \lambda\}\,.
   \end{equation}
   It is a contraction of \eqref{Paper1-2pcase3} by imposing $2+\lambda_1-\lambda_2=2\lambda_2-2\Rightarrow\lambda_1=3\lambda_2-4$. This crystal contains only (++$-$) couplings.
   \item The seed $(1,1,2-\lambda ,\lambda)$ generates a crystal with $8$ fields:
   \begin{equation}
       \pm\{0,\lambda -1,2 \lambda -2\}\cup\{1,\lambda,2-\lambda\}\,.
   \end{equation}
   It is a contraction of \eqref{Paper1-2pcase4} by imposing $2\lambda_2-2=2-2\lambda_2\Rightarrow\lambda_2=1$.
   \item The seed $(\lambda ,\lambda ,2-\lambda ,2-\lambda)$ generates a crystal with $5$ fields:
   \begin{equation}
       \pm\{0,2 \lambda -2\}\cup\{\lambda,2-\lambda\}\,.
   \end{equation}
    It is a contraction of \eqref{Paper1-2pcase5} by imposing $\lambda_1+\lambda_2-2=2-\lambda_1-\lambda_2\Rightarrow\lambda_2=2-\lambda_1$. It coincides with the small crystal.
   \item The seed $(4-3\lambda ,2-\lambda ,\lambda ,3\lambda-2)$ generates a crystal with $11$ fields:
   \begin{equation}\label{Paper1-example4}
       \pm\{0,2 \lambda-2 ,4 \lambda-4 ,6 \lambda-6 \}\cup\{ 4-3\lambda ,2-\lambda ,\lambda ,3\lambda-2\}\,.
   \end{equation}
   It is a contraction of \eqref{Paper1-2pcase4} by imposing $\lambda_2-\lambda_1=2\lambda_1-2\Rightarrow\lambda_2=3\lambda_1-2$.
   \item The seed $(4-3\lambda ,\lambda ,\lambda ,\lambda)$ generates a crystal with $4$ fields:
   \begin{equation}
       \pm\{2\lambda-2 \}\cup\{\lambda ,4-3\lambda \}\,.
   \end{equation}
   It is a contraction of \eqref{Paper1-2pcase5} by imposing $2\lambda_1-2=\lambda_1+\lambda_2-2\Rightarrow\lambda_2=\lambda_1$. This crystal contains only (++$-$) couplings. It coincides with the small crystal.
\end{enumerate}
We also present $5$ additional one-parameter families of infinite crystals which, although arising as specific contractions of \eqref{Paper1-2inf}, also originate from the two-parameter families of finite crystals. As such, they are important for the complete classification.
\begin{enumerate}
    \item The seed $(0,0,4-\lambda,\lambda)$ generates a crystal with the following fields:
    \begin{equation}
        \pm\{0,2,(2-\lambda)k,(2-\lambda)k\pm 2\},\qquad k\in \mathbb{Z}_{\geq 0}\,.
    \end{equation}
    It comes from \eqref{Paper1-2pcase1} by imposing $\lambda_1=0$.

    \item The seed $(0,4,-\lambda,\lambda)$ generates a crystal with the following fields:
    \begin{subequations}
    \begin{align}\label{Paper1-HSSDGRseed}
        &\pm\{0,2,k\}, &&k\in \mathbb{Z}_{\geq 0}\,,\qquad \lambda\;\;\text{odd}\,,\\
        & \pm\{0,2,2k\},&&k\in \mathbb{Z}_{\geq 0}\,,\qquad \lambda\;\;\text{even}\,.
    \end{align}
    \end{subequations}
    It comes from \eqref{Paper1-2pcase1} by imposing $4-\lambda_1-\lambda_2=-\lambda_1\Rightarrow \lambda_2=4$. This seed leads to HS-SDGR. In particular, starting from an even $\lambda$ yields a truncation of HS-SDGR to even helicities, while an odd $\lambda$ gives the full HS-SDGR spectrum.
    \item The seeds $(2,-2,4-\lambda,\lambda)$, $(2 \lambda-6,\lambda,4-\lambda,6-2 \lambda)$ and $(\lambda,\lambda,-\lambda,4-\lambda)$ generate a crystal with the following fields:
    \begin{equation}
        \pm\{2,(2-\lambda)k\pm 2\}\,,\qquad k\in \mathbb{Z}_{\geq 0}\,.
    \end{equation}
    It comes from \eqref{Paper1-2pcase2} by imposing $\lambda_2=-2$ and from \eqref{Paper1-2pcase3} by imposing $2-2\lambda_1=\lambda_2-\lambda_1-2\Rightarrow\lambda_1=4-\lambda_2$ and from \eqref{Paper1-2pcase5} by imposing $\lambda_2=-\lambda_1$, respectively.
    \item The seed $(2+2 \lambda,\lambda,-\lambda,2-2 \lambda)$ generates a crystal with the following fields:
    \begin{equation}
        \pm\{2,2k\lambda\pm 2,(2k+1)\lambda\}\,,\qquad k\in \mathbb{Z}_{\geq 0}\,.
    \end{equation}
    It comes from \eqref{Paper1-2pcase3} by imposing $\lambda_2=-\lambda_1$.
    \item The seed $(\lambda,2+\lambda,-\lambda,2-\lambda)$ generates a crystal with the following fields:
    \begin{equation}
        \pm\{0,2,k\lambda\pm 2,k\lambda\}\,,\qquad k\in \mathbb{Z}_{\geq 0}\,.
    \end{equation}
    It comes from \eqref{Paper1-2pcase4} by imposing $\lambda_2=-\lambda_1$.
\end{enumerate}

\paragraph{Further observations.} 
From the classification above, we find that the only seeds leading to the full HS-SDGR theory are: 
\begin{equation*}
\begin{aligned}
    &(\lambda,-\lambda,1,3)
    &&C^{\lambda,-\lambda,2}C^{-2,1,3},&&C^{\lambda,1,1-\lambda}C^{\lambda-1,-\lambda,3},&& C^{\lambda,3,-1-\lambda}C^{\lambda+1,-\lambda,1}\,,\\
    &(\lambda,-\lambda,0,4)&&
    C^{\lambda,-\lambda,2}C^{-2,0,4},&& C^{\lambda,0,2-\lambda}C^{\lambda-2,-\lambda,4},&&C^{\lambda,4,-2-\lambda}C^{\lambda+2,-\lambda,0}\,,
    &&\lambda\;\text{odd}\,,\\
    &(\lambda,-\lambda,-1,5)&&
    C^{ \lambda,- \lambda,2}C^{-2,-1,5},&& C^{ \lambda,-1,3- \lambda}C^{-3+ \lambda,- \lambda,5},&&C^{ \lambda,5,-3- \lambda} C^{3+ \lambda,- \lambda,-1}\,, &&\lambda\equiv 0\;(\text{mod}\;3)\,,\\
    &(-2,6,\lambda,-\lambda)&&
   C^{-2,6,-2}C^{2,\lambda,-\lambda},&&C^{-2,\lambda,4-\lambda}C^{-4+\lambda,6,-\lambda},&& C^{-2,-\lambda,4+\lambda}C^{-4-\lambda,6,\lambda}\,, &&\lambda\;\text{odd}\,.
\end{aligned}
\end{equation*}
Therefore, HS-SDGR is the unique solution iff at least one of these $C^2C^2$ coupling pairs is turned on.

Another noteworthy observation is that the only finite crystals containing the SDGR coupling $C^{-2,2,2}$ --- and which can thus be interpreted as chiral higher-spin gravity theories --- are those generated by \eqref{Paper1-2pcase2} and its contractions. In contrast, all infinite crystals include the SDGR coupling.

As we will see later, when solving the system of equations for the couplings associated with a crystal, some solutions may allow certain couplings to vanish. This gives rise to a sub-crystal, which corresponds to a subalgebra in the terminology of \cite{Ponomarev:2017nrr}. As shown in \cite{Ponomarev:2017nrr}, solutions to the light-cone holomorphic constraint correspond, with a few exceptions, to specific Lie algebras.

\subsection{One-derivative chiral higher-spin theories}
Here, we classify all inequivalent solutions of chiral higher-spin theories involving one-derivative interactions in the presence of a $U(N)$ gauge group. For $SO(N)$ and $USp(N)$, the classification of the crystals remains the same. The only difference is in the system for the couplings, as discussed in Appendix \ref{Paper1-AppendixD}.
Satisfying the holomorphic constraint \eqref{Paper1-Unconstraint} requires two basic rules:
\begin{itemize}
    \item Starting from a single pair of couplings $C^1C^1$, four additional ones are required. In particular, the following set of couplings must be present together:
    \begin{equation}\label{Paper1-small_crystal_1d}
        C^{\lambda_1,\lambda_2,1-\lambda_{12}}C^{\lambda_{12}-1,\lambda_3,\lambda_4}
        ,\;\; C^{\lambda_1,\lambda_3,1-\lambda_{13}}C^{\lambda_{13}-1,\lambda_2,\lambda_4}
        ,\;\; C^{\lambda_1,\lambda_4,1-\lambda_{14}}C^{\lambda_{14}-1,\lambda_2,\lambda_3}\,.
    \end{equation}
    We also impose $\sum_{i=1}^4\lambda_i=2$ to ensure that only one-derivative interactions are present. We refer to the collection of the $6$ couplings above as a small crystal. An exception is the pair $C^{\lambda_1,\lambda_1,0}C^{0,\lambda_2,\lambda_2}$, which is excluded from the small crystal, as it does not participate in the constraint \eqref{Paper1-LCholocolour}.      
    \item These couplings have to satisfy the following system of equations:
    \begin{align}\label{Paper1-coupling1d}
    \begin{split}
     &C^{\lambda_1,\lambda_2,1-\lambda_{12}}C^{\lambda_{12}-1,\lambda_3,\lambda_4}=C^{\lambda_4,\lambda_1,1-\lambda_{14}}C^{\lambda_{14}-1,\lambda_2,\lambda_3}\,,\\
    &C^{\lambda_1,\lambda_1,0}C^{0,\lambda_2,\lambda_2}=\;\text{generic}\,.
    \end{split}
    \end{align}
    The same equations also appear for the other colour-ordered constraints \eqref{Paper1-colour-ordered_terms}.
    While this system does not affect the classification below, it plays a role in determining relations among the couplings in any given theory. Notably, a solution is always guaranteed to exist --- namely the one corresponding to HS-SDYM --- in which all one-derivative couplings in the theory take the same value. However, in general, for smaller theories than HS-SDYM, some couplings remain unconstrained and correspond to free parameters. In what follows, we assume that all such free parameters are kept different from zero.
\end{itemize}

To justify the need for the small crystal in \eqref{Paper1-small_crystal_1d}, we recall that the constraint \eqref{Paper1-LCholocolour} applies only to a single colour-ordered case, specifically $[1234]$, and requires just two pairs of couplings. We make a simplifying assumption that $C^{\lambda_1,\lambda_2,\omega}\neq0$ implies $C^{\lambda_2,\lambda_1,\omega}\neq0$. As a result, we also need to satisfy the colour-ordered constraint $[1342]$, which in turn implies that all permutations must be included and all colour-ordered constraints must be fulfilled.

Apart from the change in the number of derivatives of the vertices, the classification strategy remains unchanged. Then we follow the same procedure as in the two-derivative case.
Let us give an example of a crystal generated by the seed $(1,1,-\lambda,\lambda)$, which yields
\begin{equation}\label{Paper1-example_U(N)}
\begin{aligned}
&C^{1,1,-1}C^{1,-\lambda,\lambda},&&\;\;&& C^{1,-\lambda,\lambda}C^{-\lambda,1,\lambda},&&\;\;&& C^{1,\lambda,-\lambda}C^{\lambda,1,-\lambda}\\
&C^{1,1,-1}C^{1,-1,1},&&\;\;&& C^{1,-1,1}C^{-1,1,1},&&\;\;&&C^{1,1,-1}C^{1,1,-1}\,.
\end{aligned}
\end{equation}
The above defines a specific chiral higher-spin theory with one-derivative interactions. It contains only a finite number of couplings, a finite spectrum, and the SDYM interaction $C^{-1,1,1}$.\\
The notion of equivalence between crystals is the same as in the two-derivative case.

\paragraph{Observations and simple solutions.}

First, HS-SDYM is a well-defined theory. It contains all higher-spin fields and all possible chiral one-derivative couplings, and is thus defined by the largest possible crystal. Moreover, starting from a seed containing only odd helicities, we will never generate even ones. This tells us that HS-SDYM admits a consistent truncation to the odd-helicity sector.

As in the two-derivative case, solutions that include only (++$-$) cubic couplings are allowed. The argument is the same as before, starting from the small crystal \eqref{Paper1-small_crystal_1d} and by requiring
\begin{equation}
    \lambda_1,\lambda_2,\lambda_3\geq n\geq 1\,,\qquad n\in\mathbb{Z}\,.
\end{equation}
As a consequence
\begin{align}
\begin{split}
    \lambda_4=2-\lambda_{123}\leq -n\leq -1&\,,\\
    1-\lambda_{ij}\leq -n\leq -1&\,,\qquad i,j=\{1,2,3\}\,,\\
    \lambda_{1j}-1\geq n\geq 1&\,,\qquad i,j=\{1,2,3\}\,. 
\end{split}
\end{align}
These define consistent truncations of HS-SDYM, and we can have arbitrary truncations, containing fields of helicities $|\lambda|\geq n$.
As before, there exist solutions that incorporate all possible cubic couplings constructed from the truncated spectrum.

Let us try to include a scalar field. A one-derivative cubic coupling involving a scalar field takes the form $(\lambda,1-\lambda,0)$ and generates the small crystal
\begin{equation}
    C^{\lambda_1,1-\lambda_1,0}C^{0,\lambda_2,1-\lambda_2},\;\;
    C^{\lambda_1,\lambda_2,1-\lambda_{12}}C^{\lambda_{12}-1,1-\lambda_1,1-\lambda_2},\;\;
    C^{\lambda_1,1-\lambda_2,\lambda_2-\lambda_1}C^{\lambda_1-\lambda_2,1-\lambda_1,\lambda_2}\,.
\end{equation}
This crystal contains both (++$-$) and ($--$+) couplings if at least one of the helicities satisfies $\lambda\geq 2$ or $\lambda\leq -1$. Therefore, adding a scalar necessarily introduces ($--$+) couplings as well.\\
The only exceptions occur for $\lambda=0,1$, which gives a crystal with two couplings $C^{0,0,1},C^{-1,1,1}$.

We can now ask which is the simplest theory that includes the SDYM coupling $C^{-1,1,1}$ and a generic coupling containing at least one scalar field $C^{\lambda,1-\lambda,0}$. The smaller crystal containing them is generated by the seed $(0,1,1-\lambda,\lambda)$, which gives the spectrum
\begin{equation}\label{Paper1-crystal_sdym_scalar}
    \pm\{0,1,\lambda,\lambda-1,1-2\lambda\}\,.
\end{equation}
The cubic couplings are then all the possible ones constructed from the spectrum above:
\begin{equation}
\{C^{0, 0, 1},C^{0,1-\lambda,\lambda},C^{1,-\lambda,\lambda}C^{1,1-\lambda,-1+\lambda},C^{1-\lambda,1-\lambda-1+2 \lambda}, C^{1-2 \lambda, \lambda, \lambda}, C^{-1,1, 1}, C^{1,1 - 2 \lambda, -1 + 2 \lambda}\}\,.
\end{equation}

One final observation is that we can easily recover the minimal coupling to the spin-$1$ gauge field for higher-spin fields, thanks to the small crystal
\begin{align}\label{Paper1-sdym_universality}
    &C^{1,1,-1}C^{1,\lambda,-\lambda},&
    &C^{1,\lambda,-\lambda}C^{\lambda,1,-\lambda},&
    &C^{1,-\lambda,\lambda}C^{-\lambda,1,\lambda}\,.
\end{align}
This configuration yields the following solution to the system of equations: 
\begin{align}
    &C^{1,1,-1}C^{2,\lambda,-\lambda}=C^{1,\lambda,-\lambda}C^{\lambda,1,-\lambda}&
    &\Rightarrow&
    C^{1,1,-1}=C^{1,\lambda,-\lambda}\,.
\end{align}
This result demonstrates the universality of the coupling to spin-$1$ for higher-spin fields.\\
Both the crystal \eqref{Paper1-crystal_sdym_scalar} and \eqref{Paper1-sdym_universality} will appear again in the classification we present below.

\paragraph{All one-derivative solutions.}
We follow the same procedure as in the two-derivative case. Recall that, in this case as well, the spectrum completely determines the cubic couplings.

The first solution is a three-parameter family with the following seed $(\lambda_1,\lambda_2,\lambda_3,2-\lambda_{123})$ and the crystal contains $10$ fields:
\begin{equation}\label{Paper1-1dgeneral}
    \pm\{1-\lambda_{12},1-\lambda_{13},1-\lambda_{23}\}\cup\{\lambda_1,\lambda_2,\lambda_3,2-\lambda_{123}\}\,.
\end{equation}
There are $4$ two-parameter families of finite crystals:
\begin{enumerate}
    \item The seed $(0, \lambda_1, 2 - \lambda_{12}, \lambda_2)$ generates a crystal with $22$ fields:
    \begin{align}\label{Paper1-1d2pcase1}
    \begin{split}
\pm\{&0,1-2 \lambda_1,1-\lambda_1,1-2 \lambda_2,3-2 \lambda_{12},2-\lambda_1-2 \lambda_2,1-\lambda_2,\\
&2-2 \lambda_1-\lambda_2,1-\lambda_{12},\lambda_1-\lambda_2\}\cup\{\lambda_1,2-\lambda_{12},\lambda_2\}\,.
    \end{split}
    \end{align}
    It is obtained by imposing $\lambda_3=-\lambda_3\Rightarrow \lambda_3=0$.
    \item The seed $(1, \lambda_1, 1 - \lambda_{12}, \lambda_2 )$ generates a crystal with $8$ fields:
    \begin{equation}\label{Paper1-1d2pcase2}
        \pm\{1,\lambda_1,1-\lambda_{12},\lambda_2\}\,.
    \end{equation}
    It is obtained by imposing $\lambda_{13}-1=\lambda_1\Rightarrow\lambda_3=1$.
    \item The seed $(1-2 \lambda_1,\lambda_1,1+\lambda_1-\lambda_2,\lambda_2)$ generates a crystal with $13$ fields:
    \begin{equation}\label{Paper1-1d2pcase3}
        \pm\{\lambda_1,1-2 \lambda_2,1+2 \lambda_1-2 \lambda_2,1-\lambda_{12},2 \lambda_1-\lambda_2\}\cup\{1-2 \lambda_1,1+\lambda_1-\lambda_2,\lambda_2\}\,.
    \end{equation}
    It is obtained by imposing $1-\lambda_{13}=\lambda_1\Rightarrow \lambda_3=1-2\lambda_1$.
    \item The seed $(1-\lambda_1,\lambda_1,1-\lambda_2,\lambda_2)$ generates a crystal with $13$ fields:
    \begin{equation}\label{Paper1-1d2pcase4}
        \pm\{0,1-2 \lambda_1,1-2 \lambda_2,1-\lambda_{12},\lambda_1-\lambda_2\}\cup\{1-\lambda_1,\lambda_1,1-\lambda_2,\lambda_2\}\,.
    \end{equation}
    It is obtained by imposing $1-\lambda_{13}=\lambda_{13}-1 \Rightarrow \lambda_3=1-\lambda_1$.
\end{enumerate}
There are $4$ one-parameter families of solutions:
\begin{enumerate}
     \item The seed $(0,3-4 \lambda,\lambda,3 \lambda-1)$ generates a crystal with $24$ fields:
    \begin{align}
        \begin{split}
        \pm&\{0,7-12 \lambda,5-9 \lambda,5-8 \lambda,4-7 \lambda,3-6 \lambda,3-5 \lambda,\\
        &2-4 \lambda,2-3 \lambda,1-2 \lambda,1-\lambda\} \cup\{3-4 \lambda,\lambda,-1+3 \lambda\}\,.
        \end{split}
    \end{align}
    It comes from \eqref{Paper1-1d2pcase1} by imposing $\lambda_2-\lambda_1=2\lambda_1-1\Rightarrow \lambda_2=3\lambda_1-1$.
    \item The seed $(0,2-3 \lambda,\lambda,2 \lambda)$ generates a crystal with $31$ fields:
    \begin{align}
        \begin{split}
        \pm\{&0,5-12 \lambda,4-10 \lambda,4-9 \lambda,3-8 \lambda,3-7 \lambda,2-6 \lambda,3-6 \lambda,2-5 \lambda,\\
        &1-4 \lambda,2-4 \lambda,1-3 \lambda,1-2 \lambda,1-\lambda,-\lambda\}\cup\{2-3 \lambda,2 \lambda\}\,.
        \end{split}
    \end{align}
    It comes from \eqref{Paper1-1d2pcase1} by imposing $\lambda_2-\lambda_1=\lambda_1\Rightarrow \lambda_2=2\lambda_1$.
        \item The seed  $(2-3 \lambda,1-2 \lambda,\lambda,4 \lambda-1)$ generates a crystal with $16$ fields:
    \begin{equation}
        \pm\{5-12 \lambda,4-9 \lambda,3-8 \lambda,3-6 \lambda,2-5 \lambda,1-2 \lambda,\lambda\}\cup\{4 \lambda-1,2-3 \lambda\}\,.
    \end{equation}
    It comes from \eqref{Paper1-1d2pcase3} by imposing $\lambda_1=1-2\lambda_2$.
    \item The seed $(6 \lambda-3,\lambda,2-3 \lambda,3-4 \lambda)$ generates a crystal with $13$ fields:
    \begin{equation}
        \pm\{7-12 \lambda,5-8 \lambda,4-7 \lambda,2-3 \lambda,1-2 \lambda\}\cup\{3-4 \lambda,\lambda,6 \lambda-3\}\,.
    \end{equation}
    It comes from \eqref{Paper1-1d2pcase3} by imposing $\lambda_1+\lambda_2-1=1-2\lambda_2\Rightarrow \lambda_1=2-3\lambda_2$.
\end{enumerate}
There are $2$ particular finite crystals:
\begin{enumerate}
    \item The seed $(0,0,1,1)$ generates a crystal with $3$ fields:
    \begin{equation}
        \pm\{1,0\}\,.
    \end{equation}
    This corresponds to SDYM coupled to a scalar field.
    \item The seed $(-1,1,1,1)$ generates a crystal with $2$ fields:
    \begin{equation}
        \pm\{1\}\,.
    \end{equation}
    This corresponds to SDYM.
\end{enumerate}
There is $1$ two-parameter family of infinite crystals:
\begin{enumerate}
    \item The seed $(-\lambda_1,\lambda_1,2-\lambda_2,\lambda_2)$ generates a crystal with the following fields:
    \begin{equation}\label{Paper1-1inf}
        \pm\{1,(1-\lambda_2)k\pm 1,(1-\lambda_2)k\pm\lambda_1\}\,,\qquad k\in \mathbb{Z}_{\geq 0}\,.
    \end{equation}
    It is obtained by imposing $\lambda_3=-\lambda_1$.
\end{enumerate}
There is $1$ one-parameter family of infinite crystals:
\begin{enumerate}
     \item The seed $(0,3-2 \lambda,\lambda-1,\lambda)$ generates a crystal with the following fields:
    \begin{equation}
        \pm\{0,1,(1-\lambda)k,(1-\lambda)k\pm 1\}\,,\qquad k\in \mathbb{Z}_{\geq 0}\,.
    \end{equation}
    It comes from \eqref{Paper1-1d2pcase1} by imposing $\lambda_1=\lambda_2-1$.
\end{enumerate}
There are some particular infinite solutions:
\begin{itemize}
    \item The following seeds generate all integer helicities:
    \begin{align}
        \begin{split}
        \{&(-1,0,1,2), (0,0,0,2), (-2,0,2,2), (-1,0,0,3), (-3,0,2,3),\\
        &(-5,0,2,5), (-4,0,2,4), (-2,-1,0,5), (-2,-1,2,3)\}\,.
         \end{split}
    \end{align}
    \item The following seeds generate all odd helicities:
    \begin{equation}
        \{(-1,-1,1,3), (-5,-1,3,5), (-3,-1,3,3)\}\,.
    \end{equation}
\end{itemize}
In the following, we present additional crystals obtained through specific contractions of the spectra of the previously discussed crystals.\\
There is $1$ two-parameter family of finite crystals:
\begin{enumerate}
    \item The seed $(\lambda_1,\lambda_1,2-2 \lambda_1-\lambda_2,\lambda_2)$ generates a crystal with $7$ fields:
    \begin{equation}\label{Paper1-1d2pcase5}
        \pm\{ 1 - 2\lambda_1, 1 - \lambda_{12} \}\cup
\{ \lambda_1, 2 - 2\lambda_1 - \lambda_2, \lambda_2\}\,. 
    \end{equation}
    It is a contraction of \eqref{Paper1-1dgeneral} by imposing $1-\lambda_{12}=1-\lambda_{23}\Rightarrow \lambda_3=\lambda_1$. It coincides with the small crystal.
\end{enumerate}
There are $11$ one-parameter families of finite crystals:
\begin{enumerate}
    \item The seed $(0,1,1-\lambda,\lambda)$ generates a crystal with $9$ fields:
    \begin{equation}
        \pm\{ 0,1,\lambda, 2\lambda-1, \lambda-1\}\,.
    \end{equation}
    It is a contraction of \eqref{Paper1-1d2pcase1} by imposing $2\lambda_1-1=\lambda_1\Rightarrow \lambda_1=1$.
    \item The seeds $(0,\lambda,\lambda,2-2 \lambda)$ and $(2 \lambda-1,\lambda,1-\lambda,2-2 \lambda)$ generate a crystal with $11$ fields:
    \begin{equation}
        \pm\{0,3-4 \lambda,2-3 \lambda,1-2 \lambda,1-\lambda\}\cup\{2-2 \lambda,\lambda\}\,.
    \end{equation}
    It is a contraction of \eqref{Paper1-1d2pcase1} by imposing $\lambda_1-\lambda_2=\lambda_2-\lambda_1\Rightarrow \lambda_2=\lambda_1$ and of \eqref{Paper1-1d2pcase3} by imposing $1-2\lambda_1=2\lambda_2-1\Rightarrow \lambda_1=1-\lambda_2$, respectively.
    \item The seed $(0,3-3 \lambda,\lambda,2 \lambda-1)$ generates a crystal with $17$ fields:
    \begin{equation}
       \pm\{0, 5 - 6 \lambda, 4 - 5 \lambda, 3 - 4 \lambda, 
  2 - 3 \lambda, 1 - 2 \lambda, 2 - 2 \lambda, 
  1 - \lambda\}\cup\{3 - 3 \lambda, \lambda\}\,.
    \end{equation}
    It is a contraction of \eqref{Paper1-1d2pcase1} by imposing $\lambda_2=2\lambda_1-1$.
    \item The seed $(1,1,-\lambda,\lambda)$ generates a crystal with $4$ fields:
    \begin{equation}
        \pm\{1,\lambda\}\,.
    \end{equation}
    It is a contraction of \eqref{Paper1-1d2pcase2} by imposing $\lambda_2=1$. This crystal contains only (++$-$) couplings. It coincides with the small crystal.
    \item The seed $(1,1-2 \lambda,\lambda,\lambda)$ generates a crystal with $6$ fields:
    \begin{equation}
        \pm\{1,1-2 \lambda,\lambda\}\,.
    \end{equation}
    It is a contraction of \eqref{Paper1-1d2pcase2} by imposing $\lambda_2=\lambda_1$. This crystal contains only (++$-$) couplings.
    \item The seed $(3-4 \lambda,\lambda,\lambda,2 \lambda-1)$ generates a crystal with $6$ fields:
    \begin{equation}
        \pm\{2-3 \lambda,1-2 \lambda\}\cup\{3-4 \lambda,\lambda\}\,.
    \end{equation}
    It is a contraction of \eqref{Paper1-1d2pcase3} by imposing $\lambda_1=2\lambda_2-1$. This crystal contains only (++$-$) couplings. It coincides with the small crystal.
    \item The seed $(1-2 \lambda,1-2 \lambda,\lambda,3 \lambda)$ generates a crystal with $8$ fields:
    \begin{equation}
        \pm\{1-6 \lambda,1-4 \lambda,\lambda\}\cup\{3 \lambda,1-2 \lambda\}\,.
    \end{equation}
    It is a contraction of \eqref{Paper1-1d2pcase3} by imposing $\lambda_2=1-2\lambda_1$.
    \item The seed $(5-6 \lambda,\lambda,2 \lambda-1,3 \lambda-2)$ generates a crystal with $10$ fields:
    \begin{equation}
        \pm\{4-5 \lambda,3-4 \lambda,2-3 \lambda,1-2 \lambda\}\cup\{5-6 \lambda,\lambda\}\,.
    \end{equation}
    It is a contraction of \eqref{Paper1-1d2pcase3} by imposing $1+\lambda_1-\lambda_2=2\lambda_2-1\Rightarrow \lambda_1=3\lambda_2-2$. This crystal contains only (++$-$) couplings.
    \item The seed $(\lambda,\lambda,1-\lambda,1-\lambda)$ generates a crystal with $5$ fields:
    \begin{equation}
        \pm\{0,1-2 \lambda\}\cup\{1-\lambda,\lambda\}\,.
    \end{equation}
    It is a contraction of \eqref{Paper1-1d2pcase5} by imposing $\lambda_1+\lambda_2-1=1-\lambda_1-\lambda_2\Rightarrow \lambda_2=1-\lambda_1$. It coincides with the small crystal.
    \item The seed $(3 \lambda-1,\lambda,1-\lambda,2-3 \lambda\}$ generates a crystal with $11$ fields:
    \begin{equation}
    \pm\{0,3-6 \lambda,2-4 \lambda,1-2 \lambda\}\cup\{2-3 \lambda,1-\lambda,\lambda,-1+3 \lambda\}\,.
    \end{equation}
    It is a contraction of \eqref{Paper1-1d2pcase4} by imposing $\lambda_2-\lambda_1=2\lambda_1-1\Rightarrow \lambda_2=3\lambda_1-1$.
    \item The seed $(2-3 \lambda,\lambda,\lambda,\lambda)$ generates a crystal with $4$ fields:
    \begin{equation}
        \pm\{1-2 \lambda\}\cup\{2-3 \lambda,\lambda\}\,.
    \end{equation}
    It is a contraction of \eqref{Paper1-1d2pcase5} by imposing $2\lambda_1-1=\lambda_1+\lambda_2-1\Rightarrow \lambda_2=\lambda_1$. This crystal contains only (++$-$) couplings. It coincides with the small crystal.
    \end{enumerate}
There are $7$ one-parameter families of infinite crystals which, although arising as specific contractions of \eqref{Paper1-1inf}, also originate from the two-parameter families of finite crystals:
\begin{enumerate}
    \item The seed $(0,0,2-\lambda,\lambda)$ generates a crystal with the following fields: 
    \begin{equation}
        \pm\{0,1,(1-\lambda)k,(1-\lambda)k\pm 1\}\,,\qquad k\in \mathbb{Z}_{\geq 0}\,.
    \end{equation}
    It comes respectively from \eqref{Paper1-1d2pcase1} by imposing $\lambda_1=0$.
    \item The seed $(0,2,-\lambda,\lambda)$ generates a crystal with the following fields: 
    \begin{equation}
        \pm\{0,1,k\}\,,\qquad k\in \mathbb{Z}_{\geq 0}\,.
    \end{equation}
    It comes from \eqref{Paper1-1d2pcase1} by imposing $2-\lambda_1-\lambda_2=-\lambda_1\Rightarrow \lambda_2=2$. This seed leads to HS-SDYM.
    \item The seeds $(1,-1,2-\lambda,\lambda)$ and $(\lambda,\lambda,-\lambda,2-\lambda)$ generate a crystal with the following fields:
    \begin{equation}
        \pm\{1,(1-\lambda)k\pm 1\}\,,\qquad k\in \mathbb{Z}_{\geq 0}\,.
    \end{equation}
    It comes from \eqref{Paper1-1d2pcase2} by imposing $\lambda_2=-1$ and from \eqref{Paper1-1d2pcase5} by imposing $\lambda_2=-\lambda_1$, respectively.
    \item The seed $(1-2 \lambda,2-\lambda,\lambda,2 \lambda-1)$ generates a crystal with the following fields:
    \begin{equation}
        \pm\{1,(1+\lambda)k\pm 1\}\,,\qquad k\in \mathbb{Z}_{\geq 0}\,.
    \end{equation}
    It comes from \eqref{Paper1-1d2pcase3} by imposing $1+\lambda_1-\lambda_2=2\lambda_1-1\Rightarrow\lambda_2=2-\lambda_1$.
    \item The seed $(1+2 \lambda,\lambda,-\lambda,1-2 \lambda)$ generates a crystal with the following fields:
    \begin{equation}
        \pm\{1,2k\lambda\pm 1,(2k+1)\lambda\}\,,\qquad k\in \mathbb{Z}_{\geq 0}\,.
    \end{equation}
    It comes from \eqref{Paper1-1d2pcase3} by imposing $\lambda_2=-\lambda_1$.
    \item The seed $(-1,3,-\lambda,\lambda)$ generates a crystal with the following fields: 
    \begin{subequations}
    \begin{align}
        &\pm\{0,1,k\}\,,&&k\in \mathbb{Z}_{\geq 0}\,,\qquad \lambda\;\;\text{even}\,,\\
        & \pm\{1,2k+1\}\,,&&k\in \mathbb{Z}_{\geq 0}\,,\qquad \lambda\;\;\text{odd}\,.
    \end{align}
     \end{subequations}
    It comes from \eqref{Paper1-1d2pcase3} by imposing $\lambda_2=\lambda_2-\lambda_1-1\Rightarrow\lambda_1=-1$. This leads to HS-SDYM. In particular, starting from an odd $\lambda$ yields a truncation of HS-SDYM to odd helicities, while an even $\lambda$ gives the full HS-SDYM spectrum.
    \item The seed $(\lambda,1+\lambda,-\lambda,1-\lambda)$ generates a crystal with the following fields:
    \begin{equation}
        \pm\{0,1,k\lambda\pm 1,k\lambda\}\,,\qquad k\in \mathbb{Z}_{\geq 0}\,.
    \end{equation}
    It comes from \eqref{Paper1-1d2pcase4} by imposing $\lambda_2=-\lambda_1$.
\end{enumerate}
\paragraph{Further observations.}
From the classification above, we find that the only seeds leading to HS-SDYM are:
\begin{equation*}
\begin{aligned}
    &(\lambda,-\lambda,2,0)&&
    C^{\lambda,-\lambda,1}C^{-1,2,0},&& C^{\lambda,2,-1-\lambda}C^{1+\lambda,-\lambda,0},&& C^{\lambda,0,1-\lambda}C^{-1+\lambda,-\lambda,2}\,,\\
    &(\lambda,-\lambda,3,-1)&& C^{\lambda,-\lambda,1}C^{-1,3,-1},&& C^{\lambda,3,-2-\lambda}C^{2+\lambda,-\lambda,-1},&&C^{\lambda,-1,2-\lambda}C^{-2+\lambda,-\lambda,3}\,,&& \lambda\;\text{even}\,,\\
    &(\lambda,-\lambda,-2,4)&& C^{\lambda,-\lambda,1}C^{-1,-2,4},&& C^{\lambda,-2,3-\lambda}C^{-3+\lambda,-\lambda,4},&&C^{\lambda,4,-3-\lambda}C^{3+\lambda,-\lambda,-2} \,, &&\lambda\equiv 0\;(\text{mod}\;3)\,,\\
    &(-2,-1,0,5)&& C^{-2,-1,4}C^{-4,0,5},&& C^{-2,0,3}C^{-3,-1,5},&& C^{-2,5,-2}C^{2,-1,0}\,.
\end{aligned}
\end{equation*}
Therefore, HS-SDYM is the unique solution iff at least one of these pairs of $C^1C^1$ couplings is turned on.

Another interesting observation is that the only finite crystals containing the SDYM coupling $C^{-1,1,1}$ are those arising from \eqref{Paper1-1d2pcase2} and its contractions. In contrast, all infinite crystals do contain the SDYM coupling.
\subsection{Higher-derivative}
Here, we explore the various features that emerge in the higher-derivative case, which also makes its classification more involved.
We focus on the case without a gauge algebra. When a gauge algebra is present, one simply needs to include odd-derivative couplings as well; the rest of the analysis remains unchanged. Satisfying the holomorphic constraint \eqref{Paper1-holo2}, requires respecting two basic rules:
\begin{itemize}
    \item Starting from a single pair of couplings $C^nC^m$, with $n+m=k+2$ being the total number of derivatives, the following set of $3k$ couplings must be present together:\footnote{Sometimes, for convenience, we indicate the number of derivatives of a coupling on top of it.}
    \begin{equation*}
    \begin{aligned}\label{Paper1-symHD}
&\overset{k}{C^{\lambda_1,\lambda_2,k-\lambda_{12}}}\overset{2}{C^{\lambda_{12}-k,\lambda_3,\lambda_4}},&&\;\;&&\overset{k}{C^{\lambda_1,\lambda_3,k-\lambda_{13}}}\overset{2}{C^{\lambda_{13}-k,\lambda_2,\lambda_4}},&&\;\;&& \overset{k}{C^{\lambda_2,\lambda_3,k-\lambda_{23}}}\overset{2}{C^{\lambda_{23}-k,\lambda_1,\lambda_4}}\\
    &\overset{k-2}{C^{\lambda_1,\lambda_2,k-2-\lambda_{12}}}\overset{4}{C^{\lambda_{12}+2-k,\lambda_3,\lambda_4}},&&\;\;&&\overset{k-2}{C^{\lambda_1,\lambda_3,k-2-\lambda_{13}}}\overset{4}{C^{\lambda_{13}+2-k,\lambda_2,\lambda_4}},&&\;\;&& \overset{k-2}{C^{\lambda_2,\lambda_3,k-2-\lambda_{23}}}\overset{4}{C^{\lambda_{23}+2-k,\lambda_1,\lambda_4}}\\
    &\cdots,&&\;\;&&\cdots,&&\;\;&&\cdots\\
    &\overset{2}{C^{\lambda_1,\lambda_2,2-\lambda_{12}}}\overset{k}{C^{\lambda_{12}-2,\lambda_3,\lambda_4}},&&\;\;&& \overset{2}{C^{\lambda_1,\lambda_3,2-\lambda_{13}}}\overset{k}{C^{\lambda_{13}-2,\lambda_2,\lambda_4}},&&\;\;&& \overset{2}{C^{\lambda_2,\lambda_3,2-\lambda_{23}}}\overset{k}{C^{\lambda_{23}-2,\lambda_1,\lambda_4}}\,.
    \end{aligned}
 \end{equation*}
We also impose $\sum_{i=1}^4\lambda_i=k+2$ to ensure that only couplings with up to $k$ derivatives are present. We refer to the collection of the $3k$ couplings above as a small crystal. An exception is the pair $C^{\lambda_1,\lambda_1,0}C^{0,\lambda_2,\lambda_2}$, which is excluded from the small crystal, as it does not participate in the constraint \eqref{Paper1-LCholo}.
\item These couplings have to satisfy the following system of equations:
    \begin{align}
\begin{split} 
     &C^{\lambda_1,\lambda_2,\omega}C^{-\omega,\lambda_3,\lambda_4}=\frac{k^{1234}_+(\Lambda-2)!}{2^{\Lambda-3}(\lambda_{12}+\omega-1)!(\lambda_{34}-\omega-1)!}\qquad 
    \forall\;\omega\,,\\
    &\text{same for $(1324)$ and $(1423)$}\,,\\
    &k_+^{1234}= k_+^{1324}= k_+^{1423}\,,\qquad C^{\lambda_1,\lambda_1,0}C^{0,\lambda_2,\lambda_2}=\;\text{generic}\,.
\end{split}
    \end{align}
    While this system does not affect the classification, it plays a role in determining relations among the couplings in any given theory. Notably, a solution is always guaranteed to exist --- namely the one corresponding to full chiral higher-spin theory, with the Metsaev solution \eqref{Paper1-alleven}.
\end{itemize}

Let us examine a specific example to highlight the key differences from the lower-derivative case. The seed $(6,-4,2,2)$, with $\Lambda=6$, generates the higher-derivative crystal
\begin{equation}\label{Paper1-L=6_crystal}
\begin{aligned}
    &C^{6,-4,[2,0]}C^{[-2,0],2,2},&&\;\;&& C^{6,2,[-4,-6]}C^{[4,6],-4,2},&&\;\;&& C^{6,2,[-4,-6]}C^{[4,6],-4,2}\\
    &C^{2,6,-6}C^{6,-6,2},&&\;\;&& C^{2,-6,6}C^{-6,6,2},&&\;\;&& C^{2,2,-2}C^{2,6,-6}\\
    &C^{2,6,-6}C^{6,-4,0},&&\;\;&& C^{2,-4,4}C^{-4,6,0},&&\;\;&&C^{2,0,0}C^{0,6,-4}\\
    &C^{-4,6,0}C^{0,-4,6},&&\;\;&&C^{-4,-4,10}C^{-10,6,6},&&\;\;&&C^{-4,6,0}C^{0,6,-4}\\
    &C^{-10,6,6}C^{-6,2,6},&&\;\;&&C^{-10,2,10}C^{-10,6,6},&&\;\;&& C^{-10,6,6}C^{-6,6,2}\\
    &C^{2,10,-10}C^{10,-10,2},&&\;\;&& C^{2,-10,10}C^{-10,10,2},&&\;\;&& C^{2,2,-2}C^{2,10,-10}\\
    &C^{2,10,-10}C^{10,-4,-4},&&\;\;&&C^{2,-4,4}C^{-4,10,-4},&&\;\;&& C^{2,-4,4}C^{-4,10,-4}\\
    &C^{2,4,-4}C^{4,-4,2},&&\;\;&& C^{2,-4,4}C^{-4,4,2},&&\;\;&& C^{2,2,-2}C^{2,4,-4}\\
    &C^{2,2,-2}C^{2,-2,2},&&\;\;&& C^{2,-2,2}C^{-2,2,2},&&\;\;&& C^{2,2,-2}C^{2,2,-2}\\
    &C^{2,2,-2}C^{2,0,0},&&\;\;&& C^{2,0,0}C^{0,2,0},&&\;\;&& C^{2,0,0}C^{0,2,0}\\
    &C^{2,2,[0,-2]}C^{[0,2],0,2},&&\;\;&&C^{2,0,[2,0]}C^{[-2,0],2,2},&&\;\;&& C^{2,2,[0,-2]}C^{[0,2],2,0}\,,
\end{aligned}
\end{equation}
where the square bracket $[-,-]$ is a notation to indicate the range of exchanged helicities, incremented by steps of $2$.\footnote{If we want to include both even- and odd-derivative interactions and/or in the presence of a gauge group, the step would be of $1$.}

The above crystal contains the following $10$ couplings:
\begin{equation}
    \{\overset{2}{C^{-2, 2, 2}}, \overset{2}{C^{-4, 2, 4}}, \overset{2}{C^{-4, 0, 6}}, \overset{2}{C^{-6, 2, 6}}, \overset{2}{C^{0, 0, 2}}, \overset{2}{C^{-4, -4, 10}}, \overset{2}{C^{-10, 6, 6}}, \overset{2}{C^{-10, 2, 10}},\overset{4}{C^{0, 2, 2}},\overset{4}{C^{-4, 2, 6}}\}\,.
\end{equation}
It includes SDGR, but does not contain all possible couplings that can be constructed from the spectrum up to $4$-derivatives. In particular, it omits the following $12$ couplings:
\begin{align}
    \begin{split}
    &\{\overset{2}{C^{-6, -2, 10}},\overset{2}{C^{-6, 4,4}}, \overset{2}{C^{-2, -2, 6}}, \overset{2}{C^{-2, 0, 
  4}},\overset{4}{C^{-10, 4, 10}}, \overset{4}{C^{-6, 0, 10}},\\
  &\overset{4}{C^{-6, 4, 6}}, \overset{4}{C^{-4, -2, 10}}, \overset{4}{C^{-4,4, 4}}, \overset{4}{C^{-2, 0, 6}}, \overset{4}{C^{-2, 2, 4}}, \overset{4}{C^{0, 0, 4}}\}\,.
  \end{split}
\end{align}
The solution for the couplings in the crystal \eqref{Paper1-L=6_crystal} is:
\begin{align}
    \begin{split}
&\{C^{-2, 2, 2}=C^{0, 0, 2}=C^{-4, 2, 4}=C^{-6, 2, 6}=C^{-10, 2, 10}\,,\\
    &C^{-4, -4, 10}=\frac{(C^{-4, 0, 6})^2}{C^{-10, 6, 6}},\qquad 
C^{0, 2, 2}=\frac{C^{-2, 2, 2} C^{-4, 2, 6}}{C^{-4, 0, 6}}\}\,.
    \end{split}
\end{align}
The first line is a manifestation of the universality of gravitational interactions.

It is worth stressing that, without the freedom of having an arbitrary value for the product of couplings $C^{\lambda_1,\lambda_1,0}C^{0,\lambda_2,\lambda_2}$, we would get an additional constraint:
\begin{equation}
    C^{2, 2, 2}=\frac{10}{3} \frac{(C^{0,2,2})^2}{C^{-2,2,2}}\,.
\end{equation}
This is the same constraint encountered in \eqref{Paper1-wrong_condition} for the higher-derivative theory for lower spins, and also here it would lead to a vanishing amplitude.

The example above illustrates all the new features of the higher-derivative case:
\begin{itemize}
    \item The couplings are no longer fully determined by the spectrum. In the example above, some $2$- and $4$-derivative couplings are missing. This complicates the classification of higher-derivative chiral theories.
    \item The presence or absence of a gauge group becomes more relevant. In its presence, odd-derivative vertices must also be included, leading to potentially very different solutions, unlike the similarity observed in the lower-derivative case.
    \item As already seen in  \eqref{Paper1-HDcolourtheory} and in the example above, the freedom to choose the product $C^{\lambda_1,\lambda_1,0}C^{0,\lambda_2,\lambda_2}$ becomes relevant. It is also tied to the possibility of obtaining non-vanishing amplitudes.
\end{itemize}

\paragraph{Simple observations.}
Full chiral higher-spin theory is a well-defined theory. It contains all higher-spin fields and all possible chiral even-derivative cubic couplings, and is thus defined by the largest possible crystal. Moreover, starting from a seed containing only even helicities, we will never generate odd ones. This tells us that full chiral higher-spin admits a consistent truncation to the even-helicity sector.

As in the low-derivative cases, we have solutions that include only (++$-$) cubic couplings. Starting from a small crystal, we impose that the newly generated couplings never produce any ($--$+) terms, by requiring
\begin{equation}
    \lambda_1,\lambda_2,\lambda_3\geq n\geq k\,,\qquad n\in\mathbb{Z}\,.
\end{equation}
As a consequence
\begin{align}
    \begin{split}
    \lambda_4=k+2-\lambda_{123}\leq -n\leq -k&\,,\\
    \ell-\lambda_{ij}\leq -n\leq -k&\,,\qquad i,j=\{1,2,3\}\,,\quad \ell=\{2,4,...,k\}\,,\\
    \lambda_{ij}-\ell\geq n\geq k&\,,\qquad i,j=\{1,2,3\}\,,\quad \ell=\{2,4,...,k\}\,.
    \end{split}
\end{align}
These define consistent truncations of the full chiral higher-spin theory and can have arbitrary truncations containing fields of helicities $|\lambda|\geq k$. There also exist solutions that include all possible cubic couplings constructed from the truncated spectrum.

Note that a theory that includes both higher-derivative and the SDGR coupling $C^{-2,2,2}$, or one that includes a scalar field, inevitably leads to the presence of both types of couplings (++$-$) and ($--$+).

\paragraph{Metsaev solution.}
Including the self-interacting higher-derivative coupling $C^{-\lambda,\lambda,\lambda}$, for $|\lambda|>2$, uniquely leads to the Metsaev solution. We now present a proof of this statement.

We begin by using the coupling $C^{-\lambda,\lambda,\lambda}$ to form the small crystal
\begin{equation}
    \overset{2\lambda}{C^{\lambda,-\lambda,[2\lambda-2,2]}C^{[2-2\lambda,-2],\lambda,\lambda}},\;\;\overset{2\lambda}{C^{\lambda,\lambda,[-2,2-2\lambda]}C^{[2,2\lambda-2],-\lambda,\lambda}},\;\;\overset{2\lambda}{C^{\lambda,\lambda,[-2,2-2\lambda]}C^{[2,2\lambda-2],-\lambda,\lambda}}\,.
\end{equation}
From this point onward, we write only the necessary
permutations of the small crystal and, for clarity, indicate the total number of derivatives above each pair.

Starting from the small crystal above, we can generate further ones, forming a chain of small crystals:
\begin{subequations}\label{Paper1-Metsaev_chain}
\begin{align}
&\overset{2\lambda}{C^{\lambda,-\lambda,[2\lambda-2,2]}C^{[2-2\lambda,-2],\lambda,\lambda}},\;\;\overset{2\lambda}{C^{\lambda,\lambda,[-2,2-2\lambda]}C^{[2,2\lambda-2],-\lambda,\lambda}}\,,\\
&\overset{4\lambda-4}{C^{\lambda,2\lambda-2,[\lambda-4,4-3\lambda]}C^{[4-\lambda,3\lambda-4],-\lambda,2\lambda-2}},\;\;\overset{4\lambda-4}{C^{\lambda,-\lambda,[4\lambda-6,2]}C^{[6-4\lambda,-2],2\lambda-2,2\lambda-2}}\,,\\
&\overset{8\lambda-12}{C^{\lambda,4\lambda-6,[3\lambda-8,8-5\lambda]}C^{[8-3\lambda,5\lambda-8],-\lambda,4\lambda-6}},\;\;\overset{8\lambda-12}{C^{\lambda,-\lambda,[8\lambda-14,2]}C^{[14-8\lambda,-2],4\lambda-6,4\lambda-6}}\,,\\
\nonumber
&\cdots,\;\;\cdots\,.
\end{align}
\end{subequations}
Since $|\lambda|>2$,\footnote{Indeed, for $\lambda=2$, we have the SDGR coupling $C^{-2,2,2}$, which does not lead to the full chiral higher-spin theory.} we can reach a pair of arbitrary numbers of total derivatives, and generate the coupling $C^{\lambda,-\lambda,n}$ for arbitrary $n\geq 2$. Taking two such couplings, we can generate the small crystal
\begin{equation}
    \overset{n_{12}}{C^{\lambda,n_1,[n_2-\lambda-2,2-n_1-\lambda]}C^{[\lambda-n_2+2,\lambda+n_1-2],-\lambda,n_2}},\;\;\overset{n_{12}}{C^{\lambda,-\lambda,[n_{12}-2,2]}C^{[2-n_{12},-2],n_1,n_2}}\,.
\end{equation}
The coupling $C^{[2-n_{12},-2],n_1,n_2}$ has now $n_1,n_2$ as arbitrary helicities. Looking again at \eqref{Paper1-Metsaev_chain}, we see that couplings of the form $C^{\lambda,n_3,2}$ can always be generated. In particular, they can be generated from $C^{\lambda,-\lambda,n}$ and $C^{[2-n_{23},-2],n_3,n_2}$ in this way:
\begin{equation}
    \overset{n_{23}}{C^{\lambda,-\lambda,[n_{23}-2,2]}C^{[2-n_{23},-2],n_3,n_2}},\;\;\overset{n_{23}}{C^{\lambda,n_3,[n_2-\lambda-2,2-n_3-\lambda]}C^{[2+\lambda-n_2,\lambda+n_3-2],-\lambda,n_2}}\,.
\end{equation}
Then by suitably choosing $n_2$, we can always generate $C^{\lambda,n_3,2}$ with arbitrary $n_3$.\\
Now we can pair $C^{\lambda,n_3,2}$ with $C^{n_1,n_2,[2-n_{12},-2]}$ by choosing $n_3=k+2-\lambda-n_{12}$ and generate
\begin{equation}
    \overset{k+2}{C^{n_1,n_2,[k-n_{12},2-n_{12}]}C^{[n_{12}-k,n_{12}-2],\lambda,n_3}},\;\;\overset{k+2}{C^{n_1,n_3,[k-n_{13},2-n_{13}]}C^{[n_{13}-k,n_{13}-2],n_2,\lambda}}\,.
\end{equation}
Then $C^{n_1,n_2,[k-n_{12},2-n_{12}]}$ is a generic cubic coupling $C^{n_1,n_2,n_3}$ involving up to $k$ derivatives with arbitrary $k$. This proves that all couplings must be present.

Therefore, following Appendix A of \cite{Ponomarev:2016lrm} we conclude that the Metsaev solution is the unique one in the presence of at least one higher-derivative coupling of the form $C^{-\lambda,\lambda,\lambda}$.
While this result has long been known in the higher-spin literature, to the best of our knowledge, it has not been rigorously proven before.

As a consequence of the result above, we can make two additional observations.
Firstly, in the presence of a coupling of the type $C^{\lambda,-\lambda,n}$ with $2n-\lambda\geq 2$, one has to add $C^{\lambda,-\lambda,\lambda}$, thereby recovering the Metsaev solution. Secondly, if we assume the existence of at least one nonabelian higher-derivative coupling, the following reasoning applies. Higher derivatives imply that many values of $\omega$ can appear in the exchange. Therefore, starting from a generic pair (assuming $\lambda_1\geq \lambda_2$), we can always recover the specific type of coupling mentioned above:
\begin{align}
    &C^{\lambda_1,\lambda_2,\omega}C^{-\omega,\lambda_3,\lambda_4}&
    &\Longrightarrow &
    &C^{\lambda_1,\lambda_2,-\lambda_2}C^{-\lambda_2,\lambda_3,\lambda_4}\,.
\end{align}
This again leads to the Metsaev solution. In this sense, one can view the emergence of the Metsaev solution as more tied to the presence of high-derivative vertices rather than to the presence of higher-spin fields themselves.

Note that in the presence of a gauge group, we can follow the same logic as above by replacing the condition on $C^{-\lambda,\lambda,\lambda}$ with the requirement $|\lambda|>1$. Consequently, a colour-graviton can self-interact through the coupling $C^{-2,2,2}$ only if all cubic couplings are present.

\subsection{Solutions for the couplings}
Now that we have the classification, we can solve the system for the couplings for each case. This also provides information about the sub-crystals contained within each inequivalent crystal, as well as explicit solutions for the couplings. The couplings generally have more freedom than in HS-SDGR, HS-SDYM, and the full chiral higher-spin theory.
We do not present all cases here --- only some examples --- to clarify the logic and display explicit solutions.

We begin by noting that the crystal in \eqref{Paper1-2d_lower_spin} correctly reproduces the result of \eqref{Paper1-lower_der_lower_spin}. The simplest example involving higher-spin fields is given by \eqref{Paper1-simplestcase}, which admits the following solutions for the couplings:
\begin{subequations}
\begin{align}
    &\{C^{2,\lambda,-\lambda}=C^{2,2,-2}\}\,,\\
    & \{C^{-2,2,2}\}\,,
\end{align}
\end{subequations}
where we list only the nonzero couplings. The first line corresponds to the full crystal, and the second to its sub-crystal, which coincides with SDGR.

The crystal \eqref{Paper1-2pcase2} admits the solutions:
\begin{subequations}
\begin{align}
    &\{C^{2,-\lambda_2,\lambda_2}=C^{2,2-\lambda_1-\lambda_2,-2+\lambda_1+\lambda_2}=C^{2,-\lambda_1,\lambda_1}=C^{-2,2,2},\quad C^{\lambda_1,\lambda_2, 2 - \lambda_1 - \lambda_2}\}\,,\\
    &\{C^{2,2,-2}=C^{2,-\lambda_2,\lambda_2}=C^{2,-\lambda_1,\lambda_1}\}\,,\\
    &\{C^{2,2,-2}=C^{2,-\lambda_1,\lambda_1}=C^{2,2-\lambda_1-\lambda_2,-2+\lambda_1+\lambda_2}\}\,,\\
    &\{C^{2,2,-2}=C^{2,-\lambda_2,\lambda_2}=C^{2,2-\lambda_1-\lambda_2,-2+\lambda_1+\lambda_2}\}\,,\\
    &\{C^{2,2,-2}=C^{2,-\lambda_1,\lambda_1}\}\,,\\
    &\{C^{2,2,-2}=C^{2,-\lambda_2,\lambda_2}\}\,,\\
    &\{C^{2,2,-2}=C^{2,2-\lambda_1-\lambda_2,-2+\lambda_1+\lambda_2}\}\,,\\
    &\{C^{2,2,-2}, C^{\lambda_1, \lambda_2, 2 - \lambda_1 - \lambda_2}\}\,.
\end{align}
\end{subequations}
Thus, both \eqref{Paper1-simplestcase} and SDGR are contained as sub-crystals. Note that the coupling $C^{\lambda_1, 2 - \lambda_1 - \lambda_2, \lambda_2}$ is free and is what makes this crystal more than just a sum of \eqref{Paper1-simplestcase}.

We give two other examples, where the couplings differ from the standard Metsaev-type solution. The crystal \eqref{Paper1-example3} has the solution:
\begin{align}
\begin{split}
\{&C^{-2, 2, 2}=C^{0, 0, 2}=C^{2, 2 - 2 \lambda, -2 + 2 \lambda}=C^{2, 2 - \lambda, -2 + \lambda}=C^{2, -\lambda, \lambda},\\
&C^{2 - \lambda, 2 - \lambda, -2 + 2 \lambda}=\frac{(C^{0, 2 - \lambda, \lambda})^2}{C^{2 - 2 \lambda, \lambda, \lambda}},C^{0, 2 - \lambda, \lambda},C^{2 - 2  \lambda, \lambda, \lambda}\}\,.
\end{split}
\end{align}
Here, we only illustrate the solution involving all the couplings, not all the possible truncations. As we can see, there is more freedom compared to HS-SDGR, where all couplings must be equal.

The crystal \eqref{Paper1-example4} has the solution:
\begin{align}
\begin{split}
    &\{C^{4 - 3 \lambda, 4 - 3 \lambda, -6 + 6 \lambda}=\frac{(C^{0, 4 - 3 \lambda, -2 + 3 \lambda})^2}{C^{6 - 6 \lambda, -2 + 3 \lambda, -2 + 3 \lambda}},
C^{4 - 3 \lambda, 2 - \lambda, -4 + 4 \lambda}=\frac{C^{0, 4 - 3 \lambda, -2 + 3 \lambda} C^{0, 
   2 - \lambda, \lambda}}{C^{4 - 4 \lambda, \lambda, -2 + 3 \lambda}},\\
&C^{2 - 2 \lambda, 2 - \lambda, -2 + 3 \lambda}=\frac{C^{0, 4 - 3 \lambda, -2 + 3 \lambda}, C^{0, 
   2 - \lambda, \lambda}}{C^{4 - 3 \lambda, \lambda, -2 + 2 \lambda}}, 
C^{2 - \lambda, 2 - \lambda, -2 + 2 \lambda}=\frac{(C^{0, 2 - \lambda, \lambda})^2}{ C^{2 - 2 \lambda, \lambda, \lambda}},\\
&C^{4 - 4 \lambda, \lambda, -2 + 3 \lambda}, C^{
 4 - 3 \lambda, \lambda, -2 + 2  \lambda},C^{0, 4 - 3 \lambda, -2 + 3 \lambda}, C^{0, 
 2 - \lambda, \lambda},C^{6 - 6 \lambda, -2 + 3 \lambda, -2 + 3  \lambda},C^{2 - 2 \lambda, \lambda, \lambda}\}\,.
 \end{split}
\end{align}

\section{Amplitudes}\label{Paper1-section5}
In \cite{Skvortsov:2020wtf, Skvortsov:2020gpn}, it was shown that the tree-level and one-loop amplitudes of the full chiral higher-spin theory constructed from the Metsaev couplings vanish. A more general statement about solutions of the holomorphic constraints was made in \cite{Ponomarev:2017nrr}, see also \cite{Monteiro:2022lwm,Monteiro:2022xwq}. Here, we show that this result can be extended to all chiral higher-spin theories, with a caveat.

Let us compute the $4$-pt amplitude for a generic chiral higher-spin theory. We use identities from Appendix \ref{Paper1-AppendixB}, which hold on-shell $(H_2=0)$. The amplitude can be written as
\begin{align}\label{Paper1-OPE_Associativity}
\begin{split}
\mathcal{A}&=\mathcal{A}_s+\mathcal{A}_t+\mathcal{A}_u=
    \sum_{\omega}\mathcal{C}^{1234\omega}\frac{\PPb_{12}^{\lambda_{12}+\omega}}{\beta_1^{\lambda_1}\beta_2^{\lambda_2}}\frac{1}{(q_1+q_2)^2}\frac{\PPb_{34}^{\lambda_{34}-\omega}}{\beta_3^{\lambda_3}\beta_4^{\lambda_4}}+2\leftrightarrow 4+2\leftrightarrow 3\\
    &=\frac{\PPb_{12}\PPb_{34}}{(q_1+q_2)^2\prod_{i=1}^4\beta_i^{\lambda_i}}\sum_{\omega}\Big(\mathcal{C}^{1234\omega}\PPb_{12}^{\lambda_{12}+\omega-1}\PPb_{34}^{\lambda_{34}-\omega-1}-\mathcal{C}^{1423\omega}\PPb_{14}^{\lambda_{14}+\omega-1}\PPb_{32}^{\lambda_{23}-\omega-1}\\
    &\;\;\;\;-\mathcal{C}^{1324\omega}\PPb_{13}^{\lambda_{13}+\omega-1}\PPb_{24}^{\lambda_{24}-\omega-1}\Big)=\frac{\PPb_{12}\PPb_{34}}{(q_1+q_2)^2\prod_{i=1}^4\beta_i^{\lambda_i}}\sum_{\omega}\Big(\mathcal{C}^{1234\omega}\PPb_{12}^{\lambda_{12}+\omega-1}\PPb_{34}^{\lambda_{34}-\omega-1}\\
&\;\;\;\;+\mathcal{C}^{2314\omega}\PPb_{23}^{\lambda_{23}+\omega-1}\PPb_{14}^{\lambda_{14}-\omega-1}
+\mathcal{C}^{1324\omega}\PPb_{31}^{\lambda_{13}+\omega-1}\PPb_{24}^{\lambda_{24}-\omega-1}\Big)=0\,,
\end{split}
\end{align}
where in the last equality we assume only even couplings, so the sum is over even or odd values of $\omega$ only.
Plugging in the general solution \eqref{Paper1-symfinalsystem}, we obtain
\begin{align}
\begin{split}
\mathcal{A}=&\frac{k^{1234}_+\PPb_{12}\PPb_{34}}{2^{\Lambda-2}\prod_{i=1}^4\beta_i^{\lambda_i}(q_1+q_2)^2}\Big((\PPb_{12}+\PPb_{34})^{\Lambda-2}-(\PPb_{34}-\PPb_{12})^{\Lambda-2}\\
&+(\PPb_{23}+\PPb_{14})^{\Lambda-2}-(\PPb_{14}-\PPb_{23})^{\Lambda-2}+(\PPb_{31}+\PPb_{24})^{\Lambda-2}-(\PPb_{24}-\PPb_{31})^{\Lambda-2}\Big)\,,
\end{split}
\end{align}
which, using the variables $A,B,C$, becomes
\begin{equation}
    \mathcal{A}=\frac{k^{1234}_+\PPb_{12}\PPb_{34}}{\prod_{i=1}^4\beta_i^{\lambda_i}(q_1+q_2)^2}\Big((A)^{\Lambda-2}-(B)^{\Lambda-2}+(C)^{\Lambda-2}-(-A)^{\Lambda-2}+(-B)^{\Lambda-2}-(-C)^{\Lambda-2}\Big)=0\,.
\end{equation}
The same conclusion holds if we include odd-derivative couplings, as long as the conditions \eqref{Paper1-case4system} are satisfied.

There is a caveat: in the case discussed above, we did not account for the fact that the product $C^{\lambda_1,\lambda_1,0}C^{0,\lambda_2,\lambda_2}$ can be arbitrary. When this is properly taken into account, certain exceptions to the vanishing of the amplitudes may arise.
For example, we can examine the amplitude in the simplest higher-derivative theory that includes the abelian $R^2\phi$ and/or $R^3$ terms \eqref{Paper1-HDlowerspin}, and compute the $(2,2,2,2)$ scattering amplitude
\begin{align}
\begin{split}
    \mathcal{A}&(2,2,2,2)=\frac{\PPb_{12}\PPb_{34}}{(q_1+q_2)^2\prod_{i=1}^4\beta_i^2}\sum_{\omega}C^{2,2,\omega}C^{-\omega,2,2}\Big(\PPb_{12}^{3+\omega}\PPb_{34}^{3-\omega}+\PPb_{23}^{3+\omega}\PPb_{14}^{3-\omega}+\PPb_{31}^{3+\omega}\PPb_{24}^{3-\omega}\Big)\\
    &=\frac{\PPb_{12}\PPb_{34}}{(q_1+q_2)^2\prod_{i=1}^4\beta_i^2}(C^{2,2,0}C^{0,2,2}-\frac{10}{3}C^{2,2,-2}C^{2,2,2})\Big(\PPb_{12}^3\PPb_{34}^3+\PPb_{23}^3\PPb_{14}^3+\PPb_{31}^3\PPb_{24}^3\Big)\,.
    \end{split}
\end{align}
Here the freedom in choosing $C^{2,2,0}C^{0,2,2}$ allows for a non-vanishing amplitude. Notice that this does not contradict the Weinberg no-go theorem \cite{Weinberg:1964ev}, since these are abelian (Born-Infeld type) vertices, not constrained by the Weinberg low-energy theorem.\\
Note that in a larger theory (e.g. one including all higher-spin fields), the products $C^{\lambda_1,\lambda_1,0}C^{0,\lambda_2,\lambda_2}$ may be subject to constraints.

We can also demonstrate the vanishing of amplitudes in the presence of a gauge group. Remember that in this case we can separate the colour factor and rewrite it in the form
\begin{equation}
    \mathcal{A}(12\cdots n)=\sum_{\sigma\in S_n/\mathbb{Z}_n}\mathrm{Tr}(T^{a_{\sigma_1}}T^{a_{\sigma_2}}\cdots T^{a_{\sigma_n}})\tilde{\mathcal{A}}(\sigma_1\sigma_2\cdots\sigma_n)\,.
\end{equation}
By considering a single colour-ordered amplitude, namely $[1234]$, since the same is valid for the others, we obtain
\begin{align}
\begin{split}
    \tilde{\mathcal{A}}&(1234)=\frac{\PPb_{12}\PPb_{34}}{(q_1+q_2)^2\prod_{i=1}^4\beta_i^{\lambda_i}}\sum_{\omega}\Big(\mathcal{C}^{1234\omega}\PPb_{12}^{\lambda_{12}+\omega-1}\PPb_{34}^{\lambda_{34}-\omega-1}-\mathcal{C}^{2341\omega}\PPb_{23}^{\lambda_{23}+\omega-1}\PPb_{41}^{\lambda_{41}-\omega-1}\Big)\\
    &=\frac{\PPb_{12}\PPb_{34}}{(q_1+q_2)^2\prod_{i=1}^4\beta_i^{\lambda_i}}\sum_{\omega}\Big(\mathcal{C}^{1234\omega}\PPb_{12}^{\lambda_{12}+\omega-1}\PPb_{34}^{\lambda_{34}-\omega-1}+(-)^{\omega}\mathcal{C}^{1423\omega}\PPb_{14}^{\lambda_{14}+\omega-1}\PPb_{23}^{\lambda_{23}-\omega-1}\Big)\\
    &=\frac{\PPb_{12}\PPb_{34}}{2^{\Lambda-2}(q_1+q_2)^2\prod_{i=1}^4\beta_i^{\lambda_i}}\Big((\PPb_{12}+\PPb_{34})^{\Lambda-2}+(\PPb_{14}-\PPb_{23})^{\Lambda-2}\Big)\sim (A^{\Lambda-2}-A^{\Lambda-2})=0\,.
\end{split}
\end{align}
Here, we include both even- and odd-derivative vertices in the sum.

Again, examples exist where the amplitude does not vanish. For instance, in the lower-spin theory with an $F^2\phi$ and/or $F^3$ interactions, the amplitude $A(1111)$ can be nonzero, thanks to the freedom of the product $C^{1,1,0}C^{0,1,1}$.

In principle, there seems to be no obstruction to extending the analysis of \cite{Skvortsov:2018jea,Skvortsov:2020wtf,Skvortsov:2020gpn} to show that all tree-level amplitudes vanish, up to the small caveat we discussed, for any generic $n$-pt scattering in all chiral higher-spin theories.

\section{Conclusions and discussion}\label{Paper1-section6}
One of the main results of the paper is that we have shown that, contrary to common belief, there are nontrivial higher-spin theories with finite spectra of fields. We have found a great number of nontrivial solutions to the quartic consistency condition in the light-cone gauge, which enriches the space of higher-spin theories in $4d$. However, we have not found all of them since we omitted the higher-derivative interactions, and this is the first obvious direction for future work. In addition, the choice of representations for the colour cases has been very restrictive and we have not studied theories that have both coloured and singlet fields under some gauge algebra as well. Lastly, we have completely omitted fermions and, as a result, supersymmetric theories.

So far, we have directly analysed the quartic consistency condition in the light-cone gauge. It was shown in \cite{Ponomarev:2017nrr} that it entails the Jacobi identity for a Lie algebra, save for some degenerate cases. Therefore, it seems plausible that the problem can be solved by the algebraic tools and one can find a class of finite- and infinite-dimensional Lie algebras that lead to consistent higher-spin theories.\footnote{Not all of the new theories feature a spin-two field, i.e. not all of them are ``gravities''.} The latter would make the parallel to the case $3d$ higher-spin gravities even more pronounced.

It is worth discussing to what extent the new theories are ``nontrivial''. Obviously, one can take any number of abelian couplings and add them up, which imposes no constraints in the light-cone gauge. One can also activate some $C^{\lambda_1,\lambda_2,\lambda_3}$ that do not ``talk'' to each other, i.e. no exchange diagram can be formed. These cases are trivial. Whenever an exchange diagram exists, there is a nontrivial equation for the couplings, which quite often forces us to introduce further couplings that contribute to the same exchange diagram to make it Poincaré invariant. From what we can see, there does not seem to be any way to discriminate such cases further, i.e. to divide couplings into ``more nonabelian'' and ``less nonabelian''. Nevertheless, the final results, i.e. the set of couplings that need to be turned on, can be rather different. For example, $C^{+s,+s,-s}$, $s>2$ leads to the full chiral higher-spin gravity and activates all couplings save for $C^{0,0,0}$. Also, the presence of scalar couplings $C^{\lambda_1,\lambda_2,0}$ is very important.

It is also important to note that in the classification we presented we assume that all coupling constants, if more than just one, are nonvanishing. First of all, this gives a simple way to list all cubic vertices that belong to a given theory, which is what we call ``spectrum determines couplings/cubic vertices''. Secondly, the equations for the couplings $C_i$ have the schematic form $C_1C_2+C_3C_4+...=0$ with many degenerate points. Nevertheless, if some of the coupling constants are set to zero, we will have to end up with another theory that belongs to the classification. It would be interesting to study the ``degeneration tree'' that determines the relation between various theories in the classification when some of the coupling constants are set to zero.

There are a number of natural future directions triggered by the results obtained in the paper. 
\begin{description}
    \item[Covariantization.] So far, it has always been possible to covariantize the results obtained in the light-cone gauge, e.g. \cite{Krasnov:2021nsq}, and we expect this to be true for the new theories. Any cubic interaction can easily be written in a covariant way, e.g. one can extract such couplings from the equations of motion of chiral theory in flat space \cite{Sharapov:2022faa,Sharapov:2022wpz,Sharapov:2022awp,Sharapov:2022nps,Sharapov:2023erv}. However, the covariantization of chiral theory forces one to introduce higher-order interactions. Therefore, the covariantization of the new theories may not be straightforward.

    We also note that the covariant form of HS-SDYM and HS-SDGR obtained in \cite{Krasnov:2021nsq} does not contain vertices of type ($--$+) and the scalar field as in \cite{Ponomarev:2017nrr}, see also \cite{Monteiro:2022xwq}.

    \item[Twistor space.] Self-dual theories admit a simple description on twistor space. This applies to HS-SDYM and HS-SDGR as well, \cite{Tran:2021ukl,Herfray:2022prf,Tran:2022tft,Adamo:2022lah}, see also \cite{Adamo:2016ple} for the self-dual conformal higher-spin gravity. However, the twistor formulation of chiral higher-spin gravity was lacking, see \cite{Tran:2022tft} for the first steps and \cite{Mason:2025pbz} for the very recent update. 

    \item[(Anti)-de Sitter space.] All known higher-spin gravities are smooth in the cosmological constant, and one could expect this to be true for the new solutions. However, the light-cone analysis becomes more complicated, see \cite{Metsaev:2018xip}. In particular, there seem to be more constraints, because $AdS_4$ cubic vertices have a more complicated structure compared to their flat space limits, but some interesting results can still be obtained \cite{Skvortsov:2018uru, Neiman:2023bkq, Neiman:2024vit, Lipstein:2023pih, Chowdhury:2024dcy}. Therefore, covariantization of the new theories seems a necessary step to proceed. 

    \item[Quantum corrections.] One can quite generally show that most holomorphic solutions of the quartic light-cone consistency condition have vanishing tree level amplitudes, \cite{Ponomarev:2017nrr} and also the present paper. This implies that there are no UV-divergences in these theories at least at one-loop. However, for the theories that have a finite spectrum, one cannot expect the additional mechanism (regularized sum over the spectrum) that would make them vanish \cite{Skvortsov:2018jea,Skvortsov:2020wtf,Skvortsov:2020gpn}. After all, the scattering of higher-spin fields might have nontrivial S-matrix \cite{Tran:2022amg}. For a more recent analysis of anomalies for higher-spin theories in twistor space, see \cite{Tran:2025uad}.

    \item[Celestial holography.] It was shown in \cite{Ren:2022sws} that chiral higher-spin gravity solves the celestial OPE associativity condition. We expect that all of the theories constructed here (and other, higher derivative, solutions that are yet to be found) do solve the same constraints. On the other hand, the study of flat space holography for chiral higher-spin gravity has just begun in  \cite{Ponomarev:2022atv,Ponomarev:2022ryp,Ponomarev:2022qkx,Monteiro:2022xwq}. It would be interesting to see if similar statements can be made about the new theories.

    \item[Unitary higher-spin theory.] It would be interesting to investigate whether some of these chiral higher-spin theories --- especially the higher-derivative ones --- admit a local unitary completion, given that they display more freedom in their couplings. In any case, studying their unitary completions could provide valuable insights and better control over the non-localities. We have also found that there exist some exotic low-spin couplings, e.g.\footnote{If fermions are included, there is also $C^{0,-1/2,3/2}$. } $C^{-2,1,2}$ that pass certain consistency checks and, in general, more freedom (holomorphic and anti-holomorphic vertices can be considered independent) is available in the choice of vertices as compared to covariant approaches \cite{Bengtsson:1986kh,Metsaev:1991nb,Metsaev:1991mt,Metsaev:1993ap,Bengtsson:2014qza}. Therefore, it is worth investigating if more consistent theories can be constructed with the usual, low-spin fields.

    \item[Higher dimensions.] Recently, some of the chiral higher-spin ``magic'' has been extended to $6d$ and, more generally, to all even dimensions \cite{Basile:2024raj}. One can hope that some of the results of the present paper can be extended to even dimensions.
\end{description}

\section*{Acknowledgments}
\label{Paper1-sec:Aknowledgements}
I would like to thank Zhenya Skvortsov for the guidance and numerous discussions during this project.  I am grateful to Dmitry Ponomarev for the many useful comments on the preliminary version of the paper. This project has received funding from the European Research Council (ERC) under the European Union’s Horizon 2020 research and innovation programme (grant agreement No 101002551).

\refstepcounter{section}
\section*{2.A \hspace{1mm} Standard results and notations}
\addcontentsline{toc}{section}{2.A \hspace{1mm} Standard results and notations}
\label{Paper1-AppendixA}
The $4d$ Poincaré algebra $iso(3,1)$ is defined as
\begin{subequations}
\begin{align}
    [P^A,P^B]=&\,0\,,\\
    [J^{AB},P^C]=&\,P^A\eta^{BC}-P^B\eta^{AC}\,,\\
    [J^{AB},J^{CD}]=&\,J^{AD}\eta^{BC}-J^{BD}\eta^{AC}-J^{AC}\eta^{BD}+J^{BC}\eta^{AD}\,,
\end{align}
\end{subequations}
where $P^A$ are the generators of translations and $J^{AB}$ of Lorentz transformations.

To perform canonical quantisation in quantum field theory, one has to choose a hypersurface. A standard approach is to select equal time slices, $t=t_0$, and define the Hamiltonian $H=P^0$ as the operator responsible for evolving the system in the time direction.

In this work, however, we adopt light-front quantisation, where we fix a light-like hypersurface and treat $x^+$ as our time direction. The corresponding Hamiltonian is then given by $H\equiv P^-$.\footnote{Unlike standard equal time slices of the spacetime $\mathcal{M}$ at fixed $x^0$, we consider surfaces of constant $x^+$, and for simplicity we can fix $x^+=0$. Choosing initial data on a light-like surface changes the standard Cauchy problem, requiring careful treatment of boundary conditions. A detailed discussion can be found in \cite{Neville:1971zk,Heinzl:2000ht}.}

We consider general massless higher-spin fields, which in $4d$ possess two degrees of freedom. These are denoted $\phi^{\pm s}$, representing helicities $\pm s$, and are complex conjugates of each other.

We use both light-cone coordinates and the light-cone gauge.
In flat spacetime, we adopt the $4d$ Minkowski metric with mostly plus signature as
\begin{equation}
    ds^2=-(dx^0)^2+\sum^{3}_{i=1}\,(dx^i)^2\,.
\end{equation}
We define light-cone coordinates as
\begin{align}
    &x^+=\frac{x^3+x^0}{\sqrt{2}}\,,&
    &x^-=\frac{x^3-x^0}{\sqrt{2}}\,,&
    &z=\frac{x^1-ix^2}{\sqrt{2}}\,,&
    &\bar{z}=\frac{x^1+ix^2}{\sqrt{2}}\,,
\end{align}
and the metric becomes
\begin{equation}
    ds^2=2\,dx^+ dx^- + 2\,dz d\bar{z}\,.
\end{equation}

We work with Fourier transformed fields with respect to $x^-$ and the transverse directions. The Dirac bracket is given by
\begin{equation}
    [\phi_p^{\mu}(x^+),\phi_q^{\nu}(x^+)]=\delta^{\mu,-\nu}\frac{\delta^3(p+q)}{2p^+}\,.
\end{equation}
The free field realisation of the kinematical Poincaré generators is\footnote{Following standard notations in the light-cone, we rename $\beta=p^+$.}
\begin{subequations}
\begin{align}
    &P^+=\beta\,,&
    &P=q\,,&
    &\bar{P}=\bar{q}\,,\\
    &J^{z+}=-\beta\frac{\partial}{\partial \bar{q}}\,,&
    &J^{\bar{z}+}=-\beta\frac{\partial}{\partial q}\,,&
    &J^{-+}=-N_{\beta}-1\,,\\
    &J^{z\bar{z}}=N_q-N_{\bar{q}}-\lambda\,,
\end{align}
\end{subequations}
where $N_q=q\partial_q$ is the Euler operator.
The dynamical generators are
\begin{align}
    &H_2=-\frac{q\bar{q}}{\beta}\,,&
    &J_2^{z-}=\frac{\partial}{\partial\bar{q}}\frac{q\bar{q}}{\beta}+q\frac{\partial}{\partial\beta}+\lambda\frac{q}{\beta}\,,&
    &J_2^{\bar{z}-}=\frac{\partial}{\partial q}\frac{q\bar{q}}{\beta}+\bar{q}\frac{\partial}{\partial\beta}-\lambda\frac{\bar{q}}{\beta}\,.
\end{align}
We now deform the dynamical generators using a local ansatz:\footnote{Only the dynamical generators are deformed. One advantage of working in light-front quantisation is the reduced number of dynamical generators. Out of $10$ Poincaré generators only $3$ are dynamical ($H,J^{z-},J^{\bar{z}-}$), compared to $4$ in the usual equal-time quatisation ($H^0,J^{0a}$).}
\begin{align}\label{Paper1-hamiltonian}
    H=&\,H_2+\sum_n\int d^{3n}q\;\delta\Big(\sum_i q_i\Big)h^{q_1,...,q_n}_{\lambda_1,...,\lambda_n}\phi^{\lambda_1}_{q_1}\cdots\phi^{\lambda_n}_{q_n}\,,\\\label{Paper1-boostz}
    J^{z-}=&\,J_2^{z-}+\sum_n\int d^{3n}q\;\delta\Big(\sum_i q_i\Big)\Big[j^{q_1,...,q_n}_{\lambda_1,...,\lambda_n}-\frac{1}{n}\,h^{q_1,...,q_n}_{\lambda_1,...,\lambda_n}\Big(\sum_j\frac{\partial}{\partial \bar{q}_j}\Big)\Big]\phi^{\lambda_1}_{q_1}\cdots\phi^{\lambda_n}_{q_n}\,,\\ \label{Paper1-boostzbar}
    J^{\bar{z}-}=&\,J_2^{\bar{z}-}+\sum_n\int d^{3n}q\;\delta\Big(\sum_i q_i\Big)\Big[\bar{j}^{q_1,...,q_n}_{\lambda_1,...,\lambda_n}-\frac{1}{n}\,h^{q_1,...,q_n}_{\lambda_1,...,\lambda_n}\Big(\sum_j\frac{\partial}{\partial q_j}\Big)\Big]\phi^{\lambda_1}_{q_1}\cdots\phi^{\lambda_n}_{q_n}\,.
\end{align}
Let us define the momentum combinations
\begin{align}
    &\PP_{ij}=q_i\beta_j-q_j\beta_i\,,&
    &\PPb_{ij}=\bar{q}_i\beta_j-\bar{q}_j\beta_i\,,
\end{align}
where $\PPb_{ij}=-\PPb_{ji}$ and $\PP_{ij}=-\PP_{ji}$. Solving order by order all constraints required by the closure of the Poincaré algebra \cite{Ponomarev:2016lrm}, leads to the classification of cubic vertices:
\begin{align}
h_{\lambda_1,\lambda_2,\lambda_3}=&\,C^{\lambda_1,\lambda_2,\lambda_3}\frac{\PPb^{\lambda_{123}}}{\beta_1^{\lambda_1}\beta_2^{\lambda_2}\beta_3^{\lambda_3}}+\bar{C}^{-\lambda_1,-\lambda_2,-\lambda_3}\frac{\PP^{-\lambda_{123}}}{\beta_1^{-\lambda_1}\beta_2^{-\lambda_2}\beta_3^{-\lambda_3}}\,,\\
    j_{\lambda_1,\lambda_2,\lambda_3}=&\,\frac{2}{3}\,C^{\lambda_1,\lambda_2,\lambda_3}\frac{\PPb^{\lambda_{123}-1}}{\beta_1^{\lambda_1}\beta_2^{\lambda_2}\beta_3^{\lambda_3}}\Lambda^{\lambda_1,\lambda_2,\lambda_3}\,,\\
    \bar{j}_{\lambda_1,\lambda_2,\lambda_3}=&\,-\frac{2}{3}\,\bar{C}^{-\lambda_1,-\lambda_2,-\lambda_3}\frac{\PP^{-\lambda_{123}-1}}{\beta_1^{-\lambda_1}\beta_2^{-\lambda_2}\beta_3^{-\lambda_3}}\Lambda^{\lambda_1,\lambda_2,\lambda_3}\,,
\end{align}
with $\lambda_{123}=\lambda_1+\lambda_2+\lambda_3$ and where we define
\begin{align}
    &\PP^a_{12}=\PP^a_{23}=\PP^a_{31}=\PP^a=\frac{1}{3}\,\Big[(\beta_1-\beta_2)q_3^a+(\beta_2-\beta_3)q_1^a+(\beta_3-\beta_1)q_2^a\Big]\,,\\
    &\Lambda^{\lambda_1,\lambda_2,\lambda_3}=\,\beta_1(\lambda_2-\lambda_3)+\beta_2(\lambda_3-\lambda_1)+\beta_3(\lambda_1-\lambda_2)\,.
\end{align}
Note that due to momentum conservation $\PP$ is cyclic invariant, then $\sigma_{123}\PP=\PP\,,$ same for $\PPb$.

Here $C^{\lambda_1,\lambda_2,\lambda_3}$ and $\bar{C}^{\lambda_1,\lambda_2,\lambda_3}$ are a priori independent coupling constants.
For dimensional reasons, we can introduce a length parameter to compensate for the powers of momenta. We denote it by $\ell_P$, which can naturally be associated with the Planck length. The couplings take the form
\begin{align}
    &C^{\lambda_1,\lambda_2,\lambda_3}=(\ell_P)^{\lambda_{123}-1}c^{\lambda_1,\lambda_2,\lambda_3}\,,&
    &\bar{C}^{\lambda_1,\lambda_2,\lambda_3}=(\ell_P)^{\lambda_{123}-1}\bar{c}^{\lambda_1,\lambda_2,\lambda_3}\,.
\end{align}
If we demand unitarity, in the case with no internal symmetry, we have to impose
\begin{equation}
    \bar{C}^{\lambda_1,\lambda_2,\lambda_3}=(C^{\lambda_1,\lambda_2,\lambda_3})^*\,.
\end{equation}
If we include a gauge group, unitarity imposes
\begin{align}
     \bar{C}^{\lambda_1,\lambda_2,\lambda_3}&= (-)^{\lambda_{123}}(C^{\lambda_1,\lambda_2,\lambda_3})^*&&\text{for:}\;SO(N),USp(N)\,.
\end{align}
\refstepcounter{section}
\section*{2.B \hspace{1mm} Useful relations}
\addcontentsline{toc}{section}{2.B \hspace{1mm} Useful relations}
\label{Paper1-AppendixB}
In $4d$ and light-cone gauge, for a given $N$-pt function, there are $N-2$ independent $\PPb_{ij}$ variables.
In particular, for $4$-pt scattering, we have $2$ independent $\PPb$ variables, which, for example, can be chosen to be $\PPb_{12}$, $\PPb_{34}$. Additionally, there are three independent $\beta$'s. All other $\PPb_{ij}$ can be expressed as
\begin{subequations}
\begin{align}
    \PPb_{13}&=\frac{\beta_3 \PPb_{12}+\beta_1 \PPb_{34}}{\beta_1+\beta_2}\,,
    &&\PPb_{14}=-\frac{\PPb_{12} (\beta_1+\beta_2+\beta_3)+\beta_1 \PPb_{34}}{\beta_1+\beta_2}\,,\\
    \PPb_{23}&=\frac{\beta_2 \PPb_{34}-\beta_3 \PPb_{12}}{\beta_1+\beta_2}\,,
    &&\PPb_{24}=\frac{\PPb_{12} (\beta_1+\beta_2+\beta_3)-\beta_2 \PPb_{34}}{\beta_1+\beta_2}\,.
\end{align}
\end{subequations}
Below, we collect some useful relations for $n > 0$ that are employed in the main text:
\begin{subequations}
\begin{align}
    \sum_{k=0}^{n}\begin{pmatrix}
        n\\
        k
\end{pmatrix}k\,(k-1)\cdots (k-\ell+1)=&\,n\,(n-1)\cdots (n-\ell+1)\,2^{n-\ell}\,,\\
\sum_{k\in\text{even/odd}}^{n}\begin{pmatrix}
        n\\
        k
\end{pmatrix}k\,(k-1)\cdots (k-\ell+1)=&\,
n\,(n-1)\cdots (n-\ell+1)\,2^{n-\ell-1}\,.
\end{align}
\end{subequations}
These identities can be derived by taking $\ell$ derivatives of the binomial theorem. In particular, the following relations will also prove useful:
\begin{align}
    \sum_{k=0}^n
    &\begin{pmatrix}
        n\\
        k
    \end{pmatrix}
    (-)^k(A-B)^k(A+B)^{n-k}=2^nB^n\,,\\
 \sum_{k=0}^n
    &\begin{pmatrix}
        n\\
        k
    \end{pmatrix}
    (A-B)^k(A+B)^{n-k}=2^nA^n\,,\\
    \sum_{k\in\text{even}}^n
    &\begin{pmatrix}
        n\\
        k
    \end{pmatrix}
    (A-B)^k(A+B)^{n-k}=2^{n-1}(A^n+B^n),\\
    \sum_{k\in\text{odd}}^n
    &\begin{pmatrix}
        n\\
        k
    \end{pmatrix}
    (A-B)^k(A+B)^{n-k}=2^{n-1}(A^n-B^n)\,,
\end{align}
\vspace{-2em}  
\begin{align}
    \begin{split}
    \sum_{k=0}^n
    \begin{pmatrix}
        n\\
        k
    \end{pmatrix}
    (-)^k k\,(k-&1)\cdots (k-\ell+1)(A-B)^k(A+B)^{n-k}\\
    &=n\,(n-1)\cdots (n-\ell+1)(B-A)^{\ell}(2B)^{n-\ell}\,,
    \end{split}
\end{align}
\vspace{-2em}  
\begin{align}
\begin{split}
    \sum_{k=0}^n
    \begin{pmatrix}
        n\\
        k
    \end{pmatrix}
    k\,(k-1)\cdots& (k-\ell+1)(A-B)^k(A+B)^{n-k}\\
    &=n\,(n-1)\cdots (n-\ell+1)(A-B)^{\ell}(2A)^{n-\ell}\,.
     \end{split}
\end{align}
Finally, for $4$-pt scattering, the following relations hold on-shell:
\begin{align}
    &(q_i+q_j)^2=-\frac{2}{\beta_i\beta_j}\PP_{ij}\PPb_{ij}\,,& 
    &\sum_{j=1}^4\frac{\PP_{ij}\PPb_{jk}}{\beta_j}=0\,.
\end{align}
From these, we deduce
\begin{equation}
    \frac{\PPb_{12}\PPb_{34}}{(q_1+q_2)^2}=\frac{\PPb_{31}\PPb_{24}}{(q_1+q_3)^2}=\frac{\PPb_{14}\PPb_{23}}{(q_1+q_4)^2}\,.
\end{equation}
\refstepcounter{section}
\section*{2.C \hspace{1mm} Low-derivative constraints}
\addcontentsline{toc}{section}{2.C \hspace{1mm} Low-derivative constraints}
\label{Paper1-AppendixC}
We begin by analysing the cases $\Lambda = 2$ and $\Lambda = 3$ for the constraint \eqref{Paper1-LCholo}. For simplicity, we adopt the following notation in both cases:
\begin{equation}\label{Paper1-simple_notation}
    C^{1234}_i\equiv C^{\lambda_1,\lambda_2,i-\lambda_{12}}C^{\lambda_{12}-i,\lambda_3,\Lambda-\lambda_{123}}\,,
\end{equation}
with analogous definitions for the orderings $(1324)$ and $(1423)$.

\paragraph{Case $\Lambda=2$.} In this case, only one-derivative interactions are present, and the constraint is
\begin{subequations}
\begin{align}
    &(\lambda_{13}-1)(C^{1234}_1-C^{1324}_1+C^{1423}_1)=0\,,\\
    &(1-\lambda_{23})(C^{1234}_1-C^{1324}_1+C^{1423}_1)=0\,,\\
    &(\lambda_{12}-1)(C^{1234}_1-C^{1324}_1+C^{1423}_1)=0\,.
\end{align}
\end{subequations}
These equations are consistent with the antisymmetry of the couplings, and they admit the solution
\begin{equation}\label{Paper1-L=2_constraint}
\boxed{
    C^{1234}_1=C^{1324}_1-C^{1423}_1\,.
    }
\end{equation}
Therefore, in one-derivative theories, it is possible to find solutions even when only odd-derivative vertices are present. This stands in contrast to the general case, where odd-derivative vertices must be accompanied by even-derivative ones for consistent solutions to exist.

As expected, this solution does not lead to a self-interacting photon. For example, assuming a theory with the coupling $C^{-1,1,1}$, substituting it into \eqref{Paper1-L=2_constraint} gives $C^{-1,1,1}=0$. This result also follows directly from symmetry arguments --- in particular, $C^{\lambda,\lambda,\lambda'} \equiv 0$ for odd-derivative vertices, with the sole exception of $C^{0,0,1}$, which gives rise to the self-dual sector of scalar QED. In this case, the symmetry argument no longer applies because the two scalar fields need not be identical; rather, they can be complex conjugates of one another. Accordingly, it is more appropriate to write the coupling as $C^{0,\bar{0},1}$ corresponding to the cubic interaction $\phi\bar{\phi}A^+$. Furthermore, the constraint \eqref{Paper1-L=2_constraint} leaves $C^{0,\bar{0},1}$ unconstrained.

\paragraph{Case $\Lambda=3$.} In this case, the sum runs over two values of $\omega$ and includes two-derivative interactions. The constraint becomes
\begin{subequations}
\begin{align}
    (2\lambda_{13}-3)((-)^{\lambda_{12}}C_1^{1234}+(-)^{\lambda_{12}+1}C_2^{1234}+(-)^{\lambda_{14}}C_1^{1423}+(-)^{\lambda_{14}}C_2^{1423})=&\,0\,,\\
    (3-2\lambda_{23})((-)^{\lambda_{12}}C_1^{1234}+(-)^{\lambda_{12}}C_2^{1234}+(-)^{\lambda_{13}}C_1^{1324}+(-)^{\lambda_{13}}C_2^{1324})=&\,0\,,\\
    (2\lambda_{12}-3)((-)^{\lambda_{13}}C_1^{1324}+(-)^{\lambda_{13}+1}C_2^{1324}+(-)^{\lambda_{14}}C_1^{1423}+(-)^{\lambda_{14}+1}C_2^{1423})=&\,0\,.
\end{align}
\end{subequations}
Using the definitions in \eqref{Paper1-definitions1}, this can be rewritten as
\begin{subequations}
\begin{align}
    (2\lambda_{13}-3)(k^{1234}_--k^{1423}_+)=&\,0\,,\qquad
    (3-2\lambda_{23})(k^{1234}_++k^{1324}_+)=\,0\,,\\
    (2\lambda_{12}-3)(k^{1324}_-+k^{1423}_-)=&\,0\,.
\end{align}
\end{subequations}
These are the same conditions found for the general case \eqref{Paper1-case4system}.
\paragraph{Case $\Lambda=2$ with a gauge group.} We now consider the constraint \eqref{Paper1-LCholocolour} at $\Lambda=2$ with a $U(N)$ gauge group. The constraint for the $[1234]$ colour-ordering takes the form
\begin{subequations}
\begin{align}
    (\lambda_{13}-1)((-)^{1-\lambda_{12}}\theta_{1-\lambda_{12}}C^{1234}_1+(-)^{\lambda_{14}}\theta_{1-\lambda_{14}}C^{4123}_1)=&\,0\,,\\
    (1-\lambda_{23})((-)^{1-\lambda_{12}}\theta_{1-\lambda_{12}}C^{1234}_1+(-)^{1-\lambda_{14}}\theta_{\lambda_{14}}C^{4123}_1)=&\,0\,,\\
    (\lambda_{12}-1)((-)^{1-\lambda_{12}}\theta_{1-\lambda_{12}}C^{1234}_1+(-)^{\lambda_{14}}\theta_{1-\lambda_{14}}C^{4123}_1)=&\,0\,,
\end{align}
\end{subequations}
and gives the solution
\begin{align}\label{Paper1-U(N)_1d_constraint}
    &(-)^{1-\lambda_{12}}\theta_{1-\lambda_{12}}C^{1234}_1=(-)^{1-\lambda_{14}}\theta_{1-\lambda_{14}}C^{4123}_1&
    &\Rightarrow&
    &C^{1234}_1=(-)^{\lambda_{24}}\theta_{1-\lambda_{12}}\theta_{1-\lambda_{14}}C^{4123}_1\,.
\end{align}
Meanwhile, the constraint corresponding to the colour ordering $[2341]$ leads to
\begin{equation}
    C^{2341}_1=(-)^{\lambda_{13}}\theta_{1-\lambda_{23}}\theta_{1-\lambda_{14}}C^{1234}_1\,.
\end{equation}
Using the identity $C^{4123}_1=C^{2341}_1$, we find $\theta_{1-\lambda_{23}}\theta_{1-\lambda_{14}}=1$, which implies four possible choices
\begin{align}
    \theta_{\lambda}=+1,-1,(-)^{\lambda},(-)^{\lambda+1}\,.
\end{align}
The cases $\theta_{\lambda}=+1,-1$ and $\theta_{\lambda}=(-)^{\lambda},(-)^{\lambda+1}$ yield, respectively
\begin{equation}
\boxed{
    C_1^{1234} = (-)^{\lambda_{24}} C_1^{4123}\,,
}
\qquad
\boxed{
    C_1^{1234} = C_1^{4123}\,.
}
\end{equation}
In the main text, we consider $C_1^{1234}=C^{4123}_1$, in analogy with the higher-derivative case. Moreover, for the classification of crystals, the specific values of the couplings are not important.

As we will see in Appendix \ref{Paper1-AppendixE}, the special solutions with $\theta_{\omega}=\pm 1$ are closely tied to the existence of solutions in which all fields transform in the adjoint representation. 

For the gauge groups $SO(N)$ and $USp(N)$, by using the symmetry properties given in \eqref{Paper1-SO(n)_sym} and \eqref{Paper1-USp(n)_sym} of Appendix \ref{Paper1-AppendixD}, we arrive at the same constraint \eqref{Paper1-U(N)_1d_constraint} as in the $U(N)$ case. However, here the solutions to the various colour-ordered constraints mix due to these symmetry properties, which also determine the symmetry of the couplings. In particular, we find the following solutions.

For $\theta_{\lambda}=(-)^{\lambda},(-)^{\lambda+1}$ the solution is given by
\begin{equation}
\boxed{
    C_1^{1234}=C^{4123}_1=C^{1324}_1\,,
    }
\end{equation}
and for $SO(N)$ the symmetry of the couplings is
\begin{align}
&C^{\lambda_1,\lambda_2,\lambda_3}=C^{\lambda_{\sigma_1},\lambda_{\sigma_2},\lambda_{\sigma_3}}\quad(\theta_{\omega}=(-)^{\lambda})\,,&
    &C^{\lambda_1,\lambda_2,\lambda_3}=-C^{\lambda_{\sigma_1},\lambda_{\sigma_2},\lambda_{\sigma_3}}\quad(\theta_{\omega}=(-)^{\lambda+1})\,,
\end{align}
while for $USp(N)$ the symmetry of the couplings is
\begin{align}
&C^{\lambda_1,\lambda_2,\lambda_3}=C^{\lambda_{\sigma_1},\lambda_{\sigma_2},\lambda_{\sigma_3}}\quad(\theta_{\omega}=(-)^{\lambda+1})\,,&
    &C^{\lambda_1,\lambda_2,\lambda_3}=-C^{\lambda_{\sigma_1},\lambda_{\sigma_2},\lambda_{\sigma_3}}\quad(\theta_{\omega}=(-)^{\lambda})\,.
\end{align}
For $\theta_{\omega}=+1,-1$ the solution is given by
\begin{equation}
\boxed{
    C_1^{1234}=(-)^{\lambda_{24}} C^{4123}_1=(-)^{\lambda_{14}} C^{1324}_1\,,
    }
\end{equation}
and for $SO(N)$ the symmetry of the couplings is
\begin{align}
&C^{\lambda_1,\lambda_2,\lambda_3}=C^{\lambda_{\sigma_1},\lambda_{\sigma_2},\lambda_{\sigma_3}}\quad(\theta_{\omega}=-1)\,,&
    &C^{\lambda_1,\lambda_2,\lambda_3}=-C^{\lambda_{\sigma_1},\lambda_{\sigma_2},\lambda_{\sigma_3}}\quad(\theta_{\omega}=1)\,,
\end{align}
while for $USp(N)$ the symmetry of the couplings is
\begin{align}
&C^{\lambda_1,\lambda_2,\lambda_3}=C^{\lambda_{\sigma_1},\lambda_{\sigma_2},\lambda_{\sigma_3}}\quad(\theta_{\omega}=1)\,,&
    &C^{\lambda_1,\lambda_2,\lambda_3}=-C^{\lambda_{\sigma_1},\lambda_{\sigma_2},\lambda_{\sigma_3}}\quad(\theta_{\omega}=-1)\,.
\end{align}
\refstepcounter{section}
\section*{2.D \hspace{1mm} SO(N) and USp(N) gauge groups}
\addcontentsline{toc}{section}{2.D \hspace{1mm} SO(N) and USp(N) gauge groups}
\label{Paper1-AppendixD}
Without diving into explicit computation, we examine the $SO(N)$ and $USp(N)$ cases. For $SO(N)$, the Poisson bracket takes the form
\begin{equation}\label{Paper1-SO(N)poisson_bracket}
[(\phi^{\lambda}_p)_{AB},(\phi^{\mu}_q)_{CD}]=\frac{\delta^{\lambda,-\mu}\delta^3(p+q)}{2p^+}\,(\delta_{AC}\delta_{BD}+\theta_{\lambda}\delta_{AD}\delta_{BC})\,,
\end{equation}
and also here, due to the antisymmetry of the Poisson bracket, we have $\theta_\lambda = \theta_{-\lambda}$.
Moreover, in this case, the phase can be related to the symmetry properties of the field under transposition
\begin{equation} 
[(\phi^{\lambda}_p)_{BA},(\phi^{\mu}_q)_{CD}]=\frac{\delta^{\lambda,-\mu}\delta^3(p+q)}{2p^+}\,(\delta_{BC}\delta_{AD}+\theta_{\lambda}\delta_{BD}\delta_{AC})= \theta_\lambda [(\phi^{\lambda}_p)_{AB},(\phi^{\mu}_q)_{CD}]\,,
\end{equation} 
where we used the identity $\theta_{\lambda}^2 = \theta_{\lambda} \theta_{-\lambda} = 1$ and we find the property $(\phi^{\lambda}_q)_{BA}=\theta_{\lambda}(\phi^{\lambda}_q)_{AB}$. This implies the following symmetry for the couplings:
\begin{align}\label{Paper1-SO(n)_sym}
    &(\phi^{\lambda}_q)_{BA}=\theta_{\lambda}(\phi^{\lambda}_q)_{AB}&\Rightarrow&
    &C^{\lambda_1,\lambda_2,\lambda_3}=(-)^{\lambda_{123}}\theta_{\lambda_1}\theta_{\lambda_2}\theta_{\lambda_3}C^{\lambda_{\sigma_1},\lambda_{\sigma_2},\lambda_{\sigma_3}}\,,
\end{align}
where $\sigma\in\Sigma_3$ is an odd permutation. Given this, the Poisson bracket yields
\begin{equation}
    [\mathrm{Tr}(\phi^{\lambda_1}_{q_1}\phi^{\lambda_2}_{q_2}\phi^{\omega}_{q_{\omega}}),\mathrm{Tr}(\phi^{\lambda_3}_{q_3}\phi^{\lambda_4}_{q_4}\phi^{-\omega}_{-q_{\omega}})]=\frac{1}{2q^+_{\omega}}\Big(\theta_{\lambda_3}\theta_{\lambda_4}\mathrm{Tr}(\phi^{\lambda_1}_{q_1}\phi^{\lambda_2}_{q_2}\phi^{\lambda_4}_{q_4}\phi^{\lambda_3}_{q_3})+\theta_{\omega}\mathrm{Tr}(\phi^{\lambda_1}_{q_1}\phi^{\lambda_2}_{q_2}\phi^{\lambda_3}_{q_3}\phi^{\lambda_4}_{q_4})\Big)\,,
\end{equation}
where we contract only $\phi^{\omega}_{q_{\omega}}$, and the constraint becomes
\begin{align}\label{Paper1-SOnconstraint}
\begin{split}
\sum_{\omega}(-)^{\omega}\text{Cycl}\Big[&\Big(\theta_{\omega}\mathcal{C}^{1234\omega}+(-)^{\lambda_{34}+\omega}\theta_{\lambda_3}\theta_{\lambda_4}\mathcal{C}^{1243\omega}\Big)\times\\
&\frac{(\lambda_1+\omega-\lambda_2)\beta_1-(\lambda_2+\omega-\lambda_1)\beta_2}{2(\beta_1+\beta_2)}\,
    \PPb_{12}^{\lambda_{12}+\omega-1}\PPb_{34}^{\lambda_{34}-\omega}\Big]=0\,.
\end{split}
\end{align}
Using \eqref{Paper1-SO(n)_sym} we find the relation
\begin{equation}
    \mathcal{C}^{1243\omega}=(-)^{\lambda_{34}+\omega}\theta_{\lambda_3}\theta_{\lambda_4}\theta_{\lambda_\omega}\mathcal{C}^{1234\omega}\,.
\end{equation}
The constraint \eqref{Paper1-SOnconstraint} can then be solved in the same way as for $U(N)$. If we additionally require that there be no relative signs among the various couplings, as in Metsaev's solution, we obtain the condition
\begin{align}
   &\theta_{\omega}(1+\theta_{\lambda_3}^2\theta_{\lambda_4}^2)=2(-)^{\omega}&
   &\Rightarrow&
   &\theta_{\omega}=(-)^{\omega}\,.
\end{align}
For $USp(N)$, the Poisson bracket is
\begin{equation}\label{Paper1-USp(N)poisson_bracket}
[(\phi^{\lambda}_p)_{AB},(\phi^{\mu}_q)_{CD}]=\frac{\delta^{\lambda,-\mu}\delta^3(p+q)}{2p^+}\,(C_{AC}C_{BD}+\theta_{\lambda}C_{AD}C_{BC})\,,
\end{equation}
where $C_{AB}$ is the antisymmetric invariant tensor
\begin{align}
    &C_{AB}=-C_{BA}\,,&
    &C_{AB}C^{CB}=\delta_A^C\,.
\end{align}
The matrices $C$ raise and lower indices as follows: $V^A=C^{AB}V_B$ and $V^BC_{BA}=V_A$. As before, we can find the properties
\begin{align}\label{Paper1-USp(n)_sym}
    &\theta_\lambda = \theta_{-\lambda}\,,&
    &(\phi^{\lambda}_q)_{BA}=\theta_{\lambda}(\phi^{\lambda}_q)_{AB}\,,&
    &C^{\lambda_1,\lambda_2,\lambda_3}=(-)^{\lambda_{123}+1}\theta_{\lambda_1}\theta_{\lambda_2}\theta_{\lambda_3}C^{\lambda_{\sigma_1},\lambda_{\sigma_2},\lambda_{\sigma_3}}\,,
\end{align}
where $\sigma\in\Sigma_3$ is an odd permutation. Given this, the Poisson bracket yields
\begin{equation}
    [\mathrm{Tr}(\phi^{\lambda_1}_{q_1}\phi^{\lambda_2}_{q_2}\phi^{\omega}_{q_{\omega}}),\mathrm{Tr}(\phi^{\lambda_3}_{q_3}\phi^{\lambda_4}_{q_4}\phi^{-\omega}_{-q_{\omega}})]=-\frac{1}{2q^+_{\omega}}\Big(\theta_{\lambda_3}\theta_{\lambda_4}\mathrm{Tr}(\phi^{\lambda_1}_{q_1}\phi^{\lambda_2}_{q_2}\phi^{\lambda_4}_{q_4}\phi^{\lambda_3}_{q_3})+\theta_{\omega}\mathrm{Tr}(\phi^{\lambda_1}_{q_1}\phi^{\lambda_2}_{q_2}\phi^{\lambda_3}_{q_3}\phi^{\lambda_4}_{q_4})\Big)\,,
\end{equation}
where we contract only $\phi^{\omega}_{q_{\omega}}$, and the constraint becomes
\begin{align}\label{Paper1-USpnconstraint}
\begin{split}
\sum_{\omega}(-)^{\omega+1}\text{Cycl}\Big[&\Big(\theta_{\omega}\mathcal{C}^{1234\omega}-(-)^{\lambda_{34}+\omega}\theta_{\lambda_3}\theta_{\lambda_4}\mathcal{C}^{1243\omega}\Big)\times\\
&\frac{(\lambda_1+\omega-\lambda_2)\beta_1-(\lambda_2+\omega-\lambda_1)\beta_2}{2(\beta_1+\beta_2)}\, \PPb_{12}^{\lambda_{12}+\omega-1}\PPb_{34}^{\lambda_{34}-\omega}\Big]=0\,.
\end{split}
\end{align}
This constraint can then be solved as before. If we additionally require that there be no relative signs among the various couplings, as in Metsaev's solution, we obtain the condition
\begin{align}
   &\theta_{\omega}(1+\theta_{\lambda_3}^2\theta_{\lambda_4}^2)=2(-)^{\omega+1}&
   &\Rightarrow&
   \theta_{\omega}=(-)^{\omega+1}\,.
\end{align}
Using the values of $\theta_{\lambda}$ found for the two cases, we get the properties
\begin{align}\label{Paper1-sym_gauge}
    SO(N):\quad
    (\phi^\lambda_q)_{AB}=&\;(-)^{\lambda}(\phi^\lambda_q)_{BA}\quad
    &&\Rightarrow\quad
    C^{\lambda_1,\lambda_2,\lambda_3}= C^{\lambda_{\sigma_1},\lambda_{\sigma_2},\lambda_{\sigma_3}}\,,\\
    USp(N):\quad
    (\phi^\lambda_q)_{AB}=&\;(-)^{\lambda+1}(\phi^\lambda_q)_{BA}\quad
    &&\Rightarrow\quad
    C^{\lambda_1,\lambda_2,\lambda_3}=C^{\lambda_{\sigma_1},\lambda_{\sigma_2},\lambda_{\sigma_3}}\,,
\end{align}
where $\sigma\in\Sigma_3$ is an odd permutation.
These results are consistent with the analysis in \cite{Skvortsov:2020wtf}, which was carried out under the assumption of Metsaev-type couplings only.
Notably, in all cases, fields with odd helicity always take values in the adjoint representation of the respective gauge group. This matches the Chan-Paton structure appearing in open string theory \cite{Marcus:1982fr}. 

Note also that the couplings are fully symmetric, which is consistent with the fact that, for example, Yang-Mills theory exists for all compact simple gauge groups, thus including both $SO(N)$ and $USp(N)$. This would not have been possible if the couplings turned out to be antisymmetric. 

Due to the symmetry of the couplings in \eqref{Paper1-sym_gauge}, the solution to the constraint for both $SO(N)$ and $USp(N)$ --- once all colour-ordered constraints are taken into account --- takes the form
\begin{equation}
\boxed{
\begin{aligned}\label{Paper1-symfinalsystem_SO(N)}
    &\mathcal{C}^{1234\omega}=\frac{k^{1234}_-(\Lambda-2)!}{2^{\Lambda-2}(\lambda_{12}+\omega-1)!(\lambda_{34}-\omega-1)!}\quad 
    \forall\;\omega\,,\quad\text{same for $(1324)$ and $(1423)$}\,,\\
    &k_-^{1234}=k_-^{1324}=k_-^{1423}\,,\qquad C^{\lambda_1,\lambda_1,0}C^{0,\lambda_2,\lambda_2}=\;\text{generic}\,.
\end{aligned}
}
\end{equation}
\refstepcounter{section}
\section*{2.E \hspace{1mm} All fields in the adjoint}
\addcontentsline{toc}{section}{2.E \hspace{1mm} All fields in the adjoint}
\label{Paper1-AppendixE}
Following \cite{Ponomarev:2016lrm}, we assume that all fields take values in the adjoint representation of some Lie algebra of internal symmetry with structure constant $f^{a_1a_2a_3}$. In this setup, the space-time part of the cubic vertices remains unchanged, while the coupling constant is multiplied by the structure constant
\begin{equation}\label{Paper1-newcouplings}
    C^{\lambda_1,\lambda_2,\lambda_3}\quad\rightarrow\quad C^{\lambda_1,\lambda_2,\lambda_3}f^{a_1a_2a_3}\,.
\end{equation}
Since the structure constant is antisymmetric, even-derivative couplings become antisymmetric, and vice versa; the symmetry properties of the vertices are effectively reversed.
As a result, the new symmetry property for the coupling is
\begin{equation}\label{Paper1-new_sym}
    C^{\lambda_1,\lambda_2,\lambda_3}=(-)^{\lambda_{123}+1}C^{\lambda_{\sigma_1},\lambda_{\sigma_2},\lambda_{\sigma_3}}\,.
\end{equation}
where $\sigma\in\Sigma_3$ represents an odd permutation.
This implies that odd-derivative vertices involving two identical fields can be nonzero. For example, the Yang-Mills vertex $C^{1,1,-1}\neq 0$ is allowed. We now rewrite Eq.~\eqref{Paper1-LCholo} using the shorthand notation
\begin{equation}
    F_{1234}(A,B)+F_{1324}(B,C)+F_{1423}(C,A)=0\,,
\end{equation}
where each term represents a different line of Eq.~\eqref{Paper1-LCholo}. Upon substituting \eqref{Paper1-newcouplings} into this expression, we obtain
\begin{equation}
    F_{1234}(A,B) f_{a_1a_2b}f^b_{\phantom{b}a_3a_4}+ F_{1324}(B,C) f_{a_1a_3b}f^b_{\phantom{b}a_2a_4}+ F_{1423}(C,A) f_{a_1a_4b}f^b_{\phantom{b}a_2a_3}=0\,,
\end{equation}
and using the Jacobi identity, we find two constraints:\\
$F_{1234}(A,B)=F_{1423}(C,A)$ :
\begin{align}\label{Paper1-fabc1}
    \begin{split}
    &\sum_{\omega} (-)^{\omega}\big[((\lambda_{13}-\lambda_{24})A+(\lambda_{14}-\lambda_{23})B-2\omega C)\,\mathcal{C}^{1234\omega}(A-B)^{\lambda_{12}+\omega-1}(A+B)^{\lambda_{34}-\omega-1}\\
    &-((\lambda_{12}-\lambda_{34})C+(\lambda_{13}-\lambda_{24})A-2\omega B)\,\mathcal{C}^{1423\omega}(C-A)^{\lambda_{14}+\omega-1}(C+A)^{\lambda_{23}-\omega-1}\big]=0\,.
    \end{split}
\end{align}
$F_{1234}(A,B)=-F_{1324}(B,C)$ : 
\begin{align}\label{Paper1-fabc2}
    \begin{split}
    &\sum_{\omega} \big[(-)^{\omega}((\lambda_{13}-\lambda_{24})A+(\lambda_{14}-\lambda_{23})B-2\omega C)\,\mathcal{C}^{1234\omega}(A-B)^{\lambda_{12}+\omega-1}(A+B)^{\lambda_{34}-\omega-1}\\
    &+(-)^{\lambda_{24}}((\lambda_{14}-\lambda_{23})B+(\lambda_{12}-\lambda_{34})C-2\omega A)\,\mathcal{C}^{1324\omega}(B-C)^{\lambda_{13}+\omega-1}(B+C)^{\lambda_{24}-\omega-1}\big]=0\,.
    \end{split}
\end{align}
To solve the constraints \eqref{Paper1-fabc1} and \eqref{Paper1-fabc2}, we can follow the same procedure applied previously for Eq.~\eqref{Paper1-LCholocolour}.  Using \eqref{Paper1-Unconstraint}, we rewrite here the colour-ordered constraints for $[1234]$ and $[1243]$, respectively:
\begin{align}
\nonumber
    &\sum_{\omega} \big[(-)^{\omega}\theta_{\omega}((\lambda_{13}-\lambda_{24})A+(\lambda_{14}-\lambda_{23})B-2\omega C)\,
    \mathcal{C}^{1234\omega}(A-B)^{\lambda_{12}+\omega-1}(A+B)^{\lambda_{34}-\omega-1}\\
    &+(-)^{\lambda_{14}}\theta_{\omega}((\lambda_{12}-\lambda_{34})C+(\lambda_{13}-\lambda_{24})A-2\omega B)\,
    \mathcal{C}^{4123\omega}(C-A)^{\lambda_{14}+\omega-1}(C+A)^{\lambda_{23}-\omega-1}\big]=0\,,\\
\nonumber
    &\sum_{\omega} \big[(-)^{\lambda_{34}}\theta_{\omega}((\lambda_{13}-\lambda_{24})A+(\lambda_{14}-\lambda_{23})B-2\omega C)\,
    \mathcal{C}^{1243\omega}(A-B)^{\lambda_{12}+\omega-1}(A+B)^{\lambda_{34}-\omega-1}\\
    &+(-)^{\Lambda+\omega}\theta_{\omega}((\lambda_{14}-\lambda_{23})B+(\lambda_{12}-\lambda_{34})C-2\omega A)\,
    \mathcal{C}^{3124\omega}(B-C)^{\lambda_{13}+\omega-1}(B+C)^{\lambda_{24}-\omega-1}\big]=0\,.
\end{align}
By setting $\theta_{\omega}=-1$ and applying the modified symmetry relation \eqref{Paper1-nocolour_coupling_sym}, we match them to the constraints \eqref{Paper1-fabc1} and \eqref{Paper1-fabc2}, respectively.

From the previous analysis, we already know that a solution with $\theta_{\omega}=-1$ exists only for $\Lambda=2$, that is, in the presence of one-derivative interactions only. This corresponds to one of the special solutions found at the end of Appendix \ref{Paper1-AppendixC}.
The solution is then given by
\begin{equation}
\boxed{
    C_1^{1234}=(-)^{\lambda_{24}} C^{4123}_1=(-)^{\lambda_{14}} C^{1324}_1\,.
    }
\end{equation}
If we further assume all helicities to be odd, we obtain
\begin{equation}
\boxed{
    C_1^{1234}=C^{4123}_1=C^{1324}_1\,.
    }
\end{equation}
In particular, the HS-SDYM theory truncated to only odd helicities constitutes a consistent solution. This agrees with the findings of \cite{Ponomarev:2016lrm}, although the specific truncations and possible extensions involving both odd and even helicities were not explicitly identified there. Nevertheless, the conclusion that consistent solutions involving fields all in the adjoint representation of the same internal symmetry algebra exist only for one-derivative theories is confirmed.

\renewcommand{\thesection}{\thechapter.\arabic{section}}
\setcounter{section}{0}

\newpage
\clearpage

\chapter[Associativity of celestial OPE, higher-spins and self-duality]{Associativity of celestial OPE, higher-spins and self-duality}
\chaptermark{OPE associativity, higher-spins and self-duality}

\vspace{5mm}

\paragraph{Abstract:} We\footnote{The content of this chapter is identical to the content of the paper.} highlight and clarify the connection between several ideas and self-dual theories: (a) the operator product expansion (OPE) associativity in celestial conformal field theory (CCFT); (b) the vanishing of tree-level amplitudes; (c) the Jacobi identity for the ``gauge'' algebra; (d) the light-cone holomorphic constraints. Naturally, (b), (c), or (d) are closely related to self-duality. In particular, the recently classified \cite{Serrani:2025owx} chiral higher-spin theories with one- and two-derivative interactions (i.e. with gauge and gravitational interactions, which are extensions of self-dual Yang-Mills and self-dual gravity) also satisfy the OPE associativity constraint. We discuss the OPE associativity constraint and the holomorphic constraint for the most general class of cubic vertices.

\clearpage

\section{Introduction}\label{Paper2-section1}

In the paper, we elaborate on the links between several recent and old developments. (1) The requirement of associativity of the celestial OPE \cite{Pate:2019lpp,Himwich:2021dau,Strominger:2021mtt,Costello:2022upu,Bittleston:2022jeq} leads to certain constraints for couplings \cite{Mago:2021wje} that relate low- and higher-derivative ones. Such constraints have nothing to do with Lorentz invariance in the bulk and are puzzling at first sight. (2) It was shown in \cite{Ren:2022sws,Monteiro:2022lwm} that, remarkably, Chiral higher-spin gravity passes the celestial OPE associativity test. (3) Chiral higher-spin gravity is, under certain assumptions, a unique perturbatively local higher-spin gravity in $4d$ \cite{Metsaev:1991mt,Metsaev:1991nb,Ponomarev:2016lrm,Ponomarev:2017nrr}. (4) It was shown in \cite{Ponomarev:2017nrr} that the equations of Chiral higher-spin gravity can be formulated as a self-duality constraint for a certain ``gauge'' algebra. Two new theories were also found in \cite{Ponomarev:2017nrr} that can be thought of as higher-spin extensions of self-dual Yang-Mills (SDYM) and self-dual gravity (SDGR), which are contractions of Chiral higher-spin gravity that feature only gauge and gravitational interactions,\footnote{As will be made clearer later, by gauge and gravitational interactions, we mean those that have one- and two-derivative cubic interactions, respectively, when counted in the light-cone gauge.} respectively. (5) It is a folklore statement that self-dual theories have vanishing tree-level amplitudes, which is one of the definitions of classical integrability. (6) Very recently, it was shown \cite{Serrani:2025owx} that there are more ``chiral'' theories with higher-spin fields, all of which activate different subsets of the Chiral higher-spin gravity couplings. In the paper, we aim to clarify the relation between all these statements, which can be summarised by the following diagram 
$$ 
\begin{tikzcd}[row sep=1.5cm, column sep=1.5cm]
\parbox{4.5cm}{Holomorphic constraint in the light-cone gauge} \arrow[dr, leftrightarrow, "\text{minor caveat}", sloped, above] & & \parbox{4.5cm}{OPE associativity \\ in (Celestial) CFT} \arrow[dl, leftrightarrow] \\
& \textbf{self-duality} & \\
\parbox{4.5cm}{Vanishing of (tree-level) amplitudes} \arrow[ur, leftrightarrow] & & \parbox{4.5cm}{Jacobi identity of \\ the ''gauge'' algebra} \arrow[ul, leftrightarrow]
\end{tikzcd} 
$$
In more detail, let us begin with higher-spin gravity \cite{Bekaert:2022poo,Ponomarev:2022vjb}, which was the initial motivation for this work. Chiral higher-spin gravity and its contractions \cite{Metsaev:1991mt,Metsaev:1991nb,Ponomarev:2016lrm,Ponomarev:2017nrr} were first constructed in the light-cone gauge, relying on the earlier works by Bengtsson, Bengtsson, and Brink \cite{Bengtsson:1983pd,Bengtsson:1983pg}, who were the first to construct cubic interactions of massless higher-spin fields. The light-cone gauge has always been a very useful tool for initial developments, with the main idea being to directly construct generators of the Poincaré algebra order by order in terms of the physical degrees of freedom, putting aside any gauge redundancy a covariant formulation can lead to. A remarkable feature \cite{Metsaev:1991mt,Metsaev:1991nb} of the light-cone gauge in $4d$ is that the quartic constraint, which is equivalent to the Poincaré invariance of the quartic amplitude, contains the (anti)holomorphic parts that are insensitive to the quartic Hamiltonian itself, but still heavily constrain the cubic couplings. The (anti)holomorphic constraints involve only the vertices whose cubic amplitudes are made of (square)angle brackets. Solving the holomorphic constraint leads to Chiral higher-spin gravity, whose cubic amplitudes are 
\begin{align}
    \frac{1}{\Gamma[\lambda_1+\lambda_2+\lambda_3]}[12]^{\lambda_1+\lambda_2-\lambda_3}[23]^{\lambda_2+\lambda_3-\lambda_1}[13]^{\lambda_1+\lambda_3-\lambda_2}\,, \qquad \sum\lambda_i>0\,.
\end{align}
The Gamma-function is what the holomorphic constraint fixes the helicity dependence of couplings to be. This is a unique solution provided one starts with at least one nonabelian higher-spin self-interaction, as proven in \cite{Serrani:2025owx}. It should be noted that self-dual Yang-Mills (SDYM) and self-dual gravity (SDGR) are also solutions to the holomorphic constraint. As found in \cite{Ponomarev:2016lrm,Ponomarev:2017nrr,Monteiro:2022xwq,Serrani:2025owx}, there are other solutions to the holomorphic constraint that feature only gauge or gravitational interactions (HS-SDGR and HS-SDYM) and can even have finitely many higher-spin fields. In fact, all solutions of the holomorphic constraint immediately give consistent theories by setting all higher orders and anti-holomorphic couplings to zero. In this sense, Chiral higher-spin gravity is the maximal completion of self-dual theories that activate all spins and all couplings. All of these theories --- with a minor caveat to be discussed --- have vanishing tree-level amplitudes, which can also be understood as being closely related to self-duality. 

The relation of the above to self-duality can be clarified with the help of the kinematic and ``gauge'' (Lie) algebra \cite{Monteiro:2011pc,Ponomarev:2017nrr,Ponomarev:2024jyg}.\footnote{In \cite{Monteiro:2011pc}, the kinematical algebra of SDYM is identified as the Lie algebra defined by the SDYM cubic vertex after stripping off the colour factor. By contrast, in \cite{Ponomarev:2017nrr,Ponomarev:2024jyg}, the “gauge” algebra is defined as the Lie algebra obtained by factoring out the SDYM vertex from the cubic vertices.} It is a nontrivial statement that there is a Lie algebra behind any of the theories above, which amounts to checking the Jacobi identity. The gauge algebra allows one to rewrite the equations of motion as a self-duality constraint. Remarkably, all chiral higher-spin theories (with a minor caveat), including the new ones \cite{Serrani:2025owx}, have a gauge algebra, i.e. are self-dual, as we show in the paper. 

We also show that all chiral higher-spin theories  --- with a minor caveat ---  pass the celestial OPE associativity test with the first example given in \cite{Ren:2022sws,Monteiro:2022lwm}. We also find a general solution to the celestial OPE associativity and provide some explicit low-spin examples. The same condition can also be related to the vanishing of the four-point amplitudes. 

The paper is organised as follows: In Section \ref{Paper2-section2}, we review the light-front approach to higher-spin interactions, and in particular the quartic holomorphic constraint. In Section \ref{Paper2-section3}, we review the associativity of the celestial OPE and extend it to the case where fields live in some representation of a gauge group $G$. In Section \ref{Paper2-section4}, we solve the celestial OPE associativity constraint in various cases by rewriting it in terms of the
light-cone variables. In Section \ref{Paper2-section5}, we discuss in detail lower-spin theories that satisfy the OPE associativity constraint. We also match the results with those in \cite{Ren:2022sws} and identify additional solutions. In Section \ref{Paper2-section6}, we discuss the relations between: (a) the celestial OPE associativity; (b) the vanishing of tree-level amplitudes; (c) the Jacobi identity for the gauge algebra; (d) the light-cone holomorphic constraints. Finally, in Section \ref{Paper2-section7}, we conclude and outline possible future directions for extending the connection between the light-cone constraints and the OPE associativity in CCFT.

We include four appendices. Appendix \ref{Paper2-AppendixA} introduces our light-cone notation. Appendix \ref{Paper2-AppendixB} presents the CCFT notation for massless fields. In Appendix \ref{Paper2-AppendixC}, we present an on-shell map between the $4d$ light-cone and $2d$ CCFT formalisms via their relation to spinor-helicity. In Appendix \ref{Paper2-AppendixD}, we solve the OPE associativity constraint for the most general class of cubic vertices.

\section{Light-cone holomorphic constraint}
\label{Paper2-section2}

We begin by briefly reviewing the essentials of the light-front approach to interactions and the derivation of the \textit{quartic holomorphic constraint}. For further details and background, see \cite{Ponomarev:2016lrm,Ponomarev:2017nrr,Ponomarev:2016cwi,Serrani:2025owx,Ponomarev:2022vjb}. The conventions for the light-cone notation adopted are detailed in Appendix~\ref{Paper2-AppendixA}.

The light-front deformation procedure relies on two key ingredients: the Hamiltonian formulation of classical/quantum field theory in light-cone gauge and the non-linear realisation of the Poincaré algebra through the inclusion of higher-order corrections to the free (quadratic) generators of the Poincaré algebra. Accordingly, we begin with the $4d$ Poincaré algebra $iso(3,1)$, defined as
\begin{subequations}\label{Paper2-Poincare1}
\begin{align}
    [P^A,P^B]=&\,0\,,\\
    [J^{AB},P^C]=&\,P^A\eta^{BC}-P^B\eta^{AC}\,,\\
    [J^{AB},J^{CD}]=&\,J^{AD}\eta^{BC}-J^{BD}\eta^{AC}-J^{AC}\eta^{BD}+J^{BC}\eta^{AD}\,,
\end{align}
\end{subequations}
where $P^A$ are the generators of translations and $J^{AB}$ of Lorentz transformations.

In $4d$, massless spinning fields have two degrees of freedom; i.e. they are effectively represented by two ``scalars'' --- except for the scalar field. We denote $\phi^{\lambda}$ and $\phi^{-\lambda}$ as the helicity $+\lambda$ and $-\lambda$ fields, respectively. In Lorentzian signature, the one adopted here, they are complex fields and complex conjugates of each other $\phi^{-\lambda}=(\phi^{\lambda})^*$. The free action is $S=\tfrac{1}{2}\int d^4x\, \phi^{-\lambda}\Box\phi^{\lambda}$ and is real in any signature. 

We work with Fourier transformed fields with respect to $x^-$ and the transverse directions $x$ and $\bar{x}$. The Dirac bracket is given by
\begin{equation}
    [\phi_q^{\lambda}(x^+),\phi_p^{s}(x^+)]=\delta^{\lambda,-s}\frac{\delta^3(q+p)}{2q^+}\,.
\end{equation}
In the Hamiltonian approach, it is crucial to distinguish between the kinematical and dynamical generators of the Poincaré algebra \eqref{Paper2-Poincare1}. The kinematical generators correspond to the stability group of the codimension-one hypersurface on which quantisation is performed (i.e. in the light-front at $x^+=0$), and they remain unaffected by the introduction of interactions. In contrast, the dynamical generators are deformed once interactions are included, as they govern the evolution from one hypersurface to another. The allowed deformations are determined by solving the dynamical constraints order by order in the deformation procedure while carefully imposing locality at each order.

The free field realisation of the kinematical Poincaré generators reads\footnote{Following standard notations in the light-cone, we rename $\beta=p^+$.}
\begin{subequations}
\begin{align}
    &P^+=\beta\,,&
    &P=q\,,&
    &\bar{P}=\bar{q}\,,\\
    &J^{x+}=-\beta\frac{\partial}{\partial \bar{q}}\,,&
    &J^{\bar{x}+}=-\beta\frac{\partial}{\partial q}\,,&
    &J^{-+}=-N_{\beta}-1\,,\\
    &J^{x\bar{x}}=N_q-N_{\bar{q}}-\lambda\,,
\end{align}
\end{subequations}
where $N_q=q\partial_q$ is the Euler operator.
The (free) dynamical generators are
\begin{align}
    &H_2=-\frac{q\bar{q}}{\beta}\,,&
    &J_2^{z-}=\frac{\partial}{\partial\bar{q}}\frac{q\bar{q}}{\beta}+q\frac{\partial}{\partial\beta}+\lambda\frac{q}{\beta}\,,&
    &J_2^{\bar{z}-}=\frac{\partial}{\partial q}\frac{q\bar{q}}{\beta}+\bar{q}\frac{\partial}{\partial\beta}-\lambda\frac{\bar{q}}{\beta}\,.
\end{align}
We now deform the dynamical generators using a local ansatz:\footnote{Only the dynamical generators are deformed. One advantage of working in light-front quantisation is that the number of dynamical generators attains its minimum. Out of $10$ Poincaré generators, only $3$ are dynamical ($H,J^{x-},J^{\bar{x}-}$), compared to $4$ in the usual equal-time quantisation ($H^0,J^{0a}$).}
\begin{align}\label{Paper2-hamiltonian_P2}
    H=&\,H_2+\sum_n\int d^{3n}q\;\delta\Big(\sum_i q_i\Big)h_n\,\phi^{\lambda_1}_{q_1}\cdots\phi^{\lambda_n}_{q_n}\,,\\\label{Paper2-boostz_P2}
    J^{x-}=&\,J_2^{x-}+\sum_n\int d^{3n}q\;\delta\Big(\sum_i q_i\Big)\Big[j_n\,-\frac{1}{n}\,h_n\,\Big(\sum_j\frac{\partial}{\partial \bar{q}_j}\Big)\Big]\phi^{\lambda_1}_{q_1}\cdots\phi^{\lambda_n}_{q_n}\,,\\ 
    J^{\bar{x}-}=&\,J_2^{\bar{x}-}+\sum_n\int d^{3n}q\;\delta\Big(\sum_i q_i\Big)\Big[\bar{j}_n\,-\frac{1}{n}\,h_n\,\Big(\sum_j\frac{\partial}{\partial q_j}\Big)\Big]\phi^{\lambda_1}_{q_1}\cdots\phi^{\lambda_n}_{q_n}\,,
\end{align}
where we used the shorthand notation
\begin{align}
    &h_n\equiv h^{q_1,...,q_n}_{\lambda_1,...,\lambda_n}\,,&
    &j_n\equiv j^{q_1,...,q_n}_{\lambda_1,...,\lambda_n}\,,&
    &\bar{j}_n\equiv \bar{j}^{q_1,...,q_n}_{\lambda_1,...,\lambda_n}\,.
\end{align}
Let us define the momentum combinations
\begin{align}
    &\PP_{ij}=q_i\beta_j-q_j\beta_i\,,&
    &\PPb_{ij}=\bar{q}_i\beta_j-\bar{q}_j\beta_i\,,
\end{align}
where $\PPb_{ij}=-\PPb_{ji}$ and $\PP_{ij}=-\PP_{ji}$. Solving all cubic constraints required by the closure of the Poincaré algebra \cite{Bengtsson:1983pg,Bengtsson:1983pd,Bengtsson:1986kh,Metsaev:1991mt,Metsaev:1991nb} leads to the classification of cubic vertices:
\begin{align}\label{Paper2-cubic_hamiltonian}
h_3=&\,C^{\lambda_1,\lambda_2,\lambda_3}\frac{\PPb^{\lambda_{123}}}{\beta_1^{\lambda_1}\beta_2^{\lambda_2}\beta_3^{\lambda_3}}+\bar{C}^{-\lambda_1,-\lambda_2,-\lambda_3}\frac{\PP^{-\lambda_{123}}}{\beta_1^{-\lambda_1}\beta_2^{-\lambda_2}\beta_3^{-\lambda_3}}\,,\\
    j_3=&\,\frac{2}{3}\,C^{\lambda_1,\lambda_2,\lambda_3}\frac{\PPb^{\lambda_{123}-1}}{\beta_1^{\lambda_1}\beta_2^{\lambda_2}\beta_3^{\lambda_3}}\Lambda^{\lambda_1,\lambda_2,\lambda_3}\,,\\
    \bar{j}_3=&\,-\frac{2}{3}\,\bar{C}^{-\lambda_1,-\lambda_2,-\lambda_3}\frac{\PP^{-\lambda_{123}-1}}{\beta_1^{-\lambda_1}\beta_2^{-\lambda_2}\beta_3^{-\lambda_3}}\Lambda^{\lambda_1,\lambda_2,\lambda_3}\,,
\end{align}
with $\lambda_{123}=\lambda_1+\lambda_2+\lambda_3$ and where we defined
\begin{align}\label{Paper2-PP_cyclic}
    &\PP^a_{12}=\PP^a_{23}=\PP^a_{31}=\PP^a=\frac{1}{3}\,\Big[(\beta_1-\beta_2)q_3^a+(\beta_2-\beta_3)q_1^a+(\beta_3-\beta_1)q_2^a\Big]\,,\\
    &\Lambda^{\lambda_1,\lambda_2,\lambda_3}=\,\beta_1(\lambda_2-\lambda_3)+\beta_2(\lambda_3-\lambda_1)+\beta_3(\lambda_1-\lambda_2)\,.
\end{align}
Eq.~\eqref{Paper2-PP_cyclic} follows from momentum conservation, which also implies the cyclic invariance of $\PP$ and $\PPb$: $\sigma_{123}\PP=\PP$, $\sigma_{123}\PPb=\PPb$. These expressions can be used to construct the Hamiltonian $P^-=H$ and the dynamical boost generators $J^{x-}$ and $J^{\bar{x}-}$ via Eqs.~\eqref{Paper2-hamiltonian_P2} and \eqref{Paper2-boostz_P2}.

Written in this form, the cubic vertices exhibit a clear separation between holomorphic and anti-holomorphic components, corresponding respectively to the terms involving $\PPb$ and $\PP$ in the equations above. This can be reached through field redefinitions, which at the cubic order correspond to the freedom of adding terms proportional to powers of the free Hamiltonian $H_2\sim \PP\PPb$, as explained in \cite{Metsaev:2005ar,Ponomarev:2016lrm}. 

Each cubic vertex in \eqref{Paper2-cubic_hamiltonian} comes with an independent coupling constant, $C^{\lambda_1,\lambda_2,\lambda_3}$ or $\bar{C}^{-\lambda_1,-\lambda_2,-\lambda_3}$, which remains unfixed by the deformation procedure at the cubic level. To uncover relations among these couplings and constrain the spectrum of the theory, one must analyse the quartic consistency conditions (at the very least). In particular, it turns out that only a specific subset of these conditions --- namely, the (anti-)holomorphic constraint --- is sufficient for this purpose. On the other hand, if we are interested in theories that do not close at the cubic level and thus require a genuine quartic interaction, the full set of quartic constraints needs to be studied.

We stress that in Lorentzian signature, holomorphic and anti-holomorphic vertices are related by complex conjugation. Consequently, the reality of the action --- and hence unitarity --- requires including both sectors with cubic couplings satisfying $C^{\lambda_1,\lambda_2,\lambda_3}=(\bar{C}^{-\lambda_1,-\lambda_2,-\lambda_3})^*$. In contrast, parity invariance would require $C^{\lambda_1,\lambda_2,\lambda_3}=\bar{C}^{-\lambda_1,-\lambda_2,-\lambda_3}$. In what follows, we focus on the holomorphic sector alone, in which case the action is complex.

It is important to note that the cubic constraints, which determine the structure of the cubic vertices in \eqref{Paper2-cubic_hamiltonian}, treat all cubic vertices as independent. This allows us to extend the field content by assigning to them an additional index, indicating that they belong to some representation of a gauge group $G$ with structure constants $f_{abc}$. Possible choices for $G$ include $G=U(N),SO(N)$ and $USp(N)$; for further details, see \cite{Metsaev:1991nb,Skvortsov:2020wtf,Serrani:2025owx}. One can assume that a field of helicity $\lambda$ takes values in a representation $V_\lambda$ of $G$. We will loosely denote the tensor that specifies a cubic coupling $\fA_{abc}^{\lambda_1,\lambda_2,\lambda_3}$, i.e. use the same Latin indices for all modules $V_\lambda$. Therefore, with a gauging turned on, the most general form of the cubic vertices is
\begin{equation}\label{Paper2-cubic_vertex_general}
H_3^{\lambda_1,\lambda_2,\lambda_3}=\fA_{abc}^{\lambda_1,\lambda_2,\lambda_3}\int d^9q\;\delta\Big(\sum_i q_i\Big)\frac{\PPb^{\lambda_{123}}}{\beta_1^{\lambda_1}\beta_2^{\lambda_2}\beta_3^{\lambda_3}}(\phi^{\lambda_1}_{q_1})^a(\phi^{\lambda_2}_{q_2})^b(\phi^{\lambda_3}_{q_3})^c\,.
\end{equation}
It is convenient to sum over all $\lambda_{1,2,3}$ instead of all distinct triplets $\lambda_{1,2,3}$. To do so correctly, we need to impose a specific symmetry on $\fA_{abc}^{\lambda_1,\lambda_2,\lambda_3}$ for the coupling to be non-zero. Following \cite{Serrani:2025owx}, given any permutation $\sigma\in\Sigma_3$, we assume
\begin{equation}\label{Paper2-coupling_sym}
    \fA_{a_{\sigma_1}a_{\sigma_2}a_{\sigma_3}}^{\lambda_{\sigma_1},\lambda_{\sigma_2},\lambda_{\sigma_3}}=(-)^{\lambda_{123}}\fA_{a_1 a_2 a_3}^{\lambda_1,\lambda_2,\lambda_3}\,.
\end{equation}
This will impose a symmetry property on the cubic couplings, but only in the case of identical fields, such as 
\begin{align}
    &\fA_{a_2a_1a_3}^{\lambda,\lambda,s}=(-)^{2\lambda+s}\fA_{a_1 a_2 a_3}^{\lambda,\lambda,s}&
    &\implies&
    &\fA_{a_2a_1a_3}=(-)^{2\lambda+s}\fA_{a_1 a_2 a_3}\,.
\end{align}
In contrast, when the fields do not carry any group indices, we must always impose a specific symmetry on the coupling constants $C^{\lambda_1,\lambda_2,\lambda_3}$. In particular, we require
\begin{equation}\label{Paper2-nocolour_coupling_sym_P2}
    C^{\lambda_1,\lambda_2,\lambda_3}=(-)^{\lambda_{123}}C^{\lambda_{\sigma_1},\lambda_{\sigma_2},\lambda_{\sigma_3}}\,,
\end{equation}
where $\sigma\in\Sigma_3$. Therefore, we can assume that the coupling constants are symmetric for even-derivative interactions and antisymmetric for odd-derivative ones. This observation also implies that odd-derivative couplings involving at least two identical fields vanish by symmetry  $C^{\lambda,\lambda,\lambda'}\equiv 0$. To further simplify calculations, we can sometimes introduce generators $T_a^\lambda$ such that  $\phi^\lambda\equiv \phi^\lambda_a T^a_\lambda$ and the generators can depend on the helicity, which we often omit. This does not, of course, give the most general coupling tensor $\fA_{abc}^{\lambda_1,\lambda_2,\lambda_3}$.

Assuming the symmetry above, we can take the following form for the generators:
\begin{align}
    H_3&=\sum_{\lambda_1,\lambda_2,\lambda_3}\fA_{abc}^{\lambda_1,\lambda_2,\lambda_3}\int d^9 q\;\delta\Big(\sum_i q_i\Big)h_3\,(\phi^{\lambda_1}_{q_1})^{a}(\phi^{\lambda_2}_{q_2})^{b}(\phi^{\lambda_3}_{q_3})^{c}\,,\\
    J^{x-}_3&=\sum_{\lambda_1,\lambda_2,\lambda_3}\fA_{abc}^{\lambda_1,\lambda_2,\lambda_3}\int d^9q\;\delta\Big(\sum_i q_i\Big)\Big[j_3\,-\frac{1}{3}\,h_3\,\Big(\sum_j\frac{\partial}{\partial \bar{q}_j}\Big)\Big](\phi^{\lambda_1}_{q_1})^{a}(\phi^{\lambda_2}_{q_2})^{b}(\phi^{\lambda_3}_{q_3})^{c}\,,
\end{align}
and for the case with no gauge group:
\begin{align}
    H_3&=\sum_{\lambda_1,\lambda_2,\lambda_3}C^{\lambda_1,\lambda_2,\lambda_3}\int d^9 q\;\delta\Big(\sum_i q_i\Big)h_3\,\phi^{\lambda_1}_{q_1}\phi^{\lambda_2}_{q_2}\phi^{\lambda_3}_{q_3}\,,\\
    J^{x-}_3&=\sum_{\lambda_1,\lambda_2,\lambda_3}C^{\lambda_1,\lambda_2,\lambda_3}\int d^9q\;\delta\Big(\sum_i q_i\Big)\Big[j_3\,-\frac{1}{3}\,h_3\,\Big(\sum_j\frac{\partial}{\partial \bar{q}_j}\Big)\Big]\phi^{\lambda_1}_{q_1}\phi^{\lambda_2}_{q_2}\phi^{\lambda_3}_{q_3}\,,
\end{align}
i.e. to sum over all triplets of helicities instead of all distinct (up to permutation) triplets. 

\paragraph{Quartic consistency.} The quartic dynamical constraint takes the form
\begin{align}
    &[H,J^{a-}]\Big|_4=[H_4,J_2^{a-}]+[H_3,J_3^{a-}]-[J_4^{a-},H_2]\,,&
    &a=\{x,\bar{x}\}\,.
\end{align}
We can observe, by examining the degree of homogeneity of $q$, that the following two conditions must be satisfied independently \cite{Metsaev:1991mt,Metsaev:1991nb}
\begin{align}
    &[H_3(\PPb),J^{x-}_3]=0,&
    &[H_3(\PP),J^{\bar{x}-}_3]=0\,.
\end{align}
We refer to them, respectively, as \textit{holomorphic} and \textit{anti-holomorphic} quartic constraints. Explicitly, first the Poisson bracket between two cubic couplings is computed, then we substitute the explicit form of the generators, and once the dust has settled, we obtain
\begin{align}\label{Paper2-holo_commuting_fields}
    \begin{split}
    [H_3,J_3^{x-}]=&\sum_{\lambda_i,\omega}\int d^{12}q\;\delta \left(\sum_i q_i\right)\frac{9}{2}\Big[(-)^{\omega}\frac{(\lambda_1+\omega-\lambda_2)\beta_1-(\lambda_2+\omega-\lambda_1)\beta_2}{(\beta_1+\beta_2)\beta_1^{\lambda_1}\beta_2^{\lambda_2}\beta_3^{\lambda_3}\beta_4^{\lambda_4}}\,\times\\
    &C^{\lambda_1,\lambda_2,\omega}C^{-\omega,\lambda_3,\lambda_4}\PPb_{12}^{\lambda_{12}+\omega-1}\PPb_{34}^{\lambda_{34}-\omega}\,\phi^{\lambda_1}_{q_1}\phi^{\lambda_2}_{q_2}\phi^{\lambda_3}_{q_3}\phi^{\lambda_4}_{q_4}\Big]\,,
    \end{split}
\end{align}
where $\lambda_{ij}\equiv\lambda_i+\lambda_j$. Recall that the couplings $C^{\lambda_1,\lambda_2,\lambda_3}$ obey the symmetry property \eqref{Paper2-nocolour_coupling_sym_P2}. In the presence of a gauge group, we obtain
\begin{align}\label{Paper2-holo_gauge_group}
    \begin{split}
    [H_3,J_3^{x-}]=&\sum_{\lambda_i,\omega}\int d^{12}q\;\delta \left(\sum_i q_i\right)\frac{9}{2}\Big[(-)^{\omega}\frac{(\lambda_1+\omega-\lambda_2)\beta_1-(\lambda_2+\omega-\lambda_1)\beta_2}{(\beta_1+\beta_2)\beta_1^{\lambda_1}\beta_2^{\lambda_2}\beta_3^{\lambda_3}\beta_4^{\lambda_4}}\,\times\\
    &\fA_{a_1a_2c}\fA^c_{\phantom{c}a_3a_4}C^{\lambda_1,\lambda_2,\omega}C^{-\omega,\lambda_3,\lambda_4}\PPb_{12}^{\lambda_{12}+\omega-1}\PPb_{34}^{\lambda_{34}-\omega}\,(\phi^{\lambda_1}_{q_1})^{a_1}(\phi^{\lambda_2}_{q_2})^{a_2}(\phi^{\lambda_3}_{q_3})^{a_3}(\phi^{\lambda_4}_{q_4})^{a_4}\Big]\,.
    \end{split}
\end{align}
In particular, it is worth noting that if we assume the theory to be purely (anti-)holomorphic and to satisfy the (anti-)holomorphic quartic constraints, it will be a well-defined theory at any order and will contain only cubic interactions.\footnote{We stress once again the chronological development of these results. In 1983, Bengtsson, Bengtsson, and Brink \cite{Bengtsson:1983pd,Bengtsson:1983pg} determined the form of the cubic interactions by solving the light-cone constraints at cubic order, and the complete classification of cubic vertices was later provided in \cite{Bengtsson:1986kh}. Several years later, in 1991, Metsaev began the analysis of the quartic constraints in \cite{Metsaev:1991mt, Metsaev:1991nb}. He observed that a simpler condition --- the holomorphic quartic constraint --- was already sufficient to fix all the cubic couplings (under certain assumptions), leading to what we call the Metsaev solution. Twenty-five years later, Ponomarev and Skvortsov revisited the problem in \cite{Ponomarev:2016lrm}. Remarkably, they found that if the theory is truncated to the (anti-)holomorphic sector, it becomes a consistent theory with only cubic vertices at all orders, and the couplings are the same as those found by Metsaev. More recently, in \cite{Serrani:2025owx}, it was shown that the holomorphic constraint admits multiple consistent truncations, even allowing for theories with a finite spectrum that still involve higher-spin fields.} 

\section{OPE associativity in 2d CCFT}\label{Paper2-section3}
Recently, considerable attention has been devoted to the study of the \textit{OPE associativity constraint}, see \cite{Mago:2021wje,Ren:2022sws,Monteiro:2022lwm,Costello:2022upu,Bittleston:2022jeq,Ball:2022bgg,Ball:2023sdz,Ball:2023qim,Ball:2024oqa,Fernandez:2024qnu,Guevara:2024ixn,Bhattacharyya:2025nfp}. Here, we briefly review the basics of $2d$ CCFT for massless particles (see Appendix \ref{Paper2-AppendixB} for notations), to provide context for the OPE associativity constraint. For more details and reviews on the topic, see \cite{Pasterski:2021rjz,Raclariu:2021zjz}.

In CCFT, of fundamental importance is the isomorphism $SO^+(1,3)\simeq SL(2,\mathbb{C})/\mathbb{Z}_2$, between the $4d$ (connected) Lorentz group and the set of conformal transformations of the $2d$ celestial sphere $\mathbb{C}\PP^1$\cite{oblak2018lorentzgroupcelestialsphere}. 

This is what allows the S-matrix in asymptotically flat spacetime to be recast as a celestial amplitude. The idea is to perform a change of basis: from the standard energy-momentum eigenstate basis, with definite energy and momentum, to a new basis of boost eigenstates. In this new basis, the S-matrix is interpreted as a correlation function of operators inserted at points on the celestial sphere at null infinity. These operators are labelled by the conformal dimension and spin associated with the $2d$ global conformal group of the celestial sphere $SL(2,\mathbb{C})/\mathbb{Z}_2$.

The $S$-matrix element for the scattering of $n$ massless fields in the standard energy-momentum eigenstate basis is 
\begin{equation}
    \mathcal{A}_n(\omega_j,z_j,\bar{z}_j)=\langle \text{out}|S|\text{in}\rangle\,.
\end{equation}
The integral operation that allows one to perform the change of basis to the boost eigenstate one and then trades the energy\footnote{To be precise, the energy is given by $k^0=\omega q^0=\omega(1+z\bar{z})$.} $\omega$ for the conformal dimension $\Delta$ is the Mellin transform
\begin{equation}
    M(\cdot)=\int_0^{\infty}\frac{d\omega}{\omega}\omega^{\Delta}(\cdot)\,.
\end{equation}
Then the $S$-matrix in the boost eigenstates (i.e. the celestial amplitude) is given by the Mellin-transformed amplitude
\begin{equation}\label{Paper2-Mellin_tranform_amplitude}
\mathcal{M}_n(\Delta_j,z_j,\bar{z}_j)=\prod_{j=1}^n\int_0^{\infty}\frac{d\omega_j}{\omega_j}\omega_j^{\Delta_j}\mathcal{A}_n(\omega_j,z_j,\bar{z}_j)\,,
\end{equation}
where, by definition, we have
\begin{equation}
    \mathcal{M}_n(\Delta_j,z_j,\bar{z}_j)\equiv \prescript{}{boost}{\langle} \text{out}|S|\text{in}\rangle_{boost}=\langle\mathcal{O}^{\pm}_{\Delta_1,s_1}(z_1,\bar{z}_1)\cdots \mathcal{O}^{\pm}_{\Delta_n,s_n}(z_n,\bar{z}_n)\rangle_{CCFT}\,,
\end{equation} 
where the operator $\mathcal{O}^{\pm}_{\Delta_n,s_n}$ represents outgoing $(+)$ or incoming $(-)$ celestial conformal primary operators with $2d$ conformal (or boost) weight $\Delta_n$ and spin $s_n$ that crosses the celestial sphere at a point $(z,\bar{z})$. In particular \cite{Pasterski:2017kqt}, when $\Delta_n\in 1+i\mathbb{R}$, corresponding to the principal continuous series of the Lorentz group, the transformation above can be inverted using the inverse Mellin transform 
\begin{equation}
    \mathcal{A}_n(\omega_j,z_j,\bar{z}_j)=\prod_{i=1}^n\int_{1-i\infty}^{1+i\infty}\frac{d\Delta_j}{2\pi i}\omega_j^{-\Delta_i}\mathcal{M}_n(\Delta_j,z_j,\bar{z}_j)\,.
\end{equation}
From this point on, we omit the $\pm$ sign and adopt the all–outgoing convention, as in Appendix \ref{Paper2-AppendixB}.

One of the central ingredients in any conformal field theory is the operator product expansion (OPE), which provides a systematic way to compute products of local operators and correlation functions. The holomorphic OPE of two conformal primary operators with conformal weights $(h_i,\bar{h}_i)$, where $s_i=h_i-\bar{h}_i$ and $\Delta_i=h_i+\bar{h}_i$, in celestial CFT can be obtained by the holomorphic collinear limit (i.e. $z_1\rightarrow z_2$) of scattering amplitudes \cite{Fan:2019emx,Pate:2019lpp,Himwich:2021dau} and can be expressed as
\begin{align}\label{Paper2-CelestialOPE}
   & \mathcal{O}_{h_1,\bar{h}_1}(z_1,\bar{z}_1)\mathcal{O}_{h_2,\bar{h}_2}(z_2,\bar{z}_2)\sim\frac{1}{z_{12}}\sum_p\sum_{m=0}^{\infty}C_p^{(m)}(\bar{h}_1,\bar{h}_2)\bar{z}_{12}^{p+m}\bar{\partial}^m\mathcal{O}_{h_{12}-1,\bar{h}_{12}+p}(z_2,\bar{z}_2)\,,\\
    & z_{ij}=z_i-z_j\,,\quad
    \bar{z}_{ij}=\bar{z}_i-\bar{z}_j\,,\quad
    h_{ij}=h_i+h_j\,,\quad
    \bar{h}_{ij}=\bar{h}_i+\bar{h}_j\,.\quad
    s_{ij}=s_i+s_j\,,
\end{align}
where $C_p^{(m)}(\bar{h}_1,\bar{h}_2)$ denotes the OPE coefficient determining the contribution of the $m$th right-moving descendant with weights $(h_{12}-1,\bar{h}_{12}+p+m)$ and is given by
\begin{equation}
    C_p^{(m)}(\bar{h}_1,\bar{h}_2)=-\frac{1}{2}C^{s_1,s_2,s_3}\frac{1}{m!}B(2\bar{h}_1+p+m,2\bar{h}_2+p)\,,
\end{equation}
where $B(a,b)$ is the Euler beta function and $C^{s_1,s_2,s_3}$ is the coupling constant appearing in the flat bulk $3$-pt function of massless (higher-spin) particles with spins $(s_1,s_2,s_3=p+1-s_{12})$. Therefore, there is a one-to-one correspondence between holomorphic bulk $3$-pt vertices and holomorphic celestial OPE.

Notice that the OPE expansion \eqref{Paper2-CelestialOPE} represents only a partial result, as it captures solely the holomorphic sector at tree level. In fact, one-loop contributions already introduce double poles $\sim \frac{1}{z_{12}^2}$, arising from massless loops, as well as logarithms such as $\sim\log z_{12}$ \cite{Costello:2022upu, Bittleston:2022jeq, Ball:2023qim, Bhardwaj:2022anh, Krishna:2023ukw, Bhardwaj:2024wld, Bissi:2024brf}. Moreover, taking into account the anti-holomorphic sector leads to a more complete, ``all-order'' celestial OPE that involves series expansions in powers of both $\bar{z}$ and $z$. This has so far been achieved only in the maximally helicity-violating (MHV) sector at tree level, via twistor string theory in \cite{Adamo:2022wjo} and through collinear expansions using on-shell recursion relations in \cite{Ren:2023trv}.

A crucial consistency condition --- expected to be necessary for the existence of a well-defined celestial dual theory of gravity --- is the associativity of the OPE. We now review the (anti-)holomorphic OPE associativity constraint in $2d$ CCFT as derived in \cite{Ren:2022sws}, which implies the Jacobi identity for the charges studied in \cite{Mago:2021wje}.

A standard way to check OPE associativity is to introduce a mode expansion of the operators involved and compute the commutators using their OPE, see \cite{Guevara:2021abz}. Then OPE associativity implies the Jacobi identity for the modes. This is of particular interest when applied to conformally soft currents, see \cite{Mago:2021wje}. The relation is made possible thanks to the commutators for holomorphic objects \cite{Raclariu:2021zjz,Strominger:2021mtt} determined by a contour integral as
\begin{equation}
    [A,B](z_1)=\oint_{z_1}\frac{d z_2}{2\pi i}A(z_2)B(z_1)=\underset{z_2\rightarrow z_1}{\text{Res}}A(z_2)B(z_1)\,.
\end{equation}
We start with the following identity between correlators of conformal primaries involving contour integrals\footnote{This was the method used in \cite{Ren:2022sws}. A related one–loop analysis for self-dual Yang–Mills was already carried out in \cite{Costello:2022upu}, where extra powers of $z_k$ in the Jacobi identity are needed to extract higher–order poles coming from the one-loop OPE. At tree level, no such terms are required.}  
\begin{align}
    \begin{split}
    \Bigg(\oint_{|z_{13}|=2}dz_1\oint_{|z_{23}|=1}dz_2\,-&\oint_{|z_{23}|=2}dz_2\oint_{|z_{13}|=1}dz_1\,-\oint_{|z_{23}|=2}dz_2\oint_{|z_{12}|=1}dz_1\Bigg)\times\\
    &\langle\mathcal{O}^{a_1}_{\Delta_1,s_1}(z_1,\bar{z}_1)\cdots \mathcal{O}^{a_n}_{\Delta_n,s_n}(z_n,\bar{z}_n)\rangle=0\,,
    \end{split}
\end{align}
where we allow the presence of internal indices, enabling the operators to live in specific representations of an internal gauge group. This can be rewritten in terms of a ``double residue condition'' as
\begin{equation}\label{Paper2-OPE_associativity}
\left(\underset{z_1\rightarrow z_3}{\text{Res}}\,\underset{z_2\rightarrow z_3}{\text{Res}}-\underset{z_2\rightarrow z_3}{\text{Res}}\,\underset{z_1\rightarrow z_3}{\text{Res}}-\underset{z_2\rightarrow z_3}{\text{Res}}\,\underset{z_1\rightarrow z_2}{\text{Res}}\right)\langle\mathcal{O}^{a_1}_{\Delta_1,s_1}(z_1,\bar{z}_1)\cdots \mathcal{O}^{a_n}_{\Delta_n,s_n}(z_n,\bar{z}_n)\rangle=0\,.
\end{equation}
For further discussion of this constraint and possible ambiguities in its definition, see \cite{Ball:2022bgg,Ball:2024oqa}. To express \eqref{Paper2-OPE_associativity} in terms of the standard momentum space amplitude, we use the Mellin transform \eqref{Paper2-Mellin_tranform_amplitude} and compute the residues of the amplitude $\mathcal{A}_n(\omega_j,z_j,\bar{z}_j)$ as
\begin{equation}\label{Paper2-OPE_amplitude}
\left(\underset{z_1\rightarrow z_3}{\text{Res}}\,\underset{z_2\rightarrow z_3}{\text{Res}}-\underset{z_2\rightarrow z_3}{\text{Res}}\,\underset{z_1\rightarrow z_3}{\text{Res}}-\underset{z_2\rightarrow z_3}{\text{Res}}\,\underset{z_1\rightarrow z_2}{\text{Res}}\right)\mathcal{A}_n(1\cdots n)=0\,.
\end{equation}
At the amplitude level, the residue can be efficiently computed by taking the appropriate collinear limits. Schematically, proceeding step by step, we first take the collinear limit $q_1^{\mu}\parallel q_2^{\mu}$, which gives
\begin{equation}
    \lim_{z_{12}\rightarrow 0}\mathcal{A}_n(12\cdots n)=\sum_{\lambda_i}\frac{\mathcal{A}_3(12i)\mathcal{A}_{n-1}(i34\cdots n)}{\langle 12\rangle [12]}\,,
\end{equation}
where we focus on the leading order in $\frac{1}{z_{12}}$ and possible other channels decouple in the limit. Then it is followed by the collinear limit $q_2^{\mu}\parallel q_3^{\mu}$, and we get
\begin{equation}
    \lim_{z_{23}\rightarrow 0}\,\lim_{z_{12}\rightarrow 0}\mathcal{A}_n(12\cdots n)=\sum_{\lambda_i}\sum_{\lambda_j}\frac{\mathcal{A}_3(12i)\mathcal{A}_3(i3j)\mathcal{A}_{n-2}(j4\cdots n)}{\langle 12\rangle [12]\langle i3\rangle [i3]}\,.
\end{equation}
Then we can extract the residue of the last expression as
\begin{align}\label{Paper2-computing_residues}
    \begin{split}
    &\underset{z_2\rightarrow z_3}{\text{Res}}\,\underset{z_1\rightarrow z_2}{\text{Res}}\mathcal{A}_n(12\cdots n)=\sum_{\lambda_i}\sum_{\lambda_j}\frac{\mathcal{A}_3(12i)\mathcal{A}_3(i3j)\mathcal{A}_{n-2}(j4\cdots n)}{\omega_1\omega_2\omega_i\omega_3 \bar{z}_{12} \bar{z}_{i3}}\,,\\
    &\bar{z}_i=\frac{\omega_1\bar{z}_1+\omega_2\bar{z}_2}{\omega_1+\omega_2}\,,\qquad
    \omega_i=\omega_1+\omega_2\,,
    \end{split}
\end{align}
where we used the notation introduced in Appendix \ref{Paper2-AppendixC}. Using the expression for the double residue \eqref{Paper2-computing_residues} inside \eqref{Paper2-OPE_amplitude}, we get\footnote{From now on, we change notation and denote the spin $s_n$ of the massless fields by the helicity $\lambda_n$, in line with the light-cone higher-spin literature.}
\begin{align}\label{Paper2-OPE_Ass_commuting}
    \begin{split}
\sum_{\lambda_i}\Bigg[&\frac{\mathcal{A}_3(p_1^{\lambda_1},p_2^{\lambda_2},p_i^{\lambda_i})\mathcal{A}_3(p_3^{\lambda_3},p_4^{\lambda_4},p_i^{-\lambda_i})}{\bar{z}_{12}\bar{z}_{i3}(\omega_1+\omega_2)\omega_1\omega_2\omega_3}+\frac{\mathcal{A}_3(p_2^{\lambda_2},p_3^{\lambda_3},p_i^{\lambda_i})\mathcal{A}_3(p_1^{\lambda_1},p_4^{\lambda_4},p_i^{-\lambda_i})}{\bar{z}_{23}\bar{z}_{i1}(\omega_2+\omega_3)\omega_1\omega_2\omega_3}\\
    &+\frac{\mathcal{A}_3(p_3^{\lambda_3},p_1^{\lambda_1},p_i^{\lambda_i})\mathcal{A}_3(p_2^{\lambda_2},p_4^{\lambda_4},p_i^{-\lambda_i})}{\bar{z}_{31}\bar{z}_{i2}(\omega_1+\omega_3)\omega_1\omega_2\omega_3}\Bigg]=0\,,
    \end{split}
\end{align}
where we dropped the sum over $\lambda_j$ because the OPE must vanish for a generic $\mathcal{A}_{n-2}$, so the constraint must hold separately for each $\lambda_j$. For simplicity, we relabel $\lambda_j$ as $\lambda_4$.

If we consider fields transforming in a matrix representation of a Lie algebra, the associated generators $T^{a_n}$ are carried along throughout the computation. As shown above, the procedure remains essentially the same, with the only difference being that we can factor out the trace over the generators. The clearest way to illustrate this is to begin with the colour amplitudes, written in terms of the colour-ordered ones as
\begin{equation}\label{Paper2-colour_ordered_ampl}
    \mathcal{A}(12\cdots n)=\sum_{\sigma\in S_n/\mathbb{Z}_n}\mathrm{Tr}(T^{a_{\sigma_1}}T^{a_{\sigma_2}}\cdots T^{a_{\sigma_n}})\tilde{\mathcal{A}}(\sigma_1\sigma_2\cdots\sigma_n)\,.
\end{equation}
Consequently, the constraint \eqref{Paper2-OPE_amplitude} takes the form
\begin{equation}\label{Paper2-OPE_Ass_U(N)}
    \sum_{\sigma\in S_n/\mathbb{Z}_n}\sum_{\lambda_i}\Bigg[\frac{\tilde{\mathcal{A}}_3(p_{\sigma_1}^{\lambda_{\sigma_1}},p_{\sigma_2}^{\lambda_{\sigma_2}},p_{\sigma_i}^{\lambda_{\sigma_i}})\tilde{\mathcal{A}}_3(p_{\sigma_3}^{\lambda_{\sigma_3}},p_{\sigma_4}^{\lambda_{\sigma_4}},p_{\sigma_i}^{-\lambda_{\sigma_i}})}{\bar{z}_{\sigma_1\sigma_2}\bar{z}_{\sigma_i\sigma_3}(\omega_{\sigma_1}+\omega_{\sigma_2})\omega_{\sigma_1}\omega_{\sigma_2}\omega_{\sigma_3}}\mathrm{Tr}(T^{a_{\sigma_1}}T^{a_{\sigma_2}}T^{a_{\sigma_3}}T^{a_{\sigma_4}})\Bigg]=0\,.
\end{equation}
These two constraints \eqref{Paper2-OPE_Ass_commuting} and \eqref{Paper2-OPE_Ass_U(N)} are the celestial OPE associativity constraints, which we will solve in full generality. A similarity with the previously discussed light-cone constraint is already apparent (in the sense that both consist of two three-point vertices brought together times some additional kinematical factors), and we will aim to establish a direct connection between the two in the following sections.

Even more in general, we can assume, as considered for the holomorphic constraint, fields belonging to some representation of a gauge group $G$ with generic structure constants $f_{abc}$, and arrive at
\begin{equation}
    \sum_{\sigma\in S_n/\mathbb{Z}_n}\sum_{\lambda_i}\Bigg[\frac{\tilde{\mathcal{A}}_{a_{\sigma_1}a_{\sigma_2}}^{\phantom{a_{\sigma_1}a_{\sigma_2}}c}(p_{\sigma_1}^{\lambda_{\sigma_1}},p_{\sigma_2}^{\lambda_{\sigma_2}},p_{\sigma_i}^{\lambda_{\sigma_i}})\tilde{\mathcal{A}}_{c\,a_{\sigma_3}a_{\sigma_4}}(p_{\sigma_3}^{\lambda_{\sigma_3}},p_{\sigma_4}^{\lambda_{\sigma_4}},p_{\sigma_i}^{-\lambda_{\sigma_i}})}{\bar{z}_{\sigma_1\sigma_2}\bar{z}_{\sigma_i\sigma_3}(\omega_{\sigma_1}+\omega_{\sigma_2})\omega_{\sigma_1}\omega_{\sigma_2}\omega_{\sigma_3}}T^{a_{\sigma_1}}T^{a_{\sigma_2}}T^{a_{\sigma_3}}T^{a_{\sigma_4}}\Bigg]=0\,,
\end{equation}
where we denoted $\mathcal{A}_3=\tilde{\mathcal{A}}_{abc}T^aT^bT^c=\tilde{\mathcal{A}}_3\fA_{abc}T^aT^bT^c$. We solve this more general case in Appendix \ref{Paper2-AppendixD}.

\section{Celestial OPE associativity constraint}\label{Paper2-section4}
In this section, we present an exact solution to the OPE associativity constraint by switching to the light-front approach\footnote{We will use the terms light-front and light-cone interchangeably throughout this work.}. It turns out to be very useful to rewrite the OPE associativity constraint in terms of the light-cone variables. This can be done by following Appendix \ref{Paper2-AppendixC}, where we explain how to relate the two formalisms through the spinor-helicity one. The result is that, on-shell, we have the following identity:
\begin{equation}
    \sum_{\lambda_i}\frac{\mathcal{A}_3(p_1^{\lambda_1},p_2^{\lambda_2},p_i^{\lambda_i})\mathcal{A}_3(p_3^{\lambda_3},p_4^{\lambda_4},p_i^{-\lambda_i})}{\bar{z}_{12}\bar{z}_{i3}(\omega_1+\omega_2)\omega_1\omega_2\omega_3}=\sum_{\lambda_i}(-)^{\lambda_i}C^{\lambda_1,\lambda_2,\lambda_i}C^{-\lambda_i\lambda_3,\lambda_4}\frac{\PPb_{12}^{\lambda_{12}+\lambda_i-1}\PPb_{34}^{\lambda_{34}-\lambda_i-1}}{\beta_1^{\lambda_1}\beta_2^{\lambda_2}\beta_3^{\lambda_3}\beta_4^{\lambda_4}}\,.
\end{equation}
The right-hand side is not the same term present in the quartic holomorphic constraint, but, as we will see in the final section, the OPE associativity constraint is closely related to it. The quartic holomorphic constraint was solved in \cite{Serrani:2025owx} in two specific cases: for singlet fields and for fields living in some matrix representation of a gauge group $G$.

If we wish to reproduce all lower-spin constraints arising from the OPE associativity \eqref{Paper2-OPE_Ass_commuting} and \eqref{Paper2-OPE_Ass_U(N)}, as found in \cite{Ren:2022sws}, we need to consider the following types of cubic couplings:
\begin{align}\label{Paper2-h3_1}
&h_3^{(1)}\sim\phi^{\lambda_1}_{q_1}\phi^{\lambda_2}_{q_2}\phi^{\lambda_3}_{q_3}\,,\\\label{Paper2-h3_2}
    &h_3^{(2)}\sim\mathrm{Tr}(T_{a_1}T_{a_2})(\phi^{\lambda_1}_{q_1})^{a_1}(\phi^{\lambda_2}_{q_2})^{a_2}\phi^{\lambda_3}_{q_3}\,,\\\label{Paper2-h3_3}
    &h_3^{(3)}\sim\mathrm{Tr}(T_{a_1}T_{a_2}T_{a_3})(\phi^{\lambda_1}_{q_1})^{a_1}(\phi^{\lambda_2}_{q_2})^{a_2}(\phi^{\lambda_3}_{q_3})^{a_3}\,.
\end{align}
For semisimple compact Lie algebras, we can write
\begin{align}\label{Paper2-semisimple_a}
    &\mathrm{Tr}(T_{a_1}T_{a_2})=\frac{1}{2}\delta_{a_1a_2}\,,\\
    &\mathrm{Tr}(T_{a_1}T_{a_2}T_{a_3})=\frac{1}{2}\big(\mathrm{Tr}(T_{a_1}\{T_{a_2},T_{a_3}\})+\mathrm{Tr}(T_{a_1}[T_{a_2},T_{a_3}])\big)=\frac{1}{4}(d_{a_1a_2a_3}+if_{a_1a_2a_3})\,,
\end{align}
where the first line is a normalisation that can always be achieved, $f_{a_1a_2a_3}$ denotes the antisymmetric structure constant, and $d_{a_1a_2a_3}$ the symmetric $d$-symbol, both defined as
\begin{align}
    &[T_{a_1},T_{a_2}]=if_{a_1a_2a_3}T^{a_3}\,,&
    &\{T_{a_1},T_{a_2}\}=\frac{1}{N}\delta_{a_1a_2}\mathbb{I}+d_{a_1a_2a_3}T^{a_3}\,,\\\label{Paper2-semisimple_b}
    &if_{a_1a_2a_3}\equiv 2\,\mathrm{Tr}(T_{a_1}[T_{a_2},T_{a_3}])\,,&
    &d_{a_1a_2a_3}\equiv 2\,\mathrm{Tr}(T_{a_1}\{T_{a_2},T_{a_3}\})\,.
\end{align}
Looking at the various couplings in \eqref{Paper2-h3_1}--\eqref{Paper2-h3_3}, we can identify five possible structures that can occur at the quartic order:
\begin{align}\label{Paper2-normal_case_1}
&(h_3^{(1)},h_3^{(1)})&&\phi^{\lambda_1}_{q_1}\phi^{\lambda_2}_{q_2}\phi^{\lambda_3}_{q_3}\phi^{\lambda_4}_{q_4}\,,\\\label{Paper2-normal_case_2}
&(h_3^{(3)},h_3^{(3)})&&\mathrm{Tr}(T_{a_1}T_{a_2}T_{a_3}T_{a_4})(\phi^{\lambda_1}_{q_1})^{a_1}(\phi^{\lambda_2}_{q_2})^{a_2}(\phi^{\lambda_3}_{q_3})^{a_3}(\phi^{\lambda_4}_{q_4})^{a_4}\,,\\\label{Paper2-mixed_case_1}
&(h_3^{(2)},h_3^{(2)})\,,(h_3^{(1)},h_3^{(2)})&&\mathrm{Tr}(T_{a_1}T_{a_2})(\phi^{\lambda_1}_{q_1})^{a_1}(\phi^{\lambda_2}_{q_2})^{a_2}\phi^{\lambda_3}_{q_3}\phi^{\lambda_4}_{q_4}\,,\\\label{Paper2-mixed_case_2}
&(h_3^{(2)},h_3^{(2)})&&\mathrm{Tr}(T_{a_1}T_{a_2})\mathrm{Tr}(T_{a_3}T_{a_4})(\phi^{\lambda_1}_{q_1})^{a_1}(\phi^{\lambda_2}_{q_2})^{a_2}(\phi^{\lambda_3}_{q_3})^{a_3}(\phi^{\lambda_4}_{q_4})^{a_4}\,,\\\label{Paper2-mixed_case_3}
&(h_3^{(2)},h_3^{(3)})&&\mathrm{Tr}(T_{a_1}T_{a_2}T_{a_3})(\phi^{\lambda_1}_{q_1})^{a_1}(\phi^{\lambda_2}_{q_2})^{a_2}(\phi^{\lambda_3}_{q_3})^{a_3}\phi^{\lambda_4}_{q_4}\,,
\end{align}
where on the left of each structure, we indicate the type of pair of cubic vertices responsible for generating each distinct type of constraint. In particular, we observe that while the first two involve the same type of cubic vertices, the others mix and intertwine different types of cubic vertices. As we will see, there are two main mechanisms by which the constraint can be solved. Every type of constraint will fall into one of these two cases.

\subsection{Singlet fields}
We start by solving the OPE associativity constraint \cite{Ren:2022sws}, for singlet fields \eqref{Paper2-normal_case_1}, then
\begin{align}\label{Paper2-OPE_Ass}
    \begin{split}
\sum_{\lambda_i}\Bigg[&\frac{\mathcal{A}_3(p_1^{\lambda_1},p_2^{\lambda_2},p_i^{\lambda_i})\mathcal{A}_3(p_3^{\lambda_3},p_4^{\lambda_4},p_i^{-\lambda_i})}{\bar{z}_{12}\bar{z}_{i3}(\omega_1+\omega_2)\omega_1\omega_2\omega_3}+\frac{\mathcal{A}_3(p_2^{\lambda_2},p_3^{\lambda_3},p_i^{\lambda_i})\mathcal{A}_3(p_1^{\lambda_1},p_4^{\lambda_4},p_i^{-\lambda_i})}{\bar{z}_{23}\bar{z}_{i1}(\omega_2+\omega_3)\omega_1\omega_2\omega_3}\\
    &+\frac{\mathcal{A}_3(p_3^{\lambda_3},p_1^{\lambda_1},p_i^{\lambda_i})\mathcal{A}_3(p_2^{\lambda_2},p_4^{\lambda_4},p_i^{-\lambda_i})}{\bar{z}_{31}\bar{z}_{i2}(\omega_1+\omega_3)\omega_1\omega_2\omega_3}\Bigg]=0\,,
    \end{split}
\end{align}
and we rewrite it in the light-cone formalism (see Appendix \ref{Paper2-AppendixC}) as
\begin{align}\label{Paper2-OPE_Ass_LC}
\begin{split}
\sum_{\lambda_i}(-)^{\lambda_i}\Big(&C^{\lambda_1,\lambda_2,\lambda_i}C^{-\lambda_i,\lambda_3,\lambda_4}\PPb_{12}^{\lambda_{12}+\lambda_i-1}\PPb_{34}^{\lambda_{34}-\lambda_i-1}+C^{\lambda_1,\lambda_4,\lambda_i}C^{-\lambda_i,\lambda_2,\lambda_3}\PPb_{14}^{\lambda_{14}+\lambda_i-1}\PPb_{23}^{\lambda_{23}-\lambda_i-1}\\
&+C^{\lambda_3,\lambda_1,\lambda_i}C^{-\lambda_i,\lambda_2,\lambda_4}\PPb_{31}^{\lambda_{13}+\lambda_i-1}\PPb_{24}^{\lambda_{24}-\lambda_i-1}\Big)=0\,,\qquad \lambda_{ij}=\lambda_i+\lambda_j\,.
\end{split}
\end{align}
This structure is reminiscent of the quartic holomorphic constraint studied in \cite{Ponomarev:2016lrm,Serrani:2025owx}. As we will see in the final section, the two are closely related. Moreover, our strategy to solve the constraint follows the same ideas used to address the quartic holomorphic constraint in \cite{Serrani:2025owx}. To avoid unnecessary minus signs between products of couplings and extra $i$ factors between couplings, we assign an additional factor of $i$ to odd-helicity fields. This is equivalent to taking even-helicity fields to be Hermitian and odd-helicity fields to be anti-Hermitian matrices (with singlets treated as $1\times 1$ matrices). With this convention, the factor $(-)^{\lambda_i}$ in \eqref{Paper2-OPE_Ass_LC} is removed. We begin by introducing three independent variables:
\begin{align}\label{Paper2-ABC_variables}
\begin{split}
    2A&=\PPb_{12}+\PPb_{34}=\PPb_{23}-\PPb_{14}\,,\qquad
    2B=\PPb_{13}-\PPb_{24}=\PPb_{34}-\PPb_{12}\,,\\
    2C&=\PPb_{14}+\PPb_{23}=-\PPb_{13}-\PPb_{24}\,,
    \end{split}
\end{align}
with the following transformation properties:
\begin{subequations}\label{Paper2-ABC_sym}
\begin{align}
    &A\overset{1\leftrightarrow 2}{\xleftrightarrow{\hspace{5mm}}} B\,,&
    &A\overset{1\leftrightarrow 3}{\xleftrightarrow{\hspace{5mm}}} -A\,,&
    &A\overset{1\leftrightarrow 4}{\xleftrightarrow{\hspace{5mm}}} C\,,&
    &A\overset{2\leftrightarrow 3}{\xleftrightarrow{\hspace{5mm}}} -C\,,&
    &A\overset{2\leftrightarrow 4}{\xleftrightarrow{\hspace{5mm}}} -A\,,&
    &A\overset{3\leftrightarrow 4}{\xleftrightarrow{\hspace{5mm}}} -B\,,\\
    &C\overset{1\leftrightarrow 2}{\xleftrightarrow{\hspace{5mm}}} -C\,,&
    &B\overset{1\leftrightarrow 3}{\xleftrightarrow{\hspace{5mm}}} C\,,&
    &B\overset{1\leftrightarrow 4}{\xleftrightarrow{\hspace{5mm}}} -B\,,&
    &B\overset{2\leftrightarrow 3}{\xleftrightarrow{\hspace{5mm}}} -B\,,&
    &B\overset{2\leftrightarrow 4}{\xleftrightarrow{\hspace{5mm}}} -C\,,&
    &C\overset{3\leftrightarrow 4}{\xleftrightarrow{\hspace{5mm}}} -C\,,
\end{align}
\end{subequations}
and define
\begin{align}\label{Paper2-definitions}
\nonumber
    &\mathcal{C}^{1234\lambda_i}\definition C^{\lambda_1,\lambda_2,\lambda_i}C^{-\lambda_i,\lambda_3,\lambda_4}\,,\qquad
        k_+^{1234}\definition (-)^{\lambda_{12}}\sum_{\lambda_i}(-)^{\lambda_i}\mathcal{C}^{1234\lambda_i}\,,\qquad
        k_-^{1234}\definition \sum_{\lambda_i}\mathcal{C}^{1234\lambda_i}\,,\\
    &f_+^{1234}(A,B)\definition (-)^{\lambda_{34}}\sum_{\lambda_i}(-)^{\lambda_i}\mathcal{C}^{1234\lambda_i}(A-B)^{\lambda_{12}+\lambda_i-1}(A+B)^{\lambda_{34}-\lambda_i-1}\,,\\
    \nonumber
    &f_-^{1234}(A,B)\definition \sum_{\lambda_i}\mathcal{C}^{1234\lambda_i}(A-B)^{\lambda_{12}+\lambda_i-1}(A+B)^{\lambda_{34}-\lambda_i-1}\,,
\end{align}
where the upper index $1234$ denotes the order of the external helicities. We note immediately that $f_-^{1234}(A,B)=f_+^{1234}(B,A)(-)^{\Lambda-1}$, where $\Lambda=\lambda_1+\lambda_2+\lambda_3+\lambda_4$, but it is convenient to keep both of them. The OPE associativity constraint can be recast as
\begin{align}\label{Paper2-AssOPE}
\nonumber
\sum_{\lambda_i}&\Big(\mathcal{C}^{1234\lambda_i}(A-B)^{\lambda_{12}+\lambda_i-1}(A+B)^{\lambda_{34}-\lambda_i-1}+(-)^{\Lambda}\mathcal{C}^{3124\lambda_i}(B-C)^{\lambda_{13}+\lambda_i-1}(B+C)^{\lambda_{24}-\lambda_i-1}\\
&+\mathcal{C}^{1423\lambda_i}(C-A)^{\lambda_{14}+\lambda_i-1}(C+A)^{\lambda_{23}-\lambda_i-1}\Big)=0\,,
\end{align}
and plugging in the definitions \eqref{Paper2-definitions} for the functions $f^{\cdot\cdots}_-$, we have
\begin{equation}\label{Paper2-AssOPEv2}
f^{1234}_-(A,B)+(-)^{\Lambda}f^{3124}_-(B,C)+f^{1423}_-(C,A)=0\,.
\end{equation}
The most general polynomial form of the functions $f^{\cdot\cdots}_-$ that is compatible with the constraint \eqref{Paper2-AssOPEv2}, is the following:\footnote{For instance, choosing $f^{1234}_-(A,B)\sim A^iB^j$, with $i,j>0$ and $i+j=\Lambda-2$, would never satisfy \eqref{Paper2-AssOPEv2}, since neither $f^{3124}_-(B,C)$ nor $f^{1423}_-(C,A)$ could cancel such contributions.}
\begin{subequations}\label{Paper2-eq_all}
\begin{align}
    f^{1234}_-(A,B)&=k^{1234}_-A^{\Lambda-2}-k^{1234}_+B^{\Lambda-2}\,,\\
    f^{3124}_-(B,C)&=k^{3124}_-B^{\Lambda-2}-k^{3124}_+C^{\Lambda-2}\,,\\
    f^{1423}_-(C,A)&=k^{1423}_-C^{\Lambda-2}-k^{1423}_+A^{\Lambda-2}\,.
\end{align}
\end{subequations}
where $k^{\cdots\cdot}$ are some free coefficients (that we identify with those in \eqref{Paper2-definitions}).
Substituting these back into \eqref{Paper2-AssOPEv2} gives the constraints
\begin{align}\label{Paper2-condition_eq1}
    &k^{1234}_-=k^{1423}_+\,,&
    &k^{3124}_-=(-)^{\Lambda}k^{1234}_+\,,&
    &k^{1423}_-=(-)^{\Lambda}k^{3124}_+\,.
\end{align}
The form of the functions in \eqref{Paper2-eq_all} plus the conditions in \eqref{Paper2-condition_eq1} gives the solution to the OPE associativity constraint \eqref{Paper2-AssOPEv2}. To express the solution in terms of the cubic couplings, we need to solve \eqref{Paper2-eq_all} in terms of the product of couplings $\mathcal{C}^{\bullet\bullet\bullet\bullet\bullet}$. This can be done by expressing \eqref{Paper2-eq_all} using the previous variables $\PP_{34}=A+B$ and $\PP_{12}=A-B$, as follows
\begin{equation}
    \sum_{\lambda_i}\mathcal{C}^{1234\lambda_i}\PP_{12}^{\lambda_{12}+\lambda_i-1}\PP_{34}^{\lambda_{34}-\lambda_i-1}=k^{1234}_-\left(\frac{\PP_{34}+\PP_{12}}{2}\right)^{\Lambda-2}-k^{1234}_+\left(\frac{\PP_{34}-\PP_{12}}{2}\right)^{\Lambda-2}\,.
\end{equation}
By expanding the Binomial on the RHS and matching the monomials in terms of $\PP_{34}$ and $\PP_{12}$ we get
\begin{equation}\label{Paper2-Ass_system_eq}
    \mathcal{C}^{1234\lambda_i}=\frac{(k^{1234}_-+(-)^{\lambda_i+\lambda_{12}}k^{1234}_+)(\Lambda-2)!}{2^{\Lambda-2}(\lambda_{12}+\lambda_i-1)!(\lambda_{34}-\lambda_i-1)!}\quad
    \forall\;\lambda_i\,,\quad \text{same for $(3124)$ and $(1423)$}\,.\\
\end{equation}
Summing over $\lambda_i$, we can check consistency:
\begin{equation}\label{Paper2-consistency_1}
   \sum_{\lambda_i=-\lambda_{12}+1}^{\lambda_{34}-1}\frac{(\Lambda-2)!}{2^{\Lambda-2}(\lambda_{12}+\lambda_i-1)!(\lambda_{34}-\lambda_i-1)!}=\frac{1}{2^{\Lambda-2}}\sum_{n=0}^{\Lambda-2}
   \begin{pmatrix}
       \Lambda-2\\
       n
   \end{pmatrix}
   =1\,.
\end{equation}
Notice that, if we consider only even-derivative vertices (i.e. with $k^{\cdots\cdot}_-=k^{\cdots\cdot}_+$), we get
\begin{equation}
    k^{1234}_-=k^{3124}_-=k^{1423}_-=k^{1234}_+=k^{3124}_+=k^{1423}_+\,.
\end{equation}
This case makes the classification of solutions to the OPE associativity constraint easier. 

\paragraph{Summary.}
The general solution to the OPE associativity constraint \eqref{Paper2-OPE_Ass_LC} is
\begin{equation}
\boxed{
\begin{aligned}\label{Paper2-commuting_case}
&\mathcal{C}^{1234\lambda_i}=\frac{(k^{1234}_- +(-)^{\lambda_i+\lambda_{12}}k^{1234}_+)(\Lambda-2)!}{2^{\Lambda-2}(\lambda_{12}+\lambda_i-1)!(\lambda_{34}-\lambda_i-1)!}\qquad 
    \forall\;\lambda_i\,,\quad \text{same for $(3124)$ and $(1423)$}\\ 
    & k^{1234}_-=k^{1423}_+,\qquad
    k^{3124}_-=(-)^{\Lambda}k^{1234}_+,\qquad
    k^{1423}_-=(-)^{\Lambda}k^{3124}_+\,.
\end{aligned}
}
\end{equation}
When only even-derivative vertices are present, the solution is
\begin{equation}
\boxed{
\begin{aligned}\label{Paper2-commuting_even_case}
    &\mathcal{C}^{1234\lambda_i}=\frac{k^{1234}_+(\Lambda-2)!}{2^{\Lambda-3}(\lambda_{12}+\lambda_i-1)!(\lambda_{34}-\lambda_i-1)!}\qquad 
    \forall\;\lambda_i\,,\\
    &\;\text{same for $(3124)$ and $(1423)$}\qquad k^{1234}_+=k^{3124}_+=k^{1423}_+\,.
\end{aligned}
}
\end{equation}

\subsection{Colour case}
In the presence of a colour factor --- such as when fields live in representations of a gauge group $G=U(N), SO(N)$ and $USp(N)$ --- we only need to consider colour-ordered amplitudes, as discussed above. For instance, we focus here on the ordering $[1234]$, then \eqref{Paper2-normal_case_2}. The associativity constraint gets modified into
\begin{align}\label{Paper2-OPE_colour}
    \begin{split}
\sum_{\lambda_i}(-)^{\lambda_i}\theta_{\lambda_i}\Bigg[&\frac{\mathcal{A}_3(p_1^{\lambda_1},p_2^{\lambda_2},p_i^{\lambda_i})\mathcal{A}_3(p_3^{\lambda_3},p_4^{\lambda_4},p_i^{-\lambda_i})}{\bar{z}_{12}\bar{z}_{i3}(\omega_1+\omega_2)\omega_1\omega_2\omega_3}+\frac{\mathcal{A}_3(p_4^{\lambda_4},p_1^{\lambda_1},p_i^{\lambda_i})\mathcal{A}_3(p_2^{\lambda_2},p_3^{\lambda_3},p_i^{-\lambda_i})}{\bar{z}_{41}\bar{z}_{2i}(\omega_4+\omega_1)\omega_4\omega_1\omega_2}\Bigg]=0\,,
    \end{split}
\end{align}
where we introduce a factor $\theta_{\lambda_i}$ which can be justified by interpreting the amplitudes as arising from a Poisson bracket, with an associated $\theta_{\lambda_i}$ factor. For instance, in the $u(N)$ case, we have
\begin{equation}\label{Paper2-Poisson_U(N)}
  [(\phi^{\lambda}_p)^A_{\;B},(\phi^{\mu}_q)^C_{\;D}]=\frac{\delta^{\lambda,-\mu}\delta^3(p+q)}{2p^+}\,\theta_{\lambda}\delta^C_{\;B}\delta^A_{\;D}\,,
\end{equation}
where $\theta_{\lambda}=e^{ix\lambda}$ is a phase factor reflecting a possible ambiguity in the Poisson bracket. Following the same conventions used in \cite{Serrani:2025owx,Skvortsov:2020wtf}, and in analogy with the singlet case, we fix $\theta_{\lambda_i}=(-)^{\lambda_i}$. This choice is again equivalent to taking even-helicity fields to be Hermitian and odd-helicity fields to be anti-Hermitian matrices.\footnote{In the case of Yang-Mills, we can choose to work with hermitian matrices $(T^a)^{\dag}=T^a$, then an $i$ factor would appear in the commutator $[T^a,T^b]=if^{abc}T^c$, or work with anti-hermitian matrices $(T^a)^{\dag}=-T^a$ and then $[T^a,T^b]=f^{abc}T^c$.} This choice will become particularly useful when we later search for consistent lower-spin theories satisfying the OPE associativity constraint.
Rewriting \eqref{Paper2-OPE_colour} using light-cone variables, we find
\begin{equation}\label{Paper2-OPE_Ass_color}
\sum_{\lambda_i}\Big(\mathcal{C}^{1234\lambda_i}\PPb_{12}^{\lambda_{12}+\lambda_i-1}\PPb_{34}^{\lambda_{34}-\lambda_i-1}-\mathcal{C}^{4123\lambda_i}\PPb_{41}^{\lambda_{14}+\lambda_i-1}\PPb_{23}^{\lambda_{23}-\lambda_i-1}\Big)=0\,.
\end{equation}
Using the variables defined in \eqref{Paper2-ABC_variables}, we get 
\begin{align}\label{Paper2-Ass_color}
    \begin{split}
    \sum_{\lambda_i}\Big(&\mathcal{C}^{1234\lambda_i}(A-B)^{\lambda_{12}+\lambda_i-1}(A+B)^{\lambda_{34}-\lambda_i-1}\\
    &+(-)^{\lambda_{14}+\lambda_i}\mathcal{C}^{4123\lambda_i}(C-A)^{\lambda_{14}+\lambda_i-1}(C+A)^{\lambda_{23}-\lambda_i-1}\Big)=0\,,
    \end{split}
\end{align}
and plugging in the definitions \eqref{Paper2-definitions} for the functions $f^{\cdot\cdots}_-$, we have
\begin{equation}\label{Paper2-Ass_color_f}
f^{1234}_-(A,B)+(-)^{\Lambda}f^{4123}_+(C,A)=0\,.
\end{equation}
As before, we can determine the form of these functions to be 
\begin{subequations}\label{Paper2-eq_color}
\begin{align}\label{Paper2-eq1_color}
    f^{1234}_-(A,B)&=k^{1234}_-A^{\Lambda-2}\,,\\
    f^{4123}_+(C,A)&=(-)^{\Lambda+1}k^{4123}_-A^{\Lambda-2}\,,
\end{align}
\end{subequations}
where in the definition of $k^{\cdots}_-$ we also include the $\theta_{\lambda_i}$ factor. Substituting them into \eqref{Paper2-Ass_color_f} leads to the constraint 
\begin{equation}
    k_-^{1234}=k_-^{4123}\,.
\end{equation}
From \eqref{Paper2-eq_color}, we find the system for the couplings:
\begin{align}
    \begin{split}
    &\mathcal{C}^{1234\lambda_i}=\frac{k_-^{1234}(\Lambda-2)!}{2^{\Lambda-2}(\lambda_{12}+\lambda_i-1)!(\lambda_{34}-\lambda_i-1)!}\quad\forall\;\lambda_i\,,\quad\text{same for $(4123)$}\,,\\
    &k^{1234}_-=\sum_{\lambda_i}\mathcal{C}^{1234\lambda_i}\,,
    \end{split}
\end{align}
Let us conclude with a comment on the possibility of using $SU(N)$ as the gauge group. In this case, the Poisson bracket in \eqref{Paper2-Poisson_U(N)} acquires an additional term of the form $\sim \frac{1}{N}\delta^A_{\;B}\delta^C_{\;D}$. This contribution induces, in the holomorphic constraint, a term of the type \eqref{Paper2-mixed_case_2}. As a result, the solution to the constraint becomes more involved. However, in special cases, such as for the cubic vertices of Yang–Mills theory, where the couplings are proportional to the fully antisymmetric structure constants $f_{abc}$ this additional term vanishes. Indeed, one would obtain an extra contribution of the form $\sim \mathrm{Tr}([T_{a_1},T_{a_2}])\mathrm{Tr}([T_{a_3},T_{a_4}])(\phi^{\lambda_1}_{q_1})^{a_1}(\phi^{\lambda_2}_{q_2})^{a_2}(\phi^{\lambda_3}_{q_3})^{a_3}(\phi^{\lambda_4}_{q_4})^{a_4}$, which vanishes due to the cyclicity of the trace. This mechanism is sometimes referred to as photon decoupling: at tree level, gluon scattering amplitudes are identical whether one works with $U(N)$ or $SU(N)$.

\paragraph{Summary.} The solution to the associativity constraint in the colour-ordered case is
\begin{equation}\label{Paper2-colour_case}
    \boxed{
    \begin{aligned}
    &\mathcal{C}^{1234\lambda_i}=\frac{k_-^{1234}(\Lambda-2)!}{2^{\Lambda-2}(\lambda_{12}+\lambda_i-1)!(\lambda_{34}-\lambda_i-1)!}\qquad\forall\;\lambda_i\,,\\
    &\;\text{same for $(4123)$}
    \qquad k^{1234}_-=k^{4123}_-\,.
    \end{aligned}
    }
\end{equation}

\subsection{Main mechanism}
As we have seen above, there is a simple procedure to solve the constraints. By looking at the most general case presented in Appendix \ref{Paper2-AppendixD}, one can see that this is always the case.
We can divide the procedure into two parts. First, we see that the constraint reduces to setting a polynomial in three variables to zero. The admissible functions available for this purpose are, as indicated in \eqref{Paper2-possible_functions}, the following:
\begin{align}\label{Paper2-fAB}
    f(A,B)&= a_1 A^{\Lambda-2}+ b_1 B^{\Lambda-2}\,,& \mathcal{C}^{1234\lambda_i},\,\mathcal{C}^{2134\lambda_i},\,\mathcal{C}^{1243\lambda_i},\,\mathcal{C}^{2143\lambda_i}\,,\\\label{Paper2-fBC}
    f(B,C)&= b_2 B^{\Lambda-2}+ c_1 C^{\Lambda-2}\,,& \mathcal{C}^{1342\lambda_i},\,\mathcal{C}^{3142\lambda_i},\,\mathcal{C}^{1324\lambda_i},\,C^{3124\lambda_i}\,,\\\label{Paper2-fCA}
    f(C,A)&= c_2 C^{\Lambda-2}+ a_2 A^{\Lambda-2}\,,& \mathcal{C}^{1423\lambda_i},\,\mathcal{C}^{4123\lambda_i},\,\mathcal{C}^{1432\lambda_i},\,\mathcal{C}^{4132\lambda_i}\,,
\end{align}
where on the right of each function, we wrote the product of couplings that can produce them. The constraint they have to satisfy is 
\begin{equation}\label{Paper2-cocycle?_}
    f(A,B)+f(B,C)+f(C,A)=0\,.
\end{equation}
These polynomials can cancel each other in different ways to ensure the total sum vanishes:
\begin{enumerate}
    \item If all coefficients are non-zero, consistency requires setting $a_1=-a_2$, $b_1=-b_2$, and $c_1=-c_2$. This corresponds to the case of singlet fields, where only even-derivative vertices are present, as shown in \eqref{Paper2-commuting_even_case}. In this case, we need at least one product of couplings in \eqref{Paper2-fAB}, in \eqref{Paper2-fBC}, and in \eqref{Paper2-fCA} to be present.

    \item If only two terms are non-zero --- for instance, $b_1=b_2=c_1=c_2=0$ and $a_1=-a_2$ --- we obtain a solution that corresponds to the colour case \eqref{Paper2-colour_case}; the same holds under permutations of $a,b,c$. In this case, we need at least one product of couplings in \eqref{Paper2-fAB} and \eqref{Paper2-fCA} to be present.

    \item One can consider configurations with two vanishing terms and four non-zero coefficients, for example, $c_1=c_2=0$, $a_1=-a_2$, and $b_1=-b_2$. This type of cancellation may appear in the singlet case as well, as in \eqref{Paper2-commuting_case}, but it requires the inclusion of odd-derivative vertices. In this case, we need at least one product of couplings in \eqref{Paper2-fAB}, \eqref{Paper2-fBC}, and \eqref{Paper2-fCA} to be present.
\end{enumerate}
Once one of these cases is realised, by looking at the form of a single function, for instance, $f(A,B)= a_1 A^{\Lambda-2}+ b_1 B^{\Lambda-2}$, and recalling the definitions \eqref{Paper2-definitions}, we can extract the solutions for the couplings and find two distinct cases:
\begin{enumerate}
    \item If we have $a_1\neq 0$ and $b_1\neq 0$, to reproduce the correct form of the function, we must sum over all even or odd integer values of $\lambda_i$.

    \item If we have $a_1\neq 0$ and $b_1=0$, or vice versa, to reproduce the correct form of the function, we must sum over all integer values of $\lambda_i$.
\end{enumerate}

\subsection{Mixed cases}
Here we analyse the three remaining cases \eqref{Paper2-mixed_case_1}, \eqref{Paper2-mixed_case_2}, and \eqref{Paper2-mixed_case_3}. We start by looking at \eqref{Paper2-mixed_case_1} then
\begin{equation}
    \mathrm{Tr}(T_{a_1}T_{a_2})(\phi^{\lambda_1}_{q_1})^{a_1}(\phi^{\lambda_2}_{q_2})^{a_2}\phi^{\lambda_3}_{q_3}\phi^{\lambda_4}_{q_4}\sim \delta_{a_1a_2}(\phi^{\lambda_1}_{q_1})^{a_1}(\phi^{\lambda_2}_{q_2})^{a_2}\phi^{\lambda_3}_{q_3}\phi^{\lambda_4}_{q_4}\,.
\end{equation}
In this case, all fields ``commute'', as in the case of singlet fields. Therefore, the solution to the constraint coincides with the one found in the singlet case \eqref{Paper2-commuting_case} and \eqref{Paper2-commuting_even_case}.
Regarding \eqref{Paper2-mixed_case_2} we have
\begin{equation}
    \mathrm{Tr}(T_{a_1}T_{a_2})\mathrm{Tr}(T_{a_3}T_{a_4})(\phi^{\lambda_1}_{q_1})^{a_1}(\phi^{\lambda_2}_{q_2})^{a_2}(\phi^{\lambda_3}_{q_3})^{a_3}(\phi^{\lambda_4}_{q_4})^{a_4}\sim \delta_{a_1a_2}\delta_{a_3a_4}(\phi^{\lambda_1}_{q_1})^{a_1}(\phi^{\lambda_2}_{q_2})^{a_2}(\phi^{\lambda_3}_{q_3})^{a_3}(\phi^{\lambda_4}_{q_4})^{a_4}
\end{equation}
In this case, we have to consider the following four exchanges
\begin{align}
    &(1234)+(2134)+(1243)+(2143)\,,
\end{align}
where for each $(1234)$ ordering we also include the associated $(3412)$ exchange. As shown above, in the presence of these exchanges, the holomorphic quartic constraint, reducing to the equation \eqref{Paper2-cocycle?_}, forces $f(A,B)=0$, which cannot be satisfied. This conclusion holds for generic vertices, and in the absence of additional vertices that would contribute extra terms to \eqref{Paper2-cocycle?_}. We now consider the last case, \eqref{Paper2-mixed_case_3}. We have
\begin{equation}
    \mathrm{Tr}(T_{a_1}T_{a_2}T_{a_3})(\phi^{\lambda_1}_{q_1})^{a_1}(\phi^{\lambda_2}_{q_2})^{a_2}(\phi^{\lambda_3}_{q_3})^{a_3}\phi^{\lambda_4}_{q_4}\,.
\end{equation}
In this case, using the cyclic property of the trace, and that $\phi^{\lambda_4}_{q_4}$ ``commutes'' with the others, we have to consider the following exchanges
\begin{align}
    &(123)4+(312)4+(231)4&\rightarrow&
    &(1234)+(1243)+(3142)+(3124)+(1423)+(4123)\,,
\end{align}
where again for each $(1234)$ we also have the associated $(3412)$ exchange. To read off the solution, we can look at the general case in Appendix \ref{Paper2-AppendixD}, and we get 
\begin{subequations}\label{Paper2-mixed_constraint}
\begin{align}
    &(\mathcal{F}^{1234}+\mathcal{F}^{1243})k^{1234}_- - (\mathcal{F}^{1423}+\mathcal{F}^{4123})k^{1423}_+=0\;\implies\;\mathcal{F}^{1234}k^{1234}_- = \mathcal{F}^{1423}k^{1423}_+\,,\\
    &(\mathcal{F}^{1342}+\mathcal{F}^{1342})k^{1342}_- - (\mathcal{F}^{1234}+\mathcal{F}^{1243})k^{1234}_+=0\;\implies\;\mathcal{F}^{1342}k^{1342}_- = \mathcal{F}^{1234}k^{1234}_+\,,\\
    &(\mathcal{F}^{1423}+\mathcal{F}^{4123})k^{1423}_- - (\mathcal{F}^{1342}+\mathcal{F}^{1342})k^{1342}_+=0\;\implies\;\mathcal{F}^{1423}k^{1423}_- = \mathcal{F}^{1342}k^{1342}_+\,,
\end{align}
\end{subequations}
where in the second expression we identified the various $\mathcal{F}$ terms since they are equal in this case, like $\mathcal{F}^{1234}=f_{a_1a_2a_3}=\mathcal{F}^{1243}$, and the value of the coupling product $\mathcal{C}^{1234}$, as well as for $(1342)$ and $(1423)$ is given in \eqref{Paper2-usual_couplings}.
If we consider only even-derivative vertices (i.e. with $k^{\cdots\cdot}_-=k^{\cdots\cdot}_+$), we get
\begin{equation}\label{Paper2-mixed_only_even}
   \mathcal{F}^{1234}k^{1234}_-=\mathcal{F}^{1342}k^{1342}_-=\mathcal{F}^{1423}k^{1423}_-=\mathcal{F}^{1234}k^{1234}_+=\mathcal{F}^{1342}k^{1342}_+=\mathcal{F}^{1423}k^{1423}_+\,.
\end{equation}
Another particular solution to the constraint in \eqref{Paper2-mixed_constraint} arises when we start with a pair of couplings $C^{\lambda_1,\lambda_2,\lambda_i}C^{-\lambda_i,\lambda_3,\lambda_4}$  where one is an even-derivative and the other an odd-derivative vertex. Borrowing ideas from \cite{Serrani:2025owx}, we can rewrite the solution in a more convenient form using
\begin{align}\label{Paper2-new_variables}
    k^{1234}_E\definition \frac{1}{2}(k^{1234}_-+k^{1234}_+)&\equiv\sum_{(\lambda_{12}+\lambda_i)\in\,\text{even}}\mathcal{C}^{1234\lambda_i}\,,\\
    k^{1234}_O\definition\frac{1}{2}(k^{1234}_--k^{1234}_+)&\equiv \sum_{(\lambda_{12}+\lambda_i)\in\,\text{odd}}\mathcal{C}^{1234\lambda_i}\,.
\end{align}
We can then rewrite the system governing the couplings in the form
\begin{align}\label{Paper2-even_first_P2}
    \mathcal{C}^{1234\lambda_i}&=\frac{k^{1234}_E (\Lambda-2)!}{2^{\Lambda-3}(\lambda_{12}+\lambda_i-1)!(\lambda_{34}-\lambda_i-1)!}\qquad 
    \forall\;(\lambda_{12}+\lambda_i)\in\text{even}\,,\\ \label{Paper2-odd_first_P2}
    \mathcal{C}^{1234\lambda_i}&=\frac{k^{1234}_O (\Lambda-2)!}{2^{\Lambda-3}(\lambda_{12}+\lambda_i-1)!(\lambda_{34}-\lambda_i-1)!}\qquad 
    \forall\;(\lambda_{12}+\lambda_i)\in\text{odd}\,.
\end{align}
If we rewrite the constraint \eqref{Paper2-mixed_constraint} in terms of these new variables, we find 
\begin{align}\label{Paper2-mixed_odd_even}
    \nonumber
   &\mathcal{F}^{1234}(k^{1234}_E+k^{1234}_O) = \mathcal{F}^{1423}(k^{1423}_E-k^{1423}_O)\,,\quad
    \mathcal{F}^{1342}(k^{1342}_E+k^{1342}_O) = \mathcal{F}^{1234}(k^{1234}_E-k^{1234}_O)\,,\\
    &\mathcal{F}^{1423}(k^{1423}_E+k^{1423}_O) = \mathcal{F}^{1342}(k^{1342}_E-k^{1342}_O)\,,
\end{align}
and becomes easier to identify the two special solutions. The first occurs when $k^{\cdots}_O=0$, leaving only $k^{\cdots}_E$, which corresponds to the purely even-derivative case discussed in \eqref{Paper2-mixed_only_even}, and gives
\begin{equation}\label{Paper2-mixed_1}
\mathcal{F}^{1234}k^{1234}_E=\mathcal{F}^{1342}k^{1342}_E=\mathcal{F}^{1423}k^{1423}_E\,.
\end{equation}
The second special case arises when two of the $k^{\cdots}_E$ coefficients and one $k^{\cdots}_O$ term vanish. For instance, setting $k^{1234}_E=k^{1342}_E=k^{1423}_O=0$ we find
\begin{equation}\label{Paper2-mixed_2}
    \mathcal{F}^{1234}k^{1234}_O =-\mathcal{F}^{1342}k^{1342}_O=\mathcal{F}^{1423}k^{1423}_E\,.
\end{equation}
Notice that the same solution also applies to the constraint found in the singlet case \eqref{Paper2-commuting_case}, provided we drop the $\mathcal{F}$ terms. On the other hand, if we look for solutions involving only odd-derivative vertices (i.e. with $k^{\cdots\cdot}_-=-k^{\cdots\cdot}_+$), we encounter a contradiction. This is evident from the structure of \eqref{Paper2-mixed_constraint}, and was already shown for the singlet case in \cite{Serrani:2025owx}.

\section{Solutions to the OPE associativity}\label{Paper2-section5}

In the section above, we solved the OPE associativity constraint in several specific cases, with the most general case addressed in Appendix \ref{Paper2-AppendixD}. However, finding a consistent theory composed of holomorphic cubic couplings that simultaneously solves all possible OPE associativity constraints arising among them is a related but distinct problem.

Fortunately, this issue was already addressed in \cite{Serrani:2025owx} for the holomorphic light-cone constraint, and the same approach applies here. The key idea is that, starting from a specific pair of cubic couplings $C^{\lambda_1,\lambda_2,\lambda_i}C^{-\lambda_i,\lambda_3,\lambda_4}$ involving at least one opposite helicity field,\footnote{Note that the OPE associativity constraint becomes non-trivial only if at least two of the couplings involved contain fields of opposite helicity, i.e. $\lambda_i = -\lambda_j$. In principle, one could consider very generic couplings that never generate any non-trivial constraint, though such theories would be rather trivial. Indeed, recall that to obtain an exchange amplitude, we need at least two cubic couplings involving opposite helicity fields.} requiring the constraint to be satisfied may --- and typically does --- generate new cubic couplings. These, in turn, can give rise to further constraints, and the process continues recursively. For a detailed discussion, see \cite{Serrani:2025owx}.

We briefly explain here the procedure to find consistent solutions for various cases. In the case of a singlet field \eqref{Paper2-commuting_even_case} and the mixed case \eqref{Paper2-mixed_1} with only even-derivative vertices, satisfying the OPE associativity requires that each time one of the pairs of couplings below is present, all the others must follow:
\begin{align}\label{Paper2-singlet_theory}
    \begin{split}
    &C^{\lambda_1,\lambda_2,[2-\lambda_{12},k-\lambda_{12}]}C^{[\lambda_{12}-2,\lambda_{12}-k],\lambda_3,\lambda_4}\,,\qquad
    C^{\lambda_1,\lambda_3,[2-\lambda_{13},k-\lambda_{13}]}C^{[\lambda_{13}-2,\lambda_{13}-k],\lambda_2,\lambda_4}\,,\\
    &C^{\lambda_1,\lambda_4,[2-\lambda_{14},k-\lambda_{14}]}C^{[\lambda_{14}-2,\lambda_{14}-k],\lambda_2,\lambda_3}\,,
    \end{split}
\end{align}
where we write them in pairs to emphasize that each involves an opposite-helicity field, $k$ is an even integer, the square bracket $[-,-]$ is a notation to indicate the range of exchanged helicities, incremented by steps of $2$, and $\Lambda=\lambda_1+\lambda_2+\lambda_3+\lambda_4$ --- the total number of derivatives --- must be even. For a better explanation of the notation used, we refer to \cite{Serrani:2025owx}. For instance, the following 
\begin{align}
    \begin{split}
    &C^{\lambda_1,\lambda_2,[2-\lambda_{12},4-\lambda_{12}]}C^{[\lambda_{12}-2,\lambda_{12}-4],\lambda_3,\lambda_4}\,,\qquad
    C^{\lambda_1,\lambda_3,[2-\lambda_{13},4-\lambda_{13}]}C^{[\lambda_{13}-2,\lambda_{13}-4],\lambda_2,\lambda_4}\,,\\
    &C^{\lambda_1,\lambda_4,[2-\lambda_{14},4-\lambda_{14}]}C^{[\lambda_{14}-2,\lambda_{14}-4],\lambda_2,\lambda_3}\,,
    \end{split}
\end{align}
would correspond to the following $6$ pairs of coupling
\begin{align}
    \begin{split}
    &C^{\lambda_1,\lambda_2,2-\lambda_{12}}C^{\lambda_{12}-2,\lambda_3,\lambda_4}\,,\qquad
    C^{\lambda_1,\lambda_3,2-\lambda_{13}}C^{\lambda_{13}-2,\lambda_2,\lambda_4}\,,\qquad
    C^{\lambda_1,\lambda_4,2-\lambda_{14}}C^{\lambda_{14}-2,\lambda_2,\lambda_3}\,,\\
    &C^{\lambda_1,\lambda_2,4-\lambda_{12}}C^{\lambda_{12}-4,\lambda_3,\lambda_4}\,,\qquad
    C^{\lambda_1,\lambda_3,4-\lambda_{13}}C^{\lambda_{13}-4,\lambda_2,\lambda_4}\,,\qquad
    C^{\lambda_1,\lambda_4,4-\lambda_{14}}C^{\lambda_{14}-4,\lambda_2,\lambda_3}\,.
    \end{split}
\end{align}
For the colour case \eqref{Paper2-colour_case} with both even- and odd-derivative vertices, we need the following set of couplings:
\begin{align}\label{Paper2-colour_theory}
    \begin{split}
    &C^{\lambda_1,\lambda_2,[1-\lambda_{12},k-\lambda_{12}]}C^{[\lambda_{12}-1,\lambda_{12}-k],\lambda_3,\lambda_4}\,,\qquad
    C^{\lambda_1,\lambda_3,[1-\lambda_{13},k-\lambda_{13}]}C^{[\lambda_{13}-1,\lambda_{13}-k],\lambda_2,\lambda_4}\,,\\
    &C^{\lambda_1,\lambda_4,[1-\lambda_{14},k-\lambda_{14}]}C^{[\lambda_{14}-1,\lambda_{14}-k],\lambda_2,\lambda_3}\,,
    \end{split}
\end{align}
where $k$ is an integer, and the square bracket $[-,-]$ is incremented by steps of $1$. For the mixed case \eqref{Paper2-mixed_2}, we need the following set of couplings:
\begin{align}\label{Paper2-mixed2_theory}
    \begin{split}
    &C^{\lambda_1,\lambda_2,[1-\lambda_{12},k-\lambda_{12}]}C^{[\lambda_{12}-1,\lambda_{12}-k],\lambda_3,\lambda_4}\,,\qquad
    C^{\lambda_1,\lambda_3,[1-\lambda_{13},k-\lambda_{13}]}C^{[\lambda_{13}-1,\lambda_{13}-k],\lambda_2,\lambda_4}\,,\\
    &C^{\lambda_1,\lambda_4,[k-\lambda_{14},1-\lambda_{14}]}C^{[\lambda_{14}-k,\lambda_{14}-1],\lambda_2,\lambda_3}\,,
    \end{split}
\end{align}
where $k$ is an even integer, the square bracket $[-,-]$ is incremented by steps of $2$, and $\Lambda$ must be odd.

Once all the couplings required to satisfy the constraint are included, an explicit solution relating the various couplings can always be extracted from \eqref{Paper2-commuting_even_case}, \eqref{Paper2-colour_case}, \eqref{Paper2-even_first_P2}, and \eqref{Paper2-odd_first_P2}. Below, we write down some explicit solutions for the couplings. In particular, we match the result of \cite{Ren:2022sws} for lower-spin theories and pinpoint the difference with the solutions to the light-cone quartic holomorphic constraint. 

\subsection{Lower-spin solutions}
Following ideas from \cite{Metsaev:1991mt,Metsaev:1991nb,Ponomarev:2016lrm,Ponomarev:2017nrr,Serrani:2025owx} we look for the most general lower-spin cubic vertices (i.e. involving lower-spins $|\lambda|=0,1,2$) that solve the OPE associativity constraint.

\paragraph{Singlet fields.} This presentation closely follows that of \cite{Serrani:2025owx}, with the necessary modifications. All possible even-derivative couplings are given by
\begin{equation}
    \left\{C^{-2,2,2},C^{-1,1,2},C^{0,0,2},C^{0,1,1},C^{0,2,2},C^{1,1,2},C^{2,2,2}\right\}\,.
\end{equation}
In total, we have $7$ couplings: $5$ abelian and $2$ non-abelian.\footnote{The non-abelian vertices are the most interesting ones, as they deform the gauge algebra in a covariant formulation. The abelian ones do not. Nevertheless, it is worth noting that, in general, abelian vertices can also be constrained and may play an important role in ensuring the consistency of the theory. Since we work in the light-cone gauge and with holomorphic theories, we define abelian couplings as those with $\lambda_i\geq0$ and $\sum_i\lambda_i>0$. Note that $(0,s,s)$ is still considered abelian this way, which is justified since the spin-zero exchange disappears from the constraint. Such couplings are, obviously, consistent on their own unless there are non-abelian couplings that ``talk to them''. } We can now search for all possible solutions to the OPE associativity, following \eqref{Paper2-singlet_theory}.

We begin with two-derivative couplings, then $\left\{C^{-2,2,2},C^{-1,1,2},C^{0,0,2},C^{0,1,1}\right\}$.
Solutions to the OPE associativity are
\begin{subequations}
\begin{align}
        &\{C^{-2,2,2}\}\,,&&\text{graviton coupling}\,,\\
        &\{C^{1,1,0}\}\,,&&\text{photons coupled to scalars via } \phi F_{\mu\nu}^2\,,\\
        &\{C^{-2,2,2}=C^{0,0,2}\}\,,&&\text{scalars coupled to graviton}\,,\\
        &\{C^{-2,2,2}=C^{-1,1,2}\}\,,&&\text{graviton coupled to photons}\,,\\
        &\{C^{-2,2,2}=C^{-1,1,2}=C^{0,0,2},C^{0,1,1}\}\,,&& \text{graviton, photons and scalars}\,.
\end{align}
\end{subequations}
We use the following notation: within $\{\cdots\}$, we list the active (free) couplings. When a coupling appears alone, it is considered unconstrained --- that is, it can appear in the theory’s action with a free coefficient. It is important to specify the active couplings before solving the constraints. Note that the relations above are consistent with the universality of gravitational interactions.

If we allow for higher-derivative terms, the most general solution we obtain is the following:
\begin{equation}\label{Paper2-HDlowerspin_P2}
    \left\{C^{-2,2,2}=C^{-1,1,2}=C^{0,0,2},C^{1, 1, 2}=\frac{C^{0, 1, 1} C^{0, 2, 2}}{C^{-2, 2, 2}},C^{2, 2, 2}=\frac{3}{10}\frac{(C^{0, 2, 2})^2}{C^{-2, 2, 2}}\right\}\,.
\end{equation}
The constraint above implies that a theory containing  $R^3$ and $RF^2$ terms with arbitrary coefficients would not admit a celestial dual, due to the failure of OPE associativity. Instead, the coefficients must be fixed to the specific values given above. These results correctly reproduce those of \cite{Ren:2022sws}. The constraint imposed to $C^{0, 1, 1}$ in \eqref{Paper2-HDlowerspin_P2} for the higher-derivative theories
\begin{equation}
    C^{1, 1, 2}=\frac{C^{0, 1, 1} C^{0, 2, 2}}{C^{-2, 2, 2}}\,,
\end{equation}
is novel, because in \cite{Ren:2022sws} the $C^{0, 1, 1}$ abelian coupling between photons and scalars was not considered.

\paragraph{Colour case.} This presentation closely follows that of \cite{Serrani:2025owx}, with the necessary modifications. We now examine all solutions in the presence of a gauge group, following \eqref{Paper2-colour_theory}. The complete set of even- and odd-derivative couplings that can be constructed is given by
\begin{align}
\begin{split}
    \{&C^{-2,1,2},\textcolor{blue}{C^{-2,2,2}},\textcolor{blue}{C^{-1,0,2}},C^{-1,1,1},\textcolor{blue}{C^{-1,1,2}},\textcolor{red}{C^{-1,2,2}},C^{0,0,1},\textcolor{blue}{C^{0,0,2}},\\
&C^{0,1,1},\textcolor{blue}{C^{0,1,2}},\textcolor{blue}{C^{0,2,2}},C^{1,1,1},\textcolor{red}{C^{1,1,2}},\textcolor{red}{C^{1,2,2}},\textcolor{red}{C^{2,2,2}}\}\,.
\end{split}
\end{align}
In total, we have $15$ couplings: $9$ abelian and $6$ non-abelian. We highlight the lower-spin couplings as follows: blue indicates those which, on their own, require higher-spin couplings for consistency; red denotes those which, when combined with any other coupling, also require higher-spin ones; uncoloured couplings are, as we will show in more detail below, the only consistent lower-spin couplings. We now search for possible theories that satisfy the OPE associativity constraint. We begin by considering one-derivative interactions, involving the subset of couplings
\begin{equation}\label{Paper2-one_derivative_colour}
    \{C^{-2,1,2},C^{-1,0,2},C^{-1,1,1},C^{0,0,1}\}\,.
\end{equation}
Theories that satisfy the OPE associativity constraint are given by
\begin{subequations}
\begin{align}
    &\{C^{-1,1,1}\}\,, && \text{Yang-Mills coupling}\,,\\
    &\{C^{-1,1,1}=C^{0,0,1}\}\,,&& \text{Yang-Mills coupled to scalars}\,,\\\label{Paper2-coloured_graviton}
    &\{C^{-2,1,2}=C^{-1,1,1}\}\,,&& \text{coloured graviton}\,,
\end{align}
\end{subequations}
where we do not have any sign factor appearing, thanks to the use of the $\theta_{\lambda_i}=(-)^{\lambda_i}$ term in \eqref{Paper2-OPE_colour}. This also matches standard definitions for lower-spin theories and with the results in \cite{Ren:2022sws}.
Note that including $C^{-1,0,2}$ and at least one other of the couplings in \eqref{Paper2-one_derivative_colour} leads to a non-associative OPE.

Another interesting option is the coloured graviton. It is well known that, under the usual assumptions, gravitons cannot carry colour \cite{Boulanger:2000rq}. However, we find that it is possible to have multi-graviton theories that satisfy the OPE associativity, provided we restrict to the self-dual case. The coupling $C^{-2,1,2}$ is the ``usual'' current interaction of one helicity of the Yang-Mills field with a current built from a multiplet of gravitons.

Including higher-derivative couplings, the most general consistent solution is
\begin{equation}\label{Paper2-HDcolortheory}
\Big\{C^{-1,1,1}=C^{-2,1,2}=C^{0,0,1},C^{1,1,1}=\frac{1}{2}\frac{(C^{0, 1, 1})^2}{C^{-1, 1, 1}}\Big\}\,.
\end{equation}
In this case as well, the associativity constraint implies that a theory containing an $F^3$ term with an arbitrary coefficient would not admit a celestial dual, due to the failure of OPE associativity. Any attempt to include other couplings breaks OPE associativity, forcing us to introduce higher-spin fields. This is consistent with expectations from the covariant formalism. For example, the coupling $C^{-2,2,2}$ would lead to ``colour gravity'', which is known not to be consistent, at least when including only lower spins. These results correctly reproduce those of \cite{Ren:2022sws}, once we use notation valid for any semisimple Lie algebra given in \eqref{Paper2-semisimple_a}--\eqref{Paper2-semisimple_b}. 
The possibility of having OPE associativity in the presence of the non-abelian $C^{-2, 1, 2}$ coupling \eqref{Paper2-coloured_graviton} is novel because in \cite{Ren:2022sws} the exotic $C^{-2, 1, 2}$ non-abelian coupling between gravitons and spin-$1$ fields was not included.

\paragraph{Other solutions for lower-spin.} Some vertices are still missing to fully reproduce the lower-spin results obtained in \cite{Ren:2022sws}. In particular, we need to study the OPE associativity constraints imposed on the following couplings, which were not included in the analysis above
\begin{equation}
    \left\{\delta_{ab}\frac{\tilde{C}^{0,1,1}\PPb^2}{\beta_2\beta_3}\phi^0_{q_1}(\phi^1_{q_2})^a(\phi^1_{q_3})^b,\delta_{ab}\frac{\tilde{C}^{1,1,2}\PPb^4}{\beta_1^2\beta_2\beta_3}\phi^2_{q_1}(\phi^1_{q_2})^a(\phi^1_{q_3})^b,\delta_{ab}\frac{\tilde{C}^{-1,1,2}\PPb^2}{\beta_1^2\beta_2^{-1}\beta_3}\phi^2_{q_1}(\phi^{-1}_{q_2})^a(\phi^1_{q_3})^b\right\}\,,
\end{equation}
where we denote the coupling by $\tilde{C}$ to distinguish it from the other $C$'s.
To treat these cases and match the result in \cite{Ren:2022sws}, we need to use the solutions found for the mixed cases. For the mixed cases \eqref{Paper2-mixed_case_1}, following \eqref{Paper2-singlet_theory} and using \eqref{Paper2-commuting_even_case} we can find
\begin{align}
    &(C^{-2,2,2}-\tilde{C}^{-1,1,2})\tilde{C}^{-1,1,2}=0\,,&
    &(C^{-2,2,2}-\tilde{C}^{-1,1,2})\tilde{C}^{1,1,2}=0\,,\\
    &(C^{0,0,2}-\tilde{C}^{-1,1,2})\tilde{C}^{0,1,1}=0\,,&
    &\tilde{C}^{-1,1,2}\tilde{C}^{1,1,2}=C^{0,2,2}\tilde{C}^{0,1,1}\,.
\end{align}
While for the mixed cases \eqref{Paper2-mixed_case_3}, following \eqref{Paper2-mixed2_theory} and using \eqref{Paper2-mixed_2} we can find
\begin{align}
\begin{split}
    \frac{1}{2!}f_{a_1a_2c}C^{1,1,-1}\delta_{c\,a_3}\tilde{C}^{1,1,2}=-\frac{1}{2!}f_{a_1a_3c}C^{1,1,-1}\delta_{c\,a_2}\tilde{C}^{1,1,2}=\frac{1}{2!}\delta_{a_1c}\tilde{C}^{1,2,1}f_{c\,a_2a_3}C^{-1,1,1}\\
    =\frac{1}{3!}f_{a_1a_2c}C^{1,1,1}\delta_{c\,a_3}\tilde{C}^{-1,1,2}=-\frac{1}{3!}f_{a_1a_3c}C^{1,1,1}\delta_{c\,a_2}\tilde{C}^{-1,1,2}=\frac{1}{3!}\delta_{a_1c}\tilde{C}^{1,2,-1}f_{c\,a_2a_3}C^{1,1,1}\,.
    \end{split}
\end{align}
Then, using the full antisymmetry of $f_{abc}$ and the fact that the spin-$2$ field is a singlet, we get
\begin{equation}
    f_{a_1a_2a_3}\tilde{C}^{-1,1,2}C^{1,1,1}=3f_{a_1a_2a_3}\tilde{C}^{1,1,2}C^{-1,1,1}\implies\tilde{C}^{-1,1,2}C^{1,1,1}=3\,\tilde{C}^{1,1,2}C^{-1,1,1}\,.
\end{equation}
In principle, there may be --- and likely are --- other admissible lower-spin theories that satisfy OPE associativity, but we do not analyse them further here.

We conclude this section by noting that this method of solving the OPE associativity constraints substantially simplifies the analysis compared to \cite{Mago:2021wje,Ren:2022sws}. Furthermore, it enables extensions to higher-spin fields and to solutions with higher-derivative cubic couplings, as discussed in \cite{Serrani:2025owx}.

\section{Holomorphic constraint vs OPE associativity}\label{Paper2-section6}

In this section, we make the connection between the quartic light-cone holomorphic constraint and the holomorphic OPE associativity more explicit. In particular, we borrow the idea of a \textit{``gauge'' algebra} \cite{Ponomarev:2017nrr} constructed starting from the cubic chiral vertices. First, we review and define this Lie algebra. Then we relate the following four objects: the Jacobi identity for the gauge algebra, the four-point tree-level amplitude constructed out of the cubic chiral vertices, the OPE associativity in CCFT, and the quartic light-cone holomorphic constraint.

\subsection{A gauge algebra from cubic vertices}
In \cite{Ponomarev:2017nrr}, the author proposed a gauge algebra built from the holomorphic vertices (including the chiral higher-spin theories). This idea was originally introduced in the context of self-dual Yang-Mills (SDYM) and self-dual gravity (SDGR) in \cite{Monteiro:2011pc}, to derive the BCJ relations for these theories. In particular, they identified the relevant gauge algebra of SDGR as the Poisson algebra of area-preserving diffeomorphisms of $S^2$, and related it to the kinematical algebra of SDYM.

Let us consider the most general cubic couplings, but omit $\fA_{abc}$ whenever it is not crucial. Starting from a chiral vertex in \eqref{Paper2-cubic_hamiltonian} (specifically, the holomorphic one), we associate to it a structure constant as follows:
\begin{align}
    &h_{A_1A_2A_3}=-C^{\lambda_1,\lambda_2,\lambda_3}\frac{\PPb_{12}^{\lambda_{123}}}{\beta_1^{\lambda_1}\beta_2^{\lambda_2}\beta_3^{\lambda_3}}\delta\Big(\sum_i q_i\Big)&
    &\rightarrow&
    &\fB_{A_1A_2A_3}\definition \frac{1}{2}h_{A_1A_2A_3}\frac{\beta_2\beta_3}{\beta_1\PPb_{12}}\,,
\end{align}
where capital Latin letters such as $A_i$ collectively denote helicity, momentum, and possible other internal labels. Let us note that this procedure inverts the parity of the original coupling. Effectively, it does send a cubic vertex $H_3^{\lambda_1,\lambda_2,\lambda_3}$ to a shifted one $H_3^{\lambda_1+1,\lambda_2-1,\lambda_3-1}$, thereby reversing the parity of the coupling. This way, the structure constants read
\begin{equation}\label{Paper2-structure_const}
\fB_{A_1A_2A_3}=-\frac{1}{2}C^{\lambda_1,\lambda_2,\lambda_3}\frac{\PPb_{12}^{\lambda_{123}-1}}{\beta_1^{\lambda_1+1}\beta_2^{\lambda_2-1}\beta_3^{\lambda_3-1}}\delta\Big(\sum_i q_i\Big)\,. 
\end{equation}
We raise and lower indices using the natural inner product
\begin{equation}
    (\phi_1,\phi_2)\definition \sum_{\lambda,s}\int d^4q\, d^4p\,\delta^4(q+p)\,\delta_{ab}\,\delta^{\lambda,-s}(\phi^{\lambda}_q)^a(\phi^{s}_p)^b\,,
\end{equation}
where $\phi$ are the fields of the theory. Notice that, according to the definition of the inner product, raising and lowering indices correspond to flipping the sign of both helicity and momentum. Instead, nothing happens to internal indices, if present, due to the metric normalized to $\delta_{ab}$. For any chiral theory, the cubic chiral action can be written as
\begin{equation}
    S=\frac{1}{2}(\phi,\Box\phi)-h_{A_1A_2A_3}\phi^{A_1}\phi^{A_2}\phi^{A_3}\,.
\end{equation}
The idea to define $\fB_{ABC}$ is to factor out the kinematical part of the self-dual Yang-Mills vertex, and interpret other theories, e.g. the Chiral higher-spin theories, as generalisations of SDYM \cite{Ponomarev:2017nrr} with a certain gauge algebra defined by $\fB_{ABC}$. If we introduce the Lie bracket $[\bullet,\bullet]$ 
\begin{equation}
    [T_{A_1},T_{A_2}]=\fB^{A_3}_{\;\;\;\;A_1A_2}T_{A_3}\,,
\end{equation}
that corresponds to $\fB_{ABC}$, we can rewrite the action as the ``SDYM'' where the colour Lie algebra is replaced by the gauge algebra.\footnote{We write ``SDYM'' since theories can have fields of various helicities. It is also important that there is no scalar self-coupling for such an interpretation to work (this is the only coupling from which $\PPb$ cannot be factored out). One can also consider SDYM/SDGR with these gauge algebras. For example, the Moyal deformation of SDGR was discussed recently in \cite{Bu:2022iak}. Such theories would break Lorentz invariance since the number of $\PPb$ in a vertex is strongly correlated with the helicities. }   

It is interesting to see whether $\fB_{A_1A_2A_3}$ defines a Lie algebra. We need to check the Jacobi identity, which gives a constraint on the couplings. As a warm-up example, let us review what we get for SDYM \cite{Ponomarev:2017nrr}. We could start from the Chalmers-Siegel action \cite{Chalmers:1996rq} as in \cite{Ponomarev:2017nrr} and assume to activate the SDYM coupling $C^{-1,1,1}$.
Using the following SDYM vertex:
\begin{equation}
    h_{A_1A_2A_3}=2f_{a_1a_2a_3}\frac{\PPb_{12}\beta_1}{\beta_2\beta_3}\delta_{\lambda_1,-1}\delta_{\lambda_2,1}\delta_{\lambda_3,1}\delta\Big(\sum_i q_i\Big)\,.
\end{equation}
From \eqref{Paper2-structure_const} we can read the structure constant for the gauge algebra
\begin{equation}
\fB_{A_1A_2A_3}=f_{a_1a_2a_3}\big(\delta_{\lambda_1,-1}\delta_{\lambda_2,1}\delta_{\lambda_3,1}+\frac{\beta_2^2}{\beta_1^2}\delta_{\lambda_1,1}\delta_{\lambda_2,-1}\delta_{\lambda_3,1}+\frac{\beta_3^2}{\beta_1^2}\delta_{\lambda_1,1}\delta_{\lambda_2,1}\delta_{\lambda_3,-1}\big)\delta\Big(\sum_i q_i\Big)\,,
\end{equation}
and the Jacobi identity is a natural consequence of the one for the structure constant $f^{a_1a_2a_3}$.

We can do the same for SDGR starting from
\begin{equation}
    h_{A_1A_2A_3}=2\frac{\PPb_{12}^2\beta_1^2}{\beta_2^2\beta_3^2}\delta_{\lambda_1,-2}\delta_{\lambda_2,2}\delta_{\lambda_3,2}\delta\Big(\sum_i q_i\Big)\,,
\end{equation}
then the structure constant is
\begin{equation}
\fB_{A_1A_2A_3}=\frac{\PPb_{12}\beta_1}{\beta_2\beta_3}\big(\delta_{\lambda_1,-2}\delta_{\lambda_2,2}\delta_{\lambda_3,2}+\frac{\beta_2^4}{\beta_1^4}\delta_{\lambda_1,2}\delta_{\lambda_2,-2}\delta_{\lambda_3,2}+\frac{\beta_3^4}{\beta_1^4}\delta_{\lambda_1,2}\delta_{\lambda_2,2}\delta_{\lambda_3,-2}\big)\delta\Big(\sum_i q_i\Big)\,.
\end{equation}
The Jacobi identity is then verified thanks to the identity
\begin{equation}
    \PPb_{12}\PPb_{34}+\PPb_{23}\PPb_{14}+\PPb_{31}\PPb_{24}=A^2-B^2+C^2-A^2+B^2-C^2=0\,,
\end{equation}
that on-shell coincides with a special case of the Schouten identity. In \cite{Ponomarev:2016lrm,Serrani:2025owx} it was shown that a theory with the following couplings $C^{2,-2,2}=C^{\lambda,-\lambda,2}=\ell$ is consistent. Therefore, we could search for its gauge algebra:
\begin{align}
\begin{split}
\fB_{A_1A_2A_3}=&\,\ell\Big(\frac{\PPb_{12}\beta_1}{\beta_2\beta_3}(\delta_{\lambda_1,-2}\delta_{\lambda_2,2}\delta_{\lambda_3,2}+\left(\frac{\beta_2}{\beta_1}\right)^4\delta_{\lambda_1,2}\delta_{\lambda_2,-2}\delta_{\lambda_3,2}+\left(\frac{\beta_3}{\beta_1}\right)^4\delta_{\lambda_1,2}\delta_{\lambda_2,2}\delta_{\lambda_3,-2})\\
&+ \frac{\PPb_{12}\beta_1^{\lambda-1}}{\beta_2^{\lambda-1}\beta_3}(\delta_{\lambda_1,-\lambda}\delta_{\lambda_2,\lambda}\delta_{\lambda_3,2}+\left(\frac{\beta_2}{\beta_1}\right)^{2\lambda-2}\delta_{\lambda_1,\lambda}\delta_{\lambda_2,-\lambda}\delta_{\lambda_3,2})\\
&+\frac{\PPb_{12}\beta_1^{\lambda-1}}{\beta_3^{\lambda-1}\beta_2}(\delta_{\lambda_1,-\lambda}\delta_{\lambda_2,2}\delta_{\lambda_3,\lambda}+\left(\frac{\beta_3}{\beta_1}\right)^{2\lambda-2}\delta_{\lambda_1,\lambda}\delta_{\lambda_2,2}\delta_{\lambda_3,-\lambda})\\
&+\frac{\PPb_{12}\beta_2^{\lambda-1}}{\beta_3^{\lambda-1}\beta_1}(\delta_{\lambda_1,2}\delta_{\lambda_2,-\lambda}\delta_{\lambda_3,\lambda}+\left(\frac{\beta_3}{\beta_2}\right)^{2\lambda-2}\delta_{\lambda_1,2}\delta_{\lambda_2,\lambda}\delta_{\lambda_3,-\lambda})\Big)\delta\Big(\sum_i q_i\Big)\,,
\end{split}
\end{align}
and also here, the Jacobi identity is then verified thanks to the identity
\begin{equation}\label{Paper2-JI_gauge_algebra}
    \PPb_{12}\PPb_{34}+\PPb_{23}\PPb_{14}+\PPb_{31}\PPb_{24}=A^2-B^2+C^2-A^2+B^2-C^2=0\,.
\end{equation}
Now that we understood the general idea, we can try to solve the Jacobi identity for a general gauge algebra.

\subsection{Jacobi identity and celestial OPE associativity}
The Jacobi identity for a generic gauge algebra is
\begin{equation}
    \fB_{A_1A_2B}\fB^B_{\;\;A_3A_4}+\fB_{A_2A_3B}\fB^B_{\;\;A_1A_4}+\fB_{A_3A_1B}\fB^B_{\;\;A_2A_4}=0\,.
\end{equation}
Using the definition \eqref{Paper2-structure_const} we get
\begin{align}\label{Paper2-JI}
\begin{split}
\sum_{\lambda_i}(-)^{\lambda_i}\Big(\mathcal{C}^{1234\lambda_i}\PPb_{12}^{\lambda_{12}+\lambda_i-1}&\PPb_{34}^{\lambda_{34}-\lambda_i-1}+\mathcal{C}^{2314\lambda_i}\PPb_{23}^{\lambda_{23}+\lambda_i-1}\PPb_{14}^{\lambda_{14}-\lambda_i-1}\\
&+\mathcal{C}^{3124\lambda_i}\PPb_{31}^{\lambda_{13}+\lambda_i-1}\PPb_{24}^{\lambda_{24}-\lambda_i-1}\Big)=0\,,
\end{split}
\end{align}
and it coincides with the OPE associativity constraint \eqref{Paper2-OPE_Ass_LC}! Similar observations had already been made in \cite{Guevara:2022qnm,Guevara:2021abz}.

If we assume that the cubic vertices used to construct the gauge algebra belong to some representation of a gauge group $G$, and then, for example, we assume they are accompanied by the structure constants $f_{abc}$, we get 
\begin{align}\notag
\sum_{\lambda_i}(-)^{\lambda_i}\Big(\mathcal{C}^{1234\lambda_i}\PPb_{12}^{\lambda_{12}+\lambda_i-1}&\PPb_{34}^{\lambda_{34}-\lambda_i-1}f_{a_1a_2c}f^c_{\;\;a_3a_4}+\mathcal{C}^{2314\lambda_i}\PPb_{23}^{\lambda_{23}+\lambda_i-1}\PPb_{14}^{\lambda_{14}-\lambda_i-1}f_{a_2a_3c}f^c_{\;\;a_1a_4}+\\
&+\mathcal{C}^{3124\lambda_i}\PPb_{31}^{\lambda_{13}+\lambda_i-1}\PPb_{24}^{\lambda_{24}-\lambda_i-1}f_{a_3a_1c}f^c_{\;\;a_2a_4}\Big)=0\,.
\end{align}
Now using the Jacobi identity for $f_{abc}$, we obtain two independent constraints
\begin{align}\label{Paper2-JI_colour}
&\sum_{\lambda_i}(-)^{\lambda_i}\Big(\mathcal{C}^{1234\lambda_i}\PPb_{12}^{\lambda_{12}+\lambda_i-1}\PPb_{34}^{\lambda_{34}-\lambda_i-1}+\mathcal{C}^{2314\lambda_i}\PPb_{23}^{\lambda_{23}+\lambda_i-1}\PPb_{14}^{\lambda_{14}-\lambda_i-1}\Big)=0\,,\\
&\sum_{\lambda_i}(-)^{\lambda_i}\Big(\mathcal{C}^{1234\lambda_i}\PPb_{12}^{\lambda_{12}+\lambda_i-1}\PPb_{34}^{\lambda_{34}-\lambda_i-1}+\mathcal{C}^{3124\lambda_i}\PPb_{31}^{\lambda_{13}+\lambda_i-1}\PPb_{24}^{\lambda_{24}-\lambda_i-1}\Big)=0\,,
\end{align}
which after moving the external helicity using the symmetries \eqref{Paper2-coupling_sym}, the first corresponds to the colour ordering $[1234]$ and then to \eqref{Paper2-OPE_Ass_color}, while the second to the colour ordering $[1243]$. Again, a term $\theta_{\lambda_i}=(-)^{\lambda_i}$ can be added to match notations with the OPE associativity case. Moreover, by assuming a more general gauge algebra including all the possible generic couplings \eqref{Paper2-cubic_vertex_general} contributing, we get the same expression as in Appendix \ref{Paper2-AppendixD}.

\subsection{Jacobi identity and vanishing of the four-point amplitude}
Another interesting observation is the relation between the Jacobi identity and the vanishing of the four-point tree-level amplitude for a generic chiral theory. Using identities from Appendix \ref{Paper2-AppendixC} valid on-shell, the four-point amplitude takes the form
\begin{equation}
\begin{aligned}\label{Paper2-4_pt_amplitude}
\mathcal{A}&=\mathcal{A}_s+\mathcal{A}_t+\mathcal{A}_u=
    \sum_{\lambda_i}(-)^{\lambda_i}\mathcal{C}^{1234\lambda_i}\frac{\PPb_{12}^{\lambda_{12}+\lambda_i}}{\beta_1^{\lambda_1}\beta_2^{\lambda_2}}\frac{1}{(q_1+q_2)^2}\frac{\PPb_{34}^{\lambda_{34}-\lambda_i}}{\beta_3^{\lambda_3}\beta_4^{\lambda_4}}+2\leftrightarrow 4+2\leftrightarrow 3\\
    &=\frac{\PPb_{12}\PPb_{34}}{(q_1+q_2)^2\prod_{i=1}^4\beta_i^{\lambda_i}}\sum_{\lambda_i}(-)^{\lambda_i}\Big(\mathcal{C}^{1234\lambda_i}\PPb_{12}^{\lambda_{12}+\lambda_i-1}\PPb_{34}^{\lambda_{34}-\lambda_i-1}+\mathcal{C}^{1423\lambda_i}\PPb_{23}^{\lambda_{23}+\lambda_i-1}\PPb_{14}^{\lambda_{14}-\lambda_i-1}\\
    &\qquad\qquad\qquad\qquad\qquad\qquad\qquad+\mathcal{C}^{3124\lambda_i}\PPb_{31}^{\lambda_{13}+\lambda_i-1}\PPb_{24}^{\lambda_{24}-\lambda_i-1}\Big)=0\,.
\end{aligned}
\end{equation}
Therefore, the vanishing of the four-point amplitude for a chiral theory gives the same expression as the Jacobi identity \eqref{Paper2-JI} and the OPE associativity \eqref{Paper2-OPE_Ass_LC}.

In the presence of colours, the four-point tree-level amplitude  can be written in terms of the colour-ordered ones \eqref{Paper2-colour_ordered_ampl} and by considering a single colour-ordered amplitude, here $[1234]$, since the same is valid for the others, we obtain
\begin{align}
    \begin{split}
    \tilde{\mathcal{A}}(1234)=\frac{\PPb_{12}\PPb_{34}}{(q_1+q_2)^2\prod_{i=1}^4\beta_i^{\lambda_i}}&\sum_{\lambda_i}(-)^{\lambda_i}\Big(\mathcal{C}^{1234\lambda_i}\PPb_{12}^{\lambda_{12}+\lambda_i-1}\PPb_{34}^{\lambda_{34}-\lambda_i-1}\\
    &-\,\mathcal{C}^{2341\lambda_i}\PPb_{23}^{\lambda_{23}+\lambda_i-1}\PPb_{41}^{\lambda_{41}-\lambda_i-1}\Big)=0\,,
    \end{split}
\end{align}
where we can always introduce the $\theta_{\lambda_i}$ factors and correspond to the Jacobi identity for the gauge algebra in \eqref{Paper2-JI_colour}, and to the OPE associativity in \eqref{Paper2-OPE_Ass_color}. Moreover, by considering even more generic couplings \eqref{Paper2-cubic_vertex_general} contributing to the amplitude, we get the same expression as in Appendix \ref{Paper2-AppendixD} with the same prefactor as in the expressions above.

\subsection{Relation to the light-cone quartic holomorphic constraint}
We revisit the statement made in \cite{Ponomarev:2017nrr} about the relation between the light-cone consistency condition and the Jacobi identity for the gauge algebra. We point out that, on the energy-shell,\footnote{That is, we impose the energy conservation (i.e. with $\mathcal{H}=0$) for the external fields.} the two conditions are the same up to a specific product of couplings, and we elaborate on the implications of such a difference.

Let us recall the form of the total free Hamiltonian $\mathcal{H}$ and the boost operator $\mathcal{J}$:
\begin{align}
    &\mathcal{H}=\sum_{i=0}^nh_2^{\lambda_i}(q_i)=\sum_{i=0}^n-\frac{q_i\bar{q}_i}{\beta_i}\,,&
    &\mathcal{J}=\sum_{i=0}^nj_2^{\lambda_i}(q_i)=\sum_{i=0}^n\left(-\frac{q_i\bar{q}_i}{\beta_i}\frac{\partial}{\partial\bar{q}_i}-q_i\frac{\partial}{\partial\beta_i}+\lambda_i\frac{q_i}{\beta_i}\right)\,.
\end{align}
The light-cone (holomorphic) constraint is equivalent to the Lorentz invariance of the S-matrix. This can be rewritten as a set of Ward identities for all $n$-point off-shell amplitudes \cite{Ponomarev:2016cwi}.\footnote{For a definition of the off-shell amplitudes, we refer to \cite{Ponomarev:2016cwi}.} The statement is the following:
\begin{equation}
    [H,J]=0\qquad
    \Longleftrightarrow\qquad
    [\mathcal{A},J_2]=0\,,
\end{equation}
where $\mathcal{A}$ is the off-shell amplitude.
In particular, for chiral theories, where only cubic couplings are present, we can focus on the four-point amplitude\footnote{Indeed, the quartic holomorphic constraint is the only constraint for a cubic theory. It is equivalent to the Poincaré invariance of the four-point amplitude. Iteratively, one can show that higher-point amplitudes have vanishing cuts (e.g. a cut of a five-point amplitude is the product of the (vanishing) four-point and a three-point), i.e. they should vanish as well.}
\begin{equation}\label{Paper2-H3J3=A4J2}
    [H_3,J_3]=0\qquad
    \Longleftrightarrow\qquad
    [\mathcal{A}_4,J_2]=0\,.
\end{equation}
We can rewrite the condition above as follows
\begin{equation}\label{Paper2-Amplitude4}
\delta\left(\sum_{i=1}^4q_i\right)\mathcal{J}\left(\sum_{\lambda_i}h_3^{\lambda_1,\lambda_2,\lambda_i}(q_1,q_2,q_{\lambda_i})\frac{1}{s}h_3^{\lambda_3,\lambda_4,-\lambda_i}(q_3,q_4,-q_{\lambda_i})+ (t,u)\text{-channels}\right)=0\,,
\end{equation}
where $\lambda_i$ represents the helicities of the particles exchanged in the quartic tree-level diagram. The equivalence \eqref{Paper2-H3J3=A4J2} may be established by a direct computation and arises as a consequence of the following identity:
\begin{align}
\begin{split}
&\mathcal{J}\left(h_3^{\lambda_1,\lambda_2,\lambda_i}(q_1,q_2,q_{\lambda_i})\frac{1}{s}h_3^{\lambda_3,\lambda_4,-\lambda_i}(q_3,q_4,-q_{\lambda_i})\right)=\\
&\left((-)^{\lambda_i}\frac{(\lambda_1+\lambda_i-\lambda_2)\beta_1-(\lambda_2+\lambda_i-\lambda_1)\beta_2}{\beta_1+\beta_2}\PPb_{12}^{\lambda_{12}+\lambda_i-1}\PPb_{34}^{\lambda_{34}-\lambda_i}+(1,2)\leftrightarrow (3,4)\right)\,,
\end{split}
\end{align}
where $s=\tfrac{1}{2}([12]\langle 12\rangle+[34]\langle 34\rangle)$, and the use of momentum conservation and energy conservation (i.e. with $\mathcal{H}=0$) for the external fields is required. As we have reviewed above, and following \cite{Ponomarev:2016cwi}, to isolate the gauge algebra structure, we can factor out the SDYM kinematical part of the vertices as
\begin{align}
    &h_3^{\lambda_1,\lambda_2,\lambda_i}=\fB_{A_1A_2B}h^{1,1,-1}_\text{YM}\,,&
    &h_3^{\lambda_3,\lambda_4,-\lambda_i}=\fB^B_{\phantom{B}A_3A_4}h^{1,-1,1}_\text{YM}\,.
\end{align}
Then, we rewrite \eqref{Paper2-Amplitude4} in the following form:
\begin{align}
    \begin{split}
    \delta\left(\sum_{i=1}^4q_i\right)\Bigg(&\Big(h^{1,1,-1}_\text{YM}\frac{1}{s}h^{1,-1,1}_\text{YM}\Big)\mathcal{J}^{\lambda_1-1,\lambda_2-1,\lambda_3-1,\lambda_4+1}[\fB_{A_1A_2B}\fB^B_{\phantom{B}A_3A_4}]\\
    &+\mathcal{J}^{1,1,1,-1}[h^{1,1,-1}_\text{YM}\frac{1}{s}h^{1,-1,1}_\text{YM}](\fB_{A_1A_2B}\fB^B_{\;\;A_3A_4})+(t,u)\text{-channels}\Bigg)\,.
    \end{split}
\end{align}
By introducing the following operator
\begin{align}
    \mathcal{O}_s=\Big(h^{1,1,-1}_\text{YM}\frac{1}{s}h^{1,-1,1}_\text{YM}\Big)\mathcal{J}^{\lambda_1-1,\lambda_2-1,\lambda_3-1,\lambda_4+1}+\mathcal{J}^{1,1,1,-1}[h^{1,1,-1}_\text{YM}\frac{1}{s}h^{1,-1,1}_\text{YM}]\,,
\end{align}
and the corresponding one for the $t$ and $u$ channels. The equation above has the following schematic form:
\begin{equation}\label{Paper2-Jacobi_O}
    \mathcal{O}_s[\fB_{A_1A_2B}\fB^B_{\phantom{B}A_3A_4}]+\mathcal{O}_t[\fB_{A_2A_3B}\fB^B_{\;\;A_1A_4}]+\mathcal{O}_u[\fB_{A_3A_1B}\fB^B_{\phantom{B}A_2A_4}]=0\,.
\end{equation}
The idea of \cite{Ponomarev:2016cwi} was to show that
\begin{equation}
    \mathcal{O}=\mathcal{O}_s=\mathcal{O}_t=\mathcal{O}_u\,.
\end{equation}
By employing momentum conservation inside the Jacobi identity, this can be proven for $\mathcal{H}=0$. Indeed, the three operators will differ only by terms proportional to the total free Hamiltonian $\mathcal{H}$.

We have thus established that any solution of the Jacobi identity must also satisfy the light-cone holomorphic constraint. However, proving the converse requires first determining the kernel of the operator $\mathcal{J}$. To this end, we apply $\mathcal{J}$ to the most general chiral polynomial function $K(\PPb_{12},\PPb_{34},\beta_1,\beta_2,\beta_3,\beta_4)$ and we obtain
\begin{align}
    &\mathcal{J}^{\lambda_1,\lambda_2,\lambda_3,\lambda_4}\cdot K=0\,,&
    &K=\beta_1^{\lambda_1-n}\beta_2^{\lambda_2-n}\beta_3^{\lambda_3-m}\beta_4^{\lambda_4-m}\PPb_{12}^n\PPb_{34}^m\,.
\end{align}
This can be translated to the kernel of $\mathcal{O}_s$, $\mathcal{O}_t$, and $\mathcal{O}_u$ and gives
\begin{subequations}
\begin{align}
    &\mathcal{O}_s\cdot K_s=0\,,&
    &K_s=\beta_1^{\lambda_1-n}\beta_2^{\lambda_2-n}\beta_3^{\lambda_3-m}\beta_4^{\lambda_4-m-2}\PPb_{12}^n\PPb_{34}^m\,,\\
    &\mathcal{O}_t\cdot K_t=0\,,&
    &K_t=\beta_2^{\lambda_2-n}\beta_3^{\lambda_3-n}\beta_1^{\lambda_1-m}\beta_4^{\lambda_4-m-2}\PPb_{23}^n\PPb_{14}^m\,,\\
    &\mathcal{O}_u\cdot K_u=0\,,&
    &K_u=\beta_3^{\lambda_3-n}\beta_1^{\lambda_1-n}\beta_2^{\lambda_2-m}\beta_4^{\lambda_4-m-2}\PPb_{31}^n\PPb_{24}^m\,.
\end{align}
\end{subequations}
At this point, we can look if the operators in \eqref{Paper2-Jacobi_O} contain terms in the kernel of $\mathcal{O}_s$, $\mathcal{O}_t$, and $\mathcal{O}_u$. The explicit form of the product of two structure constants is
\begin{equation}
    \fB_{A_1A_2B}\fB^B_{\phantom{B}A_3A_4}\sim C^{\lambda_1,\lambda_2,\lambda_i}C^{-\lambda_i,\lambda_3,\lambda_4}\frac{\PPb_{12}^{\lambda_{12}+\lambda_i-1}\PPb_{34}^{\lambda_{34}-\lambda_i-1}}{\beta_1^{\lambda_1-1}\beta_2^{\lambda_2-1}\beta_3^{\lambda_3-1}\beta_4^{\lambda_4+1}}\,.
\end{equation}
This term is in the kernel of the operator $\mathcal{O}_s$ only for the specific case of $C^{\lambda_1,\lambda_1,0}C^{0,\lambda_2,\lambda_2}$, where
\begin{equation}
    \mathcal{O}_s\cdot \frac{\PPb_{12}^{2\lambda_1-1}\PPb_{34}^{2\lambda_2-1}}{\beta_1^{\lambda_1-1}\beta_2^{\lambda_1-1}\beta_3^{\lambda_2-1}\beta_4^{\lambda_2+1}}=0\,,
\end{equation}
and the same, with obvious modifications, holds for $\mathcal{O}_t$ and $\mathcal{O}_u$.
These products of couplings coincide with those left unconstrained by the quartic holomorphic constraint, as shown in \cite{Serrani:2025owx}. We then identified the transformation relating the two constraints as that induced by the operators $\mathcal{O}_s$, $\mathcal{O}_t$, and $\mathcal{O}_u$. 

To conclude, we showed that the solutions to the quartic holomorphic constraint and to the OPE associativity constraint coincide, up to the products of couplings $C^{\lambda_1,\lambda_1,0}C^{0,\lambda_2,\lambda_2}$ which remain unconstrained in the holomorphic constraint.\footnote{Notice that this can also be checked directly by solving both constraints. The solution to the OPE associativity found here, and the solution to the holomorphic constraint in \cite{Serrani:2025owx}, indeed coincide, up to the special product $C^{\lambda_1,\lambda_1,0}C^{0,\lambda_2,\lambda_2}$.}

This distinction is significant, as already emphasised in \cite{Serrani:2025owx}. In particular, it not only allows for terms such as $\sim F^3$ and $\sim R^3$ to appear with free coupling constants --- consistent with the covariant formulation of lower-spin theories --- but also permits non-vanishing amplitudes. Indeed, even in chiral theories and with higher-spin interactions, where the Weinberg low-energy theorem would suggest a vanishing amplitude, this argument does not apply to abelian terms (also referred to as Born–Infeld type interactions), which can give rise to non-vanishing contributions.

\section{Conclusions}\label{Paper2-section7}

After a review of both the light-cone quartic holomorphic constraint in $4d$ higher-spin theories \cite{Ponomarev:2016lrm,Metsaev:1991mt,Metsaev:1991nb}, and the OPE associativity in $2d$ Celestial CFT \cite{Ren:2022sws,Mago:2021wje}, we have solved the latter using the same ideas employed to solve the former \cite{Serrani:2025owx}. We have also clarified their relation.

This is inspired by \cite{Ren:2022sws,Monteiro:2022lwm}, where it was argued that the Metsaev couplings of the full chiral higher-spin theory are solutions to the OPE associativity constraint. In this work, we established an explicit relation between the two constraints. We identify the operator that relates the two and provide a method to classify all solutions, including those involving higher-spin fields and higher-derivative vertices. Moreover, we also uncover an interesting relation with the vanishing of the scattering amplitude and the Jacobi identity of a gauge algebra  constructed from the cubic vertices, as discussed in \cite{Ponomarev:2017nrr}. 

In \cite{Serrani:2025owx}, we initiated the programme of classifying all possible chiral higher-spin theories, which, for lower-derivative interactions, also leads to an OPE associative Celestial CFTs. Differences can arise for higher-derivative theories due to the unconstrained product $C^{\lambda_1,\lambda_1,0}C^{0,\lambda_2,\lambda_2}$, as we showed for certain low-spin examples.

It would be interesting to provide a geometrical interpretation of the gauge algebras constructed from the holomorphic vertices \eqref{Paper2-structure_const}, which give rise to OPE-associative celestial CFTs. A partial step was taken in \cite{Ponomarev:2017nrr}, where the gauge algebra of the full chiral higher-spin theory and its truncations to HS-SDYM and HS-SDGR were described. In particular, dual celestial chiral algebras exist for SDYM and SDGR \cite{Monteiro:2022lwm}. Extending this construction to all OPE-associative celestial CFTs would be natural, since we have shown that each bulk gauge algebra gives rise to an associative OPE on the boundary. Consequently, OPE associativity guarantees the existence of an underlying celestial chiral algebra, which could also admit a twistor interpretation, as in SDGR \cite{Adamo:2021lrv}.

As shown in \cite{Ponomarev:2017nrr}, with the help of the gauge algebra, one can reformulate the equations of motion of the Chiral higher-spin gravities as the self-duality constraint. Therefore, one can also refer to Chiral higher-spin gravities as self-dual ones. It is then tempting to place self-duality at the centre of the picture and interpret the OPE associativity, the vanishing of tree-level amplitudes, and the holomorphic light-cone constraint as consequences of self-duality. 

A natural question that arises is whether some of the results of the paper can be extended to the full quartic level by identifying the complete quartic constraint with the OPE associativity constraint of the “all-order” celestial OPE \cite{Adamo:2022wjo,Ren:2023trv}. More broadly, it would be worthwhile to extend this correspondence in two directions: first, by incorporating the classification of massive cubic vertices \cite{Metsaev:2005ar}, and second, by investigating how the relation might persist --- or be modified --- at the loop level. Loop corrections for  Chiral higher-spin gravity were studied in \cite{Skvortsov:2018jea,Skvortsov:2020wtf,Skvortsov:2020gpn,Tsulaia:2022csz}. It would be interesting to see what happens to the theories \cite{Serrani:2025owx} with finitely many higher-spin fields, where some of the results of \cite{Skvortsov:2018jea,Skvortsov:2020wtf,Skvortsov:2020gpn,Tsulaia:2022csz} do not apply directly.

Another important direction for future work is the classification of higher-derivative cases, both for the OPE associativity and for the light-cone holomorphic constraint. Here, the relation to twistor methods should be fruitful \cite{Tran:2021ukl,Herfray:2022prf,Tran:2022tft,Adamo:2022lah,Mason:2025pbz,Tran:2025xbt,Tran:2025uad}. It would also be important to clarify the relation to the results of \cite{Ponomarev:2022atv,Ponomarev:2022ryp,Ponomarev:2022qkx} where a ``flat space'' analogue of the singleton representation was proposed and amplitudes of Chiral higher-spin gravity were computed.

At least the theories of HS-SDYM type \cite{Ponomarev:2016lrm,Ponomarev:2017nrr,Krasnov:2021nsq,Monteiro:2022xwq,Serrani:2025owx}, fully classified in \cite{Serrani:2025owx}, are conformally invariant. Therefore, none of the main conclusions of the paper should change once a conformal transformation is applied to map them to anti-de Sitter space. It remains to be seen what role the celestial OPE would play in AdS/CFT correspondence (see \cite{Skvortsov:2018uru,Sharapov:2022awp,Jain:2024bza,Aharony:2024nqs} for the discussion of Chiral higher-spin gravity in the AdS/CFT context) and whether it can be extended beyond the theories of (HS)-SDYM type that are conformal, where covariant formulations \cite{Sharapov:2022faa,Sharapov:2022wpz,Sharapov:2022awp,Sharapov:2022nps,Sharapov:2023erv,Skvortsov:2024rng,Tran:2025yzd} of higher-spin theories should play an important role. 

We conclude by noting that the holomorphic OPE in CCFT can be determined using Poincaré invariance \cite{Himwich:2021dau}, which also allows fixing the cubic vertices in the bulk, interpreted as OPE coefficients in the CCFT. This parallels the role of the cubic light-cone constraint. In contrast, OPE associativity appears unrelated to Poincaré invariance. However, its close connection to the quartic holomorphic light-cone constraint in the bulk suggests otherwise. The slight discrepancy --- namely, the fact that the product $C^{\lambda_1,\lambda_1,0}C^{0,\lambda_2,\lambda_2}$ remains unconstrained by the light-cone holomorphic constraint --- deserves further investigation and may have a deeper explanation.
We hope to clarify these aspects in future work by trying to extend this correspondence to the full quartic level.

\section*{Acknowledgments}
\label{Paper2-sec:Aknowledgements1}
This project has received funding from the European Research Council (ERC) under the European Union’s Horizon 2020 research and innovation programme (grant agreement No 101002551). I am grateful to Evgeny Skvortsov for suggesting to elaborate on \cite{Ren:2022sws} and for many useful discussions. I am grateful to Akshay Yelleshpur Srikant and Dmitry Ponomarev for useful comments on the draft. I thank the anonymous Referee for the valuable suggestions and insightful questions.

\refstepcounter{section}
\section*{3.A \hspace{1mm} Light-cone notations}
\addcontentsline{toc}{section}{3.A \hspace{1mm} Light-cone notations}
\label{Paper2-AppendixA}

We use both light-cone coordinates and the light-cone gauge (in fact, double-null).
In flat spacetime, we adopt the $4d$ Minkowski metric with ``mostly plus'' signature
\begin{align}
    &x^{\mu}=(x^0,x^1,x^2,x^3)\,,&
    &\eta^{\mu\nu}=\text{diag}(-,+,+,+)\,,\\
    &ds^2=-(dx^0)^2+(dx^1)^2+(dx^2)^2+(dx^3)^2\,.
\end{align}
We define the light-cone coordinates as
\begin{align}
    x^+&=\frac{x^3+x^0}{\sqrt{2}}\,,&
    x^-&=\frac{x^3-x^0}{\sqrt{2}}\,,&
    x&=\frac{x^1-ix^2}{\sqrt{2}}\,,&
    \bar{x}&=\frac{x^1+ix^2}{\sqrt{2}}\,,\\
    \partial_-=\partial^+ &= \frac{\partial^3+\partial^0}{\sqrt{2}}\,, &
    \partial_+=\partial^- &= \frac{\partial^3-\partial^0}{\sqrt{2}}\,, &
    \partial &= \frac{\partial^1+i\partial^2}{\sqrt{2}}\,, &
    \bar{\partial} &= \frac{\partial^1-i\partial^2}{\sqrt{2}}\,,
\end{align}
where the metric and the wave-operator become
\begin{align}
    &x^{\mu}=(x^+,x^-,x,\bar{x})\,,&
    &ds^2=2\,dx^+ dx^- + 2\,dx d\bar{x}\,,\\
    &\Box=\partial_{\mu}\partial^{\mu}=2(\partial^+\partial^-+\partial\bar{\partial})\,,
\end{align}
the light-front scalar product becomes
\begin{align}
    A_{\mu}B^{\mu}=A_+B^++A_-B^-+A\bar{B}+\bar{A}B=A^-B^++A^+B^-+A\bar{B}+\bar{A}B\,,
\end{align}
and our definitions for the derivatives imply
\begin{align}
    \partial^+ x^-=\partial^-x^+=\partial x=\bar{\partial}\bar{x}=1\,.
\end{align}
In the light-front, $x^+$ is taken to be the time variable, and $\partial^-$ is the time derivative. Moreover, one assumes that $\partial^+$ is always non-zero and therefore can be inverted as an operator.

\refstepcounter{section}
\section*{3.B \hspace{1mm} Celestial CFT notations for massless fields}
\addcontentsline{toc}{section}{3.B \hspace{1mm} Celestial CFT notations for massless fields}
\label{Paper2-AppendixB}
In CCFT, to capture outgoing radiative data near future null infinity $\mathscr{I}^+$, we parametrise massless momenta (i.e. $q^2=0$) as 
\begin{align}
    \begin{split}
     q^{\mu}(\omega,z,\bar{z},\epsilon)&=\epsilon\,\omega(1+z\bar{z},z+\bar{z},i(\bar{z}-z),1-z\bar{z})\\
     &=\epsilon\,\omega(1+|z|^2,2\,\text{Re}(z),2\,\text{Im}(z),1-|z|^2)\,,
     \end{split}
\end{align}
where $\epsilon=\pm$ distinguishes outgoing ($+$) and incoming\footnote{We use the all-outgoing convention for amplitudes; then all particles are considered to be outgoing, and a minus sign is introduced for incoming particles.} ($-$) particles. The energy of a massless particle is given by $q^0=\omega(1+z\bar{z})$, with $\omega>0$. The complex variables $z$ and $\bar{z}$ serve as coordinates on the celestial sphere. While they are independent in complexified Minkowski space, under the Lorentzian signature with the ``mostly plus'' convention $\eta_{\mu\nu}=\text{diag}(-,+,+,+)$, they are taken to be complex conjugates of each other.

A scattering process involving $n$ massless fields with momenta $q^{\mu}_i$ and spins $s_i$ can, once expressed in a basis of boost eigenstates, be reinterpreted as a correlation function of $n$ conformal primary operators in the dual $2d$ CCFT. Each operator is characterised by conformal weights $(h_i,\bar{h}_i)$, from which one can obtain the spin $s_i=h_i-\bar{h}_i$ and the conformal dimension $\Delta_i=h_i+\bar{h}_i$.

\refstepcounter{section}
\section*{3.C \hspace{1mm} Spinor-helicity in both formalism}
\addcontentsline{toc}{section}{3.C \hspace{1mm} Spinor-helicity in both formalism}
\label{Paper2-AppendixC}

For our spinor-helicity conventions for massless particles, we follow \cite{Elvang:2013cua}:
\begin{align}
    &q_{a\dot{b}}= q_{\mu}(\sigma^{\mu})_{a\dot{b}}\,,&
    &\text{det}(q_{a\dot{b}})=-q^{\mu}q_{\mu}=m^2\,,&
    &q^2=0\;\; \Rightarrow\;\; q_{a\dot{b}}=-|q]_a\langle q|_{\dot{b}}\equiv -\lambda_a\tilde{\lambda}_{\dot{b}}\,,
\end{align}
\begin{align}
        &\langle ij\rangle\definition \langle q_i|_{\dot{a}}|q_j\rangle^{\dot{a}}\equiv\tilde{\lambda}_{i\dot{a}}\tilde{\lambda}_j^{\dot{a}}\,,&
    &[ij]\definition [q_i|^a|q_j]_a\equiv \lambda_i^{a}\lambda_{ja}\,\,,&
    &\lambda^{a}=\epsilon^{ab}\lambda_{b}\,,&
&\tilde{\lambda}^{\dot{a}}=\epsilon^{\dot{a}\dot{b}}\tilde{\lambda}_{\dot{b}}\,,
\end{align}
\begin{align}
&\sigma^0=
    \begin{pmatrix}
        1 & 0\\
        0 & 1
    \end{pmatrix}\,,&
    &\sigma^1=
    \begin{pmatrix}
        0 & 1\\
        1 & 0
    \end{pmatrix}\,,&
    &\sigma^2=
    \begin{pmatrix}
        0 & -i\\
        i & 0
    \end{pmatrix}\,,&
    &\sigma^3=
    \begin{pmatrix}
        1 & 0\\
        0 & -1
    \end{pmatrix}\,,
\end{align}
\begin{align}
    &\epsilon^{ab}=-\epsilon_{ab}=
    \begin{pmatrix}
        0 & 1\\
        -1 & 0
    \end{pmatrix}\,,&
    &\epsilon^{\dot{a}\dot{b}}=-\epsilon_{\dot{a}\dot{b}}=
    \begin{pmatrix}
        0 & 1\\
        -1 & 0
    \end{pmatrix}\,.
\end{align}
Notice that for complex momenta $q^{\mu}$ the two spinors $(\lambda_a,\tilde{\lambda}_{\dot{b}})$ are independent two-dimensional complex vectors. In Minkowski space and for real momenta $q_{a\dot{b}}$ is hermitian, and the two spinors become complex conjugate $\tilde{\lambda}_{\dot{a}}=\pm(\lambda_a)^*$ (where the sign depends on whether the energy is taken to be positive or negative, then on the convention we use on the background flat metric).

Spinor-helicity variables $(\lambda_i,\tilde{\lambda}_i)$ are defined up to little group scaling $(\lambda_i,\tilde{\lambda}_i)\sim (t_i\lambda_i,t_i^{-1}\tilde{\lambda}_i)$ for $t_i\in\mathbb{C}^*$. In CCFT, we can choose the following adapted coordinates for massless particles, following \cite{Himwich:2021dau}:
\begin{align}
    &|q_i]_a\equiv\lambda_{ai}=\sqrt{\omega_i}\begin{pmatrix}
        -\bar{z}_i\\
        1
    \end{pmatrix}\,,&
    &\langle q_i|_{\dot{a}}\equiv\tilde{\lambda}_{\dot{a}i}=\epsilon_i\sqrt{\omega_i}\begin{pmatrix}
         -z_i & 1
    \end{pmatrix}\,,&
    &z_i,\bar{z}_i\in\mathbb{C}\,,&
    &\omega_i\in\mathbb{C}^*\,.
    \end{align}
Here, $(z_i,\bar{z}_i)$ are generally taken to be complex and independent of each other,\footnote{They become complex conjugates in Lorentzian signature, where they act as coordinates in the celestial sphere $\mathbb{C}\PP^1$.} $\omega_i$ specifies the energy of the particle and $\epsilon_i=\pm 1$ stands for outgoing $(+)$ and incoming $(-)$ particles. In the following, we consider all-outgoing particles, so we fix $\epsilon_i\cdot\epsilon_j=1$. Therefore, we get the following:
\begin{align}
[ij]=
    \sqrt{\omega_i\omega_j}\bar{z}_{ij},\quad
    \langle ij \rangle=-\sqrt{\omega_i\omega_j}z_{ij},\quad
z_{ij}= z_i-z_j,\quad 
\bar{z}_{ij}=\bar{z}_i-\bar{z}_j\,.
\end{align}
In the context of the light-cone Hamiltonian approach for massless higher-spin fields, we adopt the following notation \cite{Ponomarev:2016cwi}:
\begin{equation}
    q_{a\dot{b}}=q_{\mu}(\sigma^{\mu})_{a\dot{b}}=\sqrt{2}
    \begin{pmatrix}
        q^- & \bar{q}\\
        q & - \beta
    \end{pmatrix}\approx\sqrt{2}
    \begin{pmatrix}
        -\frac{q\bar{q}}{\beta} & \bar{q}\\
        q & - \beta
    \end{pmatrix}= -|q]_a\langle q|_{\dot{b}}=-\lambda_a\tilde{\lambda}_{\dot{b}}\,,
\end{equation}
\begin{align}
&\tilde{\lambda}_{\dot{a}i}=\frac{2^{\frac{1}{4}}}{\sqrt{\beta_i}}\begin{pmatrix}
    q_i & -\beta_i
    \end{pmatrix}\,,&
    &\langle ij\rangle=-\sqrt{\frac{2}{\beta_i\beta_j}}\PP_{ij}\,,&
    &\lambda_{ai}=\frac{2^{\frac{1}{4}}}{\sqrt{\beta_i}}
    \begin{pmatrix}
       \bar{q}_i\\
        -\beta_i
    \end{pmatrix}\,,&
    &[ij]=\sqrt{\frac{2}{\beta_i\beta_j}}\PPb_{ij}\,.
\end{align}
The two formalisms are related by
\begin{equation}
    z_{ij}=\frac{\PP_{ij}}{\beta_i\beta_j},\quad
    \bar{z}_{ij}=\frac{\PPb_{ij}}{\beta_i\beta_j},\quad
    z_i=\frac{q_i}{\beta_i},\quad
    \bar{z}_i=\frac{\bar{q}_i}{\beta_i},\quad
    \sqrt{\omega_i}=-2^{\frac{1}{4}}\sqrt{\beta_i}\,.
\end{equation}
The relation between the cubic amplitude and the cubic Hamiltonian density follows:
\begin{align}\label{Paper2-AmplitudeandHamiltonian_holo}
    \mathcal{A}_3&=C^{\lambda_1,\lambda_2,\lambda_3}[12]^{d_{12}} [23]^{d_{23}} [31]^{d_{31}}=C^{\lambda_1,\lambda_2,\lambda_3}\frac{\sqrt{2}^{\lambda_{123}}\PPb^{\lambda_{123}}}{\beta_1^{\lambda_1}\beta_2^{\lambda_2}\beta_3^{\lambda_3}}=\sqrt{2}^{\lambda_{123}}h_3\,,\\\label{Paper2-AmplitudeandHamiltonian_antiholo}
    \bar{\mathcal{A}}_3&=\bar{C}^{-\lambda_1,-\lambda_2,-\lambda_3}\langle 12\rangle^{-d_{12}} \langle 23\rangle^{-d_{23}} \langle 31\rangle^{-d_{31}}=\bar{C}^{-\lambda_1,-\lambda_2,-\lambda_3}\frac{\sqrt{2}^{\lambda_{123}}\PP^{-\lambda_{123}}}{\beta_1^{-\lambda_1}\beta_2^{-\lambda_2}\beta_3^{-\lambda_3}}=\sqrt{2}^{\lambda_{123}}\bar{h}_3\,,
\end{align}
where $\mathcal{A}_3$ is valid for $\lambda_{123}>0$ and $\bar{\mathcal{A}}_3$ for $\lambda_{123}<0$, and we have defined
\begin{align}
    &d_{12}=\lambda_{12}-\lambda_3\,,&
    &d_{23}=\lambda_{23}-\lambda_1\,,&
    &d_{31}=\lambda_{31}-\lambda_2\,.
\end{align}
We also have some standard relations for the $n$-point scattering: 
\begin{align}
   &\langle ij\rangle [ij]=-\frac{2}{\beta_i\beta_j}\PP_{ij}\PPb_{ij}=2\,q_i \cdot q_j=(q_i+q_j)^2\,,\\
   &\sum^{n}_{j=1}q^{\mu}_j=0\quad\Rightarrow\quad\sum^{n}_{j=1}\langle ij\rangle[jk]=\sum^{n}_{j=1}\frac{\PP_{ij}\PPb_{jk}}{\beta_j}=0\,.
\end{align}
In particular, using the relation above for the four-point scattering, we find
\begin{align}
    &\frac{\PPb_{12}\PPb_{34}}{(q_1+q_2)^2}=\frac{\PPb_{31}\PPb_{24}}{(q_1+q_3)^2}=\frac{\PPb_{14}\PPb_{23}}{(q_1+q_4)^2}\,,&
    &s_{ij}=-(q_i+q_j)^2\,,
\end{align}
where $s=s_{12}$, $t=s_{14}$, $u=s_{13}$ are the standard Mandelstam variables for massless four-point scattering. 

The relations above, in the light-cone approach, hold only when the external particles are on-shell.  Indeed, one of the main differences between the cubic Hamiltonian density and the amplitudes in \eqref{Paper2-AmplitudeandHamiltonian_holo}-\eqref{Paper2-AmplitudeandHamiltonian_antiholo} is that the latter are inherently on-shell objects, while the former contains off-shell information. 

\refstepcounter{section}
\section*{3.D \hspace{1mm} OPE associativity: most general solution}
\addcontentsline{toc}{section}{3.D \hspace{1mm} OPE associativity: most general solution}
\label{Paper2-AppendixD}

It is straightforward to generalise the solution of the OPE associativity constraint to the most general cubic vertices \eqref{Paper2-cubic_vertex_general}. We make use of the following definitions:
\begin{align}\label{Paper2-definitions_2}
    \begin{split}
    &\mathcal{F}^{1234}\definition \fA\fdu{a_1a_2}{c}\fA_{c\,a_3a_4}=\delta^d_c\,\fA\fdu{a_1a_2}{c}\fA_{d\,a_3a_4}\,,\\
    &k_{+}^{1234}\definition(-)^{\lambda_{12}}\sum_{\lambda_i}(-)^{\lambda_i}\mathcal{C}^{1234\lambda_i}\,,\qquad
    k_{-}^{1234}\definition \sum_{\lambda_i}\mathcal{C}^{1234\lambda_i}\,,\\
    &f_-^{1234}(A,B)\definition \sum_{\lambda_i}\mathcal{C}^{1234\lambda_i}(A-B)^{\lambda_{12}+\lambda_i-1}(A+B)^{\lambda_{34}-\lambda_i-1}\,,
    \end{split}
\end{align}
where the upper index $1234$ denotes the order of the external helicities. Moreover, $\fA\fdu{a_1a_2}{c}\fA_{c\,a_3a_4}$ indicates the natural pairing between the positive and negative helicity fields $(\phi^{+\lambda})^a$ and $(\phi^{-\lambda})_b$, where the negative helicity fields take values in the dual vector space.\footnote{For example, if $f^a_{\phantom{a}bc}$ are the structure constants of some Lie algebra $\mathfrak{g}$, and $(\phi^{+\lambda})^a$ transforms in the adjoint representation, then $(\phi^{-1})_b$ transforms in the canonical dual to the adjoint, i.e. in the coadjoint one.} The contraction is performed with the standard Poisson bracket for fields decorated with vector space indices and reads
\begin{equation}
    [(\phi^{\lambda}_q)^a,(\phi^{s}_p)_b ]= \delta^{\lambda,-s}\delta^a_b\frac{\delta^3(q+p) }{2q^+}\,.
\end{equation}  
For general vertices, the holomorphic OPE associativity constraint takes the form
\begin{align}\label{Paper2-OPE_Ass_general}
\nonumber
\sum_{\lambda_i}&\Big(\mathcal{C}^{1234\lambda_i}(\mathcal{F}^{1234}+\mathcal{F}^{2134}+\mathcal{F}^{1243}+\mathcal{F}^{2143})\PPb_{12}^{\lambda_{12}+\lambda_i-1}\PPb_{34}^{\lambda_{34}-\lambda_i-1}\\
&+\mathcal{C}^{3124\lambda_i}(\mathcal{F}^{3124}+\mathcal{F}^{1324}+\mathcal{F}^{3142}+\mathcal{F}^{1342})\PPb_{31}^{\lambda_{13}+\lambda_i-1}\PPb_{24}^{\lambda_{24}-\lambda_i-1}\\
\nonumber
&+\mathcal{C}^{1423\lambda_i}(\mathcal{F}^{1423}+\mathcal{F}^{4123}+\mathcal{F}^{1432}+\mathcal{F}^{4132})\PPb_{14}^{\lambda_{14}+\lambda_i-1}\PPb_{23}^{\lambda_{23}-\lambda_i-1}\Big)=0\,,
\end{align}
where we used the symmetry property \eqref{Paper2-coupling_sym}, that of $\PPb$ and an additional minus sign due to the antisymmetry of $\langle 12\rangle$ and $\langle i3\rangle$ appearing in the denominators when computing the residue in \eqref{Paper2-computing_residues}; each exchange of spinor labels introduces an extra minus sign.
Writing \eqref{Paper2-OPE_Ass_general} in terms of the independent variables $A,B,C$, we find
\begin{align}
\begin{split}
\sum_{\lambda_i}&\Big(\mathcal{C}^{1234\lambda_i}(\mathcal{F}^{1234}+\mathcal{F}^{2134}+\mathcal{F}^{1243}+\mathcal{F}^{2143})(A-B)^{\lambda_{12}+\lambda_i-1}(A+B)^{\lambda_{34}-\lambda_i-1}\\
&+\mathcal{C}^{1342\lambda_i}(\mathcal{F}^{1342}+\mathcal{F}^{3142}+\mathcal{F}^{1324}+\mathcal{F}^{3124})(B-C)^{\lambda_{13}+\lambda_i-1}(B+C)^{\lambda_{24}-\lambda_i-1}\\
&+\mathcal{C}^{1423\lambda_i}(\mathcal{F}^{1423}+\mathcal{F}^{4123}+\mathcal{F}^{1432}+\mathcal{F}^{4132})(C-A)^{\lambda_{14}+\lambda_i-1}(C+A)^{\lambda_{23}-\lambda_i-1}\Big)=0\,,
\end{split}
\end{align}
where we have used the symmetry properties of $A,B,C$ in \eqref{Paper2-ABC_sym}. In terms of $f^{\cdot\cdots}_-$ it becomes 
\begin{align}\label{Paper2-generalOPE}
\begin{split}
&(\mathcal{F}^{1234}+\mathcal{F}^{2134}+\mathcal{F}^{1243}+\mathcal{F}^{2143})f^{1234}_-(A,B)\\
&+(\mathcal{F}^{1342}+\mathcal{F}^{3142}+\mathcal{F}^{1324}+\mathcal{F}^{3124})f^{1342}_-(B,C)\\
&+(\mathcal{F}^{1423}+\mathcal{F}^{4123}+\mathcal{F}^{1432}+\mathcal{F}^{4132})f^{1423}_-(C,A)=0\,.
\end{split}
\end{align}
We can determine the polynomial form of the functions $f^{\cdot\cdots}_-$ to be
\begin{subequations}\label{Paper2-possible_functions}
\begin{align}
    f^{1234}_-(A,B)&=k^{1234}_-A^{\Lambda-2}-k^{1234}_+B^{\Lambda-2}\,,\\
    f^{1342}_-(B,C)&=k^{1342}_-B^{\Lambda-2}-k^{1342}_+C^{\Lambda-2}\,,\\
    f^{1423}_-(C,A)&=k^{1423}_-C^{\Lambda-2}-k^{1423}_+A^{\Lambda-2}\,.
\end{align}
\end{subequations}
By substituting these forms of the functions in \eqref{Paper2-generalOPE}, we find
\begin{equation}
    A^{\Lambda-2}:(\mathcal{F}^{1234}+\mathcal{F}^{2134}+\mathcal{F}^{1243}+\mathcal{F}^{2143})k^{1234}_--(\mathcal{F}^{1423}+\mathcal{F}^{4123}+\mathcal{F}^{1432}+\mathcal{F}^{4132})k^{1423}_+=0\,,
\end{equation}
\begin{equation}
    B^{\Lambda-2}:(\mathcal{F}^{1342}+\mathcal{F}^{3142}+\mathcal{F}^{1324}+\mathcal{F}^{3124})k^{1342}_--(\mathcal{F}^{1234}+\mathcal{F}^{2134}+\mathcal{F}^{1243}+\mathcal{F}^{2143})k^{1234}_+=0\,,
\end{equation}
\begin{equation}
    C^{\Lambda-2}:(\mathcal{F}^{1423}+\mathcal{F}^{4123}+\mathcal{F}^{1432}+\mathcal{F}^{4132})k^{1423}_--(\mathcal{F}^{1342}+\mathcal{F}^{3142}+\mathcal{F}^{1324}+\mathcal{F}^{3124})k^{1342}_+=0\,.
\end{equation}
As usual, the solution for the couplings takes the form
\begin{equation}\label{Paper2-usual_couplings}
    \mathcal{C}^{1234\lambda_i}=\frac{(k^{1234}_-+(-)^{\lambda_i+\lambda_{12}}k^{1234}_+)(\Lambda-2)!}{2^{\Lambda-2}(\lambda_{12}+\lambda_i-1)!(\lambda_{34}-\lambda_i-1)!}\quad
    \forall\;\lambda_i\,,\quad
    \text{same for $(1342)$ and $(1423)$}\,.
\end{equation}
The one presented here is a general solution, in the sense that it includes all possible non-zero terms and uses the most general form for the cubic vertices \eqref{Paper2-cubic_vertex_general} in the OPE associativity constraint, with generic $\fA_{abc}$. 

As shown in the main text, the same solutions also satisfy the light-cone holomorphic constraint. The only difference is that this constraint leaves the products $C^{\lambda_1,\lambda_1,0}C^{0,\lambda_2,\lambda_2}$ unconstrained.

For instance, to recover the singlet case, we simply drop the $\mathcal{F}$ factors, which leads to the solution \eqref{Paper2-commuting_case}. The sign differences arise from the different ordering of the external fields. Here, we have chosen the ordering that avoids additional signs.

Instead, if we want to recover the solutions for the colour case, for the colour-ordering $[1234]$, we need to consider the terms $\mathcal{F}^{1234}=\mathcal{F}^{4123}$ and will match with \eqref{Paper2-colour_case}.

A complete analysis of the theories that satisfy the most general form of the constraint, along with a full classification, appears feasible but lies beyond the scope of this paper. As an application of the expression above, we solve the OPE associativity constraint in the mixed cases \eqref{Paper2-mixed_case_1} and \eqref{Paper2-mixed_case_2}. This allows us to identify certain admissible lower-spin chiral theories that would otherwise have been missed if one focused only on the singlet and colour cases.

\renewcommand{\thesection}{\thechapter.\arabic{section}}
\setcounter{section}{0}

\newpage
\clearpage

\chapter[Massless spinning fields on the
Light-Front]{Massless spinning fields on the Light-Front (quartic vertices and amplitudes)}

\vspace{5mm}

\paragraph{Abstract:} Within\footnote{The content of this chapter is identical to the content of the paper.} the light-front approach in flat space, we study the closure of the Poincaré algebra at the quartic order, specifically the non-holomorphic constraint involving both MHV and anti-MHV vertices. We first recover some well-established results: the existence of Yang-Mills theory and gravity, as well as the inconsistency of interacting multi-graviton theories. We explicitly construct several lower-derivative and lower-spin quartic vertices. We then turn to theories involving massless higher-spin fields. It becomes evident that the quartic constraint does not allow many cubic interactions to survive, in accordance with the well-known no-go results. Nevertheless, once higher-derivative cubic vertices are included, we find nontrivial solutions to the full quartic constraint and determine the corresponding quartic vertices. On this basis, we conjecture the complete set of quartic vertices that solve the light-cone consistency conditions. Exploiting this, we find all allowed unitary local higher-spin theories and identify new families of local quasi-chiral higher-spin theories. We then determine all local higher-spin four-point amplitudes using the spinor-helicity formalism together with locality in the form of consistent factorization. We conclude with a short discussion on non-locality.

\clearpage

\section{Introduction}
\label{Paper3-section1}

The light-cone or light-front approach to dynamics is perhaps the most powerful tool to search for new theories within the (perturbative) field theory approach or for proving no-go theorems against particular types of theories. For example, string theory was first quantised in the light-cone gauge \cite{Goddard:1973qh}, and the finiteness of $\mathcal{N}=4$ super-Yang-Mills theory was established in the light-cone gauge \cite{Mandelstam:1982cb,Brink:1982wv}. The first cubic interactions of massless higher-spin fields were also constructed in the light-cone gauge \cite{Bengtsson:1983pd}. The complete classification of cubic interactions of massless higher-spin fields was obtained in the light-cone gauge \cite{Bengtsson:1983pg,Bengtsson:1983pd,Bengtsson:1986kh,Metsaev:1991nb,Metsaev:1991mt,Fradkin:1991iy,Metsaev:1993ap,Metsaev:2005ar,Metsaev:2007rn}, as well as the construction of the first example of a perturbatively local higher-spin theory \cite{Metsaev:1991mt,Metsaev:1991nb,Ponomarev:2016lrm}.\footnote{... with massless propagating fields; otherwise, there are plenty of ``topological'' theories in $3d$ \cite{Blencowe:1988gj,Bergshoeff:1989ns,Campoleoni:2010zq,Henneaux:2010xg,Grigoriev:2020lzu,Pope:1989vj,Fradkin:1989xt,Grigoriev:2019xmp} and the conformal higher-spin gravity \cite{Segal:2002gd,Tseytlin:2002gz,Bekaert:2010ky, Basile:2022nou}. The only nontopological theory with massless fields that has not been constructed via the light-cone approach is \cite{Sharapov:2024euk}, but it owes its existence to the covariant form \cite{Sharapov:2022faa,Sharapov:2022wpz,Sharapov:2022awp,Sharapov:2022nps,Sharapov:2023erv} of the chiral higher-spin gravity \cite{Metsaev:1991mt,Metsaev:1991nb,Ponomarev:2016lrm}, which was first found in the light-cone gauge. } It is beyond the scope of the paper, but the light-cone gauge also allows one to construct off-shell formulations of supersymmetric theories.

There are several advantages to the light-cone gauge. Firstly, it is a unitary gauge, and all redundant degrees of freedom associated with gauge symmetries are absent (not even present to begin with). Secondly, the light-cone approach operates with physical degrees of freedom only and, hence, allows one to find unambiguous results concerning the (non)existence of particular types of interactions. By contrast, within any covariant approach, some results can depend on the type of Lorentz covariant field chosen to contain the physical degrees of freedom, e.g. the dual graviton does not exhibit gravitational interactions \cite{Bekaert:2002uh}. Thirdly, within the perturbative field-theory approach, the aim is to construct the charges of the Poincaré algebra, including the Hamiltonian, which is the minimal requirement for obtaining a Poincaré-invariant S-matrix. The light-cone approach does precisely this and requires nothing more.

In $4d$, the structure of interactions exhibits additional features that are absent in higher dimensions. There are significantly more cubic interactions (for a given set of three spins) in the light-cone gauge \cite{Metsaev:1991mt,Metsaev:1991nb,Bengtsson:2014qza, Conde:2016izb} or in the spinor-helicity language \cite{Benincasa:2011pg,Benincasa:2007xk} than the standard description in terms of a totally-symmetric (Fronsdal) tensor field can give, see e.g. \cite{Boulanger:2006gr,Zinoviev:2008ck,Manvelyan:2010je}. The light-cone form itself exhibits several peculiar features. Firstly, the cubic vertices split into holomorphic (h or MHV) and anti-holomorphic (ah or anti-MHV), which is difficult to see in a covariant approach. Secondly, the closure of the Poincaré algebra at the quartic order has a remarkable structure: there are two sectors that contain the h--h and ah--ah vertices and receive no contribution from quartic generators (we call them holomorphic constraints), but these equations are already powerful enough to fix the spectrum and coupling constants \cite{Metsaev:1991mt,Metsaev:1991nb,Ponomarev:2017nrr,Serrani:2025owx}. In this way, upon setting the ah-vertices to zero, one obtains consistent chiral/self-dual theories.  

The present paper is divided into two parts, which can be read largely independently. The first part, up to Section \ref{Paper3-section6}, is devoted to the study of the non-holomorphic quartic constraint in the light-cone gauge. This completes the analysis of higher-spin interactions at quartic order within the light-cone framework. The second part, presented in Section \ref{Paper3-section7}, adopts a complementary perspective by studying four-point amplitudes in the spinor-helicity formalism that satisfy the factorisation constraints. To systematically explore all possible cases, we employ an efficient version of the factorisation procedure. The two approaches, based respectively on the light-cone formulation and on spinor-helicity methods, lead to fully consistent results. This agreement is expected, since it was shown in \cite{Ponomarev:2016cwi} that the light-cone deformation procedure can be reformulated entirely in terms of the spinor-helicity formalism at all orders in the interactions.\footnote{For earlier works exploring the relation between the spinor-helicity formalism and the light-cone formulation, see \cite{Ananth:2012un,Bengtsson:2016alt,Bengtsson:2016hss}.} While the light-cone approach provides explicit off-shell expressions for the local quartic vertices, the spinor-helicity approach allows us to compute all local higher-spin four-point amplitudes explicitly.

The first part of the paper is a systematic study of the quartic light-cone non-holomorphic constraint, i.e. of type h--ah. In a previous paper, we derived a general solution to the (anti)-holomorphic constraints \cite{Serrani:2025owx} and explicitly classified theories with gauge and gravitational interactions (one- and two-derivative, respectively). However, parity and unitarity require both h- and ah-vertices to be present at the same time and to have equal couplings. Therefore, the non-holomorphic constraint becomes essential, and this is also where most of the no-go theorems hide. The first important steps were already taken in \cite{Metsaev:1991nb}. 

We will warm up with Yang-Mills theory and gravity and proceed to numerous cases involving higher-spin fields. We strengthen some no-go results and find some yes-go options. We also consider quasi-chiral cases, a term coined in \cite{Adamo:2022lah}, where we allow for both h- and ah-vertices, but in an asymmetric fashion. 

In the second part, we use the on-shell spinor-helicity formalism for massless particles to determine all local higher-spin four-point amplitudes consistent with factorisation and compute them explicitly. Similar ideas have been explored in several previous works. The pioneering study of higher-spin amplitudes from the perspective of factorisation is \cite{Benincasa:2007xk} (see also \cite{Schuster:2008nh,Fotopoulos:2010ay,Dempster:2012vw}),\footnote{For related results in the covariant approach, see \cite{Taronna:2017wbx,Roiban:2017iqg}.} which initiated a systematic search for constructible theories using the BCFW recursion relations \cite{Britto:2004ap,Britto:2005fq} through a four-particle test starting from cubic amplitudes. However, the BCFW construction requires amplitudes to exhibit sufficiently good behaviour under large complex deformations of the external momenta. In particular, it requires the amplitude to vanish in the limit $z\to \infty$, where $z$ is the complex deformation parameter, so that the contribution from the contour integral at infinity can be neglected. While this condition is satisfied in Yang--Mills theory and gravity, it generally fails for higher-spin amplitudes, making the BCFW approach less suitable for studying higher-spin interactions.

In subsequent works \cite{Benincasa:2011kn,Benincasa:2011pg}, a generalised notion of constructibility was introduced, allowing for amplitudes that do not vanish in the limit $z\to\infty$ and, consequently, for the presence of boundary contributions. These contributions are directly related to self-consistent quartic vertices, which correspond to solutions of the homogeneous quartic constraint in the light-cone formulation. The resulting generalised four-particle test can therefore be consistently applied to higher-spin amplitudes. These works provided important partial results, including the identification of several local four-point amplitudes constructed from cubic higher-spin amplitudes. However, a complete and systematic classification of local higher-spin four-point amplitudes was still lacking.

Another very interesting work addressing higher-spin four-point amplitudes consistent with factorisation is \cite{McGady:2013sga}. The philosophy adopted there is closer to the one pursued in the present work, as it does not rely on a BCFW complex deformation but instead derives the constraints directly from four-point factorisation.

The main difference is that the analysis of \cite{McGady:2013sga} assumes the four-point amplitude to be constructed from a given three-point amplitude $\mA_3$ together with its parity-conjugate amplitude $\bar{\mA}_3$, then assuming unitarity. In contrast, we do not make this assumption and derive the factorisation constraints in full generality. Their analysis therefore corresponds to the unitary case that will be discussed in Section \ref{Paper3-section6}. A second difference is that we do not just derive the conditions for the existence of local four-point amplitudes, but determine their explicit form in all cases. Our motivation is to classify all local four-point amplitudes, rather than only those arising in local unitary theories. Indeed, assuming parity invariance would, for example, exclude quasi-chiral higher-spin theories, which are instead local higher-spin theories at the quartic order and may be consistent to all orders. 

Lastly, we will also observe the emergence of color-kinematics duality and the double-copy structure at the level of non-holomorphic higher-spin four-point amplitudes. For similar results in the chiral sector, see \cite{Ponomarev:2017nrr,Ponomarev:2024jyg}.

\subsection*{Summary of the main results}
We summarise here the main results of the paper. 
The first results are presented in Section \ref{Paper3-section5}, where we study the quartic light-cone constraint in detail for the cubic vertices of Yang–Mills theory and gravity. We find that self-dual Yang–Mills theory is described by a complex Lie algebra with a structure constant $\fA^c_{[ab]}$; while self-dual gravity is described by a complex commutative and associative algebra with a structure constant $g^c_{(ab)}$. Moreover, upon assuming CPT symmetry or unitarity, we show that the full Yang–Mills theory is governed by a Lie algebra with imaginary and fully antisymmetric structure constants $\fA_{[abc]}$. Gravity, instead, is governed by a commutative, symmetric, and associative algebra with a real structure constant $g_{(abc)}$. We also derive the explicit form of the local quartic vertices for Yang–Mills theory, given in \eqref{Paper3-h4_1_YM} and \eqref{Paper3-h4_2_YM}, and for gravity in \eqref{Paper3-GR}. We verified the results by matching the four-point amplitudes obtained by summing the exchange diagrams and the quartic vertex. An analogous analysis can be readily extended to other classes of lower-spin vertices, as well as to vertices involving higher-spin fields.

In Section \ref{Paper3-section6}, all local quartic vertices, including higher-spin fields, that solve the quartic light-cone constraint are determined. Given two cubic vertices, one holomorphic $C^{\lambda_1,\lambda_2,\omega}$ ($\lambda_{12}+\omega>0$) and the other anti-holomorphic $\bar{C}^{-\omega,\lambda_3,\lambda_4}$ ($\lambda_{34}-\omega<0$), forming the exchange as shown in Figure \ref{Paper3-fig1}, a local quartic vertex $C^{\lambda_1,\lambda_2,\lambda_3,\lambda_4}$ exists if the following sets of inequalities are satisfied:\footnote{We adopt the notation $\lambda_{ij}=\lambda_i+\lambda_j$ and $\lambda_{ijk}=\lambda_i+\lambda_j+\lambda_k$.}
\begin{equation}\label{Paper3-conditionIntro}
\begin{aligned}
&\lambda_1 \leq \lambda_2+\omega+k-1\,,\qquad &
&\lambda_2 \leq \lambda_1+\omega+k-1\,,\qquad &
&\lambda_3 \leq \lambda_4+\omega+k-1\,,\\
&\lambda_4 \leq \lambda_3+\omega+k-1\,,\qquad &
&\lambda_{12} \geq \lambda_{34}\,,\qquad &
&k=0,1,2,3\,.
\end{aligned}
\end{equation}
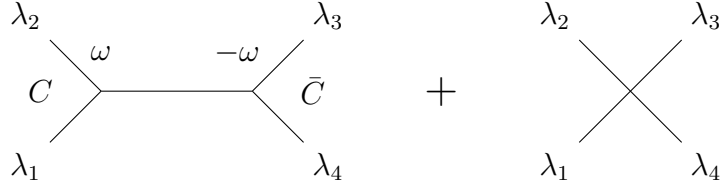
\begin{figure}[H]
    \centering
    \begin{tikzpicture}
        \begin{feynman}
            \vertex (i1) at (-6, 1) {\(\lambda_2\)};
            \vertex (i2) at (-6,-1) {\(\lambda_1\)};
            \vertex (i3) at (-2, 1) {\(\lambda_3\)};
            \vertex (i4) at (-2,-1) {\(\lambda_4\)};

            \vertex (ii1) at (1, 1) {\(\lambda_2\)};
            \vertex (ii2) at (1,-1) {\(\lambda_1\)};
            \vertex (ii3) at (3, 1) {\(\lambda_3\)};
            \vertex (ii4) at (3,-1) {\(\lambda_4\)};

            \vertex (v1) at (-5, 0);
            \vertex (v3) at (-3, 0);

            \node at (-5.8, 0) {\(C\)};
            \node at (-2.2, 0) {\(\bar{C}\)};
            \node at (-0.5,0) {\Large $+$};
            
            \vertex at (-5, 0.5) {\(\omega\)};
            \vertex at (-3.2, 0.5) {\(-\omega\)};
            
            \diagram* {
                (i1) -- (v1),
                (i2) -- (v1),
                (v1) -- [plain] (v3),
                (v3) -- (i3),
                (v3) -- (i4),
            };
            \diagram* {
                (ii1) -- (ii4),
                (ii2) -- (ii3),
            };

        \end{feynman}
    \end{tikzpicture}
    \caption{Generic $C\bar{C}$ exchange with contact term $C^{\lambda_1,\lambda_2,\lambda_3,\lambda_4}$.}
    \label{Paper3-fig1}
\end{figure}
\noindent
When \eqref{Paper3-conditionIntro} is satisfied only for $k=1,2,3$,\footnote{By ``satisfied only for'' we mean that at least one of the inequalities appearing in \eqref{Paper3-conditionIntro} is saturated, i.e. it holds as an equality.} additional exchange diagrams are required, as discussed in the main text. Accordingly, local quartic vertices fall into four classes: self-consistent quartic vertices ($k=0$);\footnote{In this case, $\omega$ no longer represents an exchange field; instead, it parametrizes the number of derivatives carried by the quartic vertex, $D=\lambda_{12}-\lambda_{34}+2\omega-2$.} vertices requiring a single-channel exchange ($k=1$); vertices requiring exchange in two channels ($k=2$), as in Yang–Mills theory; and vertices requiring $s$-, $t$-, and $u$-channel exchanges ($k=3$), as in gravity.

We used these results to determine all local unitary theories up to quartic order. We found the following possibilities:
\begin{itemize}
    \item Theories consisting exclusively of abelian couplings, both holomorphic (\(C\)) and anti-holomorphic (\(\bar{C}\)), that satisfy the triangular inequalities. These form unitary theories that can involve higher-spin fields.

    \item Theories that contain at least a parity-related pair of lower-spin non-abelian couplings. In this case, only additional lower-spin couplings are allowed, while all higher-spin couplings, whether abelian or non-abelian, are ruled out.
\end{itemize}
We further identify a class of local ``quasi-chiral'' higher-spin theories that extend both self-dual Yang–Mills theory and self-dual gravity. These theories allow for both holomorphic and anti-holomorphic cubic couplings and are therefore distinct from the chiral HS-SDYM and HS-SDGR theories \cite{Ponomarev:2017nrr, Krasnov:2021nsq}, while remaining non-parity-invariant and non-unitary.

In Section~\ref{Paper3-section7}, we determine all local four-point higher-spin amplitudes in flat space by exploiting spinor-helicity properties such as little-group scaling and the locality of the amplitudes. We also explicitly check the results in several cases by summing the exchange diagrams together with the local quartic contact vertex. When the conditions \eqref{Paper3-conditionIntro} are satisfied for $k=0$, self-consistent local quartic vertices are allowed. These correspond to solutions of the homogeneous quartic constraint and lead to the following amplitudes
\begin{equation}
    \mA^{(D)}_{\text{homo}}(1_{\lambda_1}2_{\lambda_2}3_{\lambda_3}4_{\lambda_4})=\sum_{i=0}^{\frac{D-d}{2}}\left(c_i\,s^it^{\frac{D-d}{2}-i}\right)[12]^{2\lambda_2}\langle 34\rangle^{\lambda_1-\lambda_{234}}[13]^{\lambda_{13}-\lambda_{24}}[14]^{\lambda_{14}-\lambda_{23}}\,,
\end{equation}
where $D$ denotes the total number of derivatives carried by the quartic vertex, or equivalently the mass dimension of the amplitude, while $d=3\lambda_1-\lambda_{234}$ is the total sum of the powers of spinor brackets, and $c_i$ are $\frac{D-d}{2}+1$ free coefficients. In particular, we need to associate the helicities $\lambda_1,\lambda_2,\lambda_3,\lambda_4$ with the fields $1,2,3,4$ in such a way as to maximise $d$, namely $d=\max\limits_{i\neq j\neq k\neq \ell}\{3\lambda_i-\lambda_{jk\ell},-3\lambda_i+\lambda_{jk\ell}\}$, where $i,j,k,\ell=1,2,3,4$. We can also exchange $\langle\;\rangle$ and $[\;]$ when necessary.

When the conditions \eqref{Paper3-conditionIntro} are satisfied for $k=1$, amplitudes that include single-channel exchange diagrams are allowed and are given by
\begin{subequations}
\begin{align}
    \mA^{(D)}_{s}(1_{\lambda_1}2_{\lambda_2}3_{\lambda_3}4_{\lambda_4})&=k_s\frac{t^{\frac{D-d+2}{2}}}{s}[12]^{2\lambda_2}\langle 34\rangle^{\lambda_1-\lambda_{234}}[13]^{\lambda_{13}-\lambda_{24}}[14]^{\lambda_{14}-\lambda_{23}}+\mA^{(D)}_{\text{homo}}\,,\\
    \mA^{(D)}_{t}(1_{\lambda_1}2_{\lambda_2}3_{\lambda_3}4_{\lambda_4})&=k_t\frac{u^{\frac{D-d+2}{2}}}{t}[12]^{2\lambda_2}\langle 34\rangle^{\lambda_1-\lambda_{234}}[13]^{\lambda_{13}-\lambda_{24}}[14]^{\lambda_{14}-\lambda_{23}}+\mA^{(D)}_{\text{homo}}\,,\\
    \mA^{(D)}_{u}(1_{\lambda_1}2_{\lambda_2}3_{\lambda_3}4_{\lambda_4})&=k_u\frac{s^{\frac{D-d+2}{2}}}{u}[12]^{2\lambda_2}\langle 34\rangle^{\lambda_1-\lambda_{234}}[13]^{\lambda_{13}-\lambda_{24}}[14]^{\lambda_{14}-\lambda_{23}}+\mA^{(D)}_{\text{homo}}\,,
\end{align}
\end{subequations}
where $k_{\bullet}\sim C\bar{C}$ is the product of cubic couplings in the $s$-, $t$-, and $u$-channels. In particular, when $D=d-2$, we have a unique solution corresponding to
\begin{subequations}
\begin{align}
    \mA^{(d-2)}_{s}(1_{\lambda_1}2_{\lambda_2}3_{\lambda_3}4_{\lambda_4})&=\frac{k_s}{s}[12]^{2\lambda_2}\langle 34\rangle^{\lambda_1-\lambda_{234}}[13]^{\lambda_{13}-\lambda_{24}}[14]^{\lambda_{14}-\lambda_{23}}\,,\\
    \mA^{(d-2)}_{t}(1_{\lambda_1}2_{\lambda_2}3_{\lambda_3}4_{\lambda_4})&=\frac{k_t}{t}[12]^{2\lambda_2}\langle 34\rangle^{\lambda_1-\lambda_{234}}[13]^{\lambda_{13}-\lambda_{24}}[14]^{\lambda_{14}-\lambda_{23}}\,,\\
    \mA^{(d-2)}_{u}(1_{\lambda_1}2_{\lambda_2}3_{\lambda_3}4_{\lambda_4})&=\frac{k_u}{u}[12]^{2\lambda_2}\langle 34\rangle^{\lambda_1-\lambda_{234}}[13]^{\lambda_{13}-\lambda_{24}}[14]^{\lambda_{14}-\lambda_{23}}\,.
\end{align}
\end{subequations}
When the conditions \eqref{Paper3-conditionIntro} are satisfied only for $k=2$, and we have $D=d-4$, Yang-Mills-like (YM-like) amplitudes are allowed and are given by
\begin{subequations}
\begin{align}
    \mA^{(d-4)}_{st}(1_{\lambda_1}2_{\lambda_2}3_{\lambda_3}4_{\lambda_4})&=\frac{k_{st}}{st}[12]^{2\lambda_2}\langle 34\rangle^{\lambda_1-\lambda_{234}}[13]^{\lambda_{13}-\lambda_{24}}[14]^{\lambda_{14}-\lambda_{23}}\,,\\
    \mA^{(d-4)}_{us}(1_{\lambda_1}2_{\lambda_2}3_{\lambda_3}4_{\lambda_4})&=\frac{k_{us}}{us}[12]^{2\lambda_2}\langle 34\rangle^{\lambda_1-\lambda_{234}}[13]^{\lambda_{13}-\lambda_{24}}[14]^{\lambda_{14}-\lambda_{23}}\,,\\
    \mA^{(d-4)}_{tu}(1_{\lambda_1}2_{\lambda_2}3_{\lambda_3}4_{\lambda_4})&=\frac{k_{tu}}{tu}[12]^{2\lambda_2}\langle 34\rangle^{\lambda_1-\lambda_{234}}[13]^{\lambda_{13}-\lambda_{24}}[14]^{\lambda_{14}-\lambda_{23}}\,.
\end{align}
\end{subequations}
These are not independent since $\mA^{(d-4)}_{st}=\frac{u}{t}\mA^{(d-4)}_{us}=\frac{u}{s}\mA^{(d-4)}_{tu}$. As we will show in the main text, this reflects the existence of a Jacobi identity, which implies the existence of two color-ordered amplitudes and the BCJ amplitude relations for non-holomorphic higher-spin amplitudes. When the conditions \eqref{Paper3-conditionIntro} are satisfied only for $k=3$, and we have $D=d-6$, gravity-like (GR-like) amplitudes are allowed and are given by
\begin{align}
    &\mA^{(d-6)}_{stu}(1_{\lambda_1}2_{\lambda_2}3_{\lambda_3}4_{\lambda_4})=\frac{k_{stu}}{stu}[12]^{2\lambda_2}\langle 34\rangle^{\lambda_1-\lambda_{234}}[13]^{\lambda_{13}-\lambda_{24}}[14]^{\lambda_{14}-\lambda_{23}}\,,
\end{align}
with $k_{stu}=k_s=k_t=k_u$. In the case of gravity, this reproduces the MHV four-point graviton amplitude. As we will show in the main text, GR-like amplitudes can be obtained through a ``double copy'' construction from YM-like ones.

\subsection*{Outline of the paper}

The outline of the paper is as follows:
In Section \ref{Paper3-section2}, we review various no-go theorems to appreciate the no-/yes-go results in the paper.
In Section \ref{Paper3-section3}, we briefly recall the light-front approach to massless higher-spin interactions through the study of the light-cone constraints.
In Section \ref{Paper3-section4}, we focus on the quartic constraint, highlighting some of its features and explaining our strategy for studying it. We describe the two main methods employed to study the system of PDEs arising from it.
In Section \ref{Paper3-section5}, we solve the quartic constraint for lower-derivative quartic vertices, recovering known results such as the existence of Yang–Mills and gravity vertices. We compute the four-point amplitudes by summing exchange contributions and quartic contact terms.
In Section \ref{Paper3-section6}, we extend the analysis to higher-derivative quartic vertices. We conjecture the complete set of quartic vertices that solve the light-cone quartic constraints. We use them to find all unitary local higher-spin theories in flat space and identify new families of local quasi-chiral higher-spin theories. In Section \ref{Paper3-section7}, we find all local higher-spin four-point amplitudes.
In Section \ref{Paper3-section8}, we discuss the possibility of quartic vertices with exchange-type non-localities.
Finally, in Section \ref{Paper3-section8}, we summarise our conclusions and comment on possible future directions.

We also include six appendices. 
In Appendix \ref{Paper3-AppendixA}, we explain how unitarity and parity are implemented in the light-cone formalism and, in particular, how they act on the quartic densities. 
In Appendix \ref{Paper3-AppendixB}, we collect some useful formulas and relations.  
In Appendix \ref{Paper3-AppendixC}, we comment on the quartic constraint in the presence of a cubic scalar vertex. 
Appendix \ref{Paper3-AppendixD} collects some explicit solutions for quartic vertices.
Appendix \ref{Paper3-AppendixE} contains a review of the on-shell relation between the light-cone and spinor-helicity formalisms. 
Appendix \ref{Paper3-AppendixF} collects some of the tables we used to conjecture the complete set of quartic vertices that solve the light-cone quartic constraints.

\section{No-go's}\label{Paper3-section2}

To better understand the role of the no-go and, more importantly, the yes-go results in this paper and \cite{Metsaev:1991mt,Metsaev:1991nb,Ponomarev:2016lrm,Ponomarev:2017nrr,Serrani:2025owx}, it is necessary to review the most well-known no-go-type arguments against nontrivial interactions of massless higher-spin fields. 

Various no-go theorems or other constraints implied by higher spin symmetry can be grouped into several categories: (A) global vs. (B) local; flat space (I) vs. de Sitter or anti-de Sitter (II). In the latter category, one may also add results for general gravitational background (III), or backgrounds that are less restrictive than (I-II), e.g. pp-wave or self-dual. Under (A), we group the results that constrain observables at `infinity', such as the S-matrix or the holographic S-matrix. Under (B), we collect results that impose restrictions on interactions within the perturbative approach to field theory, which is usually understood as some form of the Noether procedure.

Some important distinctions between (A) and (B) include: (1) simplicity or even triviality of the asymptotic observables from (A) does not necessarily imply that (B) is empty, e.g. interactions can conspire as to produce a simple (or trivial) S-matrix \cite{Ponomarev:2016lrm,Ponomarev:2017nrr,Skvortsov:2018jea,Skvortsov:2020wtf,Skvortsov:2020gpn}; (2) any results of (A)-type do not imply that there is a local quantum field theory that can deliver them. In fact, in some cases, it is quite clear that a given $S$-matrix that is a solution of the higher-spin symmetry constraints cannot result from a field theory. Lastly, it is never too late to rethink some of the explicit and hidden assumptions underlying the no-go results. We do not attempt to do this below and, instead, just state what every result means in ``plain terms'', i.e. what kind of conclusions are usually drawn.

\paragraph{A-category: global constraints of higher spin symmetry.} Global constraints can be derived by requiring $S$-matrix type observables to be invariant under higher spin transformations. Depending on the situation, these can be just constraints to ensure the masslessness of external states or an assumption of extended global symmetries.

{\bf AI.} In flat space, the canonical observable is the $S$-matrix and free massless fields (which are the only ones relevant for the asymptotic states) can be described, for instance, by (gauge-fixed) Fronsdal tensor $\Phi_{a_1...a_s} $ \cite{Fronsdal:1978rb} with gauge symmetry of the form
\begin{align}
    \delta \Phi_{a_1...a_s} &= \pl_{(a_1}\xi_{a_2...a_s)}\,.
\end{align}
The most powerful result in this category is Weinberg's low-energy theorem \cite{Weinberg:1964ew}. Roughly speaking, it implies that the S-matrix for any theory with massless higher-spin fields has to be the trivial one, $S=1$. There are at least two important remarks. Firstly, the conclusion is true once the higher-spin interactions survive in the low energy limit, which corresponds to the most interesting nonabelian interactions. Such interaction vertices are not trivially gauge-invariant and are the ones to produce the low-energy constraints. If one takes the abelian interactions only, no interesting constraints arise. Secondly, Weinberg's theorem is about the ``number of derivatives in the nonabelian interactions''. For one- and two-derivative interactions (gauge and gravitational ones), one gets the charge conservation and the universality of the gravitational coupling. For higher-derivative interactions, the constraints are just too strong. However, the number of derivatives depends on how the physical degrees of freedom are embedded into a covariant Lorentz tensor. In the Fronsdal formulation ``gauge'' and ``gravitational'' interactions turn out to have a higher-derivative form. By contrast, in the chiral formulation \cite{Krasnov:2021nsq}, which follows from twistor theory \cite{Penrose:1965am,Hughston:1979tq,Eastwood:1981jy,Woodhouse:1985id}, the gauge and gravitational interactions have the usual number of derivatives. This allows for an interesting loophole \cite{Tran:2022amg}. 

Another powerful constraint is the Coleman-Mandula theorem \cite{Coleman:1967ad} that implies that $S$-matrix cannot have ``extended symmetries'', where the latter means the existence of some charges $Q_{a_1...a_{s-1}}$ that transform in a higher-spin representation of the Lorentz group, i.e. with $s>2$.  

An obvious way to avoid all of the above arguments is to have a theory with nontrivial interactions that conspire to give a trivial S-matrix, as it happens in chiral theory \cite{Ponomarev:2016lrm,Ponomarev:2017nrr,Skvortsov:2018jea,Skvortsov:2020wtf,Skvortsov:2020gpn,Krasnov:2021nsq} and, more generally, in all self-dual theories. The nontriviality of interactions can still reveal itself in more nontrivial backgrounds.

{\bf AII.} In the case of anti-de Sitter space, the requirement of masslessness of external states is indistinguishable from the requirement to have extended symmetries, realized by some higher spin charges. Massless higher-spin fields are AdS/CFT dual to conserved higher-spin currents
\begin{align}
    \delta \Phi_{a_1...a_s} &= \nabla_{(a_1}\xi_{a_2...a_s)} && \longleftrightarrow && \pl^k J_{k n_2...n_s}=0\,.
\end{align}
There is an extension of the Weinberg-Coleman theorems to the AdS/CFT setting: all CFTs in $d\geq3$ that have at least one higher-spin current are free CFTs (possibly, in disguise) that have, in fact, infinitely many higher-spin currents \cite{Maldacena:2011jn,Boulanger:2013zza,Alba:2013yda,Alba:2015upa}. Therefore, the AdS/CFT analogue of $S=1$ is $S=$ free CFT on the boundary. 

An interesting deviation from this ``trivial'' S-matrix is that one can consider different boundary conditions and, eventually, a higher-spin gravity can be dual to something interesting, e.g. to the critical vector model \cite{Sezgin:2002rt,Klebanov:2002ja,Sezgin:2003pt,Leigh:2003gk} and, more generally, to Chern-Simons matter theories, see e.g. \cite{Giombi:2011kc}. A recent development \cite{Skvortsov:2018uru,Sharapov:2022awp,Jain:2024bza,Aharony:2024nqs} is the duality between chiral higher-spin gravity and a subsector of Chern-Simons matter theories.

\paragraph{B-category: local constraints of higher spin symmetry.} What falls into this category are various attempts to construct higher-spin theories as (reasonably) local field theories. Here, the main players are the Noether procedure, the light-front approach, and various AdS/CFT-related ideas. 

{\bf BI. Flat. } The Noether procedure is a way to search for theories by fixing some free field content and then deforming the action and gauge symmetries with higher-order terms while using gauge invariance and locality to constrain interactions. 
\begin{align}
    \delta \Phi_{a_1...a_s} &= \pl_{(a_1}\xi_{a_2...a_s)} +\mathcal{O}(\Phi\xi)\,.
\end{align}
Gauge symmetry is important in maintaining the correct number of degrees of freedom. While some cubic interactions are available \cite{Bengtsson:1986kh,Metsaev:1991mt,Metsaev:1991nb,Berends:1984rq,Boulanger:2006gr,Zinoviev:2008ck,Manvelyan:2010je}, there is an obstruction at the quartic order \cite{Bekaert:2010hp,Roiban:2017iqg,Ponomarev:2017nrr}. The obstruction is that one cannot find a local quartic vertex that is compatible with any ``interesting'' cubic interaction (note that abelian cubic interactions are consistent, but they do not induce any deformation of gauge symmetries and do not contain interactions such as gravitational and gauge). One might think of relaxing the standard definition of locality, but it is not clear how (if one completely abandons locality, then anything can be made into a theory \cite{Barnich:1993vg}). Nevertheless, the best one can do at present is to construct a chiral/self-dual subsector of a hypothetical theory that has some degree of nonlocality. 

{\bf BII. (A)dS. } Not surprisingly, the same problems persist in $(A)dS$. Thanks to the AdS/CFT duality, one can get a better handle on the problem since one can just take a free CFT on the boundary and try to ``reconstruct'' the interactions in $AdS$ that would give the same correlation functions back. It happens that such interactions are too nonlocal \cite{Bekaert:2015tva,Sleight:2017pcz,Ponomarev:2017qab}: the degree of nonlocality is the same as that of an exchange diagram, i.e. one cannot separate classical from quantum, roughly speaking.   

Note that the initial excitement about massless higher-spin fields in AdS was based on \cite{Vasiliev:1986bq} where ``gravitational interactions'' of higher-spin fields were constructed in $AdS_4$, while no such result was available in flat space in terms of Fronsdal fields. However, the complete classification of the cubic interactions in flat space \cite{Bengtsson:1986kh} has both gauge and gravitational interactions on the list, but in the light-cone gauge. Therefore, \cite{Vasiliev:1986bq} revealed that, while gauge and gravitational interactions cannot be constructed in terms of Fronsdal fields in flat space, one can do so in $AdS_4$, i.e. it revealed some singularity of the Fronsdal description. Objectively, there is a one-to-one correspondence between cubic interactions in flat space and $AdS_4$, \cite{Metsaev:2018xip,Nagaraj:2018nxq}.

{\bf BIII. More general backgrounds. } It is easy to show that the Fronsdal equations, with partial derivatives replaced by covariant ones, cease to be gauge invariant on general gravitational backgrounds \cite{Aragone:1979hx}. Adding various curvature-dependent (non-minimal) terms does not improve the situation. The situation does not improve either if one restricts to Einstein backgrounds. The reason is simple: in checking gauge invariance, one has to commute covariant derivatives, and the commutator brings the four-index Riemann tensor, $[\nabla,\nabla] \xi\sim R\,\xi$. Its traceless component, the Weyl tensor, survives on Einstein backgrounds and destroys the gauge invariance. However, in view of the discussion above, one should phrase this result as the impossibility of putting Fronsdal fields on backgrounds more general than maximally symmetric space-times.

It has been known since \cite{Hughston:1979tq,Aragone:1979hx,Woodhouse:1985id}, see also \cite{Krasnov:2021nsq}, that higher-spin fields can propagate on self-dual backgrounds; see e.g.  \cite{Skvortsov:2025ohi}. However, to achieve this, one has to choose an appropriate Lorentz covariant field description, which originates from twistor theory, not the Fronsdal fields. It is also known that higher-spin fields can propagate on pp-wave backgrounds \cite{Metsaev:1997ut}, see also \cite{Tran:2025yzd}.

\section{Living on the Light-Front}\label{Paper3-section3}

The light-front (or light-cone) approach to field dynamics dates back to Dirac \cite{Dirac:1949cp} and has been instrumental in making the first important steps in various directions. We refer to \cite{Ponomarev:2022vjb} for a pedagogical exposition of the subject. The present paper is based on \cite{Bengtsson:1986kh}, where the complete classification of cubic interactions of massless spinning fields was obtained, and on \cite{Metsaev:1991mt, Metsaev:1991nb, Ponomarev:2016lrm}, where the first quartic analysis of the higher spin problem was conducted. 

The main idea of the approach is to construct the charges of the Poincaré algebra, $P^A$ and $J^{AB}$, directly in terms of physical degrees of freedom (the light-cone gauge is unitary):\footnote{We choose the mostly plus convention for $\eta_{AB}$ and $A,B,...=0,...,d-1$. In fact, $d=4$. The light-front coordinates are $x^\pm=(x^3\pm x^0)/\sqrt{2}$, so that $\eta^{+-}=\eta^{-+}=1$. In $4d$, we replace $x^{1,2}$ with two complex conjugate variables $z$ and $\bar{z}$, so that $x^Ax^B \eta_{AB}=2x^+x^-+2z\bar{z}$. }
\begin{align}
[P^A,P^B]&=0\,,\\
[J^{AB},P^C]&=P^A\eta^{BC}-P^B\eta^{AC}\,,\\
[J^{AB},J^{CD}]&=J^{AD}\eta^{BC}-J^{BD}\eta^{AC}-J^{AC}\eta^{BD}+J^{BC}\eta^{AD}\,.
\end{align}
It is also convenient to choose $x^+$ as the light-cone time.\footnote{This choice may not be free of troubles, in principle, since $x^\pm$ lies along the characteristic surfaces. Various subtleties are discussed, e.g. in \cite{Neville:1971zk,Heinzl:2000ht,Heinzl:1993px,Barnich:2024aln}. As far as we can see, none of them matters for the present paper.} Most of the generators turn out to be kinematical, i.e. they do not receive any corrections due to interactions (or receive simple ones, e.g. $J^{+-}$, $J^{+a}$) and stay as in the free theory. The choice of the light-cone time minimises the number of generators that need to be deformed. It is well-known that the only relations to worry about are\footnote{The Latin indices $a,b,...=z,\bar{z}$ are reserved for the transverse directions.}
\begin{align}
    [J^{a-},H]&=0\,, &[J^{a-},J^{b-}]&=0\,,
\end{align}
where $H=P^-$ is the light-cone Hamiltonian. It can further be shown that, at least classically, the second relation is a consequence of the first one, see \cite{Ponomarev:2016lrm} for more details. 

One of the crucial simplifications of the light-cone gauge\footnote{It is worth mentioning that we do not have to impose any gauge. It is true that, given one or another covariant description, we can go into the light-cone gauge and get what is in the text. However, there is no general statement that allows one to uplift any light-cone result to a given covariant description, e.g. to express the result as a local interaction in terms of Fronsdal fields. The existence of the $S$-matrix relies on $J^{AB}$, $P^A$ rather than on having Lorentz-covariant fields with such indices.} in $4d$ is that a massless spin-$s$ field is represented by a complex scalar field $\phi(x)=\phi^{+ \lambda}(x)$ and its conjugate $\bar\phi(x)=\phi^{-\lambda}(x)$ that are helicity eigenstates. In Lorentzian signature, the one adopted here, they are complex conjugates of each other $\phi^{-\lambda}=(\phi^{+\lambda})^*$. The free action is $S=\tfrac{1}{2}\int d^4x\, \phi^{-\lambda}\Box\phi^{\lambda}$ and is real in any signature. 

It is convenient to work with Fourier-transformed fields
\begin{align}
    \phi(p,x^+)&=(2\pi)^{-\tfrac{d-1}2}  \int e^{-i(x^-p^++p\cdot x)} \phi(x,x^+)\, d^{d-1}x\,.
\end{align}
One more ingredient is the Poisson bracket:\footnote{When seen as coming from the covariant Fronsdal free theory as a result of imposing the light-cone gauge, this turns out to be a Dirac bracket.}
\begin{align}\label{Paper3-equaltime}
    [\phi^{\mu}(p,x^+),\phi^{\lambda}(q,x^+)]&=\delta^{\mu,-\lambda}\frac{\delta^{3}(p+q)}{2p^+}\,.
\end{align}
Once generators $J^{AB}$ and $P^A$ satisfy the Poincaré algebra relations at $x^+=0$, they do so at any $x^+$. Therefore, another important simplification is to set $x^+=0$, which explains why we omit it from all formulas below. The free field realisation reads
\begin{align}\label{Paper3-FreeFieldReal}
    h_2&=-\frac{p\pb}{\beta}\,, &&
    \begin{aligned}
        j^{z-}_2&= \pfrac{\pb} \frac{ p\pb}{\beta} +p \pfrac{\beta} +\lambda\frac{p}{\beta}\,,\\
         j^{\zb-}_2&= \pfrac{p} \frac{ p\pb}{\beta} +\pb \pfrac{\beta} -\lambda\frac{\pb}{\beta}  \,.     
    \end{aligned}
\end{align}
The corresponding Poincaré charges are
\begin{align}
Q_2&= \int p^+\, d^{3}p\,\phi^{-\mu}_{-p} O(p,\pl_p)\phi^{\mu}_p\,, && O=h_2\,, j^{z-}_2\,,j^{\zb-}_2\,.
\end{align}
To take into account interactions, $J^{a-}$ and $H$ are assumed to have local expansions in $\phi$. On general grounds, with some simple kinematical relations taken into account, we should have
{\allowdisplaybreaks\besubeqs\begin{align}
    H&=H_2+\sum_n\int d^{3n}q\,\deltas{\sum q_i} h_{\lambda_1,...,\lambda_n}^{q_1,...,q_n}\, \phi^{\lambda_1}_{q_1}...\,\phi^{\lambda_n}_{q_n}\,,\\
    J^{z-}&=J^{z-}_2+\sum_n\int d^{3n}q\, \deltas{\sum q_i}\left[ j_{\lambda_1,...,\lambda_n}^{q_1,...,q_n}-\frac{1}{n}h_{\lambda_1,...,\lambda_n}^{q_1,...,q_n}\left(\sum_j \pfrac{\bar{q}_j}\right)\right]\, \phi^{\lambda_1}_{q_1}...\,\phi^{\lambda_n}_{q_n}   \,,\\ 
    J^{\zb-}&=J^{\zb-}_2+\sum_n\int d^{3n}q\, \deltas{\sum q_i}\left[ \jb_{\lambda_1,...,\lambda_n}^{q_1,...,q_n}-\frac{1}{n} h_{\lambda_1,...,\lambda_n}^{q_1,...,q_n}\left(\sum_j \pfrac{q_j}\right)\right]\, \phi^{\lambda_1}_{q_1}...\,\phi^{\lambda_n}_{q_n}   \,,  
\end{align}\esubeqs}\noindent
where we have not made any assumptions regarding the spectrum of states, and all helicities are allowed. The kinematical generators constrain further the form of densities $h$, $j$ and $\jb$ to depend on 
\begin{align}
   \PP_{km}&=q_k\beta_m-q_m\beta_k\,, & \PPb_{km}&=\qb_k\beta_m-\qb_m\beta_k \,.
\end{align}
Here $k,m=1,...,N$ label fields $\phi^{\lambda_k}_{q_k}$ at order $N$. Due to momentum conservation, there are $N-2$ such independent variables among $\PP$ and the same for $\PPb$. Finally, the densities have to be
\begin{subequations}
\begin{align}
    h_{\lambda_1,...,\lambda_n}(q_1,...,q_n)&=h_{\lambda_1,...,\lambda_n}(\PP_{km},\PPb_{km},\beta_k)\,,\\
    j_{\lambda_1,...,\lambda_n}(q_1,...,q_n)&=j_{\lambda_1,...,\lambda_n}(\PP_{km},\PPb_{km},\beta_k)\,, \qquad \text{same for } \jb\,.
\end{align}
\end{subequations}
The remaining kinematical constraints (together with their origin indicated on the left) can be rewritten as
{\allowdisplaybreaks\besubeqs\label{Paper3-kinematics}\begin{align}
    J^{z\zb}&: &&\left[\NPP -\NPPb+\sum_k \lambda_k\right]h_{\lambda_1,...,\lambda_n}^{q_1,...,q_n}\sim0\,,\\
    J^{-+}&: && \left[\NPP +\NPPb+\sum_k \beta_k \pfrac{\beta_k}\right] h_{\lambda_1,...,\lambda_n}^{q_1,...,q_n}\sim0\,,\\
    J^{-+}&: && \left[\NPP +\NPPb+\sum_k \beta_k \pfrac{\beta_k}\right]j_{\lambda_1,...,\lambda_n}^{q_1,...,q_n}\sim0\,,\\
    J^{-+}&: && \left[\NPP +\NPPb+\sum_k \beta_k \pfrac{\beta_k}\right]\jb_{\lambda_1,...,\lambda_n}^{q_1,...,q_n}\sim0\,,\\
    J^{z\zb}&: &&\left[\NPP -\NPPb+\sum_k \lambda_k-1\right]j_{\lambda_1,...,\lambda_n}^{q_1,...,q_n}\sim0\,,\\
    J^{z\zb}&: &&\left[\NPP -\NPPb+\sum_k \lambda_k+1\right]\jb_{\lambda_1,...,\lambda_n}^{q_1,...,q_n}\sim0\,,
\end{align}\esubeqs}\noindent
where $\NPP=\sum \PP \pfrac{\PP}$ (sum over $\PP$'s that the densities depend on), idem. for $\NPPb$. 
\paragraph{Cubic vertices.} The complete basis of cubic interactions is given by \cite{Bengtsson:1986kh,Metsaev:1991mt,Metsaev:1991nb} and has a very neat form of (there are no negative powers of $\PP$, $\PPb$, as explained below)
\besubeqs\label{Paper3-famouscubic}\begin{align}
    h^{\lambda_i}_3\equiv h_{\lambda_1,\lambda_2,\lambda_3}&= C^{\lambda_1,\lambda_2,\lambda_3} \frac{\PPb^{\lambda_{123}}}{\beta_1^{\lambda_1}\beta_2^{\lambda_2}\beta_3^{\lambda_3}}+\bar{C}^{-\lambda_1,-\lambda_2,-\lambda_3} \frac{\PP^{-\lambda_{123}}}{\beta_1^{-\lambda_1}\beta_2^{-\lambda_2}\beta_3^{-\lambda_3}}\,,\\
    j^{\lambda_i}_3\equiv j_{\lambda_1,\lambda_2,\lambda_3}&=+\frac23 C^{\lambda_1,\lambda_2,\lambda_3} \frac{\PPb^{\lambda_{123}-1}}{\beta_1^{+\lambda_1}\beta_2^{+\lambda_2}\beta_3^{+\lambda_3}}\Lambda^{\lambda_1,\lambda_2,\lambda_3}\,,\\
    \jb^{\lambda_i}_3\equiv \jb_{\lambda_1,\lambda_2,\lambda_3}&=-\frac23\bar{C}^{-\lambda_1,-\lambda_2,-\lambda_3} \frac{\PP^{-\lambda_{123}-1}}{\beta_1^{-\lambda_1}\beta_2^{-\lambda_2}\beta_3^{-\lambda_3}}\Lambda^{\lambda_1,\lambda_2,\lambda_3}\,,
\end{align}\esubeqs
where
\begin{subequations}
\begin{align}
    &\lambda_{ijk\ell}=\lambda_i+\lambda_j+\lambda_k+\lambda_{\ell}\,,\qquad
    \lambda_{ijk}=\lambda_i+\lambda_j+\lambda_k\,,\qquad
    \lambda_{ij}=\lambda_i+\lambda_j\,,\\
    &\Lambda=\beta_1(\lambda_2-\lambda_3)+\beta_2(\lambda_3-\lambda_1)+\beta_3(\lambda_1-\lambda_2)\,.
\end{align}
\end{subequations}
There is just one independent $\PP$ at this order, $\PP_{12}=\PP_{23}=\PP_{31}$ (momentum conservation). It is convenient to define democratic variables, both for $\PP$ and $\PPb$,
\begin{align}
    \PP^a_{12}&=...=\PP^a=\frac13\left[ (\beta_1-\beta_2)q^a_3+(\beta_2-\beta_3)q^a_1+(\beta_3-\beta_1)q^a_2\right]\,.
\end{align}
The dynamical information about the theory is encoded in coupling constants $C^{\lambda_1,\lambda_2,\lambda_3}$ and $\bar{C}^{-\lambda_1,-\lambda_2,-\lambda_3}$. Note that $\lambda_1+\lambda_2+\lambda_3>0$ for $C^{\lambda_1,\lambda_2,\lambda_3}$ unless $\lambda_i=0$ and $\lambda_1+\lambda_2+\lambda_3<0$ for $\bar{C}^{-\lambda_1,-\lambda_2,-\lambda_3}$ unless $\lambda_i=0$. There is a unique $\lambda_i=0$ vertex, which corresponds to $(\phi_0)^3$. It is a matter of choice where to put this scalar cubic self-interaction, into $C$ or $\bar{C}$. To simplify notation, we assume that all coupling constants are packed in $\mathcal{C}^{\lambda_1,\lambda_2,\lambda_3}$ and it is equal to $C^{\lambda_1,\lambda_2,\lambda_3}$ for $\lambda_1+\lambda_2+\lambda_3>0$, to $\bar C^{\lambda_1,\lambda_2,\lambda_3}$ for $\lambda_1+\lambda_2+\lambda_3<0$ and $C^{0,0,0}$ is the coefficient of $(\phi_0)^3$. To have the right dimension, we should set 
\begin{align}
    C^{\lambda_1,\lambda_2,\lambda_3}&= (\ell_P)^{\lambda_{123}-1} c^{\lambda_1,\lambda_2,\lambda_3}\,,
\end{align}
where $c$ are dimensionless and $\ell_P$ is some constant of length dimension. Any choice of $c^{\lambda_1,\lambda_2,\lambda_3}$ leads to a theory that is consistent up to the cubic order. At this order, all interactions do not feel each other's presence since the main constraint
\begin{align}
    [H,J^{a-}]\Big|_3=[H_3,J^{a-}_2]-[J^{a-}_3,H_2]&=0\,,
\end{align}
is a linear equation with respect to $H_3$, $J_3$, i.e. any linear combination of solutions is a solution again. The basis is provided by \eqref{Paper3-famouscubic}. 
An interesting feature of the cubic interactions, which is hard to see in any covariant description, is that the vertices split into holomorphic, $\PPb$-dependent, and anti-holomorphic, $\PP$-dependent. Indeed, $H_2$ at the three-particle level is proportional to $\PP \PPb$ and, hence, any non-holomorphic contribution can be redefined away (via an appropriate canonical transformation). Covariant descriptions usually lead to some $\PP\PPb$-terms by default.
\paragraph{Higher orders. } The first serious constraint comes at the quartic order. It is at the quartic order that the spectrum of a theory usually gets fixed, as well as most of the cubic couplings. In general, at order $n$, we have to solve
\begin{align}\label{Paper3-LFNoether}
    [J^{a-}_2,H_n]-[H_2,J^{a-}_n]&=\sum_{\substack{i,j>2\\i+j-2=n}}[H_i^{\vphantom{a}},J^{a-}_j]\,.
\end{align}
Rewriting it in terms of the densities leads to 
\begin{align}\label{Paper3-maineq}
  \mathbf{H}_2\,  j^{a-}_n&=\mathbf{J}^{a-}_2[ h_n]+\sum_{\substack{i,j>2\\i+j-2=n}}[H_i^{\vphantom{a}},J^{a-}_j]\,,
\end{align}
where the boldface operators are sums of the one-particle ones:
\begin{align}
\label{Paper3-defHJ}
\mathbf{J}^{a-}_2&=\sum_i \tilde{j}^{a-}_2(q_i)\,, &
\mathbf{H}_2&=\sum_i h_2(q_i)\,.
\end{align}
Here $\tilde{j}^{a-}_2$ commutes with $\delta^3(\sum q_i)$ and reads
\begin{align}
\tilde{j}^{a-}_2(q_i)=\tilde{j}_2^{a-}(q_i,\pl_{q_i})= j^{a-}_2(q_i)^T-h_2(q_i)\frac{1}{n}\left(\sum_j \pfrac{q^a_j}\right)
\,,
\end{align}
where $h_2$ can be found in \eqref{Paper3-FreeFieldReal} and 
\begin{align}\label{Paper3-J2T}
    (j^{z-}_2)^T&= -\frac{ q\qb}{\beta}\pfrac{\qb} -q \pfrac{\beta} +\lambda\frac{q}{\beta}\,, &
    (j^{\zb-}_2)^T&= - \frac{ q\qb}{\beta}\pfrac{q} -\qb \pfrac{\beta} -\lambda\frac{\qb}{\beta}  \,.      
\end{align}
At first sight, it may seem puzzling how a single equation, \eqref{Paper3-LFNoether}, involving two free functions, $j_n$ and $h_n$, could admit unique solutions. Indeed, as is evident from \eqref{Paper3-maineq}, $\mathbf{H}_2$ represents merely a number (the total energy). Thus, one can always divide by $\mathbf{H}_2$ to determine $j_n$ for any given choice of $h_n$: 
\begin{align}\label{Paper3-maineqB_full}
    j^{a-}_n&=\frac{1}{\mathbf{H}_2}\Big(\mathbf{J}^{a-}_2[ h_n]+\sum_{\substack{i,j>2\\i+j-2=n}}[H_i^{\vphantom{a}},J^{a-}_j]\Big)\,.
\end{align}
In doing so, we will find transverse momenta in the denominator, which makes $j_n$ non-local at this order and will induce a similar non-locality in $h_n$ at the next order. The proof of Poincaré invariance of $S$-matrix relies on $H$ and $J^{a-}$ being local, i.e. not having transverse momenta in denominators. It is this locality assumption that allows one to find unique (or almost unique) solutions for two functions from a single equation. If locality is abandoned, one can literally take any ``interaction'' $h_n$ at each order and prolong it to a ``formally consistent'' generator of Poincaré algebra.\footnote{As a side remark, let us note that the same effect also appears in the Noether procedure \cite{Barnich:1993vg}; i.e. one can always find a solution by abandoning locality. The same reasoning applies to theories in $(A)dS_d$, with the additional subtlety that the denominators take the form $p^2+...+\Lambda$, where $\Lambda$ is the cosmological constant. This allows one to expand the denominators and observe infinite derivative tails in powers of $p^2$, rather than an explicit non-locality in the denominator, which can make one erroneously feel that the situation in $(A)dS$ is somewhat better.}

\section{Quartic analysis}\label{Paper3-section4}
A general procedure for bootstrapping theories goes as follows. One begins by making certain assumptions about the spectrum and cubic interactions, which translates into requiring some $C^{\lambda_1,\lambda_2,\lambda_3}$ and/or $\bar C^{\lambda_1,\lambda_2,\lambda_3}$ not to vanish.
The equation to be solved at the quartic order is
\begin{align}\label{Paper3-maineqB}
  \mathbf{H}_2\,  j^{a-}_4&=\mathbf{J}^{a-}_2[ h_4]+[H_3^{\vphantom{a}},J^{a-}_3]\,.
\end{align}
This is understood as a quadratic equation for the couplings. Due to the (anti)-holomorphic factorization of the cubic vertices, the equation splits into $CC$, $\bar C\bar C$ (we call these chiral), and $C \bar C$ sectors. For each of these sectors, we find that $[H_3^{\vphantom{a}},J^{a-}_3]$ contributes as
\begin{align}
    \deltas{\sum q_i}C^{\lambda_1,\lambda_2,\omega}C^{-\omega,\lambda_3,\lambda_4} F_{\lambda_1,\lambda_2,\lambda_3,\lambda_4}(\PP_{12},\PP_{34},\PPb_{12},\PPb_{34},\beta_i)\, \phi^{\lambda_1}_{q_1}\phi^{\lambda_2}_{q_2}\phi^{\lambda_3}_{q_3}\phi^{\lambda_4}_{q_4}\,,
\end{align}
where the function $F$ depends on the sector and will be given below. An important technical aspect is whether some of $\phi$'s have the same helicity and, hence, the density should take this symmetry, i.e. bosonic symmetrisation (or anti-symmetry in the case of fermions), into account. For a given helicity $\lambda$, the field $\phi^\lambda$ may also come in several species, e.g. as in the case of Yang-Mills theory.  

\paragraph{Chiral theories.} A remarkable property of \eqref{Paper3-maineqB} observed in \cite{Metsaev:1991mt,Metsaev:1991nb}, see also \cite{Ponomarev:2016lrm}, is that the chiral sectors receive no contribution from $H_4$ and $J_4$. In other words, there are two closed subsystems of equations, for $CC$ and for $\bar C\bar C$, that are independent of higher orders:
\begin{align}\label{Paper3-HOLO}
     &[H_3(\PPb),J^{z-}_3(\PPb)]=0\,,&
    &[H_3(\PP),J^{\bar{z}-}_3(\PP)]=0\,.
\end{align}
These equations are highly constraining, particularly for higher-spin fields. Solving the holomorphic ($CC$) and anti-holomorphic ($\bar C\bar C$) quartic constraints allows one to determine the allowed spectrum of cubic vertices, by fixing the various couplings $C^{\lambda_1,\lambda_2,\lambda_3}$ and $\bar{C}^{\lambda_1,\lambda_2,\lambda_3}$. In particular, a general solution to the holomorphic quartic constraint was recently found in \cite{Serrani:2025owx}, where a complete classification of lower-derivative chiral theories was also achieved. 

It is worth noting that if one allows at least one higher-spin particle to self-interact via a coupling of the type $C^{s,s,-s}$, with $s>2$ for singlet fields and $s>1$ in the color case, the holomorphic constraint forces the inclusion of all possible vertices. This leads to the so-called Metsaev solution, which uniquely determines all cubic couplings to be
\begin{align}\label{Paper3-Metsaev}
&C^{\lambda_1,\lambda_2,\lambda_3}=\frac{k(\ell_p)^{\lambda_{123}-1}}{\Gamma(\lambda_1+\lambda_2+\lambda_3)}\,,&
\bar C^{\lambda_1,\lambda_2,\lambda_3}=\frac{\bar{k}(\ell_p)^{\lambda_{123}-1}}{\Gamma(\lambda_1+\lambda_2+\lambda_3)}\,.&
\end{align}
Moreover, if we search for a unitary theory, the couplings above are related by\footnote{The reality condition for the fields is $(\phi^\lambda_q)^*=\phi^{-\lambda}_{-q}$. Complex conjugation swaps $\PP$ or $\PPb$. When the $(-q)$ is eliminated in the vertex, after the complex conjugation $\PP$ and $\PPb$ remain unchanged. In addition, the product of $\beta$'s yields $(-)^{\lambda_{123}}$. Therefore, to have a Hermitian action, every $\PP$ or $\PPb$ needs to be accompanied by $i$. This $i$ can be absorbed into $\ell_p$, and we do not write it anywhere below. After putting this $i$ under the rug, the reality conditions for the couplings are the ones written down.}
\begin{equation}
    \bar C^{-\lambda_1,-\lambda_2,-\lambda_3}=(C^{\lambda_1,\lambda_2,\lambda_3})^*\,.
\end{equation} 
In the analysis below, it is not necessary to assume Metsaev couplings; it suffices to follow the general solution to the holomorphic constraint presented in \cite{Serrani:2025owx}. Further details will be provided in the following paragraph.

\paragraph{$\mathbf{[H_3,J_3]}$ commutator.} We now analyse the quartic constraint \eqref{Paper3-maineqB} in the general case,\footnote{We will focus on the quartic constraint \eqref{Paper3-maineqB} for $a=z$. The complex conjugate constraint will be analysed in a few pages.} assuming that the starting cubic theory already satisfies both the holomorphic and anti-holomorphic constraints. Let us compute the commutator explicitly, both in the presence of an external gauge group $G$ and in its absence. Expressed in terms of the densities \eqref{Paper3-famouscubic}, the commutator takes the following form:
\begin{equation}\label{Paper3-constraint}
\begin{split}
[H_3,J_3^{z-}]= & \sum_{\lambda_i,\alpha_j}\int d^9p\,d^9q\, \delta\left(\sum_i q_i\right) \left[j_3^{\lambda_i}(q_i)-\frac{h_3^{\lambda_i}(q_i)}{3}\left(\sum_k\frac{\partial}{\partial \bar{q}_k}\right)\right] \times \\
 & \delta\left(\sum_j p_j\right)h_3^{\alpha_j}(p_j)\left[\mathrm{Tr}\prod_{i=1}^3\phi_{q_i}^{\lambda_i},\mathrm{Tr}\prod_{j=1}^3\phi_{p_j}^{\alpha_j}\right]\,,
 \end{split}
\end{equation}
where to keep the analysis general, we assume that fields take values in the matrix algebra (e.g. $\phi^\lambda=\phi^\lambda_a T^a$ with some generators $T_a$) --- analogous to what is done in Yang-Mills theory. As a result, the fields acquire a Lie algebra structure, and we apply the trace to ensure the appearance of an invariant tensor. The non-colored case corresponds to the trivial algebra of singlet fields, in which case the trace can be safely omitted.

We now choose to contract the fields $\phi_{q_3}^{\lambda_3}$ and $\phi_{p_3}^{\alpha_3}$, which correspond to the internal (exchanged) lines in the Feynman diagrams, and we get
\begin{equation}
    \sum_{\lambda_1,\lambda_2,\lambda_3}\sum_{\alpha_1,\alpha_2,\alpha_3}[\mathrm{Tr}(\phi^{\lambda_1}_{q_1}\phi^{\lambda_2}_{q_2}\phi^{\lambda_3}_{q_3}),\mathrm{Tr}(\phi^{\alpha_1}_{p_1}\phi^{\alpha_2}_{p_2}\phi^{\alpha_3}_{p_3})]=9\sum_{\lambda_1,\lambda_2}\sum_{\alpha_1,\alpha_2}\sum_{\omega}\mathrm{Tr}(\phi^{\lambda_1}_{q_1}\phi^{\lambda_2}_{q_2}\phi^{\alpha_1}_{p_1}\phi^{\alpha_2}_{p_2})[\phi^{\omega}_{q_3},\phi^{-\omega}_{p_3}]\,.
\end{equation}
We can now take a closer look at Eq.~\eqref{Paper3-constraint} and observe that the first derivatives $\pl_{\bar{q}_1}$, $\pl_{\bar{q}_2}$ still act on the fields, while the derivative $\partial_{\bar{q}_3}$ acts on the delta function associated with momentum conservation. It is therefore natural to integrate by parts the derivatives acting on the fields, but not the one acting on the delta function.
Then, using the Poisson bracket and the symmetry under the exchange $p\leftrightarrow q$ within the integral, we arrive at the following expression:
\begin{equation}\label{Paper3-constraint_rewritten}
\begin{split}
[H_3,J_3^{z-}]= & \sum_{\lambda_i,\alpha_j}\int d^9p\,d^9q\,\delta\left(\sum_i q_i\right) \delta\left(\sum_j p_j\right)9\,\delta^{\lambda_3,-\alpha_3}\frac{\delta(q_3+p_3)}{2q_3^+}\phi^{\lambda_1}_{q_1}\phi^{\lambda_2}_{q_2}\phi^{\alpha_1}_{p_1}\phi^{\alpha_2}_{p_2}\times\\
&\left(j_3^{\lambda_i}(q_i)+\sum_{k\neq 3}\frac{\partial}{\partial \bar{q}_{k}}\frac{h_3^{\lambda_i}(q_i)}{3}\right)h_3^{\alpha_j}(p_j)\,.
\end{split}
\end{equation}
As expected, momentum conservation factors out. This follows directly from the Poincaré algebra, which implies the relation
\begin{subequations}
\begin{align}
    [P^a,H]=[P^+,H]=0\,,\qquad
    [P^a,J^{b-}]=\delta^{ab}P^-&\,,\qquad
    [P^+,J^{a-}]=-P^a\,,\\
    [P^a,[J^{b-},H]]+[J^{b-},[H,P^a]]+[H,[P^a,J^{b-}]]&=0
    \;\implies\;
    [P^a,[J^{b-},H]]=0\,,\\
    [P^+,[J^{b-},H]]+[J^{b-},[H,P^+]]+[H,[P^+,J^{b-}]]&=0
    \;\implies\;
    [P^+,[J^{b-},H]]=0\,,
\end{align}
\end{subequations}
where we used the definition of the Poincaré algebra and the Jacobi identity. The last relation means that the commutator (at any order) must commute with $P^a$ and $P^+$, and therefore it must be proportional to a delta function ensuring momentum conservation.
\begin{figure}[H]
    \centering
    \begin{tikzpicture}
        \begin{feynman}
            \vertex (i1) at (-6, 1) {\(\lambda_2\)};
            \vertex (i2) at (-6,-1) {\(\lambda_1\)};
            \vertex (i3) at (-2, 1) {\(\lambda_3\)};
            \vertex (i4) at (-2,-1) {\(\lambda_4\)};

            \vertex (v1) at (-5, 0);
            \vertex (v3) at (-3, 0);

            \node at (-6.8, 0) {\(K^{\text{holo}}_{1234\omega}\)};
            \node at (-5.8, 0) {\(C\)};
            \node at (-2.2, 0) {\(C\)};
            \node at (-8.5,-0.2) {\Large $\sum\limits_{\lambda_i\in S_4}\sum\limits_{\omega}$};
            \node at (0,0) {\Large $=\;0$};
            
            \vertex at (-5, 0.5) {\(\omega\)};
            \vertex at (-4.1, 1.4) {\([H_3(\PPb),J^{z-}_3(\PPb)]\)};
            \vertex at (-3.2, 0.5) {\(-\omega\)};
            
            \diagram* {
                (i1) -- (v1),
                (i2) -- (v1),
                (v1) -- [plain] (v3),
                (v3) -- (i3),
                (v3) -- (i4),
            };

        \end{feynman}
    \end{tikzpicture}
    \caption{Holomorphic constraint.}
        \label{Paper3-fig_holo}
\end{figure}
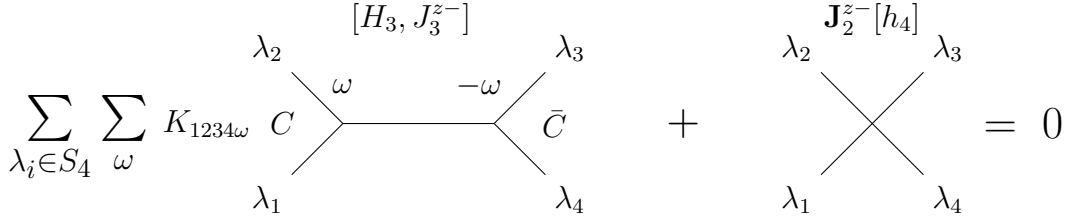
\begin{figure}[H]
    \centering
    \begin{tikzpicture}
        \begin{feynman}
            \vertex (i1) at (-6, 1) {\(\lambda_2\)};
            \vertex (i2) at (-6,-1) {\(\lambda_1\)};
            \vertex (i3) at (-2, 1) {\(\lambda_3\)};
            \vertex (i4) at (-2,-1) {\(\lambda_4\)};

            \vertex (ii1) at (1, 1) {\(\lambda_2\)};
            \vertex (ii2) at (1,-1) {\(\lambda_1\)};
            \vertex (ii3) at (3, 1) {\(\lambda_3\)};
            \vertex (ii4) at (3,-1) {\(\lambda_4\)};

            \vertex (v1) at (-5, 0);
            \vertex (v3) at (-3, 0);

            \node at (-6.8, 0) {\(K_{1234\omega}\)};
            \node at (-5.8, 0) {\(C\)};
            \node at (-2.2, 0) {\(\bar{C}\)};
            \node at (-8.5,-0.2) {\Large $\sum\limits_{\lambda_i\in S_4}\sum\limits_{\omega}$};
            \node at (-0.5,0) {\Large $+$};
            \node at (4,0) {\Large $=\;0$};
            
            \vertex at (-5, 0.5) {\(\omega\)};
            \vertex at (-4.1, 1.4) {\([H_3,J^{z-}_3]\)};
            \vertex at (-3.2, 0.5) {\(-\omega\)};
            
            \vertex at (2, 1.4) {\(\mathbf{J}^{z-}_2[ h_4]\)};
            
            \diagram* {
                (i1) -- (v1),
                (i2) -- (v1),
                (v1) -- [plain] (v3),
                (v3) -- (i3),
                (v3) -- (i4),
            };
            \diagram* {
                (ii1) -- (ii4),
                (ii2) -- (ii3),
            };

        \end{feynman}
    \end{tikzpicture}
    \caption{Non-holomorphic constraint.}
    \label{Paper3-fig_nonholo}
\end{figure}
\noindent
Starting from Eq.~\eqref{Paper3-constraint_rewritten} and using the explicit form of the densities \eqref{Paper3-famouscubic}, we obtain the following expression:
\begin{align}\label{Paper3-commutator_singlet}
     \begin{split}
       [H_3,J^{z-}_3]=&\sum_{\lambda_i,\omega}\int d^{12}q\,\delta\left(\sum_i q_i\right)\frac{9}{2}\Big[(-)^{\omega}\frac{\beta_1(\lambda_1+\omega-\lambda_2)-\beta_2(\lambda_2+\omega-\lambda_1)}{(\beta_1+\beta_2)^{2\omega+1}}\frac{\beta_3^{\lambda_3}\beta_4^{\lambda_4}}{\beta_1^{\lambda_1}\beta_2^{\lambda_2}}\times\\
       &\mathcal{C}^{1234\omega}\PPb_{12}^{\lambda_{12}+\omega-1}\PP_{34}^{-\lambda_{34}+\omega}\Big]\phi^{\lambda_1}_{q_1}\phi^{\lambda_2}_{q_2}\phi^{\lambda_3}_{q_3}\phi^{\lambda_4}_{q_4}\,,
       \end{split}
\end{align}
where we denote $\mathcal{C}^{1234\omega}\equiv C^{\lambda_1,\lambda_2,\omega}\bar{C}^{-\omega,\lambda_3,\lambda_4}$ the product of the couplings, in analogy with \cite{Serrani:2025owx}, and we have used the fact that the holomorphic part of the commutator already vanishes. A diagrammatic representation of the holomorphic and anti-holomorphic quartic constraints is shown in Figures \ref{Paper3-fig_holo} and \ref{Paper3-fig_nonholo}, respectively. In the case of matrix-valued fields, we have
\begin{align}\label{Paper3-commutator_color}
     \begin{split}
       [H_3,J^{z-}_3]=&\sum_{\lambda_i,\omega}\int d^{12}q\,\delta\left(\sum_i q_i\right)\frac{9}{2}\Big[(-)^{\omega}\frac{\beta_1(\lambda_1+\omega-\lambda_2)-\beta_2(\lambda_2+\omega-\lambda_1)}{(\beta_1+\beta_2)^{2\omega+1}}\frac{\beta_3^{\lambda_3}\beta_4^{\lambda_4}}{\beta_1^{\lambda_1}\beta_2^{\lambda_2}}\times\\
       &\mathcal{C}^{1234\omega}\PPb_{12}^{\lambda_{12}+\omega-1}\PP_{34}^{-\lambda_{34}+\omega}\Big]\mathrm{Tr}(\phi^{\lambda_1}_{q_1}\phi^{\lambda_2}_{q_2}\phi^{\lambda_3}_{q_3}\phi^{\lambda_4}_{q_4})\,.
       \end{split}
\end{align}
In order to be as general as possible, and following \cite{Serrani:2025oaw}, let us compute the commutator for the most general class of cubic vertices. We extend the field content by assigning to each field an additional index, indicating that it belongs to a representation of a gauge Lie algebra $\mathfrak{g}$ with structure constants $f_{abc}$. A field of helicity $\lambda$ is then assumed to take values in a representation $V_{\lambda}$ of $\mathfrak{g}$. We will denote by $\fA^{\lambda_1,\lambda_2,\lambda_3}_{abc}$ the tensor specifying the cubic coupling, using the same Latin indices for all modules $V_{\lambda}$. The most general form of the cubic Hamiltonian and boost generators is then given by
\begin{align}
    H_3&=\sum_{\lambda_1,\lambda_2,\lambda_3}\fA_{abc}^{\lambda_1,\lambda_2,\lambda_3}\int d^9 q\;\delta\Big(\sum_i q_i\Big)h_3^{\lambda_i}(q_i)\,(\phi^{\lambda_1}_{q_1})^{a}(\phi^{\lambda_2}_{q_2})^{b}(\phi^{\lambda_3}_{q_3})^{c}\,,\\
    J^{z-}_3&=\sum_{\lambda_1,\lambda_2,\lambda_3}\fA_{abc}^{\lambda_1,\lambda_2,\lambda_3}\int d^9q\,\delta\Big(\sum_i q_i\Big)\Big[j_3^{\lambda_i}(q_i)\,-\frac{1}{3}\,h_3^{\lambda_i}(q_i)\,\Big(\sum_j\frac{\partial}{\partial \bar{q}_j}\Big)\Big](\phi^{\lambda_1}_{q_1})^{a}(\phi^{\lambda_2}_{q_2})^{b}(\phi^{\lambda_3}_{q_3})^{c}\,,
\end{align}
where we have summed over all $\lambda_{1,2,3}$. This choice is convenient, especially for computations. However, to remain consistent, we need to take into account the appropriate symmetrisation\footnote{For bosons, and then integer helicities, this is a consequence of Bose symmetry. Fermions would get an extra minus sign.} \cite{Serrani:2025oaw} of the coupling constants:
\begin{equation}\label{Paper3-f_sym}
\fA_{a_{\sigma_1}a_{\sigma_2}a_{\sigma_3}}^{\lambda_{\sigma_1},\lambda_{\sigma_2},\lambda_{\sigma_3}}=(-)^{\lambda_{123}}\fA_{a_1 a_2 a_3}^{\lambda_1,\lambda_2,\lambda_3}\,.
\end{equation}
Note that this imposes a symmetry property on the cubic couplings, though only in the case of identical fields, such as 
\begin{align}\label{Paper3-sym_ff}
    &\fA_{a_2a_1a_3}^{\lambda,\lambda,s}=(-)^{2\lambda+s}\fA_{a_1 a_2 a_3}^{\lambda,\lambda,s}&
    &\implies&
    &\fA_{a_2a_1a_3}=(-)^{2\lambda+s}\fA_{a_1 a_2 a_3}\,.
\end{align}
Considering the most general class of cubic vertices, the commutator takes the form 
\begin{align}\label{Paper3-commutator_general}
    \begin{split}
    [H_3,J_3^{z-}]=&\sum_{\lambda_i,\omega}\int d^{12}q\;\delta \left(\sum_i q_i\right)\frac{1}{2}\Big[(-)^{\omega}\frac{\beta_1(\lambda_1+\omega-\lambda_2)-\beta_2(\lambda_2+\omega-\lambda_1)}{(\beta_1+\beta_2)^{2\omega+1}}\frac{\beta^{\lambda_3}_3\beta_4^{\lambda_4}}{\beta_1^{\lambda_1}\beta_2^{\lambda_2}}\,\times\\
    &\mathcal{F}^{1234\omega} \PPb_{12}^{\lambda_{12}+\omega-1}\PPb_{34}^{\lambda_{34}-\omega}\,(\phi^{\lambda_1}_{q_1})^{a_1}(\phi^{\lambda_2}_{q_2})^{a_2}(\phi^{\lambda_3}_{q_3})^{a_3}(\phi^{\lambda_4}_{q_4})^{a_4}\Big]\,,
    \end{split}
\end{align}
where, for convenience, we rewrite the tensor couplings as $\fA^{\lambda_1,\lambda_2,\lambda_3}_{abc}\equiv \fA_{abc}C^{\lambda_1,\lambda_2,\lambda_3}$ and define $\mathcal{F}^{1234\omega}=\fA_{a_1a_2c}\fA^c_{\phantom{c}a_3a_4}C^{\lambda_1,\lambda_2,\omega}\bar{C}^{-\omega,\lambda_3,\lambda_4}$. Therefore, in the singlet case, we have $\mathcal{F}^{1234\omega}\equiv \mathcal{C}^{1234\omega}$.\footnote{ Here, $\fA\fdu{a_1a_2}{c}\fA_{c\,a_3a_4}$ indicates the natural pairing between the positive and negative helicity fields $(\phi^{+\lambda})^a$ and $(\phi^{-\lambda})_b$, where the negative helicity fields take values in the dual vector space (note that the Poisson bracket pairs the fields of opposite helicities). For example, if $f^a_{\phantom{a}bc}$ are structure constants of some Lie algebra $\mathfrak{g}$, and $(\phi^{+\lambda})^a$ transforms in the adjoint representation, then $(\phi^{-\lambda})_b$ transforms in the canonical dual to the adjoint, i.e. in the coadjoint one. Since we consider only the case of (semi)-simple Lie algebras, one can use the invariant metric tensor $k_{ab}$ instead of $\delta^a_b$. For the case where generators $T_a$ are given, this is the Killing form $k_{ab}=\mathrm{Tr}(T_aT_b)$.} The contraction is performed with the standard Poisson bracket for fields decorated with vector space indices and reads 
\begin{equation}\label{Paper3-commutator_f}
    [(\phi^{\lambda}_q)^a,(\phi^{s}_p)_b]= \delta^{\lambda,-s}\delta^a_b\frac{\delta^3(q+p) }{2q^+}\,.
\end{equation}
Although this point is not directly relevant for our purposes, let us emphasise that, following \cite{Serrani:2025owx,Serrani:2025oaw,Skvortsov:2020wtf}, the factor $(-)^{\omega}$ appearing in $[H_3,J_3]$ can be removed by choosing even-helicity fields to be Hermitian and odd-helicity fields to be anti-Hermitian matrices (with singlets treated as $1\times1$ matrices).

At this stage, we have all the necessary tools to begin our analysis of quartic interactions. In particular, to find $h_4$, as suggested in \cite{Ponomarev:2016lrm}, we can use the following trick:
\begin{equation}\label{Paper3-quartic}
    \textbf{H}_2 j_4^{z-}=\textbf{J}_2^{z-}[h_4]+[H_3,J_3^{z-}]\implies (\textbf{J}_2^{z-}[h_4]+[H_3,J_3^{z-}])|_{\textbf{H}_2=0}=0\,.
\end{equation}
In this way, we first determine $h_4$, and by substituting it back into the first equation above, we then obtain the corresponding $j_4$. Note that this is not as strong as being on-shell, where, in addition to the full four-momentum conservation, one also has $p_i^2=0$. 

Let us highlight some features of the commutator \eqref{Paper3-commutator_general} in comparison with the holomorphic one studied in \cite{Ponomarev:2016lrm,Serrani:2025owx}. The symmetry properties of the cubic couplings remain unchanged. Moreover, the kinematical part of the commutator --- disregarding the couplings and the fields --- is, as for the holomorphic one, (anti-)symmetric under the exchange  $1\leftrightarrow 2$ or $3\leftrightarrow 4$ for (odd-)even-derivative cubic vertices. However, two important differences emerge due to the presence of both the holomorphic $\PPb$ and the anti-holomorphic $\PP$ terms:
\begin{itemize}
    \item First, once the ordering of the external helicities $\lambda_i$ is fixed, each $\omega$ in the sum contributes independently (there are more independent monomials that can be constructed from $\PP$ and $\PPb$ than from just $\PP$ as was the case for the holomorphic constraints). As a result, only the various permutations of $\lambda_i$ can contribute to the same quartic constraint, while for the (anti)holomorphic constraints, different $\omega$ could 'talk' to each other. This is precisely what makes the constraint stronger and more difficult to solve than the holomorphic one, where the sum over $\omega$ played a crucial role in solving it in the presence of higher-derivative vertices.
    \item Second, the sum over pairs such as $(1234)$ and $(3412)$ no longer leads to the clean simplifications we obtained in terms of only three independent variables $A,B,C$, as in the holomorphic constraint \cite{Ponomarev:2016lrm,Serrani:2025owx}. As we will see later, it is now possible for one of these permutations to be present while the other is absent --- a situation that did not occur before. In the holomorphic case, the presence of $\mathcal{F}^{1234\omega}$ automatically implies the presence of $\mathcal{F}^{3412\omega}$. Here, however, the two products of couplings are generally different, so summing over $(1234)$ and $(3412)$ would implicitly assume that both products are nonzero.
\end{itemize}
This is, in a sense, expected: in the present case, the commutators are not required to vanish by themselves but rather to cancel out through the inclusion of a suitable quartic interaction, as shown in \eqref{Paper3-quartic}. Let us now rewrite the quartic commutator for singlet fields \eqref{Paper3-commutator_singlet} as\footnote{As already noted, although the full commutator involves a sum over $\omega$, it decomposes into several independent constraints. Each constraint receives contributions only from terms with identical external helicities, and with the helicity $\omega$ of the exchanged field fixed by requiring not to change the powers of $\PPb$ and $\PP$, ensuring they belong to the same constraint.}
\begin{equation}\label{Paper3-sym_constraint}
    [H_3,J_3^{z-}]\sim\text{Sym}\Big[(-)^{\omega}\mathcal{C}^{1234\omega}K_{1234\omega}\Big]\,,
\end{equation}
where we have defined the following kinematical factor:
\begin{equation}\label{Paper3-kinematical_factor}
    K_{1234\omega}=\frac{\beta_1(\lambda_1+\omega-\lambda_2)-\beta_2(\lambda_2+\omega-\lambda_1)}{(\beta_1+\beta_2)^{2\omega+1}}\frac{\beta_3^{\lambda_3}\beta_4^{\lambda_4}}{\beta_1^{\lambda_1}\beta_2^{\lambda_2}}
       \PPb_{12}^{\lambda_{12}+\omega-1}\PP_{34}^{-\lambda_{34}+\omega}\,,
\end{equation}
and where Sym denotes the sum over the following $6$ distinct permutations of the external helicities:
\begin{equation}\label{Paper3-sym_terms}
    (1234)+(1324)+(1423)+(3412)+(2413)+(2314)\,.
\end{equation}
Notice that the presence of these $6$  contributions is fully analogous to the holomorphic constraint. The key difference, as already emphasised, is that all of them must be treated as independent. Summing over terms such as $(1234)$ and $(3412)$ would not only fail to simplify the expressions but would also be inconsistent if we wish to consider the most general case.

In the case where the fields take values in the matrix algebra \eqref{Paper3-commutator_color}, we have to consider color-ordered constraints. For the color ordering $[1234]$ corresponding to
\begin{equation}
    [1234]=(1234)+(2341)+(3412)+(4123)\,,
\end{equation}
we get
\begin{equation}\label{Paper3-cycl_constraint}
    [H_3,J_3^{z-}]\sim\text{Cycl}\Big[(-)^{\omega}\theta_{\omega}\mathcal{C}^{1234\omega}K_{1234\omega}\Big]\,,
\end{equation}
where Cycl denotes the sum over the cyclic permutations, and $\theta_{\omega}$ arises from the freedom to choose a phase factor in the definition of the Poisson bracket; see \cite{Skvortsov:2020wtf,Serrani:2025owx,Serrani:2025oaw}.

Assuming the most general class of cubic vertices as in \eqref{Paper3-commutator_general}, the commutator could include any of the possible $4!=24$ permutations of $(1234)$ depending on the symmetries of the generic cubic couplings $\mathfrak{f}_{a_1a_2a_3}^{\lambda_1,\lambda_2,\lambda_3}$.

In the following, our goal is to solve the quartic constraint \eqref{Paper3-quartic}, where the commutator $[H_3,J^{z-}_3]$ takes either the form given in \eqref{Paper3-sym_constraint}, the cyclic form in \eqref{Paper3-cycl_constraint}, or a more general one if we consider \eqref{Paper3-commutator_general}. To be as general as possible, we will not assume any specific form for the couplings. However, as already pointed out above, the kinematical factor \eqref{Paper3-kinematical_factor} is (anti-)symmetric under the exchange  $1\leftrightarrow 2$ or $3\leftrightarrow 4$ for (odd-)even-derivative cubic vertices. 
Therefore, when searching for the most general solutions of the quartic constraint \eqref{Paper3-maineqB}, choosing the exchange $(1234)$ or any of the other permutations $(2134)$, $(1243)$, or $(2143)$ does not make any difference. We can thus focus on the six independent kinematical factors \eqref{Paper3-sym_terms}; once a solution is found, we can further analyse the possible types of structure constants that may arise.

\paragraph{$\mathbf{[H,J^{z-}]}$ and $\mathbf{[H,J^{\zb-}]}$ constraints.} Let us recall that we need to solve the two independent constraints~\eqref{Paper3-maineqB}, for both $a=z,\zb$. Together, at the quartic order, we need to solve the following system
\begin{equation}\label{Paper3-quartic_system}
    \begin{cases}
    &\textbf{H}_2 j_4^{z-}=\textbf{J}_2^{z-}[h_4]+[H_3,J_3^{z-}]\implies (\textbf{J}_2^{z-}[h_4]+[H_3,J_3^{z-}])|_{\textbf{H}_2=0}=0\\
    &\textbf{H}_2 j_4^{\bar{z}-}=\textbf{J}_2^{\bar{z}-}[h_4]+[H_3,J_3^{\bar{z}-}]\implies (\textbf{J}_2^{\bar{z}-}[h_4]+[H_3,J_3^{\bar{z}-}])|_{\textbf{H}_2=0}=0\,.\\
    \end{cases}
\end{equation}
The operators $\textbf{J}_2^{z-}$ and $\textbf{J}_2^{\bar{z}-}$ are described in \eqref{Paper3-defHJ} and \eqref{Paper3-J2T}, while the commutator $[H_3,J_3^{\zb-}]$ is written as
\begin{equation}
\begin{split}
[H_3,J_3^{\zb-}]= & \sum_{\lambda_i,\alpha_j}\int d^9p\,d^9q\,\delta\left(\sum_i q_i\right) \delta\left(\sum_j p_j\right)9\,\delta^{\lambda_3,-\alpha_3}\frac{\delta(q_3+p_3)}{2q_3^+}\phi^{\lambda_1}_{q_1}\phi^{\lambda_2}_{q_2}\phi^{\alpha_1}_{p_1}\phi^{\alpha_2}_{p_2}\times\\
&\left(\jb_3^{\lambda_i}(q_i)+\sum_{k\neq 3}\frac{\partial}{\partial q_{k}}\frac{h_3^{\lambda_i}(q_i)}{3}\right)h_3^{\alpha_j}(p_j)\,.
\end{split}
\end{equation}
Using the explicit form of the densities \eqref{Paper3-famouscubic}, we obtain
\begin{align}
     \begin{split}
       [H_3,J^{\zb-}_3]=&\sum_{\lambda_i,\omega}\int d^{12}q\,\delta\left(\sum_i q_i\right)\frac{9}{2}\Big[(-)^{\omega}\frac{\beta_2(\lambda_2+\omega-\lambda_1)-\beta_1(\lambda_1+\omega-\lambda_2)}{(\beta_1+\beta_2)^{-2\omega+1}}\frac{\beta_1^{\lambda_1}\beta_2^{\lambda_2}}{\beta_3^{\lambda_3}\beta_4^{\lambda_4}}\times\\
       &\bar{C}^{\lambda_1,\lambda_2,\omega}C^{-\omega,\lambda_3,\lambda_4}\PP_{12}^{-\lambda_{12}-\omega-1}\PPb_{34}^{\lambda_{34}-\omega}\Big]\phi^{\lambda_1}_{q_1}\phi^{\lambda_2}_{q_2}\phi^{\lambda_3}_{q_3}\phi^{\lambda_4}_{q_4}\,.
       \end{split}
\end{align}
This corresponds to $[H_3,J_3^{z-}]$ in \eqref{Paper3-commutator_singlet} after 
\begin{align}
    &q_i\longleftrightarrow \bar{q}_i\,,&
    &\lambda_i\longleftrightarrow -\lambda_i\,,&
    &C^{\lambda_1,\lambda_2,\lambda_3}\longleftrightarrow \bar{C}^{-\lambda_1,-\lambda_2,-\lambda_3}\,.&
\end{align}
In the case of matrix-valued fields, we need to take the trace, i.e. use $\mathrm{Tr}(\cdot)$ and, for the most general class of cubic vertices, introduce the tensor couplings $\fA^{\lambda_1,\lambda_2,\lambda_3}_{abc}\equiv \fA_{abc}C^{\lambda_1,\lambda_2,\lambda_3}$ as in \eqref{Paper3-commutator_general}. Notice that these are the same transformations obtained after applying parity (see Appendix \ref{Paper3-AppendixA}). This observation will be important later when we discuss quartic vertices for specific, highly symmetric cases.

For our purposes, we would like both constraints in the system \eqref{Paper3-quartic_system} to act on the same Hamiltonian density $h_4$. It is therefore convenient to rewrite the commutator in the following equivalent form:
\begin{align}
     \begin{split}
       [H_3,J^{\zb-}_3]=&\sum_{\lambda_i,\omega}\int d^{12}q\,\delta\left(\sum_i q_i\right)\frac{9}{2}\Big[(-)^{\omega}\frac{\beta_4(\lambda_4-\omega-\lambda_3)-\beta_3(\lambda_3-\omega-\lambda_4)}{(\beta_3+\beta_4)^{2\omega+1}}\frac{\beta_3^{\lambda_3}\beta_4^{\lambda_4}}{\beta_1^{\lambda_1}\beta_2^{\lambda_2}}\times\\
       &C^{\lambda_1,\lambda_2,\omega}\bar{C}^{-\omega,\lambda_3,\lambda_4}\PPb_{12}^{\lambda_{12}+\omega}\PP_{34}^{-\lambda_{34}+\omega-1}\Big]\phi^{\lambda_1}_{q_1}\phi^{\lambda_2}_{q_2}\phi^{\lambda_3}_{q_3}\phi^{\lambda_4}_{q_4}\,,
       \end{split}
\end{align} 
where the following relabelling has been performed:
\begin{align}
    &(q_1,\qb_1,\beta_1,\lambda_1)\leftrightarrow (q_3,\qb_3,\beta_3,\lambda_3)\,,&
    &(q_2,\qb_2,\beta_2,\lambda_2)\leftrightarrow (q_4,\qb_4,\beta_4,\lambda_4)\,,&
    & C\leftrightarrow \bar{C}\,,&
    & \omega\leftrightarrow -\omega\,.
\end{align} 

\subsection{Construction of the ansätze}
Here, we construct the most general ansatz for the quartic local Hamiltonian density $h_4$ and local boost density $j_4$. To do so, we examine the quartic constraint \eqref{Paper3-quartic}, from which we can determine the required powers of $\PP$, $\PPb$, and $\beta$ to build the ansatz. We denote a quartic vertex by the pair $(n,m)$, where $n$ and $m$ correspond to the number of derivatives in the holomorphic and anti-holomorphic cubic vertices appearing in the exchange term $[H_3,J^{z-}_3]$ and $[H_3,J^{\zb-}_3]$ that participate in the same quartic constraint. Specifically:
\begin{itemize}
    \item The $n$-derivative holomorphic vertex with helicity $(\lambda_1,\lambda_2,\omega)$ and cubic coupling $C^{\lambda_1,\lambda_2,\omega}$, with $n=\lambda_{12}+\omega>0$.

    \item The $m$-derivative anti-holomorphic vertex with helicity $(\lambda_3,\lambda_4,-\omega)$ and cubic coupling $\bar{C}^{-\omega,\lambda_3,\lambda_4}$, with $m=-\lambda_{34}+\omega>0$.

\end{itemize}
In the following, we refer to lower-derivative quartic vertices as those with $n,m\leq 2$, which participate in the same constraint as the lower-derivative cubic vertices.
Conversely, higher-derivative quartic vertices are those with $n,m>2$. Note also that the total number of derivatives in a quartic vertex is $n+m-2$. From the structure of the constraint \eqref{Paper3-quartic}, we can find 
\begin{align}
        &[H_3,J^{z-}_3]\sim \beta^{\lambda_{34}-\lambda_{12}-2\omega}\bar{\PP}^{\lambda_{12}+\omega-1}\PP^{-\lambda_{34}+\omega}\sim \beta^{-n-m}\PPb^{n-1}\PP^m\,,\\\label{Paper3-h4_dependence}
        &\textbf{J}_2^{z-}\cdot h_4\sim \frac{\PP}{\beta^2}h_4
        \implies
        h_4\sim \beta^{\lambda_{34}-\lambda_{12}-2\omega+2}\bar{\PP}^{\lambda_{12}+\omega-1}\PP^{-\lambda_{34}+\omega-1}\sim \beta^{2-n-m}\PPb^{n-1}\PP^{m-1}\,,\\
        &\textbf{H}_2\cdot j_4^{z-}\sim \frac{\PP\,\PPb}{\beta^3}j_4^{z-}
        \implies
        j_4^{z-}\sim \beta^{\lambda_{34}-\lambda_{12}-2\omega+3}\bar{\PP}^{\lambda_{12}+\omega-2}\PP^{-\lambda_{34}+\omega-1}\sim \beta^{3-n-m}\PPb^{n-2}\PP^{m-1}\,.
\end{align}
We immediately observe a particular feature of the $(1,1)$ quartic vertices: they admit no contribution from $j_4$, but only $h_4$. As we will see, this is precisely the case for Yang–Mills theory. Moreover, Yang–Mills theory is complete at the quartic order, meaning $h_n=0$ and $j_n=0$ for $n\geq 5$.

We now need to identify the independent variables. As mentioned above, for a four-point scattering, we have two independent $\PP$ and two independent $\PPb$. We choose $\PP_{12}$, $\PP_{34}$, $\PPb_{12}$, and $\PPb_{34}$ as our independent variables, while the remaining ones can be expressed in terms of these using the relations collected in Appendix \ref{Paper3-AppendixB}. Additionally, there are three independent $\beta$ variables, since one can be eliminated using momentum conservation. Altogether, this gives us a total of seven independent variables. We will show how to further eliminate two of them, reducing the number of independent variables to five.

First, due to the fixed homogeneity in $\beta$ of $h_4$ and $j_4$, we can remove one further variable by performing a change of variables. Of the new variables, only one must carry homogeneity different from zero. For example, assuming $\beta_1$, $\beta_2$, and $\beta_3$ as independent variables, we can make the following change of variables:
\begin{align}
    &\tilde{x}=\frac{\beta_2}{\beta_1}\quad
    \implies\quad
    \beta_2=\tilde{x}\beta_1\,,&
    &\tilde{y}=\frac{\beta_3}{\beta_1}\quad
    \implies\quad
    \beta_3=\tilde{y}\beta_1\,.
\end{align}
Using the variables $\tilde{x}$, $\tilde{y}$ and $\beta_1$, the only one that carries non-zero homogeneity in $\beta$ is $\beta_1$. Therefore, we can drop it in the construction of $h_4$ and $j_4$. Once a solution is found, we can reconstruct the $\beta_1$ dependence from the required homogeneity.

Due to them having better symmetry properties, in the following, we use the following slightly different variables:
\begin{align}\label{Paper3-xy_variables}
    &x=\frac{\beta_1-\beta_2}{\beta_1+\beta_2}\quad
    \implies\quad
    \beta_2=\frac{1-x}{x+1}\beta_1\,,&
    &y=\frac{\beta_3-\beta_4}{\beta_3+\beta_4}\quad
    \implies\quad
    \beta_3=\frac{y+1}{1-y}\beta_4\,,
\end{align}
and upon using momentum conservation, the same comments as before apply. Second, we can also remove one $\PP$ or $\PPb$ by using, as we have already pointed out in \eqref{Paper3-quartic}, the energy conservation (i.e. with $\textbf{H}_2=0$) for the external fields. This gives us a further relation of the form
\begin{align}\label{Paper3-on_shell}
    &\textbf{H}_2=\frac{1}{\beta_1+\beta_2}\left(\frac{\PPb_{12}\PP_{12}}{\beta_1\beta_2}-\frac{\PPb_{34}\PP_{34}}{\beta_3\beta_4}\right)=0&
    &\implies&
    &\PPb_{12}=\frac{\beta_1\beta_2}{\beta_3\beta_4}\frac{\PPb_{34}\PP_{34}}{\PP_{12}}=\frac{x^2-1}{y^2-1}\frac{\PPb_{34}\PP_{34}}{\PP_{12}}\,,
\end{align}
where the variables $x$ and $y$ are those introduced in \eqref{Paper3-xy_variables}.
The most natural way to take this constraint into account is to choose as independent variables $\PP_{12}\PPb_{34}$, $\PPb_{12}\PP_{34}$ and $\frac{\PPb_{12}\PP_{12}}{\beta_1\beta_2}+\frac{\PPb_{34}\PP_{34}}{\beta_3\beta_4}$. This choice works when $n=m$, whereas for $n\ge m$ we complete the set by including the variables $\PPb_{12}$ and $\PPb_{34}$, as we see in the ansatz below. 

We now present the most general ansatz used in our analysis.
We do not write the ansatz explicitly for $j^{z-}_4$, as it coincides with that of $h_4$ in the $(n-1,m)$ case. In what follows, we denote $(\beta_1+\beta_2)^2\Big(\frac{\PPb_{12}\PP_{12}}{\beta_1\beta_2}+\frac{\PPb_{34}\PP_{34}}{\beta_3\beta_4}\Big)\equiv \mathbf{s_{12}}$ and take each function $f_i$ to depend on the variables $x$ and $y$, i.e. $f_i=f_i(x,y)$. The ansätze are
{\allowdisplaybreaks\besubeqs\begin{align}
    h_4^{(1,1)}&=f(x,y)\,,\\
    h_4^{(2,1)}&=\frac{1}{\beta_1+\beta_2}\left(\PPb_{12}f_1+\PPb_{34}f_2\right)\,,\\
    h_4^{(3,1)}&=\frac{1}{(\beta_1+\beta_2)^2}\left(\PPb_{12}^2f_1+\PPb_{12}\PPb_{34}f_2+\PPb_{34}^2f_3\right)\,,\\\
    h_4^{(n,1)}&=\frac{1}{(\beta_1+\beta_2)^{n-1}}\left(\PPb_{12}^{n-1}f_1+\PPb_{12}^{n-2}\PPb_{34}f_2+\cdots+ \PPb_{12}\PPb_{34}^{n-2}f_{n-1}+\PPb_{34}^{n-1}f_n\right)\,,\\
    h_4^{(2,2)}&=\frac{1}{(\beta_1+\beta_2)^2}\left(\PP_{12}\PPb_{34}f_1+\mathbf{s_{12}}f_2+\PPb_{12}\PP_{34}f_3\right)\,,\\
    \nonumber
     h_4^{(n,n)}&=\frac{1}{(\beta_1+\beta_2)^{2n-2}}\Big(\PP_{12}^{n-1}\PPb_{34}^{n-1}f_1+\PP_{12}^{n-2}\PPb_{34}^{n-2}\mathbf{s_{12}}f_2+\cdots+\mathbf{s_{12}}^{n-1}f_n+\cdots\\
     &\qquad\qquad\qquad\qquad+\PPb_{12}^{n-2}\PP_{34}^{n-2}\mathbf{s_{12}}f_{2n-2}+\PPb_{12}^{n-1}\PP_{34}^{n-1}f_{2n-1}\Big)\,,\\
     h_4^{(3,2)}&=\frac{1}{(\beta_1+\beta_2)^3}\left(\PP_{12}\PPb^2_{34}f_1+\PPb_{34}\mathbf{s_{12}}f_2+\PPb_{12}\mathbf{s_{12}}f_3+\PP_{34}\PPb^2_{12}f_4\right)\,,\\
     \nonumber
    h_4^{(n\geq m)}&=\frac{1}{(\beta_1+\beta_2)^{n+m-2}}\Big(\PP_{12}^{m-1}\PPb_{34}^{n-1}f_1+\PP_{12}^{m-2}\PPb_{34}^{n-2}\mathbf{s_{12}}f_2+\cdots+\PPb_{34}^{n-m}\mathbf{s_{12}}^{m-1}f_m+\cdots\\
    &\qquad\PPb_{12}^{n-m}\mathbf{s_{12}}^{m-1}f_n+\cdots+\PPb_{12}^{n-2}\PP_{34}^{m-2}\mathbf{s_{12}}f_{n+m-2}+\PPb_{12}^{n-1}\PP_{34}^{m-1}f_{n+m-1}\Big)\,.
\end{align}\esubeqs}\noindent
To improve symmetry, we consistently use $(\beta_1+\beta_2)$ as the homogeneity factor. Furthermore, the number of free functions to be determined at each order is given by:
\begin{equation}
    \#f=n+m-1\,.
\end{equation}
We have focused on the case $n\geq m$; the complementary case $n\leq m$ is easily obtained by taking the parity transformed system of constraints \eqref{Paper3-quartic_system}. This gives the same set of equations, with the products of couplings mapped to the $n\leq m$ case. This simply reflects the fact that there should be no distinction between positive and negative helicities: the two are interchangeable, and their roles can always be exchanged:
\begin{align}
    &C^{\lambda_1,\lambda_2,\omega}\bar{C}^{-\omega,\lambda_3,\lambda_4}&
    &\longleftrightarrow&
    \bar{C}^{-\lambda_1,-\lambda_2,-\omega}C^{\omega,-\lambda_3,-\lambda_4}\,.
\end{align}
The ansätze for the case $n\leq m$ are then the same after exchanging $\PP$ and $\PPb$. The same ansätze were already used by Metsaev in \cite{Metsaev:1991nb}. We will comment more on \cite{Metsaev:1991nb} in the last section of the paper when discussing non-localities.

We do not write any ansatz for $(n,0)$ quartic vertices because no local quartic vertex of this type can exist. Indeed, from \eqref{Paper3-h4_dependence} we see that it would require negative powers of $\PP$ or $\PPb$, leading to non-local quartic vertices. 

The scalar cubic coupling $C^{0,0,0}$ appears in the holomorphic quartic constraint. As shown in \cite{Serrani:2025owx}, generic solutions to the holomorphic constraint do not accommodate a scalar self-coupling. Nevertheless, in special cases --- and in particular for certain lower-spin couplings --- the scalar cubic vertex can remain consistent. We discuss these possibilities in Appendix \ref{Paper3-AppendixC}.
\subsection{Solving the quartic constraint}
The analysis of the system of quartic constraints \eqref{Paper3-quartic_system} based on the ansatz described above leads to a system of coupled first-order PDEs that can be approached using various methods. We describe two methods: the first is for finding explicit solutions; however, it becomes increasingly intractable as the number of derivatives grows.\footnote{If the system of PDEs is consistent and a solution does exist, it is fast, but if the system is inconsistent, it becomes rapidly intractable.} A second method allows us to determine whether a solution may exist by exploiting the Frobenius condition (integrability condition) on flat space $d^2=0$.\footnote{The Frobenius theorem states that a distribution (locally defined by a first-order system of PDEs) is integrable if and only if it is involutive. In flat space, involutivity of such a system reduces to the statement that successive derivatives commute, i.e. to the nilpotency $d^2=0$; in particular, in Cartesian coordinates, one has $[\partial_x,\partial_y]=0$.} We also note that, in principle, each solution can be complex and written as
\begin{align}
    &h_4(x,y)=\Re[h_4(x,y)]+ i \Im[h_4(x,y)]\,,
\end{align}
where $h_4(x,y)$ is a complex function of two real variables. However, the commutator $[H_3,J^{z-}_3]$ is real (up to the possible complex value of the product of coupling), and the only potentially non-trivial part of the vertex --- i.e. the part that contributes to solving the quartic constraint (and that may not exist) --- is its real component. The imaginary part corresponds to solutions of the homogeneous PDEs. In the following, we will therefore concentrate on the real part of $h_4(x,y)$ while separately pointing out the corresponding homogeneous solutions. 

\paragraph{Structure of PDEs.} Once we use the appropriate ansatz described before and go on the energy shell via \eqref{Paper3-on_shell}, the generic form of the PDEs arising from \eqref{Paper3-quartic_system} is
\begin{align}\label{Paper3-constraint_generic}
&G(\PP_{12},\PP_{34},\PPb_{34},x,y,\partial_x,\partial_y)=0\,,&
&x=\frac{\beta_1-\beta_2}{\beta_1+\beta_2}\,,&
&y=\frac{\beta_3-\beta_4}{\beta_3+\beta_4}\,.
\end{align}
The system of PDEs arises from isolating independent terms, corresponding to terms with different powers of $\PP_{12}$, $\PP_{34}$, and $\PPb_{34}$. For the $[H_3,J^{\bar{z}-}_3]$ constraint and a $(n,m)$ quartic vertex, this always leads to a coupled system of PDEs of the form
\begin{equation}\label{Paper3-system_of_PDEs}
    \begin{cases}
    P^{(1)}_1f_1+P^{(1)}_2\partial_xf_1+P^{(1)}_3=0\\
    P^{(2)}_1f_2+P^{(2)}_2\partial_xf_2+P^{(2)}_3f_1+P^{(2)}_4\partial_yf_1+P^{(2)}_5=0\\
    P^{(3)}_1f_3+P^{(3)}_2\partial_xf_3+P^{(3)}_3f_2+P^{(3)}_4\partial_yf_2+P^{(3)}_5=0\\
    ...\\
    P^{(k-2)}_1f_{k-2}+P^{(k-2)}_2\partial_xf_{k-2}+P^{(k-2)}_3f_{k-3}+P^{(k-2)}_4\partial_yf_{k-3}+P^{(k-2)}_5=0\\
    P^{(k-1)}_1f_{k-1}+P^{(k-1)}_2\partial_xf_{k-1}+P^{(k-1)}_3f_{k-2}+P^{(k-1)}_4\partial_yf_{k-2}+P^{(k-1)}_5=0\\
    P^{(k)}_1f_{k-1}+P^{(k)}_2\partial_yf_{k-1}+P^{(k)}_3=0\,,
    \end{cases}
\end{equation}
and another similar system of PDEs for the $[H_3,J^{z-}_3]$ constraint:
\begin{equation}\label{Paper3-system_of_PDEs_2}
    \eqref{Paper3-system_of_PDEs}\quad
    \text{with}\quad 
    P_{\bullet}^{(\bullet)}\leftrightarrow \bar{P}_{\bullet}^{(\bullet)}\quad \partial_x\leftrightarrow\partial_y\,.
\end{equation}
where $2k=2(n+m)$ is the number of PDEs, $P^{(i)}_j=P^{(i)}_j(x,y)$ and $\bar{P}^{(i)}_j=\bar{P}^{(i)}_j(x,y)$ are known rational functions of $x$, $y$, and $f_i=f_i(x,y)$, with $i=1,...,k-1$ being the free functions in the ansatz to be determined. This system of PDEs is a consequence of the structure of the linear differential operators $J^{\bar{z}-}_2$ and $J^{z-}_2$ acting on $h_4$.

The PDEs in \eqref{Paper3-system_of_PDEs} and \eqref{Paper3-system_of_PDEs_2} are linear first-order partial differential equations. The Cauchy–Kovalevskaya theorem ensures that, under mild assumptions --- such as working over the field of real $\mathbb{R}$ or complex numbers $\mathbb{C}$ and having coefficients $P^{(i)}_j(x,y)$ and $\bar{P}^{(i)}_j(x,y)$ that are analytic functions of $x$ and $y$  (in our case, they are actually rational functions) --- the system always admits a local solution with real or complex analytic functions. These can be obtained, for instance, using the method of characteristics or by direct (brute-force) integration. The system does not usually have a unique solution since it admits homogeneous solutions, which parameterize quartic vertices that are consistent on their own.

Importantly, the system consists of $2k$ PDEs for only $k-1$ unknown functions. As a result, although each equation can be solved, the existence of non-trivial solutions requires satisfying the integrability conditions. In flat space, these conditions are equivalent to the commutativity of partial derivatives, $d^2=0$, or $\partial_x\partial_yf=\partial_y\partial_xf$.

\paragraph{Solving the PDEs.} The structure of the systems \eqref{Paper3-system_of_PDEs} and \eqref{Paper3-system_of_PDEs_2} allows them to be solved straightforwardly. Starting from the first equation (and analogously from the last), we solve for $f_1(x,y)$, substitute the result into the subsequent PDEs, and then solve for $f_2(x,y)$. This procedure can be iterated to determine all functions $f_i(x,y)$.

At each step, the integration introduces a free constant. Once the final function $f_{k-1}(x,y)$ has been determined and substituted back into the last two PDEs, we are left with a pair of algebraic equations
\begin{align}\label{Paper3-algebraic_system}
&\sum_{i=1}^{k-1} A_i(x,y) a_i = 0\,,&
&\sum_{i=1}^{k-1} B_i(x,y) a_i = 0\,.
\end{align}
where $A_i$ and $B_i$ are known rational functions of $x$ and $y$, while the $a_i$ are the integration constants arising from the solution of the various PDEs. Solving the algebraic system \eqref{Paper3-algebraic_system} fixes some of these free coefficients. In particular, as we will see, all free coefficients are fixed except in the case where one considers the homogeneous system alone, i.e. in the absence of exchange contributions.

Notice that, once a solution for the density $h_4$ is found, we can use the full form of the constraint in \eqref{Paper3-quartic} to extract the boost density $j_4$. This is always possible: indeed, once the Hamiltonian of the system is constructed, we know, in principle, everything about the system. Determining the explicit expressions for the boost densities then becomes merely an exercise.

Let us emphasise that, for this procedure to work, a solution must exist. Otherwise, not only would we fail to find one, but additional complications could arise, for instance, overly cumbersome integrations, the appearance of irrational functions or square roots, or situations that render the process overly involved. For this reason, having an alternative method to check integrability without performing the explicit integrations is extremely valuable. We now proceed to describe such a method.

\paragraph{Existence of a solution.}

The method we present here allows us to check the Frobenius integrability condition and, consequently, the existence of solutions without the need to solve any of the PDEs explicitly. Starting from the initial system of PDEs, we differentiate the equations and impose the condition $\partial_x\partial_y=\partial_y\partial_x$. The idea is to take successive higher-order derivatives until the number of equations exceeds the number of variables.\footnote{Here, the variables refer to the unknown functions and their derivatives. The system effectively becomes an algebraic system of equations to be solved.} In practice, it is often necessary to go further and generate a substantially larger set of equations due to the possible presence of non-trivial relations among them. This behaviour is characteristic of overdetermined systems of differential equations subject to integrability conditions.

Starting from \eqref{Paper3-system_of_PDEs}, which we rewrite schematically, we apply both derivatives $\partial_x$ and $\partial_y$ as follows:
\begin{equation}
    \begin{cases}
    (\partial_x)^{h-1}(\partial_y)^{h-1}(f_1+\partial_xf_1)=0\\
    (\partial_x)^{h-1}(\partial_y)^{h-1}(f_2+\partial_xf_2+f_1+\partial_yf_1)=0\\
   (\partial_x)^{h-1}(\partial_y)^{h-1}(f_3+\partial_xf_3+f_2+\partial_yf_2)=0\\
    ...\\
    (\partial_x)^{h-1}(\partial_y)^{h-1}(f_{k-2}+\partial_xf_{k-2}+f_{k-3}+\partial_yf_{k-3})=0\\
    (\partial_x)^{h-1}(\partial_y)^{h-1}(f_{k-1}+\partial_xf_{k-1}+f_{k-2}+\partial_yf_{k-2})=0\\
    (\partial_x)^{h-1}(\partial_y)^{h-1}(f_{k-1}+\partial_yf_{k-1})=0\,.
    \end{cases}
\end{equation}
The same is done for the system \eqref{Paper3-system_of_PDEs_2}.
To maximise efficiency, we take the same number of $\partial_x$ and $\partial_y$ derivatives.

Once a sufficient number of steps have been performed, to concretely solve the algebraic system, we fix $x$ and $y$ to specific values (while generic rational values are sufficient, we specifically choose $(x,y)=(2,0)$, as we have observed that in practice this choice significantly speeds up the integrability method) and search for a solution to this system. If a solution is found, the original system of first-order PDEs is integrable, meaning that a solution exists; otherwise, the system of PDEs is inconsistent.\footnote{More precisely, integrability would be guaranteed if one were to consider derivatives of arbitrarily high order and check the existence of a solution. In practice, however, it is sufficient to go to a sufficiently high derivative order: if a solution exists at that stage, this ensures the actual existence of a solution to the full system. Conversely, if no solution exists already for a finite subset of equations, then the full system admits no solution.}

We will use this method to verify and formulate general conjectures about consistent and inconsistent quartic vertices, both for lower-derivative higher-spin interactions and, more interestingly, for higher-derivative ones, since we will find some yes-go results. This method is computationally far more efficient for proving no-go's than the previous approach based on explicitly solving the PDEs. Once this method establishes the existence of a solution, we can then construct it explicitly using the approach outlined above.

\section{Quartic vertices for lower-derivative theories}\label{Paper3-section5}
We begin by considering the system of quartic constraint \eqref{Paper3-quartic_system}, involving the exchange terms $[H_3,J^{z-}_3]$ and $[H_3,J^{\zb-}_3]$ with one- and two-derivative cubic vertices. We first examine the integrability of the resulting PDEs, and when a solution exists, we proceed to solve them explicitly to find the corresponding densities. 

As expected, we recover the no-go theorems for local minimal vertices of higher-spin fields, alongside the well-known positive results for lower-spin quartic vertices, such as those in Yang–Mills theory and gravity. We consider all possible lower-derivative quartic vertices, then three cases: $(1,1)$, $(2,1)$, and $(2,2)$ quartic vertices.

To cover all possible cases, we organise the analysis as follows.
As shown in \eqref{Paper3-sym_terms}, the commutator, due to its symmetries, can be organised in terms of the following relevant exchanges:
\begin{equation}\label{Paper3-various_ordering}
    \overset{s}{(1234)}+\overset{t}{(2341)}+\overset{s}{(3412)}+\overset{t}{(4123)}+\overset{u}{(1324)}+\overset{u}{(2413)}\,.
\end{equation}
We have slightly reorganised these terms so that the first four naturally correspond to the exchanges relevant to the color case, while the remaining two are additional.
This arrangement also facilitates the reading of the tables presented below, as it makes it straightforward to recognise patterns that can be traced back to the color case.

\paragraph{Quartic vertices.} Before presenting the results, it is important to understand the logic behind them. Let us briefly recall what we are doing and what we are looking for. We proceed to solve the constraint\footnote{The other constraint is analogous; thus, we restrict attention to this one.} \eqref{Paper3-quartic}. Quartic vertices are uniquely specified by the pair $(n,m)$ and the helicities of the external fields $(\lambda_1,\lambda_2,\lambda_3,\lambda_4)$, where $\lambda_4=n-m-\lambda_{123}$. Let us write down \eqref{Paper3-quartic} explicitly, as 
\begin{align}\label{Paper3-general_structure}
    \begin{split}
    &\Big(\sum_{i=1}^4\left(-\frac{ q_i\qb_i}{\beta_i}\pfrac{\qb_i} -q_i \pfrac{\beta_i} +\lambda\frac{q_i}{\beta_i}\right)h^{(n,m)}_{\lambda_1,\lambda_2,\lambda_3,\lambda_4} g^{\lambda_1,\lambda_2,\lambda_3,\lambda_4}_{a_1a_2a_3a_4}+(-)^{\omega}\Big(K_{1234\omega}\mathcal{F}_{1234\omega}\\
    &+K_{2341\omega}\mathcal{F}_{2341\omega}+K_{3412\omega}\mathcal{F}_{3412\omega}+K_{4123\omega}\mathcal{F}_{4123\omega}+K_{1324\omega}\mathcal{F}_{1324\omega}\\
    &+K_{2413\omega}\mathcal{F}_{2413\omega}\Big)\Big)(\phi^{\lambda_1}_{q_1})^{a_1}(\phi^{\lambda_2}_{q_2})^{a_2}(\phi^{\lambda_3}_{q_3})^{a_3}(\phi^{\lambda_4}_{q_4})^{a_4}=0\,,
    \end{split}
\end{align}
where $g^{\lambda_1,\lambda_2,\lambda_3,\lambda_4}_{a_1a_2a_3a_4}$ represents the coupling constant of the quartic vertex.
The system of PDEs above admits solutions whose form depends on the presence or absence of the various coefficients $\mathcal{F}_{\bullet\bullet\bullet\bullet\bullet}$. 

If we set all $\mathcal{F}_{\bullet\bullet\bullet\bullet\bullet}$ to zero (i.e. we set the cubic vertices to zero), the problem reduces to solving the homogeneous PDEs. These determine the quartic vertices, which are well-defined even in the absence of any cubic interactions.\footnote{When all cubic vertices are set to zero, the quartic constraint admits the trivial solution $h_4=0$ as well.}
Alternatively, one may look for solutions when at least one exchange term $\mathcal{F}_{\bullet\bullet\bullet\bullet\bullet}$ is non-zero. In this case, the quartic vertices become necessary to ensure the Lorentz invariance of the theory. 
We emphasise that, as already mentioned, the kinematical factors are (anti-)symmetric under the relevant permutations.\footnote{We recall that $K_{1234\omega}=(-)^{\lambda_{12}+\omega}K_{2134\omega}=(-)^{\lambda_{34}+\omega}K_{1243\omega}=(-)^{\lambda_{1234}}K_{2143\omega}$.} Consequently, once a solution exists, one can, in principle, choose which of the four possible kinematical structures $K_{\bullet\bullet\bullet\bullet\bullet}$ is present. In highly symmetric situations, this distinction disappears.\footnote{In general $K_{1234\omega}\mathcal{F}_{1234\omega}$ and $K_{2134\omega}\mathcal{F}_{2134\omega}$, are unrelated. For instance, one can have $f\fdu{a_1a_2}{c}\neq 0$ while $f\fdu{a_2a_1}{c}=0$. However, in special cases --- e.g. when $\lambda_1=\lambda_2$ --- one finds $K_{1234\omega}\mathcal{F}_{1234\omega}=K_{2134\omega}\mathcal{F}_{2134\omega}$, so the two contributions coincide and the distinction disappears.}

The main problem then reduces to determining whether \eqref{Paper3-general_structure} --- together with the $J^{\bar{z}-}$ constraint --- admits a solution with six free coefficients, each activating one of the six independent kinematical structures.

\paragraph{Integrability.} Using the integrability method, we can construct tables listing all quartic vertices that admit a solution to the quartic constraint \eqref{Paper3-quartic_system} for lower-derivative theories. The tables below were obtained as follows:
\begin{itemize}
    \item Assuming we are searching for an $(n,m)$ quartic vertex, we check the Frobenius integrability of the system of quartic constraint \eqref{Paper3-quartic_system} by evaluating the commutators $[H_3,J^{z-}_3]$ and $[H_3,J^{\zb-}_3]$ with all possible exchanges \eqref{Paper3-various_ordering} as explained above. There are six orderings that contribute to the same quartic constraint, each weighted by an independent free coefficient $k_i=(-)^{\omega}\mathcal{F}_{\bullet\bullet\bullet\bullet\bullet}$. For instance, for $(1234)$ we have $k_1=(-1)^{\omega}\mathcal{F}_{1234\omega}$ and analogously for the other orderings.

    \item The helicities $\omega$ appearing in the exchanges are automatically fixed by the requirement of having an $(n,m)$ quartic vertex. In addition, we can fix the value of $\lambda_4$ via the relation $n-m=\lambda_{1234}$. As a result, the problem reduces to searching for exchanges of the form:
    \begin{align}
        (1234)\equiv(\lambda_1,\lambda_2,\omega,\lambda_3,\lambda_4)=(\lambda_1,\lambda_2,n-\lambda_{12},\lambda_3,n-m-\lambda_{123})\,.
    \end{align}

    \item We set the maximum helicity $\lambda_{\text{max}}>|\lambda_1|,|\lambda_2|,|\lambda_3|$ and apply the integrability condition to all such cases. Whenever a solution to the Frobenius integrability condition is found, we record in the tables whether it also fixes any of the coefficients $k_i$. In particular, in all three tables presented below, we have checked the existence of a solution to the quartic constraint up to $\lambda_{\text{max}}=10$.
    
    \item The tables list only inequivalent solutions. Whenever a solution exists, permutations of the external helicities generally generate additional valid solutions. However, these are already taken into account, as they can always be deduced from the combined symmetries of the kinematical structures and the coupling constants.
    
\end{itemize}
Therefore, the tables collect the quartic couplings (``yes-go'') that admit a local solution to the quartic constraint \eqref{Paper3-quartic_system}, while at the same time indicating that all quartic vertices not listed do not satisfy it and thus fall into the category of ``no-go''.

The first entry $(\lambda_1,\lambda_2,\omega,\lambda_3,\lambda_4)$ specifies the helicities associated with the contribution from the $(1234)$ ordering in the exchange $[H_3,J^{z-}_3]$ and $[H_3,J^{\zb-}_3]$. The remaining entries correspond to the other orderings listed in \eqref{Paper3-various_ordering}. Due to the structure of the $\PP^\bullet\PPb^\bullet$ monomials, different orderings present in the quartic constraint can ``talk'' to each other, which may restrict and relate the products of the cubic couplings ($k_i$ in our shorthand notation). These constraints/relations are summarised in the tables below. If the entry is zero, it means that the corresponding product of couplings must vanish. If the entry contains $k_i$, it means that two (or more) $k_i$ turns out to be related or unconstrained. We summarise the results in Tables~\ref{Paper3-tab(1,1)}, \ref{Paper3-tab(2,1)}, and \ref{Paper3-tab(2,2)}. For example, in Table~\ref{Paper3-tab(1,1)} the second column has $k_2$ under $k_2$, which means that $k_2$ is unconstrained. The third column has $k_1$ under $k_3$, which means that $k_3=k_1$. The last column has $k_1-k_2$ under $k_6$, which means that $k_6=k_1-k_2$. We also keep $k_1$ as the initial free coupling; hence, there is no reason to include it in the tables. 

A final remark concerning the tables is the following. For exchanges of the form $(\lambda_1,\lambda_1,0,\lambda_2,\lambda_2)$, both commutators $[H_3,J^{z-}_3]$ and $[H_3,J^{\zb-}_3]$ vanish. In this case, the corresponding exchange does not enter the system of constraint \eqref{Paper3-quartic_system}, and the associated coefficient $k_i$ is therefore left unconstrained. In the tables, we therefore list the coefficient as free. However, it should be kept in mind that, for example, this coefficient will not appear in the solution for the quartic vertex. 

\begin{table}[H]
    \centering
    \begin{tabular}{|c|c|c|c|c|c|c|}
\hline
$(\lambda_1,\lambda_2,\omega,\lambda_3,\lambda_4)$ & $k_2=(2341)$ & $k_3=(3412)$ & $k_4=(4123)$ & $k_5=(1324)$ & $k_6=(2413)$ \\ \hline
$ (-1,1,1,-1,1)$  & $k_2$ & $k_1$ & $k_2$ & $0$ & $k_1-k_2$ \\ \hline
$ (-1,1,1,0,0)$ & $k_2$ & $k_1$ & $0$ & $0$ & $k_1-k_2$ \\ \hline
$(0,0,1,0,0)$  & $k_2$ & $k_1$ & $k_2$ & $k_5$ & $k_5$ \\ \hline
    \end{tabular}
    \caption{$(1,1)$ quartic vertices satisfying the quartic constraints.}
    \label{Paper3-tab(1,1)}
\end{table}

\begin{table}[H]
    \centering
    \begin{tabular}{|c|c|c|c|c|c|c|}
\hline
$(\lambda_1,\lambda_2,\omega,\lambda_3,\lambda_4)$ & $k_2=(2341)$ & $k_3=(3412)$ & $k_4=(4123)$ & $k_5=(1324)$ & $k_6=(2413)$ \\ \hline
$(-1,2,1,-1,1)$  & $k_1$ & $0$ & $0$ & $0$ & $k_1$ \\ \hline
$(-1,2,1,0,0)$  & $k_1$ & $0$ & $0$ & $0$ & $k_1$ \\ \hline
$(0,1,1,-1,1)$ & $0$ & $0$ & $k_4$ & $0$ & $k_1-k_4$ \\ \hline
$(0,1,1,0,0)$  & $k_2$ & $0$ & $0$ & $0$ & $k_6$ \\ \hline
    \end{tabular}
    \caption{$(2,1)$ quartic vertices satisfying the quartic constraints.}
    \label{Paper3-tab(2,1)}
\end{table}

\begin{table}[H]
    \centering
    \begin{tabular}{|c|c|c|c|c|c|c|}
\hline
$(\lambda_1,\lambda_2,\omega,\lambda_3,\lambda_4)$ & $k_2=(2341)$ & $k_3=(3412)$ & $k_4=(4123)$ & $k_5=(1324)$ & $k_6=(2413)$ \\ \hline
$(-2,2,2,-2,2)$  & $k_1$ & $k_1$ & $k_1$ & $0$ & $k_1$ \\ \hline
$(-2,2,2,-1,1)$  & $k_1$ & $k_1$ & $0$ & $0$ & $k_1$ \\ \hline
$(-2,2,2,0,0)$  & $k_1$ & $k_1$ & $0$ & $0$ & $k_1$ \\ \hline
$(-1,1,2,-1,1)$ & $k_2$ & $k_1$ & $k_2$ & $0$ & $k_6$ \\ \hline
$(-1,1,2,0,0)$ & $k_2$ & $k_1$ & $0$ & $0$ & $k_6$ \\ \hline
$(-1,2,1,-1,0)$  & $k_1$ & $0$ & $0$ & $0$ & $k_1$ \\ \hline
$(0,0,2,0,0)$ & $k_2$ & $k_1$ & $k_2$ & $k_5$ & $k_5$ \\ \hline
$(0,1,1,-2,1)$  & $0$ & $0$ & $k_1$ & $0$ & $k_1$ \\ \hline
    \end{tabular}
    \caption{$(2,2)$ quartic vertices satisfying the quartic constraints.}
    \label{Paper3-tab(2,2)}
\end{table}
\noindent
Several free coefficients remain, corresponding to solutions that are consistent (at this order!). For instance, in the Yang–Mills case $ (-1,1,1,-1,1)$, the presence of two free coefficients, $k_1$ and $k_2$, is related to the color structure of the theory. We have two different quartic contributions for the [++$--$] and [+$-$+$-$] sectors. Instead, for gravity $(-2,2,2,-2,2)$, there is only one free coefficient, reflecting the no-go for a multi-graviton theory (a color graviton would have two different color-ordered amplitudes). 

Let us stress once again that these tables represent consistency at the quartic order. The quintic light-cone constraint may impose further conditions. For instance, in Yang–Mills theory, the quintic Hamiltonian should vanish since the theory has no quintic vertex; therefore, the quintic constraint should be satisfied automatically without requiring a nonzero quintic vertex.

In the following, we prove the existence (at the quartic order) and uniqueness of Yang–Mills theory and gravity using the results collected in the tables above. Our strategy proceeds as follows:
\begin{itemize}
\item Firstly, we list the unique cubic vertices using \eqref{Paper3-famouscubic}.

\item Secondly, we solve the holomorphic constraints \eqref{Paper3-HOLO} that give relations among products of (anti-)holomorphic $CC$ ($\bar{C}\bar{C}$) couplings.

\item Thirdly, we use the information provided by the tables above, which establish the existence of a quartic vertex solving the quartic constraint \eqref{Paper3-quartic_system} and determine the relations among the products of $C\bar{C}$ couplings.

\item Fourthly, we use the CPT symmetry.
\end{itemize}
As we will see below, the holomorphic constraint gives us the Jacobi identity for the self-dual sectors, while the solution of the non-holomorphic quartic constraint \eqref{Paper3-quartic_system} gives us, together with the CPT symmetry, among others, a relation that in \cite{Fonseca:2025mzj} was called non-ambiguity condition (NAC). This allows us to relate the two self-dual sectors and find a unique Lie algebra.

\paragraph{Existence and uniqueness of Yang-Mills theory.}  We analyse the case of a spin-$1$ self-interacting theory. We start with the one-derivative holomorphic and anti-holomorphic spin-$1$ cubic vertices:
\begin{equation}\label{Paper3-YM_cubic}
    h^{\text{YM}}_3=\fA^{a}_{[bc]}\frac{\PPb \beta_1}{\beta_2\beta_3}(\phi^-_{q_1})_a(\phi^+_{q_2})^b(\phi^+_{q_3})^c+\bar{\fA}^{[ab]}_{c}\frac{\PP \beta_3}{\beta_1\beta_2}(\phi^-_{p_1})_a(\phi^-_{p_2})_b(\phi^+_{p_3})^c\,,
\end{equation}
where $\fA^{a}_{[bc]}=-\fA^{a}_{[cb]}$ and $\bar{\fA}^{[ab]}_{c}=-\bar{\fA}^{[ba]}_{c}$ are antisymmetric as a consequence of \eqref{Paper3-sym_ff}. In what follows, we omit the brackets $[\;]$. The two sets of indices (upper indices for $\phi^{+}$ and lower indices for $\phi^-$) belong, in general, to different vector spaces. The only requirement needed to define the kinetic term ($\sim \partial_{\mu}(\phi^{-})^a\partial^{\mu}(\phi^{+})_a$) and to allow contractions in the commutator \eqref{Paper3-commutator_f} is the existence of a pairing between these spaces. Accordingly, we assume that the positive- and negative-helicity fields $(\phi^+)^a$ and $(\phi^-)_b$ take values in vector spaces that are dual to each other.

Both couplings $\fA$ and $\bar{\fA}$ satisfy their own Jacobi identity as a direct consequence of the holomorphic quartic constraint \cite{Serrani:2025owx}. Let us briefly recall how this arises. For the cubic vertices \eqref{Paper3-YM_cubic}, only a single commutator contributes to the holomorphic constraint \eqref{Paper3-HOLO} (see \cite{Serrani:2025owx}):
\begin{equation}
    \fA^{a_1}_{a_2a_3}\fA^{b_1}_{b_2b_3}[(\phi^-_{q_1})_{a_1}(\phi^+_{q_2})^{a_2}(\phi^+_{q_3})^{a_3},(\phi^-_{p_1})_{b_1}(\phi^+_{p_2})^{b_2}(\phi^+_{p_3})^{b_3}]\,.
\end{equation}
The holomorphic constraint gives
\begin{align}
    \begin{split}
    [H_3(\PPb),J_3^{z-}]=\int d^{12}q\;\delta \left(\sum_i q_i\right)&\left(\frac{\beta_1-\beta_2}{\beta_1+\beta_2}\PPb_{34}+\frac{\beta_3+3\beta_4}{\beta_3+\beta_4}\PPb_{12}\right)\\
    &\fA^c_{a_1a_2}\fA^{a_3}_{a_4c}(\phi^+_{q_1})^{a_1}(\phi^+_{q_2})^{a_2}(\phi^-_{q_3})_{a_3}(\phi^+_{q_4})^{a_4}=0\,.
    \end{split}
\end{align}
We can use the three independent variables $A,B,C$ defined as 
\begin{align}\label{Paper3-ABC_variables}
\begin{split}
    2A&=\PPb_{12}+\PPb_{34}=\PPb_{23}-\PPb_{14}\,,\qquad
    2B=\PPb_{13}-\PPb_{24}=\PPb_{34}-\PPb_{12}\,,\\
    2C&=\PPb_{14}+\PPb_{23}=-\PPb_{13}-\PPb_{24}\,,
    \end{split}
\end{align}
and notice that Bose symmetrisation imposes that the equation above must be symmetric under the exchange of fields $1,2,4$. This leads to the form\footnote{Alternatively, we could apply the general formula in \eqref{Paper3-holo_constraint} directly, summing over all possible contributions.} 
    \begin{align}
    \begin{split}
    [H_3(\PPb),J_3^{z-}]=&\int d^{12}q\;\delta \left(\sum_i q_i\right)(A-B-C)\Big(\fA^c_{a_1a_2}\fA^{a_3}_{a_4c}+\fA^c_{a_2a_4}\fA^{a_3}_{a_1c}\\
    &+\fA^c_{a_4a_1}\fA^{a_3}_{a_2c}\Big)(\phi^+_{q_1})^{a_1}(\phi^+_{q_2})^{a_2}(\phi^-_{q_3})_{a_3}(\phi^+_{q_4})^{a_4}=0\,.
    \end{split}
\end{align}
The unique nontrivial solution is given by the Jacobi identity
\begin{equation}\label{Paper3-SDJI_YM}
    \fA^c_{a_1a_2}\fA^{a_3}_{a_4c}+\fA^c_{a_2a_4}\fA^{a_3}_{a_1c}+\fA^c_{a_4a_1}\fA^{a_3}_{a_2c}=0\,.
\end{equation}
The same steps can be repeated for $\bar{\fA}$, leading to the Jacobi identity in the dual space. Therefore, within the light-cone formalism, the Jacobi identities for both $\fA$ and $\bar{\fA}$ follow directly from the holomorphic quartic constraint. As a result, already the self-dual sectors are governed by a Lie algebra; a complex Lie algebra in the generic case.

Let us now turn to the analysis of the non-holomorphic quartic constraint. As we will see, the conditions that arise are closely related to those obtained by imposing consistent factorisation or via constructibility using BCFW. An analysis along these lines was recently carried out in \cite{Fonseca:2025mzj}. We will adopt the same terminology here for the various conditions we will find.

We describe in detail the information that can be extracted from the first line of Table \ref{Paper3-tab(1,1)}. We have $(-1,1,1,-1,1)$, corresponding to the exchange $C^{-1,1,1}\bar{C}^{-1,-1,1}$, with two free coefficients, $k_1$ and $k_2$. We find two independent quadratic conditions on the coupling constants:
\begin{align}
\nonumber
    &k_1=k_3\implies (-)^{\omega}\mathcal{F}_{1234}=(-)^{\omega}\mathcal{F}_{3412}\\\label{Paper3-NAC_YM}
    &\implies C^{-1,1,1}\bar{C}^{-1,-1,1}\fA^{a_1}_{a_2c}\bar{\fA}^{c\,a_3}_{a_4}=C^{-1,1,1}\bar{C}^{-1,-1,1}\fA^{a_3}_{a_4c}\bar{\fA}^{c\,a_1}_{a_2}\implies \boxed{\fA^{a_1}_{a_2c}\bar{\fA}^{c\,a_3}_{a_4}=\fA^{a_3}_{a_4c}\bar{\fA}^{c\,a_1}_{a_2}}\,,\\
    \nonumber
    &k_1-k_2=k_6\implies (-)^{\omega}\mathcal{F}_{1234}-(-)^{\omega}\mathcal{F}_{2341}=(-)^{\omega}\mathcal{F}_{2413}\\
    \nonumber
    &\implies C^{-1,1,1}\bar{C}^{-1,-1,1}\fA^{a_1}_{a_2c}\bar{\fA}^{c\,a_3}_{a_4}-C^{1,-1,1}\bar{C}^{-1,1,-1}\fA^{a_3}_{a_2c}\bar{\fA}^{c\,a_1}_{a_4}=C^{1,1,-1}\bar{C}^{1,-1,-1}\fA^{c}_{a_2a_4}\bar{\fA}^{a_1a_3}_{c}\\
    \nonumber
    &\implies C^{-1,1,1}\bar{C}^{-1,-1,1}\fA^{a_1}_{a_2c}\bar{\fA}^{c\,a_3}_{a_4}-C^{-1,1,1}\bar{C}^{-1,-1,1}\fA^{a_3}_{a_2c}\bar{\fA}^{c\,a_1}_{a_4}=C^{-1,1,1}\bar{C}^{-1,-1,1}\fA^{c}_{a_2a_4}\bar{\fA}^{a_1a_3}_{c}\\\label{Paper3-JI_YM}
    &\implies \boxed{\fA^{a_1}_{a_2c}\bar{\fA}^{c\,a_3}_{a_4}-\fA^{a_3}_{a_2c}\bar{\fA}^{c\,a_1}_{a_4}=\fA^{c}_{a_2a_4}\bar{\fA}^{a_1a_3}_{c}}\,,
\end{align}
where we have used the symmetry property \eqref{Paper3-f_sym} of $C^{\lambda_1,\lambda_2,\lambda_3}\fA^{a_1a_2a_3}$. We also notice that the third condition $k_2=k_4$ would lead to the same constraint as \eqref{Paper3-NAC_YM}. In analogy with \cite{Fonseca:2025mzj}, we call the condition \eqref{Paper3-JI_YM} the modified Jacobi identity and \eqref{Paper3-NAC_YM} the non-ambiguity condition (NAC).

These relations are, in principle, enough for the existence of a ``Yang-Mills'' theory, up to the quartic order. However, if we are willing to relate the holomorphic and anti-holomorphic couplings, we need a further assumption, CPT symmetry. CPT symmetry implies the following relation between $\fA$ and $\bar{\fA}$ (see \cite{Fonseca:2025mzj}):
\begin{equation}\label{Paper3-CPT_YM}
    \bar{\fA}^{ab}_{c}=(\fA^{c}_{ab})^*\,.
\end{equation}
Notice that this is also a consequence of imposing unitarity.

In \cite{Fonseca:2025mzj} it was proven that \eqref{Paper3-NAC_YM} together with \eqref{Paper3-CPT_YM} implies the existence of a basis where the $F_a=(\fA_a)^c_b$ are Hermitian $F_a=F^{\dag}_a$; then $\fA^c_{ab}=(\fA^b_{ac})^*$. We can now further use \eqref{Paper3-NAC_YM} to show that $\fA$ is indeed imaginary and fully antisymmetric $\fA_{abc}=\fA_{[abc]}$.
The Jacobi identity for $\fA$ is now a consequence either of the Jacobi identity for the self-dual sector \eqref{Paper3-SDJI_YM} or of the modified Jacobi identity \eqref{Paper3-JI_YM}. Therefore, the self-coupling of a spin $1$ massless particle has the structure of a Lie algebra, with the couplings $\fA^c_{ab}$ that play the role of the structure constants. 

The results just obtained can also be derived using the Noether procedure, starting from the free Maxwell Lagrangian in the covariant formulation. In particular, this problem can be addressed in a more systematic way within the framework of BV-BRST cohomology. For Yang–Mills theory, this analysis was carried out in \cite{Wald:1986bj, Barnich:2000zw}, and we find perfect agreement.

\paragraph{Existence and uniqueness of gravity.} We can perform the same analysis we have done for Yang-Mills with gravity. We start with the two cubic spin-$2$ vertices:
\begin{equation}\label{Paper3-GR_cubic}
    h^{\text{GR}}_3=g^{a}_{(bc)}\frac{\PPb^2 \beta^2_1}{\beta^2_2\beta^2_3}(\phi^{-2}_{q_1})_a(\phi^{+2}_{q_2})^b(\phi^{+2}_{q_3})^c+\bar{g}^{(ab)}_{c}\frac{\PP^2 \beta^2_3}{\beta^2_1\beta^2_2}(\phi^{-2}_{q_1})_a(\phi^{-2}_{q_2})_b(\phi^{+2}_{q_3})^c\,,
\end{equation}
where $g^{a}_{(bc)}=g^{a}_{(cb)}$ and $\bar{g}^{(ab)}_{c}=\bar{g}^{(ba)}_{c}$ are symmetric as a consequence of \eqref{Paper3-sym_ff}. In what follows, we omit the brackets $(\;)$.

Both couplings $g$ and $\bar{g}$ form an associative and commutative algebra, as a direct consequence of the holomorphic quartic constraint \cite{Serrani:2025owx}. Let us briefly recall how this arises. For the cubic vertices \eqref{Paper3-GR_cubic}, only a single commutator contributes to the holomorphic constraint \eqref{Paper3-HOLO} (see \cite{Serrani:2025owx}):
\begin{equation}
    g^{a_1}_{a_2a_3}g^{b_1}_{b_2b_3}[(\phi^{-2}_{q_1})_{a_1}(\phi^{+2}_{q_2})^{a_2}(\phi^{+2}_{q_3})^{a_3},(\phi^{-2}_{p_1})_{b_1}(\phi^{+2}_{p_2})^{b_2}(\phi^{+2}_{p_3})^{b_3}]\,,
\end{equation}
The holomorphic constraint gives
\begin{align}
    \begin{split}
    [H_3(\PPb),J_3^{z-}]=\int d^{12}q\;\delta \left(\sum_i q_i\right)&\left(\frac{2(\beta_2-\beta_1)}{\beta_1+\beta_2}\PPb_{12}\PPb_{34}^2-\frac{2(\beta_3+3\beta_4)}{\beta_3+\beta_4}\PPb_{34}\PPb_{12}^2\right)\\
    &g^c_{a_1a_2}g^{a_3}_{a_4c}(\phi^{+2}_{q_1})^{a_1}(\phi^{+2}_{q_2})^{a_2}(\phi^{-2}_{q_3})_{a_3}(\phi^{+2}_{q_4})^{a_4}=0\,.
    \end{split}
\end{align}
We can use the variables $A,B,C$ defined in \eqref{Paper3-ABC_variables} and notice that Bose symmetrisation imposes that the equation above must be symmetric under the exchange of fields $1,2,4$. This leads to the form
\begin{align}
    \begin{split}
    [&H_3(\PPb),J_3^{z-}]=\int d^{12}q\;\delta \left(\sum_i q_i\right)(-A+B+C)\Big((A^2-B^2)g^c_{a_1a_2}g^{a_3}_{a_4c}\\
    &+(B^2-C^2)g^c_{a_2a_4}g^{a_3}_{a_1c}+(C^2-A^2)g^c_{a_4a_1}g^{a_3}_{a_2c})\Big)(\phi^{+2}_{q_1})^{a_1}(\phi^{+2}_{q_2})^{a_2}(\phi^{-2}_{q_3})_{a_3}(\phi^{+2}_{q_4})^{a_4}=0\,.
    \end{split}
\end{align}
The unique nontrivial solution is given by
\begin{equation}\label{Paper3-holoGR_solution}
    g^c_{a_1a_2}g^{a_3}_{a_4c}=g^c_{a_2a_4}g^{a_3}_{a_1c}=g^c_{a_4a_1}g^{a_3}_{a_2c}\,.
\end{equation}
If we identify $g^a_{bc}$ with the structure constants of a finite $N$-dimensional algebra, with $V$ as internal space and define the product as
\begin{equation}
    (x\cdot y)^a= g^a_{bc} \,x^b y^c\,,\qquad
    \forall\;x,y\in V\,.
\end{equation}
Effectively, the condition \eqref{Paper3-holoGR_solution} coincides with the associativity of the algebra:
\begin{align}
    &(x\cdot y)\cdot z= x\cdot (y\cdot z)&
    &\implies&
    &g^c_{a_1[a_2}g^{a_3}_{a_4]c}=g^c_{a_1a_2}g^{a_3}_{a_4c}-g^c_{a_1a_4}g^{a_3}_{a_2c}=0\,.
\end{align}
Moreover, the condition $g^{a}_{bc}=g^{a}_{cb}$ implies that the algebra is commutative.
The same steps can be repeated for $\bar{g}$, leading to the commutativity and associativity of the dual algebra. Therefore, within the light-cone formalism, the associativity for both $g$ and $\bar{g}$ follow directly from the holomorphic quartic constraint. As a result, already the self-dual sectors are governed by a commutative and associative algebra.

We now turn to the analysis of the non-holomorphic quartic constraint. We describe in detail the information that can be extracted from the first line of Table \ref{Paper3-tab(2,2)}. We have $(-2,2,2,-2,2)$ corresponding to the exchange $C^{-2,2,2}\bar{C}^{-2,-2,2}$, with only one free coefficient, $k_1$. We find the following independent quadratic conditions on the structure constants: 
\begin{align}
\nonumber
    &k_1=k_3\implies (-)^{\omega}\mathcal{F}_{1234}=(-)^{\omega}\mathcal{F}_{3412}\\
    \nonumber
    &\implies C^{-2,2,2}\bar{C}^{-2,-2,2}g^{a_1}_{a_2c}\bar{g}^{c\,a_3}_{a_4}=C^{-2,2,2}\bar{C}^{-2,-2,2}g^{a_3}_{a_4c}\bar{g}^{c\,a_1}_{a_2}\\\label{Paper3-ASS_GR}
    &\implies \boxed{g^{a_1}_{a_2c}\bar{g}^{c\,a_3}_{a_4}=g^{a_3}_{a_4c}\bar{g}^{c\,a_1}_{a_2}}\,,\\
    \nonumber
    &k_1=k_2=k_6\implies (-)^{\omega}\mathcal{F}_{1234}=(-)^{\omega}\mathcal{F}_{2341}=(-)^{\omega}\mathcal{F}_{2413}\\
    \nonumber
    &\implies C^{-2,2,2}\bar{C}^{-2,-2,2}g^{a_1}_{a_2c}\bar{g}^{c\,a_3}_{a_4}=C^{2,-2,2}\bar{C}^{-2,2,-2}g^{a_3}_{a_2c}\bar{g}^{c\,a_1}_{a_4}=C^{2,2,-2}\bar{C}^{2,-2,-2}g^{c}_{a_2a_4}\bar{g}^{a_1a_3}_{c}\\\label{Paper3-NAC_GR}
    &\implies \boxed{g^{a_1}_{a_2c}\bar{g}^{c\,a_3}_{a_4}=g^{a_3}_{a_2c}\bar{g}^{c\,a_1}_{a_4}=g^{c}_{a_2a_4}\bar{g}^{a_1a_3}_{c}}\,.
\end{align}
The other condition, $k_1=k_4$, is a consequence of the one above. We call the condition \eqref{Paper3-ASS_GR} the modified associativity and \eqref{Paper3-NAC_GR} the non-ambiguity condition.

These relations are, in principle, sufficient to allow the existence of a theory of ``gravity'', up to the quartic order. However, if we are willing to relate the holomorphic and anti-holomorphic couplings, we need to assume CPT symmetry.  CPT symmetry implies the following relation between $g$ and $\bar{g}$, see \cite{Fonseca:2025mzj}:
\begin{equation}\label{Paper3-CPT_GR}
    \bar{g}^{ab}_{c}=(g^{c}_{ab})^*\,.
\end{equation}
Notice that this is also a consequence of unitarity. 

In \cite{Fonseca:2025mzj} was proven that  \eqref{Paper3-NAC_GR} together with \eqref{Paper3-CPT_GR} implies the existence of a basis where the $G_a=(g_a)^c_b$ are Hermitian $G_a=G^{\dag}_a$, then $g^c_{ab}=(g^b_{ac})^*$. We can now further use \eqref{Paper3-NAC_GR} to show that $g$ is indeed real and fully symmetric $g_{abc}=g_{(abc)}$.
The associativity for $g$ is now a consequence either of the associativity for the self-dual sector or of the modified associativity \eqref{Paper3-ASS_GR}. Therefore, the self-coupling of a spin $2$ massless particle has the structure of a commutative, symmetric, and associative algebra, with structure constant $g^c_{ab}$.

Once again, the same results can be derived using the Noether procedure within the framework of BV-BRST cohomology. For gravity, this analysis was carried out in \cite{Wald:1986bj, Boulanger:2000rq}, and we find perfect agreement. In \cite{Boulanger:2000rq}, it was also shown that finite-dimensional real algebras endowed with a positive-definite scalar product, which are commutative, symmetric, and associative, necessarily have a trivial structure: they decompose into a direct sum of one-dimensional ideals. Consequently, for the graviton, only self-interactions are possible.

Following the same line of reasoning applied to the cubic couplings of Yang–Mills theory and gravity, a similar analysis can be carried out for any other set of cubic couplings. For example, looking at the second line of Table \ref{Paper3-tab(1,1)}, where we have $(-1,1,1,0,0)$. In this case, we find the following condition:
\begin{align}
\nonumber
    &k_1=k_3\implies (-)^{\omega}\mathcal{F}_{1234}=(-)^{\omega}\mathcal{F}_{3412}\\\label{Paper3-YMScalar_cond1}
    &\implies \boxed{C^{-1,1,1}\bar{C}^{-1,0,0}\fA^{a_1}_{a_2c}\bar{\ell}^{c}_{a_3a_4}=C^{0,0,1}\bar{C}^{-1,-1,1}\ell^{a_3a_4}_{c}\bar{\fA}^{c\,a_1}_{a_2}}\,,\\
    \nonumber
    &k_1-k_2=k_6\implies (-)^{\omega}\mathcal{F}_{1234}-(-)^{\omega}\mathcal{F}_{2341}=(-)^{\omega}\mathcal{F}_{2413}\\
    \nonumber
    &\implies -C^{-1,1,1}\bar{C}^{-1,0,0}\fA^{a_1}_{a_2c}\bar{\ell}^{c}_{a_3a_4}-C^{1,0,0}\bar{C}^{0,0,-1}\ell^{a_3c}_{a_2}\bar{\ell}^{a_1}_{c\,a_4}=C^{1,0,0}\bar{C}^{0,-1,0}\ell^{a_4c}_{a_2}\bar{\ell}^{a_1}_{c\,a_3}\\\label{Paper3-YMScalar_cond2}
    &\implies \boxed{C^{-1,1,1}\bar{C}^{-1,0,0}\fA^{a_1}_{a_2c}\bar{\ell}^{c}_{a_3a_4}+C^{1,0,0}\bar{C}^{0,0,-1}\ell^{a_3c}_{a_2}\bar{\ell}^{a_1}_{c\,a_4}=C^{1,0,0}\bar{C}^{-1,0,0}\ell^{a_4c}_{a_2}\bar{\ell}^{a_1}_{c\,a_3}}\,.
\end{align}
While analysing the third line of Table \ref{Paper3-tab(1,1)}, where we have $(0,0,1,0,0)$, gives
\begin{align}
\nonumber
    &k_1=k_3\implies (-)^{\omega}\mathcal{F}_{1234}=(-)^{\omega}\mathcal{F}_{3412}\\\label{Paper3-Scalar_cond1}
    &\implies C^{0,0,1}\bar{C}^{-1,0,0}\ell^{a_1a_2}_{c}\bar{\ell}^{c}_{a_3a_4}=C^{0,0,1}\bar{C}^{-1,0,0}\ell^{a_3a_4}_{c}\bar{\ell}^{c}_{a_1a_2}\implies \boxed{\ell^{a_1a_2}_{c}\bar{\ell}^{c}_{a_3a_4}=\ell^{a_3a_4}_{c}\bar{\ell}^{c}_{a_1a_2}}\,.
\end{align}
Of particular interest would be to perform this analysis explicitly for the most general lower-spin cubic couplings, in order to classify all allowed lower-spin theories up to quartic order and to identify whether any new possibilities emerge. The advantage of the light-cone approach is threefold:
\begin{itemize}
\item It provides the most general framework for studying perturbation theory, allowing for a complete search in which nothing is missed.

\item It gives natural access to the self-dual sectors.

\item By treating all helicities on the same footing, as complex scalar fields, it allows us to find solutions for all higher-spin fields, as we will see in the next section.

\end{itemize}
We also emphasise that, in spirit, light-cone perturbation theory is closer to on-shell methods \cite{Benincasa:2007xk}. To study the self-dual sector within the BV-BRST cohomology framework, one would likely need to start from a chiral version of the free action and use chiral variables, along the lines of \cite{Krasnov:2021nsq}. In particular, it would be interesting to understand how the holomorphic constraint emerges in that approach.

\subsection{Lower-spin theories}

As a warm-up exercise and a consistency check of our method, we compute the quartic vertices $h_4$ for Yang-Mills theory and the quartic vertex $h_4$ along with its related boost generators $j^{z-}_4$ and $j^{\bar{z}-}_4$ for gravity. 
These are already known in light-cone gauge \cite{Brink:1982pd,Bengtsson:1983vn,Siegel:1999ew,Chakrabarti:2005ny,Chakrabarti:2006mb}, and we reproduce them in our formalism. The idea is to start from the most general ansatz described above and solve the system of quartic constraints \eqref{Paper3-quartic_system}.

We employ two different definitions for $x$ and $y$. The choice of definition is dictated by the symmetry properties of the specific example. In particular, we may use either of the following:
\begin{align}\label{Paper3-xy_1234}
    &x=\frac{\beta_1-\beta_2}{\beta_1+\beta_2}\,,&
    &y=\frac{\beta_3-\beta_4}{\beta_3+\beta_4}\,.
\end{align}
We then denote the quartic vertex using round brackets as $h^{(n,m)}_{(\lambda_1,\lambda_2,\lambda_3,\lambda_4)}$. Alternatively, we may use the following definition:
\begin{align}\label{Paper3-xy_1324}
    &x=\frac{\beta_1-\beta_3}{\beta_1+\beta_3}\,,&
    &y=\frac{\beta_2-\beta_4}{\beta_2+\beta_4}\,.
\end{align}
We then denote the quartic vertex using square brackets as $h^{(n,m)}_{[\lambda_1,\lambda_2,\lambda_3,\lambda_4]}$. This second definition is chosen because it leads to simpler transformation properties under cyclic permutations. We will adopt it in situations where color-like structures are present, as for Yang–Mills theory.

\paragraph{(1,1) quartic vertices.}
These vertices solve the system of quartic constraints \eqref{Paper3-quartic_system} when the commutators $[H_3,J_3^{z-}]$ and $[H_3,J_3^{\zb-}]$ involve solely cubic one-derivative interactions. The ansatz is
\begin{align}\label{Paper3-ansatz_YM}
    &h^{(1,1)}_{[\lambda_1,\lambda_2,\lambda_3,\lambda_4]}=f(x,y)\,,
\end{align}
where $f(x,y)$ is a real function of two real variables. 

\paragraph{(2,2) quartic vertices.}
These vertices solve the system of quartic constraints \eqref{Paper3-quartic_system} when the commutators $[H_3,J_3^{z-}]$ and $[H_3,J_3^{\zb-}]$ involve solely cubic two-derivative interactions. The ansatz is
\begin{align}\label{Paper3-ansatz_GR}
     \begin{split}
    &h^{(2,2)}_{(\lambda_1,\lambda_2,\lambda_3,\lambda_4)}=\frac{\PP_{12}\PPb_{34}}{(\beta_1+\beta_2)^2}f_1+\left(\frac{\PPb_{12}\PP_{12}}{\beta_1\beta_2}+\frac{\PPb_{34}\PP_{34}}{\beta_3\beta_4}\right)f_2+\frac{\PPb_{12}\PP_{34}}{(\beta_1+\beta_2)^2}f_3\,,\\
     &j^{z-\,(1,2)}_{(\lambda_1,\lambda_2,\lambda_3,\lambda_4)}=\frac{\PP_{12}g_1+\PP_{34}g_2}{(\beta_1+\beta_2)}\,,\qquad
     j^{\zb-\,(2,1)}_{(\lambda_1,\lambda_2,\lambda_3,\lambda_4)}=\frac{\PPb_{12}\bar{g}_1+\PPb_{34}\bar{g}_2}{(\beta_1+\beta_2)}\,,
     \end{split}
\end{align}
where $f_1(x,y)$, $f_2(x,y)$, $f_3(x,y)$, $g_1(x,y)$, $g_2(x,y)$, $\bar{g}_1(x,y)$, and $\bar{g}_2(x,y)$ are real functions of two real variables. 
\paragraph{Yang-Mills theory.} Searching for Yang-Mills theory, we take our fields to transform in the adjoint representation of a Lie algebra. In practice, we just take them to be matrix-valued. Due to the cyclicity of the trace, the two color-ordered contributions, namely [++$--$] and [+$-$+$-$], remain independent and do not mix. The quartic Hamiltonian takes the form:
\begin{equation}\label{Paper3-YM_quartic_Hamiltonian}
H_4=\int d^{12}q\,\delta\left(\sum q_i\right) \left(h_{[+,+,-,-]}^{(1,1)}\mathrm{Tr}[\phi^+_{q_1}\phi^+_{q_2}\phi^-_{q_3}\phi^-_{q_4}]+h_{[+,-,+,-]}^{(1,1)}\mathrm{Tr}[\phi^+_{q_1}\phi^-_{q_2}\phi^+_{q_3}\phi^-_{q_4}]\right)\,,
\end{equation}
It is well known that Yang-Mills theory provides the unique unitary and parity-invariant non-abelian quartic completion of the cubic vertices $C^{1,1,-1}$ and $\bar{C}^{-1,-1,1}$. Therefore, as a consistency check of our method, we expect to find a unique solution for the quartic vertex. 

Since $j^{z-}_4$ and $j^{\bar{z}-}_4$ are absent in the $(1,1)$ case, the quartic Hamiltonian densities must obey the following system of quartic constraints:
\begin{equation}
    \begin{cases}
    &\textbf{J}_2^{z-}[h_4]+[H_3,J_3^{z-}]=0\\
    &\textbf{J}_2^{\bar{z}-}[h_4]+[H_3,J_3^{\bar{z}-}]=0\,.
    \end{cases}
\end{equation}
Using the independence of $\PP_{12}$ and $\PP_{34}$, and of $\PPb_{12}$ and $\PPb_{34}$, respectively, we end up solving a system of inhomogeneous first-order PDEs. Solving the PDEs gives us the following quartic Hamiltonian densities:
\begin{align}\label{Paper3-h4_1_YM}
h^{(1,1)}_{[+,+,-,-]}(x,y)&=k_1\frac{x^2+y^2-2}{2(x+y)^2}\,,\\\label{Paper3-h4_2_YM}
h^{(1,1)}_{[+,-,+,-]}(x,y)&=k_1\frac{2 \left(2 x^2 y^2-x^2-y^2\right)}{\left(x^2-y^2\right)^2}\,,
\end{align}
where $k_1=-\,C^{1,1,-1}C^{1,-1,-1}=-\,C^{1,-1,1}C^{-1,1,-1}$. Note that $h^{(1,1)}_{[+,-,+,-]}(x,y)=h^{(1,1)}_{[+,-,+,-]}(y,x)$ follows from the symmetry $(1234)\leftrightarrow (3412)$, implied by the cyclicity of the trace
\begin{equation}
    h_{[+,-,+,-]}^{(1,1)}(x,y)\mathrm{Tr}[\phi^+_{q_1}\phi^-_{q_2}\phi^+_{q_3}\phi^-_{q_4}]=h_{[+,-,+,-]}^{(1,1)}(x,y)\mathrm{Tr}[\phi^+_{q_3}\phi^-_{q_4}\phi^+_{q_1}\phi^-_{q_2}].
\end{equation}

Moreover, we can check the parity and unitarity invariance of the quartic vertices. For the definitions of parity and unitarity, as well as their action on the densities $h_4$, we refer to Appendix \ref{Paper3-AppendixA}. In the present case, parity acts by exchanging positive and negative helicities. Thanks to the use of the smart variables \eqref{Paper3-xy_1324}, this action effectively translates into simple symmetry of the quartic vertices. Both solutions are therefore parity invariant, as we have 
\begin{align}\label{Paper3-YM1sym}
    &h^{(1,1)}_{[+,+,-,-]}(x,y)\overset{P}{=}h^{(1,1)}_{[-,-,+,+]}(x,y)&
    &\overset{(1234)\leftrightarrow (3412)}{\Longrightarrow}&
    &h^{(1,1)}_{[+,+,-,-]}(x,y)=h^{(1,1)}_{[+,+,-,-]}(-x,-y)\,,\\\label{Paper3-YM2sym1}
    &h^{(1,1)}_{[+,-,+,-]}(x,y)\overset{P}{=}h^{(1,1)}_{[-,+,-,+]}(x,y)&
    &\overset{(1234)\leftrightarrow (2341)}{\Longrightarrow}&
    &h^{(1,1)}_{[+,-,+,-]}(x,y)=h^{(1,1)}_{[+,-,+,-]}(y,-x)\,,\\\label{Paper3-YM2sym2}
    &h^{(1,1)}_{[+,-,+,-]}(x,y)\overset{P}{=}h^{(1,1)}_{[-,+,-,+]}(x,y)&
    &\overset{(1234)\leftrightarrow (4123)}{\Longrightarrow}&
    &h^{(1,1)}_{[+,-,+,-]}(x,y)=h^{(1,1)}_{[+,-,+,-]}(-y,x)\,.
\end{align}
As we can see, $h^{(1,1)}_{[+,-,+,-]}(x,y)=h^{(1,1)}_{[+,-,+,-]}(y,x)$, together with the symmetries \eqref{Paper3-YM2sym1} and \eqref{Paper3-YM2sym2} explains why $h^{(1,1)}_{[+,-,+,-]}=h^{(1,1)}_{[+,-,+,-]}(x^2,y^2)$.

We can compare \eqref{Paper3-h4_1_YM} and \eqref{Paper3-h4_2_YM} with an explicit computation of the quartic vertex of Yang–Mills theory using light-cone coordinates. The calculation is straightforward and begins with the Lagrangian of Yang–Mills theory:
\begin{equation}
    \mathcal{L}^{\text{YM}}(A_{\mu})=-\frac{1}{4}\mathrm{Tr}(F_{\mu\nu}F^{\mu\nu})\,.
\end{equation}
By imposing the light-cone gauge $A^+=0$ and rewriting the Yang–Mills Lagrangian in terms of light-cone coordinates, we obtain the following quartic interaction terms:
\begin{align}
    \begin{split}
H^{\text{YM}}_4&=\int d^{12}q\,\delta\left(\sum q_i\right) \Big(\frac{g^2}{4}\left(\frac{(\beta_2-\beta_3)(\beta_1-\beta_4)}{(\beta_2+\beta_3)(\beta_1+\beta_4)}-1\right)\mathrm{Tr}(A^+_{q_1}A^+_{q_2}A^-_{q_3}A^-_{q_4})\\
        &+\frac{g^2}{4}\left(\frac{(\beta_2-\beta_3)(\beta_1-\beta_4)}{(\beta_2+\beta_3)(\beta_1+\beta_4)}-\frac{(\beta_1-\beta_2)(\beta_3-\beta_4)}{(\beta_1+\beta_2)(\beta_3+\beta_4)}+2\right)\mathrm{Tr}(A^+_{q_1}A^-_{q_2}A^+_{q_3}A^-_{q_4})\Big)\,.
        \end{split}
\end{align}
where we denote by $g$ the coupling constant, which, upon using momentum conservation, corresponds precisely to the quartic vertices given in \eqref{Paper3-h4_1_YM} and \eqref{Paper3-h4_2_YM} upon identifying $k_1=-g^2$. This also shows that by employing suitable variables, the form of the vertices can be simplified. We have also verified our result by comparison with those obtained in \cite{Chakrabarti:2005ny, Chakrabarti:2006mb}, finding complete agreement; see also \cite{Ananth:2017pio}.
\paragraph{Gravity.} Repeating the same procedure but for gravity, using for $h^{(2,2)}_{(+2,+2,-2,-2)}$ the ansatz \eqref{Paper3-ansatz_GR}, we find
\begin{subequations}\label{Paper3-GR}
\begin{align}
    f_1(x,y)&=-\frac{1}{\left(x^2-y^2\right)^4}\big(32 x y \left(x^2 \left(\left(x^2-4\right) y^2+y^4+1\right)+y^2\right)\big)\,,\\
    \nonumber
    f_2(x,y)&=\frac{1}{\left(x^2-y^2\right)^4}\big(2 \big(x^6 \left(-4 y^4-7 y^2+5\right)+x^4 \left(-4 y^6+38 y^4-17 y^2+1\right)\\
    &\qquad\qquad\qquad +x^2 y^2 \big(-7 y^4-17 y^2+6\big)+5 y^6+y^4\big)\big)\,,\\
    f_3(x,y)&=-\frac{1}{\left(x^2-y^2\right)^4}\big(32 x y \left(x^4 \left(y^2+8\right)+x^2 \left(y^4-20 y^2+1\right)+8 y^4+y^2\right)\big)\,,\\
    \nonumber
    g_1(x,y)&=\frac{1}{\left(x^2-1\right)
   \left(x^2-y^2\right)^4}\big(16 x \big(-5 x^6+\left(3 x^4+5 x^2-3\right) y^6+\left(8 x^4+10 x^2-3\right) x^2 y^2\\
   &\qquad\qquad\qquad+\left(3 x^6-31 x^4+16 x^2-3\right) y^4\big)\big)\,,\\
   \nonumber
   g_2(x,y)&=-\frac{1}{\left(y^2-1\right)
   \left(x^2-y^2\right)^4}\big(16 y \big(x^6 \left(3 y^4+27 y^2-29\right)+x^4 \left(3 y^6-67 y^4+58 y^2+3\right)\\
   &\qquad\qquad\qquad+x^2 y^2 \left(22 y^4-12 y^2-7\right)+y^4 \left(y^2-2\right)\big)\big)\,.
\end{align}
\end{subequations}
The symmetry under the exchange $(1234)\leftrightarrow (2134)$ implies
\begin{align}
    &f_1(x,y)=-f_1\left(-x,y\right)\,,&
    &f_2(x,y)=f_2\left(-x,y\right)\,,&
    &f_3(x,y)=-f_3\left(-x,y\right)\,.
\end{align}
Similarly, under the exchange $(1234)\leftrightarrow (1243)$, it implies
\begin{align}
    &f_1(x,y)=-f_1\left(x,-y\right)\,,&
    &f_2(x,y)=f_2\left(x,-y\right)\,,&
    &f_3(x,y)=-f_3\left(x,-y\right)\,.
\end{align}
Moreover, some symmetries follow from parity invariance. In particular, parity relates
\begin{align}
    &h^{(2,2)}_{(+2,+2,-2,-2)}(x,y)\overset{P}{=}\bar{h}^{(2,2)}_{(-2,-2,+2,+2)}(x,y)\,.
\end{align}
This leads to additional relations among the $f_i$. For instance, the symmetry under the exchange $(1234)\leftrightarrow (3412)$ implies
\begin{align}
    &f_1(x,y)=f_1(y,x)\,,&
    &f_2(x,y)=f_2(y,x)\,,&
    &f_3(x,y)=f_3(y,x)\,.
\end{align}
The symmetry under the exchange $(1234)\leftrightarrow (2134)$ gives
\begin{align}
    &f_1(x,y)=-f_3\left(-x,y\right)\,,&
    &f_2(x,y)=f_2\left(-x,y\right)\,,
\end{align}
and under the exchange $(1234)\leftrightarrow (1243)$ gives
\begin{align}
    &f_1(x,y)=-f_3\left(x,-y\right)\,,&
    &f_2(x,y)=f_2\left(x,-y\right)\,.
\end{align}
Finally, note that the two constraints appearing in \eqref{Paper3-quartic_system} are related by a parity transformation. Concretely, under parity, one has
\begin{align}\label{Paper3-parity_related_constraint}
    &\mathbf{H}_2 j_4^{z-}=\mathbf{J}_2^{z-}[h_4]+[H_3,J_3^{z-}] &
    &\overset{P}{\longrightarrow}&
    &\mathbf{H}_2 j_4^{\zb-}=\mathbf{J}_2^{\zb-}[h_4]+[H_3,J_3^{\zb-}]\,,
\end{align}
where the Hamiltonians are parity invariant. Therefore, $j^{\bar{z}-\,(2,1)}_{(-2,-2,+2,+2)}=P\,j^{z-\,(1,2)}_{(+2,+2,-2,-2)}$.

\paragraph{Others lower-spin quartic vertices.}
We follow the same procedure used for Yang–Mills theory and gravity to construct quartic vertices for lower-spin theories with one- and two-derivative interactions. For clarity and readability of the main text, all results are collected in Appendix \ref{Paper3-AppendixD}.

\subsection{Four-point amplitudes}\label{Paper3-subsection5.2}

All massless cubic vertices found in the light-cone formalism can also be recovered in the spinor-helicity one (as amplitudes). This indicates the close relationship between the two approaches. When restricted to on-shell external fields, the two coincide; see \cite{Ponomarev:2016cwi}. For all known massless lower-spin theories, four-point amplitudes are well-known and can be expressed using the spinor-helicity formalism \cite{Elvang:2013cua}. 

Here, we employ the explicit on-shell map between the spinor-helicity and light-cone formalisms to compare our results and verify their agreement. This provides a strong consistency check for the quartic Hamiltonian densities we have derived. We illustrate this explicitly for Yang–Mills theory and gravity. The same analysis extends straightforwardly to the other cases and to higher spins, and additional amplitudes will be discussed in a following section.

It is essential to emphasise that our results extend beyond simply reproducing known on-shell amplitudes; they are genuinely new, as they hold off-shell. The quartic Hamiltonian densities we have constructed are off-shell quantities. For instance, our off-shell formulation allows one to write a Lagrangian and compute off-shell quantities (e.g. form factors). This is the trade-off for the more involved procedure compared to the spinor-helicity method, which builds amplitudes using little-group scaling for the cubic ones and applies on-shell recursion relations (like BCFW) for higher-point ones. While efficient, the latter, in its simpler form, assumes ``good'' high-energy behaviour in order to efficiently drop the contour integral at infinity, making it less suited for constructing vertices for higher-spin fields \cite{Benincasa:2007xk}. We review the on-shell relation between the light-cone and spinor-helicity formalisms in Appendix \ref{Paper3-AppendixE}. 

The total four-point amplitude in light-cone gauge is computed by summing all nonvanishing exchange contributions together with the quartic vertex:
\begin{align}\label{Paper3-total_A}
        \begin{split}
        \mA&(1_{\lambda_1}2_{\lambda_2}3_{\lambda_3}4_{\lambda_4})=k_1\mA^{\text{e}}(1_{\lambda_1}2_{\lambda_2}3_{\lambda_3}4_{\lambda_4})+k_2\mA^{\text{e}}(2_{\lambda_2}3_{\lambda_3}4_{\lambda_4}1_{\lambda_1})+k_3\mA^{\text{e}}(3_{\lambda_3}4_{\lambda_4}1_{\lambda_1}2_{\lambda_2})\\
        &+k_4\mA^{\text{e}}(4_{\lambda_4}1_{\lambda_1}2_{\lambda_2}3_{\lambda_3})+k_5\mA^{\text{e}}(1_{\lambda_1}3_{\lambda_3}2_{\lambda_2}4_{\lambda_4})+k_6\mA^{\text{e}}(2_{\lambda_2}4_{\lambda_4}1_{\lambda_1}3_{\lambda_3})+h^{(n,m)}_{(\lambda_1,\lambda_2,\lambda_3,\lambda_4)}\,,
        \end{split}
\end{align}
where $\mA^{\text{e}}$ denotes the exchange diagram that, in the light-cone gauge, takes the form
\begin{align}\label{Paper3-exchange_diagram}
    \begin{split}
    \mA^{\text{e}}(1_{\lambda_1}2_{\lambda_2}3_{\lambda_3}4_{\lambda_4})=&
    \frac{\PPb_{12}^{\lambda_1+\lambda_2+\omega}}{\beta^{\lambda_1}_1\beta^{\lambda_2}_2(-\beta_1-\beta_2)^{\omega}}\frac{1}{(q_1+q_2)^2}\frac{\PP_{34}^{-\lambda_3-\lambda_4+\omega}}{\beta^{-\lambda_3}_3\beta^{-\lambda_4}_4(\beta_1+\beta_2)^{\omega}}\\ 
    =&\frac{(-)^{\omega}}{2}\frac{\PPb_{12}^{\lambda_1+\lambda_2+\omega-1}\PP_{34}^{-\lambda_3-\lambda_4+\omega}}{\PP_{12}(\beta_1+\beta_2)^{2\omega}}\frac{\beta_3^{\lambda_3}\beta_4^{\lambda_4}}{\beta_1^{\lambda_1-1}\beta_2^{\lambda_2-1}}\,.
    \end{split}
\end{align}
The sum over six distinct contributions may seem unusual when compared to the standard $s$-, $t$-, and $u$-channel exchange contributions in the covariant formulation. The reason for more diagrams is that the same diagram can have $-+$ or $+-$ on the internal line, and these are different diagrams in the light-cone gauge. In the case of singlet fields (as for gravity), \eqref{Paper3-exchange_diagram} coincides with the sum over the $s$-, $t$-, and $u$-channel contributions in the covariant language:
\begin{equation}
    \mA(1_{\lambda_1}2_{\lambda_2}3_{\lambda_3}4_{\lambda_4})=\mA_s(1_{\lambda_1}2_{\lambda_2}3_{\lambda_3}4_{\lambda_4})+\mA_t(1_{\lambda_1}2_{\lambda_2}3_{\lambda_3}4_{\lambda_4})+\mA_u(1_{\lambda_1}2_{\lambda_2}3_{\lambda_3}4_{\lambda_4})\,.
\end{equation}
While for color amplitudes, it coincides with the expression in terms of color-ordered amplitudes as
\begin{equation}\label{Paper3-color_ordered}
    \mA(12\cdots n)=\sum_{\sigma\in S_n/Z_n}\mathrm{Tr} (T^{a_{\sigma_1}}T^{a_{\sigma_2}}\cdots T^{a_{\sigma_n}})\mtA(\sigma_1\sigma_2\cdots\sigma_n)\,.
\end{equation}
We now proceed to compute the amplitudes in the light-cone formulation for Yang–Mills theory and gravity.

\paragraph{Yang-Mills theory.} 
For the color ordering $[1^+2^+3^-4^-]$, the four-point amplitude obtained from \eqref{Paper3-total_A} and \eqref{Paper3-exchange_diagram}, with the coefficients $k_i$ fixed as in Table~\ref{Paper3-tab(1,1)}, reads
\begin{equation}
    \mA(1^+2^+3^-4^-)=\mA^{\text{e}}(1^+2^+3^-4^-)+\mA^{\text{e}}(2^+3^-4^-1^+)+\mA^{\text{e}}(4^-1^+2^+3^-)+h^{(1,1)}_{[+,+,-,-]}\,,
\end{equation}
where we set $k_1=1$. By substituting the values of the exchanges and the quartic vertex \eqref{Paper3-h4_1_YM}, we obtain
\begin{align}
    \mtA(1^+2^+3^-4^-)&=\frac{\langle 34\rangle^4}{\langle 12\rangle\langle 23\rangle\langle 34\rangle\langle 41\rangle}\,.
\end{align}
Similarly, for the color ordering $[1^+2^-3^+4^-]$, we have 
\begin{align}
    \begin{split}
    \mA(1^+2^-3^+4^-)&=\mA^{\text{e}}(1^+2^-3^+4^-)+\mA^{\text{e}}(2^-3^+4^-1^+)\\
    &\;\;\;+\mA^{\text{e}}(3^+4^-1^+2^-)+\mA^{\text{e}}(4^-1^+2^-3^+)+h^{(1,1)}_{[+,-,+,-]}\,.
    \end{split}
\end{align}
By substituting the values of the exchanges and the quartic vertex \eqref{Paper3-h4_2_YM}, we obtain
\begin{align}
    \mtA(1^+2^-3^+4^-)&=\frac{\langle 24\rangle^4}{\langle 12\rangle\langle 23\rangle\langle 34\rangle\langle 41\rangle}\,.
\end{align}
\paragraph{Gravity.} For gravity, the total four-point amplitude corresponds to
\begin{align}
    \begin{split}
    \mA(1^+2^+3^-4^-)&=\mA^{\text{e}}(1^+2^+3^-4^-)+\mA^{\text{e}}(1^+4^-3^-2^+)+\mA^{\text{e}}(3^-2^+1^+4^-)\\
    &\;\;\;+\mA^{\text{e}}(1^+3^-2^+4^-)+\mA^{\text{e}}(2^+4^-1^+3^-)+h^{(2,2)}_{(+2,+2,-2,-2)}\,,
    \end{split}
\end{align}
where again we set $k_1=1$. By substituting the values of the exchanges and the quartic vertex \eqref{Paper3-GR}, we obtain
\begin{align}
    \mA(1^+2^+3^-4^-)&=-\frac{[12]^4\langle 34\rangle^4}{s t u}\,.
\end{align}

\section{Quartic vertices for higher-derivative theories}\label{Paper3-section6}

We now turn to the case of higher-derivative interactions. As we will see, the solution space grows with derivatives, highlighting the deep relation between higher-spin fields and the necessity of higher derivatives.

As in the lower-derivative cases, we begin with an integrability-based analysis. Local higher-derivative quartic vertices are considerably more cumbersome, and for this reason, we avoid presenting them explicitly. Tables for $(3,1)$ \ref{Paper3-tab(3,1)}, $(3,2)$ \ref{Paper3-tab(3,2)}, and $(3,3)$ \ref{Paper3-tab(3,3)} quartic vertices satisfying the quartic constraints are given below.

\begin{table}[H]
    \centering
    \begin{tabular}{|c|c|c|c|c|c|c|}
\hline
$(\lambda_1,\lambda_2,\omega,\lambda_3,\lambda_4)$ & $k_2=(2341)$ & $k_3=(3412)$ & $k_4=(4123)$ & $k_5=(1324)$ & $k_6=(2413)$ \\ \hline
$(-1,2,2,-1,2)$ & $k_2$ & $k_1$ & $k_2$ & $0$ & $k_1-k_2$ \\ \hline
$(-1,2,2,0,1)$ & $k_2$ & $k_1$ & $0$ & $0$ & $k_1-k_2$ \\ \hline
$(0,1,2,0,1)$ & $k_2$ & $k_1$ & $k_2$ & $0$ & $k_6$ \\ \hline
$(0,2,1,0,0)$  & $k_2$ & $0$ & $0$ & $0$ & $k_1-k_2$ \\ \hline
$(1,1,1,-1,1)$  & $0$ & $0$ & $k_4$ & $0$ & $k_1-k_4$ \\ \hline
    \end{tabular}
    \caption{$(3,1)$ quartic vertices satisfying the quartic constraints.}
    \label{Paper3-tab(3,1)}
\end{table}
\begin{table}[H]
    \centering
    \begin{tabular}{|c|c|c|c|c|c|c|}
\hline
$(\lambda_1,\lambda_2,\omega,\lambda_3,\lambda_4)$ & $k_2=(2341)$ & $k_3=(3412)$ & $k_4=(4123)$ & $k_5=(1324)$ & $k_6=(2413)$ \\ \hline
$(-1,2,2,-2,2)$ & $0$ & $0$ & $k_1$ & $0$ & $k_1$ \\ \hline
$(-1,2,2,-1,1)$  & $k_2$ & $0$ & $0$ & $0$ & $k_1-k_2$ \\ \hline
$(-1,2,2,0,0)$  & $k_2$ & $0$ & $0$ & $0$ & $k_1-k_2$ \\ \hline
$(0,1,2,-2,2)$  & $0$ & $0$ & $k_1$ & $0$ & $k_1$ \\ \hline
$(0,1,2,-1,1)$  & $0$ & $0$ & $k_4$ & $0$ & $k_6$ \\ \hline
$(0,1,2,0,0)$  & $k_2$ & $0$ & $0$ & $0$ & $k_6$ \\ \hline
$(1,1,1,-2,1)$  & $0$ & $0$ & $k_1$ & $0$ & $k_1$ \\ \hline
    \end{tabular}
    \caption{$(3,2)$ quartic vertices satisfying the quartic constraints.}
    \label{Paper3-tab(3,2)}
\end{table}
\begin{table}[H]
    \centering
    \begin{tabular}{|c|c|c|c|c|c|c|}
\hline
$(\lambda_1,\lambda_2,\omega,\lambda_3,\lambda_4)$ & $k_2=(2341)$ & $k_3=(3412)$ & $k_4=(4123)$ & $k_5=(1324)$ & $k_6=(2413)$ \\ \hline
$(-2,2,3,-2,2)$ & $k_2$ & $k_1$ & $k_2$ & $0$ & $k_1-k_2$ \\ \hline
$(-2,2,3,-1,1)$ & $k_2$ & $k_1$ & $0$ & $0$ & $k_1-k_2$ \\ \hline
$(-2,2,3,0,0)$ & $k_2$ & $k_1$ & $0$ & $0$ & $k_1-k_2$ \\ \hline
$(-1,1,3,-1,1)$ & $k_2$ & $k_1$ & $k_2$ & $0$ & $k_6$ \\ \hline
$(-1,1,3,0,0)$ & $k_2$ & $k_1$ & $0$ & $0$ & $k_6$ \\ \hline
$(-1,2,2,-1,0)$ & $k_2$ & $0$ & $0$ & $0$ & $k_1-k_2$ \\ \hline
$(0,0,3,0,0)$ & $k_2$ & $k_1$ & $k_2$ & $k_5$ & $k_5$ \\ \hline
$(0,1,2,-2,1)$ & $0$ & $0$ & $k_4$ & $0$ & $k_1-k_4$ \\ \hline
    \end{tabular}
    \caption{$(3,3)$ quartic vertices satisfying the quartic constraints.}\label{Paper3-tab(3,3)}
\end{table}
\noindent
Additional tables are presented in Appendix~\ref{Paper3-AppendixF}, where we display results up to $(4,4)$. Although our explicit analysis extends to quartic vertices up to $(8,8)$ (with additional checks beyond this range), we do not include those tables here, as they are too lengthy. Nevertheless, we used them as well to identify patterns across the various cases, allowing us to infer the general behaviour of a generic $(n,m)$ quartic vertex, which we summarise below.

One possible way to present the results would be to write tables for a generic $(n,m)$ quartic vertex. Although this could be done in principle, such an approach neither provides intuition about the underlying structure nor offers a convenient formulation for further analysis. Therefore, we proceed differently. First, we describe the pattern that emerges from inspecting all tables up to $(8,8)$. Once this is established, we present another formulation of the result, which will prove useful for addressing various questions.

\begin{itemize}
    \item From the complete list of $(n,m)$ quartic vertices satisfying the quartic constraint \eqref{Paper3-quartic_system}, we focus on those containing the maximal and minimal external helicities. They correspond to the first line in the various tables we present.\footnote{This is the condition which, for $(1,1)$, gives the consistency of the Yang–Mills cubic vertices, and for $(2,2)$, ensures the consistency of the gravitational cubic vertices.} This depends on the specific type of quartic vertex, whether it is of the form $(even,even)$, $(odd,odd)$, $(even,odd)$, or $(odd,even)$. Let us write them here:
    \begin{subequations}
    \begin{align}
        &(even,even):&
        &(1234)=\left(-\frac{m+2}{2},\frac{n+2}{2},\frac{n+m}{2},-\frac{m+2}{2},\frac{n+2}{2}\right)\,,\\
        &\phantom{(even,even):}&
        &(2413)=\left(\frac{n+2}{2},\frac{n+2}{2},-2,-\frac{m+2}{2},-\frac{m+2}{2}\right)\,,\\
        &(odd,odd):&
        &(1234)=\left(-\frac{m+1}{2},\frac{n+1}{2},\frac{n+m}{2},-\frac{m+1}{2},\frac{n+1}{2}\right)\,,\\
        &\phantom{(odd,odd)}&
        &(2413)=\left(\frac{n+1}{2},\frac{n+1}{2},-1,-\frac{m+1}{2},-\frac{m+1}{2}\right)\,,\\
        &(even,odd):&
        &(1234)=\left(-\frac{m+1}{2},\frac{n+2}{2},\frac{n+m-1}{2},-\frac{m+1}{2},\frac{n}{2}\right)\,,\\
        &\phantom{(even,odd)}&
        &(2413)=\left(\frac{n+2}{2},\frac{n}{2},-1,-\frac{m+1}{2},-\frac{m+1}{2}\right)\,,\\
        &(odd,even):&
        &(1234)=\left(-\frac{m}{2},\frac{n+1}{2},\frac{n+m-1}{2},-\frac{m+2}{2},\frac{n+1}{2}\right)\,,\\
        &\phantom{(odd,even)}&
        &(2413)=\left(\frac{n+1}{2},\frac{n+1}{2},-1,-\frac{m}{2},-\frac{m+2}{2}\right)\,,
    \end{align}
    \end{subequations}
    where we omit the remaining exchanges, as they are straightforward to obtain. We display only $(1234)$ and $(2413)$, which also highlight the special role of spin-$1$ and spin-$2$.
    We now select from them the maximal and minimal external helicities, which we call respectively $M_{\text{ext}}$ and $m_{\text{ext}}$, and the maximal and minimal internal (exchanged) helicities, which we call respectively $M_{\text{int}}$ and $m_{\text{int}}$.
    \begin{align}
        &(even,even)&
        &M_{\text{ext}}=\frac{n+2}{2}\,,&
        &m_{\text{ext}}=-\frac{m+2}{2}\,,&
        &M_{\text{int}}=\frac{n+m}{2}\,,&
        &m_{\text{int}}=-2\,,\\
        &(odd,odd)&
        &M_{\text{ext}}=\frac{n+1}{2}\,,&
        &m_{\text{ext}}=-\frac{m+1}{2}\,,&
        &M_{\text{int}}=\frac{n+m}{2}\,,&
        &m_{\text{int}}=-1\,,\\
        &(even,odd)&
        &M_{\text{ext}}=\frac{n+2}{2}\,,&
        &m_{\text{ext}}=-\frac{m+2}{2}\,,&
        &M_{\text{int}}=\frac{n+m-1}{2}\,,&
        &m_{\text{int}}=-1\,,\\
        &(odd,even)&
        &M_{\text{ext}}=\frac{n+2}{2}\,,&
        &m_{\text{ext}}=-\frac{m+2}{2}\,,&
        &M_{\text{int}}=\frac{n+m-1}{2}\,,&
        &m_{\text{int}}=-1\,.
    \end{align}
    \item We now outline a set of rules that determine the specific elements appearing in the table.
    \begin{enumerate}
        \item All the exchanges $(\lambda_1,\lambda_2,\omega,\lambda_3,\lambda_4)$ involving either external or exchanged helicities that exceed the maximal value or fall below the minimal value identified in the previous point are set to zero.\footnote{This implies that no solution to the quartic constraints with a non-zero exchange contribution exists, in the sense that no local quartic vertex can satisfy the system.} Therefore, exchange diagrams must satisfy
        \begin{align}\label{Paper3-rule1}
        &m_{\text{ext}}\leq\lambda_1,\lambda_2,\lambda_3,\lambda_4\leq M_{\text{ext}}\,,&
        &m_{\text{int}}\leq \omega\leq M_{\text{int}}
        \end{align}

        \item If at least one of the external fields has either maximal or minimal helicity, the existence of a solution requires turning on all exchange diagrams allowed by the first rule. 

        \item For exchange diagrams allowed by the first rule, where none of the external helicities are extremal, the following applies. If an exchange diagram $(1234)$ is allowed and the corresponding diagram $(3412)$ is also allowed, then both must be included with the same value of $k_i$. Otherwise, $(1234)$ appears alone.

    \end{enumerate}
\end{itemize}
We now make a few interesting comments using the results above:
\begin{itemize}
    \item As the number of derivatives of the cubic vertices involved and the derivatives of the quartic vertices (i.e. $n+m$ and $n+m-2$, respectively) increases, the number of consistent quartic vertices does the same. This is a characteristic feature of higher-spin theories.

    \item The quartic consistency of the cubic non-abelian self-interaction for a field of helicity $\lambda$ of the type $C^{\lambda,\lambda,-\lambda}$ is satisfied only for helicities $\lambda=1,2$. Indeed, for $\lambda$ even, a $(\lambda,\lambda)$ quartic vertex has, as external helicities, at most $\frac{\lambda+2}{2}$, while $(\lambda,-\lambda,\lambda,\lambda-\lambda)$ has $\lambda$ as external helicity. Therefore, consistency requires $\frac{\lambda+2}{2}=\lambda\implies \lambda=2$. The same happens for $\lambda$ odd, replacing $2$ with $1$.
    
    \item For $(n,n)$ quartic vertices, there exists a two parameter family of solutions where the exchange diagrams are of the form $(-a,a,n,-b,b)$, with $a,b\leq\frac{n+2}{2}$ for $n$ even and $a,b\leq\frac{n+1}{2}$ for $n$ odd. However, even though the local quartic vertex exists to solve the quartic constraint, the cubic vertices do not form a consistent local theory. Indeed, if $a=b$ is immediate to see that we would need to satisfy a new constraint because of the exchange $(n,-a,a,a,-n)$. However, the associated quartic constraint does not admit a solution because $n$ exceeds the maximal helicity (with the exceptions of $\lambda=1,2$).
    
    If instead $a\neq b$, one must consider the holomorphic ($CC$) quartic constraint generated by the exchange $(-a,a,n,n,n)$. Following \cite{Serrani:2025owx}, solving this holomorphic constraint requires, among others, the presence of the cubic vertex $(-n,n,n)$, which would imply the spectrum of the full chiral higher-spin theory. Consequently, one could then construct arbitrary ($C\bar{C}$) exchanges using the $\bar{C}$ cubic vertex $(-n,-b,b)$, many of which fail to satisfy the quartic constraint.

    Therefore, for instance, even though Tables \ref{Paper3-tab(3,1)}--\ref{Paper3-tab(3,3)} might suggest that color gravity could exist, a full analysis of all newly generated quartic constraints shows that no local solutions are possible for a multi-graviton theory.
\end{itemize}
The last comment was made to draw attention to the meaning of the tables. 
Knowing the solutions to all quartic constraints provides information about the consistency of cubic vertices with Lorentz invariance and also tells us which quartic higher-spin amplitudes turn out to be local. However, this is not yet enough to construct a consistent local higher-spin theory at the quartic order. Given a set of cubic couplings, one must examine all the quartic constraints they generate, and every such constraint must admit at least one local solution.

This situation is reminiscent of the holomorphic constraint discussed in \cite{Serrani:2025owx}. The main difference is that, in that case, by successively introducing additional cubic vertices --- typically, but not always, leading to the full chiral higher-spin spectrum with all holomorphic cubic vertices turned on --- a solution always exists. Here, by contrast, identifying a set of cubic couplings that solves all quartic constraints is not always possible, and determining all admissible local higher-spin theories is considerably more challenging.

Moreover, recall that we must always verify that the cubic vertices that solve the non-holomorphic quartic constraint also solve the holomorphic one. A complete solution to the quartic light-cone constraint is obtained only when the cubic couplings satisfy both constraints simultaneously. We will illustrate how this works through specific examples in the next part. To approach this more intricate problem, we will reformulate the solutions to the quartic constraint in a more natural way.

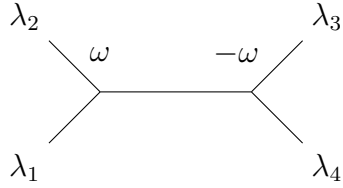
\begin{figure}[H]
    \centering
    \begin{tikzpicture}
        \begin{feynman}
            \vertex (i1) at (-6, 1) {\(\lambda_2\)};
            \vertex (i2) at (-6,-1) {\(\lambda_1\)};
            \vertex (i3) at (-2, 1) {\(\lambda_3\)};
            \vertex (i4) at (-2,-1) {\(\lambda_4\)};
            \vertex (v1) at (-5, 0);
            \vertex (v3) at (-3, 0);
            \vertex at (-5, 0.5) {\(\omega\)};
            \vertex at (-3.2, 0.5) {\(-\omega\)};
            \diagram* {
                (i1) -- (v1),
                (i2) -- (v1),
                (v1) -- [plain] (v3),
                (v3) -- (i3),
                (v3) -- (i4),
            };
        \end{feynman}
    \end{tikzpicture}
        \caption{Exchange diagrams for a generic pair $C\bar{C}$.}
    \label{Paper3-figure_generic}
\end{figure}
\noindent
\paragraph{Reformulation.} A reformulation of the solutions is as follows. Starting from an $(n,m)$ exchange diagram $(\lambda_1,\lambda_2,\omega,\lambda_3,\lambda_4)$ as in Figure \ref{Paper3-figure_generic}, where $n=\lambda_{12}+\omega$ and $m=-\lambda_{34}+\omega$. 
We start from the condition given in \eqref{Paper3-rule1}. In particular, we focus on the stronger requirement obtained by excluding both maximal and minimal external helicities. Under this assumption, we find
\begin{subequations}
\begin{align}\label{Paper3-raw_1}
    &-\frac{m}{2}\leq\lambda_1\leq \frac{n}{2}\,,&
    &-\frac{m}{2}\leq\lambda_2\leq \frac{n}{2}\,,&
    &-\frac{m}{2}\leq\lambda_3\leq \frac{n}{2}\,,&
    &-\frac{m}{2}\leq\lambda_4\leq \frac{n}{2}\,,\\\label{Paper3-raw2}
    &0\leq\omega\leq \frac{n+m}{2}\,,&
    & n,m>0\,,
\end{align}
\end{subequations}
where $\omega>0$ is a consequence of \eqref{Paper3-raw_1}. These conditions are equivalent to the following minimal set of conditions 
\begin{subequations}\label{Paper3-kinda_trinagular}
\begin{align}\label{Paper3-trinagular}
    &\lambda_1\leq \lambda_2+\omega\,,&
    &\lambda_2\leq \lambda_1+\omega\,,&
    &\lambda_3\leq \lambda_4+\omega\,,&
    &\lambda_4\leq \lambda_3+\omega\,,\\\label{Paper3-extra_condition}
    &\lambda_{12}+\omega>0\,,&
    &-\lambda_{34}+\omega> 0\,,&
    &\lambda_{12}\geq \lambda_{34}\,.
\end{align}
\end{subequations}
In this case, a solution to the system of quartic constraints \eqref{Paper3-quartic_system} does exist without the need for other exchange diagrams.\footnote{In the tables, this exchange would correspond to a free coefficient $k_i$.} In the case $\lambda_{12}=\lambda_{34}$, both $(1234)$ and $(3412)$ satisfy the condition. Indeed, equation \eqref{Paper3-kinda_trinagular} becomes symmetric under the exchange $(\lambda_1,\lambda_2)\leftrightarrow (\lambda_3,\lambda_4)$. In this situation, both exchange diagrams must therefore be included, with the same value of $k_i$. This is an analogue of the third rule discussed above. The conditions \eqref{Paper3-kinda_trinagular} show a clear difference between how the helicity in the exchange and in the external states is constrained.  
\begin{figure}[H]
    \centering
    \begin{tikzpicture}
        \begin{feynman}
            \vertex (i1) at (-6, 1) {\(\lambda_2\)};
            \vertex (i2) at (-6,-1) {\(\lambda_1\)};
            \vertex (i3) at (-2, 1) {\(\lambda_3\)};
            \vertex (i4) at (-2,-1) {\(\lambda_4\)};

            \vertex (ii1) at (1, 1) {\(\lambda_2\)};
            \vertex (ii2) at (1,-1) {\(\lambda_1\)};
            \vertex (ii3) at (3, 1) {\(\lambda_3\)};
            \vertex (ii4) at (3,-1) {\(\lambda_4\)};

            \vertex (v1) at (-5, 0);
            \vertex (v3) at (-3, 0);

            \node at (-1,0) {\Large $+$};
            
            \vertex at (-5, 0.5) {\(\omega\)};
            \vertex at (-3.2, 0.5) {\(-\omega\)};
            
            \diagram* {
                (i1) -- (v1),
                (i2) -- (v1),
                (v1) -- [plain] (v3),
                (v3) -- (i3),
                (v3) -- (i4),
            };
            \diagram* {
                (ii1) -- (ii4),
                (ii2) -- (ii3),
            };

        \end{feynman}
    \end{tikzpicture}
        \caption{A quartic vertex with a single exchange exists when the conditions \eqref{Paper3-kinda_trinagular} are satisfied.}
\end{figure}
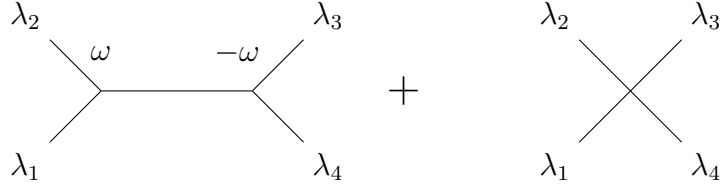
\noindent
The remaining possibilities, which follow from the first and second rules stated above, arise when some of the external helicities are allowed to be either maximal or minimal, thereby violating the conditions in \eqref{Paper3-kinda_trinagular}. This corresponds to configurations in which at least one of the external helicities (any of the four) surpasses \eqref{Paper3-kinda_trinagular}. Then, at least one of these conditions is true
\begin{equation}\label{Paper3-surpass_trinagular}
    \lambda_1=\lambda_2+\omega+n_1\,,\quad
    \lambda_2=\lambda_1+\omega+n_2\,,\quad
    \lambda_3=\lambda_4+\omega+n_3\,,\quad
    \lambda_4=\lambda_3+\omega+n_4\,,
\end{equation}
where $n_{1,2,3,4}=1,2$. In this case, the quartic vertex exists if and only if we turn on all other exchanges that satisfy \eqref{Paper3-kinda_trinagular}, with the possibility for some of the helicities to satisfy \eqref{Paper3-surpass_trinagular}.

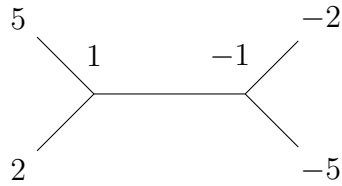
\begin{figure}[H]
    \centering
    \begin{tikzpicture}
        \begin{feynman}
            \vertex (i1) at (-6, 1) {\(5\)};
            \vertex (i2) at (-6,-1) {\(2\)};
            \vertex (i3) at (-2, 1) {\(-2\)};
            \vertex (i4) at (-2,-1) {\(-5\)};
            \vertex (v1) at (-5, 0);
            \vertex (v3) at (-3, 0);
            \vertex at (-5, 0.5) {\(1\)};
            \vertex at (-3.2, 0.5) {\(-1\)};
            \diagram* {
                (i1) -- (v1),
                (i2) -- (v1),
                (v1) -- [plain] (v3),
                (v3) -- (i3),
                (v3) -- (i4),
            };
        \end{feynman}
    \end{tikzpicture}
        \caption{Exchange diagrams between $C^{2,5,1}$ and $\bar{C}^{-2,-5,-1}$.}
        \label{Paper3-figExample}
\end{figure}

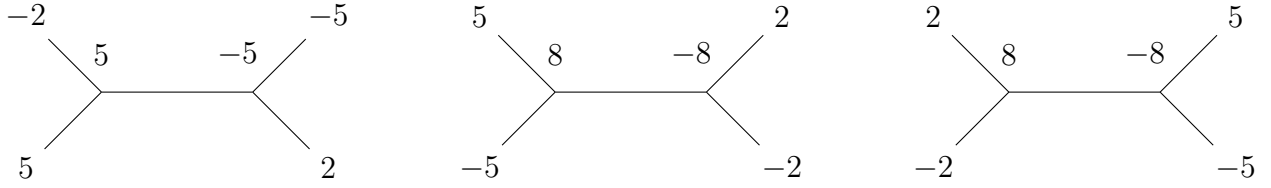
\begin{figure}[H]
    \centering
    \begin{tikzpicture}
        \begin{feynman}
            \vertex (i1) at (-6, 1) {\(-2\)};
            \vertex (i2) at (-6,-1) {\(5\)};
            \vertex (i3) at (-2, 1) {\(-5\)};
            \vertex (i4) at (-2,-1) {\(2\)};

            \vertex (ii1) at (0, 1) {\(5\)};
            \vertex (ii2) at (0,-1) {\(-5\)};
            \vertex (ii3) at (4, 1) {\(2\)};
            \vertex (ii4) at (4,-1) {\(-2\)};

            \vertex (iii1) at (6, 1) {\(2\)};
            \vertex (iii2) at (6,-1) {\(-2\)};
            \vertex (iii3) at (10, 1) {\(5\)};
            \vertex (iii4) at (10,-1) {\(-5\)};
            
            \vertex (v1) at (-5, 0);
            \vertex (v3) at (-3, 0);

            \vertex (vv1) at (1, 0);
            \vertex (vv3) at (3, 0);
            
            \vertex (vvv1) at (7, 0);
            \vertex (vvv3) at (9, 0);
            
            \vertex at (-5, 0.5) {\(5\)};
            \vertex at (-3.2, 0.5) {\(-5\)};

            \vertex at (1, 0.5) {\(8\)};
            \vertex at (2.8, 0.5) {\(-8\)};

            \vertex at (7, 0.5) {\(8\)};
            \vertex at (8.8, 0.5) {\(-8\)};

            \diagram* {
                (i1) -- (v1),
                (i2) -- (v1),
                (v1) -- [plain] (v3),
                (v3) -- (i3),
                (v3) -- (i4),
            };
            \diagram* {
                (ii1) -- (vv1),
                (ii2) -- (vv1),
                (vv1) -- [plain] (vv3),
                (vv3) -- (ii3),
                (vv3) -- (ii4),
            };
            \diagram* {
                (iii1) -- (vvv1),
                (iii2) -- (vvv1),
                (vvv1) -- [plain] (vvv3),
                (vvv3) -- (iii3),
                (vvv3) -- (iii4),
            };
        \end{feynman}
    \end{tikzpicture}
        \caption{Relevant exchange diagrams for parity invariant pairs of cubic abelian vertices.}
        \label{Paper3-three_diagrams}
\end{figure}

Let us provide an explicit example for the $(8,8)$ quartic vertex in Figure \ref{Paper3-figExample}.
\noindent
This exchange diagram respects the conditions \eqref{Paper3-kinda_trinagular}, with some of the fields having maximal helicity \eqref{Paper3-surpass_trinagular}. In particular, we have 
\begin{subequations}
\begin{align}
    &\lambda_1\leq\lambda_2+\omega\,,\quad
    \lambda_2=\lambda_1+\omega+2\,,\quad
    \lambda_3=\lambda_4+\omega+2\,,\quad
    \lambda_4\leq \lambda_3+\omega\,,\\
    &\lambda_{12}+\omega>0\,,\quad
    -\lambda_{34}+\omega>0\,,\quad
    \lambda_{12}=\lambda_{34}\,.
\end{align}
\end{subequations}
We then need to consider all other possible exchange diagrams, obtained by permuting the various external helicities, that still satisfy the conditions \eqref{Paper3-kinda_trinagular} with some maximal helicity \eqref{Paper3-surpass_trinagular}. These diagrams are represented in Figure \ref{Paper3-three_diagrams}.
\noindent
If we activate all the cubic vertices in these diagrams, we can solve the system of quartic constraints \eqref{Paper3-quartic_system}. Three possible structures arise:
\begin{enumerate}
    \item If the exchange diagram satisfies \eqref{Paper3-kinda_trinagular}, it is sufficient to consider only the $s$-channel diagram $(1234)$, and, if allowed, $(3412)$, with the same $k_i$. As we will see, the amplitudes obtained by summing the exchange contributions and the quartic vertex exhibit a $\frac{1}{s}$ pole. We refer to these as single-channel amplitudes.

    \item If the exchange diagram has at least one external helicity that surpasses \eqref{Paper3-kinda_trinagular} with at most $n_i=1$, the quartic vertices exhibit a structure similar to that of Yang–Mills theory. There are three possible quartic vertices associated with the three channel pairs $st$, $us$, and $tu$. Only two of them are independent, corresponding to the two cyclic orderings $[1234]$ and $[1324]$. In Yang–Mills theory, these are the standard color-ordered amplitudes with helicity configurations $[++--]$ and $[+-+-]$. The amplitudes obtained by summing the exchange contributions and the quartic vertex then exhibit poles of the form $\frac{1}{st}$, $\frac{1}{us}$, and $\frac{1}{tu}$, respectively. We refer to these as Yang–Mills-like (YM-like) amplitudes.

    \item If the exchange diagram has at least one external helicity that surpasses \eqref{Paper3-kinda_trinagular} with $n_i=2$, the quartic vertices exhibit a structure similar to that of gravity. In this case, there is a single quartic vertex that requires the presence of the $s$-, $t$-, and $u$-channel exchange diagrams. The amplitudes obtained by summing the exchange contributions and the quartic vertex then display a $\frac{1}{stu}$ pole. We refer to these as gravity-like (GR-like) amplitudes.
\end{enumerate}

\paragraph{Homogeneous solutions for the quartic vertices.} One aspect that has not yet been discussed is the study of the homogeneous solutions of the quartic constraint, i.e. solutions to \eqref{Paper3-quartic_system} without any exchange contributions. Once again, we can employ the integrability method. We show tables up to $(4,4)$ in Appendix \ref{Paper3-AppendixF}. Performing this analysis, we find that a homogeneous solution exists if the following conditions are satisfied:
\begin{equation}\label{Paper3-Homo_constraints}
    \lambda_1\leq \lambda_2+\omega-1\,,\quad
    \lambda_2\leq \lambda_1+\omega-1\,,\quad
    \lambda_3\leq \lambda_4+\omega-1\,,\quad
    \lambda_4\leq \lambda_3+\omega-1\,.
\end{equation}
Here, $\omega$ no longer represents an exchange field; instead, it parametrizes the number of derivatives carried by the quartic vertex, denoted by $D$. Indeed $D=n+m-2=\lambda_{12}-\lambda_{34}+2\omega-2$. Equivalently, one may eliminate $\omega$ in favour of $D$, and we get
\begin{equation}
    D-3\lambda_1+\lambda_{234}\geq 0\,,\quad
    D-3\lambda_2+\lambda_{134}\geq 0\,,\quad
    D+3\lambda_3-\lambda_{124}\geq 0\,,\quad
    D+3\lambda_4-\lambda_{123}\geq 0\,.
\end{equation}
This pattern will become clearer when we study four-point amplitudes in detail in Section \ref{Paper3-section7}.

\subsection{Initial results}
We start by collecting a series of initial results obtained using the reformulation of the full set of local solutions to the quartic constraint.

\begin{figure}[H]
    \centering
    \begin{tikzpicture}
        \begin{feynman}
            \vertex (i1) at (-6, 1) {\(\lambda_2\)};
            \vertex (i2) at (-6,-1) {\(\lambda_1\)};
            \vertex (i3) at (-2, 1) {\(-\lambda_1\)};
            \vertex (i4) at (-2,-1) {\(-\lambda_2\)};

            \vertex (ii1) at (0, 1) {\(\lambda_3\)};
            \vertex (ii2) at (0,-1) {\(\lambda_2\)};
            \vertex (ii3) at (4, 1) {\(-\lambda_3\)};
            \vertex (ii4) at (4,-1) {\(-\lambda_2\)};

            \vertex (iii1) at (6, 1) {\(\lambda_3\)};
            \vertex (iii2) at (6,-1) {\(\lambda_1\)};
            \vertex (iii3) at (10, 1) {\(-\lambda_3\)};
            \vertex (iii4) at (10,-1) {\(-\lambda_1\)};
            
            \vertex (v1) at (-5, 0);
            \vertex (v3) at (-3, 0);

            \vertex (vv1) at (1, 0);
            \vertex (vv3) at (3, 0);
            
            \vertex (vvv1) at (7, 0);
            \vertex (vvv3) at (9, 0);
            
            \vertex at (-5, 0.5) {\(\lambda_3\)};
            \vertex at (-3.2, 0.5) {\(-\lambda_3\)};

            \vertex at (1, 0.5) {\(\lambda_1\)};
            \vertex at (2.8, 0.5) {\(-\lambda_1\)};

            \vertex at (7, 0.5) {\(\lambda_2\)};
            \vertex at (8.8, 0.5) {\(-\lambda_2\)};

            \diagram* {
                (i1) -- (v1),
                (i2) -- (v1),
                (v1) -- [plain] (v3),
                (v3) -- (i3),
                (v3) -- (i4),
            };
            \diagram* {
                (ii1) -- (vv1),
                (ii2) -- (vv1),
                (vv1) -- [plain] (vv3),
                (vv3) -- (ii3),
                (vv3) -- (ii4),
            };
            \diagram* {
                (iii1) -- (vvv1),
                (iii2) -- (vvv1),
                (vvv1) -- [plain] (vvv3),
                (vvv3) -- (iii3),
                (vvv3) -- (iii4),
            };
        \end{feynman}
    \end{tikzpicture}
        \caption{$C\bar{C}$ exchange diagrams for parity invariant pairs of cubic abelian vertices.}
    \label{Paper3-figure_abelian}

    \end{figure}
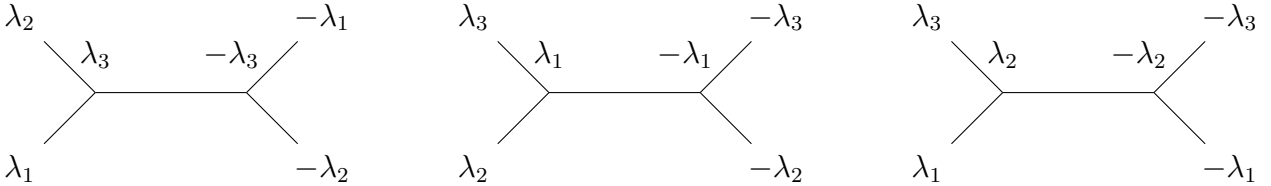

\paragraph{Abelian cubic vertices.}
Here, we study the consistency of abelian cubic vertices. As already discussed in \cite{Serrani:2025owx}, where the holomorphic quartic constraint was studied in detail, self-dual abelian cubic vertices are consistent on their own because they never generate any $CC$ exchange diagrams. In contrast, in the presence of a scalar field, $CC$ exchanges can occur, making the holomorphic constraint nontrivial. The only exception is the special case of the pair $C^{\lambda_1,\lambda_1,0}C^{0,\lambda_2,\lambda_2}$, which is unconstrained: in this case, the holomorphic constraint vanishes identically. However, once we consider the presence of both holomorphic and anti-holomorphic abelian vertices, they start to form $C\bar{C}$ diagrams, and we must then consider the non-holomorphic constraint. Let us first examine the case of a generic parity-related pair of abelian vertices, $C^{\lambda_1,\lambda_2,\lambda_3}$ and $\bar{C}^{-\lambda_1,-\lambda_2,-\lambda_3}$, with $\lambda_{1,2,3}>0$. 
The relevant exchange diagrams for the system of quartic constraints \eqref{Paper3-quartic_system} are presented in Figure \ref{Paper3-figure_abelian}.
\noindent
Following \eqref{Paper3-kinda_trinagular}, we find the conditions
\begin{align}
&\lambda_1 \leq \lambda_2+\lambda_3\,,&
&\lambda_2 \leq \lambda_1+\lambda_3\,,&
&\lambda_3 \leq \lambda_1+\lambda_2\,.
\end{align}
These are the usual triangular inequalities. Therefore, each triangle with integer unit length defines an allowed pair of abelian cubic vertices. Notice the interesting analogy with the holomorphic constraint: here as well, the only allowed scalar cubic couplings are $C^{\lambda,\lambda,0}$.

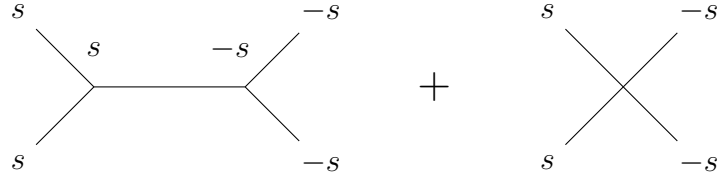
\begin{figure}[H]
    \centering
    \begin{tikzpicture}
        \begin{feynman}
            \vertex (i1) at (-6, 1) {\(s\)};
            \vertex (i2) at (-6,-1) {\(s\)};
            \vertex (i3) at (-2, 1) {\(-s\)};
            \vertex (i4) at (-2,-1) {\(-s\)};

            \vertex (ii1) at (1, 1) {\(s\)};
            \vertex (ii2) at (1,-1) {\(s\)};
            \vertex (ii3) at (3, 1) {\(-s\)};
            \vertex (ii4) at (3,-1) {\(-s\)};

            \vertex (v1) at (-5, 0);
            \vertex (v3) at (-3, 0);

            \node at (-0.5,0) {\Large $+$};
            
            \vertex at (-5, 0.5) {\(s\)};
            \vertex at (-3.2, 0.5) {\(-s\)};
            
            \diagram* {
                (i1) -- (v1),
                (i2) -- (v1),
                (v1) -- [plain] (v3),
                (v3) -- (i3),
                (v3) -- (i4),
            };
            \diagram* {
                (ii1) -- (ii4),
                (ii2) -- (ii3),
            };

        \end{feynman}
    \end{tikzpicture}
        \caption{Linearised curvature terms $(R^s)^3$ for higher-spins are consistent to the quartic level.}
    \label{Paper3-figure_Linearisedcurvature}
\end{figure}

We have just described allowed unitary higher-spin theories with abelian vertices. Particular examples of the one described above are the generalisations of linearised curvatures such as $F^3$ and $R^3$, but for higher-spin fields, see Figure \ref{Paper3-figure_Linearisedcurvature}.
\noindent

The same triangular inequalities were already found in \cite{Damour:1987vm,Damour:1987fp}. The underlying idea is to construct linearised higher-spin curvatures $R^s_{\mu_1\nu_1,...,\mu_s\nu_s}$ (de Wit–Freedman curvatures) and use them to build invariant tensors, such as $R^{s_1}R^{s_2}R^{s_3}$. By counting the number of free indices, one finds that such constructions are possible only if the spins satisfy the triangular inequalities. Let us note that imposing the light-cone gauge in a theory with only abelian cubic interactions can lead to higher order vertices in $H$, as in the process, one has to solve non-linear equations for auxiliary fields. 

\begin{figure}[H]
    \centering
    \begin{tikzpicture}
        \begin{feynman}
            \vertex (i1) at (-6, 1) {\(\lambda_2\)};
            \vertex (i2) at (-6,-1) {\(\lambda_1\)};
            \vertex (i3) at (-2, 1) {\(-\lambda_3\)};
            \vertex (i4) at (-2,-1) {\(-\lambda_4\)};
            \vertex (v1) at (-5, 0);
            \vertex (v3) at (-3, 0);
            \vertex at (-5, 0.5) {\(\omega\)};
            \vertex at (-3.2, 0.5) {\(-\omega\)};
            \diagram* {
                (i1) -- (v1),
                (i2) -- (v1),
                (v1) -- [plain] (v3),
                (v3) -- (i3),
                (v3) -- (i4),
            };
        \end{feynman}
    \end{tikzpicture}
        \caption{Exchange diagram for a pair of cubic abelian vertices with one opposite helicity.}
    \label{Paper3-figure_abelian2}
\end{figure}
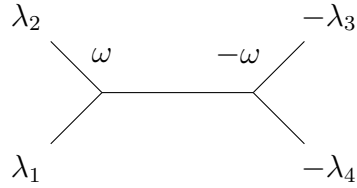

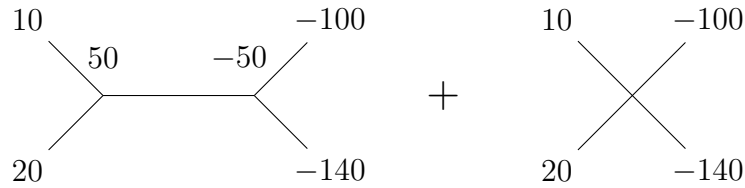
\begin{figure}[H]
    \centering
    \begin{tikzpicture}
        \begin{feynman}
            \vertex (i1) at (-6, 1) {\(10\)};
            \vertex (i2) at (-6,-1) {\(20\)};
            \vertex (i3) at (-2, 1) {\(-100\)};
            \vertex (i4) at (-2,-1) {\(-140\)};

            \vertex (ii1) at (1, 1) {\(10\)};
            \vertex (ii2) at (1,-1) {\(20\)};
            \vertex (ii3) at (3, 1) {\(-100\)};
            \vertex (ii4) at (3,-1) {\(-140\)};

            \vertex (v1) at (-5, 0);
            \vertex (v3) at (-3, 0);

            \node at (-0.5,0) {\Large $+$};
            
            \vertex at (-5, 0.5) {\(50\)};
            \vertex at (-3.2, 0.5) {\(-50\)};
            
            \diagram* {
                (i1) -- (v1),
                (i2) -- (v1),
                (v1) -- [plain] (v3),
                (v3) -- (i3),
                (v3) -- (i4),
            };
            \diagram* {
                (ii1) -- (ii4),
                (ii2) -- (ii3),
            };

        \end{feynman}
    \end{tikzpicture}
        \caption{Example of abelian cubic vertices that satisfy the quartic constraint.}
    \label{Paper3-figure_twoabelian}
\end{figure}

Let us now consider the case of two abelian couplings that do not form a parity-related pair. Let the vertices be $C^{\lambda_1,\lambda_2,\omega}$ and $\bar{C}^{-\omega,-\lambda_3,-\lambda_4}$, with $\lambda_{1,2,3,4}>0$ and $\omega>0$. In this situation, the relevant diagram is the one in Figure \ref{Paper3-figure_abelian2}.
\noindent
The conditions that must be satisfied are those in \eqref{Paper3-kinda_trinagular}, and they are less restrictive than the triangular inequalities mentioned before. An example is the one in Figure \ref{Paper3-figure_twoabelian}.
\noindent
These cubic vertices, which surpass the triangular inequalities, cannot be expressed using linearised gauge curvatures. 

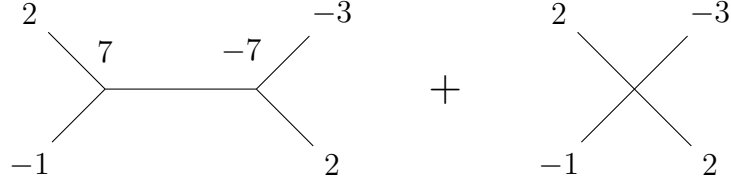
\begin{figure}[H]
    \centering
    \begin{tikzpicture}
        \begin{feynman}
            \vertex (i1) at (-6, 1) {\(2\)};
            \vertex (i2) at (-6,-1) {\(-1\)};
            \vertex (i3) at (-2, 1) {\(-3\)};
            \vertex (i4) at (-2,-1) {\(2\)};

            \vertex (ii1) at (1, 1) {\(2\)};
            \vertex (ii2) at (1,-1) {\(-1\)};
            \vertex (ii3) at (3, 1) {\(-3\)};
            \vertex (ii4) at (3,-1) {\(2\)};

            \vertex (v1) at (-5, 0);
            \vertex (v3) at (-3, 0);

            \node at (-0.5,0) {\Large $+$};
            
            \vertex at (-5, 0.5) {\(7\)};
            \vertex at (-3.2, 0.5) {\(-7\)};
            
            \diagram* {
                (i1) -- (v1),
                (i2) -- (v1),
                (v1) -- [plain] (v3),
                (v3) -- (i3),
                (v3) -- (i4),
            };
            \diagram* {
                (ii1) -- (ii4),
                (ii2) -- (ii3),
            };

        \end{feynman}
    \end{tikzpicture}
        \caption{Example of non-abelian cubic vertices that satisfy the quartic constraint.}
    \label{Paper3-figure_nonabelian}
\end{figure}

\paragraph{Non-abelian couplings.} Non-abelian couplings are subject to stronger constraints from Lorentz invariance. Nonetheless, it is possible to identify specific cubic vertices that satisfy the quartic consistency conditions. Let us take two generic vertices, $C^{\lambda_1,\lambda_2,\omega}$ and $\bar{C}^{-\omega,\lambda_3,\lambda_4}$. The constraints  \eqref{Paper3-kinda_trinagular} must still be respected. Such conditions can also be satisfied in the non-Abelian case; an example is in Figure \ref{Paper3-figure_nonabelian}.

\subsection*{Abelian vs Non-abelian cubic vertices I}
Because we have referred to both abelian and non-abelian cubic vertices, let us briefly clarify this distinction. In Figure \ref{Paper3-figure_nonabelian}, we have in fact slightly abused the terminology. At first sight, the couplings $C^{-1,2,7}$ and $\bar{C}^{-3,2,-7}$ might appear to be non-abelian, since they are of the (++$-$) and ($--$+) type, rather than the more familiar abelian (+++) and ($---$) vertices. This, however, is not the case.

The criterion distinguishing abelian from non-abelian cubic vertices was explained and found in \cite{Bekaert:2010hp}. We briefly review the relevant results for us here.

First of all, it should be emphasised that the distinction between abelian and non-abelian cubic vertices is naturally formulated in a covariant framework. Indeed, this terminology refers to the way a cubic interaction deforms the gauge algebra. Since in the light-cone formalism the gauge has already been fixed, there is no deformation to discuss.

We start by recalling that not all cubic vertices can be written in the standard (Fronsdal like) covariant formulation. If we fix $s_1\leq s_2\leq s_3$ with $s_{1,2,3}\geq 0$, covariant vertices exist only for the following helicity configurations:
\begin{align}
    &(+s_1,+s_2,+s_3)\oplus(-s_1,-s_2,-s_3) \,,&
    &(-s_1,+s_2,+s_3)\oplus (s_1,-s_2,-s_3)\,.
\end{align}
Note that in the chiral formulation, all cubic vertices can, in principle, be constructed \cite{Krasnov:2021nsq,Skvortsov:2022syz,Sharapov:2022faa}. However, here we restrict our attention to the Fronsdal formulation, since the results discussed below were derived within this framework.

The results of \cite{Bekaert:2010hp} can be summarised as follows. Vertices of the type $(s_1,s_2,s_3)$ and $(-s_1,-s_2,-s_3)$ provide the simplest examples of abelian vertices, since they do not require any deformation of the gauge transformations. There are also vertices of the type $(-s_1,s_2,s_3)$, together with their parity conjugates, satisfying $s_1+s_2\leq s_3$. These do require a deformation of the gauge transformations; however, the commutator of the deformed gauge transformations still vanishes. They are therefore also referred to as abelian, since the gauge algebra remains commutative. Non-abelian vertices are of the type $(-s_1,s_2,s_3)$ with $s_1+s_2> s_3$, together with their parity conjugates. In this case, the gauge transformations must again be deformed, but now the commutator of the deformed gauge transformations is non-vanishing. Equivalently, the gauge algebra itself is deformed into a non-abelian one.

Therefore, the cubic couplings displayed in Figure \ref{Paper3-figure_nonabelian} should more properly be referred to as abelian vertices according to the definition given above. As can already be inferred from the condition in \eqref{Paper3-surpass_trinagular}, there also exist local quartic vertices that solve the quartic light-cone constraint while involving non-abelian vertices. We will explore this point in more detail at the end of Section \ref{Paper3-section7}, where we introduce the notions of single-channel, YM-like, and GR-like amplitudes, which will provide a clearer picture and connections with amplitudes.

\subsection{Unitary local higher-spin theories in flat space}
In this section, we find all possible unitary local higher-spin theories in flat space. Starting from a cubic vertex $C^{\lambda_1,\lambda_2,\lambda_3}$, unitarity forces us to include its parity-related counterpart $\bar{C}^{-\lambda_1,-\lambda_2,-\lambda_3}$. These two vertices form a $C\bar{C}$ pair, which in turn generates quartic constraints that must be solved. 
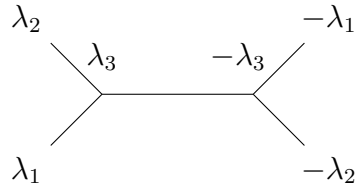
\begin{figure}[H]
    \centering
    \begin{tikzpicture}
        \begin{feynman}
            \vertex (i1) at (-6, 1) {\(\lambda_2\)};
            \vertex (i2) at (-6,-1) {\(\lambda_1\)};
            \vertex (i3) at (-2, 1) {\(-\lambda_1\)};
            \vertex (i4) at (-2,-1) {\(-\lambda_2\)};

            \vertex (v1) at (-5, 0);
            \vertex (v3) at (-3, 0);

            \vertex at (-5, 0.5) {\(\lambda_3\)};
            \vertex at (-3.2, 0.5) {\(-\lambda_3\)};
            
            \diagram* {
                (i1) -- (v1),
                (i2) -- (v1),
                (v1) -- [plain] (v3),
                (v3) -- (i3),
                (v3) -- (i4),
            };

        \end{feynman}
    \end{tikzpicture}
        \caption{Generic $C\bar{C}$ unitary exchange.}
    \label{Paper3-figure_unitaryexchange}
\end{figure}
\noindent
Considering that in this case, each of the helicities can end up on the external legs, we obtain the following conditions:
\begin{align}\label{Paper3-unitary_condition}
&\lambda_1 \leq \lambda_2+\lambda_3\,,&
&\lambda_2 \leq \lambda_1+\lambda_3\,,&
&\lambda_3 \leq \lambda_1+\lambda_2\,.&
&\lambda_1+\lambda_2+\lambda_3>0\,.
\end{align}
These conditions correspond to the triangle inequalities. As discussed above, a theory containing an arbitrary number of pairs of parity-related cubic abelian couplings that satisfy them forms a consistent and unitary theory admitting higher-spin fields.

We now start to study the cases when at least one of the external fields is allowed to be maximal. We will proceed as follows:
\begin{itemize}
    \item Firstly, we start from a generic $C\bar{C}$ unitary exchange diagram, as shown in Figure~\ref{Paper3-figure_unitaryexchange}, and consider the case where some of the external fields are maximal. There are three possible cases, which we will analyse separately.

    \item Secondly, we examine the constraints imposed by \eqref{Paper3-kinda_trinagular} and \eqref{Paper3-surpass_trinagular} and construct additional diagrams that are required to satisfy the quartic constraint.

    \item Thirdly, we notice that these new diagrams generate new cubic vertices. For each of them, we must include the corresponding parity-related pair and ensure that the triangular inequality \eqref{Paper3-unitary_condition} is satisfied. This will provide additional constraints which we use to restrict the helicities of the external fields in the original diagram.
\end{itemize}

\paragraph{First case.} By imposing that all three external fields have maximal helicities, we find
\begin{align}
\begin{split}
\lambda_1 = \lambda_{23}+n_1\,,\quad
\lambda_2 = \lambda_{13}+n_2\,,&\quad
\lambda_3 = \lambda_{12}+n_3\,,\quad
\lambda_{123}>0\\
&\implies\quad
\lambda_{123}=-(n_1+n_2+n_3)\,.
\end{split}
\end{align}
Recalling that $n_{1,2,3}=1,2$, we see that this system is inconsistent.

\paragraph{Second case.} By imposing that two external fields have maximal helicities, we find
\begin{align}\label{Paper3-consistent_1}
\begin{split}
\lambda_1 = \lambda_{23}+n_1\,,&\quad
\lambda_2 = \lambda_{13}+n_2\,\quad
\lambda_3 \leq \lambda_{12}\,,\quad
\lambda_{123}>0\\
&\implies\quad
\lambda_3=-\frac{n_1+n_2}{2}\,,\quad
\lambda_2=\lambda_1+\frac{n_2-n_1}{2}\,,\quad
\lambda_1>\frac{n_1}{2}\,.
\end{split}
\end{align}
The existence of this diagram implies the presence of additional diagrams (see Figure \ref{Paper3-figure_imply1}).\footnote{When space is limited, only $\omega$ is shown, omitting $-\omega$.} This new diagram, obtained from the original one by a permutation of the external helicities, respects the conditions \eqref{Paper3-kinda_trinagular} with the possibility of \eqref{Paper3-surpass_trinagular}. In particular, it satisfies
\begin{align}
\begin{split}
\lambda_1 \leq \lambda_2+\omega\,,&\quad
\lambda_2 = \lambda_1+\omega+n_1\,,\quad
\lambda_3=\lambda_4+\omega+n_2\,,\quad
\lambda_4 \leq \lambda_3+\omega\,,\\
&\lambda_{12}+\omega>0\,,\quad
-\lambda_{34}+\omega>0\,,\quad
\lambda_{12}=\lambda_{34}\,,
\end{split}
\end{align}
where $\lambda_{1,2,3,4}$ indicates the helicities of the diagram on the right in Figure \ref{Paper3-figure_imply1}, in the same order as in Figure \ref{Paper3-figure_generic}.
\begin{figure}[H]
    \centering
    \begin{tikzpicture}
        \begin{feynman}
            \vertex (i1) at (-10, 1) {\(\lambda_1+\frac{n_2-n_1}{2}\)};
            \vertex (i2) at (-10,-1) {\(\lambda_1\)};
            \vertex (i3) at (-6, 1) {\(-\lambda_1\)};
            \vertex (i4) at (-6,-1) {\(-\lambda_1-\frac{n_2-n_1}{2}\)};

            \vertex (arrow) at (-3,0) {\(\implies\)};

            \vertex (ii1) at (0, 1) {\(\lambda_1\)};
            \vertex (ii2) at (0,-1) {\(-\lambda_1\)};
            \vertex (ii3) at (4, 1) {\(\lambda_1+\frac{n_2-n_1}{2}\)};
            \vertex (ii4) at (4,-1) {\(-\lambda_1-\frac{n_2-n_1}{2}\)};

            \vertex (v1) at (-9, 0);
            \vertex (v3) at (-7, 0);

            \vertex (vv1) at (1, 0);
            \vertex (vv3) at (3, 0);
            
            \vertex at (-9, 0.5);
            \vertex at (-8, 0.5) {\(-\frac{n_1+n_2}{2}\)};
            \vertex at (-7.2, 0.5);

            \vertex at (1, 0.5);
            \vertex at (2, 0.5) {\(2\lambda_1-n_1\)};
            \vertex at (2.8, 0.5);

            \diagram* {
                (i1) -- (v1),
                (i2) -- (v1),
                (v1) -- [plain] (v3),
                (v3) -- (i3),
                (v3) -- (i4),
            };

            \diagram* {
                (ii1) -- (vv1),
                (ii2) -- (vv1),
                (vv1) -- [plain] (vv3),
                (vv3) -- (ii3),
                (vv3) -- (ii4),
            };

        \end{feynman}
    \end{tikzpicture}
    \caption{The left diagram, by quartic consistency, implies the presence of the right one.}
    \label{Paper3-figure_imply1}
\end{figure}
\noindent
As a consequence of unitarity, each newly generated cubic vertex, for instance the vertex $C^{\lambda_1,-\lambda_1,2\lambda_1-n_1}$, must satisfy the condition \eqref{Paper3-unitary_condition}, up to the possibility of being of maximal helicity. This additional constraint leads to
\begin{equation}
    2\lambda_1-n_1\leq 0\quad
    \implies\quad
    \lambda_1\leq \frac{n_1}{2}\,.
\end{equation}
This is inconsistent with \eqref{Paper3-consistent_1}. The only remaining possibility is to take this vertex to be maximal, which implies
\begin{equation}
2\lambda_1-n_1=n_3\quad
\implies\quad 
\lambda_1=\frac{n_1+n_3}{2}\,.
\end{equation}
At this point, we can list the cubic vertices that satisfy all the conditions derived. They are given by
\begin{equation}
(\lambda_1,\lambda_2,\lambda_3)=\left(\frac{n_1+n_3}{2},\frac{n_2+n_3}{2},-\frac{n_1+n_2}{2}\right)\,.
\end{equation}
Setting $n_{1,2,3}=1,2$, there are $2^3$ possible combinations. Among all solutions, only two contain exclusively integer helicities. Including the parity-related vertices as well, we obtain
\begin{equation}\label{Paper3-nonabelian1}
    \Big((1,1,-1) \oplus (-1,-1,1)\Big) \oplus \Big((2,2,-2) \oplus (-2,-2,2)\Big)\,.
\end{equation}
These correspond precisely to the cubic vertices of Yang-Mills theory and gravity. 

\paragraph{Third case.} By imposing that one external field has maximal helicity, we find
\begin{align}\label{Paper3-unitarycase2}
\begin{split}
\lambda_1 \leq \lambda_{23}\,,\quad
\lambda_2 =& \lambda_{13}+n_1\,,\quad
\lambda_3 \leq \lambda_{12}\,,\quad
\lambda_{123}>0\\
&\implies\quad
\lambda_1\geq-\frac{n_1}{2}\,,\quad
\lambda_3\geq-\frac{n_1}{2}\,,\quad
\lambda_{13}>-\frac{n_1}{2}\,.
\end{split}
\end{align}
This diagram implies the presence of an additional diagram (see Figure \ref{Paper3-figure_imply2}). This respects the condition \eqref{Paper3-kinda_trinagular} with the possibility of \eqref{Paper3-surpass_trinagular}. In particular, it satisfies
\begin{align}
\begin{split}
\lambda_1 \leq \lambda_2+\omega\,,&\quad
\lambda_2 \leq \lambda_1+\omega\,,\quad
\lambda_3=\lambda_4+\omega+n_1\,,\quad
\lambda_4\leq\lambda_3+\omega\,,\\
&\lambda_{12}+\omega>0\,,\quad
-\lambda_{34}+\omega>0\,,\quad
\lambda_{12}=\lambda_{34}\,,
\end{split}
\end{align}
where again by $\lambda_{1,2,3,4}$ we indicate the helicities of the diagram on the right in Figure \ref{Paper3-figure_imply2}, in the same order as in Figure \ref{Paper3-figure_generic}.
\begin{figure}[H]
    \centering
    \begin{tikzpicture}
        \begin{feynman}
            \vertex (i1) at (-10, 1) {\(\lambda_{13}+n_1\)};
            \vertex (i2) at (-10,-1) {\(\lambda_1\)};
            \vertex (i3) at (-6, 1) {\(-\lambda_{13}-n_1\)};
            \vertex (i4) at (-6,-1) {\(-\lambda_1\)};

            \vertex (arrow) at (-3,0) {\(\implies\)};

            \vertex (ii1) at (0, 1) {\(-\lambda_1\)};
            \vertex (ii2) at (0,-1) {\(\lambda_1\)};
            \vertex (ii3) at (4, 1) {\(\lambda_{13}+n_1\)};
            \vertex (ii4) at (4,-1) {\(-\lambda_{13}-n_1\)};

            \vertex (v1) at (-9, 0);
            \vertex (v3) at (-7, 0);

            \vertex (vv1) at (1, 0);
            \vertex (vv3) at (3, 0);
            
            \vertex at (-9, 0.5);
            \vertex at (-8.8, 0.5) {\(\lambda_3\)};
            \vertex at (-7.3, 0.5) {\(-\lambda_3\)};
            \vertex at (-7.2, 0.5);

            \vertex at (1, 0.5);
            \vertex at (2, 0.5) {\(2\lambda_{13}+n_1\)};
            \vertex at (2.8, 0.5);

            \diagram* {
                (i1) -- (v1),
                (i2) -- (v1),
                (v1) -- [plain] (v3),
                (v3) -- (i3),
                (v3) -- (i4),
            };

            \diagram* {
                (ii1) -- (vv1),
                (ii2) -- (vv1),
                (vv1) -- [plain] (vv3),
                (vv3) -- (ii3),
                (vv3) -- (ii4),
            };

        \end{feynman}
    \end{tikzpicture}
    \caption{The left diagram, by quartic consistency, implies the presence of the right one.}
    \label{Paper3-figure_imply2}
\end{figure}
\noindent
As a consequence of unitarity, each newly generated cubic vertex, for instance, the vertex $C^{\lambda_1,-\lambda_1,2\lambda_{13}+n_1}$, must satisfy the condition \eqref{Paper3-unitary_condition}, up to the possibility of being of maximal helicity. This additional constraint leads to
\begin{equation}
    2\lambda_{13}+n_1\leq 0\quad
    \implies\quad
    \lambda_{13}\leq -\frac{n_1}{2}\,.
\end{equation}
This is inconsistent with \eqref{Paper3-unitarycase2}. The only remaining possibility is to take this vertex to be maximal, which implies
\begin{equation}\label{Paper3-maximal2}
2\lambda_{13}+n_1=n_2\quad
\implies\quad 
\lambda_{13}=\frac{n_2-n_1}{2}\,.
\end{equation}
The constraint \eqref{Paper3-maximal2}, together with \eqref{Paper3-unitarycase2}, gives
\begin{equation}
    \lambda_{13}=\frac{n_2-n_1}{2}\,,\qquad
    \lambda_3\geq -\frac{n_1}{2}\,,\quad
    \lambda_1\geq -\frac{n_1}{2}\quad
    \implies\quad
    \lambda_1\leq \frac{n_2}{2}\,,\quad
    \lambda_3\leq \frac{n_2}{2}\,.
\end{equation}
At this point, we can list the cubic vertices that satisfy all the conditions derived. They are given by
\begin{equation}
(\lambda_1,\lambda_{13}+n_1,\lambda_3)\,,\quad
\lambda_3=\frac{n_2-n_1}{2}-\lambda_1\,,\quad
-\frac{n_1}{2}\leq \lambda_1\leq \frac{n_2}{2}\,.
\end{equation}
This gives 
\begin{equation}
\left(\lambda_1,\frac{n_2+n_1}{2},\frac{n_2-n_1}{2}-\lambda_1\right)\,,\quad
-\frac{n_1}{2}\leq \lambda_1\leq \frac{n_2}{2}\,.
\end{equation}
Setting $n_{1,2}=1,2$, there are $3$ possible combinations. Among all solutions, only two contain exclusively integer helicities. Including also the parity-related vertices, we obtain
\begin{equation}\label{Paper3-nonabelian2}
    \Big((0,1,0) \oplus (0,-1,0)\Big)\oplus\Big((0,2,0) \oplus (0,-2,0)\Big)\oplus\Big((1,2,-1) \oplus (-1,-2,1)\Big)\,.
\end{equation}
These correspond to the non-abelian cubic interactions between fields of helicities $0,1,2$; scalars, Yang-Mills fields, and gravitons.

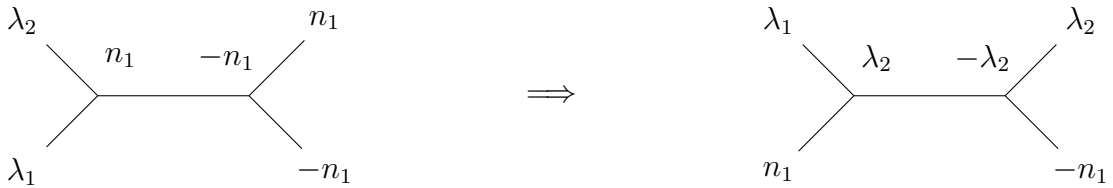
\begin{figure}[H]
    \centering
    \begin{tikzpicture}
        \begin{feynman}
            \vertex (i1) at (-10, 1) {\(\lambda_2\)};
            \vertex (i2) at (-10,-1) {\(\lambda_1\)};
            \vertex (i3) at (-6, 1) {\(n_1\)};
            \vertex (i4) at (-6,-1) {\(-n_1\)};

            \vertex (arrow) at (-3,0) {\(\implies\)};

            \vertex (ii1) at (0, 1) {\(\lambda_1\)};
            \vertex (ii2) at (0,-1) {\(n_1\)};
            \vertex (ii3) at (4, 1) {\(\lambda_2\)};
            \vertex (ii4) at (4,-1) {\(-n_1\)};

            \vertex (v1) at (-9, 0);
            \vertex (v3) at (-7, 0);

            \vertex (vv1) at (1, 0);
            \vertex (vv3) at (3, 0);
            
            \vertex at (-9, 0.5);
            \vertex at (-8.7, 0.5) {\(n_1\)};
            \vertex at (-7.3, 0.5) {\(-n_1\)};
            \vertex at (-7.2, 0.5);

            \vertex at (1, 0.5);
            \vertex at (1.3, 0.5) {\(\lambda_2\)};
            \vertex at (2.7, 0.5) {\(-\lambda_2\)};
            \vertex at (2.8, 0.5);

            \diagram* {
                (i1) -- (v1),
                (i2) -- (v1),
                (v1) -- [plain] (v3),
                (v3) -- (i3),
                (v3) -- (i4),
            };

            \diagram* {
                (ii1) -- (vv1),
                (ii2) -- (vv1),
                (vv1) -- [plain] (vv3),
                (vv3) -- (ii3),
                (vv3) -- (ii4),
            };

        \end{feynman}
    \end{tikzpicture}
    \caption{The left diagram, by quartic consistency, implies the presence of the right one.}
    \label{Paper3-figure_imply3}
\end{figure}

By applying a procedure analogous to that discussed above, one can show that imposing the presence of any of the non-abelian vertex pairs in \eqref{Paper3-nonabelian1} and \eqref{Paper3-nonabelian2} requires all higher-derivative abelian vertices to be set to zero, leaving only the lower-spin abelian interactions. We begin by observing that, due to the holomorphic constraint, as discussed in \cite{Serrani:2025owx}, the presence of the self-dual couplings $C^{0,2,0}$ and $C^{1,2,-1}$ necessarily requires the inclusion of the self-dual gravitational coupling $C^{2,2,-2}$. Similarly, the self-dual coupling $C^{0,1,0}$ requires the presence of the self-dual Yang-Mills coupling $C^{1,1,-1}$. Therefore, it is sufficient to show that the presence of either of the coupling pairs $C^{1,1,-1}\oplus \bar{C}^{-1,-1,1}$ or $C^{2,2,-2}\oplus \bar{C}^{-2,-2,2}$ implies that no additional higher-spin couplings can be consistently introduced. To this end, we start from the most general cubic vertex that forms an exchange, the one on the left of Figure \ref{Paper3-figure_imply3}, with the $\bar{C}^{n_1,-n_1,-n_1}$ cubic vertex, and require $\lambda_1$ and $\lambda_2$ to be higher-spin fields; then $\lambda_1,\lambda_2>2$. 
\noindent
This diagram implies the presence of an additional diagram (see Figure \ref{Paper3-figure_imply3}). This respects the condition \eqref{Paper3-kinda_trinagular} with the possibility of \eqref{Paper3-surpass_trinagular}, as can be easily checked.

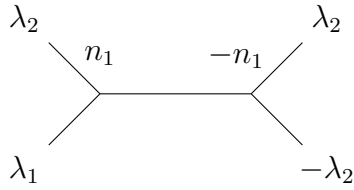
\begin{figure}[H]
    \centering
    \begin{tikzpicture}
        \begin{feynman}
            \vertex (i1) at (-6, 1) {\(\lambda_2\)};
            \vertex (i2) at (-6,-1) {\(\lambda_1\)};
            \vertex (i3) at (-2, 1) {\(\lambda_2\)};
            \vertex (i4) at (-2,-1) {\(-\lambda_2\)};

            \vertex (v1) at (-5, 0);
            \vertex (v3) at (-3, 0);

            \vertex at (-5, 0.5) {\(n_1\)};
            \vertex at (-3.2, 0.5) {\(-n_1\)};
            
            \diagram* {
                (i1) -- (v1),
                (i2) -- (v1),
                (v1) -- [plain] (v3),
                (v3) -- (i3),
                (v3) -- (i4),
            };

        \end{feynman}
    \end{tikzpicture}
        \caption{New exchange diagram between $C^{\lambda_1,\lambda_2,n_1}$ and $\bar{C}^{\lambda_2,-\lambda_2,-n_1}$.}
    \label{Paper3-figureNoAbelianHS}
\end{figure}

This new diagram generates two new cubic vertices: $C^{\lambda_1,\lambda_2,n_1}$ and $\bar{C}^{\lambda_2,-\lambda_2,-n_1}$. These can form another exchange (see Figure \ref{Paper3-figureNoAbelianHS}) that must satisfy its own non-holomorphic constraint. The constraints in \eqref{Paper3-kinda_trinagular} imply
\begin{equation}
    \lambda_2\leq n_1-\lambda_2\,,\quad
    \implies
    \lambda_2\leq\frac{n_2}{2}\,.
\end{equation}
Repeating the same steps with the roles of $\lambda_1$ and $\lambda_2$ reversed gives the constraint
\begin{equation}
    \lambda_1\leq n_1-\lambda_1\,,\quad
    \implies
    \lambda_1\leq\frac{n_1}{2}\,.
\end{equation}
Finally, we have
\begin{align}
    &\lambda_1\leq\frac{n_1}{2}\,,&
    &\lambda_2\leq\frac{n_1}{2}\,,&
    &\lambda_1+\lambda_2+n_1>0\,.
\end{align}
Therefore, no higher-spin fields are allowed.
\noindent
In the case that the external fields are maximal \eqref{Paper3-surpass_trinagular}, we find
\begin{align}
    &\lambda_1= \frac{n_1+n_2}{2}\,,&
    &\lambda_2= \frac{n_1+n_3}{2}\,.
\end{align}
In any case, no higher-spin fields are allowed.

Let us note that we have discarded possible solutions involving half-integer helicities. This is because we are considering only bosonic fields here. 
However, it seems that allowing for half-integer helicities, we would obtain additional solutions for cubic vertices involving both $\lambda=\pm\frac{1}{2}$ and $\lambda=\pm\frac{3}{2}$. This suggests that, even when massless fermions are included, an analogous structure should emerge. Moreover, it indicates that one could potentially recover --- through a similar consistency analysis at the quartic order --- the existence of interacting and consistent supergravity theories, such as $\mathcal{N}=1$ supergravity in $4d$. 
Investigating solutions to the quartic constraints in the presence of massless fermions is a natural next step, which we plan to pursue in the near future.

In conclusion, we have demonstrated that no unitary and local higher-spin theory in flat space can consistently coexist with the unitary cubic self-interactions of gravity and those of Yang–Mills theory \eqref{Paper3-nonabelian1} or with the non-abelian couplings \eqref{Paper3-nonabelian2}. The sole assumption of our analysis is that higher-spin interactions begin at the cubic order.

A potential loophole, however, is the following. One may retain the standard consistent cubic vertices for the lower-spin fields while allowing the higher-spin sector to start directly at quartic order and then attempt to close the algebra from that point onwards. In the light-cone formalism, this strategy would require a complete analysis of the quintic constraint. We leave this investigation for future work.
\subsection{On the possibility of ``quasi-chiral'' higher-spin theories}
As we have shown above, although abelian vertices are consistent among themselves, they become inconsistent once we require the presence of either of the pairs of cubic vertices in \eqref{Paper3-nonabelian1} or \eqref{Paper3-nonabelian2}. The interpretation is as follows: higher-spin abelian couplings can be consistently defined on a fixed background geometry, but they cannot interact with the geometry itself, which is represented by the cubic vertices of gravity.

A slightly weaker question is whether it is possible to couple higher-spin vertices to the self-dual sector of gravity or Yang–Mills theory in a way that includes both holomorphic and anti-holomorphic vertices, i.e. beyond the chiral theories described in \cite{Serrani:2025owx}, while satisfying both the holomorphic and non-holomorphic quartic constraints.

Here, we demonstrate the existence of such theories and describe their structure. These constructions allow coupling to one chiral sector, but not to both simultaneously. We refer to these theories as ``quasi-chiral''\footnote{This terminology was introduced in \cite{Adamo:2022lah}, where a related theory featuring a slightly different spectrum and some non-local interactions was constructed via a deformation of the Chalmers-Siegel action of self-dual Yang–Mills theory \cite{Chalmers:1996rq}.} higher-spin theories. By ``quasi-chiral'' we mean a theory that is not parity-invariant yet contains both holomorphic and anti-holomorphic quartic vertices, together with additional quartic interactions required to ensure consistency, at least up to quartic order.

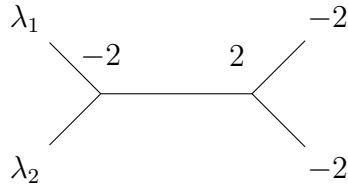
\begin{figure}[H]
    \centering
    \begin{tikzpicture}
        \begin{feynman}
            \vertex (i1) at (-6, 1) {\(\lambda_1\)};
            \vertex (i2) at (-6,-1) {\(\lambda_2\)};
            \vertex (i3) at (-2, 1) {\(-2\)};
            \vertex (i4) at (-2,-1) {\(-2\)};

            \vertex (v1) at (-5, 0);
            \vertex (v3) at (-3, 0);

            \vertex at (-5, 0.5) {\(-2\)};
            \vertex at (-3.2, 0.5) {\(2\)};
            
            \diagram* {
                (i1) -- (v1),
                (i2) -- (v1),
                (v1) -- [plain] (v3),
                (v3) -- (i3),
                (v3) -- (i4),
            };

        \end{feynman}
    \end{tikzpicture}
        \caption{Exchange diagram involving a generic vertex and the anti-MHV cubic vertex of GR.}
    \label{Paper3-figureHSGR}
\end{figure}

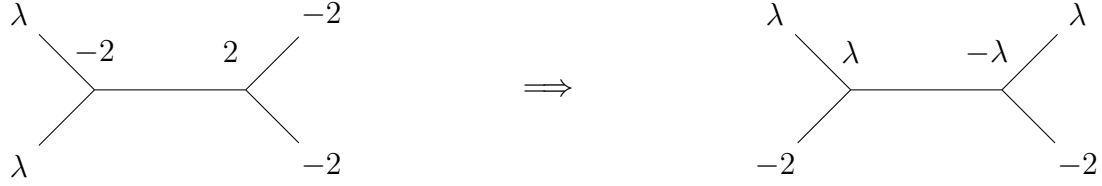
\begin{figure}[H]
    \centering
    \begin{tikzpicture}
        \begin{feynman}
            \vertex (i1) at (-10, 1) {\(\lambda\)};
            \vertex (i2) at (-10,-1) {\(\lambda\)};
            \vertex (i3) at (-6, 1) {\(-2\)};
            \vertex (i4) at (-6,-1) {\(-2\)};

            \vertex (arrow) at (-3,0) {\(\implies\)};

            \vertex (ii1) at (0, 1) {\(\lambda\)};
            \vertex (ii2) at (0,-1) {\(-2\)};
            \vertex (ii3) at (4, 1) {\(\lambda\)};
            \vertex (ii4) at (4,-1) {\(-2\)};

            \vertex (v1) at (-9, 0);
            \vertex (v3) at (-7, 0);

            \vertex (vv1) at (1, 0);
            \vertex (vv3) at (3, 0);
            
            \vertex at (-9, 0.5){\(-2\)};
            \vertex at (-7.2, 0.5){\(2\)};

            \vertex at (1, 0.5){\(\lambda\)};
            \vertex at (2.8, 0.5){\(-\lambda\)};

            \diagram* {
                (i1) -- (v1),
                (i2) -- (v1),
                (v1) -- [plain] (v3),
                (v3) -- (i3),
                (v3) -- (i4),
            };

            \diagram* {
                (ii1) -- (vv1),
                (ii2) -- (vv1),
                (vv1) -- [plain] (vv3),
                (vv3) -- (ii3),
                (vv3) -- (ii4),
            };

        \end{feynman}
    \end{tikzpicture}
        \caption{The left diagram, by quartic consistency, implies the presence of the right one.}
    \label{Paper3-figureHSGR2}
\end{figure}

\begin{figure}[H]
    \centering
    \begin{tikzpicture}
        \begin{feynman}
            \vertex (i1) at (-6, 1) {\(\lambda\)};
            \vertex (i2) at (-6,-1) {\(-2\)};
            \vertex (i3) at (-2, 1) {\(\lambda\)};
            \vertex (i4) at (-2,-1) {\(-2\)};

            \vertex (ii1) at (0, 1) {\(\lambda\)};
            \vertex (ii2) at (0,-1) {\(-\lambda\)};
            \vertex (ii3) at (4, 1) {\(-2\)};
            \vertex (ii4) at (4,-1) {\(-2\)};

            \vertex (v1) at (-5, 0);
            \vertex (v3) at (-3, 0);

            \vertex (vv1) at (1, 0);
            \vertex (vv3) at (3, 0);

            \vertex at (-5, 0.5) {\(\lambda\)};
            \vertex at (-3.2, 0.5) {\(-\lambda\)};

            \vertex at (1, 0.5) {\(-2\)};
            \vertex at (2.8, 0.5) {\(2\)};

            \diagram* {
                (i1) -- (v1),
                (i2) -- (v1),
                (v1) -- [plain] (v3),
                (v3) -- (i3),
                (v3) -- (i4),
            };
            \diagram* {
                (ii1) -- (vv1),
                (ii2) -- (vv1),
                (vv1) -- [plain] (vv3),
                (vv3) -- (ii3),
                (vv3) -- (ii4),
            };
        \end{feynman}
    \end{tikzpicture}
        \caption{Holomorphic constraints between $C^{-2,-2,2}$ and $C^{\lambda,-\lambda,-2}$.}
    \label{Paper3-figureHSGR3}
\end{figure}
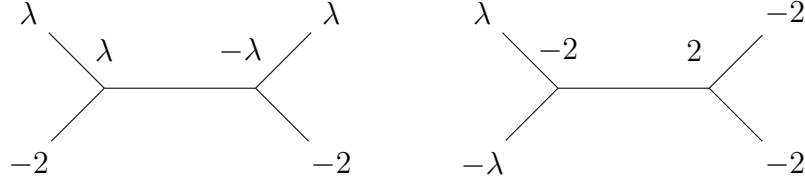

We begin by constructing a theory that couples to self-dual gravity. We start by studying the corresponding exchange diagram in Figure \ref{Paper3-figureHSGR}.
\noindent
This diagram cannot respect \eqref{Paper3-kinda_trinagular}. We have to allow the external fields to have maximal helicity \eqref{Paper3-surpass_trinagular}. The existence of quartic vertices requires
\begin{align}
    &\lambda_1=\lambda_2-2+n_1\,,&
    &\lambda_2=\lambda_1-2+n_2\,.&
    &\implies&
    &n_1=n_2=2
    &\implies&
    &\lambda_1=\lambda_2\,.
\end{align}
The existence of this diagram implies the presence of an additional diagram, as shown in Figure \ref{Paper3-figureHSGR2}.
\noindent
The newly generated cubic vertices are also required to satisfy the holomorphic constraints. In particular, one can construct the diagrams shown in Figure \ref{Paper3-figureHSGR3}.
\noindent
The holomorphic constraints fix the coefficients of the two anti-holomorphic cubic vertices $C^{-2,-2,2}$ and $C^{\lambda,-\lambda,-2}$ to be equal, implying $C^{-2,-2,2}=C^{\lambda,-\lambda,-2}$. As discussed in \cite{Serrani:2025owx}, these two anti-holomorphic cubic couplings are consistent even in the absence of additional higher-spin vertices, since they form a consistent truncation of the full chiral higher-spin gravity.

We have thus identified a fully consistent quasi-chiral higher-spin theory that we call quasi-chiral HS-GR, at least up to the quartic order, comprising a specific set of cubic and quartic vertices. The theory contains the following interactions:
\begin{equation}
\Big\{C^{-2,-2,2}=C^{\lambda,-\lambda,-2}, C^{\lambda,\lambda,-2}, h^{(2\lambda-2,2)}_{(\lambda,\lambda,-2,-2)}\Big\}\,.
\end{equation}
Notice that summing over all values of $\lambda$ gives a theory that remains consistent. Indeed, no new types of exchange diagrams are generated.

\begin{figure}[H]
    \centering
    \begin{tikzpicture}
        \begin{feynman}
            \vertex (i1) at (-6, 1) {\(\lambda\)};
            \vertex (i2) at (-6,-1) {\(-\lambda\)};
            \vertex (i3) at (-2, 1) {\(2\)};
            \vertex (i4) at (-2,-1) {\(-2\)};

            \vertex (v1) at (-5, 0);
            \vertex (v3) at (-3, 0);

            \vertex at (-5, 0.5) {\(-2\)};
            \vertex at (-3.2, 0.5) {\(2\)};

            \diagram* {
                (i1) -- (v1),
                (i2) -- (v1),
                (v1) -- [plain] (v3),
                (v3) -- (i3),
                (v3) -- (i4),
            };
        \end{feynman}
    \end{tikzpicture}
        \caption{$(2,2)$ $\bar{C}C$ exchange diagram.}
    \label{Paper3-figure_fullGR}
\end{figure}
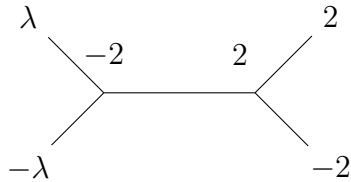

If we now attempt to include the holomorphic cubic vertex of gravity, $C^{2,2,-2}$, we can form the two-derivative exchange diagram in Figure \ref{Paper3-figure_fullGR}.
\noindent
This exchange diagram does not admit a quartic vertex that solves the quartic constraint.

Interestingly, the same phenomenon also appears in the case of Yang-Mills-like interactions. In this case, we obtain the quasi-chiral HS-YM theory --- an extension of self-dual Yang-Mills theory --- comprising the couplings
\begin{equation}\label{Paper3-quasi_chiral_YM}
    \{C^{-1,-1,1}=C^{\lambda,-\lambda,-1},C^{\lambda,\lambda,-1},h^{(2\lambda-1,1)}_{[\lambda,\lambda,-1,-1]},h^{(2\lambda-1,1)}_{[\lambda,-1,\lambda,-1]}\}\,.
\end{equation}
As before, summing over all values of $\lambda$ gives a theory that remains consistent. Attempting to include the holomorphic cubic vertex $C^{1,1,-1}$ leads to inconsistencies, similar to those encountered in the case of quasi-chiral HS-GR.

We expect that additional quasi-chiral higher-spin theories may exist that do not include the SDYM ($C^{-1,-1,1}$) or SDGR ($C^{-2,-2,2}$) cubic vertices. It would be interesting to classify all possible quasi-chiral higher-spin theories.

\paragraph{Comparison with the literature.} To our knowledge, the only other work that discusses quasi-chiral higher-spin theories is \cite{Adamo:2022lah}, in which the authors consider the possibility of extending higher-spin self-dual Yang–Mills theory (HS-SDYM) \cite{Ponomarev:2017nrr,Krasnov:2021nsq} to a quasi-chiral framework by allowing the presence of both MHV and anti-MHV cubic vertices.

The theory described in \cite{Adamo:2022lah} features, in our notation, the following spectrum of cubic vertices:
\begin{equation}
\{C^{-s,1,1},C^{-s_1,-s_2,1}\}\,,\qquad
s,s_1,s_2>0\,,
\end{equation}
where, as emphasised in \cite{Adamo:2022lah}, the positive-helicity external fields are restricted to spin-$1$ in order to preserve gauge invariance. The authors also computed four-point color-ordered amplitudes (which we will reproduce later) given by
\begin{align}
    &\tilde{\mA}(1_{s}^-2_{s}^-3_{1}^+4_{1}^+)=\frac{\langle 12\rangle^{2s+1}}{\langle 23\rangle\langle 34\rangle\langle 41\rangle}\,,&
    &\tilde{\mA}(1_{s}^-2_{1}^+3_{s}^-4_{1}^+)=\frac{\langle 13\rangle^{2s+2}}{\langle 12\rangle\langle 23\rangle\langle 34\rangle\langle 41\rangle}\,.
\end{align}
They further discuss an analogue of the Parke–Taylor formula for the $n$-point scattering of this theory. It is also noted that allowing the positive-helicity legs to carry arbitrary spin, rather than being fixed to spin-$1$, would make the amplitudes non-local.

We argue that the theory in \cite{Adamo:2022lah}, named HS-YM, does not define a local theory; rather, it defines a non-local one. As we have proven above, the simultaneous presence of both Yang–Mills cubic couplings $C^{-1,1,1}$ and $C^{-1,-1,1}$ cannot be locally consistent with any cubic higher-spin coupling. In particular, from \eqref{Paper3-quasi_chiral_YM}, the light-cone analysis indicates that the consistency of the theory requires the presence of the cubic vertex $C^{s,-s,1}$. However, together with $C^{-1,-1,1}$, this inevitably leads to non-locality.

Nevertheless, the theory remains of considerable interest, both for the way it evades various no-go results and for the peculiar structure of its non-localities, which merit further investigation. In particular, it admits a simple Lorentz-invariant formulation.

\section{Four-point higher-spin amplitudes}\label{Paper3-section7}

In the previous section, we demonstrated that a local and unitary higher-spin theory cannot consistently include parity invariant pairs of non-abelian cubic interactions. Nevertheless, we also observed the presence of many non-trivial local quartic vertices that solve the quartic constraint \eqref{Paper3-quartic_system}. Whenever such a local quartic vertex exists, one can construct a well-defined local four-point amplitude by combining the exchange contributions with the quartic contact term. Even though the local quartic vertices appear quite complicated,\footnote{While the light-cone formalism has proven extremely efficient in the self-dual sector, where only cubic vertices are present and amplitudes can be computed with remarkable simplicity \cite{Skvortsov:2018jea,Skvortsov:2020wtf,Skvortsov:2020gpn}, it is likely not the most suitable framework for computing amplitudes in the presence of higher-point vertices.} the amplitudes take an especially simple form when expressed in spinor-helicity variables.

As we will now see, contrary to the self-dual sector,\footnote{In the self-dual sector, amplitudes vanish identically \cite{Serrani:2025owx,Serrani:2025oaw}. Moreover, amplitudes can only have helicity configurations (++++) and (+++$-$), and are therefore necessarily non-MHV.} none of these local four-point amplitudes vanish, providing an explicit counterexample to the standard expectation that local higher-spin amplitudes should vanish. These amplitudes evade Weinberg's soft theorem, as shown in \cite{Tran:2022amg}. To determine the local four-point amplitudes for generic spins, one could proceed following the same strategy adopted for Yang–Mills theory and gravity in Section \ref{Paper3-subsection5.2}.

A more systematic approach is provided by the spinor-helicity formalism, which naturally incorporates the little-group scaling of massless amplitudes while implementing locality through consistent factorisation. In \cite{Ponomarev:2016cwi}, it was shown that the light-cone deformation procedure for constructing interacting massless theories in $4d$ flat space is equivalent to searching directly for spinor-helicity amplitudes satisfying the appropriate little-group scaling properties and locality constraints, the latter being reflected in the presence of only simple poles. It was further argued that this reformulation provides a more efficient framework for the search for consistent higher-spin interactions. Here we follow precisely this strategy.
In several cases, we explicitly verify the resulting amplitudes by summing the contributions from the local quartic vertices and exchange diagrams derived in the light-cone gauge.

Using this approach, we provide a complete classification and explicit expressions for all local higher-spin four-point amplitudes. To the best of our knowledge, this is the first systematic and exhaustive determination of this class of amplitudes. Partial results and specific examples had previously been obtained in \cite{Benincasa:2007xk,Benincasa:2011kn,Benincasa:2011pg,Benincasa:2012wt,McGady:2013sga,Taronna:2017wbx,Roiban:2017iqg,Ananth:2023qrf}. In addition, the quasi-chiral HS--YM amplitudes were previously derived in a covariant formulation in \cite{Adamo:2022lah}.

The recipe for determining the local amplitude $\mA(1_{\lambda_1}2_{\lambda_2}3_{\lambda_3}4_{\lambda_4})$ for generic helicities is as follows:
\begin{enumerate}
    \item We have to match the little group scaling:
    \begin{align}
        &2\lambda_i=\mathbb{N}_{|i]}-\mathbb{N}_{\langle i|}\,,&
        &\mathbb{N}_{|i]}=|i]\frac{\partial}{\partial|i]}\,,&
        &\mathbb{N}_{\langle i|}=\langle i|\frac{\partial}{\partial\langle i|}\,,&
        &i=1,2,3,4\,,
    \end{align}
    where $\mathbb{N}_{|i]}$ and $\mathbb{N}_{\langle i|}$ are the powers of $|i]$ and $\langle i|$ in the amplitude, respectively. 
    
    \item We notice that every local four-point amplitude can be written as follows: 
    \begin{equation}\label{Paper3-amplitude_base}
        \mA(1_{\lambda_1}2_{\lambda_2}3_{\lambda_3}4_{\lambda_4})=p(s,t)[12]^{x_1}\langle 34\rangle^{x_2}[13]^{x_3}[14]^{x_4}\,,
   \end{equation}
    where $x_1,x_2,x_3,x_4\in \mathbb{Z}$, and $p(s,t)$ is a meromorphic function of the Mandelstam variables $s$ and $t$,\footnote{If we require the locality of the amplitude, $p(s,t)$ must contain at most single poles. Then we can also write $p(s,t)=\frac{1}{s^at^bu^c}P(s,t)$, where $P(s,t)$ is a polynomial in $s$ and $t$, and $a,b,c=0,1$.} as $u$ can always be recovered via $s+t+u=0$. To demonstrate that every spinor-helicity expression for a four-point amplitude can be cast into this form, one needs to apply momentum conservation:
    \begin{subequations}\label{Paper3-momentum_conserv}
    \begin{align}
        &u=-s-t\,,&
        &\langle 12\rangle=-\frac{s}{[12]}\,,&
        &\langle 13\rangle=-\frac{u}{[13]}\,,&
        &\langle 14\rangle=-\frac{t}{[14]}\,,\\
        &\langle 24\rangle=-\frac{[13]\langle 34\rangle}{[12]}\,,&
        &\langle 23\rangle=\frac{[14]\langle 34\rangle}{[12]}\,,&
        &[23]=\frac{-t[12]}{[14]\langle 34\rangle}\,,&
        &[24]=\frac{u[12]}{[13]\langle 34\rangle}\,.
    \end{align}
    \end{subequations}
   In general, there are other possible structures besides \eqref{Paper3-amplitude_base}. For instance, one may replace $[13]$ with $[24]$, or $[14]$ with $[23]$. More generally, one is allowed to permute the labels $1,2,3,4$ on the right-hand side of \eqref{Paper3-amplitude_base}, as well as to exchange square and angle brackets, $[ij]\leftrightarrow \langle ij\rangle$. However, as we will explain below, configurations such as $[12]^{x_1}[34]^{x_2}$ and $\langle 12 \rangle^{x_1}\langle 34\rangle^{x_2}$ are excluded.

    \item The expression above can be constructed for arbitrary choices of the external helicities. We now impose the stronger requirement that the amplitude is generated by the exchange of two cubic vertices, as depicted in Figure \ref{Paper3-figure_generic}. This requirement enforces the factorisation properties of the amplitude and, consequently, provides the constraints necessary to ensure locality.
    
    The form of the amplitude in \eqref{Paper3-amplitude_base} requires a minimum of $D$ derivatives carried by the quartic vertex. Therefore\footnote{Here, $D$ can equivalently be understood as the momentum degree (mass dimension) of the
four-point exchange amplitude generated by two cubic vertices $A_4^{\rm exch}\sim V_3^{(n)}\frac{1}{p^2}V_3^{(m)}
\sim p^{n+m-2}\sim [M]^{n+m-2}\,.$} $D\equiv n+m-2=\lambda_{12}-\lambda_{34}+2\omega-2$,\footnote{Here, $\omega$ denotes the helicity of the exchanged field; when only a quartic vertex is present, $\omega$ is just a way to assign a number of derivatives to the specific quartic vertex.} must exceed the minimal number required by the little-group scaling $d\equiv x_1+x_2+x_3+x_4$ in \eqref{Paper3-amplitude_base}. Taking into account possible poles, then $D-d+2k\geq 0$. The integer $k$ takes the values $k=0$ for a pure quartic vertex, $k=1$ in the single-channel case ($\mA_4\sim 1/s$), $k=2$ in the YM-like case ($\mA_4\sim 1/st$), and $k=3$ in the GR-like case ($\mA_4\sim 1/stu$). By applying the constraint $D-d+2k\geq 0$ to any possible $d$ coming from all local four-point amplitude expressions such as in \eqref{Paper3-amplitude_base}, we reproduce the same algebraic constraints discussed in \eqref{Paper3-kinda_trinagular} and \eqref{Paper3-surpass_trinagular}.
\end{enumerate}
Before studying the constraints imposed on four-point amplitudes, let us justify the statement made above and prove that the following procedure is equivalent to standard factorisation. The advantage of this approach is that it can be applied systematically, given any four external helicities, and it does not require checking factorisation channel by channel.

\paragraph{``\textit{Fast factorisation}''.}
We now explain why the above procedure (which we shall refer to as \emph{fast factorisation}) is equivalent to the usual factorisation. The first step is to show that the general form \eqref{Paper3-amplitude_base} of a four-point amplitude is unique, up to permutations of the external legs $1,2,3,4$ and the exchange of square ($[\;]$) and angle ($\langle\;\rangle$) spinor brackets. This follows from an explicit construction.

Since the amplitude is gauge invariant, it can only depend on the external spinors through Lorentz-invariant spinor contractions. Without loss of generality, we first fix the little-group weight of particle $2_{\lambda_2}$ by starting with either $\mA_4\sim [12]^{2\lambda_2}$ or $\mA_4\sim \langle 12\rangle^{-2\lambda_2}$. After this choice, the remaining little-group weights must be fixed using spinor contractions involving the other external legs. There are only three independent spinor brackets left, namely $(13)$, $(14)$, and $(34)$, where $(ij)$ denotes either a square bracket $[ij]$ or an angle bracket $\langle ij\rangle$. The most general ansatz therefore takes the form
\begin{equation}
\mA(1_{\lambda_1}2_{\lambda_2}3_{\lambda_3}4_{\lambda_4})
=p(s,t)
(12)^{x_1}
(34)^{x_2}
(13)^{x_3}
(14)^{x_4}\,.
\end{equation}
Since the Mandelstam variables carry zero little-group weight, they can appear with arbitrary powers, $p(s,t)$, without affecting the little-group scaling. 
All other possible ansätze are then found by arbitrary permutations of the external legs. For example, replacing $(13)$ by $(24)$ is equivalent to the permutation $(1,2)\leftrightarrow(4,3)$.

For each factor $(ij)$ we may choose either a square or an angle bracket. However, the two configurations $[12]^{x_1}[34]^{x_2}$ and $\langle 12 \rangle^{x_1}\langle 34\rangle^{x_2}$ are excluded. There are two reasons for this. First, if the amplitude is required to arise from a factorisation channel, such structures can never be generated by gluing two three-point amplitudes. To see this explicitly, consider a factorisation channel in which the internal momentum is $k_{\omega}$:
\begin{equation}
    \mA_4\to\mA_3^{(n)}\frac{1}{s}\mA_3^{(m)}=[12]^{\lambda_{12}-\omega}[2k]^{\lambda_2+\omega-\lambda_1}[k1]^{\lambda_1+\omega-\lambda_2}\frac{1}{s}\langle 34\rangle^{-\lambda_{34}-\omega}\langle 4k\rangle^{\lambda_3+\omega-\lambda_4}\langle k3\rangle^{\lambda_4+\omega-\lambda_3}\,.
\end{equation}
The resulting four-point expression can never reduce to a structure containing $\frac{1}{s}[12]^{x_1}[34]^{x_2}$ with both $x_1,x_2\neq 0$; the same applies to $\frac{1}{s}\langle 12 \rangle^{x_1}\langle 34\rangle^{x_2}$.

There is another, more fundamental reason for excluding such configurations. Allowing them would destroy the predictive power of the pole structure, which is essential for analysing factorisation. Indeed, one would have to enlarge the class of admissible functions $p(s,t)$ by allowing arbitrary dimensionless cross-ratios of the Mandelstam variables, such as powers of ($s/t$), ($t/u$), and ($u/s$). Although these factors preserve both the little-group scaling and the total number of derivatives $D$, they can generate higher-order poles. In fact, such poles can always be removed by trading them for different spinor structures. For example
\begin{align}
&\frac{u^2}{s^2}[12]^2[34]^2=[13]^2[24]^2\,,&
&\frac{t^2}{s^2}[12]^2[34]^2=[14]^2[23]^2\,.
\end{align}
Consequently, the pole structure is no longer uniquely determined by the chosen spinor ansatz. In contrast, for the ansätze considered here, such as \eqref{Paper3-amplitude_base}, the poles contained in $p(s,t)$ cannot be eliminated by the accompanying spinor factor $[12]^{x_1}\langle34\rangle^{x_2}[13]^{x_3}[14]^{x_4}$ even after using momentum conservation. This uniqueness of the pole structure is what makes this method powerful to classify the most general four-point amplitudes satisfying factorisation.

\paragraph{\textit{Fast factorisation} = Factorisation.}

We now prove that \textit{fast factorisation} is equivalent to the usual notion of factorisation.

\begin{itemize}

\item \textbf{Factorisation $\implies$ \textit{Fast factorisation}.}

Ordinary factorisation implies that, when an internal particle goes on-shell, the four-point amplitude factorises into a product of cubic amplitudes as $\mA_4\to \mA_3^{(n)}\frac{1}{s}\mA_3^{(m)}$ (and analogously in the $t$- and $u$-channels whenever the corresponding poles are present). Since each cubic amplitude has the correct little-group scaling, the resulting four-point amplitude automatically inherits the appropriate little-group weights. It can therefore always be expressed in the form \eqref{Paper3-amplitude_base}, or, more generally, as a sum of three terms of this type.\footnote{As we will see in the single-channel case $\mA_4$ can be decomposed as $\mA_4\sim \mA_s+\mA_t+\mA_u$, where $\mA_s\sim \frac{1}{s}$, $\mA_t\sim \frac{1}{t}$, and $\mA_u\sim \frac{1}{u}$.} Moreover, by construction $D\ge d-2k$. The special case $D=d-2k$ corresponds to a constant polynomial numerator in the Mandelstam variables, for instance $P(s,t)=1$.

\item \textbf{\textit{Fast factorisation} $\implies$ Factorisation.}

Conversely, \textit{fast factorisation} states that the amplitude can be written in the form \eqref{Paper3-amplitude_base} with $D\ge d-2k$. If the amplitude contains one ($1/s$), two ($1/st$), or three ($1/stu$) physical poles, then it can be factorised in the corresponding channels.
For the $s$-channel, one may write $\mA_4\to \mA_3^{(n)}\frac{1}{s}\mA_3^{(m)}$, where $n+m-2=\lambda_{12}-\lambda_{34}+2\omega_s-2=D$, which fixes $\omega_s=\frac{1}{2}\left(D+2+\lambda_{34}-\lambda_{12}\right)$. 
Similarly, if a $t$-channel pole is present, $\mA_4\to \mA_3^{(n)}\frac{1}{t}\mA_3^{(m)}$,
with $n+m-2=\lambda_{14}-\lambda_{23}+2\omega_t-2=D$, so that $\omega_t=\frac{1}{2}\left(D+2+\lambda_{23}-\lambda_{14}\right).$
Likewise, if a $u$-channel pole is present, $\mA_4\to \mA_3^{(n)}\frac{1}{u}\mA_3^{(m)}$, with $n+m-2=\lambda_{13}-\lambda_{24}+2\omega_u-2=D$,
which gives $\omega_u=\frac{1}{2}\left(D+2+\lambda_{24}-\lambda_{13}\right)$. Therefore, every amplitude satisfying the \textit{fast factorisation} criterion admits a factorisation into appropriate three-point amplitudes in every physical channel in which a pole is present.

\end{itemize}
Let us make a few remarks. First, the degrees $n=\lambda_{12}+\omega$ (of the holomorphic cubic amplitude) and $m=-\lambda_{34}+\omega$ (of the anti-holomorphic cubic amplitude) must be identical in all three channels. Otherwise, the three factorisation channels cannot form the same four-point amplitude. This follows simply from counting angle and square brackets: momentum conservation can reshuffle spinor products ($\langle 1k\rangle[k4]=\langle 12\rangle[24]=-\langle 13\rangle[34]$) but never changes their number. This is precisely the same condition encountered in the light-cone formalism, where the powers of $\PPb$ and $\PP$ must coincide in all channels for the three exchanges to belong to the same quartic constraint. 

As a consequence, specifying the exchanged helicity $\omega$ in the $s$-channel uniquely determines the corresponding exchanges in the $t$- and $u$-channels. Indeed analysing a single exchange diagram is sufficient to determine whether an amplitude can factorise in one, two, or all three channels.

We also emphasise that there is always a representation free of spurious poles. In particular, the representation with maximal $d$ is manifestly local, with the only poles arising from the physical propagators. This follows from the fact that any denominator in \eqref{Paper3-amplitude_base} involving spinor brackets can always be converted into a numerator by exchanging angle and square brackets. 

In the following, we derive the constraints imposed by \textit{fast factorisation} and illustrate them through several examples, highlighting the properties discussed above.

Applying the first and second rules described above gives the following form of the amplitude
\begin{equation}\label{Paper3-amplitude1234}
    \mA(1_{\lambda_1}2_{\lambda_2}3_{\lambda_3}4_{\lambda_4})=p(s,t)[12]^{2\lambda_2}\langle 34\rangle^{\lambda_1-\lambda_{234}}[13]^{\lambda_{13}-\lambda_{24}}[14]^{\lambda_{14}-\lambda_{23}}\,,
\end{equation}
where $d=3\lambda_1-\lambda_{234}$. By considering the constraint $D-d+2k\geq 0$, for \eqref{Paper3-amplitude1234} we get
\begin{equation}\label{Paper3-condition1}
\lambda_1\leq \lambda_2+\omega+k-1\,.
\end{equation}
If we start from a different ansatz for the amplitude, for instance, exchanging $[ij]$ with $\langle ij\rangle$ in \eqref{Paper3-amplitude1234}, then starting from
\begin{equation}\label{Paper3-amplitude1234_inverted}
    \mA(1_{\lambda_1}2_{\lambda_2}3_{\lambda_3}4_{\lambda_4})=p(s,t)\langle 12\rangle^{-2\lambda_2}[34]^{-\lambda_1+\lambda_{234}}\langle 13\rangle^{-\lambda_{13}+\lambda_{24}}\langle 14\rangle^{-\lambda_{14}+\lambda_{23}}\,,
\end{equation}
then $d=-3\lambda_1+\lambda_{234}$; by considering again the constraint $D-d+2k\geq 0$, we obtain
\begin{equation}\label{Paper3-condition2}
\lambda_1\geq -\frac{1}{2}(-\lambda_{34}+\omega+k-1)=-\frac{1}{2}(m+k-1)\,.
\end{equation}
If, instead, we permute the RHS labels $(1234)$ in the ansatz \eqref{Paper3-amplitude1234} with $(3214)$, we have
\begin{equation}\label{Paper3-amplitude3214}
    \mA(1_{\lambda_1}2_{\lambda_2}3_{\lambda_3}4_{\lambda_4})=p(s,t)[32]^{2\lambda_2}\langle 14\rangle^{\lambda_3-\lambda_{214}}[31]^{\lambda_{31}-\lambda_{24}}[34]^{\lambda_{34}-\lambda_{21}}\,,
\end{equation}
then $d=3\lambda_3-\lambda_{214}$. The constraint does give
\begin{equation}\label{Paper3-condition3}
\lambda_3\leq \frac{1}{2}(\lambda_{12}+\omega+k-1)=\frac{1}{2}(n+k-1)\,.
\end{equation}
By further exchanging $[ij]$ with $\langle ij\rangle$ and then having
\begin{equation}\label{Paper3-amplitude3214_inverted}
    \mA(1_{\lambda_1}2_{\lambda_2}3_{\lambda_3}4_{\lambda_4})=p(s,t)\langle 32\rangle^{-2\lambda_2}[14]^{-\lambda_3+\lambda_{214}}\langle 31\rangle^{-\lambda_{31}+\lambda_{24}}\langle 34\rangle^{-\lambda_{34}+\lambda_{21}}\,,
\end{equation}
then $d=-3\lambda_3+\lambda_{214}$, we would obtain
\begin{equation}\label{Paper3-condition4}
    \lambda_4\leq \lambda_3+\omega+k-1\,.
\end{equation}
We now derive the constraints arising from all possible forms of \eqref{Paper3-amplitude_base}. It is straightforward to verify that the corresponding ansätze exhaust all possible values of $d = 3\lambda_i - \lambda_{jk\ell}$ for every permutation of $i,j,k,\ell=1,2,3,4$. Evaluating each case, we obtain the following minimal set of independent constraints:
\begin{subequations}\label{Paper3-total_conditions}
\begin{align}\label{Paper3-part12}
&\lambda_1 \leq \lambda_2+\omega+k-1\,,&
&\lambda_2 \leq \lambda_1+\omega+k-1\,,\\\label{Paper3-part34}
&\lambda_3 \leq \lambda_4+\omega+k-1\,,&
&\lambda_4 \leq \lambda_3+\omega+k-1\,,\\
&\lambda_{12} \geq \lambda_{34}\,,&
&k=0,1,2,3.
\end{align}
\end{subequations}
Notice that the condition $\lambda_{12} \geq \lambda_{34}$ is equivalent to \eqref{Paper3-condition2} and also to \eqref{Paper3-condition3} once we use \eqref{Paper3-part12} and \eqref{Paper3-part34}. These are a rewriting of the conditions found above: \eqref{Paper3-kinda_trinagular}, \eqref{Paper3-surpass_trinagular}, and \eqref{Paper3-Homo_constraints}. The additional conditions $n=\lambda_{12}+\omega> 0$ and $m=-\lambda_{34}+\omega> 0$ must be added by hand to ensure we are looking at non-holomorphic amplitude. 

The conditions \eqref{Paper3-total_conditions} can be used in two complementary ways. If we are only interested in constructing amplitudes, then for a fixed value of $k$, every choice of helicities $(\lambda_1,\lambda_2,\lambda_3,\lambda_4,\omega)$ satisfying \eqref{Paper3-total_conditions} defines an amplitude $ \mA(1_{\lambda_1}2_{\lambda_2}3_{\lambda_3}4_{\lambda_4})$ that factorises correctly into $k$ channels, whose explicit form is given by \eqref{Paper3-amplitude_base}.

Alternatively, if the goal is to recover the information encoded in the light-cone tables, one can proceed as follows. Whenever \eqref{Paper3-total_conditions} are satisfied, the exchange $(1234)$ is allowed. As a direct consequence, the exchanges $(4123)$, and $(1324)$ are always allowed as well\footnote{This differs slightly from the various tables shown because the helicities are ordered here so that $d$ is maximal.},  whereas $(3412)$ is allowed only if $\lambda_{12}=\lambda_{34}$, $(2341)$ only if $\lambda_{14}=\lambda_{23}$, and $(2413)$ only if $\lambda_{13}=\lambda_{24}$. These conditions follows directly from \eqref{Paper3-total_conditions}. For instance, if $(1234)$ is allowed, then $\lambda_{12}\geq \lambda_{34}$; demanding that $(3412)$ be allowed as well implies $\lambda_{34}\geq \lambda_{12}$, and hence $\lambda_{12}=\lambda_{34}$. The remaining cases follow analogously. Moreover, the relations among the various couplings can also be determined completely. We will illustrate this procedure below.

We can now begin to present the explicit form of all local four-point higher-spin amplitudes. Even though, as pointed out above, a given amplitude generally admits several equivalent representations, we choose a canonical one by maximising the quantity $d$, namely 
\begin{equation}\label{Paper3-max_d}
    d=\max\limits_{i\neq j\neq k\neq \ell}\{3\lambda_i-\lambda_{jk\ell},-3\lambda_i+\lambda_{jk\ell}\}\,,\qquad
    i,j,k,\ell=1,2,3,4\,.
\end{equation}
As discussed above, the representation with maximal $d$ makes locality manifest, with the only poles arising from the physical propagators.

\paragraph{Quartic amplitudes from quartic vertices.}
When the conditions \eqref{Paper3-total_conditions} are satisfied for $k=0$, local quartic vertices are allowed. These correspond to the solution of the homogeneous quartic constraint studied above and lead to the following amplitudes: 
\begin{equation}\label{Paper3-fourAmplitude_homo}
    \mA^{(D)}_{\text{homo}}(1_{\lambda_1}2_{\lambda_2}3_{\lambda_3}4_{\lambda_4})=\sum_{i=0}^{\frac{D-d}{2}}\left(c_i\,s^it^{\frac{D-d}{2}-i}\right)[12]^{2\lambda_2}\langle 34\rangle^{\lambda_1-\lambda_{234}}[13]^{\lambda_{13}-\lambda_{24}}[14]^{\lambda_{14}-\lambda_{23}}\,,
\end{equation}
where the $c_i$ are $\frac{D-d}{2}+1$ free coefficients. 

\paragraph{Single-channel amplitudes.}
When the conditions \eqref{Paper3-total_conditions} are satisfied for $k=1$, single-channel amplitudes are allowed, and the most general four-point amplitude is
\begin{equation}\label{Paper3-single-channel_stu}
    \mA^{(D)}(1_{\lambda_1}2_{\lambda_2}3_{\lambda_3}4_{\lambda_4})=\mA^{(D)}_{s}+\mA^{(D)}_{t}+\mA^{(D)}_{u}\,,
\end{equation}
where
\begin{subequations}\label{Paper3-single-channel_amplitudes}
\begin{align}
    \mA^{(D)}_{s}(1_{\lambda_1}2_{\lambda_2}3_{\lambda_3}4_{\lambda_4})&=k_s\frac{t^{\frac{D-d+2}{2}}}{s}[12]^{2\lambda_2}\langle 34\rangle^{\lambda_1-\lambda_{234}}[13]^{\lambda_{13}-\lambda_{24}}[14]^{\lambda_{14}-\lambda_{23}}+\mA^{(D)}_{\text{homo}}\,,\\
    \mA^{(D)}_{t}(1_{\lambda_1}2_{\lambda_2}3_{\lambda_3}4_{\lambda_4})&=k_t\frac{u^{\frac{D-d+2}{2}}}{t}[12]^{2\lambda_2}\langle 34\rangle^{\lambda_1-\lambda_{234}}[13]^{\lambda_{13}-\lambda_{24}}[14]^{\lambda_{14}-\lambda_{23}}+\mA^{(D)}_{\text{homo}}\,,\\
    \mA^{(D)}_{u}(1_{\lambda_1}2_{\lambda_2}3_{\lambda_3}4_{\lambda_4})&=k_u\frac{s^{\frac{D-d+2}{2}}}{u}[12]^{2\lambda_2}\langle 34\rangle^{\lambda_1-\lambda_{234}}[13]^{\lambda_{13}-\lambda_{24}}[14]^{\lambda_{14}-\lambda_{23}}+\mA^{(D)}_{\text{homo}}\,,
\end{align}
\end{subequations}
where we selected a representative four-point amplitude, noting that the addition of homogeneous solutions can always modify the explicit form of the term with the pole. The coefficients $k_{\bullet}\sim C\bar{C}$ are the products of cubic couplings in the $s$-, $t$-, and $u$-channel. In particular, we have\footnote{Here by $k_{1,2,3,4,5,6}$ we mean the products of coupling that were used in the various tables presented.} 
\begin{align}
    &k_s=k_1=k_3\,,&
    &k_t=k_2=k_4\,,&
    &k_u=k_5=k_6\,,
\end{align}
where, as discussed above for quartic vertices, coefficients such as $k_1$ and $k_3$ are equal only when both are nonzero. It may nevertheless happen that consistency requires one of them to vanish. This can be seen explicitly by checking whether a given exchange contribution satisfies the conditions in \eqref{Paper3-total_conditions}. The same reasoning applies to the pairs $(k_2, k_4)$ and $(k_5, k_6)$.

For the special case when $D=d-2$, we have a unique solution corresponding to
\begin{subequations}
\begin{align}
    \mA^{(d-2)}_{s}(1_{\lambda_1}2_{\lambda_2}3_{\lambda_3}4_{\lambda_4})&=\frac{k_s}{s}[12]^{2\lambda_2}\langle 34\rangle^{\lambda_1-\lambda_{234}}[13]^{\lambda_{13}-\lambda_{24}}[14]^{\lambda_{14}-\lambda_{23}}\,,\\
    \mA^{(d-2)}_{t}(1_{\lambda_1}2_{\lambda_2}3_{\lambda_3}4_{\lambda_4})&=\frac{k_t}{t}[12]^{2\lambda_2}\langle 34\rangle^{\lambda_1-\lambda_{234}}[13]^{\lambda_{13}-\lambda_{24}}[14]^{\lambda_{14}-\lambda_{23}}\,,\\
    \mA^{(d-2)}_{u}(1_{\lambda_1}2_{\lambda_2}3_{\lambda_3}4_{\lambda_4})&=\frac{k_u}{u}[12]^{2\lambda_2}\langle 34\rangle^{\lambda_1-\lambda_{234}}[13]^{\lambda_{13}-\lambda_{24}}[14]^{\lambda_{14}-\lambda_{23}}\,.
\end{align}
\end{subequations}
Note that if we sum single-channel amplitudes as in \eqref{Paper3-single-channel_stu}, they will admit factorisation into three channels if $k_s,k_t,k_u\neq 0$ or two channels (for example, $s$ and $t$) if we take $k_u=0$.

\paragraph{YM-like amplitudes.}
When the conditions \eqref{Paper3-total_conditions} are satisfied only for $k=2$, and we have $D=d-4$, YM-like amplitudes are allowed, and we obtain the following unique amplitudes
\begin{subequations}\label{Paper3-YM-like_amplitudes}
\begin{align}
    \mA^{(d-4)}_{st}(1_{\lambda_1}2_{\lambda_2}3_{\lambda_3}4_{\lambda_4})&=\frac{k_{st}}{st}[12]^{2\lambda_2}\langle 34\rangle^{\lambda_1-\lambda_{234}}[13]^{\lambda_{13}-\lambda_{24}}[14]^{\lambda_{14}-\lambda_{23}}\,,\\
    \mA^{(d-4)}_{us}(1_{\lambda_1}2_{\lambda_2}3_{\lambda_3}4_{\lambda_4})&=\frac{k_{us}}{us}[12]^{2\lambda_2}\langle 34\rangle^{\lambda_1-\lambda_{234}}[13]^{\lambda_{13}-\lambda_{24}}[14]^{\lambda_{14}-\lambda_{23}}\,,\\
    \mA^{(d-4)}_{tu}(1_{\lambda_1}2_{\lambda_2}3_{\lambda_3}4_{\lambda_4})&=\frac{k_{tu}}{tu}[12]^{2\lambda_2}\langle 34\rangle^{\lambda_1-\lambda_{234}}[13]^{\lambda_{13}-\lambda_{24}}[14]^{\lambda_{14}-\lambda_{23}}\,.
\end{align}
\end{subequations}
Here, by factorisation, we get the following relations among the various couplings $k_{st}=k_s=k_t$, $k_{us}=k_u=k_s$, and $k_{tu}=k_t=k_u$, then $k_{st}=k_{us}=k_{tu}=k_s=k_t=k_u=k$. There is a unique coupling. Moreover, if we sum over all of them and look for an amplitude that has three poles, we get
\begin{align}
    \begin{split}
    \mA^{(d-4)}(1_{\lambda_1}2_{\lambda_2}3_{\lambda_3}4_{\lambda_4})&=\mA^{(d-4)}_{st}+\mA^{(d-4)}_{us}+\mA^{(d-4)}_{tu}\\
    &=\frac{k(s+t+u)}{stu}[12]^{2\lambda_2}\langle 34\rangle^{\lambda_1-\lambda_{234}}[13]^{\lambda_{13}-\lambda_{24}}[14]^{\lambda_{14}-\lambda_{23}}=0\,.
    \end{split}
\end{align}
This tells us that a GR-like amplitude cannot exist. For instance, Yang–Mills theory does not admit a GR-like amplitude. More generally, however, the statement is stronger: no YM-like amplitude can ever generate a GR-like one. The same conclusion can also be reached in a more conventional way by searching for an amplitude of the form
\begin{equation}
    \mA^{(d-4)}(1_{\lambda_1}2_{\lambda_2}3_{\lambda_3}4_{\lambda_4})
    =\left(\frac{k_{st}}{st}+\frac{k_{us}}{us}+\frac{k_{tu}}{tu}\right)[12]^{2\lambda_2}\langle 34\rangle^{\lambda_1-\lambda_{234}}[13]^{\lambda_{13}-\lambda_{24}}[14]^{\lambda_{14}-\lambda_{23}}\,.
\end{equation}
Consistent factorisation (by matching the $s$, $t$, and $u$ residues) implies:
\begin{equation}\label{Paper3-system_YMHS}
    R_s=\left(\frac{k_{st}}{t}-\frac{k_{us}}{t}\right)=\frac{k_s}{t}\,,\quad
    R_t=\left(\frac{k_{tu}}{u}-\frac{k_{st}}{u}\right)=-\frac{k_t}{u}\,,\quad
    R_u=\left(\frac{k_{us}}{s}-\frac{k_{tu}}{s}\right)=-\frac{k_u}{s}\,,
\end{equation}
where $R_s$ denoted the $s$-channel residue of $\mA^{(d-4)}$,\footnote{We recall that the $s$-channel residue is obtained in the limit $\langle 12\rangle\to 0$ and $[34]\to 0$. More generally, consider an amplitude $\mA(1_{\lambda_1}2_{\lambda_2}3_{\lambda_3}4_{\lambda_4})$. If $\lambda_{12}+\omega>0$ and $\lambda_{34}-\omega<0$, the $p^2_{12}\to 0$ limit is reached via $\langle 12\rangle\to 0$ and $[34]\to 0$. Conversely, if $\lambda_{12}+\omega<0$ and $\lambda_{34}-\omega>0$, the same limit is reached via $[12]\to 0$ and $\langle 34\rangle\to 0$.} and similarly for the $t$- and $u$-channels. The system of equations \eqref{Paper3-system_YMHS}, in order to admit a solution \cite{Arkani-Hamed:2017jhn}, does require
\begin{equation}
    k_s-k_t=k_u\,.
\end{equation}
This is the modified Jacobi identity, analogous to the one encountered in Yang–Mills theory in \eqref{Paper3-JI_YM}. This observation suggests the possibility of extending color–kinematics duality and the double-copy construction to higher-spin amplitudes in the non-holomorphic sector. Color–kinematics duality and the double-copy have already been explored in the self-dual sector for higher-spin theories, where the underlying kinematic algebra is particularly tractable \cite{Ponomarev:2017nrr,Ponomarev:2024jyg}. 

In the following we will show that, at the level of four-point amplitudes, color–kinematics duality and the double-copy construction can indeed be extended beyond the self-dual sector. At higher multiplicity, one could attempt to reorganise the amplitudes into a representation better suited for the analysis of color–kinematics duality \cite{Bern:2019prr,Bern:2022wqg}.

\paragraph{GR-like amplitudes.}
When the conditions \eqref{Paper3-total_conditions} are satisfied only for $k=3$, and we have $D=d-6$, GR-like amplitudes are allowed, and we obtain the following unique amplitude
\begin{equation}\label{Paper3-GR-like_amplitude}
    \mA^{(d-6)}_{stu}(1_{\lambda_1}2_{\lambda_2}3_{\lambda_3}4_{\lambda_4})=\frac{k_{stu}}{stu}[12]^{2\lambda_2}\langle 34\rangle^{\lambda_1-\lambda_{234}}[13]^{\lambda_{13}-\lambda_{24}}[14]^{\lambda_{14}-\lambda_{23}}\,,
\end{equation}
where consistent factorisation fixes $k_{stu}=k_s=k_t=k_u=k$.

From the expression of the YM-like amplitudes in \eqref{Paper3-YM-like_amplitudes} and the GR-like amplitudes in \eqref{Paper3-GR-like_amplitude}, it is simple to see 
\begin{equation}\label{Paper3-YM-like-BCJ}
    \mA^{(d-4)}_{st}=\frac{u}{t}\mA^{(d-4)}_{us}=\frac{u}{s}\mA^{(d-4)}_{tu}\,,
\end{equation}
and 
\begin{equation}\label{Paper3-GR-like-KLT}
    \mA^{(2d-6)}_{stu}(1_{2\lambda_1}2_{2\lambda_2}3_{2\lambda_3}4_{2\lambda_4})=\frac{st}{u}\big(\mA^{(d-4)}_{st}(1_{\lambda_1}2_{\lambda_2}3_{\lambda_3}4_{\lambda_4})\big)^2\,.
\end{equation}
At four points, these structures imply the existence of color-ordered amplitudes, the BCJ amplitude relations for YM-like amplitudes, and the double-copy construction of GR-like amplitudes from YM-like ones, via KLT relations \cite{Kawai:1985xq}. We will investigate these relations in greater detail in the following, providing several explicit examples.

We stress again that all the amplitudes described above are local amplitudes. We now present some explicit four-point amplitudes for several representative examples. 

\paragraph{External scalar fields.} We start with amplitudes involving only external scalar fields, then $d=0$ and $k=0$. From \eqref{Paper3-fourAmplitude_homo}, we find
\begin{equation}
    \mA^{(D)}_{\text{homo}}(1_02_03_04_0)=\sum_{i=0}^{D/2}\left(c_i\,s^it^{D/2-i}\right)\,.
\end{equation}
In the case of $\phi^4$ theory, then for $D=0$, we have $\mA^{(D)}_{\text{homo}}(1_02_03_04_0)=c_0$, where $c_0$ corresponds to the $\phi^4$ coupling constant. 

If we now consider the single-channel amplitudes, then $d=0$, $k=1$, and $D=2\omega-2$. From \eqref{Paper3-single-channel_amplitudes} we find 
\begin{equation}
    \mA_s^{(2\omega-2)}(1_02_03_04_0)=k_s\frac{t^{\omega}}{s}+\mA^{(2\omega-2)}_{\text{homo}}(1_02_03_04_0)\,,
\end{equation}
where $k_s=C^{0,0,\omega} \bar{C}^{-\omega,0,0}$ and if we consider all three channels \eqref{Paper3-single-channel_stu} we get
\begin{equation}
    \mA^{(2\omega-2)}(1_02_03_04_0)=k_s\frac{t^{\omega}}{s}+k_t\frac{u^{\omega}}{t}+k_u\frac{s^{\omega}}{u}+\mA^{(2\omega-2)}_{\text{homo}}(1_02_03_04_0)\,.
\end{equation}
In the case of $\phi^3$ theory, then for $\omega=0$ and $k_s=k_t=k_u=k$, we get
\begin{equation}
    \mA^{(2\omega-2)}(1_02_03_04_0)=k\Big(\frac{1}{s}+\frac{1}{t}+\frac{1}{u}\Big)\,,
\end{equation}
where $k=(C^{0,0,0})^2$ is the square of the $\phi^3$ coupling constant.

It is worth noting that the four-point amplitudes constructed remain valid for the $\phi^3$ theory, even though the case $n,m=0$ was excluded from our analysis.

We also note that four-point amplitudes with external scalar fields and higher-spin exchanges were previously computed in \cite{Ponomarev:2016lrm}, where their explicit expressions in the light-cone gauge were derived.

\paragraph{External helicities $\lambda_{1,2,3,4}=+1$.}
We have again $d=0$ and $k=0$. From \eqref{Paper3-fourAmplitude_homo}, we find
\begin{equation}
    \mA^{(D)}_{\text{homo}}(1^+2^+3^+4^+)=\sum_{i=0}^{D/2}\left(c_i\,s^it^{D/2-i}\right)\frac{[12]^2}{\langle 34\rangle^2}\,.
\end{equation}
For a local amplitude we require $D\geq 4$. For $D=4$, we find 
\begin{align}
\begin{split}
    \mA^{(4)}_{\text{homo}}(1^+2^+3^+4^+)&=c_2[12]^2[34]^2+c_1[12][34][14][23]+c_0[14]^2[23]^2\\
    &=c'_2[12]^2[34]^2+c'_1[13]^2[24]^2+c'_0[14]^2[23]^2\\
    &=c'_2[12]^2[34]^2+c'_1\frac{u^2}{s^2}[12]^2[34]+c'_0\frac{t^2}{s^2}[12]^2[34]^2\,,
\end{split}
\end{align}
where we used the Schouten identity, $[12][34]+[14][23]=[13][24]$ and redefined the coefficients according to $c'_2=c_2-\frac{c_1}{2}$, $c'_1=\frac{c_1}{2}$, and $c'_0=c_0-\frac{c_1}{2}$. As already remarked, this illustrates that an ansatz containing the factor $[12][34]$ would require the function $p(s,t)$ to include contributions involving cross-ratios, such as $u/s$ and $t/s$. Consequently, the locality of the amplitude would no longer be manifest in this parametrization.

If we now consider the single-channel amplitudes, then $d=0$, $k=1$. Moreover, by the requirement that $m>0$, we get $\omega\geq 3$, then $D\geq 4$. From \eqref{Paper3-single-channel_amplitudes} we find 
\begin{equation}
    \mA_s^{(2\omega-2)}(1^+2^+3^+4^+)=k_s\frac{t^{\omega}}{s}\frac{[12]^2}{\langle 34\rangle^2}+\mA^{(2\omega-2)}_{\text{homo}}(1^+2^+3^+4^+)\,,
\end{equation}
where $k_s\sim C^{1,1,\omega} \bar{C}^{-\omega,1,1}$ and if we consider all three channels \eqref{Paper3-single-channel_stu}, we get
\begin{equation}
    \mA^{(2\omega-2)}(1^+2^+3^+4^+)=\Big(k_s\frac{t^{\omega}}{s}+k_t\frac{u^{\omega}}{t}+k_u\frac{s^{\omega}}{u}\Big)\frac{[12]^2}{\langle 34\rangle^2}+\mA_s^{(2\omega-2)}(1^+2^+3^+4^+)\,.
\end{equation}

\paragraph{Yang-Mills theory.} For Yang-Mills theory, from \eqref{Paper3-YM-like_amplitudes}, we find
\begin{subequations}\label{Paper3-YM-amplitudes}
\begin{align}
      \mA^{(0)}_{st}(1^+2^+3^-4^-)=\frac{k_{\text{YM}}}{st}[12]^2\langle 34\rangle^2&=-k_{\text{YM}}\frac{[12]^3}{[23][34][41]}\,,\\
    \mA^{(0)}_{us}(1^+2^+3^-4^-)=\frac{k_{\text{YM}}}{us}[12]^2\langle 34\rangle^2&=-k_{\text{YM}}\frac{[12]^3}{[13][34][24]}\,,\\
    \mA^{(0)}_{tu}(1^+2^+3^-4^-)=\frac{k_{\text{YM}}}{tu}[12]^2\langle 34\rangle^2&=-k_{\text{YM}}\frac{[12]^4}{[23][24][13][14]}\,,
\end{align}    
\end{subequations}
where $k_{\text{YM}}\sim C^{1,1,-1}C^{1,-1,-1}$. From this representation, it is immediate to see $\mA^{(0)}_{us}(1^+2^+3^-4^-)=\frac{t}{u}\mA^{(0)}_{st}(1^+2^+3^-4^-)=\mA^{(0)}_{st}(1^+2^+4^-3^-)$. The remaining two amplitudes correspond precisely to the two standard color-ordered amplitudes:
\begin{align}
    \tilde{\mA}(1^+2^+3^-4^-)=&-\frac{1}{k_{\text{YM}}}\mA^{(0)}_{st}(1^+2^+3^-4^-)=\frac{[12]^{3}}{[23][34][41]}\,,\\
    \tilde{\mA}(1^+3^-2^+4^-)=&-\frac{1}{k_{\text{YM}}}\mA^{(0)}_{tu}(1^+2^+3^-4^-)=\frac{[12]^{4}}{[13][32][24][41]}\,.
\end{align}
These are again related, as a consequence of $s+t+u=0$, that implies $\mA^{(0)}_{st}+\mA^{(0)}_{us}+\mA^{(0)}_{tu}=0$. Moreover, just by looking at \eqref{Paper3-YM-amplitudes} we find $\mA^{(0)}_{tu}(1^+_12^+_13^-_14^-_1)=\frac{s}{u}\mA^{(0)}_{st}(1^+_12^+_13^-_14^-_1)$. This is the simplest example of BCJ amplitude relation \cite{Bern:2019prr}.

\paragraph{Gravity.} For gravity, from \eqref{Paper3-GR-like_amplitude}, we find
\begin{equation}
    \mA^{(2)}_{stu}(1^+_22^+_23^-_24^-_2)=k_{\text{GR}}\frac{[12]^4\langle 34\rangle^4}{stu}\,.
\end{equation}
where $k_{\text{GR}}\sim C^{2,2,-2}C^{2,-2,-2}$. This can also be obtained by double-copy from the Yang-Mills quartic amplitude via the simplest of the KLT relations \cite{Bern:2019prr}
\begin{equation}
    \mA^{(2)}_{stu}(1^+_22^+_23^-_24^-_2)=\frac{st}{u}\big(\mA^{(0)}_{st}(1^+2^+3^-4^-)\big)^2\,,
\end{equation}
after the identification of $k_{\text{GR}}=k_{\text{YM}}^2$.

\paragraph{Quasi-chiral HS-YM.}
For the quasi-chiral HS-YM theories found above, four-point amplitudes are a simple generalisation of the YM ones
\begin{align}
    &\mA^{(2s-2)}_{st}(1^+_s2^+_s3^-_14^-_1)=\frac{k^{\text{HS}}_{\text{YM}}}{st}[12]^{2s}\langle 34\rangle^2=-k^{\text{HS}}_{\text{YM}}\frac{[12]^{2s+1}}{[23][34][41]}\,,\\
    &\mA^{(2s-2)}_{us}(1^+_s2^+_s3^-_14^-_1)=\frac{k^{\text{HS}}_{\text{YM}}}{us}[12]^{2s}\langle 34\rangle^2=-k^{\text{HS}}_{\text{YM}}\frac{[12]^{2s+1}}{[13][34][24]}\,,\\
    &\mA^{(2s-2)}_{tu}(1^+_s2^+_s3^-_14^-_1)=\frac{k^{\text{HS}}_{\text{YM}}}{tu}[12]^{2s}\langle 34\rangle^2=-k^{\text{HS}}_{\text{YM}}\frac{[12]^{2s+2}}{[23][24][13][14]}\,,
\end{align}
where $k^{\text{HS}}_{\text{YM}}\sim C^{s,s,-1}C^{1,-1,-1}$.
As before, we have $\mA^{(2s-2)}_{us}(1^+_s2^+_s3^-_14^-_1)=\frac{t}{u}\mA^{(2s-2)}_{st}(1^+_s2^+_s3^-_14^-_1)=\mA^{(2s-2)}_{st}(1^+_s2^+_s4^-_13^-_1)$. The remaining two amplitudes correspond to the two color-ordered amplitudes:
\begin{subequations}\label{Paper3-color-ordered-HSYM}
\begin{align}
    \tilde{\mA}(1_{s}^+2_{s}^+3_{1}^-4_{1}^-)=&-\frac{1}{k^{\text{HS}}_{\text{YM}}}\mA^{(2s-2)}_{st}(1^+_s2^+_s3^-_14^-_1)=\frac{[12]^{2s+1}}{[23][34][41]}\,,\\
    \tilde{\mA}(1_{s}^+3_{1}^-2_s^+4_{1}^-)=&-\frac{1}{k^{\text{HS}}_{\text{YM}}}\mA^{(2s-2)}_{tu}(1^+_s2^+_s3^-_14^-_1)=\frac{[12]^{2s+2}}{[13][32][24][41]}\,.
\end{align}
\end{subequations}
These amplitudes are also related as a consequence of $s+t+u=0$, that implies $\mA^{(2s-2)}_{st}+\mA^{(2s-2)}_{us}+\mA^{(2s-2)}_{tu}=0$. Moreover, just by looking at \eqref{Paper3-YM-amplitudes}, we find $\mA^{(2s-2)}_{tu}(1^+_s2^+_s3^-_14^-_1)=\frac{s}{u}\mA^{(2s-2)}_{st}(1^+_s2^+_s3^-_14^-_1)$. This is a BCJ amplitude relation for higher-spin amplitudes in the non-holomorphic sector. More generally, the same applies to all YM-like amplitudes \eqref{Paper3-YM-like-BCJ}.

The expressions \eqref{Paper3-color-ordered-HSYM} coincide with those found in \cite{Adamo:2022lah}, although we have already emphasized that the theory described there exhibits a certain degree of non-locality. Setting $s=1$, we recover the standard Yang-Mills four-point amplitudes.

\paragraph{Quasi-chiral HS-GR.} For the quasi-chiral HS-GR theories found above, the four-point amplitudes are a simple generalisation of the GR ones
\begin{equation}
    \mA^{(2s-2)}(1_{s}^+2_{s}^+3_{2}^-4_{2}^-)=-k^{\text{HS}}_{\text{GR}}\frac{[12]^{2s}\langle 34 \rangle^4 }{stu}\,,
\end{equation}
where $k^{\text{HS}}_{\text{GR}}\sim C^{s,s,-2}C^{2,-2,-2}$.
Setting $s=2$, we recover the standard MHV four-point graviton amplitude. 
Some of them can also be obtained by a ''double-copy'' for higher-spin amplitudes from the four-point amplitude of HS-YM via
\begin{equation}
    \mA^{(4s-2)}_{stu}(1^+_{2s}2^+_{2s}3^-_24^-_2)=\frac{st}{u}\big(\mA^{(2s-2)}_{st}(1^+_s2^+_s3^-_14^-_1)\big)^2\,,
\end{equation}
after the identification of $k^{\text{HS}}_{\text{GR}}=(k^{\text{HS}}_{\text{YM}})^2$.
More generally, the same applies to all GR-like amplitude \eqref{Paper3-GR-like-KLT}.

We emphasise once again that the conditions in \eqref{Paper3-total_conditions} precisely coincide with those obtained in the light-cone Hamiltonian formulation, namely \eqref{Paper3-kinda_trinagular}, \eqref{Paper3-surpass_trinagular}, and \eqref{Paper3-Homo_constraints}. This agreement is expected, as it was shown in \cite{Ponomarev:2016cwi} that the light-cone deformation procedure can be reformulated entirely in terms of the spinor-helicity formalism at all orders in the interaction.

Let us stress, however, that the existence of local four-point amplitudes does not by itself guarantee the existence of a local Lorentz-invariant theory. Indeed, while a given four-point amplitude may be local, other amplitudes in the same theory may still be non-local. Establishing Lorentz invariance of a theory requires verifying the full set of amplitudes, a task that we carry out explicitly in Section \ref{Paper3-section6}. From the light-cone perspective, this corresponds to determining whether, for a given set of cubic vertices, the quartic light-cone consistency condition admits a local solution. In some cases, although local quartic vertices solving some of the constraints do exist, additional non-local quartic vertices are nevertheless required to solve all of them.

The strategy pursued here, as already suggested in \cite{Ponomarev:2016cwi}, may provide a viable route towards solving the light-cone constraints at all orders in perturbation theory. This would require a systematic investigation of the existence and form of general $n$-point higher-spin amplitudes. Such a program would be analogous to the search for compact expressions for known classes of amplitudes, such as the Parke–Taylor formula for tree-level MHV scattering amplitudes of $n$ gluons \cite{Parke:1986gb,Berends:1987me}, the tree-level MHV graviton amplitudes \cite{Bedford:2005yy,Hodges:2012ym}, and also that of the quasi chiral HS-YM amplitudes \cite{Adamo:2022lah}.

The same ideas are likely applicable to holomorphic amplitudes, potentially reproducing the results of \cite{Ponomarev:2016lrm,Serrani:2025owx}. Furthermore, from the amplitude perspective, the extension to fermions and massive fields seems more direct \cite{Arkani-Hamed:2017jhn}.

We conclude by observing that the YM-like and GR-like amplitudes constructed above do not admit homogeneous quartic contributions. This absence suggests the possibility that a suitable complex momentum deformation, analogous to the BCFW shift, may exist such that the boundary contribution at $z\to\infty$ vanishes. Establishing the existence of such a shift would allow for the construction of higher-point amplitudes through on-shell recursion relations.

\subsection*{Abelian vs Non-abelian cubic vertices II}

Here we show that the conditions in \eqref{Paper3-total_conditions} contain information about the nature, abelian or non-abelian, of the cubic vertices entering the amplitude. It is straightforward to see that, if we require an amplitude to contain at least one cubic vertex saturating the bound in \eqref{Paper3-total_conditions} for $k=1$, then at least one of the cubic vertices must satisfy $(-s_1,s_2,s_3)$ with $s_1+s_2=s_3$, or a parity-related equivalent condition. Similarly, requiring saturation of the bound for $k=2$ implies that at least one cubic vertex satisfy $(-s_1,s_2,s_3)$ with $s_1+s_2+1=s_3$, or a parity-related equivalent condition. Finally, saturation of the bound for $k=3$ requires the presence of at least one cubic vertex that satisfy $(-s_1,s_2,s_3)$ with $s_1+s_2+2=s_3$, or a parity-related equivalent condition. These three condition are associated with the three classes of four-point amplitudes: respectively, single-channel amplitudes ($k=1$, then $s_1+s_2=s_3$), YM-like ($k=2$, then $s_1+s_2+1=s_3$) and GR-like ($k=3$, then $s_1+s_2+2=s_3$) amplitudes. 

This is consistent with the interpretation of cubic vertices as abelian or non-abelian given in Section \ref{Paper3-section6}, following \cite{Bekaert:2010hp}. Our analysis provides an on-shell characterisation of this distinction, making its relation with the light-cone formulation more explicit. Furthermore, it reveals a stronger constraint: while all abelian deformations are compatible with locality, the only non-abelian deformations allowed by locality are those satisfying $s_3=s_1+s_2+1$ corresponding to YM-like deformations, or $s_3=s_1+s_2+2$ corresponding to GR-like deformations. More generally, deformations with $s_3=s_1+s_2+\ell$, $\ell>2$ are excluded by locality.

\section{Comments on non-local quartic vertices}\label{Paper3-section8}

From the analysis above, it is clear that although the landscape of local higher-spin interactions is richer than previously thought, admitting some non-trivial higher-spin quartic vertices and, as a consequence, certain quasi-chiral higher-spin theories, it also presents a clear obstruction to achieving parity invariance and unitarity without having to abandon locality. This also helps explain why, in the covariant framework, where parity is sometimes implicit, no parity-invariant higher-spin theory has ever been found.

This inevitably leads us, if we aim to construct a parity-invariant theory, to extend our search to non-local higher-spin interactions. The difficulty with exploring non-local vertices in a perturbative Noether-like approach is that, once we allow for non-localities of the form $\frac{1}{H_2}$, the deformation procedure becomes trivial. In the light-cone, allowing for such non-localities trivialises the quartic constraint, as shown in \eqref{Paper3-maineqB_full}.

We then follow the guidance of the light-cone deformation procedure. In doing so, we should allow for certain non-localities, with the aim of constructing a unitary theory, while explicitly excluding terms of the form $\frac{1}{H_2}$. This can be referred to as \textit{mild non-localities}. Although the resulting quartic Hamiltonian will contain transverse momenta in the denominator --- and thus be effectively non-local --- it does not have to trivialise the scattering or make the amplitudes ill-defined. In general, mild non-localities extend the space of admissible quartic vertices, opening the possibility for a unitary higher-spin theory that accommodates both chiral and anti-chiral vertices.

A systematic classification and analysis of all possible mild non-localities is a highly non-trivial task. For instance, a mild non-local version of the $(1,1)$ quartic vertex could take the form
\begin{align}\label{Paper3-(n,n)-(1,1)_vertex}
\begin{split}
h^{(1,1)}_4(x,y)=&\;\frac{\PP^n_{12}}{\PP^n_{34}}f_1(x,y)+\frac{\PP^{n-1}_{12}}{\PP^{n-1}_{34}}f_2(x,y)+\cdots+\frac{\PP_{12}}{\PP_{34}}f_n(x,y)+f_{n+1}(x,y)\\
&+\frac{\PPb_{12}}{\PPb_{34}}f_{n+2}(x,y)+\cdots+\frac{\PPb^{n-1}_{12}}{\PPb^{n-1}_{34}}f_{2n}(x,y)+\frac{\PPb^n_{12}}{\PPb^n_{34}}f_{2n+1}(x,y)\,.
\end{split}
\end{align}
However, other types of mild non-localities may also arise, such as
\begin{align}
    &\frac{c_1(\beta_i)\PP_{12}}{c_2(\beta_i)\PP_{12}+c_3(\beta_i)\PP_{34}}\,,&
    &\frac{c_1(\beta_i)\PP^2_{12}}{c_2(\beta_i)\PP^2_{12}+c_3(\beta_i)\PP_{12}\PP_{34}+c_4(\beta_i)\PP^2_{34}}\,.
\end{align}
More generally, one may encounter expressions of the form
\begin{equation}
    \frac{p_1(\PP_{12},\PP_{34},\beta_i)}{p_2(\PP_{12},\PP_{34},\beta_i)}+\frac{p_1(\PPb_{12},\PPb_{34},\beta_i)}{p_2(\PPb_{12},\PPb_{34},\beta_i)}\,,
\end{equation}
where the complex conjugate has been added to ensure unitarity. Such mild non-local structures would allow us to solve the quartic constraint \eqref{Paper3-quartic_system}. However, even without trivialising the perturbative procedure, they may still give rise to a very large number of possibilities.

We can try to follow a different line of thinking. A method to address the potential non-localities was proposed in \cite{Metsaev:1991nb}. Instead of searching for a local solution to the system \eqref{Paper3-quartic_system}, the quartic Hamiltonian density is split into two parts:
\begin{equation}
h_4 = -h_4^{\text{exch}} + h_4^{\text{homo}}\,,
\end{equation}
where $h_4^{\text{exch}}$ is a non-local quartic Hamiltonian density (constructed ad hoc, as we review below) with exchange-type non-localities (these are not of the form $\sim\frac{1}{H_2}$) that solves the system \eqref{Paper3-quartic_system}, and $h_4^{\text{homo}}$ is a solution of the corresponding homogeneous system.

As emphasised in \cite{Metsaev:1991nb}, a solution with exchange-type non-localities to \eqref{Paper3-quartic_system} always exists, taking the form
\begin{equation}\label{Paper3-exchange_solution}
h_4^{\text{exch}} = \frac{9}{2s} (-)^{\omega} C^{\lambda_1,\lambda_2,\omega} \bar{C}^{-\omega,\lambda_3,\lambda_4} \frac{\PPb_{12}^{\lambda_{12}+\omega} \PP_{34}^{-\lambda_{34}+\omega} }{(\beta_1+\beta_2)^{2\omega}}\frac{\beta_3^{\lambda_3} \beta_4^{\lambda_4}}{\beta_1^{\lambda_1} \beta_2^{\lambda_2}}\,,
\end{equation}
to cancel the $(1234)$ exchange diagram \eqref{Paper3-exchange_diagram}. The same form takes the solution for the other exchanges upon permuting the helicities in \eqref{Paper3-exchange_solution} and substituting $s$ with either $t$ or $u$.

Once $h_4^{\text{exch}}$ is constructed, one can then search for the most general solution of the homogeneous system, which Metsaev also provided in terms of Jacobi polynomials. The next step is to combine the homogeneous solution $h_4^{\text{homo}}$ with the non-local $h_4^{\text{exch}}$ and determine under which conditions a local solution to \eqref{Paper3-quartic_system} exists. This is the crucial question, which, however, is not solved in \cite{Metsaev:1991nb}.

This is the question we addressed above. We have shown that a local solution to \eqref{Paper3-quartic_system} exists if and only if $D-d+2k\geq 0$, where $D \equiv \lambda_{12} - \lambda_{34} + 2\omega -2$ is the total number of derivatives carried by the quartic Hamiltonian density, $d=\max\limits_{i\neq j\neq k\neq \ell}\{\lambda_i-\lambda_{jk\ell},-\lambda_i+\lambda_{jk\ell}\}$ and $k=1,2,3$ for single-channel, YM-like, and GR-like amplitudes, respectively. This condition, however, leads to the impossibility of a consistent local higher-spin theory, except for the case containing only abelian cubic interactions. 

Therefore, the only possible way to construct an interacting unitary higher-spin theory in $4d$ is to allow for a certain degree of non-locality (while avoiding singularities of the type $\frac{1}{H_2}$). Such non-localities would necessarily modify the pole structure of the resulting amplitudes. For instance, they may introduce higher-order poles, or conversely remove poles altogether. As an example, by choosing the quartic vertex to precisely cancel the exchange-type contributions as in \eqref{Paper3-exchange_solution}, the resulting four-point amplitude can be made to vanish.

Then a minimal example of a non-local theory at quartic order, consistent with the quartic light-cone constraint, would consist of turning on all the local quartic vertices described above and then reproducing all local four-point higher-spin amplitudes, while cancelling all remaining non-local four-point amplitudes through the exchange-type non-local contributions in \eqref{Paper3-exchange_solution}.

It would be interesting to investigate the structure of all possible exchange-type non-localities, at least at quartic order, and determine whether they exhibit any special or universal properties.

Concerning higher-order light-cone constraints \eqref{Paper3-LFNoether}, we expect them to behave in a way analogous to the quartic one.\footnote{For example, in the covariant approach, it is possible to extend the triangular inequalities for abelian vertices to quadrilateral inequalities and, more generally, to $n$-point inequalities that ensure the consistency of higher-point abelian vertices, following the same line of reasoning used for cubic interactions. We can therefore expect that analogous constraints to those found in the light-cone formulation can be extended to higher-point vertices as well; where spin-$1$ and spin-$2$ will maintain a special role.}

\section{Conclusions and discussion}\label{Paper3-section9}
 
In the paper, we carried out a detailed study of the full quartic constraint \eqref{Paper3-maineqB}. We began by developing a method to consistently address the problem of determining both the local Hamiltonian quartic density $h_4$ and the corresponding boost densities $j^{z-}_4$ and $j^{\zb-}_4$. We then introduced two complementary approaches: one to construct explicit solutions and another for systematically checking the existence of consistent quartic vertices (i.e. those that solve the associated system of quartic constraints \eqref{Paper3-quartic_system}), via Frobenius integrability. The latter approach allowed us to compile several tables listing all quartic vertices that solve \eqref{Paper3-quartic_system}. We then conjectured all such $(n,m)$ quartic vertices, revealing the special role taken by the spin-$1$ and spin-$2$ fields.

Our analysis reproduced known no-go and yes-go results for lower-spin interactions. In particular, chiral higher-spin gravity \cite{Metsaev:1991mt,Metsaev:1991nb,Ponomarev:2016lrm,Sharapov:2022faa,Sharapov:2022wpz,Sharapov:2022awp,Sharapov:2022nps,Sharapov:2023erv} does not have a local completion in flat space; see also \cite{Ponomarev:2017nrr}. An important novelty of our analysis is the ability to extract quartic vertices involving higher-spin fields. Indeed, the method does not rely on any assumptions about the high-energy behaviour but comes as a solution to the quartic constraint \eqref{Paper3-quartic}, which is always well-defined. The same analysis can be carried out for fermions, as well as for massive fields, starting from the work \cite{Metsaev:2005ar,Metsaev:2007rn}.

Using the conjectured local quartic vertices, we determined all possible local unitary higher-spin theories. We find that a consistent local higher-spin theory can contain only abelian cubic vertices that satisfy the triangular inequality. Conversely, attempting to include nontrivial lower-spin cubic vertices, such as the Yang–Mills or gravitational cubic couplings, leads to a no-go result for local higher-spin interactions.

Even though the impossibility of having a local unitary higher-spin theory is confirmed, many local higher derivative higher-spin quartic vertices have been discovered. This, in turn, suggests the existence of certain consistent local ``quasi-chiral'' sectors. We have studied and commented on the existence of a quasi-chiral HS-YM theory and a quasi-chiral HS-GR theory. Quasi-chiral theories have both MHV and anti-MHV vertices, but they enter in an asymmetric way with respect to the spins involved (the symmetric combinations of type $(\lambda,\lambda,-\lambda)\oplus (-\lambda,-\lambda,\lambda)$ can exist only for $\lambda=1,2$). 

Using the spinor-helicity formalism and imposing locality in the form of factorisation, we determine all local higher-spin four-point amplitudes and identify four distinct classes: amplitudes arising from self-consistent quartic vertices, amplitudes requiring a single exchange channel, amplitudes requiring two exchange channels (YM-like), and amplitudes requiring all three channels (GR-like). The latter two classes are closely related to the standard Yang–Mills and gravity MHV four-point amplitudes. Furthermore, the YM-like amplitudes exhibit color–kinematics duality \cite{Bern:2008qj}, while the GR-like amplitudes can be obtained through a double-copy construction \cite{Bern:2010ue}, extending these ideas to higher-spin non-holomorphic amplitudes. Indeed, the holomorphic case has already been analysed in \cite{Ponomarev:2017nrr,Ponomarev:2024jyg}. See also \cite{Serrani:2026azw}, where color–kinematics duality, and in particular the self-dual higher-spin kinematic algebra, is employed to construct all $n$-point half-collinear self-dual higher-spin amplitudes as generalisations of half-collinear self-dual Yang–Mills amplitudes \cite{Guevara:2026qzd}.

A very interesting question to explore is whether colour–kinematics duality and the double-copy construction in the form of the KLT relations we have found for higher-spin non-holomorphic four-points amplitudes, extend in some way to all local $n$-point tree-level amplitudes.

One important remark is that, although it was long believed that higher-spin theories require an infinite spectrum of fields in order to be consistent, this is not necessarily the case. In fact, it was already observed in the chiral sector in \cite{Serrani:2025owx}, and also hinted at in \cite{Ponomarev:2016lrm}, that such an infinite spectrum is not strictly required. Moreover, the study of the full quartic constraint does not appear to indicate any preference for an infinite spectrum over a finite one. As we have seen, the obstruction arising from locality is the same in both cases, whether the spectrum is infinite or finite. Moreover, the ``quasi-chiral'' theories we have found seems not to require an infinite spectrum of higher-spin fields.

An interesting future direction would be to extend the idea of mild non-locality to $AdS$ and see if it improves the locality properties of the quartic vertices found via the AdS/CFT correspondence in \cite{Bekaert:2015tva}. Note that for cubic vertices, the flat-space limit --- when the cosmological constant goes to zero --- is smooth, and the two classifications (cubic vertices in flat and AdS spaces) coincide \cite{Metsaev:2018xip}. Also, there does not seem to be any mechanism that would prefer a non-vanishing cosmological constant over flat space, as far as the higher-spin problem is concerned, and we expect the same yes-go/no-go results to extend to AdS if one follows the light-cone gauge path \cite{Metsaev:2018xip}.

Perhaps most importantly, there is the search for the optimal definition of mild non-locality for higher-spin quartic vertices. Such a definition would ideally allow for the existence of a unitary higher-spin theory while preserving the desirable features of perturbative field theory. In this work, we have proposed a concrete notion of mild non-locality. The analysis of possible non-localities has also been explored from the CFT side \cite{Ponomarev:2017qab, Sleight:2017pcz, Sleight:2016xqq,Neiman:2023orj}, and it would be interesting to compare the two approaches to determine whether they are related in some way.

Finally, it would be interesting to interpret the results obtained for the local quartic vertices in the context of a putative dual 2d celestial CFT. Indeed, in \cite{Serrani:2025oaw}, building on the ideas of \cite{Ren:2022sws,Monteiro:2022lwm,Monteiro:2022xwq}, it was shown that the  holomorphic light-cone constraint is equivalent to the associativity of the celestial OPE. We expect that this relation can be extended to the full quartic light-cone constraint and beyond. Achieving this would require a study of the “all-order” celestial OPE \cite{Adamo:2022wjo, Ren:2023trv} and likely also the multi-particle celestial OPE \cite{Calkins:2026hpg}. Such a connection could provide valuable insight into the relation between the existence of a consistent gravity theory in the bulk and a well-defined celestial CFT with an associative OPE on the boundary.

\section*{Acknowledgments}
\label{Paper3-sec:Aknowledgements1}
 I am grateful to Evgeny Skvortsov for suggesting the project and for many insightful discussions, particularly for drawing my attention to the role of triangular inequalities for abelian vertices. I would also like to thank Dmitry Ponomarev for pointing out the relevance of both the $J^{z-}$ and $J^{\bar{z}-}$  quartic constraints, as well as for many useful and insightful comments. I am further indebted to Dario Francia and Evgeny Skvortsov for encouraging me to provide more detailed explanations of the various tables. Finally, I would like to thank Dmitry Ponomarev once again for his valuable comments on the first version of this paper, and for pointing out the subtle distinction between abelian and non-abelian vertices. This project has received funding from the European Research Council (ERC) under the European Union’s Horizon 2020 research and innovation (grant agreement No 101002551).

\refstepcounter{section}
\section*{4.A \hspace{1mm} Unitarity and parity in the light-cone}
\addcontentsline{toc}{section}{4.A \hspace{1mm} Unitarity and parity in the light-cone}
\label{Paper3-AppendixA}

Parity transformations are defined as
\begin{align}
    &P:\phi^{\lambda}\rightarrow\phi^{-\lambda}\,,&
    &P:C^{\lambda_i}\rightarrow \bar{C}^{-\lambda_i}\,,&    &P:\bar{q}\rightarrow q\implies P:\PPb\rightarrow \PP\,.
\end{align}
At the cubic level, parity invariance requires $C^{\lambda_1,\lambda_2,\lambda_3}=\bar{C}^{-\lambda_1,-\lambda_2,-\lambda_3}$. On the other hand, unitarity (i.e. a Hermitian Lagrangian) requires $C^{\lambda_1,\lambda_2,\lambda_3}=(\bar{C}^{-\lambda_1,-\lambda_2,-\lambda_3})^*$.\footnote{Note that if we take even-helicity fields to be Hermitian matrices and odd-helicity fields to be anti-Hermitian matrices, the conditions become $C^{\lambda_1,\lambda_2,\lambda_3}=(-)^{\lambda_{123}}\bar{C}^{-\lambda_1,-\lambda_2,-\lambda_3}$ for parity invariance and $C^{\lambda_1,\lambda_2,\lambda_3}=(-)^{\lambda_{123}}(\bar{C}^{-\lambda_1,-\lambda_2,-\lambda_3})^*$ for unitarity.} When both $C$ and $\bar{C}$ are real, the two conditions coincide, so the theory is simultaneously parity-invariant and unitary. However, if the couplings are complex, one can, in principle, obtain a theory that is unitary but not parity-invariant, or vice versa.

\paragraph{Imposing parity to $\mathbf{h_4}$.} We begin by recalling that, for a three-point vertex, parity invariance requires the presence of both its holomorphic and anti-holomorphic components. In fact, parity exchanges the two as
\begin{equation}
    C^{\lambda_1,\lambda_2,\lambda_3} \frac{\PPb^{\lambda_{123}}}{\beta_1^{\lambda_1}\beta_2^{\lambda_2}\beta_3^{\lambda_3}}\quad
    \overset{P}{\longleftrightarrow}\quad
    \bar{C}^{-\lambda_1,-\lambda_2,-\lambda_3} \frac{\PP^{\lambda_{123}}}{\beta_1^{\lambda_1}\beta_2^{\lambda_2}\beta_3^{\lambda_3}}\,.
\end{equation}
For the quartic vertices, parity acts as follows:
\begin{equation}
 H_4\sim h_{\lambda_1,\lambda_2,\lambda_3,\lambda_4}^{q_1,q_2,q_3,q_4}\, \phi^{\lambda_1}_{q_1}\phi^{\lambda_2}_{q_2}\phi^{\lambda_3}_{q_3}\phi^{\lambda_4}_{q_4}\quad
    \overset{P}{\longleftrightarrow}\quad
    \bar{H}_4\sim \bar{h}_{-\lambda_1,-\lambda_2,-\lambda_3,-\lambda_4}^{\bar{q}_1,\bar{q}_2,\bar{q}_3,\bar{q}_4}\, \phi^{-\lambda_1}_{q_1}\phi^{-\lambda_2}_{q_2}\phi^{-\lambda_3}_{q_3}\phi^{-\lambda_4}_{q_4}\,,
\end{equation}
where by $\bar{h}$ we mean the same expression as $h$, with the replacement $q \leftrightarrow \bar q$, which effectively corresponds to $\mathbb{P} \leftrightarrow \bar{\mathbb{P}}$.
We can distinguish two cases: 
\begin{itemize}
    \item If the helicities of $\bar{h}_4$ are not a permutation of those of $h_4$, no additional condition needs to be imposed on the quartic vertex. In that case, achieving parity invariance proceeds exactly as in the cubic case: we simply add $\bar{h}_4$ to the spectrum of quartic vertices. However, the presence of both vertices can introduce new constraints, possibly rendering previously consistent configurations inconsistent, exactly as it happens in the cubic case when attempting to parity-complete a chiral theory.
    
     Let us also note that for the quartic Hamiltonian density of type $(n,m)$, with $n\neq m$, that has the following dependence on transverse momenta
    \begin{equation}
    h_4\sim \PPb^{n-1}\PP^{m-1}\quad
    \overset{P}{\longleftrightarrow}\quad
    \bar{h}_4\sim \PPb^{m-1}\PP^{n-1}\,,
    \end{equation}
    parity exchanges an $(n,m)$ quartic vertex with an $(m,n)$ one, in direct analogy with the cubic case. Therefore, to obtain a parity-invariant quartic vertex, both the $(n,m)$ and $(m,n)$ must be present together. If one of them exists, it can always be ``parity completed'' by adding the other.

    \item A distinctive feature of quartic vertices is that, for an $(n,n)$ vertex, the helicities of $\bar{h}_4$ may coincide with those of $h_4$ up to a permutation. In such cases, one must impose non-trivial symmetry conditions on the quartic vertex. Specifically, any permutation symmetry acting on the helicities must be satisfied by the transverse momenta $q$. In addition, we must recall that parity exchanges $\PPb$ and $\PP$.
\end{itemize}
Some useful transformation properties of $f_i(x,y)$, used in the main text, under permutations of the external legs, are listed below. When $x$ and $y$ are defined as in \eqref{Paper3-xy_1234}, one finds
\begin{subequations}\label{Paper3-f_parity_GR}
\begin{align}
    &f_i(x,y)=f_i\left(-x,y\right)&
    &(1234)\leftrightarrow (2134)\,,\\
    &f_i(x,y)=f_i\left(x,-y\right)&
    &(1234)\leftrightarrow (1243)\,,\\
    &f_i(x,y)=f_i(y,x)&
    &(1234)\leftrightarrow (1243)\,.
\end{align}
\end{subequations}
When $x$ and $y$ are defined as in \eqref{Paper3-xy_1324}, one finds
\begin{subequations}
\begin{align}
    &f_i(x,y)=f_i(y,-x)&
    &(1234)\leftrightarrow (2341)\,,\\
    &f_i(x,y)=f_i(-x,-y)&
    &(1234)\leftrightarrow (3412)\,,\\
    &f_i(x,y)=f_i(-y,x)&
    &(1234)\leftrightarrow (4123)\,.
\end{align}
\end{subequations}

\refstepcounter{section}
\section*{4.B \hspace{1mm} Useful formulas and relations}
\addcontentsline{toc}{section}{4.B \hspace{1mm} Useful formulas and relations}
\label{Paper3-AppendixB}

For the variables $x,y$ defined in \eqref{Paper3-xy_1234}, we have the following relations
\begin{subequations}\label{Paper3-xy_1234tranf}
\begin{align}
    &\frac{\beta_1-\beta_3}{\beta_1+\beta_3}=\frac{2+x+y}{x-y}\,,&
    &\frac{\beta_1-\beta_4}{\beta_1+\beta_4}=\frac{2+x-y}{x+y}\,,\\
    &\frac{\beta_2-\beta_3}{\beta_2+\beta_3}=\frac{-2+x-y}{x+y}\,,&
    &\frac{\beta_2-\beta_4}{\beta_2+\beta_4}=\frac{-2+x+y}{x-y}\,.
\end{align}
\end{subequations}
While for the variables $x,y$ defined in \eqref{Paper3-xy_1324}, we have the following relations
\begin{subequations}\label{Paper3-xy_1324tranf}
\begin{align}
    &\frac{\beta_1-\beta_2}{\beta_1+\beta_2}=\frac{2+x+y}{x-y}\,,&
    &\frac{\beta_1-\beta_4}{\beta_1+\beta_4}=\frac{2+x-y}{x+y}\,,\\
    &\frac{\beta_2-\beta_3}{\beta_2+\beta_3}=\frac{2-x+y}{x+y}\,,&
    &\frac{\beta_3-\beta_4}{\beta_3+\beta_4}=\frac{-2+x+y}{x-y}\,.
\end{align}
\end{subequations}
For four-point scattering, we have $2$ independent $\PPb$ variables, which, for example, can be chosen to be $\PPb_{12}$ and $\PPb_{34}$. Additionally, there are three independent $\beta$'s. All other $\PPb_{ij}$ can be expressed as
\begin{subequations}\label{Paper3-PP_relations}
\begin{align}
    \PPb_{13}&=\frac{\beta_3 \PPb_{12}+\beta_1 \PPb_{34}}{\beta_1+\beta_2}\,,
    &&\PPb_{14}=-\frac{\PPb_{12} (\beta_1+\beta_2+\beta_3)+\beta_1 \PPb_{34}}{\beta_1+\beta_2}\,,\\
    \PPb_{23}&=\frac{\beta_2 \PPb_{34}-\beta_3 \PPb_{12}}{\beta_1+\beta_2}\,,
    &&\PPb_{24}=\frac{\PPb_{12} (\beta_1+\beta_2+\beta_3)-\beta_2 \PPb_{34}}{\beta_1+\beta_2}\,.
\end{align}
\end{subequations}
The same applies to $\PP$. On-shell, the following relations hold:
\begin{align}
    \begin{split}
   &\langle ij\rangle [ij]=-\frac{2}{\beta_i\beta_j}\PP_{ij}\PPb_{ij}=2\,q_i \cdot q_j=(q_i+q_j)^2\,,\\
   &\sum^{n}_{j=1}q^{\mu}_j=0\quad\Rightarrow\quad\sum^{n}_{j=1}\langle ij\rangle[jk]=\sum^{n}_{j=1}\frac{\PP_{ij}\PPb_{jk}}{\beta_j}=0\,.
   \end{split}
\end{align}
In particular, using the relation above for the four-point scattering, we find
\begin{align}
    &\frac{\PPb_{12}\PPb_{34}}{(q_1+q_2)^2}=\frac{\PPb_{31}\PPb_{24}}{(q_1+q_3)^2}=\frac{\PPb_{14}\PPb_{23}}{(q_1+q_4)^2}\,,&
    &s_{ij}=-(q_i+q_j)^2\,,
\end{align}
where $s\equiv s_{12}$, $t\equiv s_{14}$, and $u\equiv s_{13}$ are the standard Mandelstam variables for massless four-point scattering. Off-shell, the Mandelstam variables are defined as
\begin{align}
    &s\equiv\frac{\PPb_{12}\PP_{12}}{\beta_1\beta_2}+\frac{\PPb_{34}\PP_{34}}{\beta_3\beta_4}\,,&
    &t\equiv\frac{\PPb_{14}\PP_{14}}{\beta_1\beta_4}+\frac{\PPb_{23}\PP_{23}}{\beta_2\beta_3}\,,&
    &u\equiv\frac{\PPb_{13}\PP_{13}}{\beta_1\beta_3}+\frac{\PPb_{24}\PP_{24}}{\beta_2\beta_4}\,.
\end{align}
\refstepcounter{section}
\section*{4.C \hspace{1mm} Cubic scalar vertex}
\addcontentsline{toc}{section}{4.C \hspace{1mm} Cubic scalar vertex}
\label{Paper3-AppendixC}

The holomorphic constraint originally analysed in \cite{Metsaev:1991mt,Metsaev:1991nb,Ponomarev:2016lrm} was further studied in \cite{Serrani:2025owx}, whose notation we follow. Here, we examine the holomorphic constraint for the cubic scalar vertex. Although this case is not explicitly discussed in \cite{Serrani:2025owx}, it highlights several interesting features and provides a useful testing ground for the light-cone deformation procedure. As we will see, it reproduces all the correct predictions for lower-spin vertices while also providing answers for the higher-spin ones. The holomorphic constraint for generic vertices is
\begin{align}\label{Paper3-holo_constraint}
    \begin{split}
    [H_3,J_3^{z-}]=&\sum_{\lambda_i,\omega}\int d^{12}q\;\delta \left(\sum_i q_i\right)\frac{9}{2}\Big[(-)^{\omega}\frac{(\lambda_1+\omega-\lambda_2)\beta_1-(\lambda_2+\omega-\lambda_1)\beta_2}{(\beta_1+\beta_2)\beta_1^{\lambda_1}\beta_2^{\lambda_2}\beta_3^{\lambda_3}\beta_4^{\lambda_4}}\,\times\\
    &\mathcal{F}^{1234\omega}\PPb_{12}^{\lambda_{12}+\omega-1}\PPb_{34}^{\lambda_{34}-\omega}\,\,(\phi^{\lambda_1}_{q_1})^{a_1}(\phi^{\lambda_2}_{q_2})^{a_2}(\phi^{\lambda_3}_{q_3})^{a_3}(\phi^{\lambda_4}_{q_4})^{a_4}\Big]=0\,,
    \end{split}
\end{align}
where $\mathcal{F}^{1234\omega}=\fA_{a_1a_2c}\fA^c_{\phantom{c}a_3a_4}C^{\lambda_1,\lambda_2,\omega}C^{-\omega,\lambda_3,\lambda_4}$ is built from a holomorphic pair of cubic couplings (a $CC$ pair).
Following the approach of \cite{Serrani:2025owx}, we analyse the solutions in the case where one of the two couplings is the scalar cubic self-interaction $C^{0,0,0}$. As in the non-holomorphic constraint, six independent kinematical structures contribute:
\begin{equation}
    (1234)+(1324)+(1423)+(3412)+(2413)+(2314)\,.
\end{equation}
Typically, the holomorphic constraint receives contributions from both $(1234)$ and $(3412)$ (and similarly for the $t$ and $u$ channels), because they involve the same product of couplings $C^{\lambda_1,\lambda_2,\omega}C^{-\omega,\lambda_3,\lambda_4}=C^{\lambda_3,\lambda_4,-\omega}C^{\omega,\lambda_1,\lambda_2}$. However, when one of the couplings is the scalar self-coupling $C^{0,0,0}$, we notice from \eqref{Paper3-holo_constraint} that while the power of $\PPb$ is zero, the power of $\PP$ is negative, giving a non-local contribution. The resolution to this apparent issue is that Eq.~\eqref{Paper3-holo_constraint} is no longer valid in this special case. Before inserting the explicit expressions for the densities, the commutator reads 
\begin{align}
\begin{split}
[H_3,J_3^{z-}]= &\sum_{\lambda_i,\alpha_j}\int d^9p\;d^9q\; \delta\left(\sum_i q_i\right) \delta\left(\sum_j p_j\right)9\,\delta^{\lambda_3,-\alpha_3}\frac{\delta(q_3+p_3)}{2q_3^+}\,\phi^{\lambda_1}_{q_1}\phi^{\lambda_2}_{q_2}\phi^{\alpha_1}_{p_1}\phi^{\alpha_2}_{p_2}\,\times\\
&\left(j_3^{\lambda_i}(q_i)+\sum_{k\neq 3}\frac{\partial}{\partial \bar{q}_{k}}\frac{h_3^{\lambda_i}(q_i)}{3}\right)h_3^{\alpha_j}(p_j)\,.
\end{split}
\end{align}
Here, only $h_3$ can generate the $C^{0,0,0}$ contribution, since $j_3$ would vanish.\footnote{In particular, the holomorphic constraint for the pair of couplings $C^{0,0,0}C^{0,0,0}$ gives identically zero.}. The solution is then \eqref{Paper3-holo_constraint} once all contributions that would have produced non-local terms are set to zero. 

Let us start by looking at the holomorphic constraint for the pair of coupling $C^{\lambda_1,\lambda_2,0}C^{0,0,0}$. In this case, we just get one contribution
\begin{align}
    &[H_3,J_3^{z-}]\sim C^{\lambda_1,\lambda_2,0}C^{0,0,0}(\lambda_1-\lambda_2)\PP_{12}^{\lambda_{12}-1}=0&
    &\implies&
    &\lambda_1=\lambda_2\,.
\end{align}
For the pair of couplings $C^{\lambda,0,0}C^{0,0,0}$, we have three independent contributions coming from $(1234)$, $(1324)$, and $(1423)$, then we get
\begin{align}
    &[H_3,J_3^{z-}]\sim C^{\lambda,0,0}C^{0,0,0}\left(\PP_{12}^{\lambda-1}+\PP_{13}^{\lambda-1}+\PP_{14}^{\lambda-1}\right)=0&
    &\implies&
    &\lambda=2\,.
\end{align}
Lorentz consistency between $C^{1,0,0}$ and $C^{0,0,0}$ is possible in the case of a color scalar vertex; thus, we need to consider only the cyclic contribution $[1234]$, and two terms survive
\begin{align}
    &[H_3,J_3^{z-}]\sim C^{\lambda,0,0}C^{0,0,0}\left(\PP_{12}^{\lambda-1}-\PP_{41}^{\lambda-1}\right)=0&
    &\implies&
    &\lambda=1\,.
\end{align}
Interestingly, we can also make $C^{1,0,0}$ consistent with $C^{0,0,0}$ if the cubic scalar is made out of three different scalars, then we find
\begin{align}
    \begin{split}
    [H_3,J_3^{z-}]\sim \;&C^{\lambda,0_1,0_1}C^{0_1,0_2,0_3}\PP_{12}^{\lambda-1}+C^{\lambda,0_2,0_2}C^{0_2,0_3,0_1}\PP_{13}^{\lambda-1}+C^{\lambda,0_3,0_3}C^{0_3,0_1,0_2}\PP_{14}^{\lambda-1}=0\\
    &\implies\quad
    \lambda=1\; \wedge\; C^{1,0_1,0_1}+C^{1,0_2,0_2}+C^{1,0_3,0_3}=0\,,\\
    &\implies\quad
    \lambda=2\; \wedge\; C^{2,0_1,0_1}=C^{2,0_2,0_2}=C^{2,0_3,0_3}\,,
    \end{split}
\end{align}
where $0_1$, $0_2$, and $0_3$ denote three distinct scalar fields. Notice that the solution for $\lambda=1$ implies that the total charge must vanish, while the case $\lambda=2$ confirms the universality of gravitational interactions.

As we have seen, the consistency of the scalar cubic couplings $C^{0,0,0}$ with higher-spin cubic couplings is restricted to interactions of the form $C^{\lambda,\lambda,0}$.
Consequently, for example, in the full chiral higher-spin theory, where all holomorphic cubic couplings are turned on, the cubic scalar coupling is not allowed.

The above results for lower-spin couplings are consistent with what is known in the covariant formulation. Recall that the Lorentz-invariance constraint in the light-cone formalism corresponds to gauge invariance in the covariant language. Indeed, the minimal coupling of a scalar field to gravity remains consistent even in the presence of a $\phi^3$ interaction term.

\refstepcounter{section}
\section*{4.D \hspace{1mm} Explicit form of lower-derivative quartic vertices}
\addcontentsline{toc}{section}{4.D \hspace{1mm} Explicit form of lower-derivative quartic vertices}
\label{Paper3-AppendixD}

We present explicit examples of lower-derivative quartic vertices. Free coefficients arising from solutions of the homogeneous system of quartic constraints are denoted by $c_i$ when present. Throughout, we use the commutator in the form given in \eqref{Paper3-commutator_general}.

\paragraph{(1,1) quartic vertices.}
\begin{align}
    &h^{(1,1)}_{[0,0,0,0]}:&
    &f(x,y)=k_5\frac{x y}{4}+k_1\frac{x y-1}{(x-y)^2}-k_2\frac{1+x y}{(x+y)^2}+c_1\,,\\
    &h^{(1,1)}_{[1,-1,0,0]}:&
    &f(x,y)=k_1\frac{1-x y}{(x-y)^2}\,,\\
    &h^{(1,1)}_{[1,0,-1,0]}:&
    &f(x,y)=-\frac{k_1}{2}\,.
\end{align}
In the case of $h^{(1,1)}_{[0,0,0,0]}$, the quartic constraint allows for single-channel, YM-like, and GR-like vertices.\footnote{This notation is explained both in section \ref{Paper3-section6} and section \ref{Paper3-section7}.} In particular, we have
\begin{align}
    h^{(1,1)}_{[0,0,0,0]}:\qquad\text{single-channel:}\qquad f(x,y)&=k_1\frac{x y-1}{(x-y)^2}+c_1\,,\\
    \text{YM-like:}\qquad f(x,y)&=k_1\left(\frac{x y-1}{(x-y)^2}-\frac{1+x y}{(x+y)^2}\right)+c_1\,,\\
    \text{GR-like:}\qquad f(x,y)&=k_1\left(\frac{x y}{4}+\frac{x y-1}{(x-y)^2}-\frac{1+x y}{(x+y)^2}\right)+c_1\,,
\end{align}
\paragraph{(2,2) quartic vertices.}
\begin{subequations}
\begin{align}
\nonumber
    &h^{(2,2)}_{(-1,1,-1,1)}:\\
    &f_1(x,y)=\frac{(x-1) (y+1) \left(k_1 (x+y)^4-2 k_2 \left(2 x^2+3 x y+x+y (2 y-1)+1\right)\right)}{2 (x+y)^4}\,,\\
    & f_2(x,y)=\frac{\left(x^2-1\right) \left(y^2-1\right) \left(3 k_1 (x+y)^4-8 k_2 \left(2 x^2+3 x y+2 y^2-1\right)\right)}{32 (x+y)^4}\,,\\
    &f_3(x,y)=\frac{(x+1) (y-1) \left(k_1 (x+y)^4-2 k_2 \left(3 x y+x (2 x-1)+2 y^2+y+1\right)\right)}{2 (x+y)^4}\,.
\end{align}
\end{subequations}
\begin{subequations}
\begin{align}
    h^{(2,2)}_{(-2,2,0,0)}:\qquad f_1(x,y)&=\frac{1}{2} k_1 (x-2) y\,,\\
     f_2(x,y)&=\frac{1}{32} k_1 \left(3 \left(x^2-3\right) y^2-x^2+7\right)\,,\\
    f_3(x,y)&=\frac{1}{2} k_1 (x+2) y\,.
\end{align}
\end{subequations}
\begin{subequations}
\begin{align}
    h^{(2,2)}_{(-2,2,-1,1)}:f_1(x,y)&=-\frac{k_1 (y+1) (x+y-1) (x (x+y-1)-2 y+1)}{2 (x+y)^2}\,,\\
    f_2(x,y)&=-\frac{k_1 \left(y^2-1\right) \left(3 x^4+3 \left(x^2-3\right) y^2+6 \left(x^2-1\right) x y-x^2+4\right)}{32 (x+y)^2}\,,\\
    f_3(x,y)&=-\frac{k_1 (y-1) (x+y+1) \left(x^2+(x+2) y+x+1\right)}{2 (x+y)^2}\,.
\end{align}
\end{subequations}

\refstepcounter{section}
\section*{4.E \hspace{1mm} Light-cone and spinor-helicity}
\addcontentsline{toc}{section}{4.E \hspace{1mm} Light-cone and spinor-helicity}
\label{Paper3-AppendixE}

For our spinor-helicity conventions for massless particles, we follow \cite{Elvang:2013cua}:
\begin{align}
    &q_{a\dot{b}}= q_{\mu}(\sigma^{\mu})_{a\dot{b}}\,,&
    &\text{det}(q_{a\dot{b}})=-q^{\mu}q_{\mu}=m^2\,,&
    &q^2=0\;\; \Rightarrow\;\; q_{a\dot{b}}=-|q]_a\langle q|_{\dot{b}}\equiv -\lambda_a\tilde{\lambda}_{\dot{b}}\,,
\end{align}
\begin{align}
        &\langle ij\rangle\definition \langle q_i|_{\dot{a}}|q_j\rangle^{\dot{a}}\equiv\tilde{\lambda}_{i\dot{a}}\tilde{\lambda}_j^{\dot{a}}\,,&
    &[ij]\definition [q_i|^a|q_j]_a\equiv \lambda_i^{a}\lambda_{ja}\,\,,&
    &\lambda^{a}=\epsilon^{ab}\lambda_{b}\,,&
&\tilde{\lambda}^{\dot{a}}=\epsilon^{\dot{a}\dot{b}}\tilde{\lambda}_{\dot{b}}\,,
\end{align}

\begin{align}
&\sigma^0=
    \begin{pmatrix}
        1 & 0\\
        0 & 1
    \end{pmatrix}\,,&
    &\sigma^1=
    \begin{pmatrix}
        0 & 1\\
        1 & 0
    \end{pmatrix}\,,&
    &\sigma^2=
    \begin{pmatrix}
        0 & -i\\
        i & 0
    \end{pmatrix}\,,&
    &\sigma^3=
    \begin{pmatrix}
        1 & 0\\
        0 & -1
    \end{pmatrix}\,,
\end{align}
\begin{align}
    &\epsilon^{ab}=-\epsilon_{ab}=
    \begin{pmatrix}
        0 & 1\\
        -1 & 0
    \end{pmatrix}\,,&
    &\epsilon^{\dot{a}\dot{b}}=-\epsilon_{\dot{a}\dot{b}}=
    \begin{pmatrix}
        0 & 1\\
        -1 & 0
    \end{pmatrix}\,.
\end{align}
Notice that for complex momenta $q^{\mu}$ the two spinors $(\lambda_a,\tilde{\lambda}_{\dot{b}})$ are independent two-dimensional complex vectors. In Minkowski space and for real momenta $q_{a\dot{b}}$ is hermitian, and the two spinors become complex conjugate $\tilde{\lambda}_{\dot{a}}=\pm(\lambda_a)^*$ (where the sign depends on whether the energy is taken to be positive or negative, then on the convention we use on the background flat metric). Spinor-helicity variables $(\lambda_i,\tilde{\lambda}_i)$ are defined up to little group scaling $(\lambda_i,\tilde{\lambda}_i)\sim (t_i\lambda_i,t_i^{-1}\tilde{\lambda}_i)$ for $t_i\in\mathbb{C}^*$. 

In the context of the light-cone Hamiltonian approach for massless higher-spin fields, we adopt the following notation \cite{Ponomarev:2016cwi}:
\begin{equation}
    q_{a\dot{b}}=q_{\mu}(\sigma^{\mu})_{a\dot{b}}=\sqrt{2}
    \begin{pmatrix}
        q^- & \bar{q}\\
        q & - \beta
    \end{pmatrix}\approx\sqrt{2}
    \begin{pmatrix}
        -\frac{q\bar{q}}{\beta} & \bar{q}\\
        q & - \beta
    \end{pmatrix}= -|q]_a\langle q|_{\dot{b}}=-\lambda_a\tilde{\lambda}_{\dot{b}}\,,
\end{equation}
\begin{align}
&\langle i|=\frac{2^{\frac{1}{4}}}{\sqrt{\beta_i}}\begin{pmatrix}
    q_i & -\beta_i
    \end{pmatrix}\,,&
    &\langle ij\rangle=-\sqrt{\frac{2}{\beta_i\beta_j}}\PP_{ij}\,,&
    &|i]=\frac{2^{\frac{1}{4}}}{\sqrt{\beta_i}}
    \begin{pmatrix}
       \bar{q}_i\\
        -\beta_i
    \end{pmatrix}\,,&
    &[ij]=\sqrt{\frac{2}{\beta_i\beta_j}}\PPb_{ij}\,.
\end{align}
The relation between the cubic amplitude and the cubic Hamiltonian density follows:
\begin{align}\label{Paper3-AmplitudeandHamiltonian_holo}
    \mathcal{A}_3&=C^{\lambda_1,\lambda_2,\lambda_3}[12]^{d_{12}} [23]^{d_{23}} [31]^{d_{31}}=C^{\lambda_1,\lambda_2,\lambda_3}\frac{\sqrt{2}^{\lambda_{123}}\PPb^{\lambda_{123}}}{\beta_1^{\lambda_1}\beta_2^{\lambda_2}\beta_3^{\lambda_3}}=\sqrt{2}^{\lambda_{123}}h_3\,,\\\label{Paper3-AmplitudeandHamiltonian_antiholo}
    \bar{\mathcal{A}}_3&=\bar{C}^{\lambda_1,\lambda_2,\lambda_3}\langle 12\rangle^{-d_{12}} \langle 23\rangle^{-d_{23}} \langle 31\rangle^{-d_{31}}=\bar{C}^{\lambda_1,\lambda_2,\lambda_3}\frac{\sqrt{2}^{\lambda_{123}}\PP^{-\lambda_{123}}}{\beta_1^{-\lambda_1}\beta_2^{-\lambda_2}\beta_3^{-\lambda_3}}=\sqrt{2}^{\lambda_{123}}\bar{h}_3\,,
\end{align}
where $\mathcal{A}_3$ is valid for $\lambda_{123}>0$ and $\bar{\mathcal{A}}_3$ for $\lambda_{123}<0$, and we have defined
\begin{align}
    &d_{12}=\lambda_{12}-\lambda_3\,,&
    &d_{23}=\lambda_{23}-\lambda_1\,,&
    &d_{31}=\lambda_{31}-\lambda_2\,.
\end{align}
The relations above, in the light-cone approach, hold only when the external particles are on-shell.  Indeed, one of the main differences between the cubic Hamiltonian density and the amplitudes in \eqref{Paper3-AmplitudeandHamiltonian_holo}-\eqref{Paper3-AmplitudeandHamiltonian_antiholo} is that the latter are inherently on-shell objects, while the former contains off-shell information. 

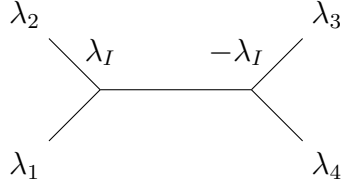
\begin{figure}[H]
    \centering
    \begin{tikzpicture}
        \begin{feynman}
            \vertex (i1) at (-6, 1) {\(\lambda_2\)};
            \vertex (i2) at (-6,-1) {\(\lambda_1\)};
            \vertex (i3) at (-2, 1) {\(\lambda_3\)};
            \vertex (i4) at (-2,-1) {\(\lambda_4\)};

            \vertex (v1) at (-5, 0);
            \vertex (v3) at (-3, 0);

            \vertex at (-5, 0.5) {\(\lambda_I\)};
            \vertex at (-3.2, 0.5) {\(-\lambda_I\)};

            \diagram* {
                (i1) -- (v1),
                (i2) -- (v1),
                (v1) -- [plain] (v3),
                (v3) -- (i3),
                (v3) -- (i4),
            };
        \end{feynman}
    \end{tikzpicture}
    \caption{Generic four-point exchange diagram.}
    \label{Paper3-gen_exch}
\end{figure}
\noindent
Given a four-point scattering, as shown in Figure \ref{Paper3-gen_exch}, momentum conservation implies
\begin{align}
    \begin{split}
     \langle 1|q_I|4]&=-\langle 1I\rangle[I4]=\langle 1|-q_1-q_2|4]=\langle 1|q_3+q_4|4]\\
     &\implies
     \langle 1I\rangle[I4]=\langle 12\rangle[24]=-\langle 13\rangle[34]\,.
     \end{split}
\end{align}

\refstepcounter{section}
\section*{4.F \hspace{1mm} Tables for higher-derivatives quartic vertices}
\addcontentsline{toc}{section}{4.F \hspace{1mm} Tables for higher-derivatives quartic vertices}
\label{Paper3-AppendixF}

Tables \ref{Paper3-tab(4,1)}--\ref{Paper3-tab(4,4)} collect quartic vertices that solve the system of quartic constraints~\eqref{Paper3-quartic_system}. These tables\footnote{We display them only up to $(4,4)$, although our analysis extended up to $(8,8)$.} have been used to conjecture the list of all quartic vertices that solve~\eqref{Paper3-quartic_system}, as discussed in the main text. Table \ref{Paper3-tabHomo} collects quartic vertices that solve the homogeneous system of quartic constraints.

\begin{table}[H]
    \centering
    \begin{tabular}{|c|c|c|c|c|c|c|}
\hline
$(\lambda_1,\lambda_2,\omega,\lambda_3,\lambda_4)$ & $k_2=(2341)$ & $k_3=(3412)$ & $k_4=(4123)$ & $k_5=(1324)$ & $k_6=(2413)$ \\ \hline
$(-1,3,2,-1,2)$ & $k_1$ & $0$ & $0$ & $0$ & $k_1$ \\ \hline
$(-1,3,2,0,1)$ & $k_1$ & $0$ & $0$ & $0$ & $k_1$ \\ \hline
$(0,2,2,-1,2)$ & $0$ & $0$ & $k_4$ & $0$ & $k_1-k_4$ \\ \hline
$(0,2,2,0,1)$ & $k_2$ & $0$ & $0$ & $0$ & $k_6$ \\ \hline
$(0,3,1,0,0)$ & $k_1$ & $0$ & $0$ & $0$ & $k_1$ \\ \hline
$(1,1,2,-1,2)$ & $0$ & $0$ & $k_4$ & $0$ & $k_1-k_4$ \\ \hline
$(1,1,2,0,1)$ & $0$ & $0$ & $k_4$ & $0$ & $k_6$ \\ \hline
    \end{tabular}
    \caption{$(4,1)$ quartic vertices satisfying the quartic constraints.}
    \label{Paper3-tab(4,1)}
\end{table}

\begin{table}[H]
    \centering
    \begin{tabular}{|c|c|c|c|c|c|c|}
\hline
$(\lambda_1,\lambda_2,\omega,\lambda_3,\lambda_4)$ & $k_2=(2341)$ & $k_3=(3412)$ & $k_4=(4123)$ & $k_5=(1324)$ & $k_6=(2413)$ \\ \hline
$(-2,3,3,-2,3)$ & $k_1$ & $k_1$ & $k_1$ & $0$ & $k_1$ \\ \hline
$(-2,3,3,-1,2)$ & $k_1$ & $k_1$ & $0$ & $0$ & $k_1$ \\ \hline
$(-2,3,3,0,1)$ & $k_1$ & $k_1$ & $0$ & $0$ & $k_1$ \\ \hline
$(-1,2,3,-1,2)$ & $k_2$ & $k_1$ & $k_2$ & $0$ & $k_6$ \\ \hline
$(-1,2,3,0,1)$ & $k_2$ & $k_1$ & $0$ & $0$ & $k_6$ \\ \hline
$(-1,3,2,-1,1)$ & $k_1$ & $0$ & $0$ & $0$ & $k_1$ \\ \hline
$(-1,3,2,0,0)$ & $k_1$ & $0$ & $0$ & $0$ & $k_1$ \\ \hline
$(0,1,3,0,1)$ & $k_2$ & $k_1$ & $k_2$ & $0$ & $k_6$ \\ \hline
$(0,2,2,-2,2)$ & $0$ & $0$ & $k_1$ & $0$ & $k_1$ \\ \hline
$(0,2,2,0,0)$ & $k_2$ & $0$ & $0$ & $0$ & $k_6$ \\ \hline
$(1,1,2,-2,2)$ & $0$ & $0$ & $k_1$ & $0$ & $k_1$ \\ \hline
$(1,1,2,-1,1)$ & $0$ & $0$ & $k_4$ & $0$ & $k_6$ \\ \hline
    \end{tabular}
    \caption{$(4,2)$ quartic vertices satisfying the quartic constraints.}
    \label{Paper3-tab(4,2)}
\end{table}

\begin{table}[H]
    \centering
    \begin{tabular}{|c|c|c|c|c|c|c|}
\hline
$(\lambda_1,\lambda_2,\omega,\lambda_3,\lambda_4)$ & $k_2=(2341)$ & $k_3=(3412)$ & $k_4=(4123)$ & $k_5=(1324)$ & $k_6=(2413)$ \\ \hline
$(-2,3,3,-2,2)$ & $k_1$ & $0$ & $0$ & $0$ & $k_1$ \\ \hline
$(-2,3,3,-1,1)$ & $k_1$ & $0$ & $0$ & $0$ & $k_1$ \\ \hline
$(-2,3,3,0,0)$ & $k_1$ & $0$ & $0$ & $0$ & $k_1$ \\ \hline
$(-1,2,3,-2,2)$ & $0$ & $0$ & $k_4$ & $0$ & $k_1-k_4$ \\ \hline
$(-1,2,3,-1,1)$ & $k_2$ & $0$ & $0$ & $0$ & $k_6$ \\ \hline
$(-1,2,3,0,0)$ & $k_2$ & $0$ & $0$ & $0$ & $k_6$ \\ \hline
$(-1,3,2,-1,0)$ & $k_1$ & $0$ & $0$ & $0$ & $k_1$ \\ \hline
$(0,1,3,-2,2)$ & $0$ & $0$ & $k_4$ & $0$ & $k_1-k_4$ \\ \hline
$(0,1,3,-1,1)$ & $0$ & $0$ & $k_4$ & $0$ & $k_6$ \\ \hline
$(0,1,3,0,0)$ & $k_2$ & $0$ & $0$ & $0$ & $k_6$ \\ \hline
$(1,1,2,-2,1)$ & $0$ & $0$ & $k_4$ & $0$ & $k_1-k_4$ \\ \hline
    \end{tabular}
    \caption{$(4,3)$ quartic vertices satisfying the quartic constraints.}
    \label{Paper3-tab(4,3)}
\end{table}

\begin{table}[H]
    \centering
    \begin{tabular}{|c|c|c|c|c|c|c|}
\hline
$(\lambda_1,\lambda_2,\omega,\lambda_3,\lambda_4)$ & $k_2=(2341)$ & $k_3=(3412)$ & $k_4=(4123)$ & $k_5=(1324)$ & $k_6=(2413)$ \\ \hline
$(-3,3,4,-3,3)$ & $k_1$ & $k_1$ & $k_1$ & $0$ & $k_1$ \\ \hline
$(-3,3,4,-2,2)$ & $k_1$ & $k_1$ & $0$ & $0$ & $k_1$ \\ \hline
$(-3,3,4,-1,1)$ & $k_1$ & $k_1$ & $0$ & $0$ & $k_1$ \\ \hline
$(-3,3,4,0,0)$& $k_1$ & $k_1$ & $0$ & $0$ & $k_1$ \\ \hline
$(-2,2,4,-2,2)$ & $k_2$ & $k_1$ & $k_2$ & $0$ & $k_6$ \\ \hline
$(-2,2,4,-1,1)$ & $k_2$ & $k_1$ & $0$ & $0$ & $k_6$ \\ \hline
$(-2,2,4,0,0)$ & $k_2$ & $k_1$ & $0$ & $0$ & $k_6$ \\ \hline
$(-2,3,3,-2,1)$ & $k_1$ & $0$ & $0$ & $0$ & $k_1$ \\ \hline
$(-2,3,3,-1,0)$ & $k_1$ & $0$ & $0$ & $0$ & $k_1$ \\ \hline
$(-1,1,4,-1,1)$ & $k_2$ & $k_1$ & $k_2$ & $0$ & $k_6$ \\ \hline
$(-1,1,4,0,0)$ & $k_2$ & $k_1$ & $0$ & $0$ & $k_6$ \\ \hline
$(-1,2,3,-3,2)$ & $0$ & $0$ & $k_1$ & $0$ & $k_1$ \\ \hline
$(-1,2,3,-1,0)$ & $k_2$ & $0$ & $0$ & $0$ & $k_6$ \\ \hline
$(-1,3,2,-1,-1)$ & $k_1$ & $0$ & $0$ & $0$ & $k_1$ \\ \hline
$(0,0,4,0,0)$ & $k_2$ & $k_1$ & $k_2$ & $k_5$ & $k_5$ \\ \hline
$(0,1,3,-3,2)$ & $0$ & $0$ & $k_1$ & $0$ & $k_1$ \\ \hline
$(0,1,3,-2,1)$ & $0$ & $0$ & $k_4$ & $0$ & $k_6$ \\ \hline
$(1,1,2,-3,1)$ & $0$ & $0$ & $k_1$ & $0$ & $k_1$ \\ \hline
    \end{tabular}
    \caption{$(4,4)$ quartic vertices satisfying the quartic constraints.}
    \label{Paper3-tab(4,4)}
\end{table}

\begin{table}[H]
    \centering
    \begin{tabular}{|c|c|c|c|c|c|c|}
\hline
$(n,m),D$ & $(\lambda_1,\lambda_2,\lambda_3,\lambda_4)$ & $(n,m),D$ & $(\lambda_1,\lambda_2,\lambda_3,\lambda_4)$ & $(n,m),D$ & $(\lambda_1,\lambda_2,\lambda_3,\lambda_4)$ \\ \hline
$(1,1),0$ & $(0,0,0,0)$ &$(4,3),5$ & $(1,0,0,0)$&$(5,3),6$ & $(2,1,0,-1)$\\ \hline
$(2,2),2$ & $(0,0,0,0)$ & $(4,3),5$ & $(1,1,0,-1)$&$(5,3),6$ & $(2,2,-1,-1)$\\ \hline
$(3,1),2$ & $(1,1,0,0)$ & $(5,2),5$ & $(1,1,1,0)$&$(6,2),6$ & $(1,1,1,1)$\\ \hline
$(3,2),3$ & $(1,0,0,0)$ &  $(5,2),5$ & $(2,1,0,0)$&$(6,2),6$ & $(2,1,1,0)$\\ \hline
$(4,1),3$ & $(1,1,1,0)$ &  $(6,1),5$ & $(2,1,1,1)$&$(6,2),6$ & $(2,2,0,0)$\\ \hline
$(3,3),4$ & $(0,0,0,0)$ & $(6,1),5$ & $(2,2,1,0)$&$(7,1),6$ & $(2,2,1,1)$\\ \hline
$(3,3),4$ & $(1,0,0,-1)$ & $(4,4),6$ & $(0,0,0,0)$&$(7,1),6$ & $(3,1,1,1)$\\ \hline
$(3,3),4$ & $(1,1,-1,-1)$ & $(4,4),6$ & $(1,0,0,-1)$&$(7,1),6$ & $(2,2,2,0)$\\ \hline
$(4,2),4$ & $(1,1,0,0)$ & $(4,4),6$ & $(1,1,-1,-1)$&$(7,1),6$ & $(3,2,1,0)$\\\hline
$(5,1),4$ & $(1,1,1,1)$ &$(5,3),6$ & $(1,1,0,0)$&$(7,1),6$ & $(3,3,0,0)$\\\hline
$(5,1),4$ & $(2,1,1,0)$ &$(5,3),6$ & $(2,0,0,0)$&...& ...\\\hline
$(5,1),4$ & $(2,2,0,0)$ &$(5,3),6$ & $(1,1,-1,-1)$& ...&... \\\hline
    \end{tabular}
    \caption{Homogeneous quartic vertices satisfying the quartic constraints.}
        \label{Paper3-tabHomo}
\end{table}

\backmatter
\footnotesize
\providecommand{\href}[2]{#2}\begingroup\raggedright\endgroup


\begin{thebibliography}{100}

\bibitem{Bengtsson:1983pd}
A.~K.~H. Bengtsson, I.~Bengtsson, and L.~Brink, ``{Cubic interaction terms for arbitrary spin},'' {\em Nucl. Phys.} {\bf B227} (1983)
31.

\bibitem{Bengtsson:1983pg}
A.~K.~H. Bengtsson, I.~Bengtsson, and L.~Brink, ``{Cubic interaction terms for arbitrarily extended Supermultiplets},'' {\em Nucl. Phys.} {\bf B227} (1983)
41.

\bibitem{Bengtsson:1986kh}
A.~K.~H. Bengtsson, I.~Bengtsson, and N.~Linden, ``{Interacting Higher Spin Gauge Fields on the Light Front},'' {\em Class. Quant. Grav.} {\bf 4} (1987)
1333.

\bibitem{Metsaev:1991mt}
R.~R. Metsaev, ``{Poincare invariant dynamics of massless higher spins: Fourth order analysis on mass shell},'' {\em Mod. Phys. Lett.} {\bf A6} (1991)
359--367.

\bibitem{Metsaev:1991nb}
R.~R. Metsaev, ``{$S$ matrix approach to massless higher spins theory. 2: The Case of internal symmetry},'' {\em Mod. Phys. Lett.} {\bf A6} (1991)
2411--2421.

\bibitem{Ponomarev:2016lrm}
D.~Ponomarev and E.~D. Skvortsov, ``{Light-Front Higher-Spin Theories in Flat Space},'' {\em J. Phys.} {\bf A50} (2017), no.~9, 095401,
\href{http://arXiv.org/abs/1609.04655}{{\tt 1609.04655}}.

\bibitem{Serrani:2025owx}
M.~Serrani, ``{On classification of (self-dual) higher-spin gravities in flat space},'' {\em JHEP} {\bf 08} (2025) 032, \href{http://arXiv.org/abs/2505.12839}{{\tt 2505.12839}}.

\bibitem{Serrani:2025oaw}
M.~Serrani, ``{Associativity of celestial OPE, higher spins and self-duality},'' {\em JHEP} {\bf 04} (2026) 047, \href{http://arXiv.org/abs/2508.16804}{{\tt 2508.16804}}.

\bibitem{Serrani:2026dbs}
M.~Serrani, ``{Massless spinning fields on the Light-Front: quartic vertices and amplitudes},'' \href{http://arXiv.org/abs/2602.12826}{{\tt 2602.12826}}.

\bibitem{Bekaert:2022poo}
X.~Bekaert, N.~Boulanger, A.~Campoleoni, M.~Chiodaroli, D.~Francia, M.~Grigoriev, E.~Sezgin, and E.~Skvortsov, ``{Snowmass White Paper: Higher Spin Gravity and Higher Spin Symmetry},'' \href{http://arXiv.org/abs/2205.01567}{{\tt 2205.01567}}.

\bibitem{Einstein:1916vd}
A.~Einstein, ``{The foundation of the general theory of relativity.},'' {\em Annalen Phys.} {\bf 49} (1916), no.~7, 769--822.

\bibitem{Weinberg:2004kv}
S.~Weinberg, ``{The Making of the standard model},'' {\em Eur. Phys. J. C} {\bf 34} (2004) 5--13, \href{http://arXiv.org/abs/hep-ph/0401010}{{\tt hep-ph/0401010}}.

\bibitem{NYT1919}
{New York Times}, ``Lights all askew in the heavens.'' \url{http://graphics8.nytimes.com/packages/pdf/arts/LightsAllAskew.pdf}, Nov., 1919.

\bibitem{Dyson:1920cwa}
F.~W. Dyson, A.~S. Eddington, and C.~Davidson, ``{A Determination of the Deflection of Light by the Sun's Gravitational Field, from Observations Made at the Total Eclipse of May 29, 1919},'' {\em Phil. Trans. Roy. Soc. Lond. A} {\bf 220} (1920) 291--333.

\bibitem{LIGOScientific:2016aoc}
{\bf LIGO Scientific, Virgo} Collaboration, B.~P. Abbott {\em et al.}, ``{Observation of Gravitational Waves from a Binary Black Hole Merger},'' {\em Phys. Rev. Lett.} {\bf 116} (2016), no.~6, 061102, \href{http://arXiv.org/abs/1602.03837}{{\tt 1602.03837}}.

\bibitem{LIGOScientific:2017vwq}
{\bf LIGO Scientific, Virgo} Collaboration, B.~P. Abbott {\em et al.}, ``{GW170817: Observation of Gravitational Waves from a Binary Neutron Star Inspiral},'' {\em Phys. Rev. Lett.} {\bf 119} (2017), no.~16, 161101, \href{http://arXiv.org/abs/1710.05832}{{\tt 1710.05832}}.

\bibitem{Fan:2022eto}
X.~Fan, T.~G. Myers, B.~A.~D. Sukra, and G.~Gabrielse, ``{Measurement of the Electron Magnetic Moment},'' {\em Phys. Rev. Lett.} {\bf 130} (2023), no.~7, 071801, \href{http://arXiv.org/abs/2209.13084}{{\tt 2209.13084}}.

\bibitem{Aoyama:2017uqe}
T.~Aoyama, T.~Kinoshita, and M.~Nio, ``{Revised and Improved Value of the QED Tenth-Order Electron Anomalous Magnetic Moment},'' {\em Phys. Rev. D} {\bf 97} (2018), no.~3, 036001, \href{http://arXiv.org/abs/1712.06060}{{\tt 1712.06060}}.

\bibitem{Noether:1918zz}
E.~Noether, ``{Invariant Variation Problems},'' {\em Gott. Nachr.} {\bf 1918} (1918) 235--257, \href{http://arXiv.org/abs/physics/0503066}{{\tt physics/0503066}}.

\bibitem{Palti:2019pca}
E.~Palti, ``{The Swampland: Introduction and Review},'' {\em Fortsch. Phys.} {\bf 67} (2019), no.~6, 1900037, \href{http://arXiv.org/abs/1903.06239}{{\tt 1903.06239}}.

\bibitem{Englert:1964et}
F.~Englert and R.~Brout, ``{Broken Symmetry and the Mass of Gauge Vector Mesons},'' {\em Phys. Rev. Lett.} {\bf 13} (1964) 321--323.

\bibitem{Higgs:1964pj}
P.~W. Higgs, ``{Broken Symmetries and the Masses of Gauge Bosons},'' {\em Phys. Rev. Lett.} {\bf 13} (1964) 508--509.

\bibitem{Guralnik:1964eu}
G.~S. Guralnik, C.~R. Hagen, and T.~W.~B. Kibble, ``{Global Conservation Laws and Massless Particles},'' {\em Phys. Rev. Lett.} {\bf 13} (1964) 585--587.

\bibitem{Barausse:2020rsu}
E.~Barausse {\em et al.}, ``{Prospects for Fundamental Physics with LISA},'' {\em Gen. Rel. Grav.} {\bf 52} (2020), no.~8, 81, \href{http://arXiv.org/abs/2001.09793}{{\tt 2001.09793}}.

\bibitem{Abac:2025saz}
A.~Abac {\em et al.}, ``{The Science of the Einstein Telescope},'' \href{http://arXiv.org/abs/2503.12263}{{\tt 2503.12263}}.

\bibitem{Maggiore:2007ulw}
M.~Maggiore, {\em {Gravitational Waves. Vol. 1: Theory and Experiments}}.
\newblock Oxford University Press, 2007.

\bibitem{Buonanno:1998gg}
A.~Buonanno and T.~Damour, ``{Effective one-body approach to general relativistic two-body dynamics},'' {\em Phys. Rev. D} {\bf 59} (1999) 084006, \href{http://arXiv.org/abs/gr-qc/9811091}{{\tt gr-qc/9811091}}.

\bibitem{Tomaselli:2024dbw}
G.~M. Tomaselli, T.~F.~M. Spieksma, and G.~Bertone, ``{Legacy of Boson Clouds on Black Hole Binaries},'' {\em Phys. Rev. Lett.} {\bf 133} (2024), no.~12, 121402, \href{http://arXiv.org/abs/2407.12908}{{\tt 2407.12908}}.

\bibitem{tHooft:1974toh}
G.~'t~Hooft and M.~J.~G. Veltman, ``{One-loop divergencies in the theory of gravitation},'' {\em Ann. Inst. H. Poincare Phys. Theor. A} {\bf 20} (1974), no.~1, 69--94.

\bibitem{Goroff:1985th}
M.~H. Goroff and A.~Sagnotti, ``{The Ultraviolet Behavior of Einstein Gravity},'' {\em Nucl. Phys.} {\bf B266} (1986)
709--736.

\bibitem{Penrose:1964wq}
R.~Penrose, ``{Gravitational collapse and space-time singularities},'' {\em Phys. Rev. Lett.} {\bf 14} (1965) 57--59.

\bibitem{Hawking:1967ju}
S.~Hawking, ``{The occurrence of singularities in cosmology. III. Causality and singularities},'' {\em Proc. Roy. Soc. Lond. A} {\bf 300} (1967) 187--201.

\bibitem{Carney:2018ofe}
D.~Carney, P.~C.~E. Stamp, and J.~M. Taylor, ``{Tabletop experiments for quantum gravity: a user{\textquoteright}s manual},'' {\em Class. Quant. Grav.} {\bf 36} (2019), no.~3, 034001, \href{http://arXiv.org/abs/1807.11494}{{\tt 1807.11494}}.

\bibitem{Weinberg:1978kz}
S.~Weinberg, ``{Phenomenological Lagrangians},'' {\em Physica A} {\bf 96} (1979), no.~1-2, 327--340.

\bibitem{Georgi:1993mps}
H.~Georgi, ``{Effective field theory},'' {\em Ann. Rev. Nucl. Part. Sci.} {\bf 43} (1993) 209--252.

\bibitem{Donoghue:1994dn}
J.~F. Donoghue, ``{General relativity as an effective field theory: The leading quantum corrections},'' {\em Phys. Rev. D} {\bf 50} (1994) 3874--3888, \href{http://arXiv.org/abs/gr-qc/9405057}{{\tt gr-qc/9405057}}.

\bibitem{Burgess_2004}
C.~P. Burgess, ``Quantum gravity in everyday life: General relativity as an effective field theory,'' {\em Living Reviews in Relativity} {\bf 7} (Apr., 2004).

\bibitem{Birrell:1982ix}
N.~D. Birrell and P.~C.~W. Davies, {\em {Quantum Fields in Curved Space}}.
\newblock Cambridge Monographs on Mathematical Physics. Cambridge University Press, Cambridge, UK, 1982.

\bibitem{Hawking:1974rv}
S.~W. Hawking, ``{Black hole explosions},'' {\em Nature} {\bf 248} (1974) 30--31.

\bibitem{Hawking:1975vcx}
S.~W. Hawking, ``{Particle Creation by Black Holes},'' {\em Commun. Math. Phys.} {\bf 43} (1975) 199--220. [Erratum: Commun.Math.Phys. 46, 206 (1976)].

\bibitem{Bekenstein:1973ur}
J.~D. Bekenstein, ``{Black holes and entropy},'' {\em Phys. Rev. D} {\bf 7} (1973) 2333--2346.

\bibitem{Bardeen:1973gs}
J.~M. Bardeen, B.~Carter, and S.~W. Hawking, ``{The Four laws of black hole mechanics},'' {\em Commun. Math. Phys.} {\bf 31} (1973) 161--170.

\bibitem{tHooft:1973alw}
G.~'t~Hooft, ``{A Planar Diagram Theory for Strong Interactions},'' {\em Nucl. Phys. B} {\bf 72} (1974) 461.

\bibitem{tHooft:1993dmi}
G.~'t~Hooft, ``{Dimensional reduction in quantum gravity},'' {\em Conf. Proc. C} {\bf 930308} (1993) 284--296, \href{http://arXiv.org/abs/gr-qc/9310026}{{\tt gr-qc/9310026}}.

\bibitem{Flato:1978qz}
M.~Flato and C.~Fronsdal, ``{One Massless Particle Equals Two Dirac Singletons: Elementary Particles in a Curved Space. 6.},'' {\em Lett.Math.Phys.} {\bf 2} (1978)
421--426.

\bibitem{Brown:1986nw}
J.~D. Brown and M.~Henneaux, ``{Central Charges in the Canonical Realization of Asymptotic Symmetries: An Example from Three-Dimensional Gravity},'' {\em Commun. Math. Phys.} {\bf 104} (1986) 207--226.

\bibitem{Susskind:1994vu}
L.~Susskind, ``{The World as a hologram},'' {\em J. Math. Phys.} {\bf 36} (1995) 6377--6396, \href{http://arXiv.org/abs/hep-th/9409089}{{\tt hep-th/9409089}}.

\bibitem{Polyakov:1997tj}
A.~M. Polyakov, ``{String theory and quark confinement},'' {\em Nucl. Phys. B Proc. Suppl.} {\bf 68} (1998) 1--8, \href{http://arXiv.org/abs/hep-th/9711002}{{\tt hep-th/9711002}}.

\bibitem{Maldacena:1997re}
J.~M. Maldacena, ``{The large N limit of superconformal field theories and supergravity},'' {\em Adv. Theor. Math. Phys.} {\bf 2} (1998) 231--252,
\href{http://arXiv.org/abs/hep-th/9711200}{{\tt hep-th/9711200}}.

\bibitem{Wess:1974tw}
J.~Wess and B.~Zumino, ``{Supergauge Transformations in Four-Dimensions},'' {\em Nucl. Phys. B} {\bf 70} (1974) 39--50.

\bibitem{Green:1987sp}
M.~B. Green, J.~H. Schwarz, and E.~Witten, {\em {SUPERSTRING THEORY. VOL. 1: INTRODUCTION}}.
\newblock Cambridge Monographs on Mathematical Physics. Cambridge University Press, 7, 1988.

\bibitem{Green:1987mn}
M.~B. Green, J.~H. Schwarz, and E.~Witten, {\em {SUPERSTRING THEORY. VOL. 2: LOOP AMPLITUDES, ANOMALIES AND PHENOMENOLOGY}}.
\newblock Cambridge University Press, 7, 1988.

\bibitem{Mandelstam:1982cb}
S.~Mandelstam, ``{Light Cone Superspace and the Ultraviolet Finiteness of the N=4 Model},'' {\em Nucl. Phys. B} {\bf 213} (1983) 149--168.

\bibitem{Brink:1982wv}
L.~Brink, O.~Lindgren, and B.~E.~W. Nilsson, ``{The Ultraviolet Finiteness of the N=4 Yang-Mills Theory},'' {\em Phys. Lett. B} {\bf 123} (1983) 323--328.

\bibitem{Grisaru:1976vm}
M.~T. Grisaru, H.~N. Pendleton, and P.~van Nieuwenhuizen, ``{Supergravity and the S Matrix},'' {\em Phys. Rev. D} {\bf 15} (1977) 996.

\bibitem{Deser:1977yyz}
S.~Deser, J.~H. Kay, and K.~S. Stelle, ``{Renormalizability Properties of Supergravity},'' {\em Phys. Rev. Lett.} {\bf 38} (1977) 527, \href{http://arXiv.org/abs/1506.03757}{{\tt 1506.03757}}.

\bibitem{VanNieuwenhuizen:1981ae}
P.~Van~Nieuwenhuizen, ``{Supergravity},'' {\em Phys. Rept.} {\bf 68} (1981)
189--398.

\bibitem{Wess:1992cp}
J.~Wess and J.~Bagger, {\em {Supersymmetry and supergravity}}.
\newblock Princeton University Press, Princeton, NJ, USA, 1992.

\bibitem{Fradkin:1986ka}
E.~S. Fradkin and M.~A. Vasiliev, ``Candidate to the role of higher spin symmetry,'' {\em Ann. Phys.} {\bf 177} (1987)
63.

\bibitem{Angelantonj:2002ct}
C.~Angelantonj and A.~Sagnotti, ``{Open strings},'' {\em Phys. Rept.} {\bf 371} (2002) 1--150, \href{http://arXiv.org/abs/hep-th/0204089}{{\tt hep-th/0204089}}. [Erratum: Phys.Rept. 376, 407 (2003)].

\bibitem{Sugimoto:1999tx}
S.~Sugimoto, ``{Anomaly cancellations in type I D-9 - anti-D-9 system and the USp(32) string theory},'' {\em Prog. Theor. Phys.} {\bf 102} (1999) 685--699, \href{http://arXiv.org/abs/hep-th/9905159}{{\tt hep-th/9905159}}.

\bibitem{Baykara:2024tjr}
Z.~K. Baykara, H.-C. Tarazi, and C.~Vafa, ``{New Non-Supersymmetric Tachyon-Free Strings},'' \href{http://arXiv.org/abs/2406.00185}{{\tt 2406.00185}}.

\bibitem{Gross:1987ar}
D.~J. Gross and P.~F. Mende, ``{String Theory Beyond the Planck Scale},'' {\em Nucl. Phys. B} {\bf 303} (1988) 407--454.

\bibitem{Gross:1987kza}
D.~J. Gross and P.~F. Mende, ``{The High-Energy Behavior of String Scattering Amplitudes},'' {\em Phys. Lett. B} {\bf 197} (1987) 129--134.

\bibitem{Gross:1988ue}
D.~J. Gross, ``{High-Energy Symmetries of String Theory},'' {\em Phys. Rev. Lett.} {\bf 60} (1988) 1229.

\bibitem{Vasiliev:1986td}
M.~A. Vasiliev, ``Free massless fields of arbitrary spin in the de sitter space and initial data for a higher spin superalgebra,'' {\em Fortsch. Phys.} {\bf 35} (1987)
741--770.

\bibitem{Vasiliev:1988sa}
M.~A. Vasiliev, ``Consistent equations for interacting massless fields of all spins in the first order in curvatures,'' {\em Annals Phys.} {\bf 190} (1989)
59--106.

\bibitem{Majorana:1932chs}
E.~Majorana, ``{Relativistic theory of particles with arbitrary intrinsic angular momentum},'' {\em Nuovo Cim.} {\bf 9} (1932) 335--344.

\bibitem{Dirac:1936tg}
P.~A.~M. Dirac, ``{Relativistic wave equations},'' {\em Proc. Roy. Soc. Lond. A} {\bf 155} (1936) 447--459.

\bibitem{Wigner:1939cj}
E.~P. Wigner, ``On unitary representations of the inhomogeneous lorentz group,'' {\em Annals Math.} {\bf 40} (1939)
149--204.

\bibitem{Bekaert:2006py}
X.~Bekaert and N.~Boulanger, ``{The unitary representations of the Poincar{\textbackslash}'e group in any spacetime dimension},'' {\em SciPost Phys. Lect. Notes} {\bf 30} (2021) 1, \href{http://arXiv.org/abs/hep-th/0611263}{{\tt hep-th/0611263}}.

\bibitem{Bekaert:2005in}
X.~Bekaert and J.~Mourad, ``The continuous spin limit of higher spin field equations,'' {\em JHEP} {\bf 01} (2006) 115,
\href{http://arXiv.org/abs/hep-th/0509092}{{\tt hep-th/0509092}}.

\bibitem{Bekaert:2017khg}
X.~Bekaert and E.~D. Skvortsov, ``{Elementary particles with continuous spin},'' {\em Int. J. Mod. Phys. A} {\bf 32} (2017), no.~23n24, 1730019, \href{http://arXiv.org/abs/1708.01030}{{\tt 1708.01030}}.

\bibitem{Bargmann:1948ck}
V.~Bargmann and E.~P. Wigner, ``{Group Theoretical Discussion of Relativistic Wave Equations},'' {\em Proc. Nat. Acad. Sci.} {\bf 34} (1948)
211.

\bibitem{Binegar:1981gv}
B.~Binegar, ``{Relativistic Field Theories in Three-dimensions},'' {\em J. Math. Phys.} {\bf 23} (1982) 1511--1517.

\bibitem{Singh:1974qz}
L.~P.~S. Singh and C.~R. Hagen, ``Lagrangian formulation for arbitrary spin. 1. the boson case,'' {\em Phys. Rev.} {\bf D9} (1974)
898--909.

\bibitem{Singh:1974rc}
L.~P.~S. Singh and C.~R. Hagen, ``Lagrangian formulation for arbitrary spin. 2. the fermion case,'' {\em Phys. Rev.} {\bf D9} (1974)
910--920.

\bibitem{Fierz:1939ix}
M.~Fierz and W.~Pauli, ``On relativistic wave equations for particles of arbitrary spin in an electromagnetic field,'' {\em Proc. Roy. Soc. Lond.} {\bf A173} (1939)
211--232.

\bibitem{Rarita:1941mf}
W.~Rarita and J.~Schwinger, ``{On a theory of particles with half integral spin},'' {\em Phys. Rev.} {\bf 60} (1941)
61.

\bibitem{Fronsdal:1978rb}
C.~Fronsdal, ``Massless fields with integer spin,'' {\em Phys. Rev.} {\bf D18} (1978)
3624.

\bibitem{Fang:1978wz}
J.~Fang and C.~Fronsdal, ``{Massless Fields with Half Integral Spin},'' {\em Phys. Rev.} {\bf D18} (1978)
3630.

\bibitem{Fronsdal:1978vb}
C.~Fronsdal, ``{Singletons and Massless, Integral Spin Fields on de Sitter Space (Elementary Particles in a Curved Space. 7.},'' {\em Phys.Rev.} {\bf D20} (1979)
848--856.

\bibitem{Labastida:1987kw}
J.~M.~F. Labastida, ``Massless particles in arbitrary representations of the lorentz group,'' {\em Nucl. Phys.} {\bf B322} (1989)
185.

\bibitem{Didenko:2014dwa}
V.~E. Didenko and E.~D. Skvortsov, ``{Elements of Vasiliev Theory},'' {\em Lect. Notes Phys.} {\bf 1028} (2024) 269--456, \href{http://arXiv.org/abs/1401.2975}{{\tt 1401.2975}}.

\bibitem{Hehl:1976kj}
F.~Hehl, P.~Von Der~Heyde, G.~Kerlick, and J.~Nester, ``{General Relativity with Spin and Torsion: Foundations and Prospects},'' {\em Rev.Mod.Phys.} {\bf 48} (1976)
393--416.

\bibitem{Palatini:1919ffw}
A.~Palatini, ``{Deduzione invariantiva delle equazioni gravitazionali dal principio di Hamilton},'' {\em Rend. Circ. Mat. Palermo} {\bf 43} (1919), no.~1, 203--212.

\bibitem{Dirac:1963ta}
P.~A.~M. Dirac, ``{A Remarkable representation of the 3 + 2 de Sitter group},'' {\em J. Math. Phys.} {\bf 4} (1963)
901--909.

\bibitem{Penrose:1965am}
R.~Penrose, ``{Zero rest mass fields including gravitation: Asymptotic behavior},'' {\em Proc. Roy. Soc. Lond.} {\bf A284} (1965)
159.

\bibitem{Hughston:1979tq}
L.~P. Hughston, R.~S. Ward, M.~G. Eastwood, M.~L. Ginsberg, A.~P. Hodges, S.~A. Huggett, T.~R. Hurd, R.~O. Jozsa, R.~Penrose, A.~Popovich, {\em et al.}, eds., {\em Advances in Twistor Theory}, vol.~37 of {\em Research Notes in Mathematics}.
\newblock Pitman, San Francisco, USA,
1979.
\newblock

\bibitem{Eastwood:1981jy}
M.~G. Eastwood, R.~Penrose, and R.~O. Wells, ``{Cohomology and Massless Fields},'' {\em Commun. Math. Phys.} {\bf 78} (1981)
305--351.

\bibitem{Woodhouse:1985id}
N.~M.~J. Woodhouse, ``{Real methods in twistor theory},'' {\em Class. Quant. Grav.} {\bf 2} (1985)
257--291.

\bibitem{Joung:2014qya}
E.~Joung and K.~Mkrtchyan, ``{Notes on higher-spin algebras: minimal representations and structure constants},'' {\em JHEP} {\bf 05} (2014) 103,
\href{http://arXiv.org/abs/1401.7977}{{\tt 1401.7977}}.

\bibitem{Dirac:1949cp}
P.~A.~M. Dirac, ``{Forms of Relativistic Dynamics},'' {\em Rev. Mod. Phys.} {\bf 21} (1949) 392--399.

\bibitem{Leutwyler:1977vy}
H.~Leutwyler and J.~Stern, ``{Relativistic Dynamics on a Null Plane},'' {\em Annals Phys.} {\bf 112} (1978) 94.

\bibitem{Fubini:1964boa}
S.~Fubini and G.~Furlan, ``{Renormalization effects for partially conserved currents},'' {\em Physics Physique Fizika} {\bf 1} (1965), no.~4, 229--247.

\bibitem{Neville1968}
R.~A. Neville, {\em Quantum Electrodynamics in a Laser Pulse}.
\newblock PhD thesis, Syracuse University, 1968.

\bibitem{Neville:1971uc}
R.~A. Neville and F.~Rohrlich, ``{Quantum electrodynamics on null planes and applications to lasers},'' {\em Phys. Rev. D} {\bf 3} (1971) 1692--1707.

\bibitem{Kogut:1969xa}
J.~B. Kogut and D.~E. Soper, ``{Quantum Electrodynamics in the Infinite Momentum Frame},'' {\em Phys. Rev. D} {\bf 1} (1970) 2901--2913.

\bibitem{Bjorken:1970ah}
J.~D. Bjorken, J.~B. Kogut, and D.~E. Soper, ``{Quantum Electrodynamics at Infinite Momentum: Scattering from an External Field},'' {\em Phys. Rev. D} {\bf 3} (1971) 1382.

\bibitem{Lepage:1980fj}
G.~P. Lepage and S.~J. Brodsky, ``{Exclusive Processes in Perturbative Quantum Chromodynamics},'' {\em Phys. Rev. D} {\bf 22} (1980) 2157.

\bibitem{Feynman:1969ej}
R.~P. Feynman, ``{Very high-energy collisions of hadrons},'' {\em Phys. Rev. Lett.} {\bf 23} (1969) 1415--1417.

\bibitem{Gross:1973id}
D.~J. Gross and F.~Wilczek, ``{Ultraviolet Behavior of Nonabelian Gauge Theories},'' {\em Phys. Rev. Lett.} {\bf 30} (1973) 1343--1346.

\bibitem{Politzer:1973fx}
H.~D. Politzer, ``{Reliable Perturbative Results for Strong Interactions?},'' {\em Phys. Rev. Lett.} {\bf 30} (1973) 1346--1349.

\bibitem{Brodsky:1997de}
S.~J. Brodsky, H.-C. Pauli, and S.~S. Pinsky, ``{Quantum chromodynamics and other field theories on the light cone},'' {\em Phys. Rept.} {\bf 301} (1998) 299--486, \href{http://arXiv.org/abs/hep-ph/9705477}{{\tt hep-ph/9705477}}.

\bibitem{Goddard:1973qh}
P.~Goddard, J.~Goldstone, C.~Rebbi, and C.~B. Thorn, ``{Quantum dynamics of a massless relativistic string},'' {\em Nucl. Phys. B} {\bf 56} (1973) 109--135.

\bibitem{Schwarz:1982jn}
J.~H. Schwarz, ``{Superstring Theory},'' {\em Phys. Rept.} {\bf 89} (1982) 223--322.

\bibitem{Ananth:2005vg}
S.~Ananth, L.~Brink, and P.~Ramond, ``{Eleven-dimensional supergravity in light-cone superspace},'' {\em JHEP} {\bf 05} (2005) 003, \href{http://arXiv.org/abs/hep-th/0501079}{{\tt hep-th/0501079}}.

\bibitem{Metsaev_2005}
R.~R. Metsaev, ``Eleven dimensional supergravity in light cone gauge,'' {\em Physical Review D} {\bf 71} (Apr., 2005).

\bibitem{Ananth:2004es}
S.~Ananth, L.~Brink, and P.~Ramond, ``{Oxidizing superYang-Mills from (N=4,d = 4) to (N=1,d = 10)},'' {\em JHEP} {\bf 07} (2004) 082, \href{http://arXiv.org/abs/hep-th/0405150}{{\tt hep-th/0405150}}.

\bibitem{Berends:1984wp}
F.~A. Berends, G.~J.~H. Burgers, and H.~Van~Dam, ``{On spin three selfinteractions},'' {\em Z. Phys.} {\bf C24} (1984)
247--254.

\bibitem{Deser:1990bk}
S.~Deser and Z.~Yang, ``{Inconsistency of Spin 4 - Spin-2 Gauge Field Couplings},'' {\em Class. Quant. Grav.} {\bf 7} (1990) 1491--1498.

\bibitem{Bekaert:2010hp}
X.~Bekaert, N.~Boulanger, and S.~Leclercq, ``{Strong obstruction of the Berends-Burgers-van Dam spin-3 vertex},'' {\em J.Phys.} {\bf A43} (2010) 185401,
\href{http://arXiv.org/abs/1002.0289}{{\tt 1002.0289}}.

\bibitem{Berends:1984rq}
F.~A. Berends, G.~J.~H. Burgers, and H.~van Dam, ``{On the theoretical problems in constructing interactions involving higher spin massless particles},'' {\em Nucl. Phys.} {\bf B260} (1985)
295.

\bibitem{Bengtsson:1983bp}
A.~K.~H. Bengtsson, ``{On Gauge Invariance for Spin 3 Fields},'' {\em Phys. Rev. D} {\bf 32} (1985) 2031.

\bibitem{Berends:1985xx}
F.~A. Berends, G.~J.~H. Burgers, and H.~van Dam, ``{Explicit construction of conserved currents for massless fields of arbitrary spin},'' {\em Nucl. Phys. B} {\bf 271} (1986) 429--441.

\bibitem{Aragone:1979hx}
C.~Aragone and S.~Deser, ``{Consistency Problems of Hypergravity},'' {\em Phys. Lett.} {\bf B86} (1979)
161.

\bibitem{Berends:1979kg}
F.~A. Berends, J.~W. van Holten, B.~de~Wit, and P.~van Nieuwenhuizen, ``{ON SPIN 5/2 GAUGE FIELDS},'' {\em J. Phys. A} {\bf 13} (1980) 1643--1649.

\bibitem{Fradkin:1986qy}
E.~S. Fradkin and M.~A. Vasiliev, ``{Cubic Interaction in Extended Theories of Massless Higher Spin Fields},'' {\em Nucl. Phys.} {\bf B291} (1987)
141.

\bibitem{Fradkin:1987ks}
E.~S. Fradkin and M.~A. Vasiliev, ``{On the Gravitational Interaction of Massless Higher Spin Fields},'' {\em Phys. Lett.} {\bf B189} (1987)
89--95.

\bibitem{Boulanger:2008tg}
N.~Boulanger, S.~Leclercq, and P.~Sundell, ``{On The Uniqueness of Minimal Coupling in Higher-Spin Gauge Theory},'' {\em JHEP} {\bf 08} (2008) 056,
\href{http://arXiv.org/abs/0805.2764}{{\tt 0805.2764}}.

\bibitem{Metsaev:2018xip}
R.~R. Metsaev, ``{Light-cone gauge cubic interaction vertices for massless fields in AdS(4)},'' {\em Nucl. Phys.} {\bf B936} (2018) 320--351,
\href{http://arXiv.org/abs/1807.07542}{{\tt 1807.07542}}.

\bibitem{Krasnov:2021nsq}
K.~Krasnov, E.~Skvortsov, and T.~Tran, ``{Actions for self-dual Higher Spin Gravities},'' {\em JHEP} {\bf 08} (2021) 076, \href{http://arXiv.org/abs/2105.12782}{{\tt 2105.12782}}.

\bibitem{Ponomarev:2017nrr}
D.~Ponomarev, ``{Chiral Higher Spin Theories and Self-Duality},'' {\em JHEP} {\bf 12} (2017) 141,
\href{http://arXiv.org/abs/1710.00270}{{\tt 1710.00270}}.

\bibitem{Krasnov:2016emc}
K.~Krasnov, ``{Self-Dual Gravity},'' {\em Class. Quant. Grav.} {\bf 34} (2017), no.~9, 095001, \href{http://arXiv.org/abs/1610.01457}{{\tt 1610.01457}}.

\bibitem{Bassetto:1991ue}
A.~Bassetto, G.~Nardelli, and R.~Soldati, {\em {Yang-Mills theories in algebraic noncovariant gauges: Canonical quantization and renormalization}}.
\newblock World Scientific, 1991.

\bibitem{Mannheim:2020rod}
P.~D. Mannheim, P.~Lowdon, and S.~J. Brodsky, ``{Comparing light-front quantization with instant-time quantization},'' {\em Phys. Rept.} {\bf 891} (2021) 1--65, \href{http://arXiv.org/abs/2005.00109}{{\tt 2005.00109}}.

\bibitem{Ponomarev:2016cwi}
D.~Ponomarev, ``{Off-Shell Spinor-Helicity Amplitudes from Light-Cone Deformation Procedure},'' {\em JHEP} {\bf 12} (2016) 117,
\href{http://arXiv.org/abs/1611.00361}{{\tt 1611.00361}}.

\bibitem{Metsaev:2005ar}
R.~R. Metsaev, ``{Cubic interaction vertices for massive and massless higher spin fields},'' {\em Nucl. Phys.} {\bf B759} (2006) 147--201,
\href{http://arXiv.org/abs/hep-th/0512342}{{\tt hep-th/0512342}}.

\bibitem{Bengtsson:2012jm}
A.~K.~H. Bengtsson, ``{Systematics of Higher-spin Light-front Interactions},''
\newblock 5, 2012.
\newblock \href{http://arXiv.org/abs/1205.6117}{{\tt 1205.6117}}.

\bibitem{Ponomarev:2022vjb}
D.~Ponomarev, ``{Basic Introduction to Higher-Spin Theories},'' {\em Int. J. Theor. Phys.} {\bf 62} (2023), no.~7, 146, \href{http://arXiv.org/abs/2206.15385}{{\tt 2206.15385}}.

\bibitem{Perry:1994kp}
R.~J. Perry, ``{Hamiltonian light front field theory and quantum chromodynamics},'' in {\em {Hadrons 94 Workshop}}.
\newblock 7, 1994.
\newblock \href{http://arXiv.org/abs/hep-th/9407056}{{\tt hep-th/9407056}}.

\bibitem{Burkardt:1995ct}
M.~Burkardt, ``{Light front quantization},'' {\em Adv. Nucl. Phys.} {\bf 23} (1996) 1--74, \href{http://arXiv.org/abs/hep-ph/9505259}{{\tt hep-ph/9505259}}.

\bibitem{6db1d5c6aea346a68475d79dd4ab667e}
N.~Ligterink, {\em Light-front hamiltonian field theory. Covariance and renormalization.}
\newblock Phd-thesis - research and graduation internal, VU, Amsterdam, 1996.
\newblock Naam instelling promotie: VU, Amsterdam Naam instelling onderzoek: VU, Amsterdam.

\bibitem{Harindranath:1996hq}
A.~Harindranath, ``{An Introduction to light front dynamics for pedestrians},'' in {\em {International School on Light-Front Quantization and Non-Perturbative QCD (To be followed by the Workshop 3-14 Jun 1996)}}.
\newblock 5, 1996.
\newblock \href{http://arXiv.org/abs/hep-ph/9612244}{{\tt hep-ph/9612244}}.

\bibitem{Heinzl:2000ht}
T.~Heinzl, ``{Light cone quantization: Foundations and applications},'' {\em Lect. Notes Phys.} {\bf 572} (2001) 55--142, \href{http://arXiv.org/abs/hep-th/0008096}{{\tt hep-th/0008096}}.

\bibitem{Henneaux:1992ig}
M.~Henneaux and C.~Teitelboim, {\em {Quantization of gauge systems}}.
\newblock Princeton University Press, 1992.

\bibitem{Gitman:1990qh}
D.~M. Gitman and I.~V. Tyutin, {\em {Quantization of Fields with Constraints}}.
\newblock Springer Series in Nuclear and Particle Physics. Springer, Berlin, Germany, 1990.

\bibitem{Siegel:1999ew}
W.~Siegel, ``{Fields},'' \href{http://arXiv.org/abs/hep-th/9912205}{{\tt hep-th/9912205}}.

\bibitem{Conde:2016izb}
E.~Conde, E.~Joung, and K.~Mkrtchyan, ``{Spinor-Helicity Three-Point Amplitudes from Local Cubic Interactions},'' {\em JHEP} {\bf 08} (2016) 040,
\href{http://arXiv.org/abs/1605.07402}{{\tt 1605.07402}}.

\bibitem{Benincasa:2007xk}
P.~Benincasa and F.~Cachazo, ``{Consistency Conditions on the S-Matrix of Massless Particles},'' \href{http://arXiv.org/abs/0705.4305}{{\tt 0705.4305}}.

\bibitem{Benincasa:2011pg}
P.~Benincasa and E.~Conde, ``{Exploring the S-Matrix of Massless Particles},'' {\em Phys. Rev. D} {\bf 86} (2012) 025007, \href{http://arXiv.org/abs/1108.3078}{{\tt 1108.3078}}.

\bibitem{Chalmers:1996rq}
G.~Chalmers and W.~Siegel, ``{The Selfdual sector of QCD amplitudes},'' {\em Phys. Rev.} {\bf D54} (1996) 7628--7633,
\href{http://arXiv.org/abs/hep-th/9606061}{{\tt hep-th/9606061}}.

\bibitem{Flato:1980zk}
M.~Flato and C.~Fronsdal, ``{On {DIS} and Racs},'' {\em Phys. Lett. B} {\bf 97} (1980) 236--240.

\bibitem{Sezgin:2002rt}
E.~Sezgin and P.~Sundell, ``{Massless higher spins and holography},'' {\em Nucl.Phys.} {\bf B644} (2002) 303--370,
\href{http://arXiv.org/abs/hep-th/0205131}{{\tt hep-th/0205131}}.

\bibitem{Klebanov:2002ja}
I.~R. Klebanov and A.~M. Polyakov, ``{AdS dual of the critical $O(N)$ vector model},'' {\em Phys. Lett.} {\bf B550} (2002) 213--219,
\href{http://arXiv.org/abs/hep-th/0210114}{{\tt hep-th/0210114}}.

\bibitem{Sezgin:2003pt}
E.~Sezgin and P.~Sundell, ``{Holography in 4D (super) higher spin theories and a test via cubic scalar couplings},'' {\em JHEP} {\bf 0507} (2005) 044,
\href{http://arXiv.org/abs/hep-th/0305040}{{\tt hep-th/0305040}}.

\bibitem{Leigh:2003gk}
R.~G. Leigh and A.~C. Petkou, ``{Holography of the N=1 higher spin theory on AdS(4)},'' {\em JHEP} {\bf 0306} (2003) 011,
\href{http://arXiv.org/abs/hep-th/0304217}{{\tt hep-th/0304217}}.

\bibitem{Giombi:2011kc}
S.~Giombi, S.~Minwalla, S.~Prakash, S.~P. Trivedi, S.~R. Wadia, and X.~Yin, ``{Chern-Simons Theory with Vector Fermion Matter},'' {\em Eur. Phys. J.} {\bf C72} (2012) 2112,
\href{http://arXiv.org/abs/1110.4386}{{\tt 1110.4386}}.

\bibitem{Weinberg:1964ew}
S.~Weinberg, ``{Photons and Gravitons in S Matrix Theory: Derivation of Charge Conservation and Equality of Gravitational and Inertial Mass},'' {\em Phys. Rev.} {\bf 135} (1964)
B1049--B1056.

\bibitem{Coleman:1967ad}
S.~R. Coleman and J.~Mandula, ``{All Possible Symmetries of the S Matrix},'' {\em Phys. Rev.} {\bf 159} (1967)
1251--1256.

\bibitem{Maldacena:2011jn}
J.~Maldacena and A.~Zhiboedov, ``{Constraining Conformal Field Theories with A Higher Spin Symmetry},'' {\em J. Phys. A} {\bf 46} (2013) 214011, \href{http://arXiv.org/abs/1112.1016}{{\tt 1112.1016}}.

\bibitem{Fitzpatrick:2012cg}
A.~L. Fitzpatrick and J.~Kaplan, ``{AdS Field Theory from Conformal Field Theory},'' {\em JHEP} {\bf 02} (2013) 054, \href{http://arXiv.org/abs/1208.0337}{{\tt 1208.0337}}.

\bibitem{Boulanger:2013zza}
N.~Boulanger, D.~Ponomarev, E.~D. Skvortsov, and M.~Taronna, ``{On the uniqueness of higher-spin symmetries in AdS and CFT},'' {\em Int. J. Mod. Phys.} {\bf A28} (2013) 1350162,
\href{http://arXiv.org/abs/1305.5180}{{\tt 1305.5180}}.

\bibitem{Alba:2013yda}
V.~Alba and K.~Diab, ``{Constraining conformal field theories with a higher spin symmetry in d=4},''
\href{http://arXiv.org/abs/1307.8092}{{\tt 1307.8092}}.

\bibitem{Alba:2015upa}
V.~Alba and K.~Diab, ``{Constraining conformal field theories with a higher spin symmetry in $d > 3$ dimensions},'' {\em JHEP} {\bf 03} (2016) 044, \href{http://arXiv.org/abs/1510.02535}{{\tt 1510.02535}}.

\bibitem{Sleight:2021iix}
C.~Sleight and M.~Taronna, ``{On the consistency of (partially-)massless matter couplings in de Sitter space},'' {\em JHEP} {\bf 10} (2021) 156, \href{http://arXiv.org/abs/2106.00366}{{\tt 2106.00366}}.

\bibitem{Roiban:2017iqg}
R.~Roiban and A.~A. Tseytlin, ``{On four-point interactions in massless higher spin theory in flat space},'' {\em JHEP} {\bf 04} (2017) 139,
\href{http://arXiv.org/abs/1701.05773}{{\tt 1701.05773}}.

\bibitem{Bekaert:2015tva}
X.~Bekaert, J.~Erdmenger, D.~Ponomarev, and C.~Sleight, ``{Quartic AdS Interactions in Higher-Spin Gravity from Conformal Field Theory},'' {\em JHEP} {\bf 11} (2015) 149,
\href{http://arXiv.org/abs/1508.04292}{{\tt 1508.04292}}.

\bibitem{Maldacena:2015iua}
J.~Maldacena, D.~Simmons-Duffin, and A.~Zhiboedov, ``{Looking for a bulk point},'' {\em JHEP} {\bf 01} (2017) 013,
\href{http://arXiv.org/abs/1509.03612}{{\tt 1509.03612}}.

\bibitem{Sleight:2017pcz}
C.~Sleight and M.~Taronna, ``{Higher-Spin Gauge Theories and Bulk Locality},'' {\em Phys. Rev. Lett.} {\bf 121} (2018), no.~17, 171604,
\href{http://arXiv.org/abs/1704.07859}{{\tt 1704.07859}}.

\bibitem{Ponomarev:2017qab}
D.~Ponomarev, ``{A Note on (Non)-Locality in Holographic Higher Spin Theories},'' {\em Universe} {\bf 4} (2018), no.~1, 2,
\href{http://arXiv.org/abs/1710.00403}{{\tt 1710.00403}}.

\bibitem{Neiman:2023orj}
Y.~Neiman, ``{Quartic locality of higher-spin gravity in de Sitter and Euclidean anti-de Sitter space},'' {\em Phys. Lett. B} {\bf 843} (2023) 138048, \href{http://arXiv.org/abs/2302.00852}{{\tt 2302.00852}}.

\bibitem{Blencowe:1988gj}
M.~Blencowe, ``{A Consistent Interacting Massless Higher Spin Field Theory in $D$ = (2+1)},'' {\em Class.Quant.Grav.} {\bf 6} (1989)
443.

\bibitem{Bergshoeff:1989ns}
E.~Bergshoeff, M.~P. Blencowe, and K.~S. Stelle, ``{Area Preserving Diffeomorphisms and Higher Spin Algebra},'' {\em Commun. Math. Phys.} {\bf 128} (1990)
213.

\bibitem{Campoleoni:2010zq}
A.~Campoleoni, S.~Fredenhagen, S.~Pfenninger, and S.~Theisen, ``{Asymptotic symmetries of three-dimensional gravity coupled to higher-spin fields},'' {\em JHEP} {\bf 1011} (2010) 007,
\href{http://arXiv.org/abs/1008.4744}{{\tt 1008.4744}}.

\bibitem{Henneaux:2010xg}
M.~Henneaux and S.-J. Rey, ``{Nonlinear $W_{\infty}$ as Asymptotic Symmetry of Three-Dimensional Higher Spin Anti-de Sitter Gravity},'' {\em JHEP} {\bf 1012} (2010) 007,
\href{http://arXiv.org/abs/1008.4579}{{\tt 1008.4579}}.

\bibitem{Grigoriev:2020lzu}
M.~Grigoriev, K.~Mkrtchyan, and E.~Skvortsov, ``{Matter-free higher spin gravities in 3D: Partially-massless fields and general structure},'' {\em Phys. Rev. D} {\bf 102} (2020), no.~6, 066003, \href{http://arXiv.org/abs/2005.05931}{{\tt 2005.05931}}.

\bibitem{Pope:1989vj}
C.~N. Pope and P.~K. Townsend, ``{Conformal Higher Spin in (2+1)-dimensions},'' {\em Phys. Lett. B} {\bf 225} (1989) 245--250.

\bibitem{Fradkin:1989xt}
E.~S. Fradkin and V.~Y. Linetsky, ``{A Superconformal Theory of Massless Higher Spin Fields in $D$ = (2+1)},'' {\em Mod. Phys. Lett. A} {\bf 4} (1989) 731.

\bibitem{Grigoriev:2019xmp}
M.~Grigoriev, I.~Lovrekovic, and E.~Skvortsov, ``{New Conformal Higher Spin Gravities in $3d$},'' {\em JHEP} {\bf 01} (2020) 059, \href{http://arXiv.org/abs/1909.13305}{{\tt 1909.13305}}.

\bibitem{Sharapov:2024euk}
A.~Sharapov, E.~Skvortsov, and A.~Sukhanov, ``{Matter-coupled higher spin gravities in 3d: no- and yes-go results},'' {\em JHEP} {\bf 04} (2025) 155, \href{http://arXiv.org/abs/2409.12830}{{\tt 2409.12830}}.

\bibitem{Alkalaev:2020kut}
K.~Alkalaev and X.~Bekaert, ``{On BF-type higher-spin actions in two dimensions},'' {\em JHEP} {\bf 05} (2020) 158, \href{http://arXiv.org/abs/2002.02387}{{\tt 2002.02387}}.

\bibitem{Segal:2002gd}
A.~Y. Segal, ``{Conformal higher spin theory},'' {\em Nucl. Phys.} {\bf B664} (2003) 59--130,
\href{http://arXiv.org/abs/hep-th/0207212}{{\tt hep-th/0207212}}.

\bibitem{Tseytlin:2002gz}
A.~A. Tseytlin, ``{On limits of superstring in $AdS_5\times S^5$},'' {\em Theor. Math. Phys.} {\bf 133} (2002) 1376--1389, \href{http://arXiv.org/abs/hep-th/0201112}{{\tt hep-th/0201112}}.
[Teor. Mat. Fiz.133,69(2002)].

\bibitem{Bekaert:2010ky}
X.~Bekaert, E.~Joung, and J.~Mourad, ``{Effective action in a higher-spin background},'' {\em JHEP} {\bf 02} (2011) 048,
\href{http://arXiv.org/abs/1012.2103}{{\tt 1012.2103}}.

\bibitem{Basile:2022nou}
T.~Basile, M.~Grigoriev, and E.~Skvortsov, ``{Covariant action for conformal higher spin gravity},'' {\em J. Phys. A} {\bf 56} (2023), no.~38, 385402, \href{http://arXiv.org/abs/2212.10336}{{\tt 2212.10336}}.

\bibitem{Ponomarev:2024jyg}
D.~Ponomarev, ``{Chiral higher-spin double copy},'' {\em JHEP} {\bf 01} (2025) 143, \href{http://arXiv.org/abs/2409.19449}{{\tt 2409.19449}}.

\bibitem{Adamo:2022lah}
T.~Adamo and T.~Tran, ``{Higher-spin Yang\textendash{}Mills, amplitudes and self-duality},'' {\em Lett. Math. Phys.} {\bf 113} (2023), no.~3, 50, \href{http://arXiv.org/abs/2210.07130}{{\tt 2210.07130}}.

\bibitem{Sperling:2017dts}
M.~Sperling and H.~C. Steinacker, ``{Covariant 4-dimensional fuzzy spheres, matrix models and higher spin},'' {\em J. Phys.} {\bf A50} (2017), no.~37, 375202,
\href{http://arXiv.org/abs/1704.02863}{{\tt 1704.02863}}.

\bibitem{deMelloKoch:2018ivk}
R.~de~Mello~Koch, A.~Jevicki, K.~Suzuki, and J.~Yoon, ``{AdS Maps and Diagrams of Bi-local Holography},'' {\em JHEP} {\bf 03} (2019) 133,
\href{http://arXiv.org/abs/1810.02332}{{\tt 1810.02332}}.

\bibitem{Metsaev:1993ap}
R.~R. Metsaev, ``{Generating function for cubic interaction vertices of higher spin fields in any dimension},'' {\em Mod. Phys. Lett.} {\bf A8} (1993)
2413--2426.

\bibitem{Metsaev:2007rn}
R.~R. Metsaev, ``{Cubic interaction vertices for fermionic and bosonic arbitrary spin fields},'' {\em Nucl. Phys. B} {\bf 859} (2012) 13--69, \href{http://arXiv.org/abs/0712.3526}{{\tt 0712.3526}}.

\bibitem{Manvelyan:2010je}
R.~Manvelyan, K.~Mkrtchyan, and W.~Ruehl, ``{A Generating function for the cubic interactions of higher spin fields},'' {\em Phys.Lett.} {\bf B696} (2011) 410--415,
\href{http://arXiv.org/abs/1009.1054}{{\tt 1009.1054}}.

\bibitem{Boulanger:2012dx}
N.~Boulanger, D.~Ponomarev, and E.~Skvortsov, ``{Non-abelian cubic vertices for higher-spin fields in anti-de Sitter space},'' {\em JHEP} {\bf 1305} (2013) 008,
\href{http://arXiv.org/abs/1211.6979}{{\tt 1211.6979}}.

\bibitem{Francia:2016weg}
D.~Francia, G.~L. Monaco, and K.~Mkrtchyan, ``{Cubic interactions of Maxwell-like higher spins},'' {\em JHEP} {\bf 04} (2017) 068,
\href{http://arXiv.org/abs/1611.00292}{{\tt 1611.00292}}.

\bibitem{Bekaert:2002uh}
X.~Bekaert, N.~Boulanger, and M.~Henneaux, ``Consistent deformations of dual formulations of linearized gravity: A no-go result,'' {\em Phys. Rev.} {\bf D67} (2003) 044010,
\href{http://arXiv.org/abs/hep-th/0210278}{{\tt hep-th/0210278}}.

\bibitem{Sharapov:2022faa}
A.~Sharapov, A.~Sharapov, E.~Skvortsov, E.~Skvortsov, A.~Sukhanov, A.~Sukhanov, R.~Van~Dongen, and R.~Van~Dongen, ``{Minimal model of Chiral Higher Spin Gravity},'' {\em JHEP} {\bf 09} (2022) 134, \href{http://arXiv.org/abs/2205.07794}{{\tt 2205.07794}}. [Erratum: JHEP 02, 183 (2023)].

\bibitem{Sharapov:2022wpz}
A.~Sharapov, E.~Skvortsov, and R.~Van~Dongen, ``{Chiral higher spin gravity and convex geometry},'' {\em SciPost Phys.} {\bf 14} (2023), no.~6, 162, \href{http://arXiv.org/abs/2209.01796}{{\tt 2209.01796}}.

\bibitem{Sharapov:2022awp}
A.~Sharapov and E.~Skvortsov, ``{Chiral higher spin gravity in (A)dS4 and secrets of Chern{\textendash}Simons matter theories},'' {\em Nucl. Phys. B} {\bf 985} (2022) 115982, \href{http://arXiv.org/abs/2205.15293}{{\tt 2205.15293}}.

\bibitem{Sharapov:2022nps}
A.~Sharapov, E.~Skvortsov, A.~Sukhanov, and R.~Van~Dongen, ``{More on Chiral Higher Spin Gravity and convex geometry},'' {\em Nucl. Phys. B} {\bf 990} (2023) 116152, \href{http://arXiv.org/abs/2209.15441}{{\tt 2209.15441}}.

\bibitem{Sharapov:2023erv}
A.~Sharapov, E.~Skvortsov, and R.~Van~Dongen, ``{Strong homotopy algebras for chiral higher spin gravity via Stokes theorem},'' {\em JHEP} {\bf 06} (2024) 186, \href{http://arXiv.org/abs/2312.16573}{{\tt 2312.16573}}.

\bibitem{Skvortsov:2024rng}
E.~Skvortsov and Y.~Yin, ``{Low spin solutions of higher spin gravity: BPST instanton},'' {\em JHEP} {\bf 07} (2024) 032, \href{http://arXiv.org/abs/2403.17148}{{\tt 2403.17148}}.

\bibitem{Tran:2025yzd}
T.~Tran, ``{Self-dual pp-wave solutions in chiral higher-spin gravity},'' {\em JHEP} {\bf 03} (2025) 041, \href{http://arXiv.org/abs/2501.06445}{{\tt 2501.06445}}.

\bibitem{Skvortsov:2018jea}
E.~D. Skvortsov, T.~Tran, and M.~Tsulaia, ``{Quantum Chiral Higher Spin Gravity},'' {\em Phys. Rev. Lett.} {\bf 121} (2018), no.~3, 031601,
\href{http://arXiv.org/abs/1805.00048}{{\tt 1805.00048}}.

\bibitem{Skvortsov:2020wtf}
E.~Skvortsov, T.~Tran, and M.~Tsulaia, ``{More on Quantum Chiral Higher Spin Gravity},'' {\em Phys. Rev.} {\bf D101} (2020), no.~10, 106001,
\href{http://arXiv.org/abs/2002.08487}{{\tt 2002.08487}}.

\bibitem{Skvortsov:2020gpn}
E.~Skvortsov and T.~Tran, ``{One-loop Finiteness of Chiral Higher Spin Gravity},'' {\em JHEP} {\bf 07} (2020) 021, \href{http://arXiv.org/abs/2004.10797}{{\tt 2004.10797}}.

\bibitem{Tsulaia:2022csz}
M.~Tsulaia and D.~Weissman, ``{Supersymmetric quantum chiral higher spin gravity},'' {\em JHEP} {\bf 12} (2022) 002, \href{http://arXiv.org/abs/2209.13907}{{\tt 2209.13907}}.

\bibitem{Neiman:2024vit}
Y.~Neiman, ``{Higher-spin self-dual General Relativity: 6d and 4d pictures, covariant vs. lightcone},'' {\em JHEP} {\bf 07} (2024) 178, \href{http://arXiv.org/abs/2404.18589}{{\tt 2404.18589}}.

\bibitem{Skvortsov:2018uru}
E.~Skvortsov, ``{Light-Front Bootstrap for Chern-Simons Matter Theories},'' {\em JHEP} {\bf 06} (2019) 058,
\href{http://arXiv.org/abs/1811.12333}{{\tt 1811.12333}}.

\bibitem{Jain:2024bza}
S.~Jain, D.~K. S, and E.~Skvortsov, ``{Hidden sectors of Chern-Simons matter theories and exact holography},'' {\em Phys. Rev. D} {\bf 111} (2025), no.~10, 106017, \href{http://arXiv.org/abs/2405.00773}{{\tt 2405.00773}}.

\bibitem{Aharony:2024nqs}
O.~Aharony, R.~R. Kalloor, and T.~Kukolj, ``{A chiral limit for Chern-Simons-matter theories},'' {\em JHEP} {\bf 10} (2024) 051, \href{http://arXiv.org/abs/2405.01647}{{\tt 2405.01647}}.

\bibitem{Monteiro:2022xwq}
R.~Monteiro, ``{From Moyal deformations to chiral higher-spin theories and to celestial algebras},'' {\em JHEP} {\bf 03} (2023) 062, \href{http://arXiv.org/abs/2212.11266}{{\tt 2212.11266}}.

\bibitem{Boulanger:2000rq}
N.~Boulanger, T.~Damour, L.~Gualtieri, and M.~Henneaux, ``{Inconsistency of interacting, multigraviton theories},'' {\em Nucl. Phys.} {\bf B597} (2001) 127--171,
\href{http://arXiv.org/abs/hep-th/0007220}{{\tt hep-th/0007220}}.

\bibitem{Arkani-Hamed:2017jhn}
N.~Arkani-Hamed, T.-C. Huang, and Y.-t. Huang, ``{Scattering amplitudes for all masses and spins},'' {\em JHEP} {\bf 11} (2021) 070, \href{http://arXiv.org/abs/1709.04891}{{\tt 1709.04891}}.

\bibitem{Monteiro:2022lwm}
R.~Monteiro, ``{Celestial chiral algebras, colour-kinematics duality and integrability},'' {\em JHEP} {\bf 01} (2023) 092, \href{http://arXiv.org/abs/2208.11179}{{\tt 2208.11179}}.

\bibitem{Weinberg:1964ev}
S.~Weinberg, ``{Feynman Rules for Any Spin. 2. Massless Particles},'' {\em Phys. Rev.} {\bf 134} (1964)
B882--B896.

\bibitem{Tran:2021ukl}
T.~Tran, ``{Twistor constructions for higher-spin extensions of (self-dual) Yang-Mills},'' {\em JHEP} {\bf 11} (2021) 117, \href{http://arXiv.org/abs/2107.04500}{{\tt 2107.04500}}.

\bibitem{Herfray:2022prf}
Y.~Herfray, K.~Krasnov, and E.~Skvortsov, ``{Higher-spin self-dual Yang-Mills and gravity from the twistor space},'' {\em JHEP} {\bf 01} (2023) 158, \href{http://arXiv.org/abs/2210.06209}{{\tt 2210.06209}}.

\bibitem{Tran:2022tft}
T.~Tran, ``{Toward a twistor action for chiral higher-spin gravity},'' {\em Phys. Rev. D} {\bf 107} (2023), no.~4, 046015, \href{http://arXiv.org/abs/2209.00925}{{\tt 2209.00925}}.

\bibitem{Adamo:2016ple}
T.~Adamo, P.~Hähnel, and T.~McLoughlin, ``{Conformal higher spin scattering amplitudes from twistor space},'' {\em JHEP} {\bf 04} (2017) 021,
\href{http://arXiv.org/abs/1611.06200}{{\tt 1611.06200}}.

\bibitem{Mason:2025pbz}
L.~Mason and A.~Sharma, ``{Chiral higher-spin theories from twistor space},'' \href{http://arXiv.org/abs/2505.09419}{{\tt 2505.09419}}.

\bibitem{Neiman:2023bkq}
Y.~Neiman, ``{Self-dual gravity in de Sitter space: Light-cone ansatz and static-patch scattering},'' {\em Phys. Rev. D} {\bf 109} (2024), no.~2, 024039, \href{http://arXiv.org/abs/2303.17866}{{\tt 2303.17866}}.

\bibitem{Lipstein:2023pih}
A.~Lipstein and S.~Nagy, ``{Self-Dual Gravity and Color-Kinematics Duality in AdS4},'' {\em Phys. Rev. Lett.} {\bf 131} (2023), no.~8, 081501, \href{http://arXiv.org/abs/2304.07141}{{\tt 2304.07141}}.

\bibitem{Chowdhury:2024dcy}
C.~Chowdhury, G.~Doran, A.~Lipstein, R.~Monteiro, S.~Nagy, and K.~Singh, ``{Light-cone actions and correlators of self-dual theories in AdS$_{4}$},'' {\em JHEP} {\bf 01} (2025) 172, \href{http://arXiv.org/abs/2411.04172}{{\tt 2411.04172}}.

\bibitem{Tran:2022amg}
T.~Tran, ``{Constraining higher-spin S-matrices},'' {\em JHEP} {\bf 02} (2023) 001, \href{http://arXiv.org/abs/2212.02540}{{\tt 2212.02540}}.

\bibitem{Tran:2025uad}
T.~Tran, ``{Anomaly-free twistorial higher-spin theories},'' \href{http://arXiv.org/abs/2505.13785}{{\tt 2505.13785}}.

\bibitem{Ren:2022sws}
L.~Ren, M.~Spradlin, A.~Yelleshpur~Srikant, and A.~Volovich, ``{On effective field theories with celestial duals},'' {\em JHEP} {\bf 08} (2022) 251, \href{http://arXiv.org/abs/2206.08322}{{\tt 2206.08322}}.

\bibitem{Ponomarev:2022atv}
D.~Ponomarev, ``{Invariant traces of the flat space chiral higher-spin algebra as scattering amplitudes},'' {\em JHEP} {\bf 09} (2022) 086, \href{http://arXiv.org/abs/2205.09654}{{\tt 2205.09654}}.

\bibitem{Ponomarev:2022ryp}
D.~Ponomarev, ``{Towards higher-spin holography in flat space},'' {\em JHEP} {\bf 01} (2023) 084, \href{http://arXiv.org/abs/2210.04035}{{\tt 2210.04035}}.

\bibitem{Ponomarev:2022qkx}
D.~Ponomarev, ``{Chiral higher-spin holography in flat space: the Flato-Fronsdal theorem and lower-point functions},'' {\em JHEP} {\bf 01} (2023) 048, \href{http://arXiv.org/abs/2210.04036}{{\tt 2210.04036}}.

\bibitem{Bengtsson:2014qza}
A.~K.~H. Bengtsson, ``{A Riccati type PDE for light-front higher helicity vertices},'' {\em JHEP} {\bf 09} (2014) 105,
\href{http://arXiv.org/abs/1403.7345}{{\tt 1403.7345}}.

\bibitem{Basile:2024raj}
T.~Basile, ``{Massless chiral fields in six dimensions},'' \href{http://arXiv.org/abs/2409.12800}{{\tt 2409.12800}}.

\bibitem{Neville:1971zk}
R.~A. Neville and F.~Rohrlich, ``{Quantum field theory off null planes},'' {\em Nuovo Cim. A} {\bf 1} (1971) 625--644.

\bibitem{Marcus:1982fr}
N.~Marcus and A.~Sagnotti, ``{Tree Level Constraints on Gauge Groups for Type I Superstrings},'' {\em Phys. Lett. B} {\bf 119} (1982) 97--99.

\bibitem{Pate:2019lpp}
M.~Pate, A.-M. Raclariu, A.~Strominger, and E.~Y. Yuan, ``{Celestial operator products of gluons and gravitons},'' {\em Rev. Math. Phys.} {\bf 33} (2021), no.~09, 2140003, \href{http://arXiv.org/abs/1910.07424}{{\tt 1910.07424}}.

\bibitem{Himwich:2021dau}
E.~Himwich, M.~Pate, and K.~Singh, ``{Celestial operator product expansions and w$_{1+\infty}$ symmetry for all spins},'' {\em JHEP} {\bf 01} (2022) 080, \href{http://arXiv.org/abs/2108.07763}{{\tt 2108.07763}}.

\bibitem{Strominger:2021mtt}
A.~Strominger, ``{$w_{1+\infty}$ Algebra and the Celestial Sphere: Infinite Towers of Soft Graviton, Photon, and Gluon Symmetries},'' {\em Phys. Rev. Lett.} {\bf 127} (2021), no.~22, 221601, \href{http://arXiv.org/abs/2105.14346}{{\tt 2105.14346}}.

\bibitem{Costello:2022upu}
K.~Costello and N.~M. Paquette, ``{Associativity of One-Loop Corrections to the Celestial Operator Product Expansion},'' {\em Phys. Rev. Lett.} {\bf 129} (2022), no.~23, 231604, \href{http://arXiv.org/abs/2204.05301}{{\tt 2204.05301}}.

\bibitem{Bittleston:2022jeq}
R.~Bittleston, ``{On the associativity of 1-loop corrections to the celestial operator product in gravity},'' {\em JHEP} {\bf 01} (2023) 018, \href{http://arXiv.org/abs/2211.06417}{{\tt 2211.06417}}.

\bibitem{Mago:2021wje}
J.~Mago, L.~Ren, A.~Y. Srikant, and A.~Volovich, ``{Deformed $w_{1+\infty}$ Algebras in the Celestial CFT},'' {\em SIGMA} {\bf 19} (2023) 044, \href{http://arXiv.org/abs/2111.11356}{{\tt 2111.11356}}.

\bibitem{Monteiro:2011pc}
R.~Monteiro and D.~O'Connell, ``{The Kinematic Algebra From the Self-Dual Sector},'' {\em JHEP} {\bf 07} (2011) 007, \href{http://arXiv.org/abs/1105.2565}{{\tt 1105.2565}}.

\bibitem{Ball:2022bgg}
A.~Ball, ``{Celestial locality and the Jacobi identity},'' {\em JHEP} {\bf 01} (2023) 146, \href{http://arXiv.org/abs/2211.09151}{{\tt 2211.09151}}.

\bibitem{Ball:2023sdz}
A.~Ball, Y.~Hu, and S.~Pasterski, ``{Multicollinear singularities in celestial CFT},'' {\em JHEP} {\bf 02} (2024) 219, \href{http://arXiv.org/abs/2309.16602}{{\tt 2309.16602}}.

\bibitem{Ball:2023qim}
A.~Ball, M.~Spradlin, A.~Yelleshpur~Srikant, and A.~Volovich, ``{Supersymmetry and the celestial Jacobi identity},'' {\em JHEP} {\bf 04} (2024) 099, \href{http://arXiv.org/abs/2311.01364}{{\tt 2311.01364}}.

\bibitem{Ball:2024oqa}
A.~Ball, ``{Currents in celestial CFT},'' {\em Mod. Phys. Lett. A} {\bf 39} (2024), no.~29n30, 2430007, \href{http://arXiv.org/abs/2407.13558}{{\tt 2407.13558}}.

\bibitem{Fernandez:2024qnu}
V.~E. Fern{\'a}ndez and N.~M. Paquette, ``{Associativity is enough: an all-orders 2d chiral algebra for 4d form factors},'' \href{http://arXiv.org/abs/2412.17168}{{\tt 2412.17168}}.

\bibitem{Guevara:2024ixn}
A.~Guevara, Y.~Hu, and S.~Pasterski, ``{Multiparticle contributions to the celestial OPE},'' {\em JHEP} {\bf 07} (2025) 178, \href{http://arXiv.org/abs/2402.18798}{{\tt 2402.18798}}.

\bibitem{Bhattacharyya:2025nfp}
A.~Bhattacharyya, S.~Ghosh, and S.~Pal, ``{The sky remembers everything: Celestial amplitude, shadow and OPE in quadratic EFT of gravity},'' {\em SciPost Phys.} {\bf 19} (2025), no.~2, 041, \href{http://arXiv.org/abs/2505.02899}{{\tt 2505.02899}}.

\bibitem{Pasterski:2021rjz}
S.~Pasterski, ``{Lectures on celestial amplitudes},'' {\em Eur. Phys. J. C} {\bf 81} (2021), no.~12, 1062, \href{http://arXiv.org/abs/2108.04801}{{\tt 2108.04801}}.

\bibitem{Raclariu:2021zjz}
A.-M. Raclariu, ``{Lectures on Celestial Holography},'' \href{http://arXiv.org/abs/2107.02075}{{\tt 2107.02075}}.

\bibitem{oblak2018lorentzgroupcelestialsphere}
B.~Oblak, ``From the lorentz group to the celestial sphere,'' 2018.

\bibitem{Pasterski:2017kqt}
S.~Pasterski and S.-H. Shao, ``{Conformal basis for flat space amplitudes},'' {\em Phys. Rev. D} {\bf 96} (2017), no.~6, 065022, \href{http://arXiv.org/abs/1705.01027}{{\tt 1705.01027}}.

\bibitem{Fan:2019emx}
W.~Fan, A.~Fotopoulos, and T.~R. Taylor, ``{Soft Limits of Yang-Mills Amplitudes and Conformal Correlators},'' {\em JHEP} {\bf 05} (2019) 121, \href{http://arXiv.org/abs/1903.01676}{{\tt 1903.01676}}.

\bibitem{Bhardwaj:2022anh}
R.~Bhardwaj, L.~Lippstreu, L.~Ren, M.~Spradlin, A.~Yelleshpur~Srikant, and A.~Volovich, ``{Loop-level gluon OPEs in celestial holography},'' {\em JHEP} {\bf 11} (2022) 171, \href{http://arXiv.org/abs/2208.14416}{{\tt 2208.14416}}.

\bibitem{Krishna:2023ukw}
H.~Krishna, ``{Celestial gluon and graviton OPE at loop level},'' {\em JHEP} {\bf 03} (2024) 176, \href{http://arXiv.org/abs/2310.16687}{{\tt 2310.16687}}.

\bibitem{Bhardwaj:2024wld}
R.~Bhardwaj and A.~Yelleshpur~Srikant, ``{Celestial soft currents at one-loop and their OPEs},'' {\em JHEP} {\bf 07} (2024) 034, \href{http://arXiv.org/abs/2403.10443}{{\tt 2403.10443}}.

\bibitem{Bissi:2024brf}
A.~Bissi, L.~Donnay, and B.~Valsesia, ``{Logarithmic doublets in CCFT},'' {\em JHEP} {\bf 12} (2024) 031, \href{http://arXiv.org/abs/2407.17123}{{\tt 2407.17123}}.

\bibitem{Adamo:2022wjo}
T.~Adamo, W.~Bu, E.~Casali, and A.~Sharma, ``{All-order celestial OPE in the MHV sector},'' {\em JHEP} {\bf 03} (2023) 252, \href{http://arXiv.org/abs/2211.17124}{{\tt 2211.17124}}.

\bibitem{Ren:2023trv}
L.~Ren, A.~Schreiber, A.~Sharma, and D.~Wang, ``{All-order celestial OPE from on-shell recursion},'' {\em JHEP} {\bf 10} (2023) 080, \href{http://arXiv.org/abs/2305.11851}{{\tt 2305.11851}}.

\bibitem{Guevara:2021abz}
A.~Guevara, E.~Himwich, M.~Pate, and A.~Strominger, ``{Holographic symmetry algebras for gauge theory and gravity},'' {\em JHEP} {\bf 11} (2021) 152, \href{http://arXiv.org/abs/2103.03961}{{\tt 2103.03961}}.

\bibitem{Bu:2022iak}
W.~Bu, S.~Heuveline, and D.~Skinner, ``{Moyal deformations, W$_{1+\infty}$ and celestial holography},'' {\em JHEP} {\bf 12} (2022) 011, \href{http://arXiv.org/abs/2208.13750}{{\tt 2208.13750}}.

\bibitem{Guevara:2022qnm}
A.~Guevara, ``{Towards Gravity From a Color Symmetry},'' \href{http://arXiv.org/abs/2209.00696}{{\tt 2209.00696}}.

\bibitem{Adamo:2021lrv}
T.~Adamo, L.~Mason, and A.~Sharma, ``{Celestial $w_{1+\infty}$ Symmetries from Twistor Space},'' {\em SIGMA} {\bf 18} (2022) 016, \href{http://arXiv.org/abs/2110.06066}{{\tt 2110.06066}}.

\bibitem{Tran:2025xbt}
T.~Tran, ``{Chiral higher-spin symmetry of the celestial twistor sphere},'' \href{http://arXiv.org/abs/2507.00340}{{\tt 2507.00340}}.

\bibitem{Elvang:2013cua}
H.~Elvang and Y.-t. Huang, ``{Scattering Amplitudes},'' \href{http://arXiv.org/abs/1308.1697}{{\tt 1308.1697}}.

\bibitem{Fradkin:1991iy}
E.~S. Fradkin and R.~R. Metsaev, ``{A Cubic interaction of totally symmetric massless representations of the Lorentz group in arbitrary dimensions},'' {\em Class. Quant. Grav.} {\bf 8} (1991)
L89--L94.

\bibitem{Boulanger:2006gr}
N.~Boulanger and S.~Leclercq, ``{Consistent couplings between spin-2 and spin-3 massless fields},'' {\em JHEP} {\bf 11} (2006) 034,
\href{http://arXiv.org/abs/hep-th/0609221}{{\tt hep-th/0609221}}.

\bibitem{Zinoviev:2008ck}
Y.~M. Zinoviev, ``{On spin 3 interacting with gravity},'' {\em Class. Quant. Grav.} {\bf 26} (2009) 035022,
\href{http://arXiv.org/abs/0805.2226}{{\tt 0805.2226}}.

\bibitem{Ananth:2012un}
S.~Ananth, ``{Spinor helicity structures in higher spin theories},'' {\em JHEP} {\bf 11} (2012) 089, \href{http://arXiv.org/abs/1209.4960}{{\tt 1209.4960}}.

\bibitem{Bengtsson:2016alt}
A.~K.~H. Bengtsson, ``{Quartic amplitudes for Minkowski higher spin},'' in {\em {International Workshop on Higher Spin Gauge Theories}}, pp.~353--370.
\newblock 2017.
\newblock \href{http://arXiv.org/abs/1605.02608}{{\tt 1605.02608}}.

\bibitem{Bengtsson:2016hss}
A.~K.~H. Bengtsson, ``{Investigations into Light-front Quartic Interactions for Massless Fields (I): Non-constructibility of Higher Spin Quartic Amplitudes},'' {\em JHEP} {\bf 12} (2016) 134, \href{http://arXiv.org/abs/1607.06659}{{\tt 1607.06659}}.

\bibitem{Schuster:2008nh}
P.~C. Schuster and N.~Toro, ``{Constructing the Tree-Level Yang-Mills S-Matrix Using Complex Factorization},'' {\em JHEP} {\bf 06} (2009) 079, \href{http://arXiv.org/abs/0811.3207}{{\tt 0811.3207}}.

\bibitem{Fotopoulos:2010ay}
A.~Fotopoulos and M.~Tsulaia, ``{On the Tensionless Limit of String theory, Off - Shell Higher Spin Interaction Vertices and BCFW Recursion Relations},'' {\em JHEP} {\bf 11} (2010) 086,
\href{http://arXiv.org/abs/1009.0727}{{\tt 1009.0727}}.

\bibitem{Dempster:2012vw}
P.~Dempster and M.~Tsulaia, ``{On the Structure of Quartic Vertices for Massless Higher Spin Fields on Minkowski Background},'' {\em Nucl. Phys.} {\bf B865} (2012) 353--375,
\href{http://arXiv.org/abs/1203.5597}{{\tt 1203.5597}}.

\bibitem{Taronna:2017wbx}
M.~Taronna, ``{On the Non-Local Obstruction to Interacting Higher Spins in Flat Space},'' {\em JHEP} {\bf 05} (2017) 026, \href{http://arXiv.org/abs/1701.05772}{{\tt 1701.05772}}.

\bibitem{Britto:2004ap}
R.~Britto, F.~Cachazo, and B.~Feng, ``{New recursion relations for tree amplitudes of gluons},'' {\em Nucl. Phys. B} {\bf 715} (2005) 499--522, \href{http://arXiv.org/abs/hep-th/0412308}{{\tt hep-th/0412308}}.

\bibitem{Britto:2005fq}
R.~Britto, F.~Cachazo, B.~Feng, and E.~Witten, ``{Direct proof of tree-level recursion relation in Yang-Mills theory},'' {\em Phys. Rev. Lett.} {\bf 94} (2005) 181602, \href{http://arXiv.org/abs/hep-th/0501052}{{\tt hep-th/0501052}}.

\bibitem{Benincasa:2011kn}
P.~Benincasa and E.~Conde, ``{On the Tree-Level Structure of Scattering Amplitudes of Massless Particles},'' {\em JHEP} {\bf 11} (2011) 074, \href{http://arXiv.org/abs/1106.0166}{{\tt 1106.0166}}.

\bibitem{McGady:2013sga}
D.~A. McGady and L.~Rodina, ``{Higher-spin massless $S$-matrices in four-dimensions},'' {\em Phys. Rev. D} {\bf 90} (2014), no.~8, 084048, \href{http://arXiv.org/abs/1311.2938}{{\tt 1311.2938}}.

\bibitem{Barnich:1993vg}
G.~Barnich and M.~Henneaux, ``{Consistent couplings between fields with a gauge freedom and deformations of the master equation},'' {\em Phys. Lett.} {\bf B311} (1993) 123--129,
\href{http://arXiv.org/abs/hep-th/9304057}{{\tt hep-th/9304057}}.

\bibitem{Vasiliev:1986bq}
M.~A. Vasiliev and E.~S. Fradkin, ``Gravitational interaction of massless high spin ($s > 2$) fields,'' {\em JETP Lett.} {\bf 44} (1986)
622--627.

\bibitem{Nagaraj:2018nxq}
B.~Nagaraj and D.~Ponomarev, ``{Spinor-Helicity Formalism for Massless Fields in AdS$_4$},'' {\em Phys. Rev. Lett.} {\bf 122} (2019), no.~10, 101602,
\href{http://arXiv.org/abs/1811.08438}{{\tt 1811.08438}}.

\bibitem{Skvortsov:2025ohi}
E.~Skvortsov and Y.~Yin, ``{Higher-spins on Taub-NUT and higher-spin Taub-NUT},'' {\em JHEP} {\bf 12} (2025) 099, \href{http://arXiv.org/abs/2508.18804}{{\tt 2508.18804}}.

\bibitem{Metsaev:1997ut}
R.~R. Metsaev, ``{Massless fields in plane wave geometry},'' {\em J. Math. Phys.} {\bf 38} (1997) 648--667, \href{http://arXiv.org/abs/hep-th/9701141}{{\tt hep-th/9701141}}.

\bibitem{Heinzl:1993px}
T.~Heinzl and E.~Werner, ``{Light front quantization as an initial boundary value problem},'' {\em Z. Phys. C} {\bf 62} (1994) 521--532, \href{http://arXiv.org/abs/hep-th/9311108}{{\tt hep-th/9311108}}.

\bibitem{Barnich:2024aln}
G.~Barnich, S.~Majumdar, S.~Speziale, and W.-D. Tan, ``{Lessons from discrete light-cone quantization for physics at null infinity: bosons in two dimensions},'' {\em JHEP} {\bf 05} (2024) 326, \href{http://arXiv.org/abs/2401.14873}{{\tt 2401.14873}}.

\bibitem{Fonseca:2025mzj}
R.~M. Fonseca, C.~Hernandez-Garcia, J.~M. Lizana, and M.~Perez-Victoria, ``{Gauge theories from scattering amplitudes with minimal assumptions},'' \href{http://arXiv.org/abs/2511.21664}{{\tt 2511.21664}}.

\bibitem{Wald:1986bj}
R.~M. Wald, ``{Spin-2 Fields and General Covariance},'' {\em Phys. Rev. D} {\bf 33} (1986) 3613.

\bibitem{Barnich:2000zw}
G.~Barnich, F.~Brandt, and M.~Henneaux, ``{Local BRST cohomology in gauge theories},'' {\em Phys. Rept.} {\bf 338} (2000) 439--569,
\href{http://arXiv.org/abs/hep-th/0002245}{{\tt hep-th/0002245}}.

\bibitem{Brink:1982pd}
L.~Brink, O.~Lindgren, and B.~E.~W. Nilsson, ``{N=4 Yang-Mills Theory on the Light Cone},'' {\em Nucl. Phys. B} {\bf 212} (1983) 401--412.

\bibitem{Bengtsson:1983vn}
I.~Bengtsson, M.~Cederwall, and O.~Lindgren, ``{LIGHT CONE ACTIONS FOR GRAVITY AND HIGHER SPINS: SOME FURTHER RESULTS},''.

\bibitem{Chakrabarti:2005ny}
D.~Chakrabarti, J.~Qiu, and C.~B. Thorn, ``{Scattering of glue by glue on the light-cone worldsheet. I. Helicity non-conserving amplitudes},'' {\em Phys. Rev.} {\bf D72} (2005) 065022,
\href{http://arXiv.org/abs/hep-th/0507280}{{\tt hep-th/0507280}}.

\bibitem{Chakrabarti:2006mb}
D.~Chakrabarti, J.~Qiu, and C.~B. Thorn, ``{Scattering of glue by glue on the light-cone worldsheet. II. Helicity conserving amplitudes},'' {\em Phys. Rev.} {\bf D74} (2006) 045018, \href{http://arXiv.org/abs/hep-th/0602026}{{\tt hep-th/0602026}}.
[Erratum: Phys. Rev.D76,089901(2007)].

\bibitem{Ananth:2017pio}
S.~Ananth, A.~Kar, S.~Majumdar, and N.~Shah, ``{Deriving spin-1 quartic interaction vertices from closure of the Poincar{\'e} algebra},'' {\em Nucl. Phys. B} {\bf 926} (2018) 11--19, \href{http://arXiv.org/abs/1707.05871}{{\tt 1707.05871}}.

\bibitem{Damour:1987vm}
T.~Damour and S.~Deser, ``{'Geometry' of Spin 3 Gauge Theories},'' {\em Ann. Inst. H. Poincare Phys. Theor.} {\bf 47} (1987) 277.

\bibitem{Damour:1987fp}
T.~Damour and S.~Deser, ``{Higher Derivative Interactions of Higher Spin Gauge Fields},'' {\em Class. Quant. Grav.} {\bf 4} (1987) L95.

\bibitem{Skvortsov:2022syz}
E.~Skvortsov and R.~Van~Dongen, ``{Minimal models of field theories: Chiral higher spin gravity},'' {\em Phys. Rev. D} {\bf 106} (2022), no.~4, 045006, \href{http://arXiv.org/abs/2204.10285}{{\tt 2204.10285}}.

\bibitem{Benincasa:2012wt}
P.~Benincasa, ``{Exploration of the Tree-Level S-Matrix of Massless Particles},'' {\em Fortsch. Phys.} {\bf 60} (2012) 889--895, \href{http://arXiv.org/abs/1201.3191}{{\tt 1201.3191}}.

\bibitem{Ananth:2023qrf}
S.~Ananth, N.~Bhave, C.~Pandey, and S.~Pant, ``{Deriving interaction vertices in higher derivative theories},'' {\em Phys. Lett. B} {\bf 853} (2024) 138704, \href{http://arXiv.org/abs/2306.05074}{{\tt 2306.05074}}.

\bibitem{Bern:2019prr}
Z.~Bern, J.~J. Carrasco, M.~Chiodaroli, H.~Johansson, and R.~Roiban, ``{The duality between color and kinematics and its applications},'' {\em J. Phys. A} {\bf 57} (2024), no.~33, 333002, \href{http://arXiv.org/abs/1909.01358}{{\tt 1909.01358}}.

\bibitem{Bern:2022wqg}
Z.~Bern, J.~J. Carrasco, M.~Chiodaroli, H.~Johansson, and R.~Roiban, ``{The SAGEX review on scattering amplitudes Chapter 2: An invitation to color-kinematics duality and the double copy},'' {\em J. Phys. A} {\bf 55} (2022), no.~44, 443003, \href{http://arXiv.org/abs/2203.13013}{{\tt 2203.13013}}.

\bibitem{Kawai:1985xq}
H.~Kawai, D.~C. Lewellen, and S.~H.~H. Tye, ``{A Relation Between Tree Amplitudes of Closed and Open Strings},'' {\em Nucl. Phys. B} {\bf 269} (1986) 1--23.

\bibitem{Parke:1986gb}
S.~J. Parke and T.~R. Taylor, ``{An Amplitude for $n$ Gluon Scattering},'' {\em Phys. Rev. Lett.} {\bf 56} (1986) 2459.

\bibitem{Berends:1987me}
F.~A. Berends and W.~T. Giele, ``{Recursive Calculations for Processes with n Gluons},'' {\em Nucl. Phys. B} {\bf 306} (1988) 759--808.

\bibitem{Bedford:2005yy}
J.~Bedford, A.~Brandhuber, B.~J. Spence, and G.~Travaglini, ``{A Recursion relation for gravity amplitudes},'' {\em Nucl. Phys. B} {\bf 721} (2005) 98--110, \href{http://arXiv.org/abs/hep-th/0502146}{{\tt hep-th/0502146}}.

\bibitem{Hodges:2012ym}
A.~Hodges, ``{A simple formula for gravitational MHV amplitudes},'' \href{http://arXiv.org/abs/1204.1930}{{\tt 1204.1930}}.

\bibitem{Bern:2008qj}
Z.~Bern, J.~J.~M. Carrasco, and H.~Johansson, ``{New Relations for Gauge-Theory Amplitudes},'' {\em Phys. Rev. D} {\bf 78} (2008) 085011, \href{http://arXiv.org/abs/0805.3993}{{\tt 0805.3993}}.

\bibitem{Bern:2010ue}
Z.~Bern, J.~J.~M. Carrasco, and H.~Johansson, ``{Perturbative Quantum Gravity as a Double Copy of Gauge Theory},'' {\em Phys. Rev. Lett.} {\bf 105} (2010) 061602,
\href{http://arXiv.org/abs/1004.0476}{{\tt 1004.0476}}.

\bibitem{Serrani:2026azw}
M.~Serrani and E.~Skvortsov, ``{Amplitudes in self-dual (higher-spin) theories},'' \href{http://arXiv.org/abs/2604.24873}{{\tt 2604.24873}}.

\bibitem{Guevara:2026qzd}
A.~Guevara, A.~Lupsasca, D.~Skinner, A.~Strominger, and K.~Weil, ``{Single-minus gluon tree amplitudes are nonzero},'' \href{http://arXiv.org/abs/2602.12176}{{\tt 2602.12176}}.

\bibitem{Sleight:2016xqq}
C.~Sleight and M.~Taronna, ``{Higher-Spin Algebras, Holography and Flat Space},'' {\em JHEP} {\bf 02} (2017) 095,
\href{http://arXiv.org/abs/1609.00991}{{\tt 1609.00991}}.

\bibitem{Calkins:2026hpg}
M.~Calkins and M.~Pate, ``{Multi-particle Celestial Operator Product Expansions from the Boundary},'' \href{http://arXiv.org/abs/2601.04329}{{\tt 2601.04329}}.

\end{thebibliography}
\end{document}